\documentclass[a4paper,palatino,11pt,twoside,unsrt,textglossaries,aas_macros,usenames,dvipsnames]{thesis}

\usepackage{hyperref}

\usepackage{import} 

\usepackage{mathtools}
\usepackage{amssymb}
\numberwithin{equation}{section}
\usepackage{tensor}

\usepackage{derivative}

\usepackage{bm}

\usepackage{suffix}
\usepackage{xparse}

\usepackage[
  noabbrev,
  capitalize,
]{cleveref}

\usepackage{subcaption}

\usepackage{siunitx}
\DeclareSIUnit\parsec{pc}
\DeclareSIUnit\year{yr}

\usepackage{booktabs}

\usepackage{mathtools}
\usepackage{tensor}
\usepackage{amsmath, amsfonts,bm}
\usepackage{derivative}
\usepackage{bm}
\usepackage[utf8]{inputenc}
\usepackage{siunitx}
\usepackage{multirow}

\usepackage{suffix}

\usepackage{amsmath, amsfonts,bm}
\usepackage{xcolor}
\usepackage{graphicx}
\usepackage{float}
\usepackage[normalem]{ulem}

\usepackage{array}
\newcommand{\PreserveBackslash}[1]{\let\temp=\\#1\let\\=\temp}
\newcolumntype{C}[1]{>{\PreserveBackslash\centering}p{#1}}
\newcolumntype{R}[1]{>{\PreserveBackslash\raggedleft}p{#1}}
\newcolumntype{L}[1]{>{\PreserveBackslash\raggedright}p{#1}}
\usepackage{diagbox}

\def\ltsima{$\; \buildrel < \over \sim \;$}
\def\simlt{\lower.5ex\hbox{\ltsima}}
\def\gtsima{$\; \buildrel > \over \sim \;$}
\def\simgt{\lower.5ex\hbox{\gtsima}}

\usepackage{tikz}

\DeclareFontFamily{OT1}{pzc}{}
\DeclareFontShape{OT1}{pzc}{m}{it}{<-> s * [1.10] pzcmi7t}{}
\DeclareMathAlphabet{\mathpzc}{OT1}{pzc}{m}{it}

\usepackage{tikz-3dplot}
\usetikzlibrary{shadings}
\usepackage{pgfplots}
\pgfplotsset{compat=1.16}
\usepgfplotslibrary{fillbetween}
\tikzset{declare function={%
f(\x,\y)=\x*\x-\y*\y;
fx(\x)=ifthenelse(\x<0,0.75*(\x+1),0.75*(-\x+1));
fy(\y)=ifthenelse(\y<0,0,ifthenelse(\y>1,-2+\y,-\y));
}}
\usetikzlibrary{backgrounds,calc,positioning}

\usepackage{acro}

\usepackage{nextpage}

\usepackage{tabularx} 

\renewcommand\arraystretch{1.5}

\usepackage{adjustbox}

\newcommand\myshade{85}
\colorlet{mylinkcolor}{RoyalPurple}
\colorlet{mycitecolor}{WildStrawberry}
\colorlet{myurlcolor}{JungleGreen}

\hypersetup{
  linkcolor  = mylinkcolor!\myshade!black,
  citecolor  = mycitecolor!\myshade!black,
  urlcolor   = myurlcolor!\myshade!black,
  colorlinks = true,
}

\usepackage{minitoc}

\newcommand\mathcomma{\,,}
\newcommand\mathperiod{\,.}

\DeclareMathAlphabet{\mathup}{OT1}{\familydefault}{m}{n}

\newcommand\expe{\mathup{e}}

\def\dd{\mathrm{d}}

\def\td{\tilde}

\newcommand\covd{\nabla}

\newcommand\hub{\mathcal{H}}

\newcommand{\be}{\begin{equation}} 
\newcommand{\ee}{\end{equation}}

\newcommand\norm[1]{\lVert #1 \rVert}
\WithSuffix\newcommand\norm*[1]{\left\lVert #1 \right\rVert}

\newcommand\numeq[1]%
  {\stackrel{\scriptscriptstyle(\mkern-1.5mu#1\mkern-1.5mu)}{=}}

\newcommand\commutator[2]{[#1, #2]}
\WithSuffix\newcommand\commutator*[2]{\left[#1, #2\right]}

\makeatletter
\g@addto@macro\@floatboxreset\centering
\makeatother



\makeatletter
\def\blfootnote{\gdef\@thefnmark{}\@footnotetext}
\makeatother

\title{Illuminating the Dark Sector}
\subtitle{Searching for new interactions between dark matter and dark energy}
\author{Elsa Maria Campos Teixeira}
\department{School of Mathematics and Statistics}
\university{The University of Sheffield}

\usepackage{minitoc}

\DeclareAcronym{cmb}{
  short=CMB,
  long=Cosmic Microwave Background,
}
\DeclareAcronym{cdm}{
  short=CDM,
  long=Cold Dark Matter,
}
\DeclareAcronym{eh}{
  short=EH,
  long=Einstein-Hilbert,
}
\DeclareAcronym{flrw}{
  short=FLRW,
  long=Friedman-Lemaître-Robertson-Walker,
}
\DeclareAcronym{gr}{
  short=GR,
  long=General Relativity,
}
\DeclareAcronym{hbb}{
  short=HBB,
  long=Hot Big Bang,
}
\DeclareAcronym{bao}{
  short=BAO,
  long=Baryon Acoustic Oscillations,
}
\DeclareAcronym{bbn}{
  short=BBN,
  long=Big Bang Nucleosynthesis,
}
\DeclareAcronym{dm}{
  short=DM,
  long=Dark Matter,
}
\DeclareAcronym{de}{
  short=DE,
  long=Dark Energy,
}

\DeclareAcronym{eos}{
  short=EoS,
  long=Equation of State,
}
\DeclareAcronym{mde}{
  short=MDE,
  long=Matter Dominated Epoch,
}
\DeclareAcronym{qft}{
  short=QFT,
  long=Quantum Field Theory,
}
\DeclareAcronym{rde}{
  short=RDE,
  long=Radiation Dominated Epoch,
}
\DeclareAcronym{svt}{
  short=SVT,
  long=Scalar-Vector-Tensor,
}

\DeclareAcronym{wimps}{
  short=WIMPs,
  long=Weakly Interacting Massive Particles,
}

\DeclareAcronym{bb}{
  short=BB,
  long=Big Bang,
}

\DeclareAcronym{gw}{
  short=GW,
  long=Gravitational Wave,
}

\DeclareAcronym{cl}{
  short=CL,
  long=Confidence Level,
}

\DeclareAcronym{dede}{
  short=DEDE,
  long=Dark Energy Dominated Epoch,
}

\DeclareAcronym{shoes}{
  short=SH0ES,
  long=Supernovae H0 for the Equation of State of Dark Energy,
}

\DeclareAcronym{ddb}{
  short=DDB,
  long=Dark D-Brane,
}

\DeclareAcronym{ddm}{
  short=DDM,
  long=Disformal Dark Matter,
}

\DeclareAcronym{lss}{
  short=LSS,
  long=Large Scale Structure,
}

\DeclareAcronym{sm}{
  short=SM,
  long=Standard Model,
}

\DeclareAcronym{lisa}{
  short=LISA,
  long=Laser Interferometer Space Antenna,
}

\DeclareAcronym{et}{
  short=ET,
  long=Einstein Telescope,
}

\DeclareAcronym{sn}{
  short=SN,
  long=Supernova,
}

\begin{document}

    \startpreamble



\vspace*{150pt}

   \epigraph{Tudo no mundo começou com um sim. Uma molécula \\ disse sim a outra molécula e nasceu a vida. Mas antes da pré-história \\ havia a pré-história da pré-história e havia o nunca e havia o sim. \\ Sempre houve. Não sei o quê, mas sei que o Universo jamais começou.\\ --- \textsc{Clarice Lispector}\ \small\textup{A Hora da Estrela}}

    \epigraph{All the world began with a yes. One molecule \\ said yes to another molecule and life was born. But before prehistory \\ there was the prehistory of prehistory and there was the never and there was the yes.  \\ It was ever so. I don’t know why, but I do know that the Universe never began. \\ --- \textsc{Clarice Lispector} in \small\textup{The Hour of the Star} }

\cleardoublepage

    \chapter*{Preface}
\chaptermark{Preface}
\addcontentsline{toc}{fmchapter}{Preface}

This dissertation is presented to the School of Mathematics and Statistics of the University of Sheffield for the degree of Doctor of Philosophy. This project has been supervised by Prof. Carsten van de Bruck, and the contents report on the results of the research work conducted over the course of the four years of my PhD, supported by the Funda\c{c}\~ao para a Ci\^encia e a Tecnologia (FCT) through the grant SFRH/BD/143231/2019. The Chapters in Part II are based partially on ongoing work and on articles published in refereed journals, as the result of collaborations with Carsten van de Bruck, Noemi Frusciante, Bruno J. Barros, Vasco M.C. Ferreira, Richard Daniel, Gaspard Poulot and Cameron Thomas. I list the published work below:

\begin{enumerate}




   
    \item \textbf{Elsa M. Teixeira}, Richard Daniel, Noemi Frusciante, and Carsten van de Bruck \\
\textsc{``Forecasts on interacting dark energy with standard sirens''},\\
\href{https://journals.aps.org/prd/abstract/10.1103/PhysRevD.108.084070}{\textit{Phys.Rev.D} 108 (2023) 8, 084070}, \href{https://arxiv.org/abs/2309.06544}{arXiv:2309.06544 [astro-ph]}.

\item Carsten van de Bruck, Gaspard Poulot, and \textbf{Elsa M. Teixeira},\\
 \textsc{``Scalar field dark matter and dark energy: a hybrid model for the dark sector''},\\
\href{https://iopscience.iop.org/article/10.1088/1475-7516/2023/07/019}{\textit{JCAP} 07 (2023), 019}, \href{https://arxiv.org/abs/2211.13653}{arXiv:2211.13653 [hep-th]}.

\item \textbf{Elsa M. Teixeira}, Bruno J. Barros, Vasco M. C. Ferreira, and Noemi Frusciante,\\
 \textsc{``Dissecting kinetically coupled quintessence: phenomenology and observational tests''},\\
\href{https://iopscience.iop.org/article/10.1088/1475-7516/2022/11/059}{\textit{JCAP} 11 (2022), 059}, \href{https://arxiv.org/abs/2207.13682}{arXiv:2207.13682 [gr-qc]}. 

\item Carsten van de Bruck, and \textbf{Elsa M. Teixeira},\\
\textsc{``Dark D--Brane Cosmology: from background evolution to cosmological perturbations''},\\
\href{https://journals.aps.org/prd/abstract/10.1103/PhysRevD.101.083506}{\textit{Phys.Rev.D} 101 (2020) 8, 083506}, \href{https://arxiv.org/abs/2007.15414}{arXiv:2007.15414 [gr-qc]} 
\end{enumerate}

This thesis is structured as follows:

\begin{itemize}
    \item \cref{chap:int} provides an introduction to the main concepts relevant for modern cosmology, including the theory of general relativity and the underlying cosmological principles.

    \item \cref{chapter:standardmodel} is an account of the fundamentals of the standard model of cosmology - the $\Lambda$CDM model - including its formulation, assumptions, background and perturbed equations, and a brief history of the Universe.

    \item \cref{chap:obs} consists of a review of the concepts in observational cosmology that are relevant for the work included in the thesis. It covers the main observables and probes pf the dark sector, followed by an account of the parameters of the $\Lambda$CDM model, and a comment on the current crisis it faces.

    \item \cref{chapter:statistics} provides a summary of the statistical methods necessary for the numerical analysis and the assessment of the support for the alternative models considered. It also includes a short introduction to the numerical tools and data sets for that purpose. 

    \item \cref{chapter:sttheories} introduces the motivation and formulation of models beyond the standard $\Lambda$CDM model, with particular focus on models with a coupling in the dark sector.

    \item \cref{chap:cquint} reports on the ongoing work with Cameron Thomas and Carsten van de Bruck. Cameron Thomas is the author of part of the numerical work conducted for \cref{sec:cq_flat}.

    \item \cref{chap:kin} draws from published work in the Journal of Cosmology and Astroparticle Physics (JCAP) \cite{Teixeira:2022sjr} in collaboration with Bruno J. Barros, Vasco M. C. Ferreira and Noemi Frusciante. Bruno J. Barros is credited with the original formulation of the model.

    \item \cref{chap:gwcons} is an account on collaborative work with Richard Daniel, Noemi Frusciante, and Carsten van de Bruck, which has been published in Physical Review D and is available at \cite{Teixeira:2023zjt}. Richard Daniel is the author of the simulated standard siren catalogue on which we report in \cref{app:mock}.

    \item \cref{chap:dbi} is based partially on published work in Physical Review D \cite{vandeBruck:2020fjo} in collaboration with Carsten van de Bruck and on ongoing work for the data analysis.

    \item \cref{chap:sfdm} reports on work published in the Journal of Cosmology and Astroparticle Physics (JCAP) \cite{vandeBruck:2022xbk} in collaboration with Gaspard Poulot and Carsten van de Bruck. Gaspard Poulot is credited with part of the theoretical background work and the fluid approximation.

    \item \cref{chapter:conclusions} concludes with a summary of the work presented in this dissertation, with an outlook on coupled and interacting quintessence models, including some general thoughts on the relevance of the research outlined in this manuscript and thoughts on the future of the field.
\end{itemize}

\cleardoublepage

     \printacronyms


    


\cleardoublepage
    

\chapter*{Abstract}
\chaptermark{Abstract}
\addcontentsline{toc}{fmchapter}{Abstract}

          \epigraph{Os factos são sonoros mas entre os factos há um sussurro. \\ É esse sussurro que me impressiona.\blfootnote{\textit{The facts are sonorous but between the facts there’s a whispering. It’s the whispering that astounds me.} --- Clarice Lispector in The Hour of the Star} \\ 
  --- \textsc{Clarice Lispector}\ \small\textup{A Hora da Estrela}}

The current standard model of cosmology - the $\Lambda$CDM model - is appropriately named after its controversial foreign ingredients: a cosmological constant ($\Lambda$) that accounts for the recent accelerated expansion of the Universe and cold dark matter needed to explain the formation and dynamics of large scale structures. Together, these form the dark sector, whose nature remains a mystery. After 25 years of withstanding confirmation and support for the $\Lambda$CDM model, enough to bypass some of its unclear theoretical issues, this paradigm is facing its biggest crisis yet. The rapid advent of technology has brought cosmology to an unprecedented observational era, with increased technical precision and the emergence of independent measures, including probes of phenomena that were thought impossible to detect or even exist, such as the gravitational ripples that propagate in the spacetime. However, such precision has unveiled cracks in the porcelain of $\Lambda$CDM, with pieces that seem glued together and difficult to reconcile. Particularly worrying is the apparent lack of compatibility between measurements of the Universe's present expansion rate based on local measurements and those based on phenomena that occurred far in the early Universe and that can only be translated into present quantities through physical propagation under a cosmological model. In this dissertation, we delve into extensions to the standard model that consider alternatives to the mysterious nature of the dark sector and any possible new interactions therein. We analyse these alternative models, hoping to identify measurable observational signatures of extra degrees of freedom in the dark sector. For this purpose, we focus on Markov Chain Monte Carlo sampling techniques by exploiting recent observational data and their combination to constrain and assess the validity of the model's parameter space. The extended frameworks investigated are only found to be marginally or not at all favoured over the $\Lambda$CDM model, ultimately falling short of resolving the tensions present in the latter.
We conclude with a forecast on the constraining power of upcoming gravitational wave detectors as standard sirens, independently and combined with current background data. By generating simulated standard siren event catalogues following the specifications predicted by the proposed missions, we discuss how an emerging avenue for independent data may be decisive in constraining alternative theories of gravity and shed light on the nature of the dark sector.


\subsubsection{Units}

Unless stated otherwise, in this dissertation I will always adopt natural units in which the speed of light, Planck's constant and Boltzmann's constant are unity, $c=\hbar = k_B = 1$. This translates into an equivalence of units for energy, mass and momentum as an inverse unit of length and time: 
\begin{equation}
    \text{[Energy]} = \text{[Mass]} = \text{[Temperature]} = \text{[Length]}^{-1} = \text{[Time]}^{-1} \mathperiod
    \end{equation}
Throughout this manuscript we denote derivatives with respect to physical time $t$ by an upper dot and derivatives with respect to conformal time $\tau$ by a prime,
\begin{align}
t = \text{cosmic time}\ \ &\Longrightarrow \ \ \frac{\dd X}{\dd t} \equiv \dot{X} \mathcomma \\
\tau = \text{conformal time}\ \ &\Longrightarrow \ \ \frac{\dd X}{\dd \tau} \equiv X' \mathperiod
\end{align}

Spatial 3-vectors are denoted an upper arrow, such as $\vec{x}$, and four-dimensional spacetime vectors are denoted in terms of their components as $x = (x^{\mu})$.
We adopt the convention for the signature of the metric as $(-,+,+,+)$.

   \dominitoc 
    { \hypersetup{hidelinks} \tableofcontents }
     { \hypersetup{hidelinks}\listoffigures}
     { \hypersetup{hidelinks}\listoftables}

    \cleardoublepage



\chapter*{Acknowledgements}
\chaptermark{Acknowledgements}
\addcontentsline{toc}{fmchapter}{Acknowledgements}


 \epigraph{A culpa é minha ou \\ A hora da estrela ou \\ Ela que se arranje ou \\ O direito ao grito ou \\ Quanto ao futuro ou \\ Lamento de um blue ou \\ Ela não sabe gritar ou \\ Uma sensação de perda ou \\ Assovio no vento escuro ou \\ Eu não posso fazer nada ou \\ Registro dos fatos antecedentes ou \\ História lacrimogênia de cordel ou \\ Saída discreta pela porta dos fundos\blfootnote{\textit{It’s all my fault - or - The hour of the star - or - Let her deal with it - or - The right to scream - or - As for the future - or - Singing the blues - or - She doesn’t know how to scream - or - A sense of loss - or - Whistling in the dark wind - or - I can’t do anything - or - Account of the preceding facts - or - Cheap tearjerker - or - Discreet exit through the back door} --- \textsc{Clarice Lispector} in The Hour of the Star} \\ --- \textsc{Clarice Lispector}\ \small\textup{A Hora da Estrela}}

I have to start by expressing my enormous gratitude for having been so lucky to have Carsten as a supervisor. During these past four years, almost half spent during a health emergency lockdown and hundreds of miles apart, you were always available to help me and keep me motivated. Your excitement and curiosity for the world are definitely not restricted to science and maths and I'm grateful to have shared that with you. I have enjoyed all the non-academic times we shared, from pandemic socially distanced walks to insisting on keeping your word and taking your two PhD ‘siblings’ to Woolsthorpe Manor. I am also thankful to Sabina for sharing some of these moments with us and for all the wild berries and mushroom picking.

 Despite being affected by the pandemic, I was still lucky to travel abundantly during my PhD. I am grateful to the School of Mathematics and Statistics of the Universe of Sheffield for providing me with multiple opportunities to do so, never denying any of my funding requests, to the European Consortium for Astroparticle Theory (EuCAPT) and the EU COST Action CosmoVerse, CA2113. Thanks to the University of Rome’ La Sapienza’ for hosting my STSM, ‘Università di Napoli’ Federico II \& INFN Sezione di Napoli for hosting the EuCAPT exchange grant, and LUPM, Université Montpellier for the introductory visit. Special thanks to Noemi Frusciante, Alessandro Melchiorri, Giuseppe Fanizza, and Vivian Poulin for welcoming me.

I am indebted to all the members of the CRAG research group, past and present, who made it such a fun, friendly, and stimulating environment to work in. I thank Steffen Gielen, Sam Dolan, Eleonora di Valentino, and Elizabeth Winstanley for being academic role models and introducing me to various corners of the Universe. I am incredibly thankful to have shared this experience with Richard, from the first day in the lift, stealing your desk and our crazy little matchbox office. I will never forget the times we spent in that squared metre, whose value is greater than the amount of tea bags consumed. To Ale for always having the greatest ground-lifting (literally) hug when it was (and also when it was not) needed. A particular mention has to be made to Lisa, who has been my absolute role model and best of friends these last two years, which feel like a decade of companionship. Thank you for always being there, for your kindness, and the occasional slice of cake. I cannot thank you enough for reading and scrutinising every line in this thesis.
I am also grateful to have crossed academic paths with some wonderful people. Thank you Rita, Matilde, Ruchika, William, Jéssica, Kerkyra, Vitor, Davide, João and Léo, and all the wonderful people that made attending conferences, workshops and seminars such an enriching experience. I want to thank my collaborators with whom I learned so much: Noemi Frusciante, Bruno J. Barros, Vasco M.C. Ferreira, Richard Daniel, Gaspard Poulot and Cameron Thomas. A special mention should be made to Ana Nunes, Nelson Nunes, Noemi Frusciante, Eleonora Di Valentino and Danielle Leonard for all the advice and mentoring, and to Bruno for everything we learned together and for remaining (in)sane with me during 1.5 years of lockdown.

I cannot overstate the equally great debt I owe to my home friends. I am especially grateful to Rita and Joana for spreading roots and sharing this adventure with me, ensuring we never took ourselves too seriously, and always sending me off with a full bag of laughs and heartfelt understanding. You have truly shown me the sun behind the English clouds.
I thank In\^{e}s for being my companion in this 10-year journey through physics, mathematics and philosophy, always with amazement and creativity. I cannot wait to keep exploring the wonders of the Universe with you.
Most of all, I owe all of it to my family, who has always supported me unconditionally with complete freedom to flourish. To my parents, who have taught me the most important things of all: never take myself too seriously, always be kind, and, above all, do what makes me happy. And in particular to my sister and niece for understating my physical absence in their lives. I dedicate this work to them, who inspire and motivate me daily. To my \textit{pardalinho}, in your fairy-tale homonymous words, I can only hope to inspire you always to remain

  \epigraph{“Curiouser and curiouser!” \\ --- \textsc{Lewis Carroll}\ \small\textup{Alice's Adventures in Wonderland}}

    
    \markboth{\nomname}{\nomname}
    
    
    \stoppreamble


    \partdivider{Part I}{Foundations of Gravity and Cosmology} \label{part:foundations}

    \cleardoublepage


 \chapter{Introduction} \label{chap:int}
 \setcounter{equation}{0}
\setcounter{figure}{0}

 \epigraph{Como começar pelo início, se as coisas acontecem antes de acontecer? \\ Se antes da pré-pré-história já havia os monstros apocalípticos? \\ Se esta história não existe, passará a existir. Pensar é um ato. \\ Sentir é um fato. Os dois juntos – sou eu que escrevo o que estou escrevendo.\blfootnote{\textit{How do you start at the beginning, if things happen before they happen? If before the pre-prehistory there were already the apocalyptic monsters? If this story doesn’t exist now, it will. Thinking is an act. Feeling is a fact. Put the two together — I am the one writing what I am writing.} --- \textsc{Clarice Lispector} in The Hour of the Star} \\ --- \textsc{Clarice Lispector}\ \small\textup{A Hora da Estrela}}

 \section{Introduction to Cosmology}

 Cosmology is a unique field of scientific inquiry that seeks to understand the nature and origins of the Universe. On what might be the only personal note in this dissertation, I like to think that cosmology parallels the work of archaeologists piecing together the fragments of ancient civilisations. Both disciplines aspire to reconstruct a larger, coherent picture from fragmented pieces - whether it be artefacts buried in the earth or distant sources of light reaching our telescopes from the vast deepness of the Universe. Each cosmic signal we analyse is a relic from a long-past event, a \textit{fossil of light}, offering clues about the history and structure of the ultimate civilisation - our Universe. This presents an epistemological challenge unique to cosmology: working to understand an entire, singular Universe based on isolated data points, inherently confined by the limits of causality and our observational reach. The task has an almost philosophical undertone; it grapples with questions about the very nature of evidence and existence, especially given that we have only one Universe to study. This singularity not only imposes a limit on our understanding but also introduces an inherent bias into any laws or theories we might derive. Thus, cosmology exists at an intersection of science and philosophy, endeavouring to build a narrative of the Universe that is both empirical and conceptually profound. It confronts us with the stark reality that our efforts to understand the cosmic journey are both an ambitious scientific undertaking and a quest that treads on deeply abstract grounds.

Metaphysics aside, as an established scientific field, cosmology aspires to comprehend and articulate the fundamental characteristics and evolution of the Universe. It devises meticulous and often speculative narratives for the composition of the Universe, delving further into the journey of those constituents as they mature into complex structures such as galaxies. It also examines the dynamics and interplay of those structures within an ever-changing landscape. These overarching frameworks are referred to as \textit{cosmological models}. Unlike the word \textit{model} might suggest, contemporary cosmological models consist mainly of abstract mathematical concepts, often devoid of physical intuition or tangible form.

While commonly referred to as the birth of the Universe, the \textit{Big Bang} might be more aptly portrayed as its conception. The period preceding the creation of the first atoms, culminating in the unimpeded propagation of light, was a vastly unfamiliar environment, an obscure cosmic fog of elementary particles and their extremely energetic interactions. When the Universe was approximately $380\ 000$ years old, it grew more familiar. It gradually unfolded into the configuration we recognise today, seemingly set on a perpetual path of expansion, according to the prevailing cosmological model.


At its core, Cosmology stands as a field deeply rooted in ancient human contemplation. It is impossible to disentangle from the quest to understand our place in the world, its origins, and ontological meaning. The foundations of modern cosmology were established in 1917 with Albert Einstein's seminal article on his theory of \ac{gr}, titled \textit{Cosmological Considerations in the General Theory of Gravity} \cite{einstein2}. Building upon Einstein's work, cosmological models were further developed by Dutch mathematician Willem de Sitter \cite{deSitter:1916zza,deSitter:1916zz,deSitter:1917zz}, German physicist Karl Schwarzschild \cite{Schwarzschild:1916ae,Schwarzschild:1916uq}, and Russian mathematician Alexander Friedmann \cite{1924ZPhy...21..326F,friedmann}. These studies demonstrated that general relativity could accommodate the concept of an expanding Universe, a theoretical hypothesis that came to be confirmed by the realisation that galaxies stood as rulers to trace astronomical distances in space and which were, in fact, receding from our own Milky Way in all directions. In 1929, Edwin Hubble's landmark discovery, now known as the Hubble law, revealed a linear relation between the recessional speed of a galaxy and its distance, culminating in the hypothesis that the Universe is expanding. This idea was not unexpected and had been largely predicted from a theoretical perspective. Indeed, it was qualitatively reinforced by redshift measurements and notoriously explored during the 1920s.

As it stands at the moment of writing this dissertation, the standard cosmological model - referred to as the $\Lambda$CDM model based on its novel components - is constructed upon six foundational principles, summarised in the following equation and which will be discussed below:
\begin{equation*}
    \Lambda\text{CDM Model}\, =\, \text{GR}\, +\, \text{FLRW}\, +\, \text{Standard Matter}\, +\, \text{Initial Conditions}\, +\, \text{CDM}\, +\, \Lambda \mathperiod
\end{equation*}
The GR, FLRW, and Standard Matter triad represent well-established and tested hypotheses based on local empirical evidence. On the other hand, although highly supported by cosmological and astrophysical observations, the last three components infuse new physics into the picture and still lack more concrete and direct confirmation. In unison, these six hypotheses orchestrate a comprehensive paradigm for understanding the Universe on large scales, as listed below:

\begin{enumerate}
    \item \ac{gr}: The theory of gravity is essential for understanding cosmological phenomena, as gravity dominates the Universe on large scales and is at the heart of cosmic evolution.
    \item \ac{flrw} metric: GR admits only a few analytical solutions as a metric theory regulated by non-linear partial differential equations. Therefore, deriving physical implications requires a simple mathematical description from the metric, with the FLRW metric being a natural choice assuming maximal spatial symmetry.
    \item Standard Matter (from the standard model of particle physics): the directly detected forms of matter in the Universe - baryons (protons, neutrons, including the less massive electrons), photons, and neutrinos - whose dynamics and interactions are based on robust physical principles - must be included in the cosmological model to recount the expansion history of the Universe successfully.
    \item Initial Conditions (from inflation): an early period of rapid cosmic expansion, inflation is invoked to resolve causality problems and account for the observed spatial curvature of the Universe. Remarkably, it also provides a mechanism to explain the Universe's initial conditions, including generating perturbations around the background metric and supporting the choice of the homogeneous and isotropic metric \textit{ansatz}.
    \item \ac{cdm}: a form of pressureless matter that does not interact with light, needed to accurately describe the dynamics and process of formation of large-scale structures, such as galaxies and galaxy clusters. The simplest resolution is to introduce a new, weakly interacting particle species.
    \item Cosmological Constant ($\Lambda$): an additional degree of freedom that accounts for the observed accelerated expansion of the Universe, which in the simplest scenario is a cosmological constant, \textit{i.e.}, some energy density contribution that remains constant over the expansion.
\end{enumerate}

For completeness of the work, we first provide a general overview of the relevant concepts of cosmology, focusing mainly on the Universe's expansion history and leading towards the physics of the late Universe. The main goal is to introduce and contextualise the original work presented in Part II of this manuscript. In that sense, Part I will consist of well-established topics in the field. Nevertheless, the organisation of the discussion and the storytelling should be original, connecting the ideas and guiding the reader through the relevant concepts. 
We will start by presenting a brief overview of the dynamics of general relativity. This lays out the basis for the introduction of cosmology, focusing mainly on the large-scale structure of spacetime and the dynamics of the expansion history. To understand the full scope of the changes introduced in each model, a complete account of the physically relevant epochs of the Universe is needed. The seeds for the evolution of structures lie in the primordial perturbations produced during an initial period of inflation. These are imprinted in the \ac{cmb} radiation and are a fundamental cosmological probe.


\section{A Comment on Observational Cosmology}

The methods of scientific pursuit must be adapted to the limitations of the field, particularly in the realms of physical cosmology and extragalactic astronomy, where empirical assessment is limited to distant observations without direct interaction or manipulation of the objects of study. Moreover, in the vast expanse of the cosmos, we cannot replicate experiments or observations. Instead, we must rely on searching for the clues and evidence left behind in fossils from the ancient past. Some of these relics are found nearby, such as extragalactic hints on the Earth or the Moon, in our Solar System or even in stars within our galaxy. Even though these local remnants give important insight, they are just pieces of a large jigsaw puzzle. Nevertheless, here, the laws of physics and the principles of general relativity play in our favour: our vision can reach far beyond, as the light from distant objects carries information from earlier epochs of the Universe, based on the fact that observed objects at increasing distances effectively corresponds to looking back in time, witnessing the earlier stages of cosmic evolution.
Our view of the cosmos is shaped by our past light cone, representing the collection of all the light that has reached us throughout history. While the light cone of human existence encompasses an incredibly thin slice of the Universe's history, the information travels towards us from all directions and epochs, spanning a vast range of time and space. Collecting the remnant pieces of cosmic history and drawing them together is just like adding the pieces of the jigsaw puzzle, except that there may be infinite pieces, some inaccessible or lost forever in the vastness of the cosmos, others waiting to be found (and most likely there are no corner pieces to start with!).

The debates and discussions surrounding cosmological issues can be met with an intensity and fervour not matched by the weight of evidence. One reason is that the observational evidence needed to settle these questions often appears to be beyond our grasp or even an impossibility. Furthermore, the study of such fundamental questions is deeply entangled with an overarching interest in understanding the nature of our own existence and the world we inhabit.
It is inevitable to ponder the Universe's origins and potential fate, and it is hard not to develop a philosophical preference for certain scientific outcomes or perspectives.

The era of precision measurements has drastically changed the scientific method. Big collaborative teams have developed to address the need to deal with unprecedentedly large amounts of data, with the natural gap between theory and observations being transposed with numerical simulations. This shift has led to heightened scrutiny of research endeavours and the need for inquiries to be more rigorously justified or supported.

The standard model of cosmology today has reached the status of an empirical science; that is, it is based on and tested according to what can be observed or measured. However, scientific advancements are often made on a theoretically intuitive basis since, while being well-tested and investigated, a cosmological theory is always incomplete and subject to unpredictability. The currently accepted $\Lambda$-Cold-Dark-Matter ($\Lambda$CDM) model relies on the extrapolation of a cosmological history from imprints left in the observables that resemble those predicted by the model. The theory is then a useful though incomplete approximate picture of the history of the cosmos, with remaining cracks and hypotheses on the borderline of established physics. Nevertheless, the abundance and diversity of empirical tests that the standard model has undergone maintain outstanding confidence that any \textit{alternative theory} that attempts to glue those cracks will portray a Universe that behaves much like $\Lambda$CDM. For this reason, the first natural step in that direction is to consider small deviations or extensions to the minimal $\Lambda$CDM model while remaining conscious that this may not be enough and the answer may require a new paradigm disconnected from the $\Lambda$CDM assumptions. 

\section{General Relativity} \label{sec:grrev}

In this section, we provide a summary of general relativity, with no intention of exhaustively introducing or reviewing its full extent. For this purpose, we refer to \text{e.g.} \cite{wald,gravitation}.

Albert Einstein's theory of General Relativity, formulated in 1915 \cite{Einstein}, presents an exceptional framework where gravitational interactions are related to the geometry of spacetime \cite{carroll2004spacetime,mukhanov_2005,ellis}. The central element is the \textit{spacetime metric}\footnote{Throughout this manuscript, we adopt Einstein's summation convention where repeated indices are summed over. The metric signature is $(-,+,+,+)$ and we use natural units with $c=\hbar = k_B =1$. Greek indices range from $0$ to $3$, while Latin indices denote spacetime components only, $1$ to $3$.} degree of freedom $g$, which determines physical distances as
\begin{equation}
    \odif{s}^2 = g_{\mu \nu} \odif{x}^{\mu} \odif{x}^{\nu} \mathcomma
    \label{eq:metricgr}
\end{equation}
and is treated as a fully dynamic variable.

The time coordinate interval between two events, $\Delta t$, is not the same as perceived by different observers, motivating the introduction of the more convenient invariant proper-time interval, $\Delta \tau$. Considering the events $A$ and $B$ which are separated by a time-like interval ($\odif{s}^2 < 0$), the curve which connects the two is denoted by $\gamma$ and parameterised by $\lambda$, such that \textit{e.g.} $\gamma(\lambda_A) = A$ and $\gamma (\lambda_B) = B$. The proper-time interval $\Delta \tau$ between $A$ and $B$ is then given by
\begin{equation}
    \Delta \tau \equiv \int_{\gamma} \sqrt{- \odif{s}^2} = \int_{\lambda_A}^{\lambda_B} \odif{\lambda} \sqrt{-g_{\mu \nu} \odv{x^{\mu}}{\lambda} \odv{x^{\nu}}{\lambda}} \mathperiod
    \label{eq:deltatau}
\end{equation}

Massive free-falling particles moving in a curved spacetime follow a \textit{geodesic trajectory}, $X^{\alpha} (\tau)$, defined as the time-like trajectory that extremises the proper-time interval $\Delta \tau$ in \cref{eq:deltatau}. The geodesic trajectory is the one that minimises the action 
\begin{equation}
    S = m \int \odif{\tau},
    \label{eq:actfp}
\end{equation}
where $m$ is the rest-frame mass of such particles. The trajectory of these particles follows the \textit{geodesic equation}:
\begin{equation}
\odv[order=2]{X^{\alpha}}{\tau} +  \Gamma^{\alpha}_{\mu \nu} \odv{X^{\mu}}{\tau} \odv{X^{\nu}}{\tau} = 0 \mathcomma
\label{eq:geoeq}
\end{equation}
where
\begin{equation}
    \Gamma^{\alpha}_{\mu \nu} \equiv \frac{1}{2} g^{\alpha \lambda} \left[ \partial_{\mu} g_{\lambda \nu} + \partial_{\nu} g_{\lambda \mu} - \partial_{\lambda} g_{\mu \nu} \right] \mathcomma
    \label{eq:gammadef}
\end{equation}
is the torsion-free ($\Gamma^{\alpha}_{\mu \nu}  = \Gamma^{\alpha}_{\nu \mu}$) connection (or Christoffel symbols) that gives a metric-compatible covariant derivative $\covd_{\mu} (\cdot)$
\begin{equation}
  \covd_{\mu} Y^{\lambda ...}_{\nu...} = \partial_{\mu} Y^{\lambda...}_{\nu...} + \Gamma^{\lambda}_{\mu \alpha} Y^{\alpha...}_{\nu...} - \Gamma^{\alpha}_{\mu \nu} Y^{\lambda...}_{\alpha...} + ... - ... \quad  \xrightarrow{\text{\cref{eq:gammadef}}}\quad \covd_{\alpha} g_{\mu \nu} = 0 \mathcomma
\end{equation}
where $Y^{\lambda ...}_{\nu...}$ is some arbitrary-rank tensor.



Einstein's gravitational theory establishes a connection between the dynamics of the metric and matter through the curvature of spacetime, with the latter embodied by the \textit{Riemann tensor}:

\begin{equation}
    R^{\alpha}_{\beta \mu \nu} \equiv \partial_{\mu} {\Gamma^{\alpha}_{\nu \beta}} - \partial_{\nu} {\Gamma^{\alpha}_{\mu \beta}} + \Gamma^{\sigma}_{\beta \nu} \Gamma^{\alpha}_{\mu \sigma} - \Gamma^{\sigma}_{\beta \mu} \Gamma^{\alpha}_{\nu \sigma} \mathperiod
\end{equation}

An action for the metric is required to endow this formulation with dynamics. Remarkably, the simplest nontrivial action that can be constructed out of a scalar from $g_{\mu \nu}$ leads to second-order equations in the effective field theory and reads:
\begin{equation}
 S = S_{\text{grav}} +  S_{\text{mat}} \mathcomma
 \label{eq:fullact}
\end{equation}
where
\begin{equation}
     S_{\text{grav}} = \int \odif[order=4]{x} \sqrt{-g} \frac{R} {16 \pi G},
     \label{eq:ehact}
\end{equation}
is the \ac{eh} action. The action $S_{\text{mat}}$ encodes the matter sector, and $g$ is the determinant of the metric. The main player in the EH action is the \textit{Ricci (or curvature) scalar}, defined in terms of the trace of the \textit{Ricci tensor}:

\begin{equation}
   R \equiv R^{\mu}_{\mu} = R_{\mu \nu} g^{\mu \nu}\quad \text{with}\quad R_{\mu \nu} \equiv R^{\alpha}_{\mu \alpha \nu} \mathperiod
\end{equation}

By employing the principle of least action, the variation of the action in \cref{eq:fullact} with respect to the metric yields the \textit{Einstein field equations}:
\begin{equation}
    G_{\mu \nu} \equiv R_{\mu \nu} - \frac{1}{2} g_{\mu \nu} R  = 8 \pi G T_{\mu \nu}\mathcomma
    \label{eq:eineqdef}
\end{equation}
where $G$ is Newton's gravitational constant, $8 \pi G= \kappa^2 = \text{M}_{\text{Pl}}^{-2}$, and $\text{M}_{\text{Pl}}$ ($\kappa$) is the (reduced) Planck mass. The left-hand side (LHS) is enclosed in the \textit{Einstein tensor} $G_{\mu \nu}$. The source term on the right-hand side (RHS), the \textit{energy-momentum tensor}, is obtained from the variation of the matter action with respect to the metric and is covariantly conserved on account of the \textit{Bianchi identities}:
\begin{equation}
   \nabla_{\mu} G^{\mu \nu} =0  \quad  \xrightarrow{\text{\cref{eq:eineqdef}}}\quad  \nabla_{\mu} T^{\mu \nu} =0\quad \text{with}\quad T^{\mu \nu} \equiv 2 \sqrt{-g} \fdv{ S_{\text{mat}}}{g_{\mu \nu}}\mathperiod
    \label{eq:grcons}
\end{equation}
By definition, $T^{\mu \nu}$ is always symmetric. The $\nu = 0$ component of $\nabla_{\mu} T^{\mu \nu}$ represents energy conservation, while the three remaining space-like components correspond to momentum conservation. The Einstein equations epitomise the foundation of general relativity: matter and geometry are cut from the same fundamental cloth, with the geometry of spacetime prescribing the dynamics of matter. At the same time, massive bodies set the curvature of spacetime itself.

The seemingly compact set of Einstein's field equations, in \cref{eq:eineqdef}, govern the evolution of the Universe and all gravitating bodies, with the concrete realisation depending on its matter content and the spacetime metric based on physical considerations.


To derive the dynamics of the Universe from GR, we can start with the example of Einstein's original proposal, based upon two main assumptions, reflecting the impossibility of filling up an infinite space with matter, which must itself be finite in nature \cite{Janzen:2014xha}:
\begin{itemize}
    \item The Universe is finite;
    \item The Universe is static.
\end{itemize}

The first assumption aims to ensure a metric completely described by the stress-energy tensor. The second one is more subtle but reasonable in the scientific context in which Einstein first proposed a cosmological solution to his equations in 1917 \cite{einstein2}. By extrapolating the assumption of a spatially closed Universe to the very largest scales, a uniform distribution of matter can only be achieved if the curvature of space is constant. This is in line with Einstein's intention, which appears to have been to depict a cosmological model wherein a finite density of matter shaped the structure of the world. During that period, only stars with minor peculiar velocities had been detected, and it was not until 1922 that nebulae were acknowledged as independent stellar systems beyond our galaxy \cite{eddington1923mathematical, 1927ASSB...47...49L,friedmann}, supporting the existence of a reference frame in which matter is at rest \cite{ORaifeartaigh:2017uct}. However, the two postulates above are not compatible with the Einstein field equations, given in \cref{eq:eineqdef}, according to which a finite density of matter collapses on itself due to the gravitational attraction, invalidating the static Universe assumption.
Einstein resolved this issue by \textit{simply} introducing a constant term, $\Lambda$, in his equations to counterbalance the gravitational pull of matter, consistent with the principles of General Relativity and with the Bianchi identities $\nabla^{\mu} T_{\mu \nu} = 0$, leading to a new set of field equations \cite{straumann2012general}: 

\begin{equation} \label{eq:einseqcc}
    G_{\mu \nu} = 8 \pi G T_{\mu \nu} + \Lambda g_{\mu \nu} \mathperiod
\end{equation}
With this relation, the $\Lambda$ term stands as a negative effect against gravity, so both effects compensate. In the context of Einstein's static Universe, and with the solution proposed by Friedmann-Lema\^{i}tre-Robertson Walker (which we assume for now but will be discussed in detail in \cref{sec:flrw}), these equations imply
\begin{equation} \label{eq:ccuns}
    8 \pi G \rho = \frac{1}{a^2R_E^2} = 3 \Lambda \mathcomma
\end{equation}
where $\rho$ stands for the total mass density of the Universe, and the parameter $R_E$ is its curvature radius\footnote{The radius of curvature of space in Einstein's static Universe is also referred to as Einstein's radius. In the context of the FLRW solution (see \cref{sec:flrw}), the definition of $R_E$ emerges from setting both the first and second-time derivatives of the scale factor, $\dot{a}$ and $\ddot{a}$, to zero in the Friedmann equations, arriving at $aR_{\text{E}} = 1/\sqrt{4\pi G\rho}$, where $G$ is Newton's gravitational constant and $\rho$ represents the spatial density of this Universe.}, for a closed spherical geometry. Hence, the cosmological constant $\Lambda$ plays the role of some repulsive force counteracting the gravitational collapse of the Universe. It could be included in the theory as a fundamental constant that determines the density of matter and the radius of cosmic space. However, the solution for a steady Universe is unstable because a change in the value of $\Lambda$ in \cref{eq:ccuns} (or equivalently a local rearrangement of the mean mass density) would cause the Universe to either expand or collapse, like a ball that is carefully balanced on the top of a mountain, thus invalidating the static assumption.
Remarkably, in 1922, Friedmann proposed a new cosmological model that abolished the static hypothesis and replaced Einstein's assumptions with \cite{friedmann}: 
\begin{itemize}
    \item The Universe is homogeneous;
    \item The Universe is isotropic;
    \item The Universe is expanding.
\end{itemize}
It was only in 1929 that evidence for the expansion of the Universe came to light with Hubble's discovery of the recession of galaxies. This led to the cosmological constant being cast aside, only to be revived at the end of the millennium.
Einstein himself was determined to dismiss the cosmological constant, as the goal for its introduction had been to represent a finite density of matter in the Universe. When Friedmann showed that this could be achieved, either with or without the $\Lambda$-term - provided that the cosmic distance scale was time-dependent, which was further substantiated by Hubble's empirical confirmation that cosmic distances indeed change with time - Einstein found himself able to reject the term. He maintained that this constant undermined the theoretical elegance and logical simplicity of his original formulation. In his own words \cite{Schilpp1949-SCHTLO-38}:

\textit{The introduction of such a constant implies considerable renunciation of the logical simplicity of theory, a renunciation which appeared to me unavoidable only so long as one had no reason to doubt the essentially static nature of space. After Hubble's discovery of the "expansion" of the stellar system and since Friedmann's discovery that the unsupplemented equations involve the possibility of the existence of an average (positive) density of matter in an expanding Universe, the introduction of such a constant appears to me, from the theoretical standpoint, at present unjustified.}

Perhaps one of the most vocal critics of Einstein's rejection of the cosmological constant was Arthur Eddington. He fundamentally disagreed with Einstein's view, especially regarding the interpretation of cosmic expansion, seeing it as a logical consequence of the physical theory \cite{eddington1933}. Eddington maintained that the general relativistic field equations expressed that length measurements should be made relative to a finite constant radius of curvature. As he famously wrote \cite{eddington1923mathematical},

\textit{From this point of view it is inevitable that the constant $\Lambda$ cannot be zero; so that empty space has a finite radius of curvature relative to familiar standards. An electron could never decide how large it ought to be unless there existed some length independent of itself for it to compare itself with.}

This disagreement between Einstein and Eddington was not rooted in scientific differences but rather philosophical ones, as both relied on the same empirical evidence to support their views. Einstein believed that the time-varying spatial scale introduced by the expansion in Hubble's law rendered the $\Lambda$-term superfluous and that his theory would reduce to a simpler statement without it. On the other hand, Eddington advocated that not only did the cosmological constant simplify the theory's formulation, but this reformed version also aligned with the observed expansion, confirming the need for a significant cosmic repulsion on large scales.

Eddington's view was exalted by the Nobel Prize-winning observation of the acceleration of the cosmic expansion in 1998 \cite{acel1,acel2} (which will be discussed in \cref{chapter:standardmodel}), which validated the cosmological constant through its predictive power. Indeed, in retrospect, the evidence gathered since Einstein first proposed $\Lambda$ seems more aligned with Eddington's interpretation, who believed that the natural gravitational repulsion embodied by $\Lambda$ could set the stage for the understanding of the Universe's evolution, referring to the cosmic expansion as \textit{Einstein's almost inadvertent prediction} \cite{eddington1933}.

\section{Homogeneity and Isotropy} \label{sec:hom}



Modern cosmology emerged from Albert Einstein's investigation of how his theory of general relativity could be employed to understand the large-scale nature of the Universe. Einstein believed that a philosophically reasonable Universe should have the same properties everywhere and in all directions (therefore justifying the proposal for a uniform distribution of matter at the boundary), with only minor inhomogeneities, such as the agglomerates of matter that make up planets or stars \cite{einstein2}. This opposed the standard scientific approach of examining structures separately and at specific hierarchical levels based on a progressive narrative, such as the study of molecules, atoms in molecules, nuclei in atoms and so on \cite{1804633}. The same premise can be applied on larger scales: the study of the nature of planets around stars is followed by stars in galaxies, galaxies in clusters, and so the story proceeds. 

Einstein's vision was that the hierarchical structure of the Universe culminates in a new ideal for modern science: \textit{large-scale homogeneity}. According to this argument, the Universe is assumed to be homogeneous and isotropic on large scales, meaning it is the same on average regardless of the observer's position or orientation. Einstein's large-scale homogeneity opened the possibility of considering and testing a theory that describes the Universe as a whole rather than focusing on theories for specific levels of structure. The power of this hypothesis is that observations from our position in the Universe can be taken to represent what the Universe looks like from any other place. Nevertheless, just like any hypothesis, it requires empirical evidence. 
The concept of large-scale homogeneity grew from philosophical reasoning, intuition, collaboration, and perhaps even some wishful thinking. It was not initially rooted in empirical evidence but in the pursuit of a unified, coherent framework to describe the Universe.

In more precise terms, the \textit{cosmological principle} is the backbone of the standard cosmological model. It states that on large scales (greater than $ \sim 100\, \text{Mpc}$), the Universe is isotropic (rotational invariance) and homogeneous (translational invariance). This assumption implies no privileged positions or directions in the Universe, limiting the vast array of possible cosmological models. Isotropy\footnote{A general theorem of geometry ensures that isotropy around every point implies homogeneity \cite{Kolb:1990vq}.} has been supported on intermediate and large scales through various observations \cite{Aluri:2022hzs}, discussed in \cref{chap:obs}, including:

\begin{itemize}
\item The distribution of galaxy clusters and super-clusters;
\item Observations of large scale structure and of the interstellar medium in radio and X-ray wavelengths;
\item The uniformity of the \ac{cmb} radiation temperature, with $\Delta T/T \sim 10^{-5}$, indicating a strongly isotropic Universe at the time of radiation emission (approximately 300 thousand years after the Big Bang).
\end{itemize}

\begin{figure}[!h]
\begin{center}
          \subfloat{\includegraphics[width=0.9\linewidth]{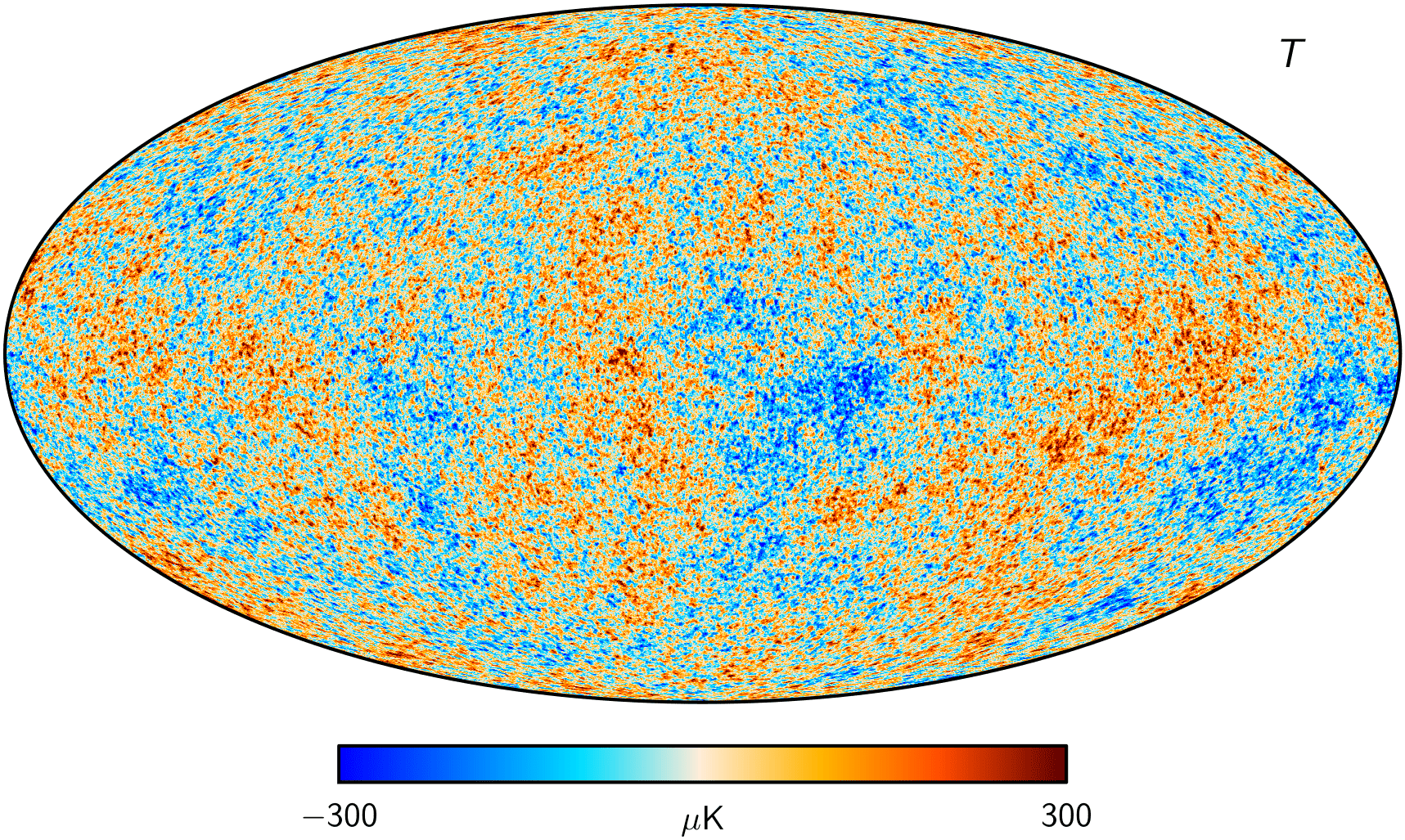}}
          \end{center}

  \caption[The \textit{Planck} 2018 temperature anisotropies map]{\label{fig:cmb} The sky map of temperature anisotropies in the Cosmic Microwave Background as measured by the \textit{Planck} Collaboration (taken from the \href{https://wiki.cosmos.esa.int/planck-legacy-archive/index.php/CMB_maps}{ \textit{Planck} Legacy Archive}) \cite{Planck:2018nkj}.}
\end{figure}

The cosmological principle is also a modern formalisation of the old \textit{Copernican principle}, in the sense that not only do we not hold a particular position in the Solar System, but also the position of the Milky Way in the Universe should not be statistically distinguishable from the spatial distribution of other galaxies. 
While the Universe contains inhomogeneous structures like galaxies and galaxy clusters on smaller scales, at early times, it looked roughly the same at every point and in all directions. This is best illustrated through \cref{fig:cmb}, where the impressively detailed map of the temperature distribution of the CMB as measured by the \textit{Planck} satellite is shown. The CMB is the monument marking the largest (and hence most ancient) scales that can be directly observed, depicting remarkable homogeneity.

We have just seen the picture of a homogeneous, isotropic and expanding Universe on large scales.
However, which scales can be classified as sufficiently large? A standard distance unit in Cosmology is the Megaparsec, $1\, \text{Mpc} = 3.2615 \times 10^6\, \text{light-years} = 3.0856 \times 10^{24}\, \text{cm}$. This is the typical distance between nearby galaxies, \textit{e.g.} the distance between Andromeda and the Milky Way is around $0.7\, \text{Mpc}$. Determining the exact scale for which the distribution of galaxies can be considered to be homogeneous is not a trivial endeavour. Even though large-scale structure observations \cite{eBOSS:2020yzd} show that the galaxy density inhomogeneities are still of the per cent order on scales of $ \sim 150\, \text{Mpc}$, the geometry of the Universe hints that these deviations must be small already on the $\text{Mpc}$ scale. The geometry can be tested through gravitational lensing, the peculiar motion of galaxies and, more importantly, the CMB \cite{Aghanim:2018eyx}.


Ultimately, the cosmological principle ensures that the laws of physics governing the Universe at large scales are identical for all observers, making it comprehensible to us.

\begin{figure}[!h]
      \subfloat{\includegraphics[width=\linewidth]{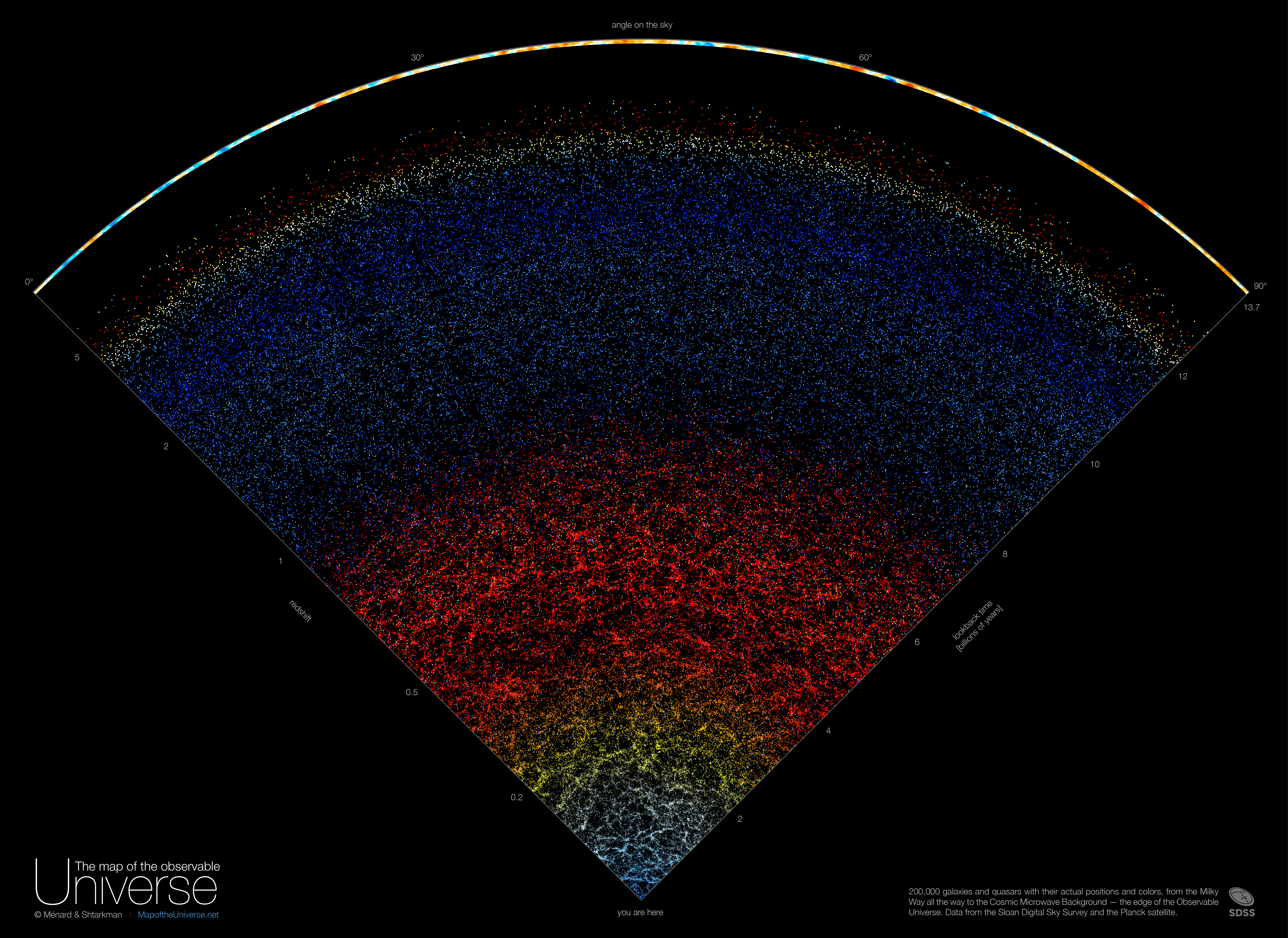}}
  \caption[The SDSS map of galaxies and quasars in the observable Universe]{\label{fig:gal} A map of galaxies and quasars in the observable Universe found by the Sloan Digital Sky Survey (SDSS) from 2000 to 2020 \cite{eBOSS:2020yzd}. This particular wedge of the full survey covers around 10 degrees. It encompasses about $200\,000$ galaxies and quasars reaching from our observable position (not special according to the cosmological principle) up to a look-back time of 12 billion years and cosmological redshift $5$, with redness indicating increasing redshift and distance. Almost every dot in the nearby lower part of the illustration represents a galaxy, and the upper part represents a distant quasar. The cosmic web structure is clearly illustrated, with the gravitational attraction between the nearby galaxies leading to a local Universe more condensed and filamentary than the distant Universe. The edge is the CMB temperature map as measured by the \textit{Planck} Collaboration \cite{Planck:2018nkj}. Visualisation created at John Hopkins University: \href{https://mapoftheUniverse.net}{https://mapoftheUniverse.net}.}
\end{figure}

Combined with the Einstein equations and a specified energy-matter distribution, the Cosmological Principle provides a framework for understanding how the Universe evolves. However, this simplistic view of the homogeneous model has inherent limitations. Specifically, it cannot account for the formation and filament-like distribution of galaxies and galaxy clusters, such as depicted in \cref{fig:gal}, nor for general variations in the matter distribution (a perfectly homogeneous and isotropic Universe would not even allow for our existence to discuss it in the first place!). Likewise, the distribution of these inhomogeneities and the variations in the temperature and polarisation of the CMB photons are intimately connected.
One can employ an approach based on \textit{linear perturbation theory} to address these limitations. If one first solves for the evolution of the background quantities (\textit{i.e.} non-perturbative), that information can be incorporated into the equations that govern the changes in these small-scale variations. The details of this method and how it connects with observations will be the focus of \cref{chapter:standardmodel,chap:obs}. 

Nevertheless, this perturbative approach is not without its limitations. During the period when matter dominates, structure formation starts taking place, invalidating the assumption of small density perturbations. Over this regime \textit{non-linear perturbation theory} must be employed, a topic which falls outside the scope of this dissertation. Nevertheless, the linear analysis of matter fluctuations and CMB anisotropies remains valuable. Its strength resides exactly in the fact that various methodologies can be complementary and mutually reinforcing.

\section{The Hubble-Lemaître Law: Cosmic Expansion}

It was Edward Arthur Milne who recognised the power of homogeneity in formulating a cosmological scenario and named it \textit{Einstein's cosmological principle} \cite{1933ZA......6....1M}. He showed that independent of general relativity, this principle explains the relation between the recession velocity $v$ of a galaxy and its distance $r$, known as \textit{Hubble's law}:
\begin{equation}
    v = H_0 r \mathperiod
\end{equation}
The proportionality constant, $H_0$, is known as Hubble's constant, and the subscript is meant to indicate that it is a measure of the present rate of expansion of the Universe. In an evolving Universe, the expansion rate is a function of time, $H(t)$. This proposal was based on the observation that light from distant galaxies is shifted towards longer wavelengths (redshifted), indicating their recession from us. A simple way to picture this in Euclidean geometry is to imagine two galaxies at arbitrary points and write their velocities as the vector relation $\vec{v} = H_0 \vec{r}$. Then an observer on galaxy $a$ sees galaxy $b$ moving away at velocity
\begin{equation}
    \vec{v}_b - \vec{v}_a = H_0 (\vec{r}_b - \vec{r}_a) \mathperiod
\end{equation}
This demonstrates that regardless of their position, all observers witness a consistent pattern of galaxies receding from one another, as expected in a homogeneous Universe.
For nonrelativistic recession velocities, the cosmological redshift is defined as the ratio $z = v/c$, akin to a Doppler shift, where $c$ is the speed of light. 

This \textit{redshift-distance} relation has recently been coined with a new name, the \textit{Hubble-Lemaître law}, in acknowledgement of Lemaître's theoretical prediction of the expanding Universe. However, the recognition could be extended to the significant contribution of other players. Vesto Melvin Slipher's redshift measurements\footnote{In fact Slipher measured the shift of spectral lines of galaxies, from which it was possible to infer their velocities using the Doppler shift. This led him to the groundbreaking conclusion that the nebulae being measured had to be outside of the Milky Way.} \cite{1913LowOB...2...56S} and Henrietta Leavitt's discovery of the Cepheid period-luminosity relation from variable stars in the Magellanic Clouds \cite{1908AnHar..60...87L} were indispensable to the production of Hubble's famous redshift-distance plot \cite{1931ApJ....74...43H}, while Milton Humason's redshift measurements \cite{1931ApJ....74...43H,1935ApJ....81..187A} and Milne's cosmological considerations\cite{1933ZA......6....1M} in the 1930s were instrumental in providing compelling and precise evidence for the nature of this result.


\section{The Friedmann-Lema\^{i}tre-Robertson-Walker Metric} \label{sec:flrw}

To extract cosmological implications from general relativity, we need to specify the components of the spacetime metric in \cref{eq:metricgr}. There are no general analytical solutions to the Einstein field equations, \cref{eq:eineqdef}, but, amongst others, Einstein realised that a coherent physical description of the Universe could only be achieved under specific assumptions. 
The fundamental underlying assumption is that the Universe is expanding and encapsulates the cosmological principle. That is to say that the metric needs to be such that it describes a time-varying Universe which is, at each time, spatially homogeneous and isotropic. 
Assembling all the geometric and symmetry arguments, the most generic expanding spacetime for a homogeneous and isotropic Universe  with matter uniformly distributed as a perfect fluid is given by the Friedmann-Lemaître-Robertson-Walker (FLRW) metric \cite{1937PLMS...42...90W,1936ApJ....83..257R,1936ApJ....83..187R,1935ApJ....82..284R,1933ASSB...53...51L,1927ASSB...47...49L,1924ZPhy...21..326F,friedmann} and has the following form:
\begin{equation} 
\label{eq:frwsph}
\odif{s}^2 =  - \odif{t}^2 + a^2(t) \gamma_{ij} \odif{x}^i \odif{x}^j  = - \odif{t}^2 + a^2(t) \odif{l}^2 \mathcomma
\end{equation}
where $\gamma_{ij}$ are the spatial components of the metric. The spatial part of the line element, $\odif{l}^2$, can be expressed in traditional Cartesian coordinates or, more informatively, in spherical polar coordinates $(x^1,x^2,x^3) = (r,\theta,\varphi)$ as
\begin{equation}
  \odif{l}^2 = \frac{\odif{r}^2}{1-K r^2/R_0^2} + r^2 \odif{\theta}^2+r^2 \sin^2(\theta) \odif{\varphi}^2 \mathcomma
  \label{eq:frwsph2}
\end{equation}
in which case
\begin{equation}
    \gamma_{11} = \left(1-K r^2/R_0^2 \right)^{-1} \mathcomma \quad \gamma_{22} = r^2 \mathcomma \quad \gamma_{33} = r^2 \sin^2(\theta) \mathperiod
\end{equation}

The scale factor $a(t)$ describes the time evolution of the expansion scale of the Universe. It is dimensionless and typically normalised at present as $a(t_0) = 1$. $r$ represents the radial comoving coordinate, while $\theta$ and $\varphi$ are the comoving angular coordinates. The whole term in \cref{eq:frwsph2} corresponds to the line element of a maximally symmetric 3D space with a constant curvature parameter $K$ and curvature scale $R_0$. The derivation of the FLRW solution can be found in \cite{weinberg,baumann_2022}.
Under the assumption of homogeneity and isotropy, the intrinsic curvature must be constant, and its character is encoded in $K$:
\begin{itemize}
    \item Flat space ($K = 0$): the simplest case corresponding to a three-dimensional Euclidean space. In such a space, parallel lines never intersect.
    \item Positively curved space ($K = +1$): known as spherical space as it can be thought of as a three-sphere embedded in a four-dimensional Euclidean space. This is the space in which all parallel lines will eventually meet.
    \item Negatively curved space ($K = -1$): called the hyperbolic space as it can be represented by a hyperboloid embedded in a four-dimensional Lorentzian space. In this space, parallel lines will always diverge.
\end{itemize}

It is readily apparent that the time and space coordinates are not scaled in the same way in the FLRW metric, \cref{eq:frwsph}, since the scale factor $a(t)$ only multiplies the spatial sector. For that purpose, the cosmic time $t$ may be replaced by the conformal time $\tau$, which is particularly useful for the study of the propagation of photons (like in the CMB), which is defined as
\begin{equation}
    \odif{\tau} \equiv \frac{\odif{t}}{a(t)} \mathperiod
    \label{eq:conftime}
\end{equation}

Moreover, without assuming a particular curvature parameter, it is convenient to redefine the radial component as $\odif{\chi} \equiv \odif{r}/\sqrt{1-K r^2/R_0^2}$ and the metric on the unitary two-sphere,
\begin{equation}
\odif{\Omega}^2 = \odif{\theta}^2+ \sin^2(\theta) \odif{\varphi}^2 \mathcomma
\end{equation}
such that the FLRW metric in \cref{eq:frwsph} becomes
\begin{equation}
 \odif{s}^2 =a^2(\tau) \left[ - \odif[order=2]{\tau} + \left(\odif{\chi}^2 + S_k^2(\chi) \odif{\Omega}^2\right) \right] \mathcomma
 \label{eq:flrwchi}
\end{equation}
where $a(\tau)$ is now the conformal scale factor and 
\begin{equation}
    S_k (\chi) \equiv R_0 
    \begin{cases}
    \chi/R_0\quad &K=0 \mathcomma \\
    \sin (\chi/R_0)\quad &K=+1 \mathcomma \\
        \sinh (\chi/R_0)\quad &K=-1 \mathcomma
    \end{cases}
\end{equation}
expresses the difference between $r$ and $\chi$ for non-flat geometries. In general we will adopt the conventions in which $r$ is normalised to $R_0$, which vanishes from the equations.
Finally, it is worth noting that the FLRW metric encapsulates all the degrees of freedom of the spacetime through a scaling function of time, $a(\tau)$, and the constant $K$. 

\section{Spacetime Geometry} \label{sec:spgeo}



Having introduced the FLRW metric, we can now compute the Einstein tensor by evaluating the left-hand side of \cref{eq:eineqdef}.

\subsection{Connections}

Substituting $g_{\mu \nu} = \diag (-1, a^2 \gamma_{ij} )$ from \cref{eq:frwsph} into the definition in \cref{eq:frwsph}, it is straightforward to compute the Christoffel symbols. All the components with two time indices vanish, that is, $\Gamma^{\mu}_{00} = \Gamma^{0}_{0\nu} = 0$. The non-vanishing components are:
\begin{align}
    \Gamma^{0}_{ij} = a \dot{a} \gamma_{ij} \mathcomma \quad  \Gamma^{i}_{0j} =  \frac{\dot{a}}{a} \delta^{i}_{j} \mathcomma \quad   \Gamma^{i}_{jk} = \frac{1}{2} \gamma^{il} \left(\partial_j \gamma_{kl} + \partial_{k} \gamma_{jl} - \partial_l \gamma_{jk} \right) \mathperiod
\end{align}

according to \cref{eq:gammadef} and the Christoffel symbols are symmetric under the exchange of the lower indices.
In spherical polar coordinates, under the normalisation $R_0=1$ over $r$ and remembering that $H=\dot{a}/a$, this becomes
 \begin{align}
         \Gamma^{0}_{ij} &= a^2 H \gamma_{ij} \mathcomma \quad  \Gamma^{i}_{0j} =  H \delta^{i}_{j} \mathcomma \quad \Gamma^1_{11} = \frac{K r}{1-Kr^2} \mathcomma \quad \Gamma^1_{22} = -r (1-K r^2)\mathcomma  \\
         \Gamma^1_{33} &= -r (1-K r^2) \sin^2\theta \mathcomma \quad \Gamma^2_{12} = \frac{1}{r} \mathcomma \quad \Gamma^2_{33} = - \sin \theta \cos \theta \mathcomma \quad \Gamma^3_{13} = \frac{1}{r} \mathcomma \quad \Gamma^3_{23} = \cot \theta \mathperiod  \nonumber
 \end{align}

\subsection{Ricci Tensor}

The time-space components vanish according to the isotropy of the spacetime: $R_{0i}=R_{i0}=0$. The non-zero components are:
\begin{align}
    R_{00} &= -3 \frac{\ddot{a}}{a} = -3 \left( H^2 +\dot{H} \right) \mathcomma \\
    R_{ij} &= a^2 \left[ \frac{\ddot{a}}{a} + 2 \left( \frac{\dot{a}}{a} \right)^2 + 2 \frac{K}{a^2} \right] \gamma_{ij} = a^2 \left[ 3H^2 +\dot{H} + 2 \frac{K}{a^2} \right] \gamma_{ij}  \mathperiod
\end{align}

\subsection{Ricci Scalar}

The Ricci scalar is computed from the components of the Ricci tensor:
\begin{equation}
    R = 6 \left[ \frac{\ddot{a}}{a} + \left( \frac{\dot{a}}{a} \right)^2 + \frac{K}{a^2}  \right] = 6 \left[ \dot{H} + 2H^2  \right]  \mathperiod
\end{equation}

\subsection{Einstein Tensor}

The non-zero components of the Einstein tensor, \cref{eq:eineqdef}, in the form $G\indices{^{\mu}_{\nu}} \equiv g^{\mu \lambda} G_{\nu \lambda}$ are:
\begin{align}
 G\indices{^0_0} &= -3 \left[ \left( \frac{\dot{a}}{a} \right)^2  + \frac{K}{a^2} \right] =  -3 \left[ H^2  + \frac{K}{a^2} \right] \mathcomma \\
 G\indices{^i_j} &= -\left[2  \frac{\ddot{a}}{a} + \left( \frac{\dot{a}}{a} \right)^2  + \frac{K}{a^2} \right] \delta\indices{^i_j} = -\left[2 \dot{H} + 3H^2  + \frac{K}{a^2} \right] \delta\indices{^i_j} \mathcomma 
\end{align}

where again isotropy ensures that the mixed components vanish.

 \section{The Hot Big Bang Model}


We close with a note on the remarkable observational evidence for what exactly is meant by the \textit{Big Bang} mentioned at the beginning of this chapter.
The advent of General Relativity (GR) sparked a paradigm shift, revolutionising our understanding of the Universe's evolution. Georges Lema\^{i}tre was the first to propose that Einstein's equations yielded a solution describing an expanding Universe that could be followed back in time to a single point \cite{lemaitre}, since $\rho \to \infty$ as $a \to 0$. Subsequently, Edwin Hubble's experimental discovery of the Universe's expansion \cite{1931ApJ....74...43H} culminated in the development of the successful \ac{hbb} model for the observable Universe (for a more detailed description and historical context about the HBB model, we refer the reader to the review in Ref.~\cite{liddleinf}).

The standard HBB model has provided a theoretical basis for explaining various experimental observations \cite{Kolb:1990vq}, namely:
\begin{enumerate}
    \item Predicting the existence and the shape of the spectrum of the cosmic microwave background;
    \item Explaining the mechanism behind the primordial abundances of light elements through nucleosynthesis;
    \item Describing the expansion of the Universe and providing a framework to understand the gravitational collapse of matter, leading to the formation of galaxies and other large-scale structures observed today.
\end{enumerate}

The Hubble parameter $H(t) = \dot{a}/a$ plays a crucial role as a physical observable, describing the expansion rate of an FLRW Universe and setting its characteristic time and distance scales. The Hubble time $t_H = H^{-1}$ and the Hubble radius $d_H = c H^{-1}$ (in units where $c=1$ we have $d_H = t_H$) define the time and length scales of the FLRW spacetime, respectively, and are strictly local quantities (\textit{i.e.} they are defined by the instantaneous expansion rate at a given time $t$). 

   \cleardoublepage


 \chapter{The Standard $\Lambda$CDM Model: Successes and Controversies} \label{chapter:standardmodel}
 \setcounter{equation}{0}
\setcounter{figure}{0}


   \epigraph{E – e não esquecer que a estrutura do átomo não é vista mas sabe-se dela. \\ Sei de muita coisa que não vi. E vós também. Não se pode dar \\ uma prova da existência do que é mais verdadeiro, o jeito é acreditar.\blfootnote{\textit{And — and don’t forget that the structure of the atom cannot be seen but is nonetheless known. I know about lots of things I’ve never seen. And so do you. You can’t show proof of the truest thing of all, all you can do is believe.} --- \textsc{Clarice Lispector} in The Hour of the Star} \\ --- \textsc{Clarice Lispector}\ \small\textup{A Hora da Estrela}}


 \nomenclature{$\Lambda$}{Cosmological Constant}
 \nomenclature{GR}{General Relativity}

In this chapter, we build upon the gravitational framework introduced in \cref{chap:int} to introduce the current standard model of cosmology, the $\Lambda$CDM model, by setting forth its underlying theoretical assumptions and predictions. Assuming the Friedmann-Lemaître-Robertson-Walker background metric as the starting point, we will explore the equations governing the Universe's expansion history in \cref{sec:fund}. Moreover, we examine the linear perturbations produced in the early Universe in \cref{sec:linpert}, which serve as the initial inhomogeneities that are the seeds of structure formation. Bringing everything together, we summarise the most relevant epochs in the Universe's evolution and the physics involved at the relevant scales in \cref{sec:infl,sec:exphis}. We conclude by contemplating the need for the cosmological constant, the simplest form of dark energy, while also discussing the conceptual challenges that arise from its inclusion in \cref{sec:theoprob}. 

\section{Fundamentals of the Background Cosmology} \label{sec:fund}


In the framework of the Hot Big Bang model for the early Universe, tracing back the evolution until the \textit{beginning} of time leads to a problematic point known as a singularity, a fundamental limitation of standard general relativity. However, it is widely expected that there exists a more comprehensive theory that resolves this singularity issue, with the concept of cosmological inflation being a prominent avenue of investigation. While inflation is a significant topic, this chapter primarily focuses on the subsequent cosmic evolution following inflation and the events in the very early Universe, deliberately setting aside the open question of what preceded (or more precisely, what was) the \ac{bb}.

\subsection{Observable Universe}

Let us start by clarifying what is meant by the observable Universe. According to GR, the speed of light $c$ (and therefore the speed of any detectable signal carrying information) is finite. It follows that not every spacetime region is within our reach. Adopting the Big Bang scenario further implies the existence of an intrinsic limit set by the age of the Universe on the maximum distance from which a point in space could have received information. This limit is known as the \textit{particle horizon} (defined as the time integral of $1/a(t)$), and its size keeps evolving. This means that whenever we make considerations about the Universe, these are based on the portion of the Universe that is causally connected to us and which lies within our particle horizon. Anything beyond this region is inaccessible and, more importantly, will not directly impact our observations. 
Nevertheless, two different spacetime regions can be accessible and causally connected to some observer while remaining disconnected from each other. All it takes is for these two regions to be contained inside the observer's particle horizon while their individual particle horizons do not overlap. It is intuitively understandable that any causally connected regions can exhibit homogeneity (without an extremely high level of fine-tuning) if they have had the chance to interact and reach some thermal equilibrium. On the other hand, if both regions are not causally connected, then local interactions cannot justify the observed homogeneity, which must only be related to some \textit{initial} homogeneity of the Universe itself.

Our observed Universe demonstrates a remarkable level of uniformity on scales larger than anticipated, given its age and the problems posed above. This challenge is known as the \textit{horizon problem} \cite{Guth:1980zm} and was first formalised in the 1960s. One possible resolution would be to consider that the Universe is significantly older. However, this would imply substantial deviations from what is estimated by observations by several orders of magnitude \cite{baumann_2022}. Consequently, under the current cosmological model, this uniformity is better explained by the inherent particularities of its initial conditions - in a brief moment at the beginning of the Universe which cosmologists believe happened and refer to as \textit{inflation}, and which will be further discussed in \cref{sec:infl}.

\subsection{Invisible Universe} \label{sec:inv}

One of the most revolutionary proposals of the standard model was the assertion that the majority of matter and energy in the Universe is in some form invisible to our detectors (non-luminous), meaning that it can only be observed through gravitational interactions or other unknown indirect signals. More precisely, the standard matter we can see and interact with amounts to around $5\%$ of the total share. The remaining $95 \%$ are in the form of the mysterious dark sector, split into two components with rather distinct nature and purpose: \textit{dark matter} (DM) and \textit{dark energy} (DE).

The term dark matter was coined by Fritz Zwicky in 1933 when he observed that galaxies in the Coma cluster moved faster than anticipated in order to remain bounded together \cite{1933AcHPh...6..110Z} given the measure of visible mass within them. This implied that either the clusters were flying apart, or they had to be held together by the gravitational attraction of mass outside the luminous parts of the cluster, or otherwise, something had to be wrong with our understanding of physics. To explain the stability of the cluster, he proposed the existence of an additional form of invisible matter. Later in the 1970s, Vera Rubin and colleagues measured the rotation speed of gas in the outer regions of galaxies \cite{1970ApJ...159..379R}, also reporting exceptionally high speeds, with nearly flat rotation curves. These and subsequent observations, based on X-rays, radio emission, \ac{bbn} or the developing studies on \ac{lss}, could be explained by hypothesising halos of some excess non-luminous matter surrounding the galaxies and in clusters of galaxies, further supporting the need for this weakly interacting matter.
Nevertheless, the most robust indirect evidence for dark matter so far comes from the gravitational lensing of the CMB. As the CMB photons travel across the Universe, they are gravitationally deflected by the nearby large-scale structure, creating a pattern of distortions in the hot and cold spots of the CMB. This phenomenon has been mapped in great detail by the \textit{Planck} satellite \cite{Planck:2018lbu}, which confirmed that its magnitude relates directly to the total amount of matter in the Universe. Observations of the abundance of light elements suggest that ordinary baryonic matter exists in relatively less abundance. The exact amount of dark matter is also necessary to explain the measured rate of gravitational clustering since the density fluctuations generated in the early Universe rely on dark matter to grow at such a rapid and efficient pace \cite{deSwart:2017heh,Bertone:2016nfn}. 

In the 1980s, even after the proposal of DM, cosmology was still at a turning point. The extrapolated age for a Universe populated only by matter appeared to be shorter than the ages of the oldest stars within it. Moreover, observations of the large-scale structure were incompatible with the predictions from inflationary cosmology of a quasi-flat Universe, suggesting that only $30\%$ of the critical density (the total energy-density contribution in the Universe to ensure spatial flatness) could be accounted for by the total matter density. As discussed in \cref{sec:grrev}, it was once suggested that Einstein's cosmological constant, introduced to create a static Universe solution, or some other form of \textit{dark energy} (a term that appears to have first been coined by Huterer and Turner \cite{Huterer:1998qv}), might provide a solution to these issues. However, lacking empirical support, this idea remained a theoretical hypothesis. Later came the simultaneous proposal by Ratra and Peebles \cite{PhysRevD.37.3406} and Wetterich \cite{Wetterich:1987fm} in 1988, positing that the cosmological constant may have been gradually approaching its natural value - zero - over an extended period.

However, this question was ultimately settled by the unanticipated discovery of the accelerating character of the Universe by two separate teams: the Supernova Cosmology Project \cite{acel2}, and the High-z Supernova Search Team \cite{acel1}, by studying standard distant supernova events and measuring the distances and recession speeds of galaxies, analogous to Hubble's analysis using local sources. These studies revealed that the expansion rate was decelerating in the early stages when the Universe was dominated by matter and transitioned to an accelerating phase quite recently (in cosmological terms). It turns out that the simplest way to account for the accelerated expansion within the framework of Einstein's theory of gravity is if the Universe has just entered a phase dominated by Einstein's cosmological constant $\Lambda$; more precisely, a source with an energy density that does not dilute with the expansion and which exerts a negative pressure symmetric to its energy density. Remarkably, the cosmological constant hypothesis accommodates the oldest stars in the estimated age of the Universe and fulfils the critical density as required in the inflation paradigm, with detailed studies of the CMB anisotropies map placing the dark energy fractional density to the missing $70\%$ of the total energy in the present-day Universe \cite{Aghanim:2018eyx}.

The addition of these two dark sector components is the \textit{genesis} of the standard model of cosmology, the Lambda-Cold-Dark-Matter ($\Lambda$CDM) model. However, and as will be one of the main focuses of this dissertation, despite addressing the problems above and fitting the data with outstanding accuracy, the $\Lambda$CDM model now faces a new crisis. On the fundamental side, the cosmological constant cannot be reconciled with predictions for the vacuum energy from quantum field theory, with several orders of magnitude of difference between the predicted and observed values. On the other hand, the increase in independent precision probes has brought to light observational incompatibilities between model-dependent measurements from the early Universe and studies within the local Universe. The origin of these inconsistencies remain one of the most significant unresolved problems in cosmology and fundamental physics \cite{Abdalla:2022yfr}.




%
\subsection{Expanding Universe}

The Hubble expansion rate is one of the protagonists of modern cosmology and is defined as:
\begin{equation}
    H \equiv \frac{\dot{a}}{a} \mathcomma
\end{equation}
with the fact that it provides information about the Universe's expansion rate now made explicit. Generally, $H$ is a function of time, and its value measured at present, denoted as $H_0$, is referred to as the Hubble constant or parameter.
As was the case for defining the conformal time, \cref{eq:conftime}, it is crucial to differentiate between the comoving distance $r$ - which is fixed to the expansion and thus \textit{oblivious} to its effect - and the physical distance $r_{\text{phys}}$, which increases proportionally to the expansion. An observer studying an object at a fixed comoving distance $r$ will measure a physical distance to it that depends on time. Therefore, the physical velocity of a point must also account for the added contribution of the expansion and is given by:
\begin{equation}
   r_{\text{phys}} \equiv a(t) r\quad \xrightarrow{\dd/\odif{t}}\quad v_{\text{phys}} \equiv \frac{\odif{r_{\text{phys}}}}{\odif{t}} = a \dot{r} + r \dot{a} \equiv v_{\text{pec}} + H r_{\text{phys}} \mathperiod
   \label{eq:physd}
\end{equation}
The first term, $v_{\text{pec}}$, is the \textit{peculiar velocity}, that is, the velocity of the object relative to the coordinate grid. The second term, $H r_{\text{phys}}$, is the \textit{Hubble flow} resulting from the expansion of the coordinate grid itself. For galaxies moving in the Hubble flow, sufficiently distant for their peculiar velocities to be insignificant in comparison to the velocity arising from the expansion, \cref{eq:physd} simplifies to:
\begin{equation}
    v_{\text{phys}} \simeq H r_{\text{phys}} \mathcomma
\end{equation}
which we have encountered before as the Hubble-Lemaître law, valid for local distances (in comparison to $H$), in which case $H \approx H_0$. In other words, the recession of galaxies away from us happens at speeds proportional to their physical distance and can be solely attributed to the expansion of the Universe. The proportionality constant measures the present rate of expansion, $H_0$. Edwin Hubble's reported observation of his homonym law in 1929 \cite{1931ApJ....74...43H} by plotting $v$ against $r$ leading to an estimated $H_0 = 500\, \text{km/s/Mpc}$, was the onset of evidence of the Universe's expansion and provided strong support for the hot Big Bang model. During the past century, the endeavour of measuring and independently confirming the value of the Hubble constant progressed. At the moment, $H_0$ is typically inferred from Cosmic Microwave Background (CMB) measurements assuming a specific cosmological model or estimated directly in the local Universe, such as the luminosity patterns of Type Ia supernovae, in a model-independent fashion. 



As is the case for the $H_0$ estimates, most observational probes in cosmology rely on detecting light emitted by astrophysical and cosmological sources.
In an expanding Universe, the momentum of particles that transverse it, as perceived by comoving observers, decays with the expansion at the rate $1/a$, resulting in an intrinsic effect in the light we observe. If the light is emitted with a specific wavelength $\lambda$, that wavelength will be stretched by the expansion proportionally to the scale factor $a$. Consequently, a photon emitted at time $t$ is detected on Earth at time $t_0$ with a wavelength given by:
\begin{equation}
    \lambda(t_0) = \frac{a(t_0)}{a(t)} \lambda (t) \mathperiod
    \label{eq:lambat}
\end{equation}

In an expanding Universe, the scale factor increases with time, meaning that $a(t_0) > a(t)$, implying that the wavelength at detection is larger than the wavelength at emission. This leads to the introduction of the redshift parameter\footnote{We had previously defined the redshift as $z=v/c$. This is valid as long as the recessional velocity of an object is much smaller than $c$ and $\lambda_0 \approx (1+v/c)\lambda$ from the Doppler effect.}, defined as the fractional change in the wavelength of a photon emitted at time $t$ and detected today on Earth:
\begin{equation}
    z \equiv \frac{\lambda (t_0) - \lambda(t)}{\lambda(t)} \mathcomma
\end{equation}
and which quantifies how much the wavelength has been stretched.
Employing \cref{eq:lambat}, we find $1 + z = a(t_0)/a(t)$, which together with the conventional normalisation $a(t_0) = 1$, yields:
\begin{equation}
    1+z \equiv \frac{1}{a(t)} \mathperiod
    \label{eq:red}
\end{equation}
Given its direct relation with the time-dependent scale factor, the redshift parameter $z$ is frequently preferred for the time-keeping of specific events during the history of the Universe.


The Hubble parameter is often expressed in terms of the dimensionless quantity $h$:

\begin{equation}
    H_0 = 100 h\, \text{km/s/Mpc} = 2.1332 h \times 10^{-42}\, \text{GeV} \mathcomma
    \label{eq:h0est}
\end{equation}

and can be used to define the average cosmological density in the Universe, also called the \textit{critical density}:

\begin{equation}
    \rho_{\text{crit},0} \equiv \frac{3 H_0^2}{8 \pi G} = 1.88 h^2 \times 10^{-29}\, \text{g}\, \text{cm}^3 \mathperiod
    \label{eq:critdens}
\end{equation}

Lastly, the Hubble rate defines a \textit{Hubble time}

\begin{equation}
    t_H \equiv \frac{1}{H_0} = 9.78 \times 10^9\, h^{-1}\, \text{years} \mathcomma 
\end{equation}

which is a crude measure of the age of the Universe, and the present \textit{Hubble radius}

\begin{equation}
    D_H \equiv \frac{c}{H_0} = 2998 \, h^{-1}\, \text{Mpc} \mathcomma
    \label{eq:horizonh}
\end{equation}

a fair approximation to the largest scales that can currently be observed.
The following two sections will address other observationally meaningful parametrisations of time and distance.

\subsection{Time Parametrisations} \label{sec:times}

 Before moving on to the details of the standard model, it is essential to summarise the various clocks that refer to the notion of time in different epochs and contexts depending on the physical system under consideration. 
The relation between the time $t$ (and equivalently $\tau$) and the redshift $z$ (and equivalently $a$) encoded in \cref{eq:red} is especially relevant when dealing with relativistic components due to the direct redshift-dependent correspondence between energy at emission and detection, encoding a valuable measure of energy loss over time which would not be evident for the proper time. 
Thus, $t$, $\tau$, $a$, and $z$ are all relevant and distinctively informative definitions of time. 

However, in the very early Universe, when particle physics processes play a crucial role, the concept of temperature $T$ constitutes a trigger for different events. For photons with energy $E$, there is a direct correspondence between temperature and redshift since $1+z \propto E \propto T$ (normalised to the temperature of the Universe today: $T_0 = 2.7255\, \text{K} \simeq 2.348 \times 10^{-4}\, \text{eV}$).

Finally, and as will be extensively discussed in the following sections, the tiny energy fluctuations in the early-Universe evolve according to a wave-like behaviour and can be decomposed into their Fourier modes $k$, related to their wavelength $\lambda$ as $k = 2 \pi a/\lambda$. Assuming the \textit{separate Universe} approximation, \textit{i.e.} that each mode evolves independently, $k$ becomes a proxy for the scales of a given event. At early times, when the Universe was smaller, only modes with small wavelengths contributed significantly to the evolution, as those with much larger wavelengths were minimally affected by processes occurring at small scales. The Fourier modes convey an idea of the size of the Universe, which, in turn, is related to the time elapsed since the beginning of its expansion. 

Therefore, all these quantities will be used interchangeably throughout the text depending on the context and the cosmological epoch.

 \subsection{Cosmological Distances}



In the same way, as for the time parametrisations considered above, different rulers can be used to measure distances, depending on which quantity is being used as a proxy of distance. Much of the evidence for dark energy comes from measurements of cosmological distances.
In order to understand how present observations relate to other processes happening in the Universe, it is essential to understand photon propagation.
Since photons have no mass, they traverse null geodesics:
\begin{equation}
    \odif{s}^2 = -  \odif{t}^2 + a^2(t)  \odif{\chi}^2 = 0 \ \ \Leftrightarrow\ \ \odif{\chi}^2 = a^{-2}(t)  \odif{t}^2 \mathcomma
\end{equation}
with the radial component $\chi$ as the direction of the photons' propagation. The last expression can be integrated to find the \textit{line-of-sight comoving distance} $d_c$ travelled by the photon in some time interval $ [ t_i, t_f ] $:
\begin{equation} 
    d_c (t) = \chi (t) = \int_{t_f}^{t_i} \frac{c\odif{t}}{a(t)}\quad \xrightarrow{\odif{t}=\odif{a}/(aH)=-\odif{z} a/H}\quad d_c (z) = \int_{z_i}^{z_f} \frac{c\odif{z}}{H(z)} \mathperiod
    \label{eq:dc}
\end{equation}
%

Another important observable in cosmology is the \textit{absolute luminosity} of astrophysical objects $L$, defined in terms of the \textit{observed flux} $F$, over a sphere of radius $\chi$ in an expanding spacetime. This sphere has an area 
%
\begin{equation}
    4 \pi a^2(t_0) d_M^2\quad \text{where}\quad d_M \equiv S_k(\chi) \mathcomma
\end{equation}
 and is called the \textit{metric or transverse comoving distance}, with $S_k (\chi)$ as defined in \cref{eq:flrwchi}, which reflects the fact that in a curved spacetime $d_M \neq \chi$. $d_c$ in \cref{eq:dc} is often used when referring to the spatially flat geometry, in which case both quantities are equivalent.
The \textit{luminosity distance} is then expressed in terms of $F$ as,
\begin{equation}
    d_L^2 = \frac{L}{4 \pi F} \quad \xrightarrow{\text{expanding space}}\quad d_L(z) = (1+z) d_M (z)  \mathcomma
    \label{eq:lumdist}
\end{equation}
%
where the multiplicative redshift factor accounts for the energy loss experienced by the photons and the reduction in their number due to cosmic expansion since

\begin{equation}
    \frac{L_i}{L_f} = \frac{\Delta E_i}{\Delta E_f} \frac{\Delta t_f}{\Delta t_i} = (1+z)^2 \mathperiod
\end{equation}

This will be important in \cref{chap:obs} where we will discuss how distance measurements can be used to infer the value of cosmological parameters such as $H_0$. 

For objects of some physical size $\Delta x$ that are observed to subtend an angle in the sky $\Delta \theta$ (perpendicular to the line of sight and $\Delta \theta \ll 1$ in radians for cosmological distances), it is more convenient to define the \textit{angular diameter distance}:
\begin{equation}
 d_A = \frac{\Delta x}{\Delta \theta} \quad \text{and} \quad \Delta x = a(t) S_k(\chi) \Delta \theta  \quad \xrightarrow{\text{expanding space}}\quad d_A(z) = d_M (z)/(1+z)  \mathcomma
\end{equation}
where we have used geometrical considerations assuming $\Delta \theta \ll 1$, and the dividing redshift factor is now included to reflect the Universe's expansion since the light was emitted until it is observed in the sky.
In summary, we have defined three different distance observables which are related as:
\begin{equation}
   d_M(z) = (1+z) d_A (z),\quad \text{and}\  \quad d_L (z) = (1+z)^{2} d_A (z) \mathperiod
   \label{eq:distz}
\end{equation}
The last expression is the distance-duality or Etherington relation \cite{Etherington}, which holds for general metrics under photon number conservation.



\subsection{Friedmann Equations}


Assuming perfect fluids coupled to Einstein's field equations, \cref{eq:eineqdef}, we can gather all the ingredients and derive the \textit{Friedmann equations} that describe the evolution of an FLRW Universe:
\begin{align}
    \left( \frac{\dot{a}}{a} \right)^2 = H^2 &= \frac{8 \pi G}{3} \rho - \frac{K}{a^2} \mathcomma \label{eq:fri1} \\
     \frac{\ddot{a}}{a} = \dot{H} + H^2 &= -\frac{4 \pi G}{3} (\rho + 3p) \mathcomma \label{eq:fri2}
\end{align}
where $\rho$ and $p$ are the generic contributions of all the fluids in the Universe, remembering that the constant $K$ represents the spatial curvature of the Universe.
It is worth remarking that \cref{eq:fri2} is also known as the \textit{Raychaudhuri equation} and is a curvature-independent measure of the expansion, as it does not depend directly on $K$.

\subsection{Cosmic Ingredients} \label{sec:ing}

The cosmological principle dictates that the background energy-momentum tensor resembles that of a perfect fluid \cite{baumann_2022}, expressed as 
\begin{equation}
    T_{\mu \nu} = (\rho + p) u_{\mu} u_{\nu} + p g_{\mu \nu} \mathcomma
\end{equation}
where $\rho$ and $p$ represent the energy density and pressure in the rest frame of the fluid, respectively, and $u^{\mu}$ is the four-velocity relative to a comoving observer, under the constraint $u^{\mu} u_{\mu} = -1$, which we take to be
\begin{equation}
    u_{\mu} = (a,\vec{0})\quad \text{and}\quad  u^{\mu} = (a^{-1},\vec{0}) \mathcomma
\end{equation}
for all cosmological components. 
The Bianchi identities, \cref{eq:grcons}, imply conservation of energy from the time component, leading to the conservation or continuity equation:
\begin{equation}
\covd^{\mu} T_{\mu \nu} =  0 \quad  \xrightarrow{\nu =0} \quad \dot{\rho} + 3H (\rho + p) = 0 \mathperiod
    \label{eq:fricon}
\end{equation}

The second Friedmann equation, introduced in \cref{eq:fri2}, is equivalent to taking the first Friedmann equation, \cref{eq:fri1}, together with the continuity equation, \cref{eq:fricon}.

Assuming a Universe dominated by a single species with an equation of state $p = w \rho$ with constant $w$ (which will no longer hold, for instance, for dynamical dark energy), \cref{eq:fricon} can be integrated to obtain 
\begin{equation}
 a \propto (t-t_i)^{\frac{2}{3(1+w)}} \quad \text{and}\quad  \rho \propto a^{-3(1+w)} \mathcomma
 \label{eq:frifl}
\end{equation}
where the subscript denotes the initial time. This result embodies how different cosmological fluids are diluted by cosmic expansion, which we list below for the three ingredients that play essential roles in the evolution of the Universe. It also shows how spatial curvature contributes to the first Friedmann equation, \cref{eq:fri1}, as a fluid with $\rho \propto a^{-2}$.

The critical energy density for a flat spatial geometry ($K=0$), defined in \cref{eq:critdens} at present, can be expressed as a time-evolving function as
\begin{equation}
    \rho_{\text{crit}}  = \frac{3 H^2}{8 \pi G}\quad \text{with}\quad \rho_{\text{crit},0} \approx 5 \times 10^{-6}\, \text{GeV}\, \text{cm}^{-3} \mathcomma
\end{equation}
where the label $0$ denotes present time, which we often omit when discussing the current matter distributions, and it is implied from the context. This can be used to express the relative energy density of the fluid components $f$ in a dimensionless manner:
\begin{equation}
    \Omega_{f} = \frac{\rho_{f}}{\rho_{crit}}\quad \text{for}\quad f=m,r,\Lambda,... \mathcomma
\end{equation}
where the subscripts $(m,r)$ denote the collective matter and radiation components, whose evolution will be detailed below, and $\Lambda$ gives the contribution from the cosmological constant. 
Defining $\Omega_K \equiv -K/(a H_0)^2$ (in which case $\Omega_K <0$ actually corresponds to $K>0$ and \textit{vice versa}), the Friedmann equation, \cref{eq:fri1}, can be rewritten as the \textit{Friedmann constraint} for the curvature density. This can be expressed according to the measured relative energy densities at present time:
\begin{equation}
    \Omega_0 \equiv \Omega_m^0 + \Omega_r^0 + \Omega_{\Lambda}^0 \quad \xrightarrow{a(t_0) \equiv 1}\quad \Omega^0_K = 1- \Omega_0 \mathperiod
    \label{eq:omega0}
\end{equation}

If the Universe's expansion is decelerating (\textit{i.e.}, $\ddot{a} < 0$), then the absolute value of the curvature term $|\Omega_K|$ will continue to rise. This is except in the particular case where the Universe has been precisely flat ($K = 0$) from its early stages. According to the Planck 2018 data \cite{Aghanim:2018eyx}, this quantity is contrained to be $\Omega_K^0 = -0.044^{+0.018}_{-0.015}$ at present at $68\%$ \ac{cl}. For $|\Omega_K|$ to decrease during the Universe's evolution, a phase of cosmic acceleration ($\ddot{a} > 0$) is necessary, or otherwise, the Universe must have emerged from an initial state extremely close to flat. To account for the current observed level of the Universe's flatness, a phase of cosmic inflation is needed before the radiation-dominated era, during which the scale factor must have increased by a factor greater than $\expe^{60}$ \cite{liddleinf}.

\begin{figure}[h]
      \subfloat{\includegraphics[width=\linewidth]{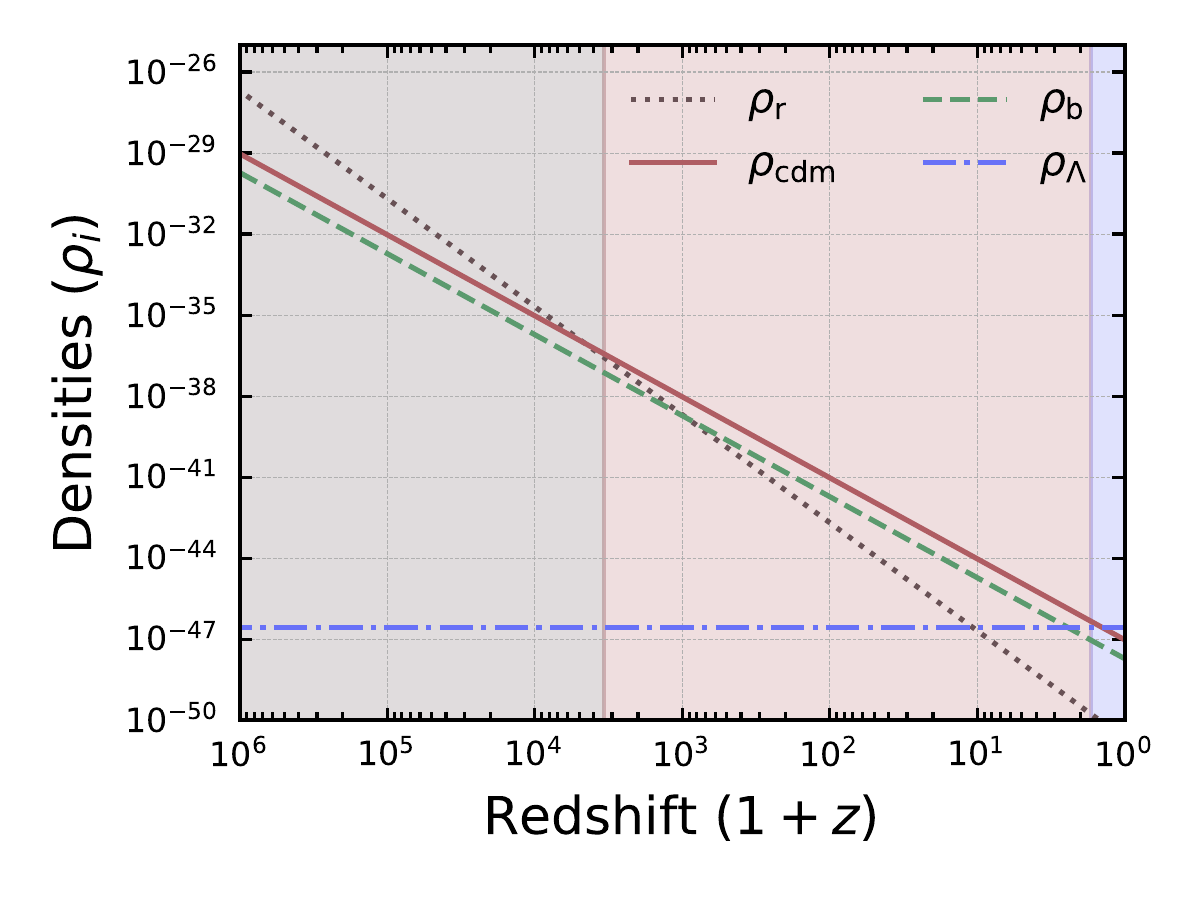}}
  \caption[The redshift evolution of the energy densities of the components of the Universe]{\label{fig:rhos} Evolution of the energy densities $\rho$ of radiation $r$ (grey dotted line), baryons $b$ (green dashed line), cold dark matter $\text{cdm}$ (red filled line), and the cosmological constant $\Lambda$ (blue dot-dashed line), as a function of redshift $z$ in the $\Lambda$CDM model.}
\end{figure}

Next, according to \cref{eq:frifl}, we list the Universe's constituents by their properties, deriving the corresponding evolution for the approximated single-component case. The evolution of each comic fluid with redshift is depicted in \cref{fig:rhos}.

\subsubsection{Matter}

The broad heading \textit{matter} refers to fluids with negligible pressure compared to the energy density (non-relativistic particles), such that $p \ll \rho\ \Rightarrow\ w \approx 0$, resulting in the dilution rate
    \begin{equation}
        a \propto (t-t_i)^{2/3}\quad \text{and}\quad \rho_m \propto a^{-3} \propto t^{-2} \quad \xrightarrow{H=\dot{a}/a} \quad H = \frac{2}{3t} \mathperiod
        \label{eq:amat}
    \end{equation}
 One way to interpret this result is to imagine that the volume of a given region of three-dimensional space increases as $V \propto a^{3}$. In contrast, the number of particles (and therefore the energy in that region) remains constant. The relevant non-relativistic matter components for the Universe's evolution are:
\begin{itemize}
    \item Baryons: refers to ordinary matter, including the less massive electrons (fermions), thoroughly comprehended by physicists as an integral part of the standard model of particle physics.

    \item  Dark Matter: approximately five times more abundant than baryons, the exact characteristics of dark matter remain a mystery, as was discussed in \cref{sec:inv}. 
    In the standard model, it is cold dark matter (\ac{cdm}). Other potential candidates include \ac{wimps}, which have yet to be detected. 
\end{itemize}

\subsubsection{Radiation}

Radiation is a gas composed of relativistic particles characterised by $p = \rho/3\ \Rightarrow\ w = 1/3$ according to statistical mechanics. As derived from \cref{eq:frifl}, this results in
\begin{equation}
     a \propto (t-t_i)^{1/2}\quad \text{and}\quad \rho_r \propto a^{-4} \propto t^{-2} \quad \xrightarrow{H=\dot{a}/a} \quad H = \frac{1}{2t} \mathperiod
    \label{eq:arad}
\end{equation}

The energy redshift of relativistic particles, expressed as $E \propto a^{-1}$, adds to the dilution due to the expansion.



The non-relativistic components are

\begin{itemize}
    \item Photons: relativistic massless particles. Photons were the dominant contribution during nucleosynthesis (formation of the first atoms) and are now detected as the cosmic microwave background.
    
    \item Light particles: In the early Universe, all the particles in the standard model had a radiation character due to the high kinetic energy relative to their mass. They behaved like matter once the temperature dropped enough for their masses to play a role.

    \item  Neutrinos: contributed to the early Universe radiation domination, and only recently have their small masses become relevant and started to act like matter.
\end{itemize}

At present, both photons and neutrinos only exist in minimal quantities. 

\subsubsection{Dark Energy}

In order to have cosmic acceleration, The Friedmann equations require $\ddot{a} > 0$ in \cref{eq:fri2}, which implies $p < -\frac{\rho}{3}$, or equivalently, $w < -\frac{1}{3}$. 
When $w = -1$ and $p = -\rho$, the relation in \cref{eq:frifl} only holds if $\rho$ is a constant. This is the case for the cosmological constant, which (in a flat Universe) implies that $H$ is also constant, in which case the scale factor evolves exponentially:

\begin{equation}
  a \propto \expe^{H_{\Lambda} t}\quad \text{and}\quad \rho_{\Lambda} = \frac{\Lambda}{4 \pi G} \propto a^0 \quad \xrightarrow{H=\dot{a}/a} \quad H = \frac{\sqrt{8\pi G}}{3} \rho_{\Lambda} \mathperiod
    \label{eq:acc}
\end{equation}

The cosmological constant cannot be responsible for inflation in the early Universe, as that would mean the accelerated expansion would never cease, contrary to the current cosmic acceleration, which may continue indefinitely.



The cosmological constant proposal can be generalised under the broader category of dark energy, referring to any fluid either with a constant $w_{\text{DE}} \approx -1$, or a time-varying equation of state parameter, $w_{\text{DE}}(t)$, in which case \cref{eq:frifl} no longer holds.


Bringing all the ingredients together for the $\Lambda$CDM model, the first Friedmann equation, \cref{eq:fri1}, can be decomposed into
\begin{equation}
  E^2(z) = \frac{H^2}{H_0^2} = \Omega_r^0 a^{-4} + \Omega_m^0 a^{-3}  + \Omega_K^0 a^{-2} + \Omega_{\Lambda}^0 \mathcomma
\end{equation}

where $E(z)$ is the normalised Hubble rate.
%



 \section{Linear Cosmological Perturbations} \label{sec:linpert}



 Up to this point, our treatment of the Universe assumed perfect homogeneity. However, to comprehend the Universe's large-scale structure, it becomes necessary to introduce inhomogeneities and study their evolution. As long as these perturbations remain relatively small, we can analyse them using linear perturbation theory. 

In practice, the quantities that can be determined observationally for a given model are not typically the exact values of the perturbation variables, subject to quantum fluctuations, but rather realisations of expectation values. 
However, this methodology has an inherent problem since we cannot precisely measure expectation values. The usual statistical considerations fail because we only have access to \textit{one Universe}, one sample of the stochastic process that generates these fluctuations. Therefore, when attempting to determine the mean square fluctuation on a given scale $\lambda$, the best approach is to average over multiple disjoint patches of size $\lambda$ and take this spatial averaging as an ensemble averaging based on an \textit{ergodic hypothesis} analogue. If the scale $\lambda$ is much smaller than the Hubble horizon $1/H$ (our observational limit), then we can compute averages over many independent volumes, and this approximation works well. However, as we approach scales of the order of the horizon, $\lambda \approx \mathcal{O} (1/H_0)$, the statistical approximation breaks down due to \textit{cosmic variance}.


The perturbations are included in the Einstein equations by taking slight deviations - denoted by $\delta$ - to the background quantities - denoted by an upper bar - in the metric (the Einstein tensor) and the matter distribution (the energy-momentum tensor):
\begin{align}
    g_{\mu \nu} (t,\vec{x}) &= \bar{g}_{\mu \nu} + \delta g_{\mu \nu} (t,\vec{x}) \mathcomma \\
     T_{\mu \nu} (t,\vec{x}) &= \bar{T}_{\mu \nu} + \delta T_{\mu \nu} (t,\vec{x}) \mathperiod
\end{align}
We can then expand the Einstein and conservation equations to linear order in the perturbations. For simplicity, we will focus on flat spacetimes only. The same derivation, including the terms dependent on the curvature scalar, can be found \textit{e.g.} in \cite{liddleinf,Kolb:1990vq,peacock_1998}.

For the study of the dynamics of the linear perturbations, it is convenient to define each relevant quantity in terms of its Fourier transform, where $k=2\pi/\lambda$ are the Fourier modes. Taking any perturbed quantity $f (\tau,\vec{x})$, the Fourier transform is
\begin{equation}
f (\tau,\vec{k}) =  \int \expe^{-i \vec{k} \cdot \vec{x}} f (\vec{x}) \odif{\vec{x}}^3\mathcomma
\end{equation}
while the inverse reads
\begin{equation}
f (\tau,\vec{x}) = \frac{1}{(2 \pi)^{3/2}} \int \expe^{-i \vec{k} \cdot \vec{x}} f (\vec{k}) \odif{\vec{k}}^3\mathcomma
\end{equation}
where $f (\vec{k})$ are the Fourier components of $f (\tau,\vec{x})$ and cosmic time has been replaced by the conformal time $\tau$, related by $\odif{t} = a \odif{\tau}$ and $\dot{(\cdot)} \mapsto (\cdot)'/a$. This implies the mapping $\nabla^2 \rightarrow -k^2$ when the spatial derivatives are applied to the relevant quantities.

The general perturbed metric containing only scalar perturbations\footnote{In general, one can perform a scalar-vector-tensor decomposition of the metric perturbations, such that the Einstein equations for scalars, vectors and tensors are independent at linear order and can be treated separately. We will only deal with scalar fluctuations and the corresponding density perturbations, so vectors and tensors may be disregarded at linear order. Vector perturbations are not produced by standard inflation but would, in any case, decay quickly over the expansion. Tensor perturbations are an important test of inflation but not significant for the scales considered in this work.} is subject to a non-unique choice of coordinates or \textit{gauge choice}. Hence, different choices of coordinates can introduce significant changes to the values of perturbed variables. Some non-physical perturbations may arise associated with that choice, the so-called \textit{gauge modes}, which can be eliminated with more convenient coordinates. Equivalently, a physical perturbation can be dropped, for instance, in comoving coordinates, but these will never truly vanish; they will instead be incorporated into additional perturbations in the metric.
 Therefore, a more physical approach to perturbation theory is one which remains invariant under changes in coordinates, allowing for a more robust and reliable study of the true perturbed cosmological degrees of freedom in cosmology. This can be achieved via the so-called Bardeen variables \cite{Bardeen:1980kt}, which encode the \textit{true spacetime perturbations} that cannot be erased by a \textit{gauge transformation}. More details on gauge invariance and transformations can be found, for example, in \cite{Mukhanov:1990me,Malik:2008im, Riotto:2002yw,Ma:1995ey}.
%
It is often convenient to fix the gauge depending on the scenario in question, at the cost of carefully keeping track of all the perturbations in the metric and the matter (there are no free lunches!). For the cosmological scenarios this dissertation considers, we will focus on two common gauge choices, primarily for consistency and cross-checking reasons: the conformal Newtonian and synchronous gauges.

 \subsection{Newtonian Gauge} \label{sec:newtgauge}

For this dissertation, we will only be interested in cosmological scenarios well-described by the FLRW background metric $\bar{g}_{\mu \nu}$, as defined in \cref{eq:flrwchi}. In the context of dealing with the Einstein and Friedmann equations, the scalar metric perturbations are most simply expressed in the \textit{Newtonian or longitudinal gauge}: 
 \begin{equation} \label{eq:pmetric}
    \odif{s}^2 = a^2 (\tau) \left[ - (1 + 2\Psi) \odif[order=2]{\tau} + (1-2\Phi) \delta_{ij} \odif{x}^i \odif{x}^j \right] \mathperiod
 \end{equation}
%
The metric perturbations $\Phi$ and $\Psi$ are also called the Bardeen potentials \cite{Bardeen:1980kt}, as they are gauge invariant and hence of physical interest. The Newtonian gauge is usually preferred for studying structure formation exactly because of its closest resemblance to the Newtonian analysis, with $\Phi$ playing the role of the gravitational potential.

\subsubsection{Metric Perturbations}

This lays out the metric of the spacetime as
\begin{equation}
    g_{\mu \nu} = a^2 \begin{pmatrix}
        -(1+2 \Psi) & 0 \\
        0 & (1-2 \Phi) \delta_{ij}
    \end{pmatrix} \mathcomma
\end{equation}
from which we can compute the evolution equations for the metric perturbations encoded in the perturbed Einstein equations.
Namely, starting from the perturbed line element, \cref{eq:pmetric}, we can compute the inverse metric and, through the definition of the connections, further calculate the perturbations of more complex objects, such as the Ricci tensor, $\delta R_{\mu \nu}$, and the Ricci scalar, $\delta R$. These are needed to write the linear order Einstein equations in terms of the perturbed Einstein tensor $\delta G_{\mu \nu}$:
\begin{equation} \label{eq:pertein}
\delta G_{\mu \nu} = \delta R_{\mu \nu} - \frac{1}{2} \left( \delta R g_{\mu \nu} + R \delta g_{\mu \nu} \right)  = 8 \pi G \delta T_{\mu \nu}\mathperiod
\end{equation}
%
For the sake of completion, in what follows, we list the main results for the perturbed quantities relevant to the Einstein equations, considering only the linear terms. Further details and a thorough derivation can be found \textit{e.g.} in Refs.~\cite{baumann_2022,gorbunov2011introduction, Riotto:2002yw}.


\subsubsection{Inverse Metric}

\begin{equation}
    g^{\mu \nu} = a^{-2} \begin{pmatrix}
        - 1+2 \Psi & 0 \\
        0 & (1 + 2 \Phi) \delta^{ij} 
    \end{pmatrix} \mathcomma
\end{equation}

\subsubsection{Connections}

\begin{align} 
   \delta \Gamma^{0}_{00} &= \Psi' \mathcomma\quad \delta \Gamma^{0}_{i0} = \partial_i \Psi \mathcomma\quad \delta \Gamma^{0}_{ij} = - \left[ \Phi' + 2 \mathcal{H} (\Phi + \Psi) \right] \delta_{ij} \mathcomma \label{eq:newtgam1} \\
    \delta \Gamma^{i}_{00} &= \partial^i \Psi \mathcomma \quad \delta \Gamma^{i}_{j0} =  - \Phi' \delta^i_j \mathcomma \quad \delta \Gamma^{i}_{jk} = -2 \delta^i_{(j} \partial_{k)} \Phi + \delta_{jk} \partial^i \Phi \mathperiod \label{eq:newtgam6}
\end{align}

\subsubsection{Ricci Tensor and Ricci Scalar}


The perturbed Ricci tensor is calculated from the connections as:

\begin{equation}
    \delta R_{\mu \nu} = \partial_{\alpha} \delta \Gamma^{\alpha}_{\mu \nu} - \partial_{\nu} \delta \Gamma^{\alpha}_{\mu \alpha} + \delta \Gamma^{\alpha}_{\mu \nu} \Gamma^{\beta}_{\alpha \beta} + \Gamma^{\alpha}_{\mu \nu} \delta \Gamma^{\beta}_{\alpha \beta} - \delta \Gamma^{\alpha}_{\mu \beta} \Gamma^{\beta}_{\alpha \nu} - \Gamma^{\alpha}_{\mu \beta} \delta \Gamma^{\beta}_{\alpha \nu} \mathperiod
\end{equation}
    
The Ricci Scalar follows from contractions of the Ricci tensor and the metric, yielding:

\begin{equation}
   \delta R = \delta g^{\mu \alpha} R_{\alpha \mu} + g^{\mu \alpha} \delta R_{\alpha \mu} \mathperiod
\end{equation}
%

%


\subsubsection{Einstein Tensor}


We now have all the ingredients to write explicitly the different components of the perturbed Einstein tensor in \cref{eq:pertein}:

\begin{equation}
  \delta  G_{\mu \nu} = \delta R_{\mu \nu} - \frac{1}{2} \delta g_{\mu \nu} R - \frac{1}{2} g_{\mu \nu} \delta R \mathcomma 
\end{equation}
%

or, equivalently, with one of the indices raised to facilitate the comparison with the matter sector:

\begin{equation}
  \delta  G\indices{^\mu _\nu} = \delta g^{\mu \alpha} G_{\alpha \nu} + g^{\mu \alpha} \delta G_{\alpha \nu} \mathperiod
\end{equation}

Writing the perturbed terms explicitly, and according to the perturbed metric in \cref{eq:pmetric}, we arrive at the different components of the perturbed Einstein tensor:

\begin{align}
      \delta  G\indices{^0_0} &= 2 a^{-2} \left[ 3 \hub (\hub \Psi + \Phi') - \nabla^2 \Phi \right] \mathcomma \\
      \delta  G\indices{^0_i} &= - 2 a^{-2} \partial_i \left( \Phi' + \hub \Psi \right) \mathcomma \\
      \delta  G\indices{^i_j} &= 2 a^{-2} \left[ (\hub^2 + 2 \hub') \Psi + \hub \Psi' + \Phi'' + 2 \hub \Phi' \right] \delta^i_j +  a^{-2} \left[ \nabla^2 (\Psi - \Phi) \delta^i_j + \partial^i \partial_j (\Phi - \Psi)\right] \mathperiod
\end{align}

\subsubsection{Matter Perturbations} \label{sec:matpert}

For the purpose of this study, we assume that the background perfect fluids remain perfect fluids when perturbed, in which case there is no anisotropic stress. Matter perturbations, that is, perturbations to the energy-momentum tensor of each species, can first be defined in terms of a perfect fluid:
\begin{align}
    T\indices{^0_0} &= -  \delta \rho \mathcomma \\
    T\indices{^0_i} &=  (1+ w) \bar{\rho} v^i \mathcomma \\
    T\indices{^i_j} &= \delta p \delta^i_j  \mathcomma 
\end{align}
where $v^i$ is the so-called scalar bulk velocity, the matter peculiar velocity with respect to the expansion, $v^i = \odv{x^i}{\tau}$. The different terms in the energy-momentum tensor from each species $f$ need to be added together to make up the perturbation of the total contribution $T_{\mu \nu} = \sum_f T_{\mu \nu}^f$:
\begin{equation}
    \delta \rho = \sum_f \delta \rho_f\mathcomma \ \ \delta p = \sum_f \delta \rho_f \mathperiod
\end{equation}
The density and velocity perturbations are often expressed in terms of the dimensionless density contrast and the velocity divergence

\begin{align}
    \delta &\equiv \delta \rho / \bar{\rho} \mathcomma    \label{eq:denscont} \\
    \theta &\equiv \partial_i v^i \mathperiod
\end{align}

Once $\delta$ becomes of the order of unity, linear perturbation theory no longer applies.

The perturbed four-velocity takes the form
\begin{equation}
u^{\mu} \equiv \bar{u}^{\mu} + \delta u^{\mu} = a^{-1} (1-\Psi, v^i) \mathcomma \ \ u_{\mu} \equiv \bar{u}_{\mu} + \delta u_{\mu} = a(-(1+\Psi), v_i ) \mathperiod 
\end{equation}
This leads to the perturbed energy-momentum tensor of a perfect fluid:
\begin{equation}
    \delta T\indices{^\mu _\nu} = \rho \left[ \delta (1 + c_s^2) \bar{u}^{\mu} \bar{u}_{\nu} + (1+w) (\delta u^{\mu} \bar{u}_{\nu} + \bar{u}^{\mu} \delta u_{\nu}) + \delta c_s^2 \delta^{\mu}_{\nu} \right] \mathperiod
\end{equation}

This equation introduces the sound velocity, $c_s^2 \equiv \delta p/\delta \rho$. For the barotropic fluids, $p$ is only a function of $\rho$, even after being perturbed, implying

\begin{equation}
    c_s^2 \equiv \frac{\delta p}{\delta \rho} = \odv{p}{\rho} = \frac{\dot{p}}{\dot{\rho}} \mathcomma
\end{equation}

where the last step only holds for the FLRW metric and represents the adiabatic sound speed, discarding any other possible internal degrees of freedom of the fluid.




\subsubsection{Einstein Equations} \label{sec:peinst}


Bringing together the expressions for the perturbed Einstein and energy-momentum tensors gives rise to the perturbed Einstein equations.

The time-time component is

\begin{equation}
    \delta G\indices{^0_0} = -2a^{-2} \left[ \nabla^2 \Phi - 3 \hub (\Phi' + \hub \Psi) \right] \mathcomma
\end{equation}
from which it follows that
\begin{equation} \label{eq:pein1}
    \nabla^2 \Phi - 3 \hub (\Phi' + \hub \Psi) = 4 \pi G a^2 \delta \rho \ \  \xrightarrow{\text{F.T.}}\ \   k^2 \Phi + 3 \mathcal{H} \left( \Phi'+ \mathcal{H} \Phi \right) = -4 \pi G a^2 \delta \rho \mathcomma
\end{equation}
in Fourier space. The second derivative of the potential hints at the fact that this is a relativistic generalisation of the Poisson equation. Inside the Hubble radius ($k \gg \hub$), the second term of the left-hand side becomes negligible, and \cref{eq:pein1} reduces to the Poisson equation in the Newtonian limit: $\nabla^2 \Phi\approx 4 \pi G a^2 \delta \rho$. Only on scales comparable to the Hubble radius do the GR corrections start to play an important role.

For the trace-free part of the spatial components, we get the following:
%
%
\begin{equation} \label{eq:pein2}
 \nabla^2 \left( \Phi - \Psi \right) = 0 \ \  \xrightarrow{\text{F.T.}}\ \ k^2 \left( \Phi - \Psi \right) = 0 \mathcomma
\end{equation}
and the approximation $\Phi \approx \Psi$ holds under the assumption of vanishing anisotropic stress considered for the scope of this work, such that:
\begin{equation}
  \partial_i(2 \Phi' + \hub \Psi) = 4 a^2 \pi G (\bar{\rho}+ \bar{p}) v^i \ \ \xrightarrow{\text{F.T.}} \ \  k^2 \left( \Phi' + \mathcal{H} \Psi \right) = 4 \pi G a^2(\bar{\rho} + \bar{p} ) \theta \mathcomma
\end{equation}
where $(\bar{\rho} + \bar{p} ) \theta \equiv \sum_f \rho_f \left( 1 + w_f \right) \theta_f$.

Finally, the trace of the space-space Einstein equation translates into the following evolution equation for the metric potential

\begin{equation} \label{eq:pein3}
 \Phi'' + \frac{1}{3} \nabla^2 (\Psi - \Phi) +  \mathcal{H} ( \Psi' +  2  \Phi') + \Psi \left( \mathcal{H}^2 + 2 \mathcal{H}' \right) = 4 \pi G a^2\delta p \mathcomma
\end{equation}
and $\delta p \equiv  \sum_f \rho_f \delta_f c_{s, f}^2$ is the total pressure perturbation. Assuming the approximation $\Psi \approx \Phi$, \cref{eq:pein2} becomes:
\begin{equation}
 \text{\cref{eq:pein3}} \ \ \xrightarrow{\text{F.T.}} \ \  \Phi'' + 3 \mathcal{H}  \Phi' + \Phi \left( \mathcal{H}^2 + 2 \mathcal{H}' \right) = 4 \pi G a^2\delta p \mathperiod
\end{equation}
This gives a closed equation for the evolution of $\Phi$ in terms of $\delta p$, and the relation between $\delta \rho$ and $\Phi$ is encoded in the Poisson-like equation, \cref{eq:pein1}. Combined with the conservation equations we derive next, this leads to a closed system of linearised equations.


Altogether, the general linearly perturbed Einstein equations in Fourier space are
\begin{align}
k^2 \Phi + 3 \mathcal{H} \left( \Phi'+ \mathcal{H} \Psi \right) &= -4 \pi G a^2 \rho \mathcomma
\label{eq:peq1} \\
k^2 \left( \Phi' + \mathcal{H} \Psi \right) &= 4 \pi G a^2 \rho \left( 1 + w \right) \theta \mathcomma
\label{eq:peq2} \\
\Phi'' + 2 \mathcal{H}  \Phi' + \hub \Psi' +  \left( \mathcal{H}^2 + 2 \mathcal{H}' \right) \Psi &= 4 \pi G a^2 \delta \rho c_{s}^2 \mathcomma \\
\Phi &\approx \Phi \mathcomma
\label{eq:peq3}
\end{align}
relating the metric perturbation $\Phi$ to the perturbed matter content. For simplicity, we have omitted the sums over each fluid $f$, running over the matter species for a generic equation of state, $w_f =\bar{p}_f/\bar{\rho}_f$, and sound speed, $c_{s,f}^2 = \delta p_f / \delta \rho_f$. 

\subsubsection{Conservation Equations} \label{sec:pertcons}

Additionally, the perturbed evolution equations for each component are derived from the corresponding conservation relations for the energy-momentum tensor:
\begin{equation} \label{eq:contpert}
 \nabla_{\mu} T\indices{^\mu_\nu} = \partial_{\mu}  T\indices{^\mu_\nu} + \Gamma^{\mu}_{\mu \sigma} T\indices{^\sigma_\nu} - \Gamma^{\sigma}_{\mu \nu} T\indices{^\mu_\sigma} = 0 \mathperiod
\end{equation}
From the expression above, employing the definitions for matter perturbations from \cref{sec:matpert} and the Christoffel symbols in the Newtonian gauge, \cref{eq:newtgam1,eq:newtgam6}, we can derive the following general perturbed conservation equations, before particularising for each fluid.

\begin{itemize}
    \item \textit{Perturbed continuity equation:} Taking the $\nu = 0$ component of \cref{eq:contpert} we find:
\begin{equation}
    \delta \rho_f' = -3 \hub (\delta \rho_f + \delta p_f) - (\bar{\rho}_f + \bar{p}_f) (\theta_f - 3  \Phi' ) \mathcomma
\end{equation}
for the evolution of the density perturbation of the fluid $f$. The different effects can be identified on the right-hand side of the equation: firstly, we have the dilution with the background expansion; secondly, the contributions from the local fluid flow; and lastly, a purely relativistic effect due to perturbations to the local expansion rate itself, encoded in $\Phi$. Employing the definition of the density contrast, $\delta \equiv \delta \rho / \bar{\rho}$, and transforming to Fourier space:
\begin{equation}
    \delta_f'+ 3 \hub \left( c_{s,f}^2-w_f \right) \delta_f = \left( 1 + w_f \right) \left( 3 \Phi' - \theta_f \right) \mathcomma
\end{equation}
recalling that $c_{s,f}^2 = \delta p_f/\delta \rho_f $ and $w_f = \bar{p}_f/\bar{\rho}_f$. 

 \item \textit{Euler equation}: If, on the other hand, we take the spatial components $\nu = i$, we obtain:
\begin{equation}
    {v_i^f}' = - \left( \hub + \frac{\bar{p}_f'}{\bar{\rho}_f + \bar{p}_f} \right) v^f_i - \frac{1}{\bar{\rho}_f + \bar{p}_f}\partial_i \delta p_f  - \partial_i \Psi \mathperiod
    \label{eq:eulint}
\end{equation}
On the right-hand side, we have a correction to the momentum components from the velocity flow and contributions from pressure and gravity, respectively. Once again, using the relation $\theta = \partial_i v^i$ and taking the derivative of \cref{eq:eulint}, leads to an equivalent equation for the velocity divergence in Fourier space:
\begin{equation}
\theta_f' + \left[ \mathcal{H} \left( 1 - 3 w_f \right) + \frac{w_f'}{1+w_f} \right] \theta_f = k^2 \left[ \frac{c_{s,f}^2}{1+w_f} \delta_f + \Psi  \right]  \mathperiod
\end{equation}
\end{itemize}
%

The conservation equations depend on the individual characteristics of the fluid $f$ whose evolution we wish to track. We remark below the special cases of matter and radiation at sub-horizon scales ($k \gg \hub$), for which the equations greatly simplify:
\begin{itemize}
    \item \textit{Matter:} Taking $p_m =0$ the continuity and Euler equations reduce to:
\begin{align}
    \delta_m' &= - \theta_m + 3 \Phi' \mathcomma \label{eq:pcontm} \\
    \theta_m' &= - \hub \theta_m + k^2 \Psi \mathperiod \label{eq:peulm}
\end{align} 
Combining derivatives of these two equations, it is possible to write a single clustering evolution equation for the matter perturbations:
\begin{equation} \label{eq:deltappm} 
    \delta_m'' + \hub \delta_m' = - k^2 \Psi + 3 \left(\Phi'' + \hub \Phi' \right) \mathperiod
\end{equation}
The second term on the left-hand side accounts for the expansion and, for this reason, is often called the Hubble friction. The $-k^2 \Psi$ accounts for the scale-dependent time-varying gravitational potential that the matter particles feel.

    \item \textit{Radiation:} For $p_r = \rho_r/3$ and no viscosity, the continuity and Euler equations become:
\begin{align}
     \delta_r' &= - \frac{4}{3} \theta_r + 4 \Phi' \mathcomma \label{eq:pcontrr} \\
    \theta_r' &= k^2 \frac{\delta_r}{4}+ k^2 \Psi \mathperiod \label{eq:peulrr}
\end{align}
Likewise, we can compute the second-order evolution equation for $\delta_r$:
\begin{equation} \label{eq:deltappr} 
    \delta_r'' + k^2 \frac{\delta_r}{3}  = - k^2 \frac{4}{3} \Psi + 4 \Phi'' \mathperiod
\end{equation}
A crucial difference from the relativistic matter case is that the radiation perturbations are not subject to the Hubble friction but, instead, feel an induced pressure contribution from the second term on the left-hand side. The interplay between the effects in both these equations, \cref{eq:deltappm,eq:deltappr}, is the genesis of the acoustic oscillation in the primordial plasma, which will be discussed in more detail in \cref{chap:obs}.

\end{itemize}

Since the equations are linear, the decomposition in Fourier modes ensures that each plane wave follows the same equations but with a different comoving wavenumber $k$. During the phase of linear evolution, the physical scale $\lambda_p= \frac{2\pi}{k} a $ of the perturbation expands along with the cosmic expansion. However, this linear treatment becomes inadequate if the perturbation enters the non-linear regime. At that point, the perturbation decouples from the Hubble expansion and starts to collapse gravitationally.

Lastly, it should be noted that this treatment of the conservation equations supposes no additional couplings between the individual fluids. For example, suppose interactions of dark matter with other components are allowed. In that case, its density perturbations will evolve according to a modified version of \cref{eq:deltappm}, including particular terms to account for such non-standard interactions, \textit{e.g.} for interacting dark energy models which we will introduce in \cref{chapter:sttheories}. The gravitational interaction of the different fluids is implicitly accounted for by the terms proportional to the expansion $\hub$ and the metric potentials $\Phi$ and $\Psi$.

 \subsection{A Few Words on the Synchronous Gauge} \label{sec:sync}


The synchronous gauge is a historically significant gauge, first introduced by Lifshitz during his groundbreaking work on cosmological perturbation theory \cite{Lifshitz:1945du}. It played a vital role in the development of early CMB codes, as will be introduced in \cref{sec:cmbcodes}. However, a drawback of the synchronous gauge is the emergence of spurious gauge degrees of freedom, which were at the origin of considerable confusion in the initial stages of cosmological perturbation theory. This was one of the driving forces behind Bardeen's development of a gauge-invariant approach \cite{Bardeen:1980kt}.

 In the previous section, we worked on the Newtonian gauge to simplify the analysis. However, the synchronous gauge is also useful to study the evolution of perturbations. The transformation between both gauges can be achieved trivially \cite{Ma:1995ey}, so it also serves as a cross-check of results. Therefore, for completeness, in this section, we provide the analogous perturbation equations in the synchronous gauge, in which the line element is written as 
\begin{equation}
\dd s^2= a^2 \left( \tau \right) \left[- \dd\tau^2 + \left(\delta_{ij} + \mathpzc{h}_{ij} \right) \dd x^{i} \dd x^{j} \right] \mathcomma
\end{equation}
 where $\mathpzc{h}_{i j}$ represents the metric perturbation. In what follows, we will adopt Ma $\&$ Bertschinger's \cite{Ma:1995ey} notation. However, we use $\mathpzc{h}$ for the scalar metric perturbation instead of the original $h$ to avoid confusion with other quantities. 
Once more, the metric perturbation can be decomposed into scalar, vector and tensor modes:
\begin{equation}
    \mathpzc{h}_{ij} = \mathpzc{h} \delta_{ij} + \mathpzc{h}^{\parallel}_{ij} + \mathpzc{h}^{\perp}_{ij} + \mathpzc{h}^{T}_{ij} \mathperiod
\end{equation}

The parallel and perpendicular components, $\mathpzc{h}^{\parallel}_{ij}$ and $\mathpzc{h}^{\perp}_{ij}$, can still be expressed in terms of a scalar $\eta$ and a divergence-free vector $\vec{A}$,

\begin{align}
    \mathpzc{h}^{\parallel}_{ij} &=\left( \partial_i \partial_j - \frac{1}{2} \delta_{ij} \nabla^2 \right) \eta \mathcomma \\
    \mathpzc{h}^{\perp}_{ij} &= \partial_i A_j + \partial_j A_i \mathperiod
\end{align}

Therefore, the scalar mode of the metric perturbations is fully characterised by $\mathpzc{h}$ and $\eta$, while the vector and tensor modes $A_i$ and $\mathpzc{h}^{T}_{ij}$ can be discarded for the purpose of this study. In what follows, we write down the relevant equations in Fourier space in terms of the Fourier transforms $\mathpzc{h}(k,\tau)$ and $\eta(k,\tau)$.

Analogously to the treatment for the Newtonian gauge, we derive the perturbed Einstein equations:

\begin{equation}
k^2 \eta - \frac{1}{2} \mathcal{H} \mathpzc{h}' = -4 \pi G a^2 \sum \delta \rho_f \mathcomma
\end{equation}
\begin{equation}
k^2  \eta' = 4 \pi G a^2 \sum \rho_f \left( 1 + w_f \right) \theta_f \mathcomma
\end{equation}
\begin{equation}
\mathpzc{h}'' + 2\mathcal{H} \mathpzc{h}' - 2 k^2 \eta = - 24 \pi G a^2 \sum \delta p_f \mathcomma
\end{equation}
\begin{equation}
\mathpzc{h}'' + 6 \eta'' + 2 \mathcal{H} \left( \mathpzc{h}' + 6 \eta' \right) - 2 k^2 \eta = 0 \mathperiod
\end{equation}

The perturbed continuity and Euler equations for the uncoupled baryonic and radiation fluids are derived directly from the perturbation of the conservation relations $ \nabla_{\mu} T^{\mu}_{\nu}=0$ and are given by 
\begin{equation}
\delta_f' + 3 \mathcal{H} \left( c_{s,f}^2-w_f \right) \delta_f = - \left(1 + w_f \right) \left( \frac{\mathpzc{h}'}{2} + \theta_f \right) \mathcomma
\label{eq:deltasyn}
\end{equation}
\begin{equation}
\theta_f' + \left[ \mathcal{H} \left( 1 - 3 w_f \right) + \frac{w_f'}{1+w_f} \right] \theta_f = k^2 \frac{c_{s,f}^2}{1 + w_f} \delta_f \mathperiod
\label{eq:thetasyn}
\end{equation}

The advantages of using the synchronous gauge become evident from when $f=\text{cdm}$. In this case, because for cold dark matter $w=p=0$, \cref{eq:thetasyn}, implies that the fluid velocity remains zero as long as it has a vanishing initial condition ($\theta_{\text{cdm}} (\tau_i)=0$), removing the gauge freedom without loss of generality. Nevertheless, this remarkable simplification is lost by introducing a coupling between cold dark matter and the other matter components (we will focus on the dark energy case), and the fluid velocity becomes non-vanishing in general.

\section{Inflation} \label{sec:infl}


The first cosmologically relevant epoch in the expansion history of the Universe is the initial inflation period, which came to a halt at around $a = 10^{-28}$ (or equivalently about $10^{-32}$ seconds). 
The mechanism of \textit{inflation} was first proposed by Guth in 1980 \cite{Guth:1980zm} as an explanation for the absence of magnetic monopoles and further developed by Linde \cite{Linde:1981mu} and Albrecht and Steinhardt \cite{PhysRevLett.48.1220}. In short, it is a phase in the very early Universe in which the comoving Hubble radius $(aH)^{-1}$ decreased dramatically through accelerated expansion, which drives $\Omega_0$ in \cref{eq:omega0} to unity, ensuring spatial flatness, which is preserved up to the present. 

Although there are some theoretical gaps and a lack of experimental proof for this inflation mechanism, it is currently accepted in the community as it addresses several fundamental issues. Namely, it provides a simple and effective solution to the \textit{horizon problem} concerning the origin of the observed homogeneity and isotropy of the Universe, enabling a full causal connection of photons in the early Universe that leads to the thermal equilibrium needed to explain the CMB measurements. Remarkably, this setting also predicts the quantum fluctuations needed to seed the small inhomogeneities that grow into the cosmic structure distribution that we observe at present.

Essentially, inflation addresses the origin of the isotropy in the distribution of photons at large angular distances. Indeed, we observe that all the photons emitted from the last scattering surface, \textit{i.e.} when photons began to freely traverse the Universe without interacting with other matter, possess the same temperature up to minor variations. If inflation had not occurred and the Universe had expanded gravitationally from an epoch dominated by radiation, then most of these photons would have been separated by distances greater than their comoving horizon at the time of their last scattering, meaning that prior to their detection, they could never have been in causal contact and thereby still exhibit the same temperature today.


The problems presented above (further discussed in \cref{app:infl}) are related to causality and, therefore, can be effectively addressed by postulating a period of extremely fast inflation before the hot Big Bang to generate the homogeneity state of the Universe and the correlation of the CMB fluctuations. 

The reason why this \textit{inflation} solves the causality problems is that, keeping in mind the importance of the Hubble radius during the standard Big Bang period, this would be equivalent to considering an initial phase of the early Universe during which the comoving Hubble radius is decreasing:
\begin{equation}
    \odv{}{t} (a H)^{-1} < 0 \Leftrightarrow \ddot{a} > 0 \mathcomma
    \label{eq:infcon}
\end{equation}
given that $H \equiv \dot{a}/a$ and the notion of an accelerated expanding period becomes evident in the last expression. The Hubble radius (a present \textit{particle} communication limit) becomes much smaller than the particle horizon (an absolute causality and particle communication limit), then this means that even though particles (or patches of the sky) are disconnected \textit{today}, they could still have been in causal contact early on. Of course, this is simply an intuitive picture of how inflation can address the causality problems, and detailed calculations of how this is achieved for particular frameworks can be found in \cite{baumann_2022}.

A possible approach to realise inflation lies in the introduction of one or more scalar fields known as the inflaton(s) $\varphi$, which dominate at the earliest stages and possess a property denoted as $w_{\varphi} < -1/3$, thereby triggering a phase of accelerated expansion in the early Universe. The entire model's degrees of freedom are determined by the shape of its potential $V(\varphi)$. As the field moves along its potential, it releases significant kinetic energy. This energy may lead to the generation of \ac{sm} particles through couplings with the inflaton, akin to \textit{Yukawa} couplings, potentially giving rise to all the particles found in the standard model, a process referred to as \textit{reheating}.
For inflation to occur and persist for an adequate duration, the inflaton field must exhibit \textit{slow-roll} behaviour. To achieve this, its potential energy must dominate over kinetic energy ($\dot{\varphi}^3 \ll V$), and the acceleration must be slow, expressed as as $\ddot{\varphi} \ll 3H \dot{\varphi}$ during inflation. These conditions ensure that the energy density of the Universe remains nearly constant with $\rho_{\varphi} \simeq V$ during inflation, resulting in exponential growth of the scale factor, similar to the $\Lambda$-dominated case. However, the balance between kinetic and potential energy must eventually change, causing inflation to cease and reheating to commence.
When treating the inflaton as a quantum field, it becomes possible to predict the perturbations generated by its quantum fluctuations. These primordial quantum fluctuations, stretched to macroscopic scales by the Universe's expansion, constitute the initial seeds for the formation of large-scale structures. Employing a perturbative approach, we can anticipate the evolution of deviations $\delta \varphi$ of the field from homogeneity. These deviations evolve as a harmonic oscillator with a time-varying mass, enabling a quantum treatment resembling that of a quantum oscillator. For a more detailed introduction to slow-roll inflation see \cref{app:infl}.

The remarkable consistency with observations and the support for its role in the early Universe has granted inflation a place as an ingredient in the standard cosmological model, even without consensus on a particular inflationary model. 

In summary, the physical predictions of inflation include:

\begin{itemize}
    \item Negligible Curvature: Inflation requires that the spatial curvature of the Universe be extremely close to zero. Observations of the CMB, combined with lensing effects and \ac{bao} measurements, put tight bounds on the curvature, $\Omega_k = 0.0007 \pm 0.0019$ \cite{Aghanim:2018eyx}, in agreement with the inflationary predictions.

    \item Adiabatic, Gaussian, and Nearly Scale Invariant Initial Perturbations: while the specific details may vary depending on the inflationary model, observations of the CMB provide strong constraints on departures from these assumptions \cite{Planck:2018jri}.
\end{itemize}

In \cref{app:infl} a more detailed account of the problems addressed by inflation is provided, along with the illustrative example of a scalar field model of \textit{slow-roll} inflation.

\section{A Brief History of the Universe in Four Epochs} \label{sec:exphis}

Bringing together the information in the previous sections, we can reconstruct the historical epochs of the Universe by tracing its expansion backwards in time \cite{Cicoli:2023opf}. Since the Universe is expanding, the scale factor would approach zero at earlier times. Under such circumstances, the Universe would have been in a drastically different physical state, significantly hotter, with its baryonic matter ionised and photons unable to propagate freely in this primordial plasma due to the extreme density of the environment.

Beginning with this early time, the Universe has undergone various epochs as part of its cooling and expansion process, in a hierarchy of gradually occuring phase transitions. We will also explore the subsequent era marked by structure formation and followed by the more recent phase of accelerated expansion. Nevertheless, all the processes described are continuous, meaning there is no absolute division between each epoch. A concise overview of the evolution of each cosmic fluid during these periods, as described in \cref{sec:ing}, is illustrated in \cref{fig:rhos}, for the $\Lambda$CDM model.

\subsection{Early Universe: The Hot Big Bang Epoch}

Indeed, according to the Hot Big Bang model, the early Universe was a much denser and high-temperature environment filled with elementary particles. As we venture back in time, we encounter periods beyond the reach of particle physics and reproduction with our present-day particle colliders. Rather than relying on these tools, we can look for the Universe's thermal evolution relics, providing an unparalleled laboratory for particle physics. Assuming conservation of entropy from thermodynamic, and as discussed briefly in \cref{sec:times}, the events in this epoch can also be pinpointed according to the cooling down temperatures:

\begin{equation}
    T(t) = T_0 \frac{a_0}{a(t)} \mathperiod
\end{equation}

A more detailed overview can be found in \cite{gorbunov2011introduction}.



\begin{itemize}
    \item The Planck Epoch ($t \sim 10^{-43}\,s$ or $T \sim 10^{19}\, \text{GeV}$): At the earliest conceivable moment after the Big Bang, the nature of the Universe was profoundly different, dominated by extreme energy conditions ($E_{\text{Pl}} \sim 10^{19}\, \text{GeV}$ is the \textit{Planck energy}). The Universe was hot, dense and governed by quantum gravity effects. The fundamental forces of electromagnetism, gravitation, weak nuclear interaction, and strong nuclear interaction are thought to have been unified, although a complete theory remains elusive. Conventional Big Bang cosmology predicts a gravitational singularity prior to this time, reflecting our inherent ignorance on the "origins" of the Universe. General relativity is assumed to break down at such scales due to the quantum effects, requiring a coherent theory of quantum gravity \cite{oriti_2009,rovelli_2004,armas_2021} or at least some quantum compatible classical treatment. 
    
    \item The Grand Unification Epoch ($10^{-43}\, s \lesssim t \lesssim 10^{-36}\,s$ or $10^{19} \, \text{GeV}\gtrsim T \gtrsim 10^{15}\, \text{GeV}$): The gravitational force separated from the other three fundamental forces. The Universe was still too hot and dense for particles as we know them to exist. This is the era where Grand Unified Theories (GUTs) aim to explain unfamiliar physics.
    The transition to the inflationary epoch is predicted to generate a vast number of magnetic monopoles \cite{Preskill:1979zi}, which are conspicuously absent in current observations. This discrepancy was one of the problems addressed and resolved by the introduction of the theory of cosmic inflation \cite{Guth:1980zm,Linde:1981mu}.

    \item Inflationary Epoch ($10^{-36}\,s \lesssim t \lesssim 10^{-32}\,s$ or $ 10^{15}\, \text{GeV}\gtrsim T \gtrsim 10^{13}\, \text{GeV}$): Described in \cref{sec:infl} and in \cref{app:infl}, during this brief moment the Universe underwent rapid exponential expansion, smoothing itself out. This explains the observed large-scale uniformity of the CMB radiation. Cosmologists generally believe that inflation happened and could be explained through a scalar field slowly rolling down a potential, called the \textit{inflaton}. Inflation led to a vast thinning of the elementary particles, which were regenerated through the symmetry breaking as the inflaton field dropped to the minimum of its potential energy, refilling the Universe with a hot and dense plasma of elementary particles in a process called \textit{reheating}. 
    
    \item The Electroweak Epoch ($10^{-32}\,s \lesssim t \lesssim 10^{-12}\,s$ or $10^{13} \, \text{GeV}\gtrsim T \gtrsim 10^{3}\, \text{GeV}$): The Universe cooled to approximately $10^{15} \, \text{GeV}$ and the strong nuclear force separated from the electroweak force, leading to distinct electromagnetic and weak nuclear forces. The Higgs field also gave particles mass in this period \cite{ATLAS:2012yve}.
    Around this period, a process known as \textit{baryogenesis} occurred ($T \sim 10^{12}\text{GeV}$), accounting for the observed predominance of matter over anti-matter in the present-day Universe.

    \item Quark Epoch ($10^{-12}\,s \lesssim t \lesssim 10^{-6}\,s$ or $ 10^{3}\, \text{GeV}\gtrsim T \gtrsim 1\, \text{GeV}$): The Universe was still too hot for quarks to combine into protons and neutrons. It was filled with a quark-gluon plasma, a hot soup of quarks and gluons.
\end{itemize}

\subsection{Radiation Dominated Epoch}

The \ac{rde} is the period extending from $10^{-10}\, s \lesssim t \lesssim 50\, \text{kyrs}$ and encompasses significant sub-phases, each characterised by cooling temperatures, ultimately reaching an energy level of approximately $10^2\, \text{GeV}$. 
The RDE is when light elements form. Since radiation decays faster with the expansion than non-relativistic matter, this epoch will eventually cease and give way to a matter-dominated regime.
The radiation-dominated epoch is depicted in grey in \cref{fig:rhos}.

\begin{itemize}
    \item Electroweak Symmetry Breaking ($t \lesssim 10^{-6}\, s$ or $ 1\, \text{GeV}\gtrsim T$): For temperature around $100\, \text{GeV}$ the forces have solidified into their current forms, and particles acquire masses through the Higgs mechanism.
    \item Neutrino Decoupling and Hadron Epoch ($10^{-6}\, s \lesssim t \lesssim 1\, s$ or $ 1\, \text{GeV}\gtrsim T \gtrsim 1\, \text{MeV}$): The temperature dropped enough for quarks to bound to create hadron and anti-hadron pairs which annihilate as the Universe cools, leaving a residue of protons and neutrons. Dark matter is expected to decouple from the primordial plasma by the end of the hadron epoch ($t \sim 10^{-5}\, \text{s}$ or $T \sim 200\, \text{MeV}$ for WIMPs), while neutrinos start to decouple around this time ($t \sim 0.2\, \text{s}$ or $T \sim 1\, \text{MeV}$).
    \item Lepton Epoch ($1\, s<t<200\, s$ or $ 1\, \text{MeV}\gtrsim T \gtrsim 0.1\, \text{MeV}$): the temperatures are still high enough for positron-electron pairs to form. For $T \approx m_e \sim 0.5\, \text{MeV}$, the photons are no longer energetic enough to keep this equilibrium and the annihilation process results in energy transfer to the thermal bath. 
    \item \ac{bbn} Epoch ($200\, s \lesssim t \lesssim 300\, \text{s}$ or $0.1\, \text{MeV} \gtrsim T \gtrsim 0.05\, \text{MeV}$): nuclear fusion starts and protons and neutrons combine into atomic nuclei. Free neutrons start fusing with protons at $T \approx 0.1\, \text{MeV}$. Around three minutes after the Big Bang, the light elements were formed, starting with deuterium, which rapidly transformed into $^4\text{He}$. 
    As the Universe's temperature and density quickly decrease to the point where nuclear fusion is no longer possible, we enter the \ac{mde}. All the neutrons have been incorporated into helium nuclei, leaving a mass ratio of roughly three times more hydrogen than $^4\text{He}$, with only trace quantities of other nuclei.
    \item Photon Epoch ($300\, \text{s} \lesssim t \lesssim t_{\text{eq}}$ or $ 0.05\, \text{MeV}\gtrsim T \gtrsim 1\, \text{eV}$): For the rest of the radiation epoch, the Universe consisted of a dense hot plasma filled with nuclei, electrons, and photons. This epoch extends through the initial stages of the \ac{mde} when $t>t_{\text{eq}}$.
\end{itemize}

\subsubsection{Matter-Radiation Equality}

At some point, the matter density dominates over radiation, which dilutes faster. In particular, the RDE ended when the densities equalled: 
\begin{equation}
    \frac{\rho_m}{\rho_r} = \frac{\rho_{m,0}}{\rho_{r,0}} \frac{a^4}{a^3} = \frac{\rho_{m,0}}{\rho_{r,0}} (1+z_{\text{eq}})^{-1} = 1 \mathcomma
\end{equation}
and thus, the redshift at the equivalence between matter and radiation was $z_{\text{eq}} \sim 3400$ or $t_{\text{eq}} \sim 50\,000$ years (for fiducial values as measured by \textit{Planck} \cite{Aghanim:2018eyx}). This is represented by the line dividing the grey and red sectors in \cref{fig:rhos}. 

\subsection{Matter Dominated Epoch}

At the beginning of the MDE, the Universe was still dominated by photons. Although neutral atoms had not formed yet, the Universe was becoming increasingly transparent. The MDE is depicted in red in \cref{fig:rhos}.

Shortly after equality, the Universe's temperature dropped to approximately $1\, \text{eV}$, equalling the binding energy of the hydrogen atom. For the first time since the Big Bang, the Universe started emancipation from the dense primordial plasma of photons, electrons, and protons. Nevertheless, it remained virtually opaque to electromagnetic radiation since photons could travel only short distances before colliding with a charged particle through Compton scattering. In this epoch it becomes relevant to also keep track of time in terms of the redshift, which for matter domination, evolves approximately as:

\begin{equation}
    z - z_0 \approx (t-t_0)^{-2/3} \mathperiod
\end{equation}

\begin{itemize}
    \item Recombination and Decoupling Epoch ($18\, \text{kyrs} \lesssim t \lesssim 380\, \text{kyrs}$ or $6000 \gtrsim z \gtrsim 1100$): As the Universe continued to expand and cool down, the formation of neutral hydrogen became energetically favoured and ionised hydrogen and helium atoms captured the free electrons. Shortly after this transitional phase, Compton scattering became inefficient, and photons decoupled from matter, leading to the Cosmic Microwave Background Radiation we can detect today. This event defines the \textit{last scattering surface}. 
    Until this point, photons and baryons had been tightly coupled, oscillating in acoustic waves generated by the compression/rarefaction cycle due to the balance between the baryons' gravitational attraction and the photons' radiation pressure. Once the two species decouple, the baryons are effectively released from the drag of the photons. This \textit{decoupling} marks the end of the \textit{drag epoch} and allowed photons to travel unimpeded through the Universe without interacting with matter, marking the earliest epoch observable today - the \textit{last scattering surface}. On the other hand, the baryons are frozen in the acoustic wave pattern at a scale determined by the size of the horizon at the drag epoch. It should be emphasised that photon decoupling and recombination are distinct events, albeit closely related. The photons present at the time of decoupling have since freely streamed through the Universe and are the same ones that we observe today in the CMB radiation, albeit significantly cooled by the expansion. 
    By the time this period ended, the Universe's composition had shifted to a dark fog containing roughly $75\%$ of hydrogen and $25\%$ of helium, with only small amounts of lithium.

    \item Dark Ages ($380\, \text{kyrs}<t<150\, \text{Myrs}$ or $ 1100\gtrsim z \gtrsim 60 $): With no stars, the Universe was in a dark state, as photons were not effectively being produced yet and hence there was no visible light. Matter consisted of a gas of atoms that kept being dragged into denser regions by the pull of gravity alone, but the first stars had not yet ignited. Slowly, the gravitational attraction pulled matter together, and overdensities started growing into clumps as dark matter started to dominate the total energy density of the Universe.

        \sloppy
    \item Cosmic Dawn ($150\, \text{Myrs}<t<200\, \text{Myrs}$ or $60 \gtrsim z \gtrsim 20$): As matter started to clump subjected to gravity's rule, the \textit{Jeans length} $\lambda_J = \sqrt{\frac{\pi c_s^2}{10 G \rho_0}}$, which determines the smallest structures that can form, began to decrease and the density perturbations started to grow in amplitude. Cold dark matter became the dominating component, setting the stage for gravitational collapse to amplify the slight inhomogeneities left by cosmic inflation. This process made dense areas denser and sparse regions more rarefied. Once the overdensities overcome the Jeans mass dictated by the Jeans length, they begin to spontaneously collapse and the first stars form

    \item Star Formation and Reionisation ($200\, \text{Myrs}<t<1\, \text{Gyrs}$ or $20 \gtrsim z \gtrsim 6$): The process of formation of the first starts heated the surrounding medium, once again ionising the hydrogen in the majority of the Universe. This moment is known as \textit{reionisation} and happened approximately one billion years after the Big Bang, marking the last phase transition in the Universe and the end of the dark ages since the star-light across much of the electromagnetic spectrum was finally able to travel unimpeded through the cosmos, eventually revealing the Universe as we see it today.

    \item Large Scale Structure Formation Epoch ($1\, \text{Gyrs}<t<5\, \text{Gyrs}$ or $6 \gtrsim z \gtrsim 1$): Subsequently, stars start assembling into galaxies, galaxy clusters, super-clusters and all the large-scale structures. The process of structure formation is dictated by non-linear gravitational effects, which thread the galactic \textit{cosmic web}, leaving behind large empty regions called \textit{cosmic voids}.  This epoch extends to the present day, even after dark energy took over the evolution at the dark energy-matter equality.
    
    \end{itemize}





\subsubsection{Dark Energy-Matter Equality}

Likewise, we can estimate the redshift value at which the transition from a \ac{mde}, decelerating expansion during which large-scale structures form, to a new regime of accelerating expansion occurs. Neglecting
radiation and for $K = 0$ we have that $\ddot{a} = 0$ when
\begin{equation}
    \frac{\rho_m}{2\rho_\Lambda} = \frac{\rho_{m,0}}{2\rho_{\Lambda,0}} \frac{1}{a^3} = \frac{\Omega_{m,0}}{2\Omega_{\Lambda,0}} (1+z_{\text{acc}})^{3} = 1 \mathcomma
\end{equation}
which corresponds to a redshift $z_{\text{acc}} \sim 0.7$, assuming fiducial \textit{Planck} values for $\Lambda$CDM. This equality epoch is depicted as the line dividing the red and blue regions in \cref{fig:rhos}.

\subsection{Dark Energy Dominated Epoch}



In the $\Lambda$CDM model, due to the cosmological constant's premise of non-dilution with the expansion, $\Lambda$ will eventually take over the Universe's evolution from matter, whose density decays with $a^{-3}$, giving rise to the current \ac{dede}. The Universe will keep expanding faster and faster, leading to a potential \textit{Big Freeze}, where stars burn out and the Universe becomes dark and cold.

\begin{itemize}

    \item Late Universe ($t \gtrsim 5\, \text{Gyrs}$ or $z \lesssim 1$): The Sun forms in the Milky Ways and the Solar System emerges in an orchestrated dance around $t \sim 9\, \text{Gyrs}$. Around the $10\, \text{Gyrs}$ mark, the primordial forms of life on Earth emerge, while our modern human ancestors came to be only about $20\,0 000$ years ago.
    
    \item Present time ($t \sim 13.8\, \text{Gyrs}$ and $z=0$): After billions of years of cosmic evolution, the Universe is as we observe it today, filled with galaxies, stars, and a variety of complex structures. We start studying the vastness of the cosmos, which now spans a radius of $46.5$ billion light-years $\sim$ $14$ Gpc $\sim$ $10^{26}$ m, larger than the naive guess from the age of the Universe since the light is carried along with the expansion of the Universe.
\end{itemize}

\section{\texorpdfstring{The Theoretical Problems with $\Lambda$}{The Theoretical Problems with the Cosmological Constant}} \label{sec:theoprob}

Throughout its twenty-year lifespan, the $\Lambda$CDM model described above has successfully withstood numerous tests under its inherent simplicity. Based on our current understanding of fundamental physics, this framework provides a detailed account of most of the Universe's expansion history with only 6 free parameters to be fixed by the observational data.
More intricate models rarely manage to achieve statistical preference from the data, as any extra model parameters must provide a sufficiently better fit to the data to justify their inclusion (see \cref{sec:stat}). 
However, the cosmological constant hypothesis still faces longstanding challenges, from motivation in quantum field theory to its role in an expanding cosmology. Moreover, more recently, the observational tensions in the parameters of $\Lambda$CDM, discussed in detail in \cref{sec:cosmotensions}, have granted extended models an extra layer of encouragement and support.

For the moment, and for what concerns the core of the standard model, we will discuss the most prominent challenges: the cosmological constant and the cosmic coincidence problems.

\subsection{The Cosmological Constant Fine Tuning Problem}


In order to realise the cosmic acceleration today, according to \cref{eq:acc}, we require that the cosmological constant $\Lambda$ is of the order of the square of the present Hubble parameter $H_0$, whose value was estimated in \cref{eq:h0est}. This corresponds to the following approximate value for the energy density (with $\text{M}_{\text{Pl}} \approx 10^{19}\, \text{GeV}$):

\begin{equation}
    \rho_{\Lambda} \approx 10^{-47}\, \text{GeV}^4 \approx 10^{-123}\, \text{M}_{\text{Pl}}^4 \mathperiod
\end{equation}

This is incompatible with the vacuum energy scales predicted for \ac{qft}, as we will illustrate next.
This issue was identified long before the discovery of the late-time acceleration or the cosmological constant proposal.
At that time, the prevailing belief was that the cosmological constant was precisely zero, and much effort was devoted to explaining why this should be the case. The vanishing of a constant in physics often hints at some underlying symmetry.

Initially referred to as the \textit{vacuum catastrophe} \cite{1995AmJPh..63..620A}, the issue lies in the \ac{qft} predictions of an inexplicably large value for the vacuum energy, arising from the sum of the zero-point energies of each degree of freedom in all quantum fields within the Universe, up to a particular cut-off scale. Such a large value should manifest itself \textit{via} significant gravitational effects which have never been observed. Observational constraints on the upper limit of this vacuum energy completely contradict the QFT result, with differences reported up to an outrageous level of $120$ orders of magnitude \cite{martin}. 
For example, the vacuum energy estimated for the extreme cut-off scales of the order of $\text{M}_{\text{Pl}}$ or of the scales predicted in particle physics (such as $\approx 0.1\, \text{GeV}$ for quantum chromodynamics (QCD)), the vacuum energy density can be estimated in limit values as:

\begin{equation}
    \rho_{\text{vac}}^{\text{M}_{\text{Pl}}} \approx 10^{74}\, \text{GeV}^4 \quad \text{and}\quad \rho_{\text{vac}}^{QCD} \approx 10^{-3}\, \text{GeV}^4 \mathperiod
\end{equation}

This shows how, even while selecting a different cut-off scale for the cumulative zero-point energy may somewhat alleviate this problem, reducing the discrepancy to around forty or sixty orders of magnitude \cite{Bernardo:2022cck}, such a solution remains highly insufficient. The observed value for $\Lambda$ is much smaller than all particle physics scales, endowing this problem with a fundamental character. 
The ultimate resolution of the cosmological constant problem remains uncertain, and whether it will emerge from QFT or cosmological arguments is unclear. The former could reflect our incomplete understanding of how to estimate the zero-point energy of quantum fields in curved spacetimes, while the latter could introduce some \textit{bare} cosmological constant to cancel out the contributions from all fields except the observed cosmological constant. Another alternative is to stick to the pattern of explaining physical mysteries by postulating a scalar field (like the Higgs field \cite{ATLAS:2012yve} or the inflaton \cite{Dimopoulos2020}). In this context, a potential approach, as we will further dissect, is to discard the cosmological constant altogether and promote it to a scalar field (or some similar effect from modified gravity) to account for the late-time acceleration \cite{Copeland:2006wr,Tsujikawa:2013fta}.

\subsection{The Cosmic Coincidence Problem}

The coincidence problem embodies the improbable fact that the observed values of the dark matter and dark energy densities are of the same order of magnitude today, given that these species evolve at entirely different rates throughout the history of the Universe, as discussed in \cref{sec:exphis}. This issue is generally less consensual than the cosmological constant problem \cite{Velten:2014nra}, often attributed to more philosophical or metaphysical considerations, out of the scope of cosmological physical theories \cite{chamcham_silk_barrow_saunders_2017}. This debate reflects the lack of consensus in the community over the relevance of anthropic arguments \cite{Carter:1974zz}.
In essence, the anthropic principle suggests that the values of physical quantities in the current state of the Universe are what they are because such conditions are necessary for the existence of observers in our condition, and it is not up to us, inevitable inhabitants of the Universe, to decide on the statistical likelihood of the \textit{present design of the Universe} \cite{chamcham_silk_barrow_saunders_2017}. This doctrine can be endorsed with the idea that such \textit{statistics} implied by the terms \textit{coincidence} or \textit{likelihood} do not make sense, to begin with, since cosmic variance implies that we have only one realisation of the Universe at our disposal \cite{sep-cosmology}.

However, the anthropic principle does not easily align with the cosmological principle, on whose pillars observational cosmology stands, and which implies that we do not occupy any preferential position in the Universe. Therefore, if we accept this incompatibility, the coincidence problem can serve as additional motivation to dismiss the cosmological constant and devise alternative dark energy models. In particular, dynamical dark energy models, where the energy density of dark energy varies over time or there is some interaction in the dark sector, stand out due to their ability to explain the seemingly coincidental similarities between the matter and dark energy densities simply through their natural dynamical character. Further discussion on this type of models will be the focus of \cref{chapter:sttheories}.




  \cleardoublepage


 \chapter{Cosmological Observational Tests} \label{chap:obs}
 \setcounter{equation}{0}
\setcounter{figure}{0}



   \epigraph{Estou me interessando terrivelmente por fatos: fatos são pedras duras. \\ Não há como fugir. Fatos são palavras ditas pelo mundo.\blfootnote{\textit{I’m becoming terribly interested in facts: facts are hard rocks. You can’t escape. Facts are words spoken by the world.} --- \textsc{Clarice Lispector} in The Hour of the Star} \\ --- \textsc{Clarice Lispector}\ \small\textup{A Hora da Estrela}}

  Cosmology differs fundamentally from most branches of physics by being an observational science rather than an experimental one. Given that the early Universe events played out without an audience (as much as we are aware!), and the dawn and subsequent evolution of the Universe cannot be replicated (we have only one sample!), our understanding is primarily based on measuring the spatial correlations between cosmological structures at late times. The fundamental endeavour of modern cosmology is no longer to map the night sky into constellations that track the seasons but rather to formulate a coherent narrative for the history of the Universe that provides motivated explanations and is compatible with these observed correlations. In this sense, cosmology is a wide-ranged effort to reconcile various observational data with predictions from theoretical models based on coherent physical principles.

The study of high-redshift Type Ia supernovae (SNe) provided the first compelling evidence for the late-time accelerated expansion of the Universe \cite{acel1,acel2}. These supernovae, which serve as \textit{standardisable candles} based on the fixed luminosity at which they occur, appeared fainter than expected in a spatially flat, matter-dominated Universe. Subsequent observations from other probes further supported this picture, with the additional assumptions that make up the $\Lambda$CDM model described in \cref{chapter:standardmodel}.

After the success of previous seminal probes and ground-based experiments \cite{durrer_2020}, the \textit{Planck} satellite has given the most precise measurement of the anisotropies in the Cosmic Microwave Background (CMB) \cite{Planck:2018nkj}. This is one of the most valuable and comprehensive sources of cosmological information from the initial perturbations that led to the formation of cosmic structures \textit{via} gravitational collapse. The CMB has also been instrumental in constraining cosmological parameters within the standard cosmological model and its extensions. Simultaneously, galaxy surveys such as the Sloan Digital Sky Survey \cite{Tegmark_2004} have extensively mapped the distribution of the Large Scale Structure across the Universe, identifying the distinct pattern known as the Baryon Acoustic Oscillation (BAO) scales within the matter correlation function. By leveraging these findings, a standard ruler is established, whose length can be quantified across various cosmological epochs, enabling scientists to monitor and trace the expansion of the Universe.

The precision and compatibility level between these observations have put the standard cosmological model under increasingly rigorous scrutiny. Verifying the standard model's hypotheses through independent means solidifies its validity, and observational support can be established from multiple sources. Upcoming missions like the space-based interferometer LISA \cite{LISACosmologyWorkingGroup:2019mwx} and the Euclid satellite mission \cite{Amendola:2012ys} will offer highly detailed maps of the cosmic microwave background and large-scale structures. These missions promise even finer measurements of the BAO scales across different cosmological epochs. At the same time, additional insight will be extracted from phenomena such as supernovae, both weak and strong gravitational lensing, galaxy clustering, and gravitational wave events, among others. These advancements will enrich our understanding of the large-scale homogeneity and any deviations therein, providing extended tests for the validity of general relativity.

This chapter is structured as follows. In \cref{sec:corr} we introduce the matter correlation function, needed to define the matter power spectrum in \cref{sec:mpk}. These two introductory sections are based on the Chapter 3 of \cite{debook}. \cref{sec:cmb} provides a description of the main properties of the cosmic microwave background and how to relate its features to the cosmological framework. In \cref{sec:otherp} we report on other relevant probes of the physics of the Universe, in particular for dark energy signatures. \cref{sec:lcdm_param} provides a complete account of the impact of the six-parameters of the $\Lambda$CDM model on the matter power spectrum and the CMB spectrum of temperature anisotropies. Finally, in \cref{sec:cosmotensions} we list the cosmological tensions and anomalies that have brought the standard model to a crisis and which motivate the exploration of alternative models.

\section{The Matter Correlation Function} \label{sec:corr}

 As elaborated in \cref{chapter:standardmodel}, the Universe's large-scale structure is not randomly distributed. Instead, cosmic objects are arranged into a tangled cosmic web with galaxies anchored along dense filaments and clustering at their intersections, surrounded by vast empty regions known as cosmic voids. The mapping of these structures encodes information about the initial state of the hot Big Bang when the small overdensities that grew into clusters are thought to have formed and the subsequent governing physical processes. Different cosmological models predict different patterns of structure formation, allowing us to infer the best-fitting values for specific cosmological parameters from the statistical characteristics of the Universe's large-scale structure.

We rely on indispensable statistics tools to assess the validity of any theoretical model and its connection to the real world. 
Because we have only one realisation of the observable Universe, spanning a wide range of physical scales, the statistical properties of the distribution of events in the sky are as vital to cosmologists as flour to a baker. 
Statistics methods in cosmology are generally categorised into two main branches: \textit{descriptive statistics}, which aims to cast the data into a compact and physically understandable way, and estimation or \textit{inferential statistics}, deals with the problem of extracting information about the model and its parameters.
This chapter will focus on the former and how observational insight can be derived from the background and perturbation theory introduced in \cref{chapter:standardmodel}. The latter will be deferred to \cref{chapter:statistics}.

The initial step is to consider a random particle distribution, which may correspond to various astrophysical objects such as galaxies or quasars. For $N$ points contained within a volume $V$, the primary statistical descriptor is the average numerical density $\rho_0 = N / V$. However, this is insufficient to distinguish between, for example, $N$ points clustered in one location and $N$ points uniformly scattered throughout the volume $V$. If, instead, we consider an infinitesimal volume $\odif{V}$ randomly selected within $V$, then the term $\rho_0 \odif{V}$ represents the mean number of particles in this infinitesimal volume. If $\odif{N}_{ab} = \langle n_a n_b\rangle$ is the average number of pairs in the infinitesimal volumes $\odif{V}_a$ and $\odif{V}_b$, separated by a distance $r_{ab}$, the next significant statistical descriptor is the \textit{two-point correlation function} $\xi(r_{ab})$, defined as:

\begin{equation}
    \odif{N}_{ab} = \langle n_a n_b\rangle = \rho_0^2 \odif{V}_a \odif{V}_b \left[ 1+ \xi (r_{ab}) \right] \mathcomma \quad \text{with}\quad r_{ab} > 0 \mathperiod
\end{equation}

In our preceding discussion, the implicit assumption is that $r_{ab} > 0$, meaning the two infinitesimal volumes do not overlap. In this context, the notion of average, denoted by $\langle \cdot \rangle$, conveys the volume or \textit{sample average}\footnote{Although we have previously introduced this notation as the average over random initial conditions (like the ones generated from inflation), the large-scale cosmic structure we observe in practice is essentially one realisation of this random process. To reconcile this with theoretical models, we rely on the principle of ergodicity, according to which, as volume approaches infinity, ensemble averages converge to spatial averages. In this context, several different patches of the Universe with various sizes can be considered separate realisations of the same random process. Thus, averaging measurements over a large volume mimics averaging over various realisations of the Universe. However, our observations are inherently limited to a finite volume, introducing what is known as \textit{sample variance}. In the case of the CMB, where this finite volume is the single whole observable Universe, this limitation is appropriately called \textit{cosmic variance}.} and considers pairs at different points within a single realisation, separated by the same distance $r_{ab}$. If the points are sufficiently far apart to be uncorrelated, the results would be similar to taking an average over many realisations of the distribution. 
Despite the limitation of being confined to observations from a single Universe, the correlation function remains a practical and meaningful descriptor.
For a completely random distribution of $N$ particles, $\odif{N}_{ab}$ should be independent of specific locations, leading to a vanishing correlation function $\xi$. In contrast, a non-vanishing $\xi$ indicates spatial correlation among the particles. The correlation function can then be expressed as a spatial average of the product of density contrasts at two distinct points:

\begin{equation}
    \xi (r_{ab}) = \frac{\odif{N}_{ab}}{\rho_0^2 \odif{V}_a \odif{V}_b} -1 = \langle \delta (r_a) \delta (r_b) \rangle \quad \text{with}\quad \delta(r_a) = n_a/(\rho_0 \odif{V}_a) - 1 \mathcomma
\end{equation}

where we used the fact that $\langle\delta(r_a)\rangle = \langle\delta(r_b)\rangle = 0$. For a sample average, $\xi$ can be computed over all possible positions as:

\begin{equation}
    \xi (\vec{r}) = \frac{1}{V} \int \delta (\vec{y}) \delta (\vec{y} + \vec{r}) \odif{V}_y \mathperiod
\end{equation}

In practical terms, calculating the correlation function is often simplified by selecting a volume $\odif{V}_a$ such that $\rho_0 \odif{V}_a = 1$ is the average particle at a distance $r$ from a test particle. In that case, the count of particle pairs reduces to the number of particles contained within the volume $\odif{V}_b$:

\begin{equation}
    \odif{N}_b = \rho_0 \odif{V}_b [1 + \xi(r_b)] \mathperiod
\end{equation}

Thus, the correlation function is typically computed as:

\begin{equation}
    \xi(r) = \frac{\odif{N}(r)}{\rho_0 \odif{V}} - 1 = \frac{\langle\rho_c\rangle}{\rho_0} - 1 \mathcomma
\end{equation}

yielding the average number of particles at a distance $r$ from any given particle, normalised by the expected number in a uniformly distributed system. This is sometimes termed the conditional density contrast. Moreover, considering a bounded volume with $N = \rho_0 V$ particles, imposes an integral constraint on $\xi(r)$:

\begin{equation}
    \int \xi(r) \odif{V} = \frac{1}{\rho_0} \int \frac{\odif{N}}{\odif{V}} \odif{V} - V = \frac{N}{\rho_0} - V = 0 \mathperiod
\end{equation}

A positive correlation $\xi(r)$ signals more particles than expected in a uniform distribution. In such instances, the distribution is termed as \textit{positively clustered}. Generally, one is more concerned with the dependency on the magnitude $r$, leading to the selection of a shell of distance $r$ surrounding each particle. 

As introduced in \cref{chapter:standardmodel}, the decomposition in Fourier modes of perturbation variables is useful for practical purposes because these modes evolve independently at the linear level. Given that the average of a perturbation variable is, by definition, zero, the next statistically meaning quantity is the average of the pairs. When expressed in Fourier space, ($ k = 2\pi/\lambda $) these yield the \textit{power spectrum}:

\begin{equation}
    P_{\delta} (k) = V |\delta_k|^2 \mathcomma
\end{equation}

where $\delta_k$ are the Fourier coefficients corresponding to the density contrast, and $V$ is the power spectrum's normalisation factor. In cases where the quadratic form involves two distinct variables, such as $|\delta_k \theta_k|$, $P(k)$ is said to be a cross-correlation power spectrum. 

The power spectrum is the most widely used measure for describing clustering in linear and mildly non-linear regimes and is one of the most important observables in the field. It serves as the Fourier-space counterpart of the correlation function, expressed as:

\begin{equation}
    P(\vec{k}) = \frac{1}{V} \int \xi (\vec{r}) \expe^{-i \vec{k} \cdot \vec{r}} \mathrm{d}V\quad \text{and}\quad \xi(\vec{r}) = \frac{V}{(2\pi)^3} \int P(\vec{k}) \expe^{i \vec{k} \cdot \vec{r}} \mathrm{d}k \mathperiod
\end{equation}

The Fourier volume factors of $1/V$ and $V$ in $P(\vec{k})$ and $\xi(\vec{r})$, respectively, are conventionally added to the Fourier transformation to guarantee that both functions have the same dimensions. When assuming spatial isotropy, that is, when the correlation function depends solely on the magnitude $r = |\vec{r}|$, the spectrum likewise depends only on $k = | \vec{k} |$.

In the following sections, we will see in more detail how the power spectrum decomposition can be used to study density fluctuations in the matter distributions.

\section{The Matter Power Spectrum} \label{sec:mpk}


As mentioned in \cref{sec:infl}, the theory of inflation is currently the leading explanation for the first generating moments of the Universe through a rapidly expanding phase in its early stages. During this period, the initial anisotropies are believed to have been generated through the quantum fluctuations of a scalar field called the \textit{inflaton}. This exponential expansion is essential as it amplifies the density perturbations responsible for the observed anisotropies in the CMB and the formation of large-scale structures such as galaxies and galaxy clusters in the Universe.
 Therefore, it is helpful to focus on what is happening at different scales to study the patterns in this matter distribution resulting from inflation. This is achieved exactly through the power spectrum decomposition of the density perturbations. The density contrast $\delta (k)$, which characterises the deviation of the density of a Fourier scale $k$ from the background density, is the Fourier counterpart of \cref{eq:denscont}.

The large-scale structure in the Universe is not distributed randomly but has interesting correlations between spatially separated points. By definition, the mean value of the density perturbations is zero, $\langle \delta \rangle = 0$. Therefore, the first non-trivial statistical measure of the density field (at a fixed time) is exactly the two-point correlation function introduced in \cref{sec:corr}.

Treating these density contrasts as variations arising from primordial quantum fluctuations, we can now formally write the power spectrum $P(k)$ in $k$-space originating from the matter distribution seeds as follows:
\begin{equation}
    \langle \delta (\vec{k}) \delta^* (\vec{k'}) \rangle = \frac{2 \pi^2}{k^3} P(k) \delta^3 (\vec{k} - \vec{k'}) \mathcomma
    \label{eq:mpk}
\end{equation}
now following the convention of defining the power spectrum with the normalisation factor $2\pi^2/k^3$, rendering $P(k)$ dimensionless. The modes are uncorrelated via the Dirac delta function $\delta^3 (\cdot)$. 
When considering scalar perturbations, we can derive the matter power spectrum for both the pre-and post-radiation-matter equality epochs, approximately resulting in \cite{baumann_2022}:
\begin{equation}
    P(k) \propto 
    \begin{cases}
    k^{n_s},\ \ k < k_{\text{eq}} \mathcomma \\
    k^{n_s-4},\ \ k > k_{\text{eq}} \mathcomma
    \end{cases}
\end{equation}
%
where $n_s$ is the \textit{spectral index} of the power spectrum produced during inflation and $k_{\text{eq}} = a_{\text{eq}} H(a_{\text{eq}})$ is the scale parametrising the distinction between large and small scales, corresponding to the mode that entered the Hubble radius exactly at the radiation–matter equality.
On large scales, $P(k)$ grows for increasing $k$, while it is a decreasing function of $k$ on small scales. This implies a peak in the power spectrum, directly related to the scale $k_{\text{eq}}$, which is sensitive to the background and perturbed expansion history.
The shape of the power spectrum changes with parameter variations is depicted for the six $\Lambda$CDM parameters in \cref{fig:varying_pk_all}.
 Furthermore, there is a subtle oscillatory signature in the tail of the spectrum at small scales (large $k$), which results from the baryonic acoustic oscillations (BAO) imprinted by cosmic evolution.

The comparison between the theoretical power spectrum of dark matter and the observed power spectrum of the galaxy distribution poses a significant challenge. Assessing how these two spectra should be related is not trivial, an issue commonly referred to as \textit{biasing}. It is commonly assumed that both spectra differ only by a constant factor, based on the premise that, on sufficiently large scales, both have their evolution dominated by the gravitational effects. This hypothesis is not necessarily universally valid as it can vary depending on the details of galaxy formation and evolution, and is assessed ultimately through \textit{N-body simulations} and observations \cite{Desjacques:2016bnm}.



This chapter introduces the observational probes relevant to this dissertation, categorises them and discusses the cosmological information they offer. In particular, we focus on characteristic signatures and effects that can be decisive in distinguishing between different models, considering the underlying assumptions and the potential for model-independent analysis. While this discussion provides a tailored overview, more comprehensive information is available in recent reviews such as \cite{Moresco:2022phi} and the sources cited within each subsection.

\section{The Cosmic Microwave Background} \label{sec:cmb}


 More than any other cosmological probe, observations of the \textit{Cosmic Microwave Background} (CMB) spectrum, mapping both intensity and polarisation, have converted cosmology into a precision science, providing a comprehensive understanding of the Universe's geometry, evolution and composition. Measurements of the temperature anisotropies over the last three decades have played a crucial role in establishing the standard cosmological model, as described in \cref{chapter:standardmodel}. These observations have confirmed the existence of primordial fluctuations, which are the seeds of structure formation consistent with the inflationary period. The power of the CMB as a cosmological probe lies in the fact that it captures these fluctuations at an early stage when they were still small and aptly described using linear perturbation theory, unlike the complex and non-linear structure formation process. The evolution of small fluctuations in the primordial plasma can be understood from fundamental physics enclosed in well-defined equations, allowing for precise theoretical predictions of the expected temperature anisotropies' distribution in the CMB.

As discussed in \cref{sec:exphis}, 
the pivotal moment in which light started to propagate freely and the Universe became transparent to light occurred at redshift $z \approx 1090$, when the Universe was around $380 000$ years old and the effective temperature of the free streaming photons was $\sim 3000\, \text{K}$, is termed \textit{photon decoupling}. Since then, the expansion of the Universe has redshifted and diluted these photons, according to the relation $T \propto (1+z)$ discussed in \cref{sec:times}. They are detected today as the CMB radiation, with a characteristic average temperature of $T_0 \approx 2.725\, \text{K}$. These photons carry imprints of their cosmic journey across the expansion history, from the early Universe until today, including the influence of gravitational wells near massive clusters. This information is encoded at different scales as deviations to this average temperature distribution. Because these spatial fluctuations break the isotropy of the CMB, they are referred to as the CMB \textit{anisotropies}. Being one of the most robust and complete cosmological probes, numerous experiments have been conducted to measure the temperature variations in the CMB radiation. Additionally, the anisotropies can be modelled for different theories rigorously and systematically using modern Einstein-Boltzmann codes, which will be the focus of \cref{sec:cmbcodes}. These solve the Boltzmann equation, which expresses how the distribution function of some particle species evolves, assuming Gaussian random fields.

\begin{figure}[!h]
      \subfloat{\includegraphics[width=\linewidth]{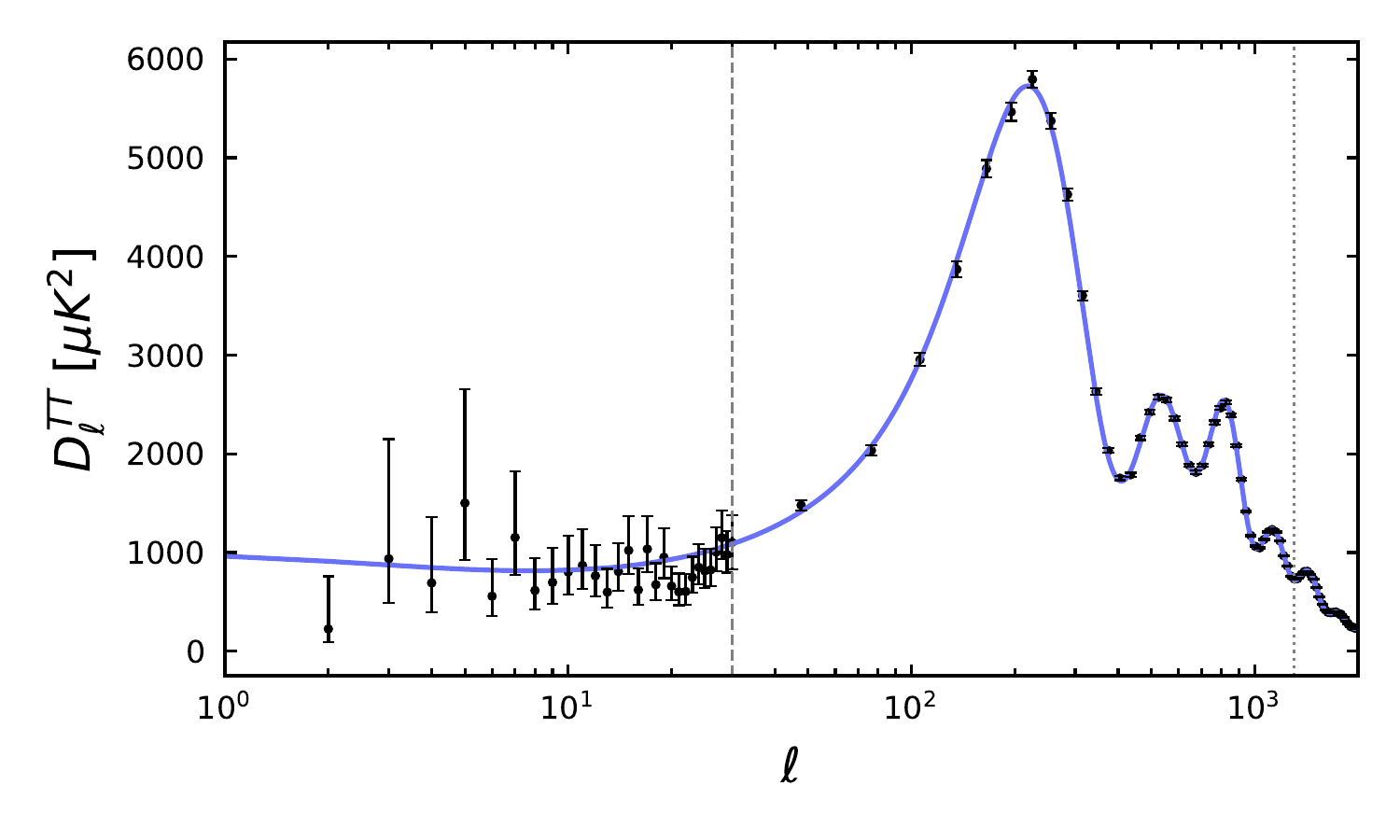}}
  \caption[The \textit{Planck} 2018 CMB temperature–temperature power spectrum]{\label{fig:planck_lcdm} Figure of the power spectrum of the CMB anisotropies (temperature). The measured spectrum $\mathcal{D}_{\ell}^{TT} = \ell (\ell+1) C_{\ell}^{TT}/(2\pi) $ has been scaled to make the smaller scale features visible. The large-scale data ($\ell>30$) is binned up by multipole intervals. The blue curve is the $\Lambda$CDM prediction for the \textit{Planck} 2018 \textit{bestfit} values, produced with the \texttt{CLASS} code \cite{Lesgourgues_2011,Blas:2011rf}. The black markers represent the \textit{Planck} data \cite{Aghanim:2018eyx} and corresponding error bars. The data is collected from the Planck legacy archive \href{http://pla.esac.esa.int/pla/}{http://pla.esac.esa.int/pla/}.}
\end{figure}

\cref{fig:cmb} presents the map of the CMB anisotropies as measured by the \textit{Planck} collaboration. This is a captivating image of the so-called last scattering surface of the Universe, with the colour scale encoding variations in the temperature of the CMB photons, which, in turn, correspond to density fluctuations at the time of photon decoupling. The study of the CMB temperature fluctuations distribution follows precise statistical techniques to find the correlations between hot and cold spots as a function of their angular separation. This results in an angular power spectrum decomposed in multipoles $\ell$, as depicted in \cref{fig:planck_lcdm}. The figure also includes the theoretical prediction of the CMB spectrum in the $\Lambda$CDM model when fitted to the observed data. The striking agreement between the theoretical prediction and the observed data highlights the remarkable success of the standard cosmological model. For a more detailed overview of the physics of the CMB, we refer the reader to Ref.~\cite{dodelson2003modern,durrer_2020}.


\subsection{The TT-Spectrum: Measuring the Anisotropies} \label{sec:anis}

As observers, we are limited to a realisation of the CMB map at a fixed point in space and time ($\vec{x}=\vec{x}_0$ and $\tau = \tau_0$), leading to the evolution equation for the temperature:
\begin{equation}
    T_{\text{CMB}} (\vec{x},\vec{p},\tau) = T_{\text{CMB},0} (\tau) \left[ 1 + \Theta  (\vec{x},\vec{p},\tau) \right] \mathcomma
\end{equation}
with $T_{\text{CMB}} \propto (1+z)$ and $\vec{p}$ is the direction of propagation of the photons. Next, we will discuss how to compare the observed temperature distribution function against the theoretical predictions.

Being a sky map, the distribution of CMB anisotropies is better expressed in the projection of the photon direction in the celestial sphere, \textit{i.e.} in polar coordinates $(\theta,\phi)$. Since our fields are assumed to be statistically homogeneous, the averages over an ensemble will be independent of the position. Together with the assumption of isotropy of the initial conditions of the fluctuations, this implies that the distribution of the temperature anisotropies will be the same for all directions. 
To compare the theoretical predictions for the temperature anisotropies with CMB observations, the temperature perturbation, $\Theta^{\text{obs}}= \Delta T/T$, is expanded in terms of spherical harmonic functions, $Y_{\ell m}$, which form an orthonormal basis for the points on a sphere, particularly suited for describing the pixel patches of the sky map:
\begin{equation}
  \Theta^{\text{obs}} (\theta, \phi) \equiv \frac{\Delta T}{T} (\theta, \phi) = \sum_{\ell = 1}^{\infty} \sum_{m=-\ell}^{+\ell} a_{\ell m}^T Y_{\ell m} (\theta, \phi) \mathcomma
\end{equation}
where $\ell$ and $m$ are conjugate to a unit vector in real space, representing the direction of propagation of the incoming photons. In particular, the relation $\theta=\pi/\ell$ in radians holds, meaning that larger scales correspond to lower multipoles.
The corresponding coefficients $a_{\ell m}$ encode the dependence on the direction of observation and hence are essential for parametrising the shape and overall distribution of the anisotropies in the CMB map in \cref{fig:cmb}. The monopole temperature is $a_{00}$ and is typically omitted from the spectra along with the dipole terms \footnote{The monopole term essentially represents a uniform temperature across the sky. It encoded the average temperature of the CMB, but it does not provide any information about fluctuations or anisotropies.
The dipole anisotropy is primarily due to our motion with respect to the CMB rest frame. It has been extensively mapped but is not particularly interesting for cosmological purposes. The dipole does not contain information about the early Universe but about the present motion. On this basis, it is usually discarded to unveil intrinsic anisotropies of cosmological interest.}.
Being statistically independent, the average value of the single harmonic coefficients vanishes, \textit{i.e.} $\langle a_{\ell m} \rangle = 0$. For the two-point correlator, similarly to the Fourier decomposition of random fields in space, the off-diagonal correlators of the expansion coefficients $a_{\ell m}$ of the expansion vanish, ensuring orthogonality. The intrinsic variance of the coefficients is defined as:
\begin{equation}
    \langle a_{\ell m} a^*_{\ell' m'} \rangle = \delta_{\ell \ell'} \delta_{m m'} C_{\ell}^{\text{obs}} \mathperiod
    \label{eq:meascl}
\end{equation}
The temperature-temperature (TT) spectrum is conventionally scaled in terms of the multipoles, resulting in

\begin{equation}
    \mathcal{D}_{\ell}^{TT} = \frac{\ell (\ell+1)}{2 \pi} C_{\ell}^{TT} \mathcomma
\end{equation}

such as the example depicted in \cref{fig:planck_lcdm}.
It is also possible to compute the three-point \textit{non-Gaussian correlations}, 
 a piece of information directly encoded by the primordial non-Gaussianity induced by some models of inflation, which CMB experiments have constrained \cite{Bartolo:2004if,Planck:2019kim} but goes beyond the scope of this work.




This method is so robustly studied according to fundamental physics that most of the statistical features had been predicted long before they were effectively observed. In particular, the TT-spectrum has been accurately mapped up to $\ell \sim 4000$ and it is largely believed that most of the available cosmological information has already been derived from it. In fact, CMB observations encode potential access to six different power spectra: temperature-temperature (TT), E-mode-E-mode polarisation (EE) and B-mode-B-mode polarisation (BB) autocorrelations and temperature-E-mode (TE), temperature-B-mode (TB), E-mode-B-mode (EB) cross-correlations. 
The BB spectra have much lower power compared to TT, and are still quite suppressed in comparison to the EE polarisation case (around 2 orders of magnitude below). So, for simplicity reasons, the work reported in Part II of this dissertation only considers CMB data from TT, EE, and TE spectra.
Nevertheless, the B-mode spectra carry unique information, and their detection and imprints are an independent research line \cite{Kamionkowski:2015yta}. This is because while scalar perturbations can only generate E-mode polarisations, primordial B-modes would unequivocally imply the existence of tensor perturbations, most likely generated during inflation. Even though the observational prospect for primordial B-modes still faces technical limitations due to secondary B-mode signals, there is a lot of potential for sensitivity improvements and their detection would significantly enrich our understanding of the early Universe. 

This work will focus on scalar perturbations only, uncorrelated with the vector and tensor modes. The vector perturbations decay, meaning their contribution is negligible (assuming that the initial conditions were set very early, as is the case in inflation), and the tensor perturbations are constant on super-horizon scales, subject to damped oscillations after entering the horizon, becoming suppressed compared to scalar perturbations.
In a broader context, scalar perturbations can be compared to sound waves propagating in the primordial plasma, leading to areas of compression and rarefaction. These variations produce the pattern of peaks and troughs in the CMB power spectrum, as seen in \cref{fig:planck_lcdm}, where the spectrum has been rescaled by $\ell (\ell+1)/(2\pi)$. Hence, the specific positions of these peaks depend non-trivially on the composition and evolution of the Universe and represent a valuable probe for testing signatures of alternative models to $\Lambda$CDM.

\subsection{Modelling the Anisotropies} \label{sec:diss}

The line of sight integration method provides an informative, straightforward way to study the physical effects behind the CMB anisotropies \cite{Seljak:1996is}. For a spatially flat Universe and adopting the conformal Newtonian gauge \cref{eq:pmetric}, the linear temperature anisotropy $\Theta \equiv \delta T/T$ due to the Fourier mode $k$ and projected on the multipole $\ell$ today can be written as \cite{durrer_2020}
%


%
\begin{equation} \label{eq:lsi}
    \Theta_{\ell}^{(s)} (k, \tau_0) = \int_{0}^{\tau_0} \odif{\tau} S_T^{(s)} (k, \tau) j_{\ell} (k (\tau_0-\tau)) \mathperiod
\end{equation}
%
The integrand in \cref{eq:lsi} is a convolution between the $\ell$-th spherical Bessel function, denoted by $j_\ell$, which gives the projection from Fourier to harmonic space. The $S_T^{(s)}$ terms in the integrand are the scalar sources:
\begin{equation} \label{eq:anis}
    S_T^{(s)} (k, \tau) \equiv g \left( \Theta_0 + \Psi \right) +  \left( g k^{-2} \theta_{b}  \right)' + \expe^{-\tau_{\text{op}}} \left( \Phi' + \Psi' \right)  \ + \ \text{polarisation terms} \mathperiod
\end{equation}
The term 
\begin{equation}
    \tau_{\text{op}} (\tau) = \int_{\tau}^{\tau_0} n_e \sigma_T a \odif{\tau'} \mathcomma
\end{equation}
represents the optical depth (not to be confused with conformal time), determined by the free electron fraction $n_e$ and the Thompson scattering cross section $\sigma_T$. An optically thin (thick) medium is characterised by $\tau_{\text{op}} \ll 1$ ($\tau_{\text{op}} \gg 1$), with $\tau_{\text{op}} = 1$ defining the optical transition. Furthermore, $g \equiv - \tau_{\text{op}}' \expe^{-\tau_{\text{op}}}$ corresponds to the visibility function, which quantifies the probability distribution that a photon last scatters at a given redshift\footnote{For simplicity, we omit the terms from the polarisation spectrum from different sources since the class of models considered does not typically affect the CMB polarisation, unless there is some specific coupling to photons.}.
In addition to the gravitational potentials $\Phi$ and $\Psi$, the source terms include the temperature monopole, the baryon velocity $\theta_b$ and subdominant polarisation modes, all dynamical functions of $\tau$ and $k$. The line-of-sight approach dramatically simplifies the computation while isolating the different contributions to the total anisotropy. The terms in the first and second brackets in \cref{eq:anis} are proportional to the visibility function, which peaks around the recombination epoch and are known as \textit{primary anisotropies} since their effect is mainly imprinted at $z \sim 1090$ when the photons decouple. In more detail, the contributions are the following:
\begin{itemize}
    \item Primary Monopole $\Theta_0$ (density perturbations): anisotropy proportional to the local photon energy density fluctuation at the last scattering surface. Overdensities within the early primordial plasma are anticipated to exhibit higher temperatures, whereas underdensities are expected to be colder. These variations in over- and under-densities inherited from the inflationary phase, will undergo an evolutionary process governed by the perturbation equations described in \cref{sec:linpert}. This evolution contimues until the epoch of decoupling, ultimately leaving a distinctive imprint in the CMB.
    
    \item Sachs-Wolfe (SW) Effect $\Psi$ (gravitational perturbations): redshift of the photons as they climbed up the initial potential wells when emitted. Regions linked to overdensities are correspondingly linked to more robust gravitational fields. Photons traversing within these perturbed gravitational fields will undergo a gravitational redshift, as elucidated by Sachs and Wolfe in \cite{Sachs:1967er}. Unlike adiabatic perturbations, this effect is anticipated to generate cooler over-densities and warmer under-densities.
    
    \item Doppler Shift (velocity perturbations): effects induced by the baryon velocity $\theta_b$ and $\theta_b'$. 
\end{itemize}
The third bracket terms are proportional to $\expe^{-\tau_{\text{op}}}$, which is active since the last scattering. This gives rise to \textit{secondary anisotropies}:
\begin{itemize}
    \item Integrated Sachs-Wolfe (ISW) Effect  $\Phi' + \Psi'$: accounts for how the photons travelling from the last scattering surface are affected by the potential wells they encounter. Comparable to the SW effect, the ISW occurs on a larger scale when photons traverse through the vast structures in the Universe. This results in an overall shift in the photon's frequency, arising from the cumulative influence of the mass encountered along its path. If the potential wells were static, the blueshift gained when entering would be compensated by the redshift necessary to escape. Nevertheless, if the potentials evolve in time, there is a net contribution to the temperature anisotropy.
\end{itemize}
The polarisation terms can be classified as tertiary contributions and discarded for simplicity.
Gravitational lensing of the CMB provides an additional secondary contribution to the temperature fluctuation in \cref{eq:anis}. However, lensing effects require bending the photon trajectories due to inhomogeneities, which are already linear effects. Hence, they constitute a higher order correction to \cref{eq:anis}.

\begin{figure}[!h]
      \subfloat{\includegraphics[width=\linewidth]{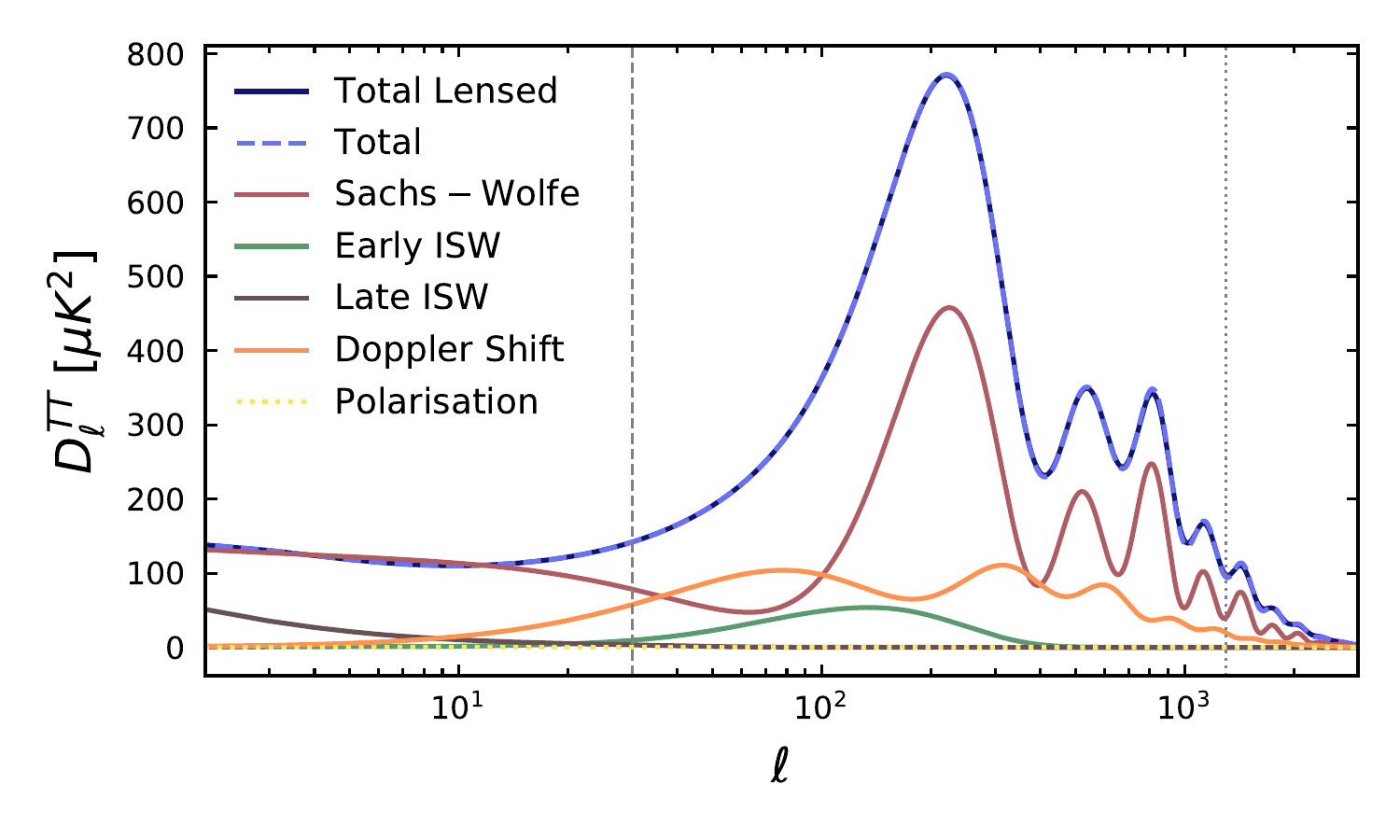}}
  \caption[Contributions to the CMB power spectrum]{\label{fig:cmb_comp} The individual main contributions to the CMB temperature power spectrum as detailed in \cref{sec:diss}, namely the total lensed and unlensed spectra, the Sachs-Wolfe, Integrated Sachs Wolfe, Doppler effect and the contributions from polarisation. This plot was generated through the Boltzmann code \texttt{CLASS} \cite{Lesgourgues:2011re,lesgourgues2011cosmic2} and has no multipole scaling compared to \cref{fig:planck_lcdm}.}
\end{figure}

The contribution of each effect to the temperature power spectrum defined by \cref{eq:lsi} is depicted in \cref{fig:cmb_comp}.



The measured value $C_{\ell}^{\text{obs}}$ in \cref{eq:meascl} can be directly compared to the the theoretical \textit{angular power spectrum} $C^{(s)}_{\ell}$, in terms of the temperature field $\Theta^{(s)}_{\ell}(k)$, as the multipole $\ell$ cross-correlation function in Fourier space $k$:
\begin{equation}
    C^{(s)}_{\ell} = \frac{2}{\pi} \int_{0}^{\infty} \odif{k} k^2\left| \Theta^{(s)}_{\ell}(k) \right|^2 \mathcomma
\end{equation}
with
\begin{equation}
    \Theta_l \equiv \frac{1}{2(-i)^{\ell}} \int_{-1}^{1} \odif{\mu} P_{\ell} (\mu) \Theta (\mu) \mathcomma
\end{equation}
in terms of the Legendre polynomials $P_{\ell} (\mu)$, where  $\mu = \hat{k}\cdot \hat{p}$ gives the direction between the direction in which the photon is propagating $\hat{p}$ and its wavenumber $\hat{k}$.

\subsection{Scale Dependence of the CMB}

Scalar perturbations can be envisioned as sound waves echoing through the primordial plasma. These waves correspond to regions of rarefaction and compression, resulting in the pattern of peaks and valleys in the CMB power spectrum, which are the dominant contribution at small scales, $\ell \gtrsim 100$. The exact location of these peaks depends on the Universe's composition and evolution, which are an important probe of dark energy.
Initial perturbations within the baryon-photon fluid can only travel a finite distance until recombination\footnote{To be more precise, during recombination, most electrons combined with the atomic nuclei formed during Big Bang nucleosynthesis. However, photons kept being scattered by the remaining free electrons. This phase should not be confused with the drag epoch, where photons \textit{fully decoupled} from the remaining electrons and could finally free stream throughout the Universe, a process which occurred about $z = 80$ later.}. The extent of this propagation distance experienced by an acoustic wave depends on its speed of sound:
\begin{equation}
    c_s = \frac{1}{\sqrt{3}} \left( 1 - \frac{3 \rho_b}{4 \rho_{\gamma}} \right)^{-1/2} \mathcomma
\end{equation}
for the oscillations imprinted in the CMB anisotropies with frequency $\omega_k (\tau) = k r_s (\tau)$, where $r_s$ is this limit distance, known as the sound horizon,
\begin{equation}
    r_s (z_{\text{dec}}) = \int_0^{(1+z_{\text{dec}})^{-1}} \frac{c_s (a)}{a^2 H(a)} \odif{a} \mathperiod
    \label{eq:soundh}
\end{equation}
The redshift at decoupling $z_{\text{dec}}$ can be deduced by intricate formulae \cite{Hu:1995en} or by taking the entire background history through fitting numerical codes, which is particularly important for non-standard scenarios with changes to the early Universe dynamics. 


As detailed in \cite{debook}, an approximate relation for the comoving wavelength of the acoustic peaks is $\lambda_c = 2\pi /k = 2 r_s/n$, where $n$ are integer numbers. The first peak is found at an angular scale of $\theta = 1$ degree which corresponds to $\ell = 180$ (according to $\theta=\pi/\ell$). It is helpful to define a characteristic angular location for these peaks' positions in terms of observable quantities:
\begin{equation}
    \theta_{\text{A}} = \frac{r_s (z_{\text{dec}})}{d_A^c (z_{\text{dec}})} \mathperiod
    \label{eq:thetacmb}
\end{equation}
%
This acoustic scale is a quasi-universal observable, as it can easily be defined across any statistically isotropic or spherically symmetric cosmology. $d_A^c$ is the comoving angular diameter distance, defined from \cref{eq:lumdist} as
\begin{equation}
  d_A^c (z) = (1+z) d_A (z) \mathperiod  
\end{equation}

The multipole associated with the angle $\theta_{\text{A}}$, is expressed as
\begin{equation}
 \ell_A = \frac{\pi}{\theta_{\text{A}}} = \frac{\pi d_A^c (z_{\text{dec}})}{r_s (z_{\text{dec}})}   \mathperiod
\end{equation}

Furthermore, the comoving angular diameter distance can be parametrised by
\begin{equation}
    d_A^c (z_{\text{dec}}) = \frac{c}{H_0} \frac{\mathcal{R}}{\sqrt{\Omega_{m,0}}} \mathcomma
\end{equation}
where $\mathcal{R}$ stands for the CMB shift parameter related to the sound horizon and given by
\begin{equation}
    \mathcal{R} = H_0 \sqrt{\Omega_{m,0}} \int_0^{z_\text{rec}} \frac{\odif{z}}{H(z)} \mathcomma
\end{equation}
implying $\ell_A \propto \mathcal{R}$, which can be bounded by CMB measurements, which also correspond to estimates of the sound horizon $r_s (z_{\text{dec}})$. The CMB shift parameter depends directly on the Universe's expansion history from recombination to the current epoch, defining the acoustic peaks' position.
This can be appreciated in \cref{fig:planck_lcdm} depicting a comparison between the theoretical prediction of the CMB temperature power spectrum for $\Lambda$CDM against the 2018 observations from \textit{Planck}. The predictions of peak locations and magnitudes in a Universe with dark energy, fully defined by a cosmological constant, align closely with the observed data.

The TT spectrum's highest amplitudes are attributed to two main components: the $\ell = 0$ component, representing the monopole (the mean value of the blackbody signal), and the $\ell = 1$ component, corresponding to the dipole as discussed in \cref{sec:diss}. Moving beyond these components, the predicted and observed higher modes ($\ell \geq 2$) are displayed in \cref{fig:planck_lcdm}, exhibiting a characteristic oscillatory pattern. Further insight into the shape of the spectrum is provided in \cref{fig:cmb_comp}, which decomposes the contributions from the sources of anisotropies introduced in \cref{sec:diss}.
The behaviour of this spectrum at different scales can be summarised as follows:

\begin{itemize}
    \item For $\ell < 100$, the Sachs-Wolf plateau emerges. These angular scales exceed the size of the causal horizon at the time of the last scattering surface. Consequently, regions separated by such scales had limited interaction and co-evolution since the end of inflation. The Sachs-Wolf plateau quantifies the primordial perturbations, primarily arising from mechanisms outlined in \cref{sec:diss}, particularly the Sachs-Wolf effect. These perturbations exhibit a distinct scale-invariant behaviour, manifesting as a flat $\mathcal{D}_{\ell}^{TT} = \ell (\ell+1) C_{\ell}^{TT}/(2\pi) $, consistent with inflationary predictions.

    \item In the range of $100 \leq \ell \leq 1000$, we enter the realm of the acoustic peaks. Scales smaller than the horizon at the last scattering surface capture causal interactions within the primordial plasma prior to decoupling. The intricate wavelike structure in this region results from the oscillations of the coupled photon/baryon fluid within the gravitational field, primarily influenced by the massive and non-interacting dark matter. Here, gravitational forces compete with fluid motion driven by photon radiation pressure. The acoustic peaks are characterised by a prominent peak around $\ell \simeq 200$, followed by two smaller peaks. These visible perturbations in the plasma resemble sound waves, justifying the term \textit{acoustic}.

    \item For $\ell \leq 1000$, we encounter the damping tail. In this region, the spectrum exhibits rapid oscillations with a progressively diminishing amplitude. This behaviour arises due to the thickness of the last scattering surface, averaging over multiple small-scale behaviours that tend to cancel each other out. This averaging process gradually erases structures at scales smaller than the thickness of the last scattering surface.
    
\end{itemize}

\subsection{CMB Lensing} \label{sec:cmblen}

\begin{figure}[!h]
      \subfloat{\includegraphics[width=\linewidth]{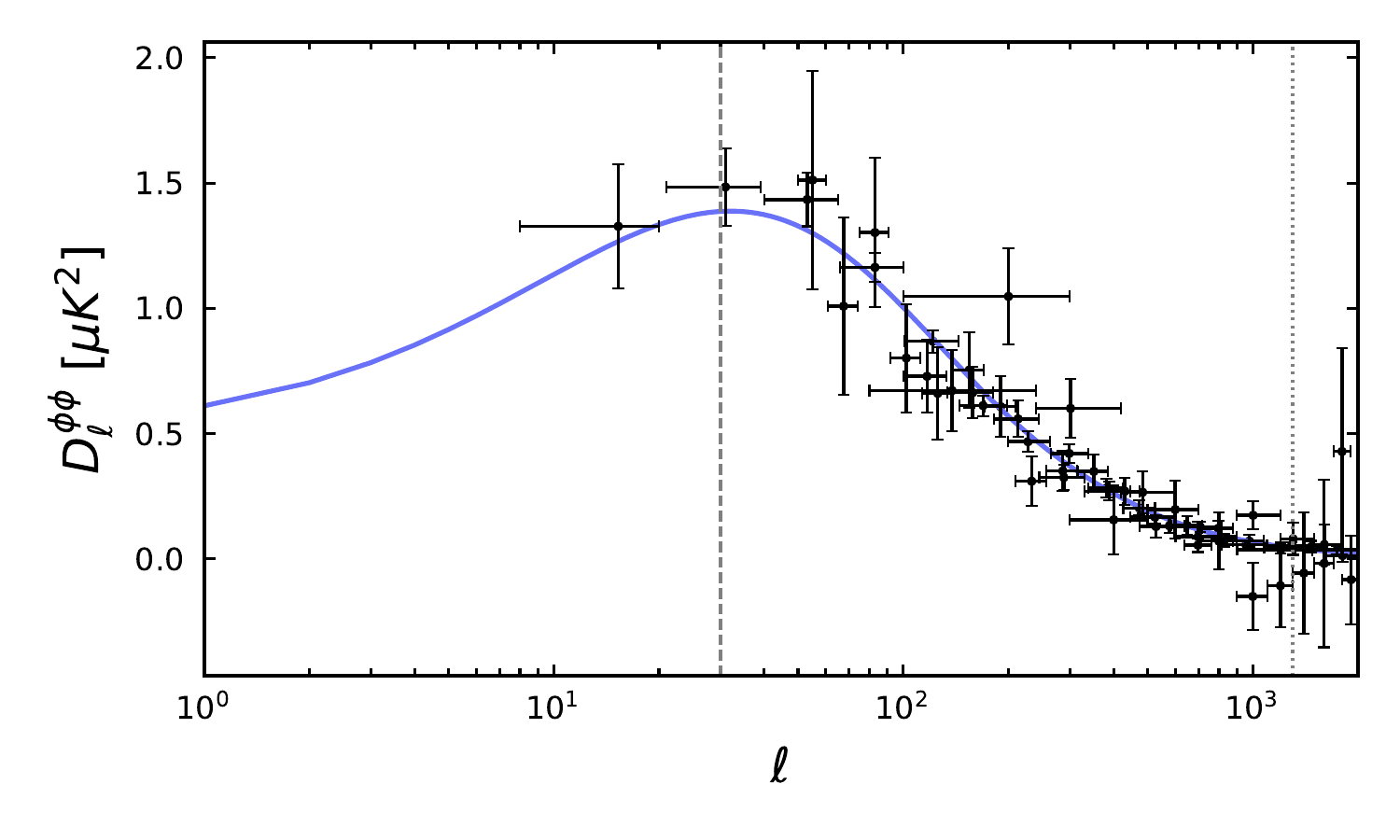}}
  \caption[The \textit{Planck} 2018 CMB gravitational lensing power spectrum]{\label{fig:planck_len} The power spectrum of the CMB gravitational lensing $\mathcal{D}^{\phi \phi}_{\ell}=10^7 \left[\ell (\ell+1)\right]^2 C_{\ell}^{\phi \phi}/(2\pi)$ factor. The blue curve is the $\Lambda$CDM prediction for the \textit{Planck} 2018 \textit{bestfit} values, produced with the \texttt{CLASS} code \cite{Lesgourgues_2011,Blas:2011rf}. The black markers represent, as an example, data from the \textit{Planck} collaboration \cite{Planck:2018lbu}, POLARBEAR 2019 \cite{POLARBEAR:2019ywi}, SPTpol 2019 \cite{Wu:2019hek}, SPT/Planck 2017 \cite{Simard:2017xtw} and ACTpol 2016 \cite{Sherwin:2016tyf}, and corresponding error bars. The data is part of a compilation from \href{https://lambda.gsfc.nasa.gov/education/lambda_graphics/more/lensing_power_source.html}{https://lambda.gsfc.nasa.gov/education/lambda$\_$graphics/more/lensing$\_$power$\_$source.html}.}
\end{figure}

In the context of general relativity, massive objects serve as gravitational lenses that bend the paths of photons. The large-scale structure between the last scattering surface and the observer gravitationally lenses both the temperature and polarisation anisotropies of the CMB. These imprints can be reconstructed as a map of the lensing potential, whose gradient determines the lensing deflections. For example, in propagating through a large, over-dense clump of matter in the line of sight, angular structures in the CMB get magnified, appearing more extensive in the sky. Essentially, by looking at how the typical size of hot and cold spots in the CMB temperature map vary across the sky, we can reconstruct the lensing deflections and, hence, the integrated distribution of dark matter.
This lensing map provides a new cosmological observable, similar to maps of cosmic shear estimated from the shapes of galaxies\footnote{Cosmic shear is induced by weak lensing as a consequence of the overall mass distribution, including non-luminous matter. Compared to galaxy counting, the advantage is that it is not affected by bias uncertainty.}. Its power spectrum, as depicted in \cref{fig:planck_len}, provides access to cosmological parameters from the CMB alone that affect the late-time expansion and geometry of the Universe and the growth of structure - parameters that have only degenerate effects in the primary CMB anisotropies.

This lensing effect offers a unique window into the Universe's structure, including its dark matter distribution. It affects various cosmological observables, such as blurring the acoustic peaks in the CMB and altering polarisation modes.

Strong lensing phenomena, often involving quasars lensed by single galaxies, offer direct ways to study cosmological parameters like the rate of the Universe's expansion. On the other hand, weak gravitational lensing, usually featuring distant galaxies lensed by closer galaxy clusters, offers insights into the growth and density of cosmic structures. Both strong and weak lensing serve as powerful tools for cosmological study, each providing complementary data that can help resolve current tensions in our understanding of parameters like the matter density and structure growth, as we will discuss in more depth in \cref{sec:s8t}.

\section{Other Cosmological Probes} \label{sec:otherp}

Despite the fact that the CMB provides a powerful and robust measure of the cosmological parameters of the $\Lambda$CDM model, these can be reinforced and independently confirmed by combining multiple surveys, ideally probing the Universe at different temporal and spatial scales. In fact, this is what allows for the breaking degeneracies in the model parameters inferred from the CMB, as will be discussed in \cref{sec:lcdm_param}. Below we detail the various observational probes relevant for the context of this dissertation which correspond mostly to probes of dark energy. Nevertheless, a multitude of other emerging independent cosmological probes could be added to this picture, such as the $21$ cm and Lymann-$\alpha$ lines, quasars, cosmic chronometers, measurements of redshift drift, surface brightness fluctuations, to name only a few. A complete review can be found in \cite{Moresco:2022phi}.

\subsection{Standard Candles: Cepheid Variables and Type IA Supernovae}


As we have briefly mentioned in the context of the seminal measurements of the accelerated expansion of the Universe \cite{acel1,acel2}, an alternative method of estimating the expansion rate draws on the knowledge associated with two particular astrophysical objects: a distance calibrator and some objected with well-known standard luminosity. The most conventional pair corresponds to Cepheid variables as calibrators for Type Ia supernovae\footnote{The explosion of supernovae results in an intense burst of radiation, making them exceedingly luminous events. These supernovae can be categorised based on the presence of chemical element absorption lines in their spectra. Specifically, if the spectrum of a supernova displays a hydrogen absorption line, it is labelled as a Type II supernova. In the absence of such a line, it is called Type I. Further sub-classifications exist within Type I supernovae: Type Ia contains an absorption line of singly ionised silicon, Type Ib features a helium line, and Type Ic lacks silicon and helium absorption lines.
The explosion characteristic of Type Ia supernovae is triggered when a white dwarf in a binary system crosses the Chandrasekhar mass limit~\cite{Chandrasekhar:1931ih} due to the accretion of gas from its companion star. One remarkable feature of Type Ia supernovae is the near-constancy of their peak absolute luminosity. Consequently, these supernovae serve as a form of \textit{standardisable} candles, enabling the observational measurement of luminosity distance by assessing their apparent brightness.}.

Cepheid variables are a vastly studied type of pulsating stars characterised by a well-established correlation between their luminosity and the period of their pulses. We can infer their intrinsic luminosity from the observed pulsation period by categorising Cepheids as Population I or Population II stars. By measuring the differences in the magnitude (a measure of a luminous object's brightness, related to the photon flux), it is possible to establish a connection with the \textit{luminosity distance}, defined in \cref{eq:lumdist}.

More precisely, defining the distance modulus, $\mu = m-M$, simply as the difference between the apparent $m$ (measure of an object's brightness when observed from the Earth) and absolute $M$ magnitudes, we find the key relation 

\begin{equation}
    \frac{\mu}{5} + 1 = \log \left( \frac{c(1+z)}{H_0} \int_0^{z} \frac{\odif{z'}}{E(z')} \right) \mathcomma
\end{equation}
where the expansion rate $E(z)$ is defined as $H^2 (z) = H_0^2 E^2(z)$, scaled by the Hubble constant $H_0$ in $\text{km}\, \text{s}^{-1}\, \text{Mpc}^{-1}$, and assuming a flat geometry ($\Omega_K =0$ and $d_M(z) = d_c(z)$). 
Since the corrected peak magnitude $M$ is the same for all the supernovae ($M \approx -19$ for type Ia), the luminosity distance can be directly obtained from $m$.
This methodology allows us to establish a connection between the apparent magnitudes of Cepheid variables and supernovae and the Hubble constant. 
The redshift associated with the supernova can be determined independently by analysing the wavelength of the light emitted and the shifts in its characteristic spectral absorption lines. With a significant sample size of supernova observations, one can establish the relationship between the observed luminosity and redshift. Large catalogues like Pantheon (1048 supernovae \cite{Scolnic:2017caz}) and its recent update Pantheon+ (1550 supernovae \cite{Brout:2022vxf}) compare these observations against calibrated distance-redshift measurements from the wavelength of the observed light based on the spatially flat $\Lambda$CDM model. From this comparison, key cosmological parameters such as $\Omega_m$ and $\Omega_{\Lambda}$ can be inferred, leaving only one free parameter: the Hubble constant. 
The SH0ES collaboration estimates $H_0$ employing Cepheid measurements as calibrators analysed using the \textit{Hubble} Space Telescope (HST) and the \textit{Gaia} mission \cite{Riess:2021jrx}. Planned surveys such as the Rubin Legacy Survey of Space and Time (Rubin LSST) are expected to provide catalogues of various transient calibrators, increasing the number of known \ac{sn} Ia by a factor of $\sim 10$ \cite{LSST:2022kad}. 

However, the measured values of the Hubble constant obtained according to this method keep showing increasing tension with estimations derived from the cosmic microwave background angular distance as the data accumulates and the results become more precise. This tension has been highlighted in the latest paper from the Pantheon+ catalogue analysis by the SH0ES collaboration \cite{Riess:2021jrx}. \cref{sec:cosmotensions} will investigate this issue more thoroughly.


\begin{figure}[!h]
      \subfloat{\includegraphics[width=0.75\linewidth]{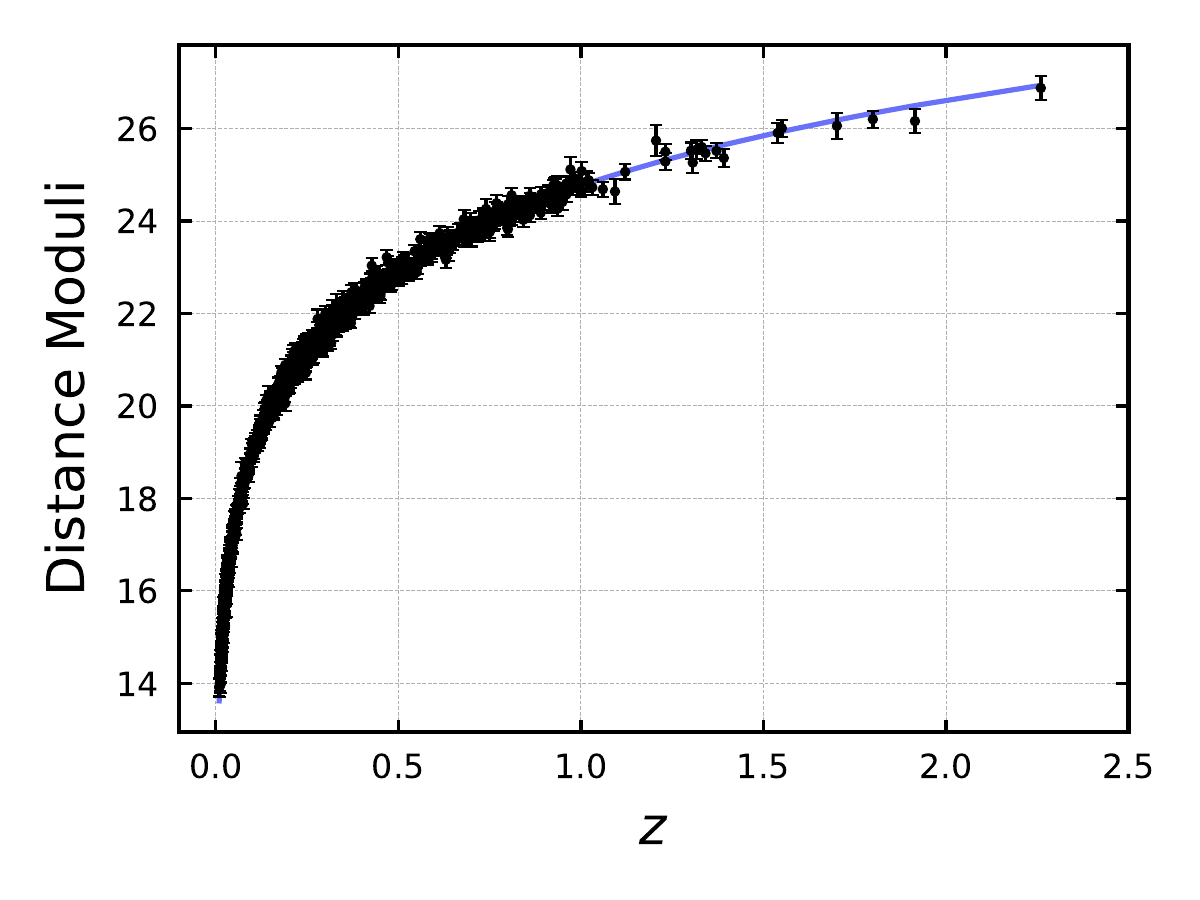}}
  \caption[The distance modulus measurements from Pantheon]{\label{fig:pantheon_lcdm} The distance modulus $\mu(z)$ as a function of redshift $z$ for the Pantheon dataset (black dots), along with the expected $\mu(z)$ predicted values according to the $\Lambda$CDM model (blue curve). The error bars show the uncertainties in the $1048$ data points used to construct the Pantheon dataset \cite{Scolnic:2017caz}.}
\end{figure}

\subsection{Standard Rulers: Baryonic Acoustic Oscillations} \label{sec:bao}


Before the recombination epoch, photons and baryons were tightly coupled, meaning that the sound waves that resulted in the temperature anisotropies in the CMB also left signatures in the baryon perturbations. The drag epoch refers to the era when photons were liberated from the baryons' drag due to Compton scattering. The oscillations produced by the acoustic waves in the CMB, as they interact with baryonic matter, also leave discernible signatures within the matter power spectrum, as it plays an instrumental role in determining the distribution of galaxies. The associated large-scale redshift-space correlation function displays a peak at the Baryon Acoustic Oscillations (BAO) scale at the sound horizon size $r_s$. The \textit{BAO peak} was initially observed at $\sim 100\,h^{-1}$ Mpc from a spectroscopic sample of luminous red galaxies from the Sloan Digital Sky Survey (SDSS) \cite{SDSS:2005xqv}. The precise location of this peak provided further evidence supporting the existence of a dominant dark energy component in the late Universe. The sound horizon at $z = z_{\text{drag}}$, which marks the baryons' release from the Compron drag of photons, can be defined as:

\begin{equation}
    r_s (\tau_{\text{drag}})= \int_0^{\tau_{\text{drag}}} c_s (\tau) \odif{\tau}\quad \xrightarrow{\odif{\tau} = \odif{a}/(a^2 H(a))}\quad   r_s (z_{\text{drag}})= \int_0^{(1+z_{\text{drag}})^{-1}} \frac{c_s (a)}{a^2 H(a)} \odif{a} \mathcomma
    \label{eq:rsdrag}
\end{equation}

where $c_s(\tau)$ denotes the sound speed of the photon–baryon plasma, given by:

\begin{equation}
c_s^2 = \frac{\delta p_{\gamma}}{\delta \rho_{\gamma} + \delta \rho_b} \mathcomma    
\end{equation}

with the subscripts $(\gamma,b)$ denoting photon and baryon quantities, respectively, and $\tau$ is the conformal time.

The power spectrum in redshift space, decomposed into wavenumbers parallel and perpendicular to the line of sight, provides the observed redshift and angular distributions of galaxies within redshift space. From this power spectrum, the following quantities can be obtained \cite{debook}, in analogy to \cref{eq:thetacmb} for the acoustic peak in the CMB:

\begin{align}
    \theta_{s} &= \frac{ r_s (z_{\text{drag}})}{(1+z)d_A(z)} \mathcomma \\
    \delta z_{s} &= \frac{ r_s (z_{\text{drag}}) H(z)}{c} \mathcomma
\end{align}

with $d_A(z)$ as previously defined in \cref{eq:distz}. The angle $\theta_{s}$ corresponds to observations perpendicular to the line of sight, and $\delta z_{s}$ corresponds to observations made along the line of sight.

Although current BAO data is still not sufficiently abundant to provide independent measurements of these two distance estimators, a combined distance scale ratio can be obtained from the spherically averaged power spectrum \cite{debook}:

\begin{equation}
    \left[ \theta^2_{s} \delta z_{s} \right]^{1/3} \equiv \frac{ r_s (z_{\text{drag}})}{\left[(1+z)^2\,d_A^2(z)c/H(z)\right]^{1/3}} \mathcomma 
\end{equation}

or the related effective distance \cite{SDSS:2005xqv}:

\begin{equation}
    d_V (z) \equiv \left[ (1+z)^2 d^2_A(z) \frac{cz}{H(z)} \right]^{1/3} \mathperiod
    \label{eq:dvbao}
\end{equation}

Thus, predictions for the distance scale $d_V (z)$ can be compared to observational data. These observations indeed reveal an amplified number of galaxies separated by a distance of roughly $150\, \text{Mpc}$ or simply $100 h^{-1}\, \text{Mpc}$, which, in turn, provides a method for calculating the expansion rate $H_0$ or its dimensionless counterpart $h$ ($H_0 = 100h\, \text{km}\,\text{s}^{-1}\, \text{Mpc}^{-1}$). 

\begin{figure}[!h]
      \subfloat{\includegraphics[width=0.75\linewidth]{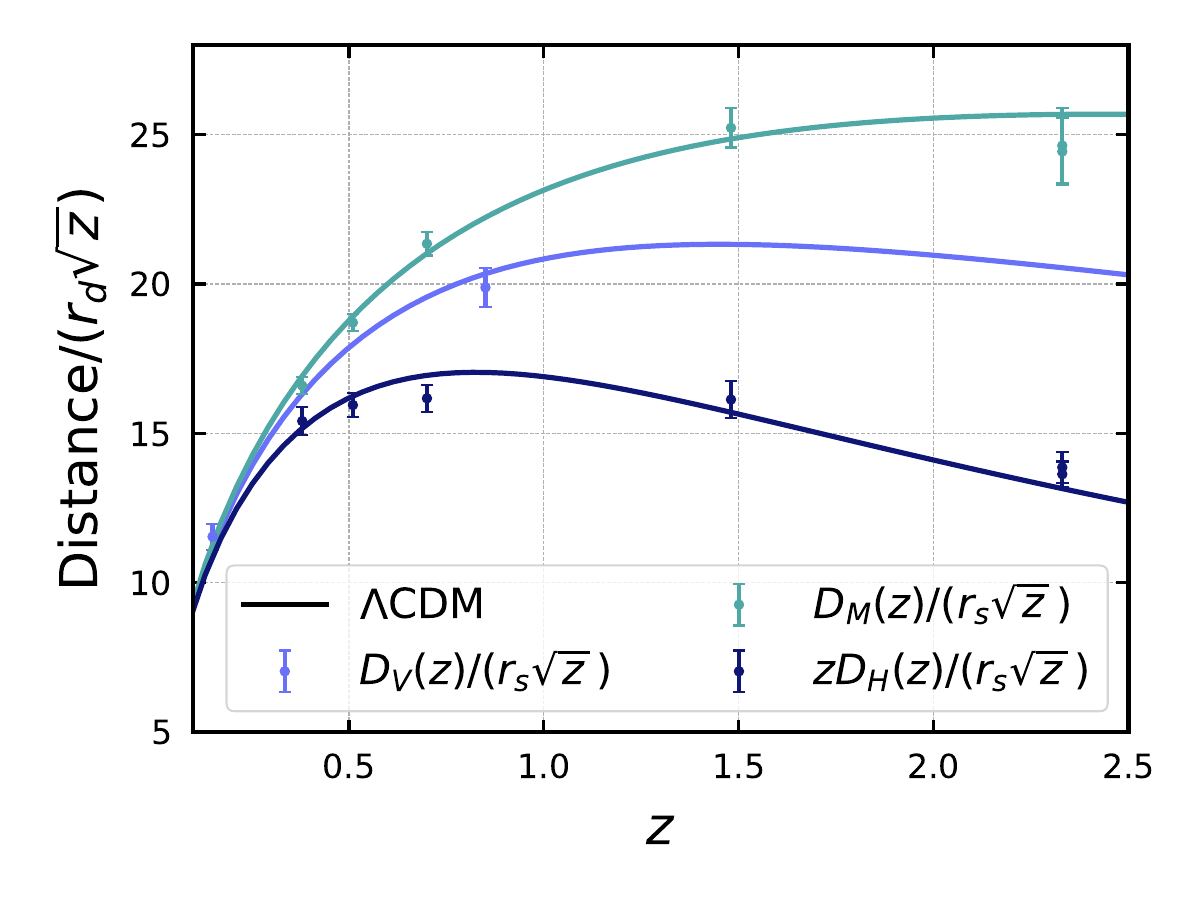}}
  \caption[BAO distance and expansion rate data from SDSS]{\label{fig:bao_lcdm} BAO distance and expansion rate data, derived from the SDSS collaboration's final results that span eight different redshift intervals, improved over two decades \cite{eBOSS:2020yzd}. The data includes isotropic BAO measurements denoted by $D_V(z)/r_s$, where $D_V(z)$ is the spherically averaged volume distance and $r_s$ is the sound horizon at baryon drag, both defined in \cref{eq:dvbao,eq:rsdrag}. It also comprises anisotropic BAO measurements, expressed as $D_M(z)/r_s$ and $D_H(z)/r_s$, where $D_M(z)$ is the comoving angular diameter distance and $D_H(z) = c/H(z)$ is the Hubble distance, both elaborated in \cref{eq:distz,eq:horizonh}. These measurements are detailed in Table III of \cite{eBOSS:2020yzd} (see references therein). The compilation of data to produce this plot was provided by William Giar\`{e}.}
\end{figure}

This probe focuses on the BAO peak in galaxy clustering, which differs from the acoustic oscillations observed in the CMB, as it primarily reflects the impact of low-redshift phenomena. The data collected from the Sloan Digital Sky Survey (SDSS) \cite{Alam:2016hwk,Ata:2017dya,BOSS:2016hvq,Ross:2014qpa,Beutler:2011hx}, amongst other sources, was instrumental in measuring this BAO peak. A recent collection of BAO measurements is depicted in \cref{fig:bao_lcdm}, along with the corresponding distance and expansion rate theoretical $\Lambda$CDM curves. It is an important probe for breaking degeneracies in the CMB data.
The large scale structures of the Universe can also be retrieved from the reciprocal weak lensing effect, which measures the response of photons to a gravitational potential, thereby mapping the mass distribution in the Universe through its effect on deflecting the light of background galaxies. For this purpose, large photometric and spectroscopic surveys are under development, including the 18th data release (DR18) from the fifth generation of the SDSS survey (SDSS-V) \cite{SDSS:2023tbz} and the Dark Energy Spectroscopic Instrument (DESI) \cite{Levi:2019ggs}. During the process of writing this thesis (July 2023), ESA's Euclid satellite was launched. This cosmological survey mission was designed to map the extragalactic sky, exploring the expansion history and the formation of structures, with the main aim of providing new insight on the nature and properties of dark energy and dark matter on universal scales \cite{Amendola:2012ys}. I am an active member of the consortium since 2022, as part of modified gravity and dark energy theory working groups.


\subsection{Standard Sirens: Gravitational Wave Detection} \label{sec:stsirens}


In the context of GR, and assuming propagation in a vacuum, gravitational waves follow the wave equation

\begin{equation}
    h_{P}'' (\tau,k) + 2 \mathcal{H} h_{P}' (\tau,k) + k^2 h_{P} (\tau,k) = 0 \mathcomma
\end{equation}

in which $h_{P} (\tau,k)$ are the Fourier modes of the gravitational wave amplitude, and the index $P = +,\, \times$ runs over the two possible polarisation states of the \ac{gw} \cite{Schutz:1986gp,zhao:2010sz}.

The groundbreaking detection of \ac{gw}s sourced by the merger of two black holes was first achieved in 2015 by the LIGO-Virgo collaboration \cite{grav}. This provided compelling evidence for the existence of gravitational waves, as remarkably predicted by GR. However, it can be argued that an equally noteworthy event for cosmology occurred two years later with the observation of a signal pointing to a binary neutron star merger along with a counterpart detection in the electromagnetic spectrum \cite{LIGOScientific:2017vwq,LIGOScientific:2017zic}. This immediately allowed the establishment of notably stringent constraints on the speed of the tensor modes, thereby ruling out theories of modified gravity that predict a speed of \ac{gw}s different from the speed of light \cite{LISACosmologyWorkingGroup:2019mwx}. 

Namely, these signals can serve as \textit{standard sirens}, which means they provide a method of constructing a Hubble diagram and derive estimations of the value of $H_0$, independent of the traditional distance ladder method \cite{Schutz:1986gp}. This complementary approach is possible because the amplitude of gravitational waves is practically inversely proportional to the luminosity distance $d_L (z)$ to the source event from which it originates \cite{Holz:2005df}. The electromagnetic counterpart signal provides redshift measurements, meaning that $H_0$ can be estimated using the relation between the comoving distance and the luminosity distance in \cref{eq:lumdist}.

Standard sirens cannot yet compete with other methods of estimating $H_0$, with the current constrained value subjected to significant errors: $H_0 = 70.0^{+12.0}_{-8.0}\, \text{km/s/Mpc}$ \cite{DiValentino:2018jbh}. However, the prospects for future observations are promising, as next-generation surveys can yield dozens or even hundreds of additional measurements of $H_0$ using standard sirens.

Presently, we have second-generation (2G) ground-based operational detectors, which include Virgo \cite{VIRGO:2014yos}, LIGO \cite{LIGOScientific:2014pky}, and the Kamioka Gravitational Wave Detector (KAGRA) \cite{Somiya:2011np}, which will be joined by the Indian Initiative in Gravitational-wave Observations (IndIGO) \cite{IndIGO}, planed for around 2030. The investment in maintaining and developing more detectors augments the potential of GW astronomy, increasing the number of detected events (covering a larger cosmic volume) and enhancing sky localisation (better triangulation to the source), which, in turn, facilitates counterpart searches.
However, their third-generation (3G) successors aim for increased sensitivity, precision, and a broader frequency range. The Einstein Telescope (ET) warrants particular attention, as it is predicted to boost the present sensitivity tenfold \cite{Punturo:2010zz}. Moreover, ET is projected to widen the redshift horizon, for instance, $z \sim 5$ for binary black holes (BBHs) in contrast to the $z\sim 0.5$ threshold of 2G instruments \cite{Sathyaprakash:2012jk}. Forecasts suggest that the number of traceable multi-messenger events might touch the count of tens of thousands of standard sirens \cite{Maggiore:2019uih}. While ground-based detectors might span a frequency spectrum of $1 \lesssim f\lesssim 10^3$ Hz \cite{Cai:2016sby}, forthcoming space-based 3G detectors, with particular focus on the Laser Interferometer Space Antenna (LISA) \cite{LISA:2017pwj}, could reach peak sensitivities close to $10^{-3}$ Hz and identify GW events beyond $z=20$. Other 3G GW detectors have been planned, such as the DECi-hertz Interferometer Gravitational Wave Observatory (DECIGO) \cite{Kawamura:2011zz}. In \cref{chap:gwcons}, we will explore a related application of standard sirens centred on ET and LISA for comparison and joint analysis of ground and space-based experiments. More precisely, we will forecast their constraining power for different models in direct comparison with current data from standard candles and rulers.

\subsection{Redshift Space Distortions} \label{sec:rsd}

As we have seen in the previous sections, the measurements of distances to galaxies rely on measuring their redshift. If the Universe were perfectly homogeneous, there would be a straightforward mapping from the radial distance (real space) to the redshift space \cite{Kaiser:1987qv}. However, we have just seen that the Universe is not homogeneous\footnote{We have assumed that there is a homogeneous distribution on cosmological scales, which is a different statement.}. 
The growth of inhomogeneous structures leads to peculiar velocities in galaxies moving in clusters that add to the expansion redshift and manifest through distortions in the mapping to redshift space. These distortions result in an apparent enhancement of large-scale clustering in the radial direction compared to the transverse direction, and the redshift distribution of galaxies appears disfigured. These \textit{redshift space distortions} (RSD) are a probe for the linear growth of structures and have been extensively reviewed in the literature, \textit{e.g.} in \cite{Hamilton:1997zq,Saito_2016}.

The linear growth rate $f(z,k)$ of structures is a differential measure of the evolution of the matter fluctuations:
\begin{equation}
    f (z,k) = \frac{\odif{\text{ln}} \delta_{\text{m}} (z,k)}{ \odif{\text{ln}\, a}} = \frac{1}{\hub} \frac{\delta_{\text{m}}' (z,k)}{\delta_{\text{m}} (z,k)} \mathcomma
\end{equation}
where the subscript $m$ stands for the collective matter component (baryons+CDM) \cite{Hamilton:2000tk}. As the name indicates, $f(z,k)$ is a measure of the rate at which the matter perturbations grow, highly correlated with the anisotropic clustering observed in redshift space distortions. Therefore, RSDs serve as a probe for the combined quantity $f \sigma_8$, which is the product of the growth rate and the root mean square mass fluctuation amplitude for spheres of size $8h^{-1}\, \text{Mpc}$, and which can be used as a normalisation factor in the matter power spectrum at that scale \cite{eBOSS:2020yzd}. More precisely, $\sigma_8$ is defined as
\begin{equation}
    \sigma_8^2 = \int_{0}^{\infty} W^2 (kR) \frac{P(k)}{k} \odif{k} \mathcomma
\end{equation}
where $W$ is a top hat filter function in Fourier space $k$
\begin{equation}
 W(kR) = 3 \left[ \frac{\sin (kR)}{(kR)^3} - \frac{\cos (kR)}{(kR)^2} \right] \mathperiod   
\end{equation}
Here, $P(k)$ denotes the matter power spectrum as defined in \cref{eq:mpk}, and $R$ represents the $\sigma_8$ scale radius $8h^{-1}\, \text{Mpc}$ \cite{liddleinf}. The combination $f \sigma_8$ becomes then
\begin{equation}
 f \sigma_8 (z,k_{\sigma_8}) = \frac{\sigma_8 (0,k_{\sigma_8})}{\hub} \frac{\delta_{\text{m}}' (z,k_{\sigma_8}) }{\delta_{\text{m}} (0,k_{\sigma_8}) } \mathcomma
\end{equation}
with $k_{\sigma_8} = 0.125 h$ Mpc$^{-1}$, and $\sigma_8$ is defined as
\begin{equation}
 \sigma_8 (z,k_{\sigma_8}) = \sigma_8 (0,k_{\sigma_8}) \frac{\delta_{\text{m}} (z,k_{\sigma_8}) }{\delta_{\text{m}} (0,k_{\sigma_8})} \mathperiod
\end{equation}
%

For observational and degeneracy-breaking purposes, a related parameter is defined as $S_8 = \sigma_8 \sqrt{\Omega_{\text{m}}/3}$. $S_8$ is conventionally employed to express the tension in measurements of $\sigma_8$ for weak lensing and CMB probes, and their correlation with measurements of $\Omega_{\text{m}}$, and will be the focus of \cref{sec:cosmotensions}.


\section{The Cosmological Parameters of the $\Lambda$CDM Model} \label{sec:lcdm_param}



\begin{figure}[!h]
      \subfloat{\includegraphics[width=\linewidth]{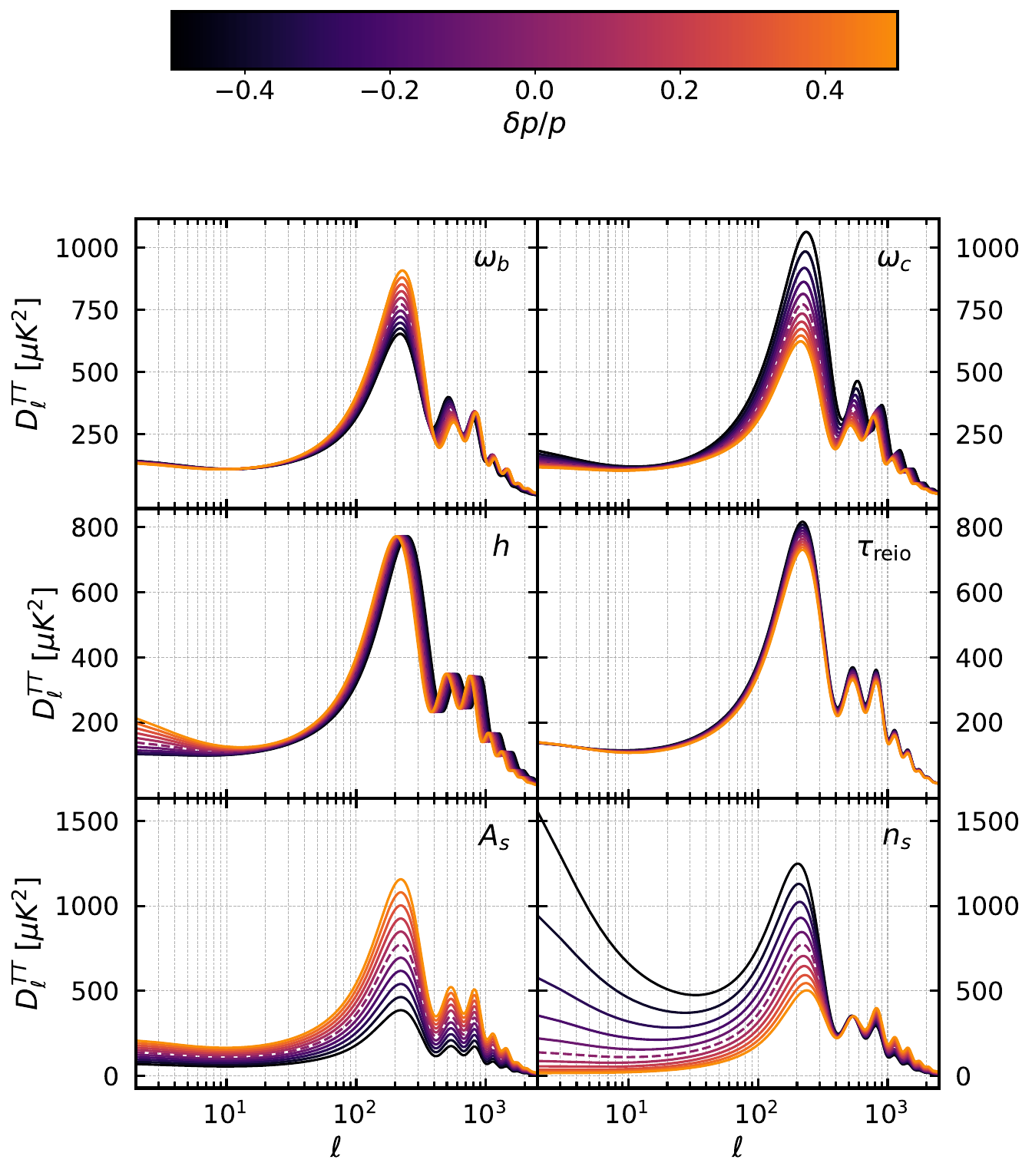}}
  \caption[The impact of varying each of the six $\Lambda$CDM parameters on the CMB temperature power spectrum]{\label{fig:varying_all} Figure with CMB temperature power spectrum variations from variations $\delta p$ of the standard $\Lambda$CDM cosmological parameters $p$. The $\Lambda$CDM fiducial case is depicted as a dashed line for reference. Variations in $\omega_b$ and $\omega_c$ implicitly hold $h$ fixed, while variations in $h$ hold the cold dark matter and baryon densities fixed. Figure produced using \texttt{CLASS} and inspired by \cite{Hart:2020mnx}.}
\end{figure}

The anisotropies of the CMB are dependent on the values of various parameters. The standard approach typically relies on six free and independent parameters chosen to avoid degeneracies and to improve the convergence of the model fit to the data.
 These parameters are $\{\omega_b, \omega_c, h, \tau_{\text{reio}},A_s,n_s \}$, which play a significant role in describing the primordial power spectrum of fluctuations, the reionisation epoch, the expansion rate, and the constituents of the Universe. In this section, we delve into the specifications of these parameters, exploring the implications of their individual variation on the power spectrum of the CMB for the $\Lambda$CDM model by keeping the remaining parameters fixed to \textit{Planck} 2018 fiducial values \cite{Aghanim:2018eyx}, given in \cref{tab:planckval}. The corresponding CMB temperature power spectrum change for variation of the six $\Lambda$CDM parameters is shown in \cref{fig:varying_all} and detailed below. 

\begin{table}[t!]
    \centering
    \begin{tabular}{l c c}
     \hline
     \multicolumn{3}{c}{Parameters of $\Lambda$CDM from \textit{Planck}}\\
    \hline
\multicolumn{2}{l}{Parameter} & Estimated value\\
\hline\hline
Baryon density & $\omega_{\text{b}}$ & $0.02236 \pm 0.00015$ \\
CDM density & $\omega_{\text{c}}$ & $0.1202 \pm 0.0014 $ \\
Dimensionless Hubble constant  & $h$ & $0.6727 \pm 0.0060 $\\
Optical depth & $\tau_{\text{reio}}$ & $0.0544^{+0.0070}_{-0.0081}$ \\
Scalar tilt & $n_s$ & $0.9649 \pm 0.0044$  \\
Scalar amplitude & $\ln \left( 10^{10}A_{s} \right)$ & $3.045 \pm 0.016 $  \\
\hline
Matter density    & $\Omega_m$ & $0.3166 \pm 0.0084 $ \\
$\Lambda$ density    & $\Omega_{\Lambda}$ & $0.6834 \pm 0.0084$ \\
Proxy angular scale of the sound horizon & $100 \theta_{\text{MC}}$ & $1.04090 \pm 0.00031 $ \\
Normalisation of matter power spectrum   & $\sigma_8$ & $0.8120 \pm 0.0073 $\\
$\sigma_8 (\Omega_m/0.3)^{1/2}$  & $S_8$ & $0.834 \pm 0.016$\\
Angular scale of the sound horizon   & $100 \theta_*$ & $1.04109 \pm 0.00030$\\
Sound horizon at recombination   & $r_{*}$ [Mpc] & $144.39 \pm 0.30$\\
Damping scale    & $k_D$ [Mpc$^{-1}$] & $0.14090 \pm 0.00032$\\
Reionisation redshift  & $z_{\text{reio}}$ & $7.68 \pm 0.79 $\\
Helium mass fraction & $Y_{P}^{BBN}$ & $0.246716^{+0.000062}_{-0.000055}$\\
Age of the Universe  & $t_0$ [Gyr] & $13.800 \pm 0.024 $\\
\hline\hline
    \end{tabular}
    \caption[Estimates of cosmological parameters in the six-parameter $\Lambda$CDM model derived from \textit{Planck}]{Estimates of the $\Lambda$CDM model cosmological parameters (upper section). The remaining parameters are either compatible with zero or can be derived (lower section) from the standard six parameters above \cite{Aghanim:2018eyx}. The results are obtained from the combination of all the \textit{Planck} spectra. The bounds on the parameters correspond to $1\sigma$ errors ($68\%$ CL).}
    \label{tab:planckval}
\end{table}


\begin{figure}[!h]
      \subfloat{\includegraphics[width=\linewidth]{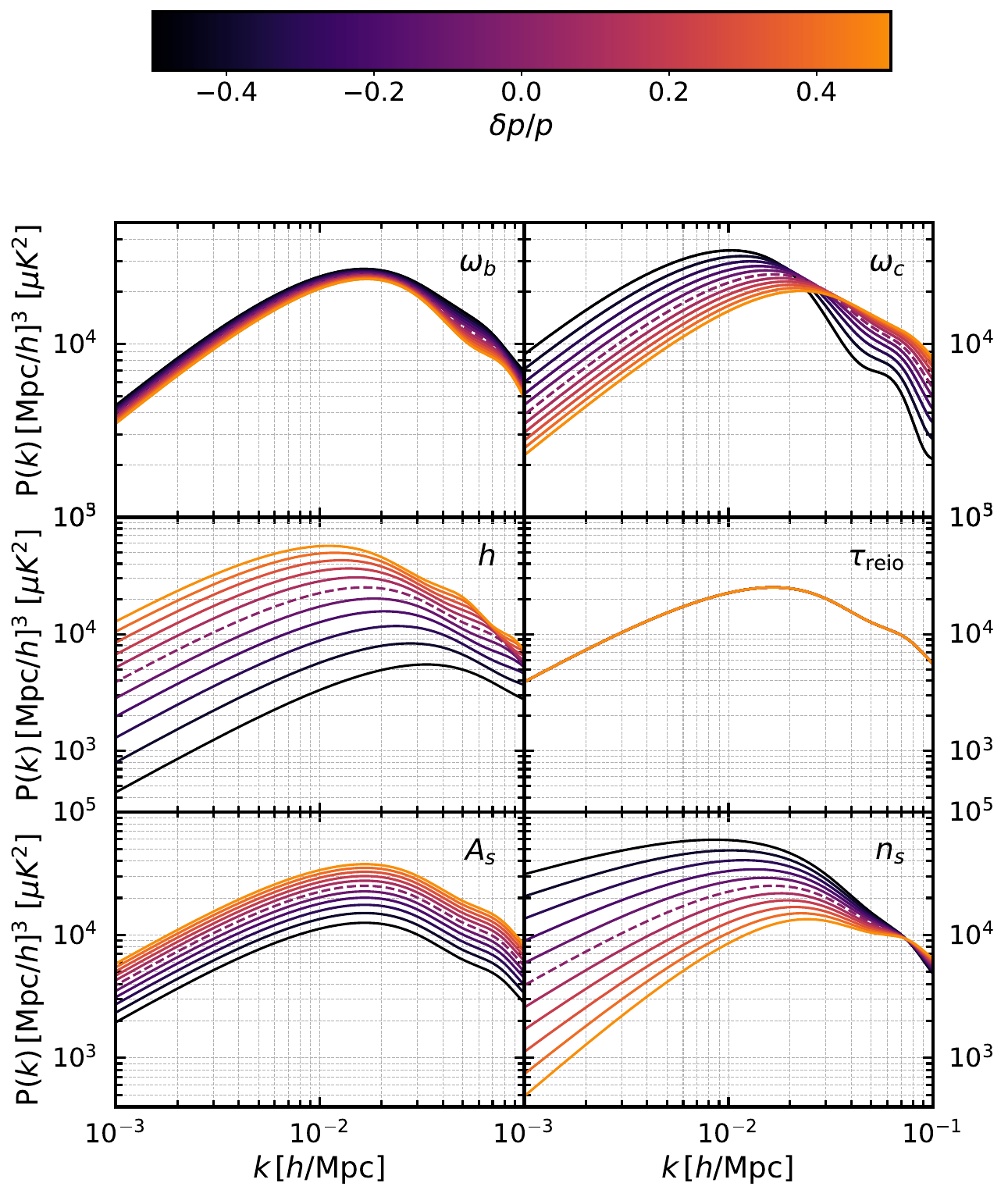}}
  \caption[The impact of varying each of the six $\Lambda$CDM parameters on the matter power spectrum]{\label{fig:varying_pk_all} Figure with variations in the matter power spectrum from variations $\delta p$ of the standard $\Lambda$CDM cosmological parameters $p$. The $\Lambda$CDM fiducial case is depicted as a dashed line for reference. Variations in $\omega_b$ and $\omega_c$ implicitly hold $h$ fixed, while variations in $h$ hold the cold dark matter and baryon densities fixed. Figure produced using \texttt{CLASS}.}
\end{figure}

\subsection{Baryon and Cold Dark Matter Densities: $\omega_{\rm b}$ and $\omega_{\rm c}$}

Having fundamentally different properties, each present matter density influences the anisotropies' imprint in distinct ways. 

The baryon density, denoted as $\omega_{\text{b}} = \Omega_{\text{b}} h^2$, alters the drag experienced by baryons within the tightly coupled baryon-photon fluid. Increasing the baryon abundance increases the drag, reducing the sound speed of the acoustic oscillations described in \cref{sec:bao}, as depicted in the upper left panel of \cref{fig:varying_all}. Consequently, larger overdensities will be generated by the cold dark matter potential wells, enhancing the odd-numbered peaks of the acoustic oscillations relative to the even-numbered ones associated with compression due to the gravitational potential. 
Moreover, while more sensitive to the total matter density $\omega_{\text{b}}+\omega_{\text{c}}$, baryons induce changes in the sound speed, which impact the size of the sound horizon, shifting all the peaks towards larger multipoles. 
While also influencing the free electron fraction $X_e$\footnote{$X_e=n_e/n_H$ represents the redshift dependence of the ionization/free electron fraction.}, higher values of $\omega_{\text{b}}$ are associated with an increase in the thickness of the last scattering surface, leading to a minor suppression of the diffusion damping at small multipoles (related to time variations in the optical depth, $\dot{\tau}_{\text{reio}}$).

In the upper left panel of \cref{fig:varying_pk_all}, we also identify an overall suppression (enhancement) of the matter power spectrum for higher (lower) values of the baryon density.
Higher baryon densities can amplify the large-scale BAO features in the matter power spectrum. Because baryons are more affected by pressure effects from structure formation or radiation compared to dark matter, it does not cluster as efficiently on small scales and higher values of $\omega_{\text{b}}$ induce a suppression of the amplitude of the spectrum.
Finally, a higher baryon density effectively increases the Jeans length (below which they do not collapse to form structure) and thus further suppresses structure formation on small scales.

Even though dark matter does not interact electromagnetically, it is responsible for the gravitational wells in which baryons and photons oscillate. This is illustrated in the upper right panel of \cref{fig:varying_all}, by an enhancement of the BAO features at large multipoles with the increase of $\omega_{\text{c}} = \Omega_{\text{c}} h^2$.
CDM has an opposite effect on variations of $\tau_{\text{reio}}$ compared to baryons. An increase in $\omega_{\text{c}}$ shifts the epoch of matter-radiation equality to larger redshifts. 
Since the cold dark matter's influence surpasses the photon-baryon fluid contribution to the gravitational potential after equality, this implies that the radiation pressure plays a less pivotal role after equality ($\tau > \tau_{\text{eq}}$). Being the motor of the baryon acoustic oscillations, the attenuation of the radiation's influence weakens the CMB anisotropies, particularly in the first peak, as the relative abundance of dark matter is increased \cite{dodelson2003modern}, a phenomenon known as \textit{radiation driving}. Dark matter is the primary driver of structure formation, with the variations in dark matter density becoming imprinted on the CMB through the ISW effect at low multipoles.

This is also reflected in the matter power spectrum depicted in the upper right panel of \cref{fig:varying_pk_all}, as a higher density of CDM will lead to more structure on all scales, effectively increasing its amplitude. Consequently, a higher $\omega_{\text{c}}$ is associated to larger values of $ \sigma_8 $. Higher CDM density results in faster structure growth, leading to more power on smaller scales and the peak of the matter power spectrum is shifted to smaller scales

\subsection{The Hubble Constant: $H_0$}

The Universe's expansion rate can be directly extrapolated from the peaks of the CMB TT-spectrum. The angular scale of the first peak, denoted as $\theta_{\text{rec}}$, sets the horizon size at the time of recombination and is parametrised as 
\begin{equation}
    \theta_{*} = \frac{s_{*}}{r_{s,*}} \approx \frac{\tau_{\text{rec}}}{\tau_0} \mathcomma \label{eq:thetarec}
\end{equation}
where $s_{*}$ is the sound horizon at recombination, approximately given by
\begin{equation}
    s_{*} = \int_0^{\tau_{\text{rec}}} c_s (\tau) \odif{\tau} \approx \frac{\tau_{\text{rec}}}{\sqrt{3}} \mathperiod
\end{equation}
The following peaks are located at an angular scale following the same pattern, that is, $\theta_{*,n} \approx n \theta_{*}$, placing the higher harmonics in the decomposition. 
A higher value of $ H_0 $ (and equivalently $h$) makes the angular diameter distance smaller, which means that the same physical scale corresponds to a larger angle on the sky. This shifts the acoustic peaks to larger angular scales (\textit{i.e.}, smaller multipoles), as illustrated in the central left panel of \cref{fig:varying_all}. 
The value of $ H_0 $ indirectly affects early-Universe physics through its relation to other cosmological parameters. 
The ISW effect, relevant at large angular scales (low $ \ell $), is sensitive to the rate of expansion related to the critical density of the Universe, which in turn determines the redshift at which matter and radiation densities become equal, with higher values of $H_0$ leading to an enhancement of the low-$\ell$ tail of the spectrum.

Analogously, in the central left panel of \cref{fig:varying_pk_all}, we identify an enhancement of the amplitude of the matter power spectrum for higher values of $h$, associated with a slight shift of the peak towards lower $k$-modes. There are different effects at play, but, more importantly, a faster expansion can suppress the growth of structure, leading to less clustering on large scales. At the same time, it also changes the maximum size of the cosmic structures by modifying the Hubble horizon, shifting the turnover peak to larger scales.

\subsection{Reionisation Optical Depth: $\tau_{\text{reio}}$}


Parameters like the optical depth to reionisation $\tau_{\text{reio}}$ capture the degree and timing of reionisation. A larger optical depth damps the CMB anisotropies on all scales because some fraction of the CMB photons have last scattered with free electrons not at the last scattering surface but more recently. This overall damping is a multiplicative effect across all multipoles $ \ell $ and is depicted in the central right panel of \cref{fig:varying_all}.
More precisely, increasing the optical depth leads to an overall suppression of the CMB power spectrum. This effect is highly degenerate with the one generated by the scalar power spectrum parameter $A_s$, as will be discussed next. This degeneracy is only broken by considering measurements of CMB polarisation \cite{Aghanim:2018eyx}. 

In the central right panel of \cref{fig:varying_pk_all}, we see that $\tau_{\text{reio}}$ has a negligible impact on the matter power spectrum for the scales considered. This is because the matter power spectrum reflects the distribution of matter at later times, long after reionisation has ceased. By this time, most of the imprints of this epoch have been \textit{washed out} or are secondary to other physical processes.

\subsection{Amplitude and Tilt of the Primordial Power Spectrum: $A_s$ and $n_s$}

The matter power spectrum, introduced in \cref{sec:mpk}, is sourced by the primordial (scalar) power spectrum, which in turn is characterised by the parameters $A_s$ and $n_s$, defined at a pivot scale of $k_* = 0.05\, \text{Mpc}^{-1}$ for \textit{Planck} observations. These determine the strength and the scale dependence of the initial density perturbations, respectively. 
A higher amplitude ($A_s$) means more pronounced density fluctuations in the matter distribution, leading to stronger clustering of matter and thus a more peaked matter power spectrum, as depicted in the lower left panel of \cref{fig:varying_all}.
The amplitude is directly related to $ \sigma_8 $ and directly scales the entire matter power spectrum, as shown in the lower left panel of \cref{fig:varying_pk_all}. 

The parameter $n_s$ accounts for the tilt of the power spectrum: for $n_s > 1$, there is more power on smaller scales; for $n_s < 1$, there is more power on larger scales, as depicted in the lower right panel of \cref{fig:varying_all}. Changing the tilt alters the relative height of the peaks and the angular scales at which they appear.
Since the tilt $n_s$ determines directly how the clustering power is distributed across different length scales, in the lower right panel of \cref{fig:varying_pk_all}, we see that larger values of $n_s$ lead to more power on smaller scales, making the matter distribution more clumpy at small scales, greatly suppressing the power on larger scales.

\subsection{Derived Parameters}

In addition to the standard six parameters mentioned above, other parameters embody more complex changes to the physics of the CMB anisotropies. These can be parameters of the standard model, which are usually kept fixed - such as the number of relativistic degrees of freedom $N_{\text{eff}}$, the fraction of primordial helium $Y_H$ or the mass of massive neutrinos $m_{\nu}$ - or parameters that account for the introduction of new physical degrees of freedom in \textit{alternative cosmological models}. The effect of some examples of both classes of modifications in the CMB anisotropies will be the focus of Part II. The estimated constraints on some of the relevant cosmological and derived parameters of $\Lambda$CDM for the \textit{Planck} 2018 analysis are reported in \cref{tab:planckval}.


\section{Cosmological Puzzles and Observational Tensions} \label{sec:cosmotensions}


 Albeit initially yielding great consistency confirmation in favour of $\Lambda$CDM, the recent advent of observational precision and techniques has revealed unexpected irreconcilable predictions from different probes, bringing to light a significant crisis for the standard model of cosmology \cite{Weinberg:1988cp, Weinberg:2000yb, martin,DiValentino:2020vvd,Abdalla:2022yfr}. These tensions arise from discrepancies between observations of the early and late stages of the Universe, such as those of the model-dependent CMB data analysis \cite{Planck:2018nkj,ACT:2020gnv,SPT-3G:2021wgf} and varied distance-ladder surveys \cite{Riess:2019cxk, Wong:2019kwg, Pesce:2020xfe,Freedman:2021ahq,Uddin:2023iob}, respectively. 
 Although some of these inconsistencies could be attributed to data errors and systematics, the statistical significance of roughly a $5\sigma$ tension suggests flaws in the standard model itself, with many extensions being proposed in the literature to ease this problem. 
Exploring alternative models may reveal new insights into the enigmatic nature of the dark sector, the physics of the early Universe, or even the fundamental assumptions of $\Lambda$CDM \cite{CANTATA:2021ktz}.

The level of concordance or discordance quantifies the tension $T_P$ (if any) on a parameter $P$ between the data sets $i$ and $j$. Considering the tension in the estimate of a single parameter with a posterior distribution that is approximately Gaussian, the difference between the mean values in the posterior $P_i$ and $P_j$ divided by the respective quadrature sum of the standard deviation of the respective data sets, $\sigma_{P_i}$ and $\sigma_{P_j}$, represents a robust tension metric:
\begin{equation}
    T_P \equiv \frac{P_i - P_j}{\sqrt{\sigma^2_{P_i} + \sigma^2_{P_j}}} \mathperiod
    \label{eq:tensiondef}
\end{equation}

But if we want to assess a tension in two or more parameters it can become inadequate, as not all of the information can be captured with one-dimensional projections. There is no unique and widely accepted metric for quantifying the tensions and for simplicity reasons we will always use the metric defined in \cref{eq:tensiondef} as a first approximation. Nevertheless we refer to \cite{DES:2020hen} for a more detailed account of this problem, including alternative metrics and illustrative examples.

\subsection{The Hubble Tension} \label{sec:hubt}



\begin{figure}[!h]
      \subfloat{\includegraphics[width=\linewidth]{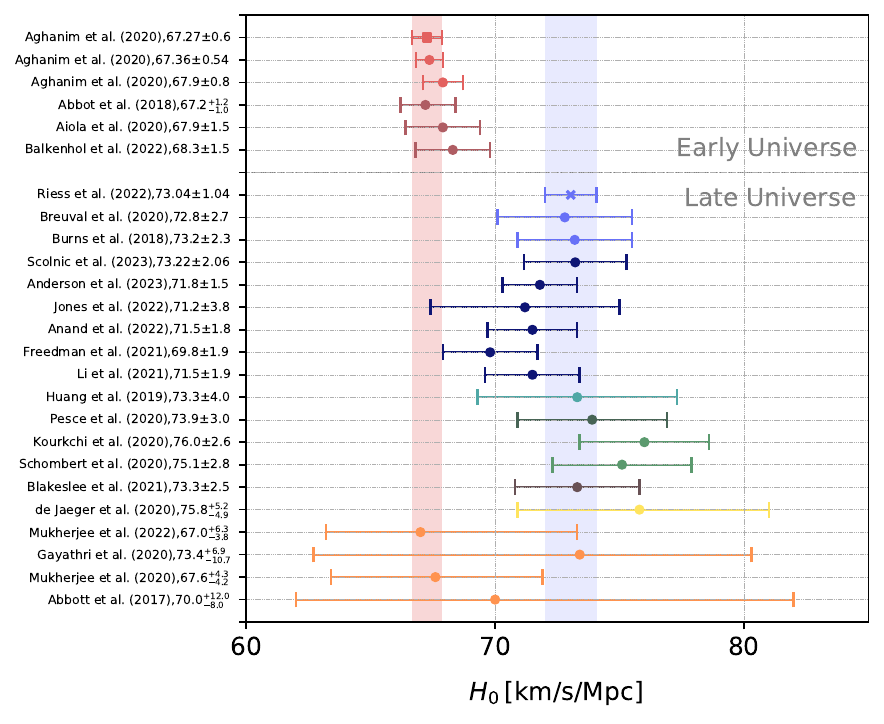}}
  \caption[Tension between estimates of $H_0$ from different probes]{\label{fig:H0WP} \sloppy Various recent estimates of $H_0$ from different probes, as detailed in \cref{sec:hubt}. The red and blue vertical bars illustrate the $68\%$ CL regions for the values measured by the \textit{Planck} \cite{Aghanim:2018eyx} and SH0ES \cite{Riess:2021jrx} collaborations. The dashed line separates measurements from probes based on the early and late Universe. The colours group the type of estimates: red for CMB with \textit{Planck}, dark red for CMB without \textit{Planck}, blue for SNIa-Cepheid, navy for SNIa-TRGB, sea green for SNIa-Miras, dark green for Masers, green for Tully Fisher, charcoal for surface brightness fluctuation, yellow for SNII, and orange for GW related measurements. Plot adapted from \href{https://github.com/lucavisinelli/H0TensionRealm}{https://github.com/lucavisinelli/H0TensionRealm} and based on \cite{DiValentino:2021izs,Abdalla:2022yfr}.}
\end{figure}

In broader terms, the Hubble tension often refers to the $\approx 5.0\, \sigma$ disagreement between the value of the Hubble constant predicted by the\textit{Planck} collaboration~\cite{Aghanim:2018eyx}, $H_0^{\textit{Planck}}=\left(67.27\pm 0.60\right)\, \text{km/s/Mpc}$ (red squared data point in the whisker plot of \cref{fig:H0WP}) at $68\%$ confidence level (CL), and the latest constraint from the \ac{shoes} collaboration R22~\cite{Riess:2021jrx}, $H_0^{\text{R22}}=(73.04 \pm 1.04)\, \text{km/s/Mpc}$ at 68\% CL (blue crossed data point in the whisker plot of \cref{fig:H0WP}), based on the supernovae calibrated by Cepheids. A CMB analysis by the Atacama Cosmology Telescope (ACT), combining data from the WMAP satellite, finds a value consistent with and independent from \textit{Planck}: $H_0^{\text{ACT+WMAP}}=(67.26 \pm 1.1)\, \text{km/s/Mpc}$ \cite{ACT:2020gnv}.
Nevertheless, it is crucial to recognise that there are multiple sets of measurements and estimates for $H_0$. These can be divided into two categories that are in agreement within their groups, reflecting the persistence of the $H_0$ tension: $(i)$ \textit{indirect model-dependent} estimates at \textit{early times} (such as CMB and Baryon Acoustic Oscillation experiments) that assume a $\Lambda$CDM cosmology, and $(ii)$ \textit{direct late-time model-independent} measurements (such as distance ladder and strong lensing). That is why the tension is said to arise between the CMB data and the direct local measurements.
This is depicted in the whisker plot of \cref{fig:H0WP} for a collection of recent $H_0$ estimates, showing the consistent trend between early and late time probes.

\subsection{The $S_8$ Tension} \label{sec:s8t}

\begin{figure}[!h]
      \subfloat{\includegraphics[width=\linewidth]{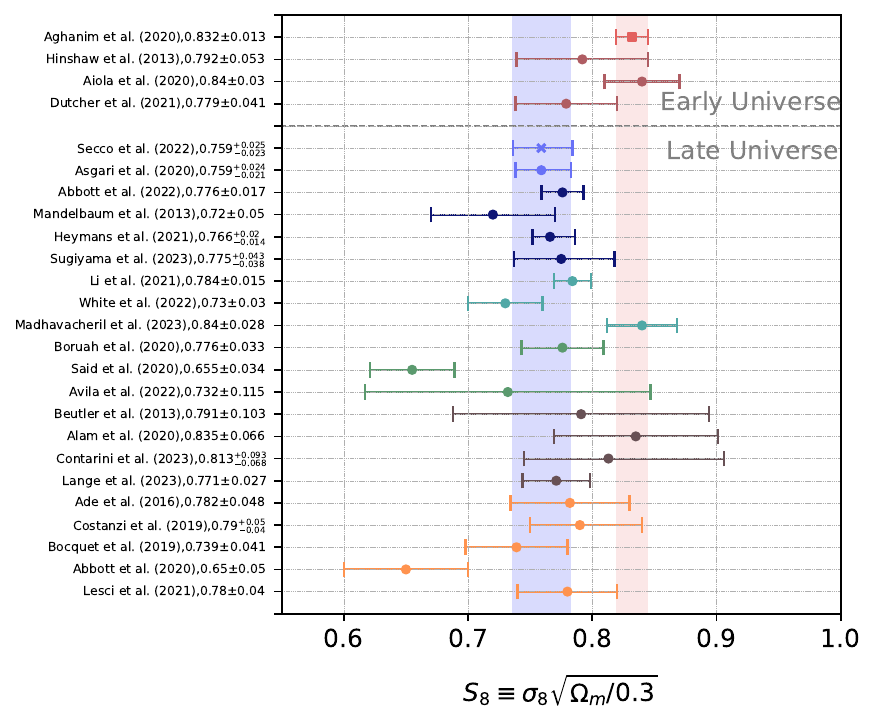}}
  \caption[Tension between estimates of $S_8$ from different probes]{\label{fig:S8WP} \sloppy Various recent estimates of $S_8$ from different probes, as detailed in \cref{sec:s8t}. The red and blue vertical bars illustrate the $68\%$ CL regions for the values measured by the \textit{Planck} \cite{Aghanim:2018eyx} and DES-Y3 \cite{DES:2021bvc,DES:2021vln} collaborations. The dashed line separates measurements from probes based on the early and late Universe. The colours group the type of estimates: red for CMB with \textit{Planck}, dark red for CMB without \textit{Planck}, blue for cosmic shear, navy for galaxy weak lensing, sea green for CMB weak lensing, green for peculiar velocities, charcoal for RSD, and orange for cluster abundance measurements. Plot adapted from \href{https://github.com/lucavisinelli/H0TensionRealm}{https://github.com/lucavisinelli/H0TensionRealm} and based on \cite{DiValentino:2021izs,Abdalla:2022yfr}.}
\end{figure}

Recent observations of the large-scale structure in the Universe have allowed us to place constraints on the clustering strength of matter. However, these constraints also differ from those reported by probes of the early Universe. More precisely, there exists a tension in the matter clustering power between the primary anisotropies of the CMB, as measured by the \textit{Planck} satellite, and lower redshift probes such as weak gravitational lensing and galaxy clustering.~\cite{Asgari:2019fkq,KiDS:2020suj,Joudaki:2019pmv, DES:2021wwk,DES:2021bvc, DES:2021vln, KiDS:2021opn,  Joudaki:2016kym,Heymans:2020gsg,Hildebrandt:2018yau,DES:2020ahh,Macaulay:2013swa,Skara:2019usd,Kazantzidis:2019nuh,Joudaki:2016mvz,Bull:2015stt,Kazantzidis:2018rnb,Nesseris:2017vor,Philcox:2021kcw} at the level of $2-3\sigma$. This tension is typically quantified in terms of the parameter $S_8$, defined as $S_8 \equiv \sigma_8 \sqrt{\Omega_{\rm m}/0.3}$, which regulates the magnitude of weak lensing measurements \cite{Li:2016bis,Gil-Marin:2016wya}.
The $S_8$ parameter is closely related to $f\sigma_8(z=0)$ measured by redshift space distortions (see \cref{sec:rsd}) and lensing probes, with $f=[\Omega_{\rm m}(z)]^{0.55}$ being a good proxy for the growth rate in GR, based simply on the matter density parameter $\Omega_{\rm m}$ at a given redshift $z$. In particular, lower redshift probes tend to favour a lower value of $S_8$ compared to the high redshift CMB estimates (red in the whisker plot of \cref{fig:S8WP}). As an example, the latest cosmic shear analysis of KiDS (KiDS-1000) and DES (DES-Y3) set $S_8^{\text{KiDS-1000}}=0.759^{+0.024}_{-0.021}$ \cite{KiDS:2020suj} and $S_8^{\text{DES-Y3}} = 0.759^{+0.025}_{-0.023}$ \cite{DES:2021bvc,DES:2021vln} (blue crossed data point in the whisker plot of \cref{fig:S8WP}), respectively, consistent with the $2-3\sigma$ tension with the fiducial \textit{Planck} analysis, which gives $S_8^{\textit{Planck}}=0.834\pm0.016$. The addition of secondary anisotropies from CMB lensing narrows the constraint to $S_8^{\textit{Planck}+\text{lens}}=0.832\pm0.013$ (red squared data point in the whisker plot of \cref{fig:S8WP}). The agreement with high and other low redshift measurements is recovered from the CMB lensing data alone. On the other hand, a combination of the high-$\ell$ data from ACT and the low-$\ell$ measurements of WMAP \cite{ACT:2020gnv} is compatible with the \textit{Planck} results $S_8^{\text{ACT+WMAP}}=0.840\pm0.030$, even though the error bars are more accommodating. 

It is important to note that unlike locally measured quantities such as the Hubble constant, $S_8$ relies on a specific cosmological model for its interpretation. In the cases reported in \cref{fig:S8WP}, the underlying model is always the standard flat $\Lambda$CDM model, which, despite providing a reasonable fit to the different data sets, yields a lower rate of structure formation compared to what observations seem to suggest. Therefore, to arrive at a fair and accurate comparison of $S_8$ values, it becomes imperative to reassess the analysis of weak lensing data while assuming a cosmological model beyond $\Lambda$CDM (as will be the focus of Part II), before quantifying any tension. This holds particularly true for cosmological models that exhibit variations in the growth of structure compared to the standard scenario. While not necessarily a trivial task, by conducting this reanalysis under different cosmological frameworks, we can gain a more comprehensive understanding of how $S_8$ is affected and whether the observed tension persists across various models. This approach promotes a more refined assessment of the discrepancies in $S_8$ values, disentangling the contributions of different cosmological parameters to the overall tension.


\subsection{Lensing Excess} \label{sec:Alens}

There is a potential indication of a systematic error in the \textit{Planck} data, known as the \textit{$A_{\rm lens}$ anomaly}. The parameter $A_{\rm lens}$ was introduced in Ref.~\cite{Calabrese:2008rt} as a rescaling factor for the effects of gravitational lensing on the CMB angular power spectra. The term \textit{anomaly} is employed since it is an \textit{unphysical} parameter computed through smoothing the peaks in the CMB damping tail. When $A_{\text{lens}}=0$, there is no lensing effect, while $A_{\rm lens}=1$ corresponds to the value expected in general relativity. In Ref.~\cite{Aghanim:2018eyx}, an analysis of $\Lambda$CDM+$A_\text{lens}$ finds $A_\text{lens} = 1.180 \pm 0.065$ at $68 \%$ CL. In fact, the \textit{Planck} CMB power spectra show a preference for $A_{\rm lens}>1$ at more than $95\%$ CL for both the standard \texttt{Plik} \cite{Planck:2019nip} and the alternative \texttt{CamSpec} \cite{Efstathiou:2019mdh,Efstathiou:2020wem} likelihoods, resulting in a significant improvement of the statistical fit. When BAO data is included, the evidence for $A_{\rm lens}>1$ strengthens to more than and approximately $99\%$ CL for \texttt{Plik} and for \texttt{CamSpec}, respectively.

Remarkably, the inclusion of the $A_{\rm lens}$ parameter in the analysis introduces a slight alleviation of the cosmological tensions; namely, it reduces the Hubble tension from $5\sigma$ to $3.9\sigma$ and mitigates the $S_8$ tension with weak lensing experiments to less than $2\sigma$, but is still insufficient to resolve the cosmic tensions fully. If $A_{\rm lens}$ is attributed to systematic errors in the \textit{Planck} data analysis, it is still valuable to understand how such systematics affect the constraints on $H_0$, $S_8$, and ultimately the cosmological tensions. Nevertheless, it should be noted that ground-based CMB experiments such as ACT and SPT, which are unaffected by the lensing excess, still find disagreement in the values of $H_0$ and $S_8$.

The evidence for $A_{\rm lens}>1$ can also be interpreted as evidence for missing physics in the theoretical frameworks. It may require considering a closed Universe, which poses challenges to several observational measurements and to the simplest inflationary models~\cite{DiValentino:2019qzk,Handley:2019tkm,DiValentino:2020hov}, or hint at alternative cosmological theories that modify GR~\cite{Planck:2015bue,DiValentino:2015bja,Aghanim:2018eyx}. If the $A_{\rm lens}$ anomaly is otherwise associated with some subtle and unobserved systematic error persistent through all releases of the \textit{Planck} data~\cite{Planck:2013pxb,DiValentino:2013mt,Planck:2015fie,Aghanim:2018eyx} then its impact on the cosmological tensions needs to be comprehended. Recent measurements from ACT, SPT-3G, and the combination of \textit{Planck} TT and SPTPol data all exhibit consistency with the standard lensing effect predicted in the $\Lambda$CDM model, \textit{e.g.} $A_{\rm lens}^{\text{ACT}}=1.01\pm0.11$~\cite{ACT:2020gnv}. This suggests that the lensing anomaly is unique to the \textit{Planck} data at small scales and not present in other CMB experiments.

\subsection{Hints for a Closed Universe from \textit{Planck} Data} \label{sec:closed}

The enhanced lensing amplitude signal ($A_{\rm lens}$ anomaly) observed in the CMB power spectrum~\cite{Calabrese:2008rt,Aghanim:2018eyx}, as indicated by various data releases from the \textit{Planck} collaboration, defies the standard $\Lambda$CDM model paradigm. One of the explanations that have been put forward is the possibility of having a closed Universe\footnote{Or something that mimics the same signal and is being expressed through the variation of the spatial curvature parameter.}, which is favoured over the flat scenario at $3.4\sigma$ level~\cite{DiValentino:2019qzk,Aghanim:2018eyx,Handley:2019tkm}. This preference for a closed Universe is ascribed to the strong degeneracy between the $A_{\rm lens}$ parameter and the spatial curvature parameter $\Omega_K$. Moreover, a closed Universe could potentially address the tension between the low and high multipoles of the angular CMB power spectrum~\cite{Addison:2015wyg,Planck:2016tof,DiValentino:2019qzk}. Nevertheless, both effects could be manifestations of hidden systematics in the \textit{Planck} data or even just statistical fluctuations.

However, while \textit{Planck} 2018~\cite{Aghanim:2018eyx} reports $\Omega_K=-0.044^{+0.018}_{-0.015}$, implying evidence for a closed Universe at about $3.4\sigma$, this evidence is reduced by replacing the baseline \texttt{Plik} likelihood~\cite{Planck:2019nip} with the alternative \texttt{CamSpec}~\cite{Efstathiou:2019mdh,Efstathiou:2020wem}. That being said, the estimated value $\Omega_K=-0.035^{+0.018}_{-0.013}$ still shows evidence above the $99\%$ CL. Other studies have also shown that taking a closed Universe in the context of $\Lambda$CDM leads to an increase in the dark matter abundance, which in turn exacerbates both the Hubble and $S_8$ tensions, leading to a drastically small prediction of $H_0 \sim 55 \text{km/s/Mpc}$~\cite{Aghanim:2018eyx, DiValentino:2019qzk, Handley:2019tkm} from \textit{Planck}.

However, alternative data combinations have been explored in the literature, leading to varying degrees of preference (or lack thereof) for a closed Universe. For example, the recent results from the ground-based experiment ACT, in an analysis including data from the WMAP experiment, are fully compatible with a flat Universe, $\Omega_K^{\text{ACT}}=-0.001^{+0.014}_{-0.010}$ \cite{ACT:2020gnv}. Bringing together the ACT data with a fraction of the \textit{Planck} data recovers some slight preference for the closed Universe scenario~\cite{ACT:2020gnv}. On the other hand, the combination of \textit{Planck} and BAO~\cite{Beutler:2011hx,Ross:2014qpa,Alam:2016hwk,Benisty:2020otr}, \textit{Planck} and CMB lensing~\cite{Planck:2018lbu} or \textit{Planck} and Pantheon data~\cite{Scolnic:2017caz}, favour a flat Universe in all cases. However, the robustness of this analysis is questionable since the combined data sets are in disagreement at more than $3\sigma$ when the curvature is left to vary~\cite{DiValentino:2019qzk,Handley:2019tkm,Vagnozzi:2020rcz}.

Moreover, while the spatial curvature prediction appears to be unaffected by local measurements of the Hubble constant ($H_0$) from the Cepheid distance ladder~\cite{Zuckerman:2021kgm}, its variation exacerbates both the $H_0$ and $S_8$ tensions~\cite{DiValentino:2019qzk}. Therefore, adding the curvature degree of freedom to the standard model cannot effectively account for the observed tensions and anomalies. 
Surveys such as the recently launched \textit{Euclid} \cite{Amendola:2012ys} have the potential to provide tighter constraints on $\Omega_K$ - optimistically of the order of $\sim 10^{-3}$ at 95\% CL - and help break parameter degeneracies~\cite{DiDio:2016ykq,Vardanyan:2009ft} and solve the anomalies in the \textit{Planck} data~\cite{Leonard:2016evk}.


   \cleardoublepage

     \chapter{Methodology} \label{chapter:statistics}
 \setcounter{equation}{0}
\setcounter{figure}{0}

 \epigraph{Vivemos exclusivamente no presente pois sempre e eternamente é o dia de hoje \\ e o dia de amanhã será um hoje, a eternidade é o estado das coisas neste momento.\blfootnote{\textit{We live exclusively in the present because it is always eternally today and tomorrow will be a today, eternity is the state of things at this very moment.} --- \textsc{Clarice Lispector} in The Hour of the Star} \\ --- \textsc{Clarice Lispector}\ \small\textup{A Hora da Estrela}}

This section outlines the relevant statistical methods and tools used in cosmology for analysing observational data, focusing on estimating model parameters. Any particular model can only make \textit{statistical predictions} about the Universe's properties, meaning that we need to optimise the amount of information that can be extracted from the available data to evaluate the validity and predictions of these models effectively. For a more detailed overview of statistical methods in cosmology, we refer to \cite{Liddle:2009xe,Trotta:2008qt,hobson_jaffe_liddle_mukherjee_parkinson_2009}.

In \cref{sec:stat} we provide a brief summary of the theory and methodology for statistical analysis and inference of parameters, including tests to assess the goodness of fit of a particular model to the data and its evidence compared with the standard $\Lambda$CDM case. In \cref{sec:cmbcodes} we present a brief review of the Einstein-Boltzmann codes in cosmology and their connection with MCMC codes for model sampling. We close with a detailed account of the data sets considered for the parameter estimates in this work.

 \section{Statistical Methods and Bayesian Inference} \label{sec:stat}




Cosmology is a unique science that fosters a strong synergy between theoretical model development, numerical analysis, and the planning and treatment of observational data. 
As new theories are put forth daily, data analysis in cosmology has become essential to test and possibly falsify their statistical support. This is part of the quest to find a framework that can handle vast amounts of data and address existing incompatibilities while navigating complexities such as theoretical biases, model independence, overfitting due to excessive parameters, and parameter space degeneracies. Ultimately, developing and improving current statistical methods linking precision data with accurate theoretical predictions is paramount. Due to its nature, analysis problems are inherently inverse problems, \textit{i.e.} from collected data, we aim to deduce something about the underlying physical processes that generated the data. For instance, we may use Planck CMB data to fit the $\Lambda$CDM model and then infer \textit{e.g.} the relative amount of dark matter or baryons from these observations.

Hence, we may categorise the data analysis process into three classes: 

\begin{enumerate}
    \item \textit{Bayesian inference} is the tool used to transform observations into constraints on theoretical models, which should balance complexity for data explanation and simplicity for concrete prediction.

    \item \textit{Parameter estimation} aims to constrain the values of the model's parameters and their uncertainties, considering errors from experimental and inherent statistical uncertainty from physical process randomness. A prominent example is the limited predictability of the CMB pattern due to quantum processes during inflation, which can nonetheless still be statistically modelled and taken into account along with measurement uncertainties. 

    \item \textit{Model selection} involves selecting among multiple models proposed to explain the observational data, each representing different physical processes and offering a unique set of parameters to be varied for data fitting.
\end{enumerate}

\subsection{Bayesian Inference}

Bayesian inference is the dish of the day in the statistical analysis methods menu. Other subjects have traditionally favoured the frequentist approach, which relies on the repeatability of the experiment.
The essence of Bayesian analysis lies in assigning probabilities to all variables, handling them based on a set of rules that embody Bayes' theorem. This theory aims to keep updating our understanding as new data comes to light. This means that while all subsequent steps follow some algorithm, a significant aspect of this approach is the need to quantify our base knowledge before gathering new data. This is called the \textit{prior probability}, which can be subject to different views among researchers.

The \textit{posterior probability}, the probability that the model's varied parameter will adopt specific values after the data is collected, is represented as $p(\theta | d)$. Here, $\theta$ represents the model's unknown parameter, while $d$ represents the observed data. Using this formulation, one can determine the expected values of the parameters and their associated uncertainties.

Usually, it is easier to calculate the reverse probability, $p(d|\theta)$. For example, we consider a Gaussian model with mean $\mu$ and variance $\sigma^2$. This model includes two parameters, $\theta=(\mu, \sigma)$, and the probability of a single variable $d$ in relation to these parameters is described by 
\begin{equation}
    p(d| \theta) = \frac{1}{\sqrt{2 \pi} \sigma} \exp \left[ - \frac{(d-\mu)^2}{2 \sigma^2}\right] \mathcomma
    \label{eq:gaussdist}
\end{equation}
which is not what we were initially looking for.
Nonetheless, we can connect this with $p(\theta, d)$ by applying the \textit{Bayes' Theorem}:
\begin{equation}
p \left(\theta | d\right) = \frac{p \left( d | \theta \right)\, p \left( \theta \right)}{p \left( d \right)} \mathcomma
\end{equation}
which expresses the confidence degree in the values of $\theta$ according to the information given by the data $d$, or in other words, the probability that the parameters $\theta$ in the theory can be explained by the data $d$. Here, $p(\theta|d)$ is the \textit{posterior probability} of the parameters, while $p(d|\theta)$ is referred to as the \textit{likelihood}, also symbolised by $\mathcal{L} (d; \theta)$. The \textit{prior probability}, $p(\theta)$, expresses our knowledge about the parameters before performing the experiment. This knowledge can come from previous experiments or theoretical understanding. In cases where no previous information exists, the prior is usually assumed to be a constant, also known as a \textit{flat prior}. 

The \textit{evidence}, represented as $p(d)$, is used to normalise the probabilities:
\begin{equation}
    p(d) = \int  p(d|\theta) p(\theta) \odif{\theta} \mathperiod
    \label{eq:evdef}
\end{equation}

For estimating parameters, the evidence does not depend on the relative probabilities of the parameters; hence, it is often overlooked. Nevertheless, it is worth noting that the evidence plays a significant part in \textit{model selection}, relating what we learn about $\theta$ after seeing the data (the posterior) to what we knew about $\theta$ before accessing the data (the likelihood and the prior). Therefore, in some way, it measures how much we have learned or how much our knowledge has been updated from the prior to the posterior. This is especially relevant when multiple theoretical models are being evaluated, and there is a need to distinguish between them, regardless of their respective parameters (this will be the focus of \cref{sec:modelsel}).

Assuming flat priors greatly simplifies the computation of the posterior: 
\begin{equation}
   p(\theta|d) \propto \mathcal{L} (d; \theta) \mathperiod
\end{equation}

Even though we might possess the complete probability distribution for the parameters $p(\theta)$, often it suffices to use the peak of the distribution as the parameter estimate. This method is known as a maximum likelihood estimate. However, if the priors are not flat, the posterior peak may not coincide with the maximum likelihood estimate.

We are generally interested in calculating the \textit{expectation value} of a parameter, or a set of derived parameters, $f(\theta)$, from the posterior distribution, also called the \textit{mean value} of $f(\theta)$:

\begin{equation}
    \langle f(\theta) \rangle = \int f(\theta) p(\theta | d) \odif{\theta} \mathperiod
\end{equation}


When dealing with multi-parameter models, $\theta= \theta_1,...,\theta_i,...,\theta_n$ (for $n>1$ parameters), it may be necessary to extract from the multivariate posterior distribution the posterior of a subset of parameters. This reduction of the posterior probability to a lower-dimensional subspace is accomplished via a process called \textit{marginalisation}:
\begin{equation}
f(\theta_i) = \int ... \int p(\theta_i | d) \odif{\theta_1} ... \odif{\theta_{i-1}}\, \odif{\theta_{i+1}} ...  \odif{\theta_{n}} \mathcomma
\end{equation}
where $\theta_i$ represents one of the elements of the $n$-dimensional subspace of parameters $\theta$. This is especially important when there are multiple \textit{nuisance parameters} related to the experiment and the method itself, which are not informative for the cosmological analysis.

The set of parameters most favoured by the data can be determined by finding the parameters that maximise the posterior probability $p(\theta|d)$:
\begin{equation}
    f(\theta)\mid_{m} = | \text{max}_{\theta}\ p(\theta | d) | \mathperiod
\end{equation}
This is usually referred to as the \textit{best-fit}, although strictly speaking, the term is used for the parameter values that maximise the likelihood and it applies strictly to $f(\theta)\mid_{m}$ in cases of uniform priors. Alternatively, the mean parameter values can be estimated by computing the posterior mean, also known as the \textit{marginalised mean}.

Lastly, the confidence level for the derived parameters can also be computed as follows:

\begin{equation}
f(\theta_i) = \int_{R(r)} p(\theta_i | d) \odif{\theta}^n = r \mathperiod
\label{eq:expval}
\end{equation}

In this equation, $r$ is a fraction associated with the likelihood that the parameters will fall within a given parameter area $R$. The fractions often analysed are $r = 0.683, 0.954, 0.997$, which correspond respectively to the $1\sigma$, $2\sigma$ and $3\sigma$ confidence levels.

  \subsection{Parameter Estimation}

The integral in \cref{eq:expval} can be computed numerically, limited to the maximum precision permitted by the computing system in use. For cases where the number of parameters, also known as dimensions, $n$, exceeds one, the most straightforward approach is to discretise the problem and sum over a uniformly distributed grid in the parameter space.
This grid must have a large enough volume to cover all the regions in which $f(\theta) p(\theta | d)$ is non-zero, with some particular width and resolution in each direction, depending on the smoothness of $f(\theta)$ and $p(\theta | d)$.
Nevertheless, when dealing with a parameter space of high dimensionality, evaluating the likelihood on a grid quickly becomes computationally expensive as the number of grid points exponentially increases with the dimension. Fortunately, there are more efficient ways to sample the likelihood surface, such as focusing on areas with a high likelihood. A popular approach in cosmology is the use of Markov Chain Monte Carlo (MCMC) methods \cite{Metropolis_1987,hobson_jaffe_liddle_mukherjee_parkinson_2009}.

Monte Carlo methods use random sampling, guided by some proposed distribution and acceptance criteria until the desired result is reached. In this case, we will obtain a chain of posterior samples with an expected number density proportional to the posterior. A notable subclass is the MCMC methods, which are characterised by their sequential nature, where each step depends solely on the one before. This sequence is known as a \textit{Markov chain}. Each step corresponds to a specific value of the parameters for which the likelihood is evaluated.

The elements of the Markov chain are structured to represent random samples from the posterior parameter distribution, meaning that each chain element shows the probability of its corresponding parameter values being correct. A straightforward method that accomplishes this is the \textit{Metropolis-Hastings} algorithm \cite{10.1063/1.1699114,b1878965-730b-33c7-b1ad-dfd3acb6f61b}, described in \cref{alg:MHalgorithm}

\begin{algorithm}[ht]
    \caption{\textbf{Metropolis-Hastings Algorithm}}
    \label{alg:MHalgorithm}
    \begin{description}
\item [Step 1.] Take a starting point in the parameter space randomly selected from the prior density.
\item [Step 2.] Propose a random candidate iteration based on some probability distribution for the length and direction of the jump, as long as it fulfils the \textit{detailed balance condition}, which states that going back to the starting point is as likely as the iteration away from it. This can be accomplished using a symmetric \textit{proposal function}, for example, a multivariate Gaussian about the current point. Calculate the likelihood at the new point and, consequently, the posterior by multiplying by the prior.
\item [Step 3.] If the probability at the iteration candidate is higher, accept the jump. If it is lower, accept the jump with a probability determined by the ratio of the posterior probabilities at the new and old points. If the jump is not accepted, remain at the same point, adding a duplicate to the chain.
\item [Step 4.] Repeat the procedure from Step 2 until the probability distribution is accurately mapped. This can be checked by comparing chains starting from different points and using convergence statistics tests.
    \end{description}
\end{algorithm}

The possibility of moving to a point with a lower probability ensures that the algorithm thoroughly explores the shape of the posterior near its maximum and does not get stranded in a local maximum.
The algorithm generally begins in a region with low likelihood and migrates toward regions with high likelihood. The initial phase, known as the \textit{burn-in}, depends on the starting point and must be eliminated during the chain's analysis. Once near the peak, most potential jumps are in areas of lower probability, and the chain roams around the peak, mapping its shape.
The proposal function should be adjusted according to the scale of variation of the likelihood near its maximum. A Gaussian function is commonly chosen, and its axes should ideally align with the principal directions of the posterior to quickly navigate along parameter degeneracies. Thus, a short initial run is often performed to map out the posterior distribution roughly, and this mapping is used to optimise the proposal function for the actual computation. It can be shown that the chain should sample the target distribution fairly once it has reached a stationary distribution, \textit{i.e.} once there are no significant differences between consecutive steps.

In practice, many chains are generated, starting from arbitrarily chosen positions in the parameter space, which should be sufficiently separated from each other. This will generate a \textit{random walk}, following an algorithm that assigns higher jumping probabilities to iterations in which the parameters contribute considerably to the distribution. On the other hand, it also implies that this algorithm is quite inefficient if the distribution is multi-modal, in which case other sampling methods may be necessary. 
An alternative is the Nested Sampling algorithm \cite{Feroz:2007kg}. This approach simplifies the problem by not sampling from the entire likelihood distribution but rather by just sampling within progressively more significant regions. 
Instead of constructing chains based on a random point from which we sample, we uniformly sample within some hard bound on the likelihood. As we sample these \textit{live} points, they are used to define a yet-to-be-determined contour of potential likelihood regions. We aim to generate a sample within this region, repeat the process based on the new sample, and continue this contraction cycle. 
The essence of the method is that instead of generating a single chain that we sample across the entire distribution, we aim to segment the distribution into numerous discrete levels or chunks. We address a simplified problem within each segment that does not sample from the full likelihood distribution but only within this unspecified region using a certain prior transform function.


The idea is that once the chains converge, the values deduced from each chain, like the sample mean and variance, should align. More specifically, this method compares the variance (whether on a single parameter or all of them) inside a chain with the variance across multiple chains. This requires running multiple chains, looking at the variance of the parameter away from its mean within each chain, and comparing it to the variance in the mean of that parameter across chains. Exactly how the Gelman-Rubin diagnosis handles these variances is complex. However, the important question for convergence is whether, as the chains get longer, these two variances asymptotically reach stable values and whether those two values agree. This is measured as the "$r-1$ value" ($r$ is expected to be $1$ for full convergence of the chains), and a usual criterion is to assume the MCMC has converged if $\max(r-1) \approx 10^{-2}$ across all the sampled parameters. We will impose this condition for all the analyses in this dissertation.

  \subsection{Model Selection Methods} \label{sec:modelsel}

Many of the questions that we wish to answer in cosmology go beyond parameter inference and fall into the model comparison realm. Examples of questions we will be asking in this dissertation are:
\begin{itemize}
    \item Does the Universe have a non-vanishing spatial curvature?
    \item Is dark energy dynamic?
    \item Is there evidence for a coupling in the dark sector?
\end{itemize}

It is important to highlight the fundamental difference between model fitting and selection. The model fitting process relies on assuming that a particular model is the true model and extracting the constraints that provide the best possible fit to the available data. On the other hand, in model selection, we are interested in assessing the level of compatibility of each model with the data. 
In exceptionally simple cases, we can select between models by comparing the maximum likelihood value, but this is not generally valid. A model with more degrees of freedom will always give an equal or larger maximum likelihood, irrespective of the data: model complexity is increased, then the maximum likelihood value will generally increase, often at the cost of over-fitting the data. 
This means an accurate model selection method should account for and penalise for the over-fitting present in models with increased complexity. The key difference is in comparing posteriors instead of likelihoods.

\subsubsection{Model Fit to Data}

If the distribution functions are Gaussian, assuming flat priors, then maximising the likelihood of a given model is essentially equivalent to the least square method of minimising the $\chi^2$ (best fit of parameters), which according to \cref{eq:gaussdist} corresponds to the exponent of the Gaussian distribution 
\begin{equation}
    \mathcal{L} (d|\theta) \propto \exp(- \chi^2/2) \mathcomma
\end{equation}
which is proportional to the peak in the posteriors for the flat-prior case. For non-Gaussian distributions (\textit{e.g.} a \textit{Poisson} distribution), minimising the $\chi^2$ does not necessarily maximise the likelihood that the fitted parameters reflect the data. In those cases the minimum $\chi^2$ might still yield a good first approximation but the best-fitting parameters will generally be found through the maximum likelihood estimation $\nabla_{\theta} \mathcal{L} (d|\theta) = 0$ instead \cite{HAUSCHILD2001384}.

For a simpler analysis, we can consider information criteria for approximate model comparison, keeping in mind that these make use of considerably strong assumptions about the posterior distribution:
\begin{itemize}
 \item Akaike Information Criterion: $\mathrm{AIC} \equiv \chi^2_{\rm eff} + 2k$;
 \item Bayesian Information Criterion: $\mathrm{BIC} \equiv \chi^2_{\rm eff} + k \ln (n)$;
 \item Deviance Information Criterion: $\mathrm{DIC} \equiv \chi^2_{\rm eff} + 2 \left[ \bar{\chi}^2_{\rm eff} - \chi^2_{\rm eff} \right]$;
 \end{itemize}
where $k$ is the number of fitted parameters, $n$ is the number of data points, the terms 

\begin{equation}
  \chi^2_{\rm eff} = \chi^2_{\rm min} \equiv -2 \ln (\mathcal{L}_\mathrm{max}) \mathcomma  
  \label{eq:stat_chieff}
\end{equation}

give the $\chi^2$ of the best-fit, and the upper bar denotes quantities computed at the average of the posterior distribution. 

In this case, the best model, the one that provides the best fit to the data, must minimise the AIC/BIC/DIC. However, each criterion penalises models differently, with the AIC being the simplest measure, only accounting for the extra parameters. The BIC includes a stronger penalty for models with a larger number of free parameters $k$ when the number of data points $n$ is sufficiently large and gives a better approximation to the full Bayesian evidence in the large $n$ limit. On the other hand, the DIC also considers whether information about the parameters is gained by evaluating the difference according to the average of the posterior distributions. Because of all these subtleties, it is always preferable to calculate the full Bayesian evidence (next section), although this may be non-trivial for more complex cases, in which the information criteria already provide some useful insight.

\subsubsection{Model Comparison} \label{sec:modelcomp}


A unique combination of variable parameters and a prior distribution for these parameters distinguishes each model in our study. A crucial aspect of Bayesian analysis is that it differentiates between a model where a quantity is fixed to a specific value and a more flexible model in which that parameter is allowed to vary, even if occasionally assuming that particular value. To exemplify this, we shall consider two competing models, labelled as $M$ and $N$, with $N$ being a simpler model containing fewer parameters ($n_N < n_M$). In addition, we assume that model $N$ is \textit{nested} within model $M$, meaning the $n_N$ parameters of model $N$ are also present in $M$, which has a total of $p = n_M - n_N$ additional parameters. In model $N$, these additional parameters are set to reference values. We use $d$ to denote the data vector, and $\theta_M$ and $\theta_N$ to denote the parameter vectors (of lengths $n_M$ and $n_N$ respectively). Bayes' theorem determines the posterior probability of each model:

\begin{equation}
    p(M|d) = \frac{p(d|M) p(M)}{p(d)} \mathcomma
\end{equation}

where $p(M)$ represents the prior information on the model $M$ and a similar equation applies for model $N$. The term $p(d|M)$ is the \textit{evidence} which we have already encountered in \cref{eq:evdef}, where the $|M$ bit was discarded since we were focusing on one model only. Accordingly, it can be written in terms of its marginalisation over the parameter space, leading to:

\begin{equation}
    p(d|M) = \int \odif{\theta_M} p(d|\theta_M;M) p(\theta_M|M) \mathcomma 
\end{equation}

which is a multi-dimensional integration. The ratio of the posterior probabilities for the two models can be written as:

\begin{equation}
    \frac{p(N|d)}{p(M|d)} = \frac{p(N)}{p(M)} \frac{ \int \odif{\theta_N} p(d|\theta_N;N) p(\theta_N|N)}{\int \odif{\theta_M} p(d|\theta_M;M) p(\theta_M|M)} \mathperiod
\end{equation}

If we do not have any preference towards either model, \textit{i.e.}, $p(N) = p(M)$, this ratio simplifies to the ratio of the evidence, which is referred to as the \textit{Bayes Factor} \cite{MOREY20166},

\begin{equation}
   B_{N;M} =  \frac{ \int \odif{\theta_N} p(d|\theta_N;N) p(\theta_N|N)}{\int \odif{\theta_M} p(d|\theta_M;M) p(\theta_M|M)} \mathperiod
\end{equation}

The more complex model $M$ inevitably yields a higher likelihood than the simpler nested model $N$. However, the evidence tends to favour the simpler model if the fit is nearly as good due to the smaller \textit{prior volume}. If we assume uniform, and therefore separable, priors for each parameter, we can express the volume as $p(\theta_M|M) = (\Delta \theta^M_1 ... \Delta \theta^M_{n_M})^{-1}$ and

\begin{equation}
   B_{N;M} =  \frac{ \int \odif{\theta_N} p(d|\theta_N;N)}{\int \odif{\theta_M} p(d|\theta_M;M)} \frac{(\Delta \theta^M_1 ... \Delta \theta^M_{n_M})}{(\Delta \theta^N_1 ... \Delta \theta^N_{n_N})} \mathperiod
   \label{eq:stat_be}
\end{equation}

We must integrate the likelihood across the parameter space to derive the evidence. While this is a standard mathematical problem, its complexity is ascribed to the most likely case that the integrand is extremely sharply peaked, and we cannot predict where this peak will occur in the parameter space. Moreover, the multi-dimensional parameter space makes individual likelihood evaluations computationally expensive. Bayesian model selection techniques rely on efficient algorithms capable of handling this type of integral. It is useful to establish a reference scale to evaluate differences in the evidence. The Jeffreys scale \cite{Jeffreys1939-JEFTOP-5} is a standard criterion that measures the strength of evidence favouring a model. A revised version proposed by Kass and Raftery \cite{Kass:1995loi,Trotta:2008qt} is presented in the following table:

\begin{table}[h]
\centering
\begin{tabular}{|c|c|c|c|}
\hline
$|\ln{B_{N;M}}|$ & Fractional Odds & Model's Probability & Evidence \\
\hline
$<1.0$ & $< 3:1$ & $<0.750$ & Inconclusive \\
$1.0$ to $2.5$ & $< 12:1$ & $0.923$ & Weak to Moderate \\
$2.5$ to $5.0$ & $< 150:1$ & $0.993$ & Moderate to Strong \\
$>5.0$ & $> 150:1$ & $> 0.993$ & Very strong or decisive \\
\hline
\end{tabular}
\caption[The Jeffreys scale ]{Jeffreys scale to evaluate the strength of the evidence of a model N over another model M, in terms of the absolute value of $\ln{B_{N; M}}$, with a positive (negative) value indicating support for model $N$ ($M$).}
\label{tab:jeff_scale}
\end{table}

In summary, the Bayes factor strikes a balance between fit quality and additional model complexity. It rewards highly predictive models whilst penalising models with unnecessary extra parameters. This principle is often referred to as Occam's razor.

\section{Einstein-Boltzmann Codes} \label{sec:cmbcodes}


In \cref{sec:linpert}, we have introduced the equations governing the evolution of gravity and the matter fields for the inhomogeneous Universe, namely the evolution of the fluctuations produced in the early Universe and that grow over time into the large-scale structures pattern we observe. However, the dynamics of this system of matter perturbations have a quite intricate character, requiring specialised numerical tools to analyse their evolution.

Various numerical codes, or Boltzmann solvers, have been developed to facilitate this complex task, starting with \texttt{COSMICS} \cite{Bertschinger:1995er}, followed by \texttt{CMBFAST} \cite{Seljak:1996is}, \texttt{CMBEASY} \cite{Doran:2003sy}, and more recently, \texttt{CAMB} \cite{Lewis:1999bs,Howlett:2012mh} and \texttt{CLASS} \cite{Lesgourgues:2011re,Blas:2011rf}. These codes are engineered to solve the Boltzmann and fluid equations of motion for every type of matter in the Universe. The work presented in Part II of this dissertation relies on a modified version of \texttt{CLASS}, which includes an implementation of general coupled dark energy. 
While both \texttt{CLASS} and \texttt{CAMB} aim to accomplish the same results, their implementations differ in several ways. \texttt{CLASS} is the more recent of the two, launched publicly in 2011, compared to CAMB, which has been public since 2000. Its design is more modular than \texttt{CAMB} 's original Fortran 90 version, making it easier to modify. Furthermore, \texttt{CLASS} is written in C, which some users might find more approachable than Fortran. Nevertheless, both codes now include a user-friendly Python wrapper, and the choice between the two boils down to individual preference.

\texttt{CLASS}\footnote{\href{https://github.com/lesgourg/class_public}{https://github.com/lesgourg/class\_public}} (Cosmic Linear Anisotropy Solving System) was originally developed on request of the \textit{Planck} science team as a tool independent from \texttt{CAMB} to check for possible bias in parameter estimation introduced by the code. Ultimately, the \texttt{CLASS}-\texttt{CAMB} comparison has led to progress in the accuracy of both codes, with the agreement established at $0.01\%$ level for CMB observables, using highest-precision settings in both codes. 
\texttt{CLASS} calculates various cosmological observables such as CMB anisotropy power spectra, matter power spectra, and the primordial power spectrum. It is especially tailored for solving Einstein-Boltzmann equations - the equations describing cosmological perturbations' evolution — and can simulate the evolution across numerous cosmological scenarios and particle types.
The software features a modular architecture. It starts by reading and initialising user-specified input parameters, followed by a one-time calculation of all vital background quantities, thereby avoiding redundant computations. These quantities are stored for future use. Next, the code calculates thermodynamic aspects of the Universe, including free electron fraction and matter temperature evolution, which are solely dependent on the background evolution. Subsequent modules calculate the primordial power spectrum and evolve perturbation equations to compute the power spectra requested by the user. If necessary, non-linear scales can be modelled using different approaches. All output data are saved in designated files. We refer to lecture notes available at \cite{Lesgourgues} for more in-depth information.

The \texttt{CLASS} code has been adapted for interacting dark energy for the purpose of this dissertation, including all the models considered in Part II. This involved several changes to existing modules, optimised for $\Lambda$CDM. Therefore, considerable adjustments had to be made to the input, background, thermodynamics and perturbation modules. Although still a private patch, the version modified from scratch during this PhD accommodates any coupling in the dark sector, including the more general disformal case (discussed in \cref{chap:gwcons,chap:dbi}), and also different types of scalar fields for both dark energy, as introduced in \cref{sec:fields}, and is flexible enough to allow for the future incorporation of similar effects, like universal coupling to the whole matter sector.

\texttt{CLASS} can be directly interfaced with the sampler \texttt{MontePython}\footnote{\href{https://github.com/brinckmann/montepython_public}{https://github.com/brinckmann/montepython\_public}} \cite{Brinckmann:2018cvx,Audren_2013}, a widely used tool for MCMC analysis in cosmology based on Python. It enables sampling various model parameter spaces, essentially analysing any cosmological data. Theoretically, observables are calculated through \texttt{CLASS} \textit{via} the Python wrapper. \texttt{MontePython} supports different sampling algorithms, such as the Metropolis-Hastings and Nested Sampling, and various convergence tests, including the \textit{Gelman-Rubin} diagnosis \cite{1992StaSc...7..457G}. This and other statistical measures can be calculated with the aid of the Python \texttt{GetDist}\footnote{\href{https://github.com/cmbant/getdist}{https://github.com/cmbant/getdist}} \cite{Lewis:2019xzd} package that takes in the chains generated by \texttt{MontePython}.
In order to determine the goodness-of-fit for a particular model, we will report the difference between the value of the minimum $\chi^2$ test in a model $M$ with respect to the $\Lambda$CDM model, $\Delta \chi^2_{\rm min} = \chi^2_{{\rm min}, M} - \chi^2_{{\rm min}, \Lambda{\rm CDM}}$, in the tables of parameter constraints. Additionally, we report on the Bayesian evidence model comparison analysis, for which we used the public \texttt{MCEvidence}\footnote{\href{https://github.com/yabebalFantaye/MCEvidence}{https://github.com/yabebalFantaye/MCEvidence}} code \cite{Heavens:2017afc}.



\section{Baseline Data Sets} \label{sec:baseline}


The baseline data set used throughout this dissertation consists of particular combinations of existing data:

\begin{enumerate}
    \item The \textit{Planck} 2018 temperature and polarisation (TT TE EE) likelihood, which includes low multipole data ($\ell < 30$) \cite{Planck:2018nkj,Planck:2019nip,Aghanim:2018eyx}. We refer to this as \textit{Planck} 2018 or \textit{Pl18}.
    \item The \textit{Planck} 2018 lensing likelihood, constructed from measurements of the lensing potential power spectrum \cite{Planck:2018lbu}. We refer to this as \textit{Planck lensing} or \textit{Pl18len}.
    \item Baryon Acoustic Oscillations measurements, collected from the 6dFGS \cite{Beutler:2011hx}, SDSS MGS \cite{Ross:2014qpa}, and BOSS DR12 \cite{Alam:2016hwk} surveys. We denote this simply as \textit{BAO}.
    \item Type Ia Supernovae (SNIa) distance moduli measurements, taken from the Pantheon sample \cite{Scolnic:2017caz}. We term this data set as \textit{Pantheon} or \textit{SN}.
\end{enumerate}

For the \textit{Planck} data, we consider both the high-multipole likelihood (including multipoles $30 \lesssim \ell \lesssim 2500$ for the TT spectrum and  $30 \lesssim \ell \lesssim 2000$ for TE and EE spectra) and the low-E polarisation likelihood (covering the multipole range $2 \leq \ell \leq 30$ for the EE spectrum). We can derive constraints on the cosmological and model-specific parameters by analysing Planck temperature anisotropies and polarisation measurements. 

We also test the differences in the constraining power in adding the \textit{Planck} lensing measurement, as introduced in \cref{sec:cmblen}, the most significant detection of CMB lensing to date \cite{Planck:2018lbu}, which helps improve constraints on cosmological parameters, with a particular focus on parameters that affect late-time expansion and the background geometry.

While \textit{Planck} lensing measurements partially break the geometric degeneracy in the $H_0$ and $\Omega_m$ parameters, incorporating BAO measurements from galaxy surveys proves to be a much more effective way to shatter this degeneracy in the geometrical sector, in particular in the relative abundance of matter. The BAO measurements, being relatively simple geometric measurements unaffected by non-linear physics, offer a robust geometrical test of cosmology due to the large scale of its acoustic peak.
We use the standard combination of measurements of $D_V /r_d$ from the 6dF survey at an effective redshift $z = 0.106$, the SDSS Main Galaxy Sample at $z= 0.15$, and the final BOSS DR12 data with separate constraints on $H(z) r_d$ and $D_M/r_d$ in three correlated redshift bins at $z = \{0.38 , 0.51 , 0.61\}$.




     \cleardoublepage
    

 \chapter{Beyond the Standard Model: Coupled Dark Sector} \label{chapter:sttheories}
 \setcounter{equation}{0}
\setcounter{figure}{0}



    \epigraph{Mas o vazio tem o valor e a semelhança do pleno. Um meio de obter é não procurar, \\ um meio de ter é o de não pedir e somente acreditar \\ que o silêncio que eu creio em mim  é resposta a meu – a meu mistério.\blfootnote{\textit{But the emptiness has the value and the appearance of plenty. One way of getting is not looking, one way of having is not asking and only believing that the silence I believe to be inside me is the answer to my — to my mystery.} --- \textsc{Clarice Lispector} in The Hour of the Star} \\ --- \textsc{Clarice Lispector}\ \small\textup{A Hora da Estrela}}

    \dominitoc

 In this chapter, we introduce the formulation of alternative theories of gravity, with a particular focus to models where dark matter and dark energy experience some non-standard interaction, as alluded to in \cref{sec:beyond}. 
 We present some examples of different dark energy models and their cosmological consequences in \cref{sec:fields}, in which a dynamically evolving scalar field replaces the cosmological constant. \cref{sec:coup} reports on models where there is a universal coupling of dark energy to the entire matter sector through conformal and disformal transformations of the metric tensor.
We close with \cref{sec:coupde} with an introduction to models of non-universally coupled dark matter and dark energy, which will be the central focus of the work reported in Part II.

\section{Beyond $\Lambda$CDM} \label{sec:beyond}

As presented in \cref{chapter:standardmodel}, the resurrection of the need for a cosmological constant was prompted by the use of the type Ia SNe distance-redshift relation to estimate the value of the Hubble constant, unveiling a Universe expanding at an accelerating rate \cite{acel1,acel2}. This observation was further cemented by simultaneous CMB anisotropy measurements, which independently found a preference for $\Lambda$CDM cosmology with a low mass content \cite{db9dc632a6894231a6a88d4d86e14413}. Nevertheless, as was also discussed in \cref{sec:theoprob,sec:cosmotensions}, the concordance model is plagued by theoretical inconsistencies and, arguably more grievous, observational tensions.
In an attempt to address these issues, a wide range of alternatives to the cosmological constant have been put forward in the literature \cite{CANTATA:2021ktz}, typically replacing the cosmological constant with another mechanism for accelerating the Universe. While providing a more elucidatory theoretical framework or vaster phenomenology, these approaches come at the cost of introducing modifications to one or more of the assumptions underlying the standard model, as briefly described in \cref{chap:obs}. While these proposals can take several forms\footnote{Extensions to GR are conventionally categorised into three classes: i) added dimensions to the spacetime; ii) higher-order derivatives of the curvature or other related invariants; iii) introduction of extra fields.}, the results reported in Part II all rely on some extension to the dark sector in the \textit{standard} $\Lambda$CDM action based on GR with a matter sector, and which we write again explicitly to make the modifications clear later in the chapter:

\begin{equation}
S_{\Lambda \text{CDM}} = \int \odif[order=4]{x} \sqrt{-g}  \left[ \frac{M_{\rm pl}^2}{2} \Big(R(g_{\mu \nu}) - 2 \Lambda \Big) +  \mathcal{L}_{\text{M}} (g_{\mu \nu}, \psi_{\text{M}}) \right] \mathperiod
\label{eq:lcdmac}
\end{equation}

The first term is the Einstein-Hilbert action, as defined in \cref{eq:ehact}, including the cosmological constant $\Lambda$, and $\mathcal{L}_{\text{M}}$ collectively denotes the matter Lagrangian for the fields $\psi_{\text{M}}$, incorporating the standard model matter and dark matter.


The models studied in Part II rely on the extensions to the action in \cref{eq:lcdmac} listed below, and how such extensions are motivated will be the focus of the referenced upcoming sections:

\begin{enumerate}
    \item Introducing a scalar field dark energy component $\phi$ that replaces the cosmological constant $\Lambda$, and can have various natures, as introduced in \cref{sec:fields} - \cref{chap:cquint,chap:kin,chap:gwcons,chap:dbi,chap:sfdm};
    \item Considering a non-universal coupling in the dark sector only, as motivated in the context of conformal and disformal transformations for universal couplings in \cref{sec:coup} - \cref{chap:cquint,chap:kin};
    \item Providing a joint scalar field origin for the dark sector through a joint geometrical or fluid description for dark matter with a scalar field origin (with the dynamics as detailed in \cref{sec:fields}), that is approximated by a coupling in the dark sector as described in \cref{sec:coupde} - \cref{chap:dbi,chap:sfdm}.
\end{enumerate}

The points listed above are encapsulated in the following extended \textit{effective action} for an \textit{interacting dark sector}

\begin{equation}
S_{\text{IDS}} = \int \odif[order=4]{x} \sqrt{-g}  \left[ \frac{M_{\rm pl}^2}{2} R(g_{\mu \nu}) + \mathcal{L}_{\phi} (g_{\mu \nu},\phi, \nabla_{\mu} \phi) +  \mathcal{L}_{\text{SM}} (g_{\mu \nu}, \psi_{\text{SM}}) + \mathcal{L}_{\text{DM}} ( g_{\mu \nu},\phi, \nabla_{\mu} \phi, \psi_{\text{DM}}) \right] \mathcomma
\end{equation}
 where $\phi$ is the dark energy scalar field, $\psi_{\text{SM}/\text{DM}}$ denote the uncoupled standard model fields and the coupled dark matter sector (which in \cref{chap:sfdm} is approximated from a scalar field description), respectively.
 Effective field theories of this kind have the advantage of being easy to implement while preserving the general covariance of the theory. We formulate toy models based on some simplistic solutions to cosmological puzzles, aiming to reproduce the effective behaviour of a more fundamental (but also more complicated) underlying dynamics. Ultimately, these stand for cosmologists as lighthouses do to sailors, signalling any effects beyond the standard model.
 As will become clear, the advantage of modifying only the dark sector is twofold: first, the theory can still be formulated at the Lagrangian level and relying only on the addition of one scalar degree of freedom; second, interactions are only allowed in the dark sector \cite{Boehmer:2008av}, whose nature is still not well understood, and is mostly probed cosmologically through the characteristic gravitational signatures left in the large scale structure and the CMB. This means that the local precision tests of general relativity can be evaded \cite{Wang:2016lxa,Sakstein:2014isa,Ip:2015qsa,Will:2014kxa,LIGOScientific:2017zic}, since the \textit{fifth-force} mediated by $\phi$ will only influence the component that is not heavily restricted by the SM of particle physics. Moreover, the interaction would also influence the dynamics of dark energy, possibly contributing to explain some of the theoretical issues that plague the dark sector, as discussed in \cref{sec:theoprob}.

\section{Scalar Field Dark Energy} \label{sec:fields}


Originally, the basic premise behind extensions to GR was the idea that the value of $\Lambda$ is measured to be so small because it has been progressively converging to its natural vanishing state, \textit{i.e.} $\Lambda \rightarrow 0$, over an extended period through the cosmic history. This evolution provides a tentative resolution of the theoretical problems discussed in \cref{sec:theoprob}. The earliest expression of this thought can be traced back to Dolgov \cite{1983veu..conf..449D}, who proposed a variation of the theory of Brans and Dicke \cite{Brans:1961sx}, in which both $\Lambda$ and the strength of the gravitational interaction are evolving to zero. In that context, Reuter and Wetterich \cite{Reuter:1986wm,Wetterich:1987fm} explored possible formulations yielding field equations where a dynamical $\Lambda$ gradually vanishes. Around the same time, Peebles and Ratra \cite{Peebles:1987ek} showed how the energy density can be driven towards zero by a scalar field self-interacting potential, mimicking an asymptotically decreasing $\Lambda$, under the assumption that the quantum vacuum energy density vanishes. The name \textit{quintessence} was introduced by Caldwell, Davé, and Steinhardt \cite{Caldwell:1997ii,1998Ap&SS.261..303C,2000BrJPh..30..215C} to refer to this dynamically changing effective cosmological constant. Inaugural considerations on the cosmological implications of this concept were further elaborated in \cite{PhysRevD.37.3406} by Ratra and Peebles.

A natural fundamental candidate for a dynamically changing entity in modern physics is a \textit{scalar field}. Relevant examples are the recently detected Higgs field \cite{ATLAS:2012yve}, responsible for the mechanism of providing mass to the particles of the standard model of particle physics, or the inflaton, considered to be the scalar field that drives inflation \cite{linde}. Both these scalar fields play a critical role in models of fundamental physics. The inflaton, in particular, gives rise to dynamics similar to dark energy since both have to be responsible for accelerated cosmic expansion periods. Additionally, there is a precedent of solving problems related to missing energy by hypothesising a new particle or field, as was the case with neutrinos and dark matter, with the latter still awaiting direct detection. Scalar field-based theories had been explored in the literature long before the discovery of the accelerated expansion, as viable alternatives to GR. 
Inspired by Weyl and Dirac's studies leading to a slowly varying gravitational constant \cite{3e8b8b57-8265-314b-aef6-91c771572f79,Weyl:1919fi,Weyl:1917gp,1938RSPSA.165..199D,1937Natur.139..323D}, in 1959 Jordan showed that by describing gravitational couplings through a scalar field, the extra degree of freedom behaves like a matter field obeying conservation relations \cite{Jordan:1959eg}. These considerations developed into a complete gravitational theory in 1961, developed by Brans and Dicke (BD) \cite{Brans:1961sx}, in which the gravitational interaction is ascribed to the metric with the addition of one scalar degree of freedom, which came to be known also as the Jordan-Fierz-Brans-Dicke theory of gravity - in recognition of Fierz's and Jordan's seminal works on the physical understanding and formulation of this framework \cite{Fierz:1956zz,Jordan:1959eg} - and is widely considered to be the inaugural alternative to GR \cite{Faraoni:2004pi}.


Under these considerations, the quintessence model has been generalised to include a dynamical scalar field playing the dark energy role, varying slowly along some potential $V(\phi)$. According to \cref{eq:omega0}, this evolving scalar field accounts for the missing energy needed to maintain the Universe's geometrical flatness. While bearing similarities to the slow-roll inflation period in the early Universe \cite{liddleinf}, the late-time acceleration mechanism differs in that non-relativistic matter (dark matter and baryons) cannot be ignored for a complete understanding of the dynamics of dark energy. In principle, this evolving scalar field could interact with other species directly through a \textit{fifth-force}, allowing for a DE component that naturally self-adjusts to reproduce the inferred present energy density of $\Lambda$. 
For instance, this mechanism can be realised with attractor-like DE solutions, which reproduce the present energy densities for a vast range of initial conditions.

An alternative approach is to consider parameterisations that can be compared against observations, disregarding the fundamental or physical origin of such extensions, hoping to understand the direction hinted at by the data. A famous example, analysed and constrained by the \textit{Planck} collaboration \cite{Aghanim:2018eyx} as well, is the \textit{wCDM model}, consisting of a constant equation of state for the dark energy fluid, that deviates from $w=-1$, as assumed in the $\Lambda$CDM model. The value of $w$ is estimated from observational data and consistency relation checks. Alternatively, the parameterisation can be generalised according to some redshift-dependent evolution $w(z)$, which provides insight into dark energy but is limited to handpicking a particular subset of functions that determine the phenomenology of the theory \cite{Escamilla:2023oce}. The opposite side of the coin is reconstructing the free function(s) in $w(z)$ without prior restrictions or considerations \cite{Escamilla:2023shf}. While parameterisations are helpful to capture physical features in a somewhat model-independent manner, self-consistent and predictive frameworks are needed to comprehend and test the model's assumptions across different regimes.

The remainder of this section focuses on prominent scalar field candidates for dark energy commonly considered in the literature and whose characteristic nature leads to distinct features in the cosmological evolution that could help alleviate the standard model's cosmological tensions and anomalies. Other possibilities have been proposed but will not be discussed in detail here, and the interested reader is referred to reviews on alternative theories of gravity such as \cite{CANTATA:2021ktz,Clifton:2011jh,debook}.

\subsection{Quintessence Field} \label{sec:quin}

In the framework of the FLRW cosmological background introduced in \cref{sec:fund}, a cosmological constant term represents a source with a constant equation of state parameter, $w_\Lambda = -1$. A similar behaviour at late times can be reproduced by a dynamically evolving scalar field, minimally coupled to gravity, of which the simplest example is the canonical scalar field $\phi$ as a QFT generalisation of the non-relativistic particle. This scalar dark energy source has been appropriately named \textit{quintessence} in a bow to Aristotle's fifth natural element. The crucial aspect of this framework is that the equation of state parameter is no longer constant and evolves according to $\phi$ and $\dot{\phi}$, naturally addressing the conceptual issues of the cosmological constant discussed in \cref{sec:theoprob}, such as the incompatibility between the small value required for the cosmological constant to match the observations and the vacuum energy predictions from QFT.
In fact, in a homogeneous Universe accurately described by GR, the second Friedmann equation, \cref{eq:fri2}, implies that cosmic acceleration can be achieved so long as the total energy-momentum tensor components meet $p < - \rho/3$. From this, a new cosmic fluid can drive the acceleration at the cost of violating the strong energy condition \cite{Wetterich:1987fm,Peebles:1987ek}.
The quintessence scalar field is a traditional example of such scenarios, and can be incorporated directly in the gravitational action alongside all the matter sources (in $S_{\text{m}}$), with a standard kinetic term and a self-interacting potential $V(\phi)$ yielding
\begin{equation} \label{eq:quinac}
    S = \int \odif[order=4]{x}\, \sqrt{-g} \left[ \frac{\text{M}_{\text{Pl}}^2}{2} R - \frac{1}{2} (\covd \phi)^2 - V(\phi) \right] + S_{\text{m}} \mathcomma
\end{equation}
where $\phi$ is the quintessence field which has dimensions of mass and, once more, the first term is simply the Einstein-Hilbert action with $\text{M}_{\text{Pl}}^2 = 1/(8\pi G)$, assuming $\Lambda = 0$. The term $(\covd \phi)^2 = g^{\mu \nu} \partial_{\mu} \phi \partial_{\nu} \phi$ encodes the dependence on the kinetic evolution of the field.

The contribution of the scalar field to the total energy-momentum tensor is given by the variation of the action in \cref{eq:quinac} according to the metric and reads
\begin{equation} \label{eq:quintmn}
    T_{\mu \nu}^{\phi}  = \partial_{\mu} \phi \partial_{\nu} \phi - g_{\mu \nu} \left[\frac{1}{2} g^{\alpha \beta} \partial_{\alpha} \phi \partial_{\beta} \phi + V(\phi)  \right] \mathperiod
\end{equation}
In the context of the flat FLRW background metric, \cref{eq:flrwchi}, the dynamical evolution of the scalar field is extracted through variation of the action in \cref{eq:quinac} with respect to $\phi$ itself and results in the Klein-Gordon equation in an expanding Universe:
\begin{equation}
    \Box \phi - \odv{V}{\phi} = 0  \ \  \xrightarrow{\text{FLRW}}\ \ \ddot{\phi} + 3 H \dot{\phi} + V'(\phi) = 0 \mathperiod
    \label{eq:covkg}
\end{equation}

The energy density and pressure of the scalar field are computed from \cref{eq:quintmn}:
\begin{equation}
        \rho_{\phi} = - T^0_0 = \frac{1}{2} \dot{\phi}^2 + V(\phi)  \mathcomma \quad \text{and}\quad p_{\phi} = T^i_i = \frac{1}{2} \dot{\phi}^2 - V(\phi) \mathperiod \label{eq:quinp}
\end{equation}
Therefore, in the case of a single-component Universe ruled by this scalar field, the Friedmann equations track its evolution as follows:
\begin{equation}
 3 \text{M}_{\text{Pl}}^2  H^2 = \frac{1}{2} \dot{\phi}^2 + V(\phi) \mathcomma  \quad \text{and}\quad 3 \text{M}_{\text{Pl}}^2  \frac{\ddot{a}}{a} = - \dot{\phi}^2 + V(\phi)  \mathperiod \label{eq:quinacc}
\end{equation}
As we can see from \cref{eq:quinacc}, the field yields a positive accelerating contribution for $\ddot{a} > 0$ as long as $\dot{\phi}^2 < V(\phi)$; therefore, in order to explain the accelerated phase without invoking a cosmological constant, the scalar field dynamics must be designed in such a way to fulfil this condition.
More precisely, the equation of state parameter is given by:
\begin{equation} \label{eq:wfquin}
    w_{\phi} = \frac{p_{\phi}}{\rho_\phi} = \frac{\dot{\phi}^2 - 2V(\phi)}{\dot{\phi}^2 + 2V(\phi)} \mathperiod
\end{equation}
It can be readily appreciated that there are two limits to this equation: if the potential is sufficiently flat at late times such that $\dot{\phi}^2 \ll V(\phi)$, then $w_\phi \simeq -1$, leading to a \textit{slow-roll} phase of cosmological constant-like accelerated expansion; if otherwise the field dynamics is dominated by its kinetic energy, that is $\dot{\phi}^2 \gg V(\phi)$, then $w_\phi \simeq 1$. The former is what makes a canonical scalar field the simplest dynamical candidate for dark energy with different models distinguished by the form of their potential $V(\phi)$, which determines how $\phi$ portrays the expansion of the Universe.

Nevertheless, quintessence fields have little impact on the growth of cosmological perturbations and are therefore challenging to probe with large-scale structure observations, producing only minor modification on the large-scale CMB through the change in the expansion history (see \cref{sec:cmb}). In addition, the potential can be chosen to satisfy practically any expansion history (with $w \geq -1$), making the general case very unpredictable without further requirements on $V$.


A sub-class of quintessence models is the tracker models. These theories exhibit a particular evolution of the energy density $\rho_{\phi}$ as it tracks the dominant component of the Universe. In this way, the coincidence problem is mitigated as the other components determine the evolution of $\rho_{\phi}$. A potential that produces such an evolution, traditionally named the \textit{Ratra-Peebles} potential \cite{PhysRevD.37.3406,Copeland:2006wr}, is
\begin{equation}
    V(\phi) = \frac{V_0}{\phi^{\alpha}} \mathcomma
\end{equation}
where $\alpha>0$ is a constant power left as a free parameter \cite{Liddle:1998xm,Zlatev:1998tr, Steinhardt:1999nw, delaMacorra:2001ay}. In this scenario, $\rho_{\phi}$ will mimic the energy density of the dominant background component $\rho_B$ in such a way that the equation of state parameter becomes
\begin{equation} \label{eq:wftrack}
    w_{\phi} = \frac{\alpha w_B - 2}{ \alpha +2} \mathperiod
\end{equation}
Imposing the acceleration condition $\ddot{a}>0$ from \cref{eq:fri2} to \cref{eq:wfquin,eq:wftrack} implies \cite{Copeland:2006wr} 
\begin{equation}
    \dot{\phi}^2 < \frac{2}{3} V(\phi) \mathcomma \quad \text{and}\quad \alpha < \frac{2}{2 w_B +1} \mathperiod
\end{equation}
Accordingly, if the dominant contribution comes from matter, $w_B = w_m =0$, accelerated expansion is achieved simply through the condition $\alpha <2$.
Without further modifications, tracking solutions lead to an overall enhancement of the Hubble rate at all times, indirectly reducing the clustering rate of ordinary matter and possibly leading to a significant \textit{early dark energy} contribution in the early Universe \cite{Poulin:2023lkg}. 

Another standard class of quintessence potentials is the exponential case \cite{Copeland:1997et, UrenaLopez:2011ur, Tamanini:2014mpa}, which leads to a power-law expansion and takes the form:

\begin{equation}
    V(\phi) = V_0 \exp \left( - \lambda \frac{\phi}{\text{M}_{\text{Pl}}} \right) \mathcomma
\end{equation}
where $V_0$ is still the scale of the potential, and the constant $\lambda$ gives the steepness of the exponential. In this simple case, the field evolves as $\phi \propto \ln t$ and accelerated expansion is achieved depending on the value of $\lambda$, namely $\lambda^2 < 2$ is required. Arguably more important is the fact that the exponential potential admits cosmological \textit{scaling solutions} \cite{Copeland:1997et,Barreiro:1999zs}, in which the field's energy density ($\rho_{\phi}$) \textit{scales} proportionally to the background fluid's energy density $\rho_B$ \cite{Billyard:1999ia,PhysRevD.61.123503,Nunes:2000yc,Ng:2001hs,PhysRevD.64.127301,Uzan:1999ch}, that is
\begin{equation} \label{eq:scale}
    \frac{\rho_\phi}{\rho_B} = r \mathcomma
\end{equation}
where $r$ is a nonzero positive constant, and the field can mimic the background even if it is sub-dominant over the radiation and matter-dominating eras. In this case, as long as the scaling solution is the attractor, then for any generic initial conditions, the field will sooner or later enter the scaling regime, thereby introducing a strategy to address the fine-tuning problem of dark energy. It is worth mentioning that scaling solutions live on the border between acceleration and deceleration. We also note that the system needs to exit from the scaling regime defined in \cref{eq:scale} to bring about an epoch of accelerated expansion. For this purpose, the field's energy density must catch up to that of the fluid, provided that the potential is shallow relative to the one corresponding to the scaling solutions.
For instance, it has been shown that the following type of double exponential potential can account for this transition \cite{Barreiro:1999zs}:
\begin{equation}
    V(\phi) = V_0 \left[ \exp \left(  - \lambda\, \frac{\phi}{\text{M}_{\text{Pl}}}\right) + \exp \left(  - \mu\, \frac{\phi}{\text{M}_{\text{Pl}}}\right) \right] \mathcomma
\end{equation}
where $\lambda$ and $\mu$ are positive constants. The requirement is that the $\lambda$-exponential term provides the scaling during the radiation- and matter-dominated eras before the $\mu$-exponential term takes over. The scalar field is then driven out of the scaling regime and towards a solution in which it dominates. An important advantage of the double exponential potential is that the scaling regime occurs for a wide range of initial conditions, followed by an accelerated expansion phase once the potential becomes shallow, evading the general fine-tuning of initial conditions of the scalar field in most quintessence models.
For a cosmological analysis of quintessence with other potentials see, for example, \cite{Barreiro:1999zs, Jarv:2004uk, Li:2005ay, UrenaLopez:2011ur, Ng:2000di, Ng:2001hs, Gong:2014dia, Fang:2008fw, Roy:2013wqa, Garcia-Salcedo:2015ora, Paliathanasis:2015gga, Zhou:2007xp, Matos:2009hf, Clemson:2008ua}.

Quintessence models will be the main focus of \cref{chap:cquint}. Beyond the technical details, it is essential to highlight the resemblance between the action of the quintessence model and that of a general scalar-tensor theory. The boundary between these two theories is somewhat blurred, and it has been shown that quintessence models can be recast into a Scalar-Tensor theory analogous to the Jordan-Brans-Dicke formulation by applying a Weyl rescaling of the metric \cite{Faraoni:1998qx,CANTATA:2021ktz}. 

\subsection{K-Essence Models} \label{sec:kessence}


A further extension of quintessence is the \textit{k-essence model} \cite{PhysRevD.62.023511,ArmendarizPicon:2000dh,ArmendarizPicon:2000ah}, in which a canonical field does not drive the accelerated expansion via a slowly varying potential, but by non-standard modifications to the kinetic energy of the scalar field instead, \textit{i.e.}
\begin{equation}
        S = \int \odif[order=4]{x}\, \sqrt{-g} \left[ \frac{\text{M}_{\text{Pl}}^2}{2} R + \mathcal{L}_K(\phi,X) \right] + S_{\text{m}} \mathperiod
    \label{eq:kess}
\end{equation}
where the Lagrangian density $\mathcal{L}_K$ encapsulates the most general scalar field action, which is a function of both $\phi$ and its kinetic energy $X = -\frac{1}{2} (\nabla \phi)^2$, and of which quintessence is a particular case. While k-essence fields have a significantly increased number of phenomenological features when compared with their quintessence ancestors, the trade-off is the fact that the action in \cref{eq:kess} is not guaranteed to be free of pathological or unstable behaviour, such as ghost degrees of freedom or superluminal propagation \cite{Sbisa:2014pzo}.  
K-essence models are motivated by low-energy effective string theory where an $\mathcal{L}$-like term of the form in \cref{eq:kess} appears for the dilaton field \cite{Armendariz-Picon:1999hyi}. The possibility of achieving an accelerated expanding period from the action in \cref{eq:kess} was suggested in the context of inflation \cite{Armendariz-Picon:1999hyi}, with the extension to the late-time dark energy scenario first proposed in \cite{PhysRevD.62.023511}, with later generalisations establishing the concept of k-essence. 


The scalar field action $\mathcal{L}_K$ is made more tractable by assuming that it is separable in $\phi$ and $X$, and in particular, the simple case of the \textit{ghost condensate model} is often assumed, in which $\mathcal{L}_K (\phi,X) = f(\phi) (X^2/M^4 -X)$, where $M$ is a mass scale and $f(\phi)$ is an arbitrary non-singular function of $\phi$. In these models, the effects of the quintessence potential are reproduced by the non-standard kinetic term, which leads to the equation of state
\begin{equation}
    w_{\phi} = \frac{1-X/M^4}{1-3X/M^4} \mathcomma
\end{equation}
where in order to have an accelerated expansion phase, that is, $w_{\phi} < -1/3$, we must require $X < 2M^4/3$. The expression above also clearly illustrates the underlying assumption of k-essence: if $X$ does not vary in time, and there is no kinetic dynamics, then $w_{\phi}$ remains a constant. In particular, for this simple k-essence scenario, the cosmological constant case corresponds to $X=M^4/2$. This analysis has been extended in multiple works, and it can be shown to address the coincidence problem of dark energy \cite{ArmendarizPicon:2000dh,ArmendarizPicon:2000ah}.
In \cref{chap:kin}, we will explore a generalised k-essence scalar field scenario derived from the framework proposed in \cite{Barros:2019rdv}.

\section{Universal Couplings to the Matter Sector} \label{sec:coup}

In the context of GR, gravity is interpreted as a fundamental geometrical property of spacetime originating from the matter sources included therein. This notion aligns with Mach's equivalence principle, which states that inertia is a product of the interaction between bodies \cite{mach_2013,einstein2014meaning}. In the proposal of Brans and Dicke \cite{Brans:1961sx}, this principle was incorporated in the form of a varying gravitational coupling according to the dynamical character of the mass distribution itself. This formalism relied on the introduction of a scalar field such that Newton's gravitational constant $G$ is actually the local value of the gravitational coupling and the weak equivalence principle - stating that in a gravitational field, all point masses follow geodesics of the same metric - still holds \cite{Brans:1961sx,wald,Clifton:2011jh}. 
Brans and Dicke chose a matter-scalar coupling acting universally on all types of matter, according to the same gravitational metric. This ensured that matter is perceived as free-falling in this gravitational metric, in agreement with the weak equivalence principle. 
Interestingly, while this scalar field was intentionally detached from the matter Lagrangian to maintain the universality of free-fall, it inevitably couples with matter in the field equations, which depend directly on the trace of the energy-momentum tensor of matter. 
Theories of gravity of this nature build upon the solid groundwork of GR plus a scalar field which is introduced in a non-trivial way through what's called as \textit{non-minimal universal coupling}. 

Two particularly relevant descriptions with an intuitive physical interpretation arise: 

\begin{itemize}
\item The \textit{Jordan frame} in which the scalar field only features explicitly in the gravitational sector (given by the the Einstein-Hilbert action in \cref{eq:ehact}) and the matter fields are said to be \textit{minimally coupled} to $\phi$, while there is a \textit{non-minimal coupling} between the scalar field and the curvature term. This means that the individual energy-momentum tensors of each source are still covariantly conserved and test particles follow the geodesics of the spacetime metric. 

\item The \textit{Einstein frame} which denotes the opposite case, where the scalar field is only explicitly included in the matter sector, which becomes \textit{non-minimally coupled}, while the gravitational sector is now minimally coupled, meaning that the EH action remains unaltered. Likewise, the set of field equations resemble the Einstein equations in GR, ensuring second-order field equations, with the caveat of losing covariant conservation of the energy-momentum tensor of the matter fields:
\begin{equation}
\nabla^{\mu} T_{\mu \nu}^{\text{M}}= \mathcal{Q}_{\nu}\quad \text{and}\quad \nabla^{\mu} T_{\mu \nu}^{\phi}= - \mathcal{Q}_{\nu} \mathperiod
\label{eq:nu0}
 \end{equation}
 The $\mathcal{Q_{\nu}}$ term encodes the non-minimal coupling between matter and the scalar field. The scalar degree of freedom is transformed into a matter source contribution, an imprint of its gravitational role in the Jordan frame. This comes with the trade-off that the matter fields no longer follow geodesics of the gravitational metric $g_{\mu \nu}$, but rather of some transformed metric including the effect of the coupling:
\begin{equation}
g_{\mu \nu} \mapsto \td{g}_{\mu \nu} = C(\phi) g_{\mu \nu} \mathcomma
\label{eq:einconf}
\end{equation}
where $C(\phi)$ is called the \textit{conformal} function.
\end{itemize}

Conformal transformations such as the one defined in \cref{eq:einconf} are a conventional and seemingly powerful tool, first exploited in the prototype Brans-Dicke (BD) model in 1961 \cite{Brans:1961sx}, which also clearly illustrated the physical meaning of such a metric transformation. The advantage is that if the theory has been shown to be invariant under these transformations, then the action can be recast between different frames (much like a change of reference frame) in which its physical meaning may be clearer or its field equations less involved\footnote{The conformal function/factor $C(\phi)$ must be a smooth non-vanishing function of the spacetime coordinates. By applying a conformal transformation, a mapping of the cosmological description is enforced, implying a modification of the structure of the spacetime. In contrast, the coordinate separations $\dd x^{\mu}$ are fixed to the spacetime itself and remain unaltered. Nevertheless, the distance between two spacetime points accompanies the structural change, meaning that a conformal transformation applies an isotropic scaling of the length and norm of time-like and space-like intervals and vectors alike.} such as the BD theory and its generalisation into the broader class of \textit{scalar-tensor theories} of modified gravity \cite{Capozziello:2010zz}. It should, however, be emphasised that the definition of these two frames is only valid when a \textit{universal transformation} is considered, that is, when the scalar field couples non-minimally to the \textit{whole matter sector}, and the equivalence principle still holds.
While in the past there was significant debate and division in the community on whether these two frames represent physically equivalent theories, the general consensus at the moment of writing this thesis, is that conformally related theories are indeed distinct representations of the same physical scenario, especially at the classical level \cite{Capozziello:2010zz,Faraoni:2006fx}.

Without introducing additional scalar degrees of freedom, this formulation can be extended by allowing the conformal factor to depend on the derivatives of the scalar field as well, through its kinetic term $ X \equiv - (1/2)\, g^{\mu \nu} \covd_{\mu} \phi \covd_{\nu} \phi$.
This transformation is more general than the traditional conformal case, introducing a dependence of the metric in the conformal factor, and a related example will be the focus of \cref{chap:kin}. 

A further extension was proposed by Bekenstein in 1992 \cite{Bekenstein:1992pj} while investigating general couplings of matter to the gravitational metric in the context of Finsler geometry \cite{Asanov_1985}. This was accomplished through the introduction of distinct geometries for each sector, restricted to be related through the so-called \textit{disformal transformation} of the metric. 
Disformal transformations have gained particular relevance in cosmology since it was shown that Horndeski theories\footnote{The Horndeski Lagrangian is a direct generalisation of the Brans-Dicke theory, based on the premise of having the most general scalar-tensor extension to GR yielding second-order equations of motion both for the metric and the scalar field in four-dimensional spacetime. This solution was put forward by Horndeski in the 1970s \cite{Horndeski:1974wa} but remained practically unnoticed for many years and was only brought to light recently in the context of Galileon theory \cite{Deffayet:2013lga}. The disformal transformation represents a symmetry of the Horndeski action, just as the conformal transformations are with respect to the Brans-Dicke action.} remain structurally invariant under the particular class of disformal transformations which depend only on $\phi$ \cite{Bettoni:2013diz}:
\begin{equation}
g_{\mu \nu} \longmapsto \td{g}_{\mu \nu}  =C (\phi)\, g_{\mu \nu} + D(\phi) \partial_{\mu} \phi \partial_{\nu} \phi \mathcomma
\label{eq:disfphii}
\end{equation}
where $C(\phi)$ and $D(\phi)$ are the conformal and disformal factors, and the conformal transformation is recovered by setting $D=0$, while the minimally coupled case requires $C=1$. It has been shown that theories in which $C$ and $D$ depend on $X$ \cite{Lin:2014jga, os, Zumalacarregui:2013pma} may be plagued by Ostrogradski's instabilities\footnote{Ostrogradski's theorem states that the Hamiltonian of any non-degenerate Lagrangian, depending on more than first-order time derivatives, is unbounded from below, leading to instability issues that result in an unphysical theory with arbitrarily high negative energy. This no-go theorem limits the space of possible scalar-tensor theories like Horndeski's. Whenever there are time derivatives of the fields of order higher than one in the Lagrangian, the theory becomes unstable and unphysical.} \cite{Woodard:2015zca,Ostrogradsky:1850fid,DeFelice:2011bh}.

Just on a qualitative basis, by exiting the conformal realm, the metric transformation can no longer be interpreted as a simple field redefinition. Disformal transformations do not represent a change in coordinates but a local change in the geometry instead, encoding a stretching (or a compression) of the metric specified according to the gradient of the scalar field, resulting in a distortion of both angles and lengths. 
There is a translation along the direction of variation of the field, implying that the new metric will also depend directly on the changing properties of the field through spacetime.

\section{Non-Universal Couplings and Interactions in the Dark Sector} \label{sec:coupde}




Finally, one may ask whether this dark energy source encodes any non-minimal interaction, that is, besides the standard gravitational coupling, to the other sources in the Universe. The idea of an \textit{interactive dark energy} scalar field $\phi$, coupled to a matter component, and its subsequent cosmological implications, were first examined in \cite{ELLIS1989264, Wetterich:1994bg, PhysRevD.48.3436, DAMOUR1994532}. 
Szydlowski introduced a similar concept of interaction between dark matter and dark energy \cite{Szydlowski:2005ph}, followed by a method of examination that was formulated later in \cite{PhysRevD.73.063516}. There's a wealth of studies exploring the dynamical implications of interactions between dark energy and matter \cite{Nunes:2000ka, Chimento:2003iea, Olivares:2007rt, Quartin:2008px, Chen:2008ca, Quercellini:2008vh, CalderaCabral:2008bx, Li:2010ju, Arevalo:2011hh, Perez:2013zya, Szydlowski:2016epq, Haba:2016swv}. 
Delving into the particular scenarios where dark energy is portrayed by a canonical scalar field, we come across the \textit{coupled quintessence models}, as direct generalisations of the models presented in \cref{sec:fields}. These were initially presented and studied in \cite{Amendola:1999er,Amendola:1999er, Zimdahl:2001ar}, based on Luca Amendola's non-minimally coupled theories \cite{Amendola:1999qq}. Various distinct models have been considered, each characterised by the universality and form of the coupling \cite{Billyard:2000bh, Holden:1999hm, Amendola:2000uh, TocchiniValentini:2001ty, Gumjudpai:2005ry, Liu:2005wga, Gonzalez:2006cj, PhysRevD.73.023502, Boehmer:2008av, Chen:2008pz, Boehmer:2009tk, LopezHonorez:2010ij, Wei:2010fz, Cicoli:2012tz, Xu:2012jf, Morris:2013hua, Tzanni:2014eja, Hossain:2014xha, Zhang:2014zfa, Roy:2014hsa, Shahalam:2015sja, Singh:2015rqa, Landim:2016gpz, Barros:2018efl}.

Furthermore, instead of imposing the coupling directly by an \textit{ad-hoc} field-dependence in the DM Lagrangian, the ideas of conformal/disformal invariance discussed in \cref{sec:coup} can be recovered, \textit{e.g.} assuming that the coupling between the scalar field and the SM species is suppressed or hidden by means of a screening mechanism \cite{PhysRevLett.126.091102,Koyama:2015vza,Joyce:2014kja,Khoury:2013yya,Babichev:2013usa,Brax:2021wcv,Brax:2013ida,Sakstein:2014isa,Brax:2004qh}.
In these mechanisms the coupling between the scalar field and matter weakens and limits the gravitational influence of the scalar field itself in dense, small-scale areas like the Earth and the solar system, where GR has been rigorously tested. Yet, on larger scales, like those generally associated with dark energy, the scalar field's gravitational effects may re-surge, in a context where the theory is less constrained. This subject is still an active research area, but is out of the scope of this thesis.

Considering an FLRW Universe, \cref{eq:flrwchi}, the continuity equations, \cref{eq:fricon}, for the scalar field and the coupled matter fluid include a symmetric extra term $\mathcal{Q}$ accounting for the interaction and preserving the covariant conservation of energy, corresponding to the $\nu = 0$ component of \cref{eq:nu0}, and take the form:

\begin{equation}
\dot{\rho}_{\rm DM} + 3 H \rho_{\rm DM} (1+w_{\rm DM})= - \mathcal{Q}, \ \ \ \  \text{\rm and}\ \ \  \ \dot{\rho}_{\phi} + 3 H \rho_{\phi} (1+w_{\phi})= \mathcal{Q},
\label{eq:contq}
\end{equation}

The sign of $\mathcal{Q}$ in \cref{eq:contq}, which is (in principle) a function of the cosmological variables, establishes the direction of the flow of energy transfer, and can itself be a dynamical quantity. 

\cleardoublepage

  \partdivider{Part II}{Research Results} \label{part:results}

\cleardoublepage


 \chapter{Coupled Quintessence in Flat and Curved Geometries} \label{chap:cquint}
 \setcounter{equation}{0}
\setcounter{figure}{0}


 \epigraph{Cada coisa é uma palavra. E quando não se a tem, inventa-se-a.\blfootnote{\textit{Each thing is a word and when there is no word it is invented.} --- \textsc{Clarice Lispector} in The Hour of the Star} \\ --- \textsc{Clarice Lispector}\ \small\textup{A Hora da Estrela}}



As discussed in Part I of this dissertation, the quest to comprehend the recent accelerated expansion of the Universe is among the most pressing open questions in the cosmological community and perhaps in the scientific endeavour. According to their inherently attractive gravitational nature, the matter fields accounted for in the standard model of particle physics cannot be the driving force behind this acceleration. When considering general relativity at the largest observable scales, this behaviour becomes important and must be accommodated by the theory, such as some form of cosmic fluid that exhibits a strongly negative pressure-to-energy density ratio — often termed dark energy — capable of balancing out the gravitational pull of ordinary matter and powering the acceleration, fitting into the observational evidence.

Over the course of \cref{chap:int,chapter:standardmodel}, we have introduced a positive cosmological constant as the most straightforward form of dark energy, a cornerstone of the well-accepted $\Lambda$CDM model. Interestingly, this constant has only recently begun to dominate the Universe's energy content. Indeed, \cref{chapter:sttheories} built up the idea of more dynamic varieties of dark energy supported by current data. For example, \cref{sec:quin} introduced quintessence models, where a scalar field drives the Universe's accelerated expansion. In these scenarios, the field's potential energy, rather than its kinetic energy, becomes the dominant factor over time, effectively mimicking the cosmological constant value.

Additionally, most studies often assume that dark energy and dark matter - classed together as the \textit{dark sector} - have independent origins and do not share any exotic interactions. However, given the enigmatic nature of these species, there is no fundamental justification for dismissing potential effective couplings between the two. For example, while solar system tests \cite{Ip:2015qsa} and experimental data severely limit any interaction between dark energy and ordinary matter, such constraints do not necessarily apply to dark matter. This subject was the focus of \cref{sec:coup,sec:coupde} and has been thoroughly explored in the literature \cite{vandeBruck:2013yxa,vandeBruck:2016hpz,vandeBruck:2016jgg,vandeBruck:2017idm,Mifsud:2017fsy,debook,Gomez-Valent:2020mqn,Amendola:1999er,Barros:2018efl,Pettorino:2012ts,Amendola:2011ie,Amendola:2014kwa,Pourtsidou:2013nha,Amendola:1999er,Pace:2013pea,PhysRevD.85.023503,Teixeira:2019hil}.

This chapter will explore various types of \textit{coupled quintessence models} and their predictive power. Namely, their potential to address evidence for a closed Universe found in the CMB data. 

The \textit{Planck} satellite's observations of the cosmic microwave background temperature fluctuations and polarisation angular power spectra \cite{Planck:2018nkj,Planck:2019nip} largely support the concordance $\Lambda$CDM cosmological model \cite{Aghanim:2018eyx}. However, as thoroughly discussed in \cref{sec:cosmotensions}, with the remarkable rise in the precision of cosmological probes in the last decade, certain anomalies require further scrutiny. 
Namely, the \textit{Planck} data leads to the prediction of a lensing amplitude above the normalised value, $A_{\text{lens}}=1$, at about $2.8$ standard deviations \cite{DiValentino:2019dzu,DiValentino:2020hov}. In minimal extensions, this so-called lensing excess implies more cold dark matter, favouring a closed Universe. This is supported by both the baseline \texttt{Plik} likelihood and the alternative \texttt{CamSpec} likelihood, with the latter even supported at a $99\%$ confidence level. These findings question the flatness assumption in the conventional $\Lambda$CDM cosmological model and have sparked significant debates. This hypothesis also accounts for other large-scale data irregularities, such as the reduced amplitude in quadrupole and octupole modes \cite{10.1046/j.1365-8711.2003.06940.x}. As a result, evidence pointing to a closed Universe has escalated to a compelling $3.4$ standard deviations \cite{DiValentino:2019qzk,Handley:2019tkm}, surpassing the anomaly reported for the lensing amplitude, sparking significant debate.

This finding is rather surprising given that other astrophysical measurements, like BAOs \cite{Beutler:2011hx,BOSS:2016hvq,Beutler:2011hx}, are found to conflict with the notion of a closed Universe when combined with the \textit{Planck} likelihood \cite{DiValentino:2019qzk}. Moreover, the now well-established and extensively investigated discrepancy between the Hubble constant as determined by the SH0ES project using Cepheid stars as calibrators for estimating Type Ia supernovae luminosity distances \cite{Riess:2021jrx} ($H_0 = 73\pm\, 1 \text{km/s/Mpc}$) and the \textit{Planck} satellite's CMB-based estimate including lensing measurements \cite{Aghanim:2018eyx} ($H_0 = 67.4 \pm 0.5\, \text{km/s/Mpc}$) have risen above the distressful $5$-sigma threshold (see also \cite{Riess:2022mme}, where the divergence reaches $5.3\sigma$), essentially diminishing the chance of this disagreement being attributed to a statistical anomaly.

Lastly, it is worth mentioning the tension in the parameter $S_8 \equiv \sigma_8 \sqrt{\Omega_m/0.3}$ between \textit{Planck} and weak lensing measurements like KiDS-1000 \cite{Heymans:2020gsg} or DES-Y3 \cite{DES:2021vln}. Again, these discrepancies dwell at a $2-3\sigma$ level under a $\Lambda$CDM assumption. Hence, there is strong support for the need for more precise measurements beyond those from \textit{Planck} for gaining clearer insights into these cosmological tensions and anomalies that allow for independently testing the fundamental assumptions behind the curtains of the modern cosmological stage.

We will introduce the coupled quintessence model for a general coupling in \cref{sec:cq:models} along with the methodology followed in this work. We include a summary of the equations governing its evolution at the background and linear perturbative levels for general spatial curvature and discuss the repercussions of this dark sector coupling on the CMB's TT spectrum.. In \cref{sec:cq_flat} we present an update on the constraints for the flat models differing in the choice of the coupling and the potential functions. In \cref{sec:cq_curved}, we will shift our focus to the changes introduced by dropping the flatness assumption and taking the curvature of the Universe to be a cosmological parameter that can be constrained and compared with other data sets in order to shed light on the cosmological tensions. 
Finally, we comment on the evidence for a closed Universe in coupled quintessence models and summarise the main conclusions and discuss the results in \cref{sec:cqc_discussion}. 


 \section{Models and Methodology} \label{sec:cq:models}

 \subsection{Background Cosmology}

 Consistent with the framework outlined in \cref{chapter:sttheories}, the interaction within the dark sector in these models is achieved through a \textit{conformal non-universal coupling}. As introduced in \cref{sec:beyond}, this setting is embodied by the following action

\begin{align}
\label{eq:cq_action}
{\cal S} =& \int d^4 x \left[ \sqrt{-g}\left( \frac{M_{\rm Pl}^2}{2} R - \frac{1}{2}g^{\mu\nu} \partial_\mu \phi \partial_\nu\phi - V(\phi) \right) + \sqrt{-g} {\cal L}_{\rm SM}(g,\Psi_i) + \sqrt{-{\tilde{g}}} {\cal L}_{\rm DM}({\tilde g},\chi) \right] \nonumber \mathperiod
\end{align}

Here, $R$ is the Ricci scalar, in terms of the metric degree of freedom $g_{\mu \nu}$, with $g$ being its determinant. We recall that $M_{\rm Pl} = (8 \pi G)^{-1/2}$ is the Planck mass and $G$ is Newton's gravitational constant. The quintessence scalar field enters the action as a canonical kinetic term plus a self-interacting potential $V(\phi)$. 
The \ac{sm} Lagrangian $\mathcal{L}_{\text{SM}}$ consists of both radiation (relativistic) and baryonic (non-relativistic) components, encapsulated by the fields $\Psi_i$, which follow geodesics defined by the gravitational metric $g_{\mu \nu}$.
Lastly, the dark matter Lagrangian $\mathcal{L}_{\text{DM}}$ yields the description of the dark matter field $\chi$, which is coupled to dark energy \textit{via} a $\phi$-dependent metric transformation. This results in the dark matter following geodesics dictated by this modified, dark-energy-dependent metric $\td{g}_{\mu \nu}$, related to the gravitational metric $g_{\mu \nu}$ via a conformal transformation:

\begin{equation}
    \td{g}_{\mu \nu} = C(\phi) {g}_{\mu \nu} \mathcomma
\end{equation}

where $C(\phi)$ is the conformal function. We will see that, in practical terms, this is equivalent to an interaction with dark matter, expressed as a modified effective Newton's constant, a dynamical function of the local scalar field value, appropriately known as conformal coupling.

 We start by recalling the general Friedmann-Lema\^{i}tre-Robertson-Walker line element with curvature signature $K$:
\begin{equation}
    \odif{s}^2 = a^2(\tau) \left[ - \odif{\tau}^2 + \frac{\odif{r}^2}{1-K r^2} + r^2 \left( \odif{\theta^2 + \sin^2(\theta) \odif{\varphi}^2} \right) \right] \mathcomma
\end{equation}
 in terms of the conformal time $\tau$ and where $a(\tau)$ is the scale factor for the expansion of the Universe and a prime will denote derivatives with respect to $\tau$ throughout this chapter. The constant $K$ is the curvature parameter, whose encodes the spacetime geometry: $K=0$ is the standard flat case, as assumed in the standard six-parameter $\Lambda$CDM model constrained by the \textit{Planck} Collaboration, while $K=+1$ and $K=-1$ correspond to the positively (closed geometry) and negatively (open geometry) curved cases. This is expressed in Cartesian coordinates as:
\begin{align}
\odif{s}^2 &\equiv g_{\mu \nu} \dd x^{\mu} \dd x^{\nu} =  a(\tau)^2 \left[ - \odif{\tau}^2 + \gamma_{i j} \odif{x}^{i} \odif{x}^{j} \right] \mathcomma \\ 
\gamma_{i j}  &= \delta_{i j} \left[1 + \frac{1}{4} K \left( x^2 + y^2 + z^2 \right) \right]^{-2} \mathperiod \label{eq:gamamet}
\end{align}

The Friedmann equations accounting for curvature, DE, DM and a radiation ($r$) and baryonic ($b$) sectors is:
 \begin{align}
\mathcal{H}^2 &= \frac{1}{3{\rm M}_{\rm Pl}^2} a^2 \left( \rho_r + \rho_b + \rho_{DM} + \rho_{\phi} \right) - K \mathcomma \label{eq:cq_fried1} \\ 
\mathcal{H}' + \mathcal{H}^2 &= - \frac{1}{6 {\rm M}_{\rm Pl}^2} a^2 \left( 2\rho_r + \rho_b + \rho_{DM} + \rho_{\phi} + 3 p_{\phi} \right) \mathperiod \label{eq:cq_fried2}
\end{align}

The Friedmann closure relation now must account for the extra contribution of curvature, encoded in the term $\Omega_K \equiv -K/ H_0^2$:

 \begin{equation}
\Omega_0 = \frac{\rho_i}{3  {\rm M}_{\rm Pl}^2 \mathcal{H}^2}\quad  \xrightarrow{\text{\cref{eq:cq_fried1}}}\quad   \Omega_r + \Omega_b + \Omega_{DM} + \Omega_{\phi} + \Omega_K = 1 \mathperiod
\end{equation}

Owing to the energy exchange, the DM conservation equation now reads 

\begin{equation}
\rho_{DM}' + 3 \mathcal{H} \rho_{DM} = \beta \rho_{DM} \phi'/ M_{\rm Pl} \mathcomma
\label{eq:cq_rhodm}
\end{equation}

where $\rho_{DM}$ is the energy density of DM. The coupling $\beta$ is related to the conformal coupling function $C(\phi)$ via

\begin{equation}
\beta = \frac{M_{\rm Pl}}{2} \frac{{\rm d} \ln C}{{\rm d} \phi} \mathperiod
\end{equation}

The energy density of dark matter particles no longer follows the scaling behaviour of a pressureless fluid. Instead, it scales as $\rho_{DM} \propto \sqrt{C(\phi)} /a^3$. It's worth noting that, when interpreted directly from the action, the mass of DM particles is field-dependent (assuming conservation of the number of particles) and scales as $m(\phi) \propto \sqrt{C(\phi)}$. The evolution of the dark energy scalar field is also influenced by this coupling, leading to a modified form of the Klein-Gordon equation, first introduced in \cref{eq:covkg}:

\begin{equation}
\phi''  + 2 \mathcal{H} \phi' +  a^2 V_{, \phi} = - a^2  \beta \rho_{\rm DM} /\text{M}_{\text{Pl}} \mathperiod
\end{equation}

\subsection{Cosmological Perturbations}

We consider scalar perturbations in the conformal Newtonian gauge (introduced in \cref{sec:newtgauge})
 
\begin{equation}
\mathrm d s^2= a^2(\tau) \left[ - \left( 1 + 2 \Psi \right) d\tau^2 + \left( 1 - 2\Phi \right)  \gamma_{i j} \delta_{ij} dx^{i} dx^{j} \right] \mathcomma
\end{equation}

with $\gamma_{i j}$ as defined in \cref{eq:gamamet}.

 According to \cref{sec:pertcons}, we can write the perturbed continuity and Euler equations for the DM component:

\begin{align}
\delta_{DM}'  &= -  \left( \theta_{DM} - 3 \Phi' \right) - \beta \delta \phi' \mathcomma \\
\theta_{DM}' + \mathcal{H}  \theta_{DM} &= k^2 \Psi +  \beta  \phi' \theta_{DM} -  \frac{\beta}{\text{M}_{\text{Pl}}^2} \delta \phi \mathperiod
\end{align}

These equations describe the clustering of matter that leads to the process of structure formation.

The perturbed KG equation reads

\begin{align}
&\delta\phi''+2\mathcal{H}\delta\phi'+\left(k2^+a^2V_{,\phi\phi}\right)\delta\phi = \\ &\left(\Psi'+3\Phi'\right)\phi'-2a^2V_{,\phi}\Psi-a^2 \beta \delta_{DM} \rho_{DM} - 2a^2  \beta \rho_{DM}\Psi \mathperiod
\end{align}

Additionally, in the subhorizon limit ($k \gg \hub$), the equation for the DM density contrast must account for the gravitational effects introduced by the coupling and becomes:

\begin{equation}
    \delta''_{\rm DM} + \mathcal{H}_{\rm eff} \delta'_{\rm DM} - \frac{3}{2} \mathcal{H}^2 \frac{G_{\rm eff}}{G} \Omega_{\rm DM} \delta_{\rm DM} \simeq \frac{3}{2} \mathcal{H}^2 \left(\Omega_b \delta_b + \Omega_r \delta_r \right) \mathcomma
\end{equation}

where $\mathcal{H}_{\rm eff} = \mathcal{H} + \beta \phi'$ is the effective Hubble parameter.

Due to the coupling in the dark sector, a long-range attractive \textit{fifth-force} between DM particles is introduced, mediated by the scalar field $\phi$. As a result, DM particles interact according to an enhanced effective gravitational constant, expressed as:

\begin{equation}
    G_{\rm eff} = G \left(  1+ 2 \beta^2   \right) \mathperiod
    \label{eq:cq_geff}
\end{equation}

\subsection{Models} \label{sec:cq_fl_ide}





This study explores three specific toy models characterised by their distinct conformal coupling and scalar field potential functions, $C(\phi)$ and $V(\phi)$.

The first model, referred to as $\text{M}1$, employs an exponential form for both the conformal coupling function and the potential \cite{Gomez-Valent:2020mqn}, more precisely:

\begin{equation}
   C(\phi) = e^{2 \alpha \phi / M_{\rm Pl}},\quad \text{and} \quad V(\phi) = V_0^{4} e^{-\lambda  \phi / M_{\rm Pl}} \mathcomma
\end{equation}

where the conformal coupling parameter $\alpha$ and the slope of the scalar field potential $\lambda >0$ are dimensionless constants, and $V_0$ is a constant mass scale of the potential. The coupling between dark energy and dark matter in this model is simply a constant, $\beta = \alpha$, resulting in a constant fifth-force experienced by DM particles, expressed through $G_{\rm eff}/G = 1+2\alpha^2$, according to \cref{eq:cq_geff}. Such an interaction is at the heart of the original coupled quintessence model, thoroughly explored in the literature for various scalar field potentials. Conventionally, studies focus on the general case where $\beta$ can take either positive or negative values. However, in this work, we choose to consider both regimes separately, distinguishing between instances where $\alpha > 0$, and the model is labelled as $\text{M}1^{+}$, and where $\alpha < 0$, and it is termed $\text{M}1^{-}$.

The second model we consider, denoted $\text{M}2$, is characterised by the same exponential conformal coupling function as in $\text{M}1$, but adopts an inverse power-law form for the potential:

\begin{equation}
   C(\phi) = e^{2 \alpha \phi / M_{\rm Pl}}, \quad \text{and} \quad V(\phi) = V_0^{4}  \left(\dfrac{\phi}{M_{\rm Pl}}\right)^{-\mu} \mathperiod
\end{equation}

In this case, the conformal coupling constant $\alpha$ and the power of the potential $\mu > 0$ are also dimensionless constants, and the scale of the potential $V_0$ is still a constant with dimensions of mass. We have already encountered this type of potential in \cref{sec:quin} in the context of quintessence scalar fields. It has the attractive feature of admitting tracker solutions, which can help alleviate the cosmic coincidence problem. Analogously, when $\alpha > 0$, this model is referred to as $\text{M}2^{+}$, and $\text{M}2^{-}$ when $\alpha < 0$.

Lastly, we examine a less conventional scenario, which we designate as $\text{M}3$. In contrast to the constant coupling present in $\text{M}1$ and $\text{M}2$, $\text{M}3$ exemplifies the case of a field-dependent dynamical coupling parameter $\beta$. Indeed, a quintessence model with a dynamical coupling remains an intriguing and relatively uncharted avenue for research. A subset of conformal couplings offering this time-varying fifth-force involves those with minimum points of the effective potential given by the combined effects of $C$ and $V$. The dark sector interaction vanishes when the field sits at this minimum, denoted as $\phi_*$. On the other hand, if the field shifts away from the minimum - possibly at later stages, influenced by the scalar field potential — a fifth-force emerges, coinciding with the onset of dark energy domination. In this work, we investigate such a scenario, accomplished through the same exponential potential as in model $\text{M}1$ and a quadratic exponential conformal function:

\begin{equation}
   C(\phi) = e^{ \gamma ( \phi - \phi_{*} )^{2} / M_{\rm Pl}^2}, \quad \text{and} \quad V(\phi) = V_0^{4} e^{-\lambda  \phi / M_{\rm Pl}} \mathperiod
\end{equation}

Here, $\gamma$ and $\lambda $ are dimensionless constants and $\phi_*$ is a constant with dimensions of mass. The conformal coupling function $C(\phi)$ reaches its minimum at $\phi_*$ when $\gamma > 0$, and the model is termed $\text{M}3^{+}$, or its maximum when $\gamma < 0$, termed $\text{M}3^{-}$. As we will see, this difference will be crucial to understand and differentiate between the two cases. This setting gives the varying dark sector coupling $\beta = \gamma ( \phi - \phi_{*} )$, implying that the fifth-force in the dark sector only switches on when the field is displaced from the minimum/maximum $\phi_*$, set to $\phi_* = 2 M_{\rm Pl}$ without loss of generality. 

Numerically, the mass scale $V_0$ for each model is determined through a shooting algorithm to avoid degeneracies in the parameter space. The initial scalar field value, $\phi_{\rm ini} = \phi(a_{\rm ini})$ with $a_{\rm ini} = 10^{-14}$, is generally allowed to vary. However, in $\text{M}1$ and $\text{M}2$, it is known not to affect the evolution equations (which we check for $\text{M}2$), and, as we will discuss in further detail, it leads to unphysical results for $\text{M}3^{+}$, since the model wants to sit at the minimum of the potential, originating an ill-defined behaviour in the sampling method. Consequently, $\text{M}1$ and $\text{M}2$ introduce two additional parameters $\{\alpha,\lambda\}$ and $\{\alpha,\mu\}$ (even though we also look at sampling $\phi_{\rm ini}$), respectively, while $\text{M}3$ introduces three extra parameters $\{\gamma,\lambda,\phi_{\rm ini}\}$. In \cref{tab:cq_models}, we summarise the differences in the three models.

Furthermore, we are interested in considering extensions to all the models with the addition of one extra degree of freedom of GR that is conventionally fixed in $\Lambda$CDM and extensions to it: the curvature background geometry, parametrised by the curvature density parameter $\Omega_K \equiv 1 - \Omega_0$, as defined in \cref{eq:omega0}. This implies relaxing the assumption of a flat Universe supported by most inflationary models \cite{Forconi:2021que}, such that negative (positive) values stand for a spatially closed (open) Universe.
Nevertheless, while the vast majority of inflationary models naturally predict spatial flatness (\textit{i.e.} $\Omega_K = 0$), inflation in a curved Universe has been extensively explored in the literature (see, \textit{e.g.}, Refs.~\cite{Guth:1980zm,Linde:1995xm,Linde:2003hc,Ratra:2017ezv,Bonga:2016iuf,Handley:2019anl,Bonga:2016cje,Ooba:2017ukj,Ellis:2001ym,Uzan:2003nk,Gordon:2020gel,Sloan:2019jyl}). Regardless of the evidence reported for a closed Universe in $\Lambda$CDM for the Planck data \cite{DiValentino:2019qzk}, for the sake of completeness, we will sample for both positive and negative values of $\Omega_K$, following the procedure considered, for instance, in Ref.~\cite{DiValentino:2020kpf} for interacting dark energy models with different phenomenological couplings.


\begin{table*}[ht!]
    \centering
    \begin{tabular}{|c|c|c|}
\hline
 \backslashbox{$V(\phi)$}{$C(\phi)$}   &  $e^{2 \alpha \phi / \text{M}_{\rm Pl}}$ & $e^{ \gamma ( \phi - \phi_{*} )^{2} / \text{M}_{\rm Pl}^2}$ \\
\hline
$ V_0^{4} e^{-\lambda  \phi / \text{M}_{\rm Pl}}$ & $\mathbf{M1}$ & $\mathbf{M3}$ \\
$V_0^{4}  \left(\phi/\text{M}_{\rm Pl}\right)^{-\mu}$ & $\mathbf{M2}$ & --- \\
    \hline
    \end{tabular}
    \caption[Coupled quintessence models]{Summary of the models considered in this work, distinguished by the conformal factor $C(\phi)$ and the potential $V(\phi)$.}
    \label{tab:cq_models}
\end{table*}

\subsection{Methodology and Observational Data}\label{sec:cq_fl_methodology_data}

To explore the extent and viability of the background and linear perturbation predictions of the scenarios presented in \cref{sec:cq_fl_ide}, we follow the strategy and numerical tools outlined in \cref{chapter:statistics}. We employ the baseline data sets listed in \cref{sec:baseline}, which, for clarity, we enumerate once more:

\begin{itemize}
    \item \textbf{Pl18}: the Cosmic Microwave Background TTTEEE+lowE likelihood from the most recent \textit{Planck} 2018 data release \cite{Aghanim:2018eyx} in all the data set combinations considered in this study. Specifically, this includes the data on the CMB temperature (TT) and polarisation (EE) anisotropies and their cross-correlations (TE) at both small (highlTTTEEE) and large angular scales (lowlTT+lowEE).
    
    \item \textbf{BAO}: a compilation of baryon acoustic oscillations distance and expansion rate measurements from BOSS DR12 \cite{BOSS:2016hvq}, the SDSS Main Galaxy Sample \cite{Ross:2014qpa}, and 6dFGS \cite{Beutler:2011hx}, aligning with the data used by the \textit{Planck} 2018 Collaboration \cite{Aghanim:2018eyx} in their analysis.
    
    \item \textbf{SN}: the distance moduli measurements from Type Ia Supernovae gathered by the Pantheon team \cite{Scolnic:2017caz}. This compilation offers 1048 data points for luminosity distance in the redshift range $z \in [0.01, 2.3]$, from which information on the late time expansion rate can be derived.
\end{itemize}

We analyse the incremental effects of adding the CMB-independent BAO and SN astrophysical data alongside the Planck \textit{likelihood} and the complete combination of these data sets. Our ultimate goal is to perform a Bayesian Monte Carlo Markov Chain analysis to establish observational constraints on the theories considered and compare their evidence against the $\Lambda$CDM model, as summarised in \cref{tab:cq_ok_lcdm_flat}-\cref{tab:cq_ok_m3n_curved}.
We recall that a larger Bayes factor $B_{M, \Lambda {\rm CDM}}$ indicates more substantial evidence of the model $M$ relative to $\Lambda$CDM. This numerical factor can be translated into a qualitative assessment of the level of statistical support for each extended model against $\Lambda$CDM through the use of Jeffreys scale \cite{Kass:1995loi}, as per the criteria listed in \cref{tab:jeff_scale}. Moreover, if $\ln B_{M, \Lambda {\rm CDM}} <0$, there is no evidence of support for that model over $\Lambda$CDM for a given data set, while the opposite holds if $\ln B_{M, \Lambda {\rm CDM}} >0$. We report the Bayes factor $\ln B_{M, \Lambda {\rm CDM}}$ and the $\Delta \chi^2_{\rm eff}$, defined in \cref{eq:stat_chieff,eq:stat_be} to assess the support and goodness of fit in each table of constraints of the corresponding models (including $\Lambda$CDM). Given the increased number of free parameters we choose to run the MCMC chains with convergence criterion based on the Gelman-Rubin statistics of $|r-1| \lesssim 0.02$, as introduced in \cref{chapter:statistics}. We present the results for each model, concluding by comparing the different scenarios and assessing their viability in light of our analysis.

We vary the conventional six $\Lambda$CDM parameters (see \cref{sec:lcdm_param} for more details): the reduced baryon $\omega_{\rm b} = \Omega_{\rm b} h^2$ and dark matter $\omega_{\rm cdm} = \Omega_{\rm cdm} h^2$ energy densities, the ratio between the sound horizon and the angular diameter distance at decoupling $\theta_s$, the reionisation optical depth $\tau_{\rm reio}$, the scalar spectral index $n_s$, and the scalar amplitude of the primordial power spectrum $A_s$, where $h$ is the reduced Hubble constant defined by $H_0 = 100 h \, {\rm km\, s^{-1}\, Mpc^{-1}}$. Additionally we also sample for the respective parameters of each coupled quintessence model: $\{\alpha,\lambda\}$ for $\text{M}1$, $\{\alpha,\mu,\phi_{\rm ini}\}$ for $\text{M}2$, and $\{\gamma,\lambda,\phi_{\rm ini}\}$ for $\text{M}3$.
From these primary parameters, we can derive and report constraints on other relevant derived quantities such as the current mass fluctuation amplitude for spheres of size $8h^{-1}\, \text{Mpc}$, $\sigma_8$, the total matter density at present time, $\Omega_m$, the $S_8$ parameter defined as $S_8 \equiv \sigma_8 \sqrt{\Omega_{m} / 0.3}$, and the Hubble expansion constant, $H_0$. 
In \cref{sec:cq_flat} we will focus on the standard six-parameter $\Lambda$CDM and the corresponding coupled quintessence extensions. These results will then be compared to the case where the flatness assumption is relaxed and $\Omega_K$ is let to vary, as reported in \cref{sec:cq_curved}.
The range of the flat priors for the sampled cosmological and model parameters is listed in \cref{tab:cq_fl_priors}.

\begin{table}[ht!]
\begin{center}
\begin{tabular}{|c|c|c|}
\hline
Model & Parameter                    & Prior \\
\hline\hline
\multirow{6}{*}{All} & $\Omega_{\rm b} h^2$                & $[0.005, 0.1]$ \\
& $\Omega_{\rm cdm} h^2$                & $[0.001, 0.99]$ \\
& $100\theta_s$                    & $[0.5,10]$ \\
& $\tau_{\rm reio}$                          & $[0.01,0.8]$ \\
& $n_s$                      & $[0.7,1.3]$ \\
& $\log \left(10^{10}A_{s} \right)$   & $[1.7, 5.0]$ \\
\hline
All with $\Omega_K = 0$ and $\Omega_K \neq 0$ & $\Omega_K$   & $[-0.3, 0.3]$ \\
\hline \hline
$\mathbf{M1}^{+}$,$\mathbf{M1}^{-}$, $\mathbf{M3}^{+}$ and $\mathbf{M3}^{-}$ & $\lambda$                            & $[0, 10]$ \\
\hline
$\mathbf{M2}^{+}$ and $\mathbf{M2}^{-}$ & $\mu$                            & $[0,10]$ \\
\hline
$\mathbf{M1}^{+}$,$\mathbf{M1}^{-}$, $\mathbf{M2}^{+}$ and $\mathbf{M2}^{-}$ & $\alpha$                            & $[0,2]$ and $[-2,0]$ \\
\hline
$\mathbf{M3}^{+}$ and $\mathbf{M3}^{-}$ & $\gamma$                            & $[0,2]$ and $[-2,0]$ \\
\hline
$\mathbf{M3}^{+}$ and $\mathbf{M3}^{-}$ & $\phi_{\rm ini}$                            & $[0,4]$ \\
\hline %
\end{tabular}
\end{center}
\caption[Priors on $\Lambda$CDM and coupled quintessence parameters]{Flat priors on the cosmological and model parameters sampled in this work, as discussed in \cref{sec:cq_fl_methodology_data}.}
\label{tab:cq_fl_priors}
\end{table}

 \section{Update on Flat Coupled Quintessence} \label{sec:cq_flat}


This section systematically reviews and interprets the findings based on the approach described in \cref{sec:cq_fl_methodology_data} for the flat $\Omega_K = 0$ case. More precisely, we allocate individual subsections for each model examined in this study. Each subsection offers a summary table detailing the cosmological parameters at a $68\%$ confidence level (CL), a graphical representation illustrating the $68\%$ and $95\%$ CL marginalised probability contours, and one-dimensional marginalised posterior distributions. The primary purpose of this analysis is to provide an update on coupled quintessence models and to set the stage for the comparison with the extended models in which the curvature parameter is free to vary and not fixed to the flat case $\Omega_K = 0$.

The results of our data analysis are listed in \cref{tab:cq_ok_lcdm_flat}-\cref{tab:cq_ok_m3n_flat} where we report the parameter constraints for the $\Lambda$CDM, $\text{M}1^{+}$ and $\text{M}1^{-}$, $\text{M}2^{+}$ and $\text{M}2^{-}$, and $\text{M}3^{-}$ models according to the Pl18, Pl18 + BAO, Pl18 + SN, and  Pl18 + BAO + SN data set combinations for each model.
 In \cref{fig:1d_m1_flat,fig:1d_m2_flat,fig:1d_m3_flat}, we show the 1D marginalised posterior distributions for the mode-specific parameters in each case, with the left and right panels corresponding to the positive and negative coupling parameter cases, respectively. In \cref{fig:rectangle_m1_flat,fig:rectangle_m2_flat,fig:rectangle_m3_flat}, we display the 2D marginalised posterior distributions for the conformal coupling parameter in each model (the absolute value) where once again the positive and negative coupling cases are displayed in the left and right panels, respectively.
We present the numerical predictions for the $\Lambda$CDM model in our methodology in \cref{tab:cq_ok_lcdm_flat} for comparison purposes and provide the main conclusions in bullet points in each section.
 



\begin{table*}[ht!]
\begin{center}
\renewcommand{\arraystretch}{1.5}
\resizebox{\textwidth}{!}{
\begin{tabular}{l c c c c c c c c c c c c c c c }
\hline\hline
\textbf{Parameter} & \textbf{ Pl18 } & \textbf{ Pl18 + BAO } & \textbf{ Pl18 + SN } & \textbf{ Pl18 + BAO + SN } \\ 
\hline\hline
$\Omega_b h^2$ & $0.02237\pm 0.00015 $ &  $0.02242\pm 0.00013   $ &  $0.02239\pm 0.00014        $&  $0.02243\pm 0.00013 $ \\
$\Omega_{cdm} h^2$ &$0.1202\pm 0.0014      $ &  $0.1193\pm 0.0010$ &  $0.1199\pm 0.0013          $& $0.11923\pm 0.00097  $  \\
$100 \theta_s$ &   $1.04188\pm 0.00029       $ & $1.04197\pm 0.00028 $ &  $1.04192\pm 0.00029          $ & $1.04196\pm 0.00028     $  \\
$\tau_{reio}$ &   $0.0547^{+0.0071}_{-0.0080} $ & $0.0557^{+0.0072}_{-0.0081} $  &  $0.0551\pm 0.0079           $ & $0.0560\pm 0.0079  $  \\
$n_s$  & $0.9654\pm 0.0044   $ & $0.9673\pm 0.0038     $  &  $0.9660\pm 0.0042          $ & $0.9676\pm 0.0037    $ \\
$\ln \left( 10^{10} A_s \right)$ & $3.046\pm 0.016     $ & $3.046\pm 0.016     $ &  $3.046\pm 0.016          $ & $3.047\pm 0.017        $  \\
\hline 
$\sigma_8$ &$0.8118\pm 0.0075       $ &   $0.8094\pm 0.0072 $ &  $0.8109\pm 0.0075           $ & $0.8093\pm 0.0073   $ \\
$\Omega_m$ & $0.3160\pm 0.0084  $ &   $0.3107\pm 0.0060  $ &  $0.3141\pm 0.0079         $ & $0.3101\pm 0.0058   $ \\
$S_8$ &  $0.833\pm 0.016$ & $0.824\pm 0.013            $ & $0.830\pm 0.016            $ & $0.823\pm 0.013            $  \\
$H_0$ &  $67.33\pm 0.60   $ &   $67.70\pm 0.44        $ &  $67.47\pm 0.57                $ & $67.75\pm 0.43      $ \\
\hline 
\hline
\end{tabular}}
\end{center}
\caption[Observational constraints for flat $\Lambda$CDM]{Observational constraints at $68 \%$ confidence level on the sampled and derived cosmological parameters for different data set combinations under the flat $\Lambda {\rm CDM}$ model.}
\label{tab:cq_ok_lcdm_flat}
\end{table*}


\subsection{\text{M}1 Flat: Exponential Conformal Factor and Exponential Potential} \label{sec:cq_m1_flat}

\begin{table*}[ht!]
\begin{center}
\renewcommand{\arraystretch}{1.5}
\resizebox{\textwidth}{!}{
\begin{tabular}{l c c c c c c c c c c c c c c c }
\hline\hline
\textbf{Parameter} & \textbf{ Pl18 } & \textbf{ Pl18 + BAO } & \textbf{ Pl18 + SN } & \textbf{ Pl18 + BAO + SN } \\ 
\hline\hline
$\Omega_b h^2$ & $0.02235\pm 0.00015$ &  $0.02239\pm 0.00015  $ &  $0.02236\pm 0.00015         $&  $0.02238\pm 0.00014$ \\
$\Omega_{cdm} h^2$ & $0.1193^{+0.0026}_{-0.0015}$ &  $0.1190\pm 0.0011  $ &  $0.1182^{+0.0022}_{-0.0015}      $& $0.1188\pm 0.0010   $  \\
$100 \theta_s$ &   $1.04185\pm 0.00030   $ & $1.04190\pm 0.00029  $ &  $1.04187\pm 0.00030    $ & $1.04190\pm 0.00029   $  \\
$\tau_{reio}$ & $0.0548^{+0.0071}_{-0.0081} $ & $0.0551^{+0.0073}_{-0.0083} $  &  $0.0553\pm 0.0080         $ & $0.0555\pm 0.0080     $  \\
$n_s$  & $0.9662\pm 0.0047  $ & $0.9670\pm 0.0040  $  &  $0.9667\pm 0.0042          $ & $0.9667\pm 0.0040 $ \\
$\ln \left( 10^{10} A_s \right)$ & $3.047^{+0.015}_{-0.016}  $ & $3.046\pm 0.017      $ &  $3.047\pm 0.016            $ & $3.047\pm 0.017 $  \\
\hline 
$\lambda$ & $< 1.26     $ & $0.81^{+0.41}_{-0.60}  $ &  $< 0.510                    $ & $< 0.477    $  \\
$\alpha$ & $< 0.0464   $ &  $0.035\pm 0.018   $  &  $0.038^{+0.017}_{-0.030}$ & $0.029\pm 0.016   $  \\
\hline 
$\sigma_8$ &  $0.801^{+0.036}_{-0.024}   $ &   $0.807^{+0.018}_{-0.016} $ &  $0.826^{+0.011}_{-0.022}           $ & $0.818^{+0.011}_{-0.015}   $\\
$\Omega_m$ &  $0.333^{+0.022}_{-0.045}   $ &   $0.3190^{+0.0098}_{-0.018}  $ &  $0.301^{+0.018}_{-0.011}         $ & $0.3068^{+0.0084}_{-0.0074}  $  \\
$S_8$ &  $0.839\pm 0.022            $ & $0.831^{+0.014}_{-0.016}   $ & $0.826\pm 0.015            $ & $0.827\pm 0.013            $  \\
$H_0$ & $65.8^{+3.8}_{-2.3} $ &   $66.8^{+1.8}_{-0.94}   $ &  $68.59^{+0.88}_{-1.6}             $ & $68.02^{+0.62}_{-0.74}   $  \\
\hline \hline 
$\Delta \chi^{2}_{\rm min} $ &  $+0.80$ & $+0.12$ & $+0.88$ & $+0.54$ \\
$\ln B_{M1^+, \Lambda {\rm CDM}}$ &  $-5.51$ & $-5.31$  & $-6.60$ & $-6.40$ \\
\hline 
\hline
\end{tabular}}
\end{center}
\caption[Observational constraints for the flat M$1^+$ model]{Observational constraints at $68 \%$ confidence level on the sampled and derived cosmological parameters for different data set combinations under the flat M$1^+$ model, studied in \cref{sec:cq_m1_flat}.}
\label{tab:cq_ok_m1p_flat}
\end{table*}

\begin{table*}[ht!]
\begin{center}
\renewcommand{\arraystretch}{1.5}
\resizebox{\textwidth}{!}{
\begin{tabular}{l c c c c c c c c c c c c c c c }
\hline\hline
\textbf{Parameter} & \textbf{ Pl18 } & \textbf{ Pl18 + BAO } & \textbf{ Pl18 + SN } & \textbf{ Pl18 + BAO + SN } \\ 
\hline\hline
$\Omega_b h^2$ &$0.02235\pm 0.00015  $ &  $0.02239\pm 0.00015  $ &  $0.02237\pm 0.00015        $&  $0.02239\pm 0.00015  $ \\
$\Omega_{cdm} h^2$ & $0.1185^{+0.0027}_{-0.0015}$ &  $0.1178^{+0.0018}_{-0.0012}$ &  $0.1179^{+0.0026}_{-0.0015}       $& $0.1179^{+0.0016}_{-0.0011} $ \\
$100 \theta_s$ & $1.04187\pm 0.00030     $ & $1.04191\pm 0.00029  $ &  $1.04189\pm 0.00030       $ & $1.04190\pm 0.00028  $  \\
$\tau_{reio}$ &  $0.0550\pm 0.0080  $ & $0.0554\pm 0.0081  $  &  $0.0547^{+0.0071}_{-0.0080}         $ & $0.0554\pm 0.0077       $  \\
$n_s$  &  $0.9656\pm 0.0044 $ & $0.9671\pm 0.0039  $  &  $0.9666\pm 0.0043         $ & $0.9670\pm 0.0040    $ \\
$\ln \left( 10^{10} A_s \right)$ & $3.047\pm 0.016    $ & $3.047\pm 0.017    $ &  $3.046\pm 0.016             $ & $3.047\pm 0.016   $\\
\hline 
$\lambda$ &$< 0.829   $ & $< 0.659          $ &  $< 0.364                       $ & $< 0.408  $  \\
$\alpha$ & $> -0.0322 $ &  $-0.028^{+0.022}_{-0.013}     $  &  $> -0.0409   $ & $-0.030\pm 0.016             $  \\
\hline 
$\sigma_8$ & $0.808\pm 0.023     $ &   $0.811\pm 0.015     $ &  $0.820^{+0.010}_{-0.018}         $ & $0.817^{+0.011}_{-0.015} $ \\
$\Omega_m$ &   $0.321^{+0.021}_{-0.024}   $ &   $0.3110^{+0.0094}_{-0.011} $ &  $0.304^{+0.016}_{-0.010}         $ & $0.3052^{+0.0087}_{-0.0079}  $  \\
$S_8$ &  $0.834\pm 0.018            $ & $0.825\pm 0.013            $ & $0.825\pm 0.016            $ & $0.824\pm 0.013            $  \\
$H_0$ & $66.5^{+2.2}_{-1.8}     $ &   $67.3^{+1.1}_{-0.80}    $ &  $68.14^{+0.77}_{-1.3}               $ & $67.98\pm 0.69   $  \\
\hline \hline 
$\Delta \chi^{2}_{\rm min} $ &  $+0.34$ & $+0.18$ & $+1.14$ & $+0.24$ \\
$\ln B_{M1^-, \Lambda {\rm CDM}}$ &  $-6.03$ & $-6.03$ & $-6.97$ & $-6.77$ \\
\hline 
\hline
\end{tabular}}
\end{center}
\caption[Observational constraints for the flat M$1^-$ model]{Observational constraints at $68 \%$ confidence level on the sampled and derived cosmological parameters for different data set combinations under the flat M$1^-$ model, studied in \cref{sec:cq_m1_flat}.}
\label{tab:cq_ok_m1n_flat}
\end{table*}

\begin{figure*}[ht!]     
\centering     
\subfloat{\includegraphics[width=0.48\textwidth]{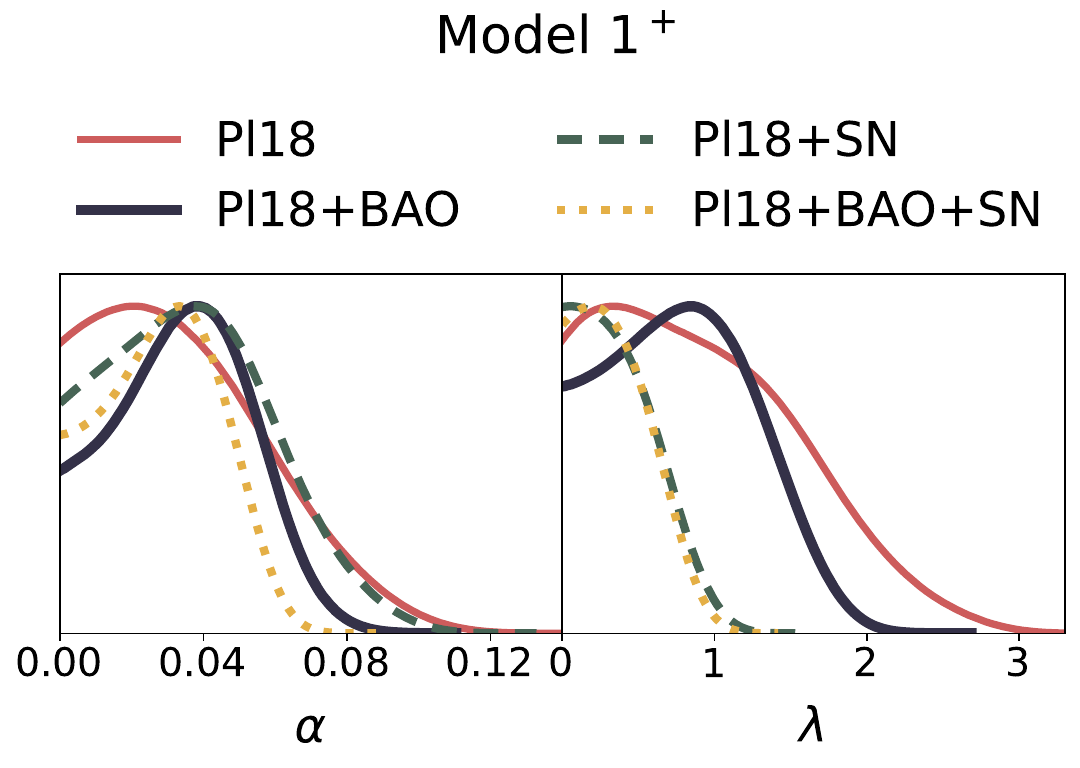}} \hfill
\subfloat{\includegraphics[width=0.48\textwidth]{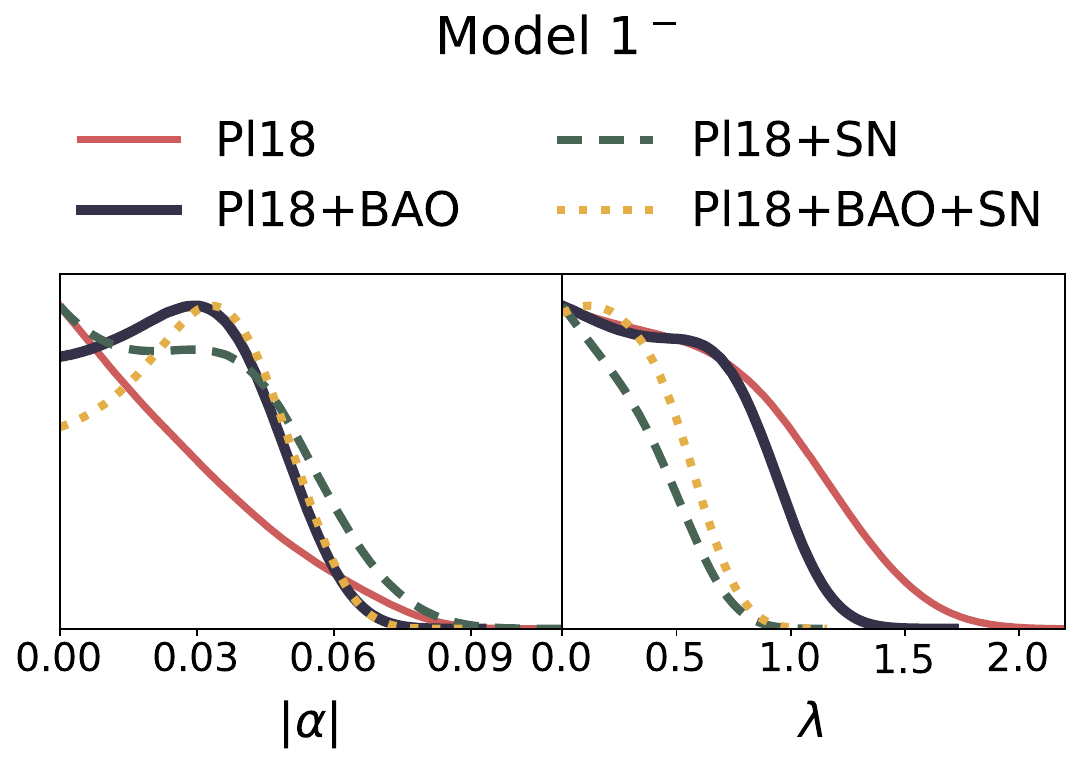}}
\caption[1D marginalised posterior distributions in the flat $\text{M}1$ models]{1D marginalised posterior distributions of the coupling parameter, $\alpha$, and slope of the potential, $\lambda$, in the $\text{M}1^{+}$ model (upper panel) and $\text{M}1^{-}$  model (lower panel) using the Pl18 (red), Pl18+BAO (blue), Pl18+SN, (green) and Pl18+BAO+SN (yellow) data set combinations.}     
\label{fig:1d_m1_flat} 
\end{figure*}

\begin{figure*}[ht!]     
\centering     
\subfloat{\includegraphics[width=\textwidth]{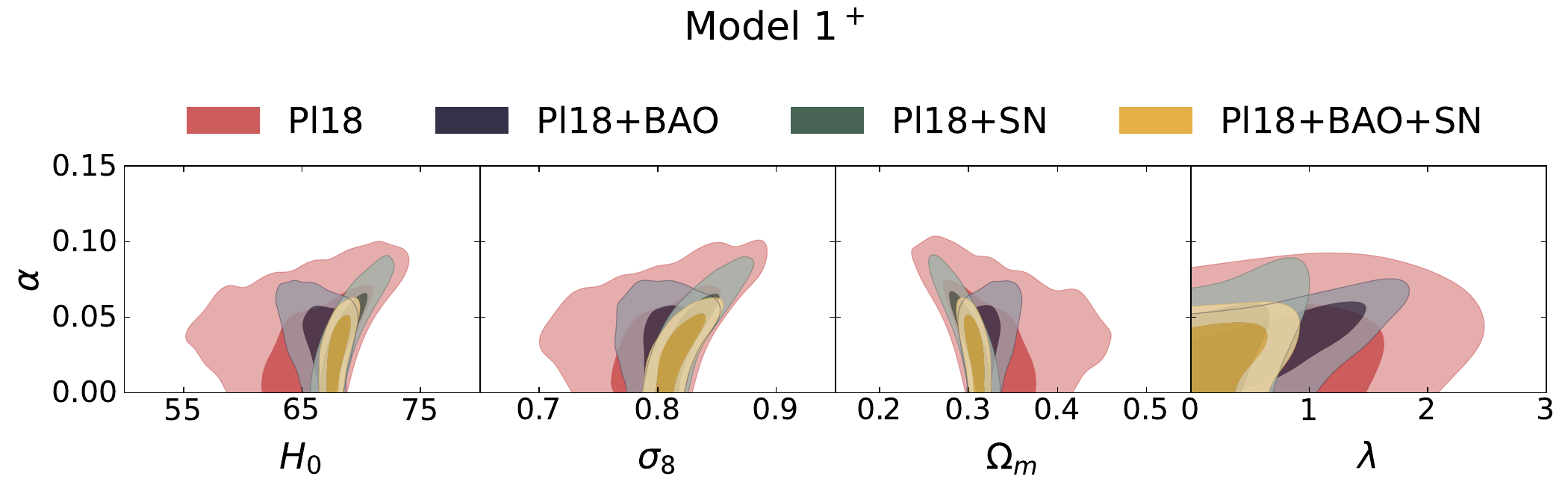}} \hfill
\subfloat{\includegraphics[width=\textwidth]{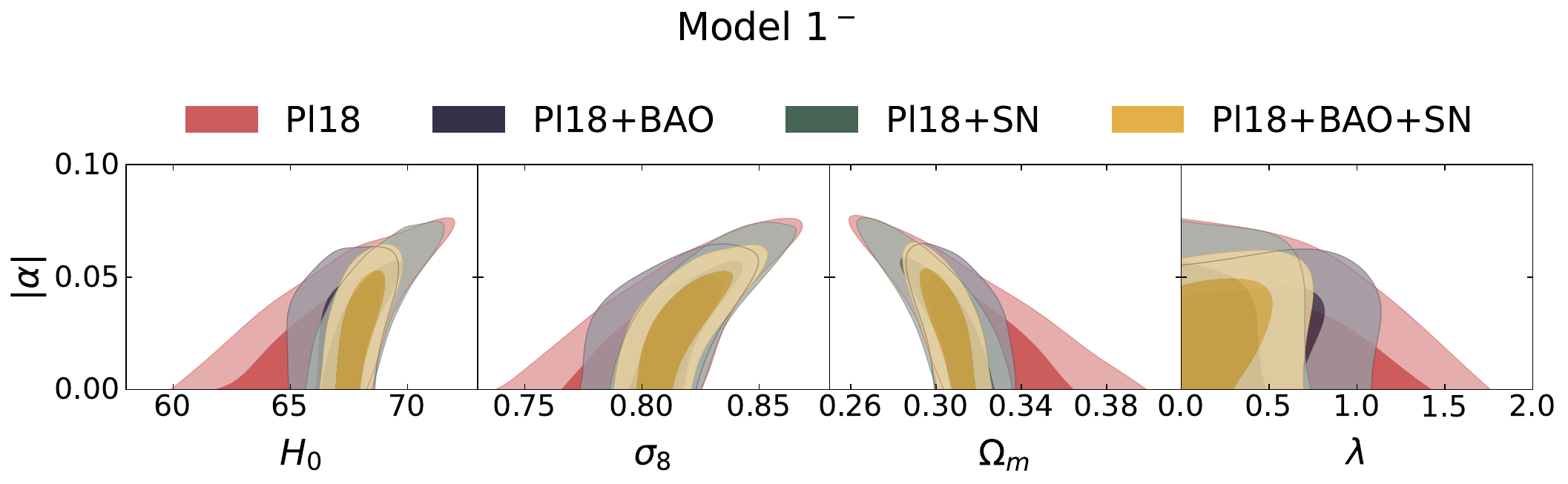}} 
\caption[2D marginalised posterior distributions in the flat $\text{M}1$ models]{2D marginalised posterior distributions of parameters in the $\text{M}1^{+}$ model (upper panel) and $\text{M}1^{-}$  model (lower panel) using the Pl18 (red), Pl18+BAO (blue), Pl18+SN, (green) and Pl18+BAO+SN (yellow) data set combinations. We plot the 2D marginalised posterior distributions of the conformal coupling parameter, $\alpha$, against the slope of the potential, $\lambda$, the Hubble constant in units ${\rm \: km \: s^{-1} \: {Mpc}^{-1}}$, $H_0$, the present-day mass fluctuation amplitude in spheres of radius $8h^{-1} {\rm Mpc}$,  $\sigma_8$, and the total matter density parameter, $\Omega_m$.  The shaded contours indicate the $1\sigma$ and $2\sigma$ confidence limits.}     
\label{fig:rectangle_m1_flat} 
\end{figure*}



\begin{itemize}
    \item Coupling: In \cref{tab:cq_ok_m1p_flat,tab:cq_ok_m1n_flat}, we observe that while Pl18 data on its own accommodates a vanishing coupling constant $\alpha$ in the $68\%$ CL region for the $\text{M}1$ models, incorporating additional individual background data sets, BAO or SN in this case, yields higher support for the dark sector interaction in both $\text{M}1^{+}$ and $\text{M}1^{-}$ models (\textit{i.e.} $|\alpha|>0$ in the $1\sigma$ region), associated with larger mean values in general. The exception is in the $\text{M}1^{-}$ model when adding SN data (Pl18+SN in \cref{tab:cq_ok_m1n_flat}), which results in tighter bounds. For the M$1^+$ model, we report bounded constraints for $\alpha$ at $1\sigma$  with all combinations involving additional background data. For M$1^-$, we find similar bounds for $|\alpha|$ with Pl18+BAO and Pl18+BAO+SN. \cref{fig:1d_m1_flat} illustrates this trend through the marginalised posterior distributions, where additional datasets result in a non-zero peak for $|\alpha|$, with BAO being the decisive driver of the peak away from zero, possibly by setting tighter constraints on $\Omega_m$. More precisely, with the Pl18 data set alone the following upper bounds are found at $68\%$ CL: $\alpha < 0.0464$ for the $\text{M}1^{+}$ model and $|\alpha| < 0.0322$ for $\text{M}1^{-}$. For the full Pl18+BAO+SN data set combination we report at $1\sigma$: $\alpha=0.029 \pm 0.016$ for the $\text{M}1^{+}$ model and $|\alpha| = 0.030 \pm 0.016$ for $\text{M}1^{-}$. In \cref{fig:rectangle_m1_flat}, we depict the 2D marginalised posterior contours for $\alpha$ against relevant cosmological and model parameters: $\{H_0,\sigma_8,\Omega_m,\lambda\}$. For both the $\text{M}1^{+}$ and $\text{M}1^{-}$ models, positive correlations are evident in the $\{|\alpha|, H_0\}$ and $\{|\alpha|, \sigma_8\}$ planes, while a negative correlation is observed in the $\{|\alpha|, \Omega_m\}$ plane. The scalar field parameters $|\alpha|$ and $\lambda$ are positively correlated for positive and negative coupling cases, except in the Pl18-only case in $\text{M}1^{-}$, where it is unclear whether any correlation exists.

    \item Potential: In \cref{tab:cq_ok_m1p_flat,tab:cq_ok_m1n_flat}, we also report on the slope of the potential $\lambda$, which is consistently constrained only from above in the M$1$ models, except when using the Pl18+BAO dataset for the $\text{M}1^{+}$ model, in which case we find a bounded $1\sigma$ constrained region. Incorporation of background datasets brings $\lambda$ closer to zero, with higher values of $\lambda$ allowed in the $\text{M}1^{+}$ model in general. This effect is depicted in \cref{fig:1d_m1_flat} and seems mostly influenced by the SN data. For Pl18 we report $\lambda < 1.26$ and $\lambda < 0.829$ at $68\%$ CL for $\text{M}1^{+}$ and $\text{M}1^{-}$, respectively, while the full Pl18+BAO+SN data set combination restricts $\lambda$ to $< 0.477$ for $\text{M}1^{+}$ and $< 0.408$ for $\text{M}1^{-}$.

    \item $H_0$ Tension: The positive correlation between $|\alpha|$ and $H_0$ aids in mitigating the $H_0$ tension of $\sim 4.8\sigma$ in $\Lambda$CDM when considering Pl18 data alone. The tension is apparently attenuated to $\sim 2.5 \sigma$ in $\text{M}1^{+}$ and to $\sim 2.9 \sigma$ in the $\text{M}1^{-}$ model. It is worth noting that the mean values of $H_0$ in the $\text{M}1$ models, as reported in the second columns of \cref{tab:cq_ok_m1p_flat,tab:cq_ok_m1n_flat} for Pl18, are smaller than in the $\Lambda$CDM model. The tension alleviation is attributed to the broader error bars for $H_0$, a direct consequence of the added number of parameters that prevent $H_0$ from being as precisely constrained as in the $\Lambda$CDM case.

    \item $S_8$ Tension: While $|\alpha|$ exhibits a positive correlation with the parameter $\sigma_8$, the mean values of $\sigma_8$ reported for the $\text{M}1$ models turn out to be smaller than their $\Lambda$CDM counterpart for Pl18 data. Nonetheless, the $68\%$ CL region is broader in the $\text{M}1$ models. A suppressed $\sigma_8$ is offset by an enhanced $\Omega_m$ when considering the Pl18 data, balancing the constraints derived for the $S_8$ parameter. For $\text{M}1^{+}$ the $1\sigma$ region for $S_8$ value is found at $0.839\pm 0.022$ and at $0.834\pm 0.018$ for $\text{M}1^{-}$. These values do not deviate far from the $\Lambda$CDM prediction of $0.833\pm 0.016$.

    \item Model Evidence: To conclude, the logarithm of the Bayes factor, $\ln B_{{\rm M1}, \Lambda {\rm CDM}}$ is the decisive criteria for any potential support in the extended models. We find a negative value for all the data combinations, suggesting no statistical evidence for the M1 coupled quintessence models over $\Lambda$CDM. M$1^+$ has a marginally less negative Bayes factor than M$1^-$, indicating a slight but unimportant preference for positive values of $\alpha$ over negative ones. The $\Delta \chi^2_{\text{min}}$ is positive in all the cases considered, suggesting a worse fit to the data, which adds to the preference over a smaller parameter space in $\Lambda$CDM.
\end{itemize}



\subsection{\text{M}2 Flat: Exponential Conformal Factor and Inverse Power-law Potential} \label{sec:cq_m2_flat}

\begin{figure*}[ht!]     
\centering     
\subfloat{\includegraphics[width=0.48\textwidth]{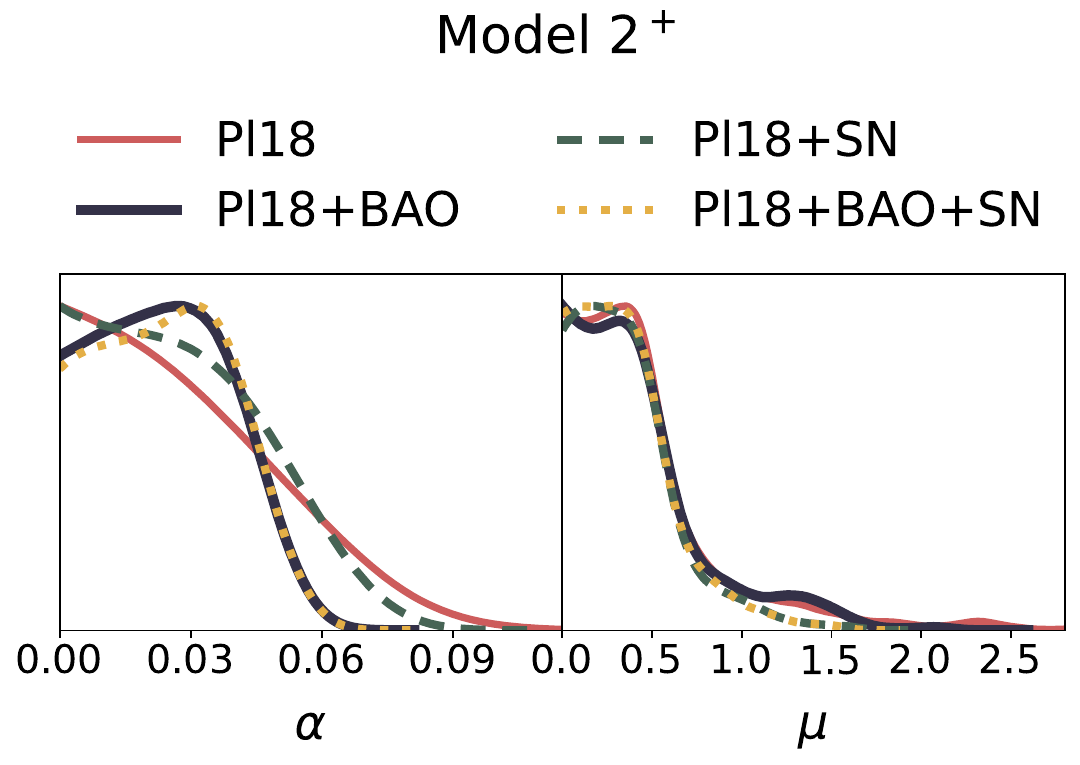}} \hfill
\subfloat{\includegraphics[width=0.48\textwidth]{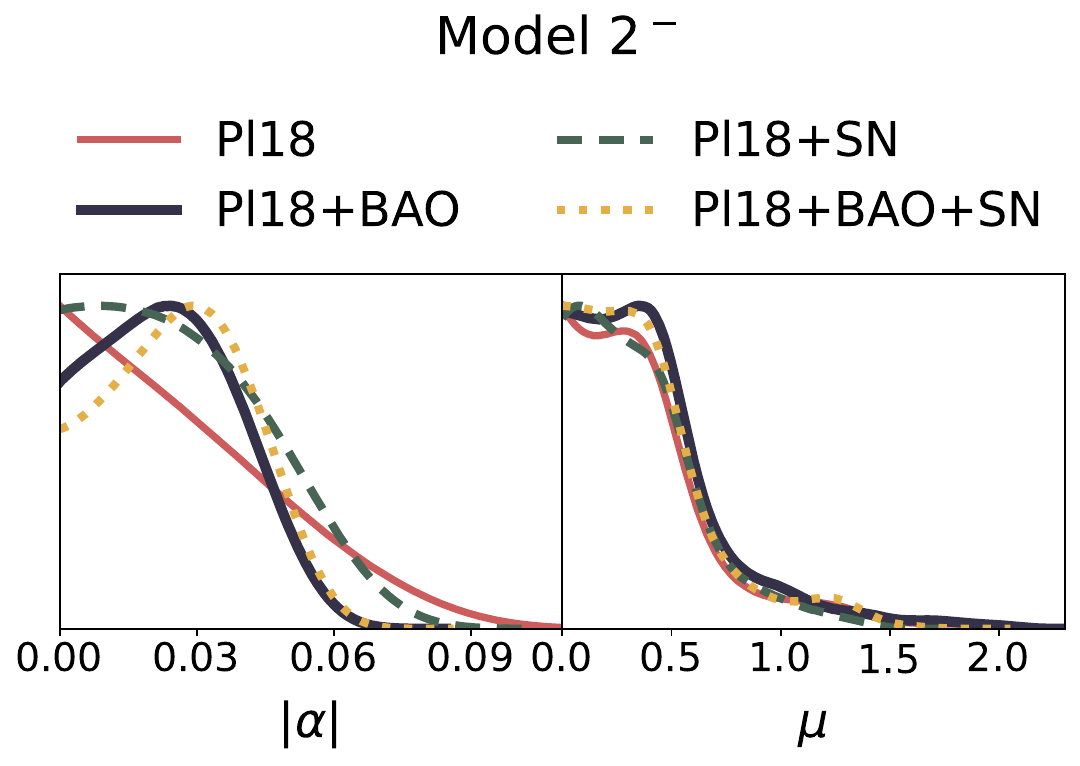}}
\caption[1D marginalised posterior distributions in the flat $\text{M}2$ models]{1D marginalised posterior distributions of the coupling parameter, $\alpha$, and slope of the potential, $\mu$, in the $\text{M}2^{+}$ model (upper panel) and $\text{M}2^{-}$  model (lower panel) using the Pl18 (red), Pl18+BAO (blue), Pl18+SN, (green) and Pl18+BAO+SN (yellow) data set combinations.}     
\label{fig:1d_m2_flat} 
\end{figure*}

\begin{figure*}[ht!]     
\centering     
\subfloat{\includegraphics[width=\textwidth]{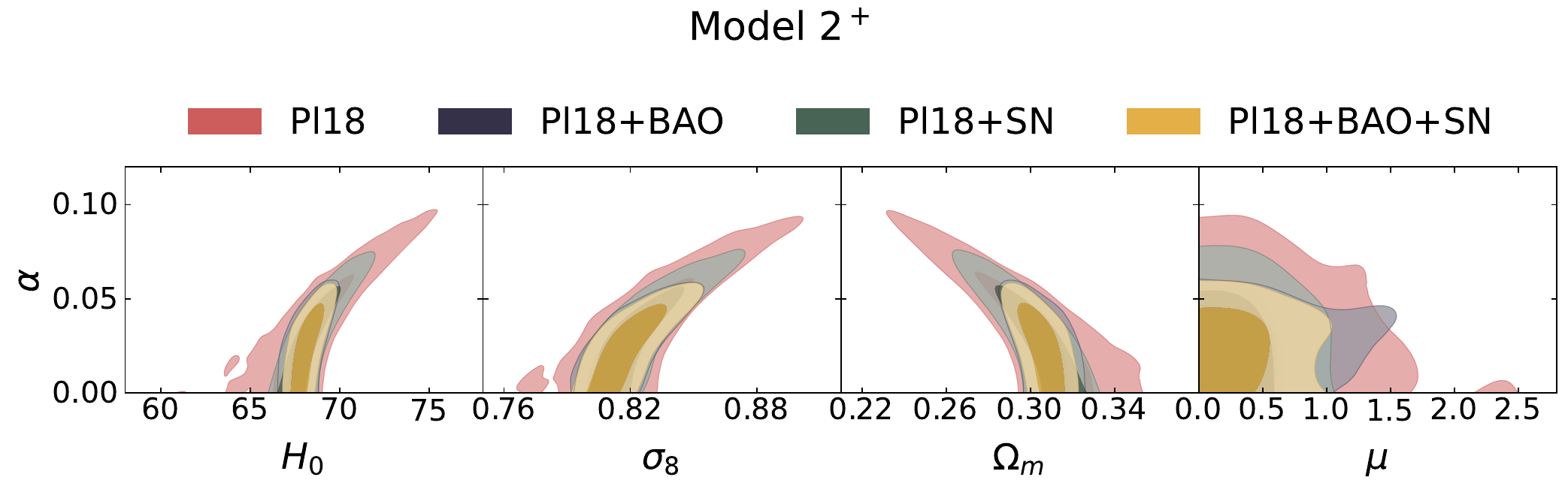}} \hfill
\subfloat{\includegraphics[width=\textwidth]{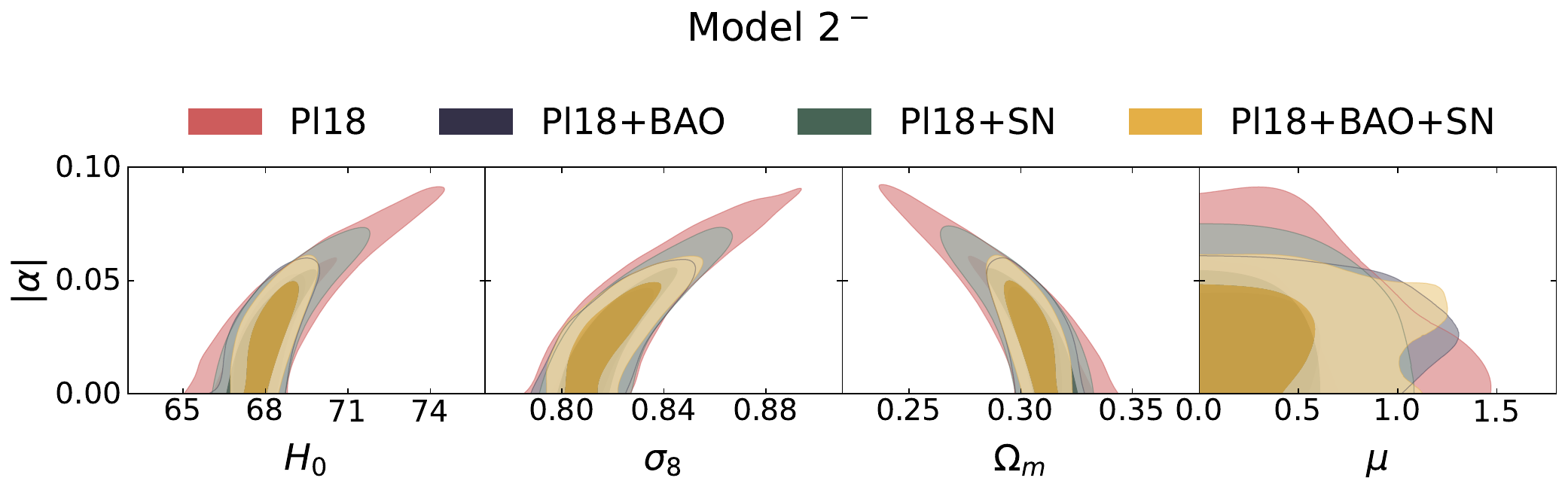}} 
\caption[2D marginalised posterior distributions in the flat $\text{M}2$ models]{2D marginalised posterior distributions of parameters in the $\text{M}2^{+}$ model (upper panel) and $\text{M}2^{-}$  model (lower panel) using the Pl18 (red), Pl18+BAO (blue), Pl18+SN, (green) and Pl18+BAO+SN (yellow) data set combinations. We plot the 2D marginalised posterior distributions of the conformal coupling parameter, $\alpha$, against the power of the potential, $\mu$, the Hubble constant in units ${\rm \: km \: s^{-1} \: {Mpc}^{-1}}$, $H_0$, the present-day mass fluctuation amplitude in spheres of radius $8h^{-1} {\rm Mpc}$,  $\sigma_8$, and the total matter density parameter, $\Omega_m$.  The shaded contours indicate the $1\sigma$ and $2\sigma$ confidence limits.}     
\label{fig:rectangle_m2_flat} 
\end{figure*}

\begin{itemize}
    \item Coupling: As evident from \cref{tab:cq_ok_m2p_flat}, for the $\text{M}2^{+}$ model, a non-zero coupling at $68\%$ CL is detected when BAO and BAO+SN data set combinations are added to the Pl18 data, more precisely $\alpha = 0.025^{+0.012}_{-0.020}$ and $\alpha = 0.026^{+0.014}_{-0.019}$, respectively. For the $\text{M}2^{+}$ model, reported in \cref{tab:cq_ok_m2n_flat}, $\alpha$ is bounded at $68\%$ CL with the addition of the BAO and BAO+SN data set combinations as well, namely $\alpha = -0.025^{+0.021}_{-0.011}$ and $\alpha =- 0.027 \pm 0.015$, respectively. While the uncoupled case $\alpha=0$ is contained in the $1\sigma$ region when Pl18 data is considered by itself and for the Pl18+SN case, the introduction of BAO and BAO+SN is again responsible for excluding a zero coupling at $1\sigma$. Close inspection of \cref{fig:1d_m2_flat} shows how the inclusion of the BAO data originates a non-null peak for $|\alpha|$. This trend is maintained in both $\text{M}2^{+}$ and $\text{M}2^{-}$ when SN data is included. The magnitude of the coupling parameter is comparable in both cases. \cref{fig:rectangle_m2_flat} shows 2D marginalised posterior distributions of the conformal coupling parameter $\alpha$ against $\{H_0,\sigma_8$,$\Omega_m,\mu\}$, just as we did for the $\text{M}1$ models,. Both $\text{M}2$ models exhibit a positive correlation in the $\{|\alpha|, H_0\}$ and $\{|\alpha|, \sigma_8\}$ planes, while showing a negative correlation in the $\{|\alpha|, \Omega_m\}$ plane instead. Unlike the $\text{M}1$ models, the coupling and potential parameters are not seemingly correlated.

    \item Potential: The power of the inverse power potential, $\mu$, is only bounded from above with $68\%$ CL in $\text{M}2$ models across any data set combinations. As shown in \cref{tab:cq_ok_m2n_flat}, adding background data sets scarcely tightens the constraint on $\mu$ in the $\text{M}2$ models. Pl18 data alone limits $\mu$ to be larger than $0.459$, while the full Pl18+BAO+SN data set sets this lower limit at $0.463$, both at $1\sigma$. A parallel constraint of $\mu < 0.502$ and $\mu < 0.451$ at $1\sigma$ is observed for the $\text{M}2^{+}$ model with Pl18 and Pl18+BAO+SN. In \cref{fig:1d_m2_flat}, the similar behaviour of $\mu$ distributions between the two $\text{M}2$ models is noticeable. The values of $|\alpha|$ and $\mu$ fall within similar ranges in both models when considering the full Pl18+BAO+SN data set.

    \item $H_0$ Tension: The $H_0$ tension is alleviated to some extent by the positive correlation between $|\alpha|$ and $H_0$, similar to the $\text{M}1$ models. Using Pl18 data, the $\text{M}2^{+}$ and $\text{M}2^{-}$ models lessen the tension from approximately $4.8\sigma$ to $\sim 4.0\sigma$ and $2.8\sigma$, respectively, which is a less significant reduction compared to the $\text{M}1$ models. Notably, contrary to the $\text{M}1$ models, the mean $H_0$ value in the $\text{M}2^{-}$ model is higher, yet with smaller error bars, making it less effective in reducing the tension, according to \cref{eq:tensiondef}.

    \item $S_8$ Tension: In contrast to $\text{M}1$, higher average values for the $\sigma_8$ parameter are obtained in the $\text{M}2$ models compared to $\Lambda$CDM for all data set combinations. We also report consistently lower mean values of $\Omega_m$. The increase in $\sigma_8$ and the decrease in $\Omega_m$ approximately balance out again, resulting in an $S_8$ value of $0.828\pm 0.018$ at $1\sigma$ for the $\text{M}2^{+}$ model, and $0.827\pm 0.018$ for $\text{M}2^{-}$, both not significantly different from the $\Lambda$CDM value assuming Pl18 data. 

    \item Model Evidence: \cref{tab:cq_ok_m2p_flat,tab:cq_ok_m2n_flat} reveal that the logarithm of the Bayes factor, $\ln B_{\text{M2}, \Lambda {\rm CDM}}$, is negative for all the data sets, suggesting a lack of evidence supporting the $\text{M}2$ models over the $\Lambda$CDM model. 
    As was the case for the $\text{M}1$ models, there is an overall trend of positive values of $\Delta \chi^2_{\text{min}}$ (except for Pl18 in $\text{M}2^+$ and Pl18+SN and Pl18+BAO+SN in $\text{M}2^-$) which is less pronounced and, consistently, the fit to the data is not sufficiently improved to justify the addition of extra parameters. Nevertheless, it should be noted that the results reported in \cref{tab:cq_ok_m2p_flat,tab:cq_ok_m2n_flat} include the sampling of $\phi_{\rm ini}$, which could in principle be fixed without loss of generality, decreasing the prior volume.
\end{itemize}


\begin{table*}[ht!]
\begin{center}
\renewcommand{\arraystretch}{1.5}
\resizebox{\textwidth}{!}{
\begin{tabular}{l c c c c c c c c c c c c c c c }
\hline\hline
\textbf{Parameter} & \textbf{ Pl18 } & \textbf{ Pl18 + BAO } & \textbf{ Pl18 + SN } & \textbf{ Pl18 + BAO + SN } \\ 
\hline\hline
$\Omega_b h^2$ & $0.02236\pm 0.00015$ & $0.02238\pm 0.00014   $ &  $0.02237\pm 0.00015        $&  $0.02238\pm 0.00015  $ \\
$\Omega_{cdm} h^2$ & $0.1185^{+0.0028}_{-0.0016}$ & $0.1190\pm 0.0011  $ &  $0.1185^{+0.0021}_{-0.0014}    $& $0.1188\pm 0.0011    $  \\
$100 \theta_s$ & $1.04186\pm 0.00030$ &  $1.04190\pm 0.00029  $ &  $1.04189\pm 0.00029           $ & $1.04191\pm 0.00029   $ \\
$\tau_{reio}$ & $0.0546\pm 0.0079$ & $0.0550^{+0.0073}_{-0.0082} $  &  $0.0546\pm 0.0082          $ & $0.0554\pm 0.0079      $ \\
$n_s$  & $0.9660\pm 0.0045$ & $0.9663\pm 0.0038   $  &  $0.9664\pm 0.0041             $ & $0.9666\pm 0.0039     $ \\
$\ln \left( 10^{10} A_s \right)$ & $3.046\pm 0.016$ &$3.047^{+0.015}_{-0.017}    $ &  $3.046\pm 0.017                 $ & $3.047\pm 0.016   $  \\
\hline 
$\mu$ & $< 0.502$ & $< 0.500      $ &  $< 0.450                  $ & $< 0.451      $  \\
$\alpha$ & $<0.0396$ &$0.025^{+0.012}_{-0.020}  $  &  $< 0.0387    $ & $0.026^{+0.014}_{-0.019}   $  \\
\hline 
$\sigma_8$ & $0.823^{+0.014}_{-0.021}$ & $0.818^{+0.010}_{-0.014}  $ &  $0.8232^{+0.0086}_{-0.018}         $ & $0.8195^{+0.0091}_{-0.014}  $  \\
$\Omega_m$ &  $0.305^{+0.020}_{-0.013}$ & $0.3068\pm 0.0083   $ &   $0.303^{+0.016}_{-0.0093} $ &  $0.3051^{+0.0079}_{-0.0070}         $ \\
$S_8$ & $0.828\pm 0.018$ & $0.827\pm 0.013            $ & $0.826\pm 0.015            $ & $0.826\pm 0.013            $  \\
$H_0$ & $68.3^{+1.0}_{-1.7}$ &$68.05\pm 0.70  $ &  $68.44^{+0.67}_{-1.4}              $ & $68.19^{+0.52}_{-0.67}     $  \\
\hline \hline 
$\Delta \chi^{2}_{\rm min} $ &  $-0.68$ & $0.12$ & $0.66$ & $0.52$ \\
$\ln B_{M2^+, \Lambda {\rm CDM}}$ &  $-6.55$ & $-6.63$ & $-6.45$ & $-6.71$ \\
\hline 
\hline
\end{tabular}}
\end{center}
\caption[Observational constraints for the flat M$2^+$ model]{Observational constraints at $68 \%$ confidence level on the sampled and derived cosmological parameters for different data set combinations under the flat M$2^+$ model, studied in \cref{sec:cq_m2_flat}.}
\label{tab:cq_ok_m2p_flat}
\end{table*}

\begin{table*}[ht!]
\begin{center}
\renewcommand{\arraystretch}{1.5}
\resizebox{\textwidth}{!}{
\begin{tabular}{l c c c c c c c c c c c c c c c }
\hline\hline
\textbf{Parameter} & \textbf{ Pl18 } & \textbf{ Pl18 + BAO } & \textbf{ Pl18 + SN } & \textbf{ Pl18 + BAO + SN } \\ 
\hline\hline
$\Omega_b h^2$ & $0.02237\pm 0.00015  $ &  $0.02238\pm 0.00014      $ &  $0.02236\pm 0.00015        $&  $0.02238\pm 0.00014 $ \\
$\Omega_{cdm} h^2$ &  $0.1183^{+0.0030}_{-0.0014}$ &  $0.1188\pm 0.0012   $ &  $0.1185^{+0.0021}_{-0.0013}       $& $0.1186^{+0.0012}_{-0.0011} $  \\
$100 \theta_s$ &   $1.04187\pm 0.00030    $ & $1.04191\pm 0.00028   $ &  $1.04187\pm 0.00029         $ & $1.04191\pm 0.00028   $ \\
$\tau_{reio}$ &  $0.0550^{+0.0072}_{-0.0084} $ & $0.0553\pm 0.0081      $  &  $0.0551\pm 0.0076           $ & $0.0547\pm 0.0074      $  \\
$n_s$  & $0.9663\pm 0.0045  $ & $0.9663\pm 0.0040   $  &  $0.9662\pm 0.0041          $ & $0.9661\pm 0.0038  $ \\
$\ln \left( 10^{10} A_s \right)$ & $3.047\pm 0.016  $ & $3.047\pm 0.016     $ &  $3.047\pm 0.016               $ & $3.046\pm 0.015  $  \\
\hline 
$\mu$ & $< 0.459   $ & $< 0.484    $ &  $< 0.458                     $ & $< 0.463    $ \\
$\alpha$ & $> -0.0385 $ &  $-0.025^{+0.021}_{-0.011}   $  &  $> -0.0379 $ & $-0.027\pm 0.015   $  \\
\hline 
$\sigma_8$ &$0.8240^{+0.0083}_{-0.021}    $ &   $0.8185^{+0.0096}_{-0.013} $ &  $0.8229^{+0.0092}_{-0.018}         $ & $0.8194^{+0.0092}_{-0.015}  $  \\
$\Omega_m$ &  $0.303^{+0.022}_{-0.010}  $ &   $0.3063\pm 0.0084  $ &  $0.304^{+0.016}_{-0.0086}        $ & $0.3045^{+0.0085}_{-0.0073}  $  \\
$S_8$ &  $0.827\pm 0.018            $ & $0.827\pm 0.013            $ & $0.827\pm 0.016            $ & $0.825\pm 0.013            $  \\
$H_0$ & $68.42^{+0.72}_{-1.8}  $ &   $68.05\pm 0.71   $ &  $68.32^{+0.62}_{-1.3}             $ & $68.21^{+0.58}_{-0.71} $ \\
\hline \hline 
$\Delta \chi^{2}_{\rm min} $ &  $+1.14$ & $+0.06$ & $-0.32$ & $-2.12$ \\
$\ln B_{M2^-, \Lambda {\rm CDM}}$ &  $-6.61$ & $-6.51$ & $-9.88$ & $-3.57$ \\
\hline 
\hline
\end{tabular}}
\end{center}
\caption[Observational constraints for the flat M$2^-$ model]{Observational constraints at $68 \%$ confidence level on the sampled and derived cosmological parameters for different data set combinations under the flat M$2^-$ model, studied in \cref{sec:cq_m2_flat}.}
\label{tab:cq_ok_m2n_flat}
\end{table*}


\subsection{\text{M}3 Flat: Coupling with Minimum/Maximum and Exponential Potential} \label{sec:cq_m3_flat}







\begin{figure*}[ht!]     
\centering     
\subfloat{\includegraphics[width=0.48\textwidth]{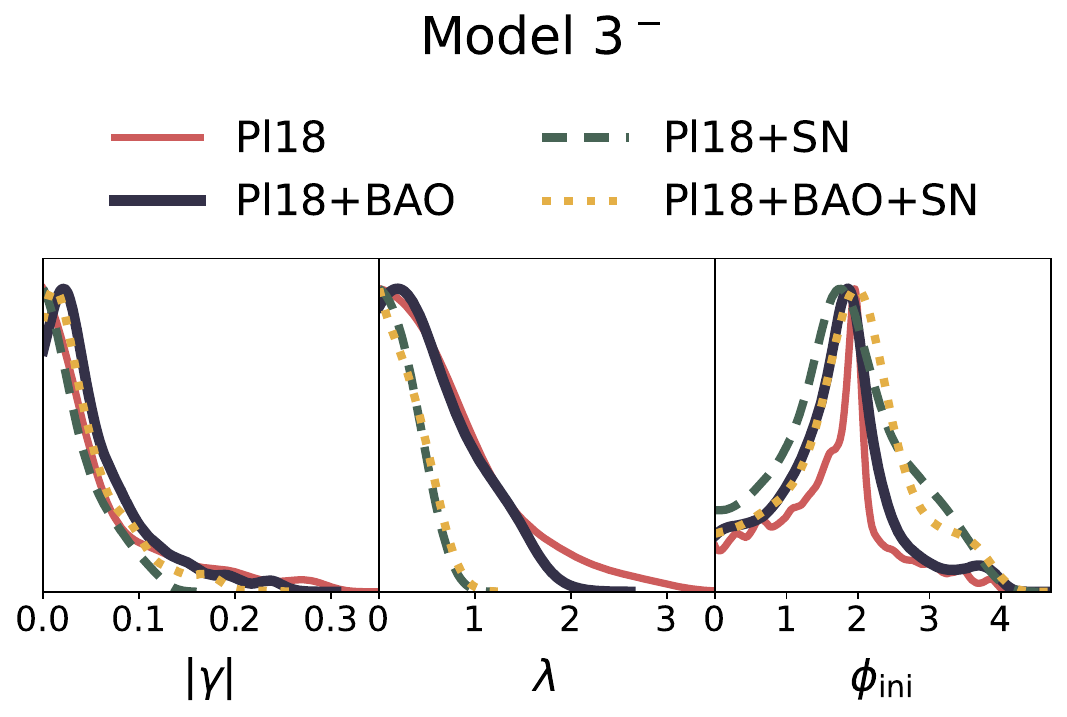}}
\caption[1D marginalised posterior distributions in the flat $\text{M}3^-$ model]{1D marginalised posterior distributions of the coupling parameter, $\gamma$, and slope of the potential, $\lambda$, in the $\text{M}3^{-}$  model using the Pl18 (red), Pl18+BAO (blue), Pl18+SN, (green) and Pl18+BAO+SN (yellow) data set combinations.}     
\label{fig:1d_m3_flat} 
\end{figure*}
\begin{figure*}[ht!]     
\centering     
\subfloat{\includegraphics[width=\textwidth]{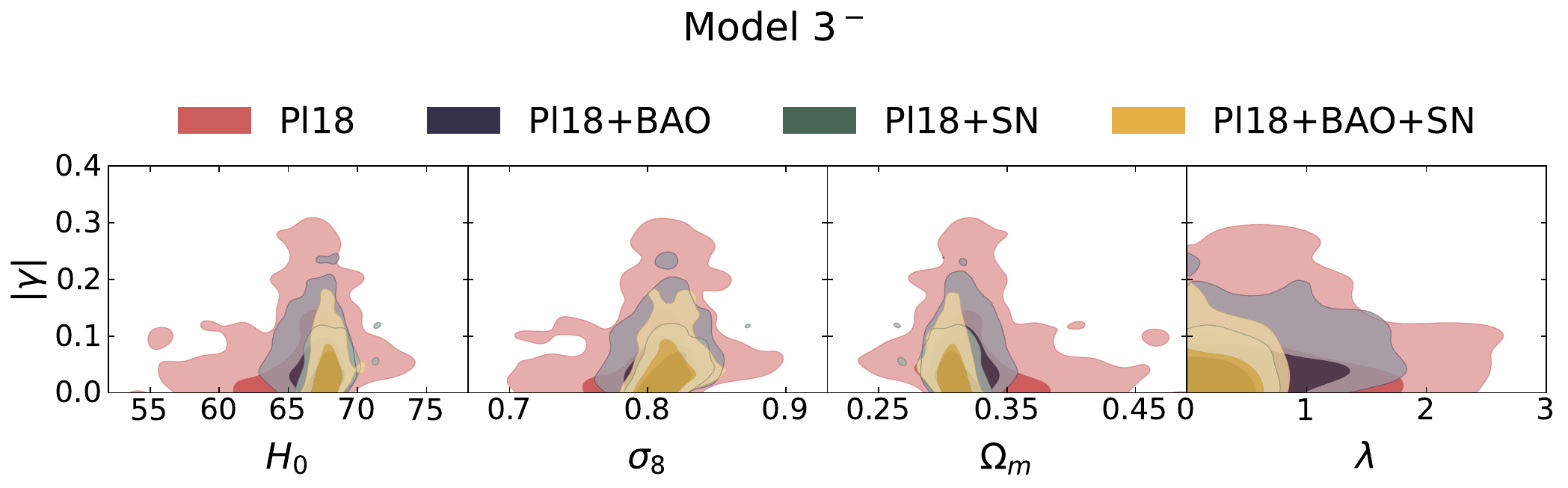}} 
\caption[2D marginalised posterior distributions in the flat $\text{M}3^-$ model]{2D marginalised posterior distributions of parameters in the $\text{M}3^{-}$  model using the Pl18 (red), Pl18+BAO (blue), Pl18+SN, (green) and Pl18+BAO+SN (yellow) data set combinations. We plot the 2D marginalised posterior distributions of the conformal coupling parameter, $\gamma$, against the slope of the potential, $\lambda$, the Hubble constant in units ${\rm \: km \: s^{-1} \: {Mpc}^{-1}}$, $H_0$, the present-day mass fluctuation amplitude in spheres of radius $8h^{-1} {\rm Mpc}$,  $\sigma_8$, and the total matter density parameter, $\Omega_m$.  The shaded contours indicate the $1\sigma$ and $2\sigma$ confidence limits.}     
\label{fig:rectangle_m3_flat} 
\end{figure*}

\begin{itemize}

    \item Initial Conditions: The M$3$ model differs more significantly from M$1$ and M$2$ due to the non-trivial form of the coupling, which is no longer a constant but depends directly on the value of the scalar field. For this reason, we chose to leave the initial condition for the scalar field $\phi_{\text{ini}}$ as a free parameter. This field dependence originates a minimum and a maximum of the effective potential at $\phi_*$ for $\text{M}3^{+}$ and $\text{M}3^{-}$, respectively. For the sampling process, this means that the scalar field will prefer to sit at the minimum of the potential in $\text{M}3^{+}$, leading to an unphysical highly peaked posterior distribution for $\phi_{\text{ini}}$ which spoils the results, namely through an unreasonably large steep peak for the coupling posterior of $\gamma$. For this reason, we have opted for presenting the results only for $\text{M}3^{-}$ with varying $\phi_{\text{ini}}$, listed in \cref{tab:cq_ok_m3n_flat}. This means that we only consider the case in which the potential has a maximum and the scalar field does not get stuck into a minimum.The features introduced by the varying initial conditions can be better appreciated by comparing both models. In \cref{fig:1d_m3_flat}, we see a clear peak close to the maximum, which is not symmetric, with values $\phi_{\text{ini}} < \phi_* = 2\, \text{M}_\text{Pl}$ being favoured. This motivates setting $\phi_{\text{ini}} = 1\, \text{M}_\text{Pl}$ for M$3^+$ in furter studies.

    \item Coupling: In \cref{tab:cq_ok_m3n_flat}, we verify that all the data combinations can accommodate a vanishing coupling constant $\gamma$ in the $68\%$ CL region for the $\text{M}3^-$ model, with only upper bounds for $|\gamma|$ being reported. \cref{fig:1d_m2_flat} illustrates this trend through the marginalised posterior distributions, where there is only a small hint of BAO data driving the peak away from zero. More precisely, with the Pl18 data set alone, the following upper bound is found at $68\%$ CL: $|\gamma| < 0.0683$ for the $\text{M}3^{-}$ model. For the full Pl18+BAO+SN data set combination, we report at $1\sigma$: $|\gamma| <0.0549$. In \cref{fig:rectangle_m3_flat}, we depict the 2D marginalised posterior contours for $|\gamma|$ against relevant cosmological and model parameters: $\{H_0,\sigma_8,\Omega_m,\lambda\}$. Due to the added number of parameters, there is less evidence of correlations, with arguably slightly positive correlations in the $\{|\gamma|, H_0\}$, $\{|\gamma|, \sigma_8\}$ and (less robust) in the $\{|\gamma|, \lambda\}$ plane, and slightly negative in the $\{|\gamma|, \Omega_m\}$ plane.

    \item Potential: In \cref{tab:cq_ok_m3n_flat}, we also report on the slope of the potential $\lambda$, which, just as in the M$1$ models, is consistently constrained only from above. Incorporation of background datasets yields tighter constraints on $\lambda$ compared to Pl18. This effect is depicted in \cref{fig:1d_m3_flat}, where there seems to be a minute shift of the peak away from zero with the addition of BAO data. For Pl18, we report $\lambda < 0.925$ at $68\%$ CL, while the full Pl18+BAO+SN data set combination is more restrictive, namely $\lambda< 0.406$.

    \item $H_0$ Tension: The $H_0$ tension is alleviated to some extent by the enlarged error bars in M$3^-$, as no significant correlation seems to exist between $|\gamma|$ and $H_0$, while a positive relation had been identified in the $\text{M}1$ and M$2$ models. Using Pl18 data, the $\text{M}3^{-}$ model lessens the tension from $\sim 4.8\sigma$ to $\sim 3.2\sigma$ which is a less significant reduction compared to the $\text{M}1$ and M$2$ models since the mean value of $H_0$ is also consistently smaller, making it less effective in reducing the tension.

    \item $S_8$ Tension: In analogy to the $\text{M}1$ models, slightly smaller values of $\sigma_8$ are reported for $\text{M}3^{-}$ compared to $\Lambda$CDM for all data set combinations, compensated by consistently lower mean values of $\Omega_m$. Again, the interplay between $\sigma_8$ and $\Omega_m$ approximately balances out, resulting in a $S_8$ value of $0.839^{+0.018}_{-0.021}$ at $1\sigma$ for the $\text{M}3^{-}$, not significantly different from the $\Lambda$CDM value assuming Pl18 data. 

    \item Model Evidence: \cref{tab:cq_ok_m2n_flat} reveals that the logarithm of Bayes factor, $\ln B_{\text{M3}, \Lambda {\rm CDM}}$, is once again negative for all the data sets, implying a lack of evidence supporting the $\text{M}3^-$ model over $\Lambda$CDM. There is a minute overall trend of positive values of $\Delta \chi^2_{\text{min}}$, with the exception being the Pl18+SN, which when compared with the Bayesian evidence, is not sufficient to justify the addition of extra parameters. 
\end{itemize}

\begin{table*}[ht!]
\begin{center}
\renewcommand{\arraystretch}{1.5}
\resizebox{\textwidth}{!}{
\begin{tabular}{l c c c c c c c c c c c c c c c }
\hline\hline
\textbf{Parameter} & \textbf{ Pl18 } & \textbf{ Pl18 + BAO } & \textbf{ Pl18 + SN } & \textbf{ Pl18 + BAO + SN } \\ 
\hline\hline
$\Omega_b h^2$ & $  0.02235\pm 0.00015 $ & $  0.02239\pm 0.00015 $ & $  0.02239\pm 0.00014 $ & $  0.02242\pm 0.00014 $ \\
$\Omega_{cdm} h^2$ & $  0.1192^{+0.0023}_{-0.0015} $ & $  0.1187^{+0.0015}_{-0.0011} $ & $  0.1187^{+0.0019}_{-0.0013} $ & $  0.1183^{+0.0016}_{-0.00099} $ \\
$100 \theta_s$ & $  1.04184\pm 0.00030 $ & $  1.04193\pm 0.00029 $ & $  1.04191\pm 0.00029 $ & $  1.04191\pm 0.00031 $  \\
$\tau_{reio}$ & $  0.0552\pm 0.0079 $ & $  0.0555\pm 0.0079 $ & $  0.0552\pm 0.0078 $ & $  0.0545\pm 0.0074 $ \\
$n_s$  & $  0.9657\pm 0.0044 $ & $  0.9666\pm 0.0042 $ & $  0.9663\pm 0.0042 $ & $  0.9673\pm 0.0039 $  \\
$\ln \left( 10^{10} A_s \right)$ &  $  3.047\pm 0.017 $ & $  3.047\pm 0.016 $ & $  3.046\pm 0.016 $ & $  3.045\pm 0.016 $ \\
\hline 
$\lambda$ & $< 0.925$ & $< 0.797$ & $< 0.369$ & $< 0.406$  \\
$\gamma$ & $> -0.0683$ & $> -0.0645$ & $> -0.0459$ & $> -0.0549$ \\
$\phi_{\rm ini}  $ & $1.64^{+0.61}_{-0.87}$ & $1.69^{+0.67}_{-0.74}$ & $1.81\pm 0.87$ & $1.92\pm 0.84$ \\
\hline 
$\sigma_8$ & $  0.805^{+0.029}_{-0.019} $ & $  0.810\pm 0.017 $ & $  0.8164^{+0.0092}_{-0.014} $ & $  0.8143^{+0.0098}_{-0.014} $  \\
$\Omega_m$ &  $0.328^{+0.012}_{-0.032}$ & $0.315^{+0.011}_{-0.016}$ & $0.308^{+0.013}_{-0.0086}$ & $0.3067^{+0.0092}_{-0.0062}$ \\
$S_8$ &  $0.839^{+0.018}_{-0.021}$ & $0.830^{+0.014}_{-0.017}$ & $0.827\pm 0.015$ & $0.823^{+0.014}_{-0.013}$  \\
$H_0$ &  $66.1^{+2.8}_{-1.0}$ & $67.1^{+1.5}_{-0.79}$ & $67.89^{+0.66}_{-1.0}$ & $67.91^{+0.54}_{-0.73}$  \\
\hline \hline 
$\Delta \chi^{2}_{\rm min} $ &  $+1.84$ & $+0.5$ & $-0.2$ & $+0.32$ \\
$\ln B_{M3^-, \Lambda {\rm CDM}}$ &  $-6.34$ & $-5.94$ & $-7.29$ & $-7.27$ \\
\hline 
\hline
\end{tabular}}
\end{center}
\caption[Observational constraints for the flat M$3^-$ model]{Observational constraints at $68 \%$ confidence level on the sampled and derived cosmological parameters for different data set combinations under the flat M$3^-$ model, studied in \cref{sec:cq_m3_flat}.}
\label{tab:cq_ok_m3n_flat}
\end{table*}

 \section{Coupled Quintessence in Non-Flat Geometries} \label{sec:cq_curved}

 While large-scale observations have historically suggested that our Universe is nearly homogeneous, isotropic, and spatially flat, recent data has called this into question. Studies on the CMB temperature and polarisation spectra, particularly from the \textit{Planck} 2018 collaboration, hint at a closed geometric structure for the Universe. Thus, observations implying a minimal prediction for the spatial curvature of the Universe cannot be taken as irrefutable evidence for spatial flatness.
In this section, we aim to thoroughly assess the validity of coupled quintessence models by treating the Universe's curvature as a variable parameter and how it influences the cosmological tensions, compared with the flat case studied in \cref{sec:cq_curved}. 

In the following sections, we analyse the different extensions of the curved $\Lambda \text{CDM}$ model. We make use of the same data set combinations as in the flat cases for all the extensions. In the \cref{tab:cq_ok_lcdm_curved,tab:cq_ok_m1p_curved,tab:cq_ok_m1n_curved,tab:cq_ok_m2p_curved,tab:cq_ok_m2n_curved,tab:cq_ok_m3p_curved,tab:cq_ok_m3n_curved} we provide the bounds at $68\%$ CL ($1\sigma$) and comment on the corresponding $95\%$ ($2\sigma$) and $99\%$ ($3\sigma$) CL in the text when relevant for the discussion.





\subsection{$\Lambda$CDM $+$ $\Omega_K$}\label{sec:cq_ok_results_lcdm}

\cref{tab:cq_ok_lcdm_curved} presents the constraints on the $\Lambda \text{CDM} + \Omega_K$ scenario for various datasets. The second column of \cref{tab:cq_ok_lcdm_curved} shows the constraints using CMB data alone, and we remark a strong evidence for a closed Universe at over $99\%$ confidence level: $\Omega_K =-0.052^{+0.041}_{-0.053}$. The Hubble constant is significantly lower in this case ($H_0 =  52.8^{+3.2}_{-4.1} \, \text{km/s/Mpc}$ at $68\%$ CL), thereby greatly exacerbating the Hubble tension. Additionally, due to a strong correlation among $H_0$, $\Omega_K$, and $\Omega_{m}$, the matter density $\Omega_{m}$ takes extremely large values, also increasing the estimate of $S_8$ in this scenario.

When BAO data are combined with CMB data, as shown in the third column of \cref{tab:cq_ok_lcdm_curved}, $\Omega_K$ aligns with a spatially flat Universe, and the Hubble constant rises to the standard flat $\Lambda \text{CDM}$ \textit{Planck} value. However, this concordance comes from a dataset combination that is in disagreement at over $3\sigma$ \cite{Handley:2019tkm,DiValentino:2019qzk,Vagnozzi:2020rcz}, calling into question the reliability of the Pl18+BAO results. The last two columns of \cref{tab:cq_ok_lcdm_curved} report on the outcomes for Pl18+SN and Pl18+BAO+SN which are fairly similar, except for a slightly lower mean value of $H_0$ in the Pl18+SN case ($H_0 = 65.7^{+1.9}_{-3.3} \, \text{km/s/Mpc}$ at $68\%$ CL), but with larger error bars. In both instances, spatial flatness is consistent with the data.

\begin{table*}[ht!]
\begin{center}
\renewcommand{\arraystretch}{1.5}
\resizebox{\textwidth}{!}{
\begin{tabular}{l c c c c c c c c c c c c c c c }
\hline\hline
\textbf{Parameter} & \textbf{ Pl18 } & \textbf{ Pl18 + BAO } & \textbf{ Pl18 + SN } & \textbf{ Pl18 + BAO + SN } \\ 
\hline\hline
$ \omega_{\rm b }  $ & $  0.02261\pm 0.00017 $ & $  0.02239\pm 0.00015 $ & $  0.02246\pm 0.00016 $ & $  0.02240\pm 0.00015 $ \\ 
$ \omega_{\rm cdm }  $ & $  0.1180\pm 0.0015 $ & $  0.1199\pm 0.0014 $ & $  0.1192\pm 0.0015 $ & $  0.1197\pm 0.0014 $ \\ 
$ 100\theta_{s }  $ & $  1.04206\pm 0.00031 $ & $  1.04193\pm 0.00030 $ & $  1.04196\pm 0.00031 $ & $  1.04194\pm 0.00030 $ \\ 
$ \tau_{\rm reio }  $ & $  0.0484\pm 0.0082 $ & $  0.0556\pm 0.0078 $ & $  0.0556\pm 0.0078 $ & $  0.0550\pm 0.0079 $ \\ 
$ n_{s }  $ & $  0.9714\pm 0.0049 $ & $  0.9662\pm 0.0045 $ & $  0.9678\pm 0.0048 $ & $  0.9664\pm 0.0045 $ \\ 
$ \ln \left( 10^{10} A_s \right)  $ & $  3.028\pm 0.017 $ & $  3.047\pm 0.016 $ & $  3.046\pm 0.016 $ & $  3.046\pm 0.016 $ \\ 
$ \Omega_K  $ & $  -0.052^{+0.020}_{-0.017} $ & $  0.0009\pm 0.0020 $ & $  -0.0051^{+0.0057}_{-0.0078} $ & $  0.0010\pm 0.0019 $ \\ 
\hline
$ \sigma_8  $ & $  0.769\pm 0.016 $ & $  0.8118\pm 0.0082 $ & $  0.8060^{+0.0094}_{-0.011} $ & $  0.8110\pm 0.0082 $ \\ 
$ \Omega_{\rm m }  $ & $  0.514^{+0.065}_{-0.073} $ & $  0.3099\pm 0.0067 $ & $  0.332^{+0.030}_{-0.021} $ & $  0.3086\pm 0.0065 $ \\ 
$ S_8  $ & $  1.003^{+0.054}_{-0.047} $ & $  0.825\pm 0.013 $ & $  0.846^{+0.035}_{-0.022} $ & $  0.822\pm 0.013 $ \\ 
$ H_0  $ & $  52.8^{+3.2}_{-4.1} $ & $  67.92\pm 0.69 $ & $  65.7^{+1.9}_{-3.3} $ & $  68.03\pm 0.67 $ \\ 
\hline \hline
\end{tabular} }
\end{center}
\caption[Observational constraints for the curved $\Lambda$CDM model]{Observational constraints at $68 \%$ confidence level on the sampled and derived cosmological parameters for different data set combinations under the curved $\Lambda\text{CDM}\, +\, \Omega_K$ model, studied in \cref{sec:cq_ok_results_lcdm}.}
\label{tab:cq_ok_lcdm_curved}
\end{table*}

\subsection{\text{M}1 $+$ $\Omega_K$: Exponential Conformal Factor and Exponential Potential}\label{sec:cq_ok_results_m1}

\begin{figure*}[ht!]     
\centering     
\subfloat{\includegraphics[width=0.48\textwidth]{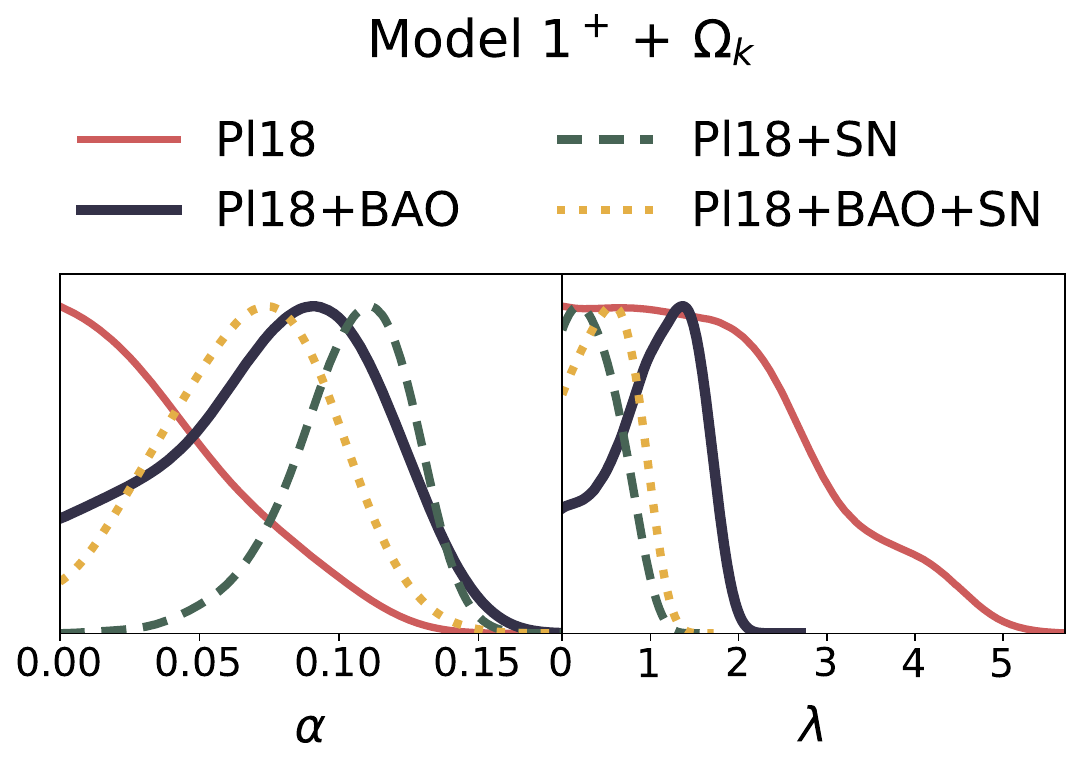}} \hfill
\subfloat{\includegraphics[width=0.48\textwidth]{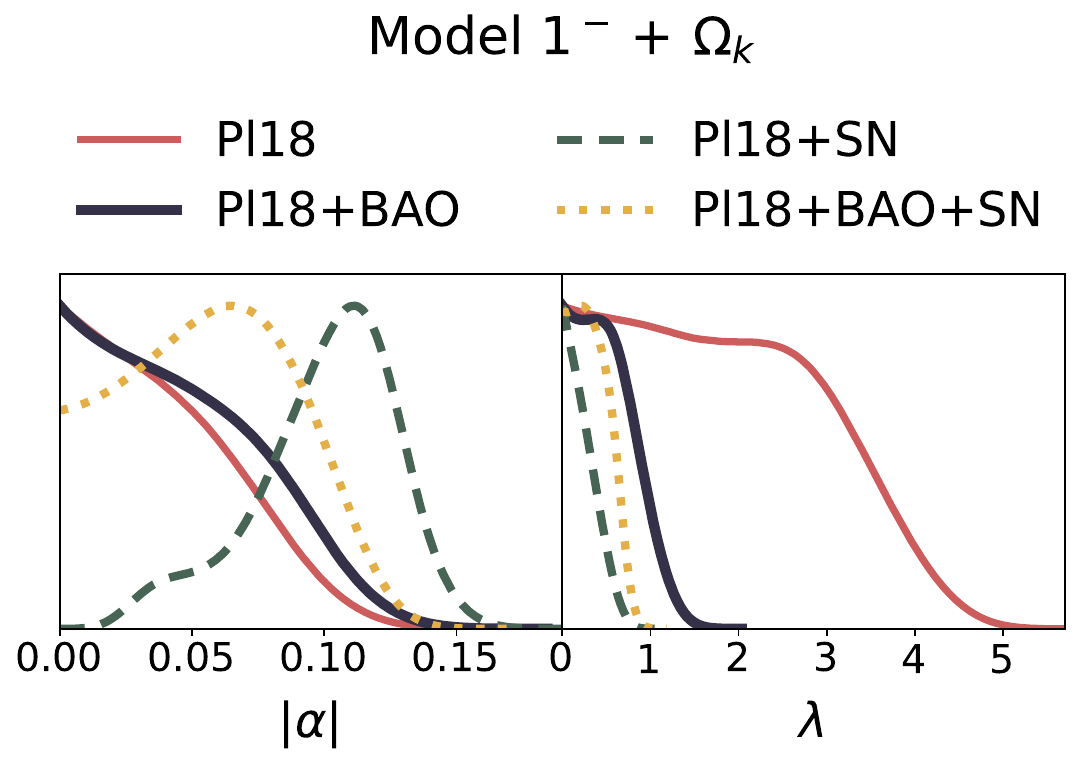}}
\caption[1D marginalised posterior distributions in the curved $\text{M}1$ models]{1D marginalised posterior distributions of the coupling parameter, $\alpha$, and slope of the potential, $\lambda$, in the $\text{M}1^{+}\, +\, \Omega_K$ model (upper panel) and $\text{M}1^{-}\, +\, \Omega_K$  model (lower panel) using the Pl18 (red), Pl18+BAO (blue), Pl18+SN, (green) and Pl18+BAO+SN (yellow) data set combinations.}     
\label{fig:1d_m1_curved} 
\end{figure*}

\begin{figure*}[ht!]     
\centering     
\subfloat{\includegraphics[width=\textwidth]{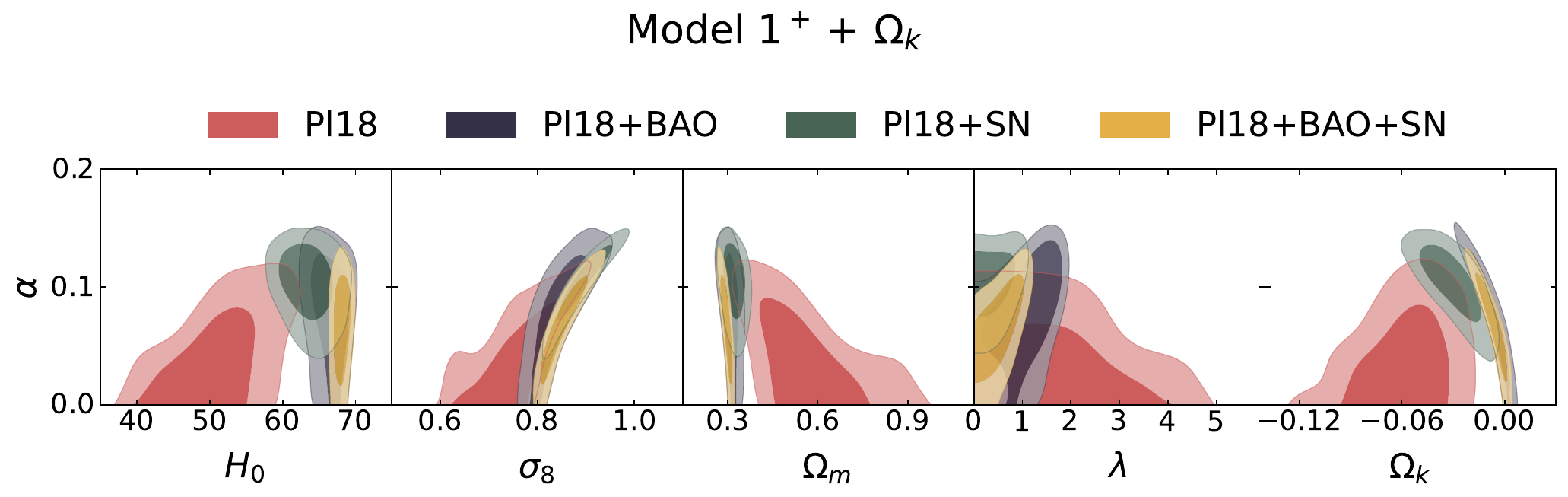}} \hfill
\subfloat{\includegraphics[width=\textwidth]{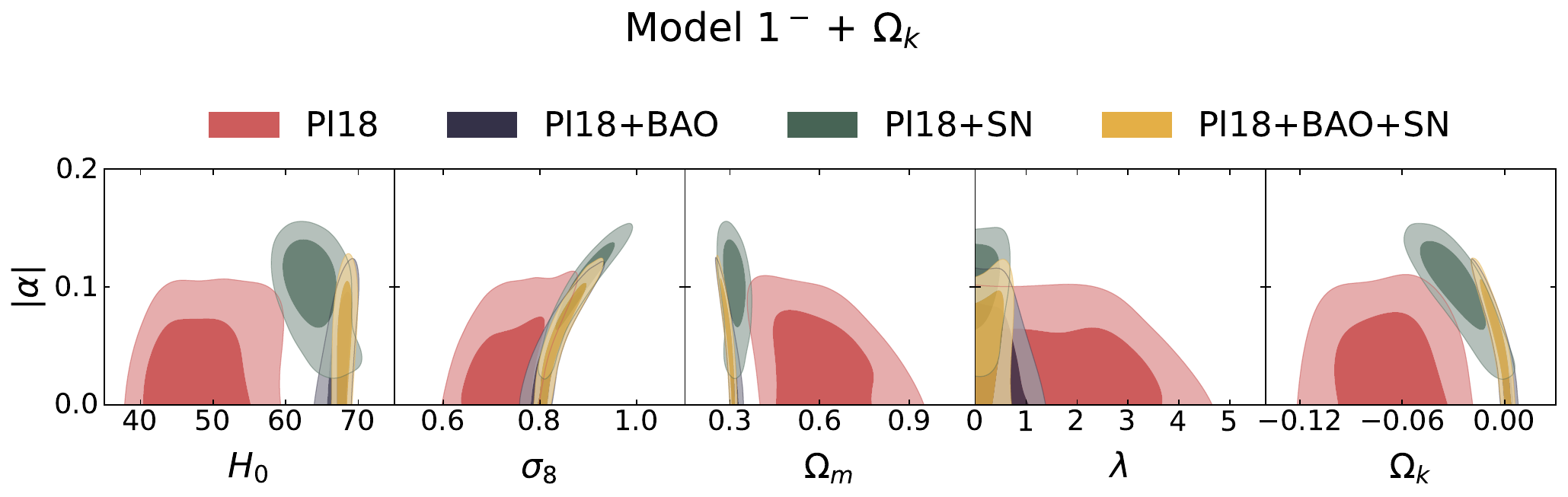}} 
\caption[2D marginalised posterior distributions in the curved $\text{M}1$ models]{2D marginalised posterior distributions of parameters in the $\text{M}1^{+}\, +\, \Omega_K$ model (upper panel) and $\text{M}1^{-}\, +\, \Omega_K$  model (lower panel) using the Pl18 (red), Pl18+BAO (blue), Pl18+SN, (green) and Pl18+BAO+SN (yellow) data set combinations. We plot the 2D marginalised posterior distributions of the conformal coupling parameter, $\alpha$, against the slope of the potential, $\lambda$, the Hubble constant in units ${\rm \: km \: s^{-1} \: {Mpc}^{-1}}$, $H_0$, the present-day mass fluctuation amplitude in spheres of radius $8h^{-1} {\rm Mpc}$,  $\sigma_8$, and the total matter density parameter, $\Omega_m$.  The shaded contours indicate the $1\sigma$ and $2\sigma$ confidence limits.}     
\label{fig:rectangle_m1_curved} 
\end{figure*}

The first extension of the non-flat $\Lambda \text{CDM}$ framework that we explore incorporates Model M$1$ as introduced in \cref{sec:cq_fl_ide}, split into the positive and negative coupling parameters variants: $\text{M}1^+ + \Omega_K$ and $\text{M}1^- + \Omega_K$. The derived parameter constraints are presented in \cref{tab:cq_ok_m1p_curved,tab:cq_ok_m1n_curved} and \cref{fig:1d_m1_curved,fig:rectangle_m1_curved} display the 1D and 2D marginalised posterior probability distributions, respectively. 

In summary, when only relying on the Planck satellite's CMB temperature and polarisation data (Pl18), a preference for a curved cosmological space emerges at over $99\% \, \text{CL}$, constraining the curvature parameter to $\Omega_K = -0.059^{+0.039}_{-0.039}$ and $\Omega_K = -0.059^{+0.045}_{-0.062}$ at $95\% \, \text{CL}$ and $99\% \, \text{CL}$ for $\text{M}1^+ + \Omega_K$ and $\Omega_K =-0.069^{+0.041}_{-0.044}$ and $\Omega_K = -0.069^{+0.053}_{-0.051}$ at $95\% \, \text{CL}$ and $99\% \, \text{CL}$ for $\text{M}1^- + \Omega_K$. This strong preference for a closed Universe vanishes when combining the Planck data with BAO, favouring spatial flatness within one standard deviation. Concerning the current expansion rate $H_0$, as depicted in \cref{fig:rectangle_m1_curved}, the mild positive correlation with the coupling parameter results in even lower preferred $H_0$ values, yielding $H_0 = 50.9\pm 5.2 \, \text{km/s/Mpc}$ and $H_0 = 48.5^{+4.5}_{-5.8} \, \text{km/s/Mpc}$ for $\text{M}1^+ + \Omega_K$ and $\text{M}1^- + \Omega_K$ at $68\% \, \text{CL}$, which sharply contrasts with the late-time independent measurements of $H_0$, diverging at approximately $4.2\sigma$ and $4.7\sigma$.

When adding the background data sets, the statistical tension persists. For instance, the $H_0$ value inferred from the Pl18 and BAO combination (third column in \cref{tab:cq_ok_m1p_curved,tab:cq_ok_m1n_curved}) is $66.5^{+2.0}_{-1.5} \, \text{km/s/Mpc}$ and $67.5^{+1.2}_{-1.0} \, \text{km/s/Mpc}$ at $68\% \, \text{CL}$ for $\text{M}1^+ + \Omega_K$ and $\text{M}1^- + \Omega_K$, closely resembling the flat $\Lambda \text{CDM}$ result. Also, the matter density parameter $\Omega_{m}$ negatively correlates with $H_0$, amplifying the tension in the $S8$ parameter. Compared to Pl18-only and Pl18+BAO+SN datasets, this tension reaches roughly $3-4\sigma$. Interestingly, the Pl18+SN case is consistent with a closed Universe at more than $95\%$ CL.

Regarding the coupling parameter, the Pl18 data alone offers an upper limit of $\alpha <  0.100$ and $|\alpha| < 0.0934$ at $95\% \, \text{CL}$ for $\text{M}1^+ + \Omega_K$ and $\text{M}1^- + \Omega_K$, which becomes less stringent, $\alpha = 0.068^{+0.052}_{-0.060}$ and $|\alpha| < 0.110$ at $95\% \, \text{CL}$ for $\text{M}1^+ + \Omega_K$ and $\text{M}1^- + \Omega_K$ when BAO and SN data are also included. 
This is in contrast to the flat case, where the upper limit is $|\alpha| < 0.0547$ and $|\alpha| < 0.0560$ for Pl18+BAO+SN at $95\% \, \text{CL}$ for $\text{M}1^+$ and $\text{M}1^-$. Thus, assuming spatial flatness imposes more restrictive constraints on the traditional coupled quintessence model.

For comparison purposes, we list below a more detailed account of each effect.

\begin{itemize}
    \item Coupling: Adding the varying curvature parameter to the M$1 + \Omega_K$ models, results in a non-vanishing prediction for $\alpha$ at $1\sigma$ for all the data sets except Pl18 in M$1^+ + \Omega_K$ and for Pl18+SN and Pl18+BAO+SN in M$1^- + \Omega_K$, as can be confirmed in \cref{tab:cq_ok_m1p_curved,tab:cq_ok_m1n_curved}. For instance, for the full combination we report $\alpha = 0.068^{+0.033}_{-0.028}$ and $\alpha = - 0.058\pm 0.32$ for M$1^+ + \Omega_K$ and M$1^- + \Omega_K$, respectively. On the other hand, for the Pl18 data alone the conditions are relaxed to upper bounds: $\alpha < 0.0515$ and $|\alpha| <0.0537 $ for M$1^+ + \Omega_K$ and M$1^- + \Omega_K$, respectively. Hence, the magnitude of the coupling parameter is comparable in both cases. As was the case in the flat scenario, the background data brings the peak in the $|\alpha|$ marginalised posterior distribution away from zero, as depicted in \cref{fig:1d_m1_curved}, with SN having the most defining impact, especially in the M$1^- + \Omega_K$ case, heavily suppressing the zero-peaked tail introduced by BAO, which pushes the maximum back for Pl18+BAO+SN. In \cref{fig:rectangle_m1_curved} we find the 2D marginalised posterior distributions of $\alpha$ against $\{H_0,\sigma_8$,$\Omega_m,\lambda,\Omega_K\}$. Both M$1 + \Omega_K$ models exhibit a positive correlation in the $\{|\alpha|, \lambda\}$ (except for the uncorrelated Pl18 case) and $\{|\alpha|, \sigma_8\}$ planes. There is also a reduced positive correlation for $\{|\alpha|, H_0\}$ with Pl18 only in M$1^+ + \Omega_K$. On the other hand, there is a minor negative correlation for $\{|\alpha|, \Omega_K\}$ in both cases excluding the Pl18-only case. This is possibly associated with the larger error bars in $\Omega_m$, which is constrained towards more conservative values when the background data is added and does not appear to be correlated with $\alpha$. Therefore, a more negative value of $\Omega_K$ must be compensated by an increased coupling, sourcing the DE component. The remaining combinations seem to not be correlated.

    \item Potential: The slope of the exponential potential $\lambda$ is only bounded from above with $68\%$ CL in the Pl18 and Pl18+SN combinations in M$1^+ + \Omega_K$ and across all cases in M$1^- + \Omega_K$. As reported in \cref{tab:cq_ok_m1p_curved,tab:cq_ok_m1n_curved}, the background data tightens the constraint on $\lambda$ in both models, in particular, due to the constraining power of BAO data on $\Omega_m$, directly influenced and positively correlated with $\lambda$. Pl18 data alone limits $\lambda$ to be smaller than $2.21$ and $2.49$, while the full Pl18+BAO+SN set brings this upper limit down to $0.54$ (with zero not included at $1\sigma$) and $0.435$, for M$1^+ + \Omega_K$ and M$1^- + \Omega_K$, respectively, both at $1\sigma$. In \cref{fig:1d_m1_curved}, we confirm the constraining power of BAO over $\lambda$, responsible for suppressing the left-hand side tail of the distribution, with SN pushing the peak towards zero. The values of $|\alpha|$ and $\lambda$ fall within similar ranges for all the data set combinations in both models, with broader error bars for Pl18+SN.

    \item $H_0$ Tension: The $H_0$ tension is mildly alleviated by the positive correlation between $|\alpha|$ and $H_0$, similar to the flat $\text{M}1$ case. Using Pl18 data, the M$1^+ + \Omega_K$ and M$1^- + \Omega_K$ models lessen the tension from approximately $5.3\sigma$ to $\sim 4.2\sigma$ and $4.7\sigma$, respectively, which is still almost double the corresponding values for $\Omega_K =0$. As happened in the flat case, the mean $H_0$ value in the M$1 + \Omega_K$ models is lower than in $\Lambda \text{CDM} + \Omega_K$, yet with larger error bars, resulting in an apparent alleviation of the tension due to the larger posterior space. Adding background data (mostly the effect of BAO) brings the $H_0$ value closer to the more reasonable values reported in the flat cases. This is consistent with these data sets' power in bringing $\Omega_K$ close to zero. Nevertheless, it should be noted that this represents yet another tension between the CMB and background data sets, which give predictions that are incompatible at $99\%$ CL.

    \item $S_8$ Tension: Even though the prediction of a closed Universe when considering Pl18 alone helps bring the value of $\sigma_8$ down, the $S_8$ tension remains due to the large values predicted for $\Omega_m$ in both the $\Lambda\text{CDM} + \Omega_K$ and M$1 + \Omega_K$ models. By introducing the background data, both $\sigma_8$ and $\Omega_m$ are restored to values compatible with the flat case. For Pl18 this results in an $S_8$ value of $1.014^{+0.055}_{-0.043}$ at $1\sigma$ for the M$1^+ + \Omega_K$ model, and $1.026^{+0.052}_{-0.037}$ for M$1^- + \Omega_K$, both not significantly different from the $\Lambda\text{CDM} + \Omega_K$ value $S_8 = 1.003^{+0.054}_{-0.047}$. 

    \item Curvature: In line with what happens in the $\Lambda\text{CDM} + \Omega_K$ case, there is a prediction for a closed Universe at more than $99\%$ CL for the M$1 + \Omega_K$ models with Pl18 and Pl18+SN. In particular, at $68\%$ CL we find $\Omega_K = -0.059^{+0.023}_{-0.018}$ for M$1^+ + \Omega_K$ and $\Omega_K = -0.069^{+0.024}_{-0.022}$ for M$1^- + \Omega_K$, in contrast with $\Omega_K = -0.052^{+0.020}_{-0.017}$ for $\Lambda\text{CDM} + \Omega_K$. These values are themselves in tension with the Pl18+BAO+SN counterparts, which report $\Omega_K = -0.0068^{+0.0071}_{-0.0047}$ for M$1^+ + \Omega_K$ and $\Omega_K = -0.0048^{+0.0069}_{-0.0042}$ for M$1^- + \Omega_K$. This is in contrast with $\Omega_K = 0.0010 \pm 0.0019$ for $\Lambda\text{CDM} + \Omega_K$, all consistent with $\Omega_K =0$ but with the mean value for $\Lambda$CDM actually depicting an open Universe. This shows the incompatibility of predictions between the different data sets in tension at $\sim 2.9\sigma$ for $\Lambda\text{CDM} + \Omega_K$, $\sim 2.5\sigma$ for M$1^+ + \Omega_K$ and $\sim 2.7\sigma$ level for M$1^- + \Omega_K$, already admitting that the data sets have been combined. This shows how care should be taken when deriving conclusions from their combination. 

    \item Model Evidence: The lower sections of \cref{tab:cq_ok_m1p_curved,tab:cq_ok_m1n_curved} report on the $\Delta \chi^2_{\text{min}}$ value for the goodness of fit comparison and the logarithm of the Bayes factor, $\ln B_{\text{M1}, \Lambda {\rm CDM}}$ for the curved cases. We see that both model comparison criteria are significantly negative (except for $\Delta \chi^2_{\text{min}}>0$ for M$1^- + \Omega_K$ with Pl18+BAO+SN), yielding no support for the M$1 + \Omega_K$ models over $\Lambda\text{CDM} + \Omega_K$, but hinting at a better fit to the data.
\end{itemize}

\begin{table*}[ht!]
\begin{center}
\renewcommand{\arraystretch}{1.5}
\resizebox{\textwidth}{!}{ 
\begin{tabular}{l c c c c c c c c c c c c c c c }
\hline\hline
\textbf{Parameter} & \textbf{ Pl18 } & \textbf{ Pl18 + BAO } & \textbf{ Pl18 + SN } & \textbf{ Pl18 + BAO + SN } \\ 
\hline\hline
$ \omega_{\rm b }  $ & $  0.02261\pm 0.00017 $ & $  0.02241\pm 0.00016 $ & $  0.02255\pm 0.00016 $ & $  0.02241\pm 0.00016 $ \\ 
$ \omega_{\rm cdm }  $ & $  0.1163^{+0.0041}_{-0.0019} $ & $  0.1137^{+0.0062}_{-0.0040} $ & $  0.1060^{+0.0049}_{-0.0056} $ & $  0.1141^{+0.0051}_{-0.0034} $ \\ 
$ 100\theta_{s }  $ & $  1.04205\pm 0.00030 $ & $  1.04191\pm 0.00030 $ & $  1.04197\pm 0.00030 $ & $  1.04191\pm 0.00030 $ \\ 
$ \tau_{\rm reio }  $ & $  0.0488\pm 0.0084 $ & $  0.0553\pm 0.0079 $ & $  0.0521\pm 0.0083 $ & $  0.0550\pm 0.0078 $ \\ 
$ n_{s }  $ & $  0.9729\pm 0.0051 $ & $  0.9718^{+0.0059}_{-0.0067} $ & $  0.9774\pm 0.0057 $ & $  0.9704^{+0.0053}_{-0.0059} $ \\ 
$ \ln \left( 10^{10} A_s \right)  $ & $  3.029\pm 0.017 $ & $  3.047\pm 0.016 $ & $  3.038\pm 0.017 $ & $  3.046\pm 0.016 $ \\ 
$ \Omega_K  $ & $  -0.059^{+0.023}_{-0.018} $ & $  -0.0083^{+0.0094}_{-0.0066} $ & $  -0.030\pm 0.012 $ & $  -0.0068^{+0.0071}_{-0.0047} $ \\ 
\hline
$ \lambda  $ & $ < 2.21 $ & $ 1.05^{+0.64}_{-0.37} $ & $ < 0.549 $ & $ 0.54^{+0.28}_{-0.39} $ \\ 
$ \alpha  $ & $  < 0.0515 $ & $  0.078^{+0.046}_{-0.031} $ & $ 0.104^{+0.026}_{-0.016} $ & $ 0.068^{+0.033}_{-0.028} $ \\ 
\hline
$ \sigma_8  $ & $  0.754^{+0.061}_{-0.049} $ & $  0.845^{+0.034}_{-0.048} $ & $  0.893\pm 0.037 $ & $  0.852^{+0.025}_{-0.038} $ \\ 
$ \Omega_{\rm m }  $ & $  0.559^{+0.079}_{-0.15} $ & $  0.310\pm 0.021 $ & $ 0.322\pm 0.023 $ & $ 0.296^{+0.014}_{-0.011} $ \\ 
$ S_8  $ & $  1.014^{+0.055}_{-0.043} $ & $ 0.856^{+0.021}_{-0.033}$ & $ 0.923\pm 0.039 $ & $  0.845^{+0.017}_{-0.023} $ \\ 
$ H_0  $ & $  50.9\pm 5.2 $ & $  66.5^{+2.0}_{-1.5} $ & $  63.5^{+2.2}_{-2.6} $ & $  68.10\pm 0.79 $ \\ 
\hline\hline 
$\Delta \chi^{2}_{\rm min} $ & $-1.60$  & $-3.68$ & $-7.66$ & $-1.36$ \\
$\ln B_{M1^+, \Lambda {\rm CDM}}$ & $-4.32$  & $-4.2$  & $-4.04$ & $-5.89$ \\
\hline \hline
\end{tabular} }
\end{center}
\caption[Observational constraints for the curved M$1^+$ model]{Observational constraints at $68 \%$ confidence level on the sampled and derived cosmological parameters for different data set combinations under the curved M$1^+\, +\, \Omega_K$ model, studied in \cref{sec:cq_ok_results_m1}.}
\label{tab:cq_ok_m1p_curved}
\end{table*}

\begin{table*}[ht!]
\begin{center}
\renewcommand{\arraystretch}{1.5}
\resizebox{\textwidth}{!}{
\begin{tabular}{l c c c c c c c c c c c c c c c }
\hline\hline
\textbf{Parameter} & \textbf{ Pl18 } & \textbf{ Pl18 + BAO } & \textbf{ Pl18 + SN } & \textbf{ Pl18 + BAO + SN } \\ 
\hline\hline

$ \omega_{\rm b }  $ & $  0.02261\pm 0.00017 $ & $  0.02239\pm 0.00016 $ & $  0.02252\pm 0.00017 $ & $  0.02239\pm 0.00015 $ \\ 
$ \omega_{\rm cdm }  $ & $  0.1130^{+0.0060}_{-0.0028} $ & $  0.1148^{+0.0062}_{-0.0030} $ & $  0.1045^{+0.0063}_{-0.0073} $ & $  0.1134^{+0.0068}_{-0.0038} $ \\ 
$ 100\theta_{s }  $ & $  1.04204\pm 0.00030 $ & $  1.04191\pm 0.00030 $ & $  1.04197\pm 0.00029 $ & $  1.04190\pm 0.00030 $ \\ 
$ \tau_{\rm reio }  $ & $  0.0485^{+0.0084}_{-0.0074} $ & $  0.0556\pm 0.0081 $ & $  0.0526\pm 0.0081 $ & $  0.0552\pm 0.0077 $ \\ 
$ n_{s }  $ & $  0.9730\pm 0.0049 $ & $  0.9686^{+0.0049}_{-0.0057} $ & $  0.9768\pm 0.0063 $ & $  0.9696\pm 0.0053 $ \\ 
$ \ln \left( 10^{10} A_s \right)  $ & $  3.029\pm 0.017 $ & $  3.047\pm 0.017 $ & $  3.039\pm 0.017 $ & $  3.047\pm 0.016 $ \\ 
$ \Omega_K  $ & $  -0.069^{+0.024}_{-0.022} $ & $  -0.0028^{+0.0067}_{-0.0036} $ & $  -0.028^{+0.012}_{-0.014} $ & $  -0.0048^{+0.0069}_{-0.0042} $ \\ 
\hline
$ \lambda  $ & $ < 2.49 $ & $ < 0.653 $ & $ < 0.305 $ & $ < 0.435 $ \\ 
$ \alpha  $ & $ > -0.0537 $ & $ > -0.0627 $ & $ -0.100^{+0.018}_{-0.032} $ & $ -0.058\pm 0.032 $ \\ 
\hline
$ \sigma_8  $ & $  0.736^{+0.058}_{-0.051} $ & $  0.829^{+0.022}_{-0.041} $ & $  0.887\pm 0.041 $ & $  0.843^{+0.022}_{-0.039} $ \\ 
$ \Omega_{\rm m }  $ & $ 0.60^{+0.10}_{-0.15} $ & $ 0.303^{+0.020}_{-0.014}$ & $ 0.315\pm 0.023$ & $ 0.294^{+0.018}_{-0.011} $ \\ 
$ S_8  $ & $ 1.026^{+0.052}_{-0.037} $ & $ 0.831^{+0.014}_{-0.018} $ & $ 0.909^{+0.045}_{-0.035} $ & $ 0.834^{+0.015}_{-0.018}$ \\ 
$ H_0  $ & $  48.5^{+4.5}_{-5.8} $ & $  67.5^{+1.2}_{-1.0} $ & $  63.7^{+2.2}_{-2.7} $ & $  68.11\pm 0.76 $ \\ 
\hline\hline 
$\Delta \chi^{2}_{\rm min} $ & $-2.46$  & $-1.60$ & $-6.38$ & $1.66$ \\
$\ln B_{M1^-, \Lambda {\rm CDM}}$ & $-4.24$ & $-5.61$  & $-4.31$ & $-6.77$ \\
\hline \hline
\end{tabular} }
\end{center}
\caption[Observational constraints for the curved M$1^-$ model]{Observational constraints at $68 \%$ confidence level on the sampled and derived cosmological parameters for different data set combinations under the curved M$1^-\, +\, \Omega_K$ model, studied in \cref{sec:cq_ok_results_m1}.}
\label{tab:cq_ok_m1n_curved}
\end{table*}

\subsection{\text{M}2 $+$ $\Omega_K$: Exponential Conformal Factor and Inverse Power-law Potential}\label{sec:cq_ok_results_m2}

\begin{figure*}[ht!]     
\centering     
\subfloat{\includegraphics[width=0.48\textwidth]{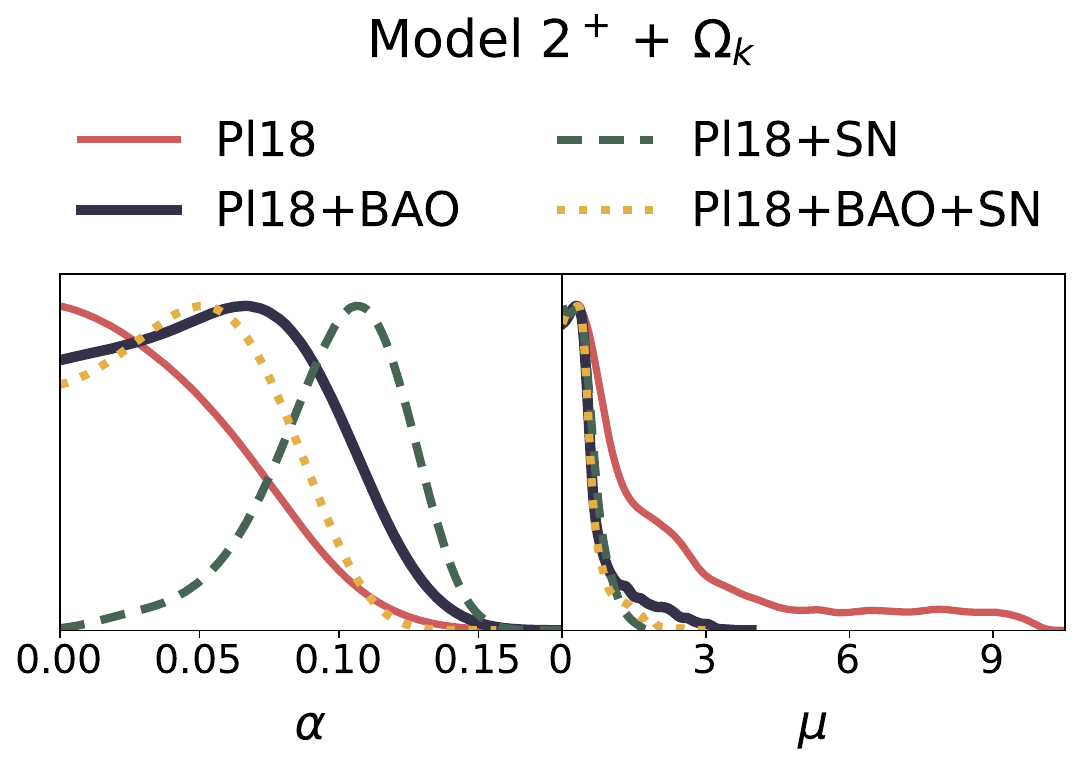}} \hfill
\subfloat{\includegraphics[width=0.48\textwidth]{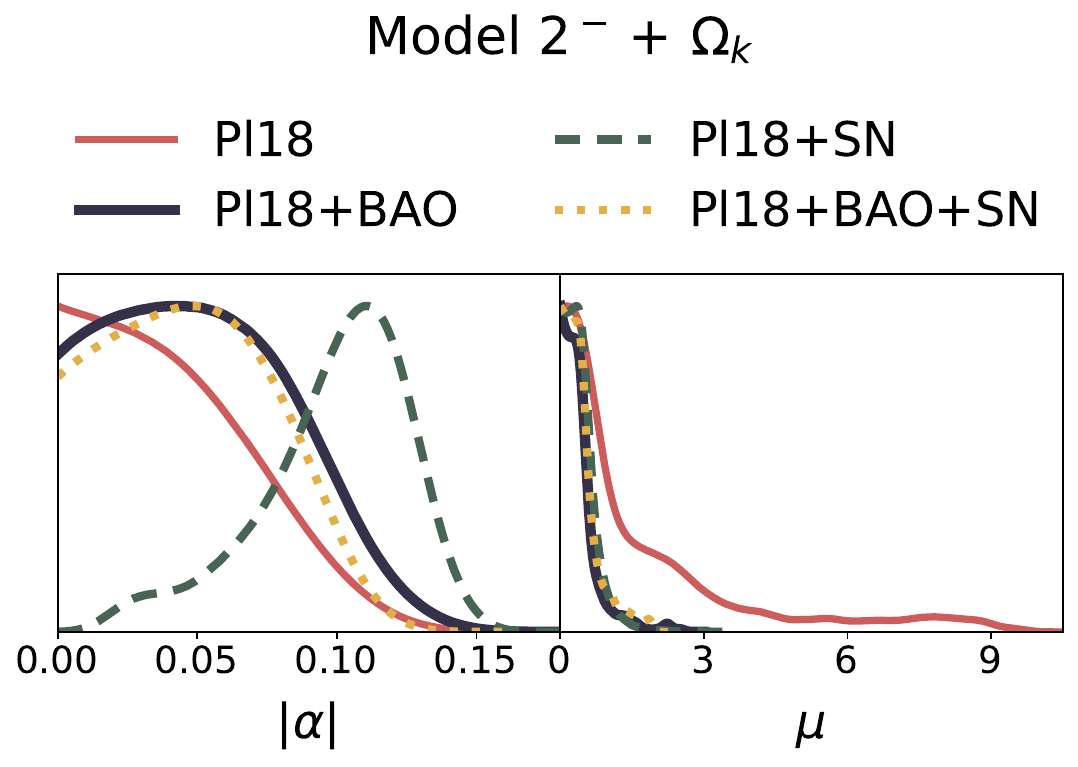}}
\caption[1D marginalised posterior distributions in the curved $\text{M}2$ models]{1D marginalised posterior distributions of the coupling parameter, $\alpha$, and slope of the potential, $\mu$, in the $\text{M}2^{+}\, +\, \Omega_K$ model (upper panel) and $\text{M}2^{-}\, +\, \Omega_K$  model (lower panel) using the Pl18 (red), Pl18+BAO (blue), Pl18+SN, (green) and Pl18+BAO+SN (yellow) data set combinations.}     
\label{fig:1d_m2_curved} 
\end{figure*}

\begin{figure*}[ht!]     
\centering     
\subfloat{\includegraphics[width=\textwidth]{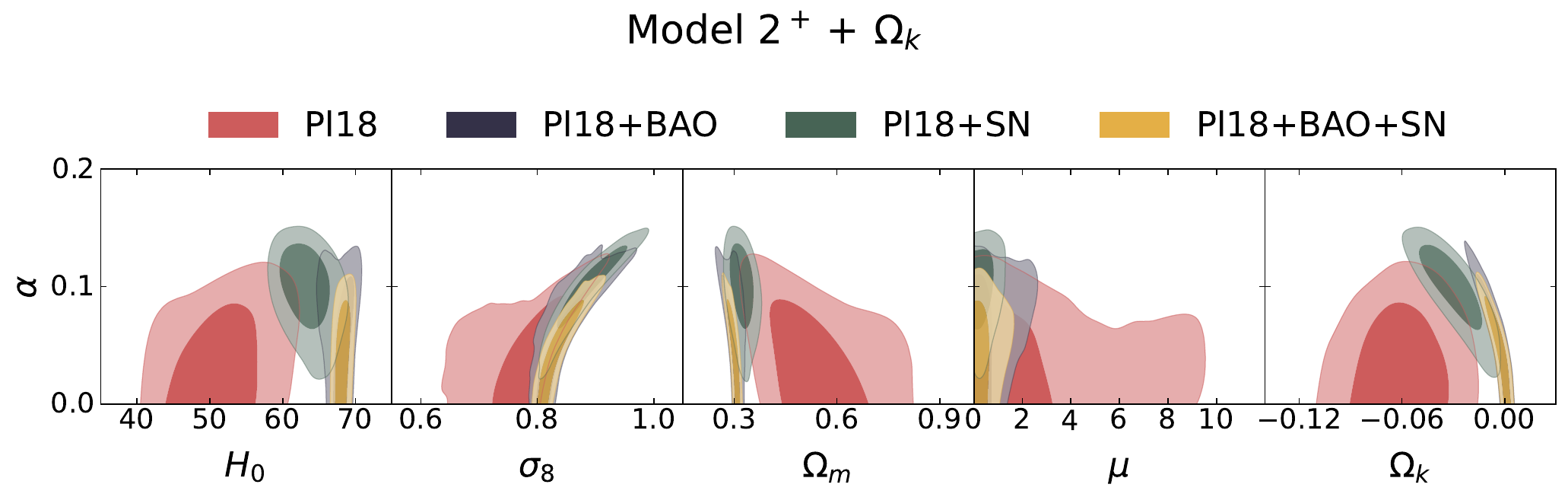}} \hfill
\subfloat{\includegraphics[width=\textwidth]{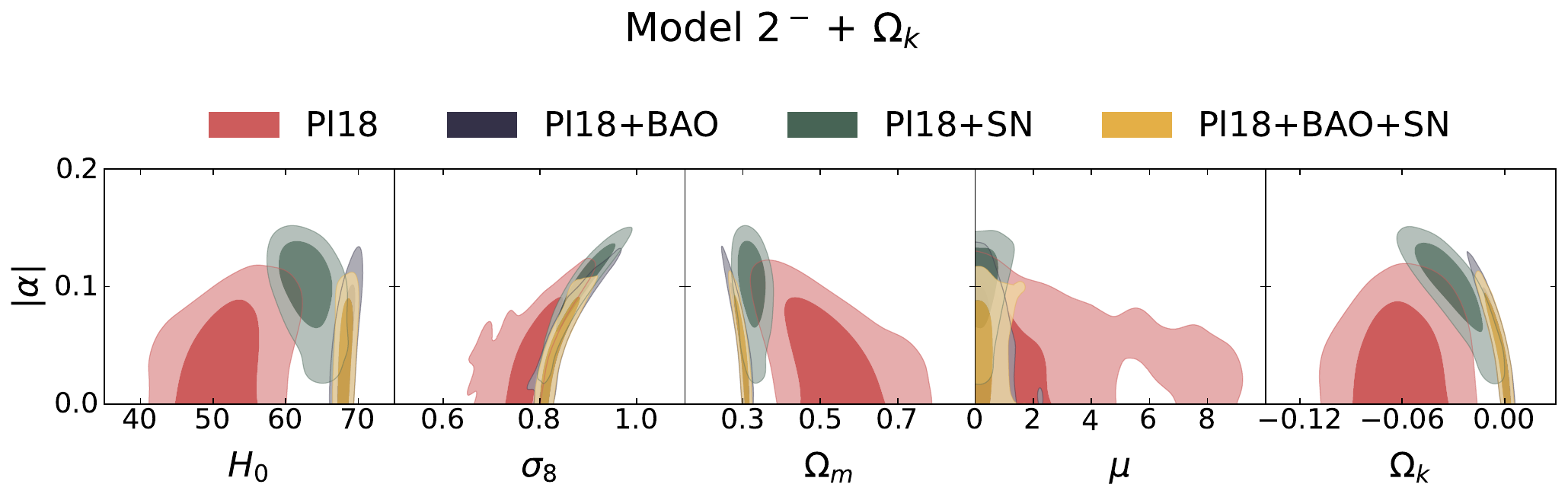}} 
\caption[2D marginalised posterior distributions in the curved $\text{M}2$ models]{2D marginalised posterior distributions of parameters in the $\text{M}2^{+}\, +\, \Omega_K$ model (upper panel) and $\text{M}2^{-}\, +\, \Omega_K$  model (lower panel) using the Pl18 (red), Pl18+BAO (blue), Pl18+SN, (green) and Pl18+BAO+SN (yellow) data set combinations. We plot the 2D marginalised posterior distributions of the conformal coupling parameter, $\alpha$, against the power of the potential, $\mu$, the Hubble constant in units ${\rm \: km \: s^{-1} \: {Mpc}^{-1}}$, $H_0$, the present-day mass fluctuation amplitude in spheres of radius $8h^{-1} {\rm Mpc}$,  $\sigma_8$, and the total matter density parameter, $\Omega_m$.  The shaded contours indicate the $1\sigma$ and $2\sigma$ confidence limits.}     
\label{fig:rectangle_m2_curved} 
\end{figure*}

In \cref{tab:cq_ok_m2p_curved,tab:cq_ok_m2n_curved}, we compile the parameter constraints for the M$2 + \Omega_K$ models, and \cref{fig:1d_m2_curved,fig:rectangle_m2_curved} illustrate the corresponding marginalised 1D and 2D posterior distributions. Observing the Pl18 data in isolation, evidence for a closed Universe manifests at more than $ 99\% \, \text{CL}$, namely $\Omega_K = -0.061^{+0.048}_{-0.053}$ and $\Omega_K = -0.062^{+0.047}_{-0.054}$ for M$2^+ + \Omega_K$ and M$2^- + \Omega_K$, respectively. The coupling parameter is bounded to similar values across both models and all data sets; namely, we report $ \alpha < 0.0563$ and $|\alpha| <0.0566$ at $ 68\% \, \text{CL} $ for Pl18 data alone, looser than in the corresponding flat scenario. The $H_0$ tension in the CMB data is maintained at the same level as for $\Lambda\text{CDM} + \Omega_K$ and M$1 + \Omega_K$, displaying mean values of $ H_0 = 51.6^{+4.2}_{-2.8} \, \text{km/s/Mpc} $ and $ H_0 = 51.9^{+3.7}_{-4.7} \, \text{km/s/Mpc} $ at $ 68\% \, \text{CL} $. Compared with the flat case, we conclude that this worsening of the tension is attributed to the curvature effects.

Incorporating BAO data into the analysis considerably shifts the constraints due to their conflicting implications in a curved Universe, drawing this model closer to the spatially flat case. Nevertheless, the combination of Pl18+SN yields intriguing outcomes. We detect a closed Universe at above $ 95\% \, \text{CL} $ ($ \Omega_K = -0.030 \pm 0.023 $ and $ \Omega_K = -0.031 \pm 0.026$) with a coupled dark sector ($ \alpha = 0.099^{+0.047}_{-0.061}$ and $ |\alpha| =0.098^{+0.046}_{-0.054} $ at $ 95\% \, \text{CL} $). The Hubble constant is suppressed in relation to the Pl18+BAO case, leading to an enlarged $\Omega_{m}$ due to its correlation with $ H_0 $.

Lastly, considering the Pl18+BAO+SN combination (see the last column of \cref{tab:cq_ok_m2p_curved,tab:cq_ok_m2n_curved}), the BAO data's tension with Planck results reinforces spatial flatness. Meanwhile, $ \alpha $ becomes more constrained and aligns with an uncoupled and a cosmological constant scenario ($\alpha =0$ and $\mu =0$) at $ 95\% \, \text{CL} $. It is noteworthy that $ H_0 $ sees a mild enhancement compared to the Pl18 flat and curved $ \Lambda\text{CDM} $ predictions, yet the tension with distance ladder measurements persists at $ 4-5\sigma $.

Below we list the results in more detail, with a focus on comparing the effects of the introduction of spatial curvature.

\begin{itemize}
    \item Coupling: In contrast with the flat case, a non-vanishing prediction for $\alpha$ at $68\%$ can be derived for all data set combinations except Pl18 in both M$2 + \Omega_K$ models, as reported in \cref{tab:cq_ok_m2p_curved,tab:cq_ok_m2n_curved}. For instance, for the full combination we report $\alpha = 0.049^{+0.026}_{-0.035}$ and $\alpha = -0.050^{+0.040}_{-0.023}$ for M$2^+ + \Omega_K$ and M$2^- + \Omega_K$, respectively. On the other hand, for the Pl18 data alone, these constraints are once again relaxed to upper bounds: $\alpha < 0.0563$ and $|\alpha| <0.0566 $ for M$2^+ + \Omega_K$ and M$2^- + \Omega_K$, respectively. Hence, the magnitude of the coupling parameter is comparable in both cases, in agreement with the symmetry found in the constant coupling case for change of signs in the pair $\{\alpha,\dot{\phi}\}$, according to \cref{eq:cq_rhodm}. In \cref{fig:1d_m2_curved}, we find a similar pattern to the flat case, with the background data driving the peak in the $|\alpha|$ marginalised posterior distribution away from zero. SN data has the greatest impact in suppressing the low-$\alpha$ tail of the distribution for Pl18+SN, which is restored with the introduction of BAO into the combination. In \cref{fig:rectangle_m2_curved} we show the 2D marginalised posterior distributions of $\alpha$ against $\{H_0,\sigma_8$,$\Omega_m,\mu,\Omega_K\}$. Both M$2 + \Omega_K$ models exhibit a positive correlation in the $\{|\alpha|, \sigma_8\}$. For Pl18 there is also a slight positive correlation for $\{|\alpha|, H_0\}$ and negative for $\{|\alpha|, \Omega_m\}$, and both vanish once background data sets are included. On the other hand, there is a negative correlation for $\{|\alpha|, \Omega_K\}$ for all cases excluding Pl18 alone. This is once again associated with the larger error bars in $\Omega_m$, which is constrained towards more conservative values when the background data is added and does not appear to be correlated with $\alpha$. Therefore, a more negative value of $\Omega_K$ must be compensated by an increased coupling, sourcing the DE component. The remaining combinations seem to be negligibly correlated. We find the scalar field parameters $\{|\alpha|, \mu\}$ to be generally uncorrelated. We highlight the consistent results between the positive and negative coupling parameter cases due to the model's symmetry and show a clear preference for the trend set by the data on the parameters.

    \item Potential: The power of the inverse potential $\mu$ is only bounded from above with $68\%$ CL for all the data combinations in both M$2 + \Omega_K$ models. These are reported in \cref{tab:cq_ok_m2p_curved,tab:cq_ok_m2n_curved}, with the background data consistently shortening the $1\sigma$ parameter region and, consequently, the upper bound. Indeed, this is illustrated in \cref{fig:1d_m2_curved}, with the background data leading to the collapse of the high-$\mu$ tail of the distribution that is present in both cases for Pl18, bringing the peak close to zero. Pl18 data alone limits $\mu$ to be smaller than $2.17$ and $1.74$, while the full Pl18+BAO+SN sets brings this upper limit down to $0.499$ for M$2^+ + \Omega_K$ and $0.484$ for M$2^- + \Omega_K$, respectively at $1\sigma$. The symmetry of the model ensures that the values of $|\alpha|$ and $\mu$ fall within similar ranges in both models for the same data set combinations.

    \item $H_0$ Tension: The $H_0$ tension is alleviated to some extent by the positive correlation between $|\alpha|$ and $H_0$, similar to the $\text{M}1 + \Omega_K$ models. In the Pl18-only case, the M$2^+ + \Omega_K$ and M$2^- + \Omega_K$ models negligibly reduce the tension in $\Lambda\text{CDM} + \Omega_K$ from approximately $5.3\sigma$ to $\sim 4.6\sigma$ and $4.9\sigma$, respectively, which is still considerably aggravated with respect to the corresponding values for M$2$ with $\Omega_K =0$. Analogously to the $\text{M}1 + \Omega_K$ case, the mean $H_0$ value in the M$2 + \Omega_K$ models is lower than in $\Lambda \text{CDM} + \Omega_K$, yet with larger error bars. Once more, adding background data (dominated by the effect of BAO) brings the $H_0$ value closer to the ones reported in the flat cases, consistent with the constraints set on $\Omega_K$ close to zero. Nevertheless, it stresses once more the incompatibility of the CMB and background data sets when $\Omega_K$ is added as a free parameter.

    \item $S_8$ Tension: We see that considering the inverse power law potential does not change the predictions significantly compared with the exponential case $\text{M}1 + \Omega_K$. The evidence for a close Universe when considering Pl18 alone drives the value of $\sigma_8$ down. However, the $S_8$ parameter remains too high when compared with the one reported by studies of weak lensing predictions, as a consequence of the large values predicted for $\Omega_m$ in both the $\Lambda\text{CDM} + \Omega_K$ and M$2 + \Omega_K$ models. By introducing the background data, both $\sigma_8$ and $\Omega_m$ are restored to values compatible with the flat case. For Pl18 this results in an $S_8$ value of $1.021^{+0.057}_{-0.040}$ at $1\sigma$ for the M$2^+ + \Omega_K$ model, and $1.019^{+0.056}_{-0.038}$ for M$2^- + \Omega_K$, both not significantly different from the $\Lambda\text{CDM} + \Omega_K$ value $S_8 = 1.003^{+0.054}_{-0.047}$. 

    \item Curvature: We report evidence for a curved Universe at more than $99\%$ CL for the M$2 + \Omega_K$ models with CMB data and with the addition of SN. In particular, at $68\%$ CL we find $\Omega_K = -0.061\pm 0.020$ for M$2^+ + \Omega_K$ and $\Omega_K = -0.062\pm 0.019$ for M$2^- + \Omega_K$, in contrast with $\Omega_K = -0.052^{+0.020}_{-0.017}$ for $\Lambda\text{CDM} + \Omega_K$, with similar sizes for the CL regions. These values are consistently in tension with the Pl18+BAO+SN counterparts, which report $\Omega_K = -0.0037^{+0.0058}_{-0.0033}$ for M$2^+ + \Omega_K$ and $\Omega_K = -0.0038^{+0.0059}_{-0.0034}$ for M$2^- + \Omega_K$, also in contrast with $\Omega_K = 0.0010 \pm 0.0019$ for $\Lambda\text{CDM} + \Omega_K$, all including $\Omega_K = 0$ in the $68\%$ CL region. This corroborates the incompatibility between the data sets of different nature in tension at $\sim 2.9\sigma$ for $\Lambda\text{CDM} + \Omega_K$, $\sim 2.8\sigma$ for M$2^+ + \Omega_K$ and $\sim 3.0\sigma$ level for M$2^- + \Omega_K$, already after the data sets have been combined.

    \item Model Evidence: The lower sections of \cref{tab:cq_ok_m2p_curved,tab:cq_ok_m2n_curved} report on the $\Delta \chi^2_{\text{min}}$ value for the goodness of fit comparison and the logarithm of the Bayes factor, $\ln B_{\text{M2}, \Lambda {\rm CDM}}$. We see that both model comparison criteria are significantly negative (except for the negligible $\Delta \chi^2_{\text{min}}>0$ for the M$2 + \Omega_K$ models with Pl18+BAO+SN), yielding no support for the M$2 + \Omega_K$ models over $\Lambda\text{CDM} + \Omega_K$ but revealing a better fit to the data.
\end{itemize}

\begin{table*}[ht!]
\begin{center}
\renewcommand{\arraystretch}{1.5}
\resizebox{\textwidth}{!}{
\begin{tabular}{l c c c c c c c c c c c c c c c }
\hline\hline
\textbf{Parameter} & \textbf{ Pl18 } & \textbf{ Pl18 + BAO } & \textbf{ Pl18 + SN } & \textbf{ Pl18 + BAO + SN } \\ 
\hline\hline

$ \omega_{\rm b }  $ & $  0.02264\pm 0.00017 $ & $  0.02240\pm 0.00015 $ & $  0.02255\pm 0.00016 $ & $  0.02240\pm 0.00016 $ \\ 
$ \omega_{\rm cdm }  $ & $  0.1150^{+0.0045}_{-0.0025} $ & $  0.1146^{+0.0061}_{-0.0034} $ & $  0.1057^{+0.0056}_{-0.0064} $ & $  0.1159^{+0.0047}_{-0.0026} $ \\ 
$ 100\theta_{s }  $ & $  1.04208\pm 0.00030 $ & $  1.04191\pm 0.00030 $ & $  1.04198\pm 0.00030 $ & $  1.04191\pm 0.00030 $ \\ 
$ \tau_{\rm reio }  $ & $  0.0478^{+0.0086}_{-0.0076} $ & $  0.0553\pm 0.0078 $ & $  0.0523\pm 0.0083 $ & $  0.0550\pm 0.0078 $ \\ 
$ n_{s }  $ & $  0.9739\pm 0.0051 $ & $  0.9697^{+0.0052}_{-0.0060} $ & $  0.9770\pm 0.0059 $ & $  0.9685^{+0.0048}_{-0.0054} $ \\ 
$ \ln \left( 10^{10} A_s \right)  $ & $  3.026\pm 0.018 $ & $  3.047\pm 0.016 $ & $  3.038\pm 0.017 $ & $  3.046\pm 0.016 $ \\ 
$ \Omega_K  $ & $  -0.061\pm 0.020 $ & $  -0.0054^{+0.0076}_{-0.0044} $ & $  -0.030\pm 0.012 $ & $  -0.0037^{+0.0058}_{-0.0033} $ \\ 
\hline
$ \mu  $ & $ < 2.17 $ & $ < 0.626$ & $ < 0.525 $ & $ < 0.499 $ \\ 
$ \alpha  $ & $ < 0.0563 $ & $0.059^{+0.031}_{-0.045} $ & $ 0.098^{+0.030}_{-0.018}$ & $ 0.049^{+0.026}_{-0.035} $ \\ 
\hline
$ \sigma_8  $ & $  0.774^{+0.041}_{-0.036} $ & $  0.846^{+0.023}_{-0.043} $ & $  0.887\pm 0.040 $ & $  0.838^{+0.017}_{-0.032} $ \\ 
$ \Omega_{\rm m }  $ & $0.533^{+0.078}_{-0.12} $ & $0.297^{+0.018}_{-0.014} $ & $ 0.323\pm 0.022 $ & $0.298^{+0.014}_{-0.0085}$ \\ 
$ S_8  $  & $ 1.021^{+0.057}_{-0.040} $ & $0.841^{+0.015}_{-0.024} $ & $ 0.919\pm 0.039 $ & $ 0.835^{+0.015}_{-0.019}$ \\ 
$ H_0  $ & $  51.6^{+4.2}_{-4.8} $ & $  68.1^{+1.3}_{-0.85} $ & $  63.3^{+2.1}_{-2.3} $ & $  68.29\pm 0.75 $ \\ 
\hline\hline 
$\Delta \chi^{2}_{\rm min} $ &  $-2.04$ & $-3.12$ & $-5.62$ & $0.04$ \\
$\ln B_{M2^+, \Lambda {\rm CDM}}$ & $-5.14$ & $-5.36$  & $-4.26$ & $-6.33$ \\
\hline \hline
\end{tabular} }
\end{center}
\caption[Observational constraints for the curved M$2^+$ model]{Observational constraints at $68 \%$ confidence level on the sampled and derived cosmological parameters for different data set combinations under the curved M$2^+\, +\, \Omega_K$ model, studied in \cref{sec:cq_ok_results_m2}.}
\label{tab:cq_ok_m2p_curved}
\end{table*}

\begin{table*}[ht!]
\begin{center}
\renewcommand{\arraystretch}{1.5}
\resizebox{\textwidth}{!}{
\begin{tabular}{l c c c c c c c c c c c c c c c }
\hline\hline
\textbf{Parameter} & \textbf{ Pl18 } & \textbf{ Pl18 + BAO } & \textbf{ Pl18 + SN } & \textbf{ Pl18 + BAO + SN } \\ 
\hline\hline

$ \omega_{\rm b }  $ & $  0.02265\pm 0.00017 $ & $  0.02240\pm 0.00015 $ & $  0.02254\pm 0.00017 $ & $  0.02239\pm 0.00015 $ \\ 
$ \omega_{\rm cdm }  $ & $  0.1137^{+0.0052}_{-0.0025} $ & $  0.1146^{+0.0062}_{-0.0031} $ & $  0.1052^{+0.0058}_{-0.0072} $ & $  0.1152^{+0.0054}_{-0.0027} $ \\ 
$ 100\theta_{s }  $ & $  1.04208\pm 0.00031 $ & $  1.04191\pm 0.00030 $ & $  1.04199\pm 0.00030 $ & $  1.04191\pm 0.00030 $ \\ 
$ \tau_{\rm reio }  $ & $  0.0478\pm 0.0085 $ & $  0.0549\pm 0.0079 $ & $  0.0522\pm 0.0082 $ & $  0.0553\pm 0.0081 $ \\ 
$ n_{s }  $ & $  0.9741\pm 0.0050 $ & $  0.9691^{+0.0048}_{-0.0059} $ & $  0.9771\pm 0.0062 $ & $  0.9688\pm 0.0050 $ \\ 
$ \ln \left( 10^{10} A_s \right)  $ & $  3.027\pm 0.018 $ & $  3.046\pm 0.016 $ & $  3.038\pm 0.017 $ & $  3.047^{+0.015}_{-0.017} $ \\ 
$ \Omega_K  $ & $  -0.062\pm 0.019 $ & $  -0.0045^{+0.0069}_{-0.0037} $ & $  -0.031\pm 0.013 $ & $  -0.0038^{+0.0059}_{-0.0034} $ \\
\hline
$ \mu  $ & $< 1.74 $ & $< 0.450$ & $ < 0.496 $ & $ < 0.484 $ \\ 
$ \alpha  $ & $ > -0.0566 $ & $ -0.053^{+0.046}_{-0.022} $ & $-0.099^{+0.017}_{-0.032} $ & $-0.050^{+0.040}_{-0.023}$ \\ 
\hline
$ \sigma_8  $ & $  0.781^{+0.033}_{-0.040} $ & $  0.844^{+0.022}_{-0.042} $ & $  0.887\pm 0.041 $ & $  0.839^{+0.019}_{-0.033} $ \\ 
$ \Omega_{\rm m }  $ & $ 0.520\pm 0.090$ & $ 0.295^{+0.019}_{-0.011} $ & $ 0.323^{+0.021}_{-0.024} $ & $ 0.297^{+0.015}_{-0.0094}$ \\ 
$ S_8  $ & $  1.019^{+0.056}_{-0.038}$ & $ 0.836^{+0.015}_{-0.019}$ & $ 0.920\pm 0.041$ & $ 0.834^{+0.014}_{-0.017}$ \\ 
$ H_0  $ & $  51.9^{+3.7}_{-4.7} $ & $  68.35\pm 0.94 $ & $  63.1\pm 2.3 $ & $  68.25\pm 0.76 $ \\
\hline\hline 
$\Delta \chi^{2}_{\rm min} $ & $-1.06$  & $-0.94$ & $-6.30$ & $0.28$ \\
$\ln B_{M2^-, \Lambda {\rm CDM}}$ & $-4.84$ & $-5.56$  & $-3.51$ & $-6.72$ \\
\hline \hline
\end{tabular} }
\end{center}
\caption[Observational constraints for the curved M$2^-$ model]{Observational constraints at $68 \%$ confidence level on the sampled and derived cosmological parameters for different data set combinations under the curved M$2^-\, +\, \Omega_K$ model, studied in \cref{sec:cq_ok_results_m2}.}
\label{tab:cq_ok_m2n_curved}
\end{table*}

    
\subsection{\text{M}3 $+$ $\Omega_K$: Coupling with Minimum/Maximum and Exponential Potential}\label{sec:cq_ok_results_m3}

\begin{figure*}[ht!]     
\centering     
\subfloat{\includegraphics[width=0.48\textwidth]{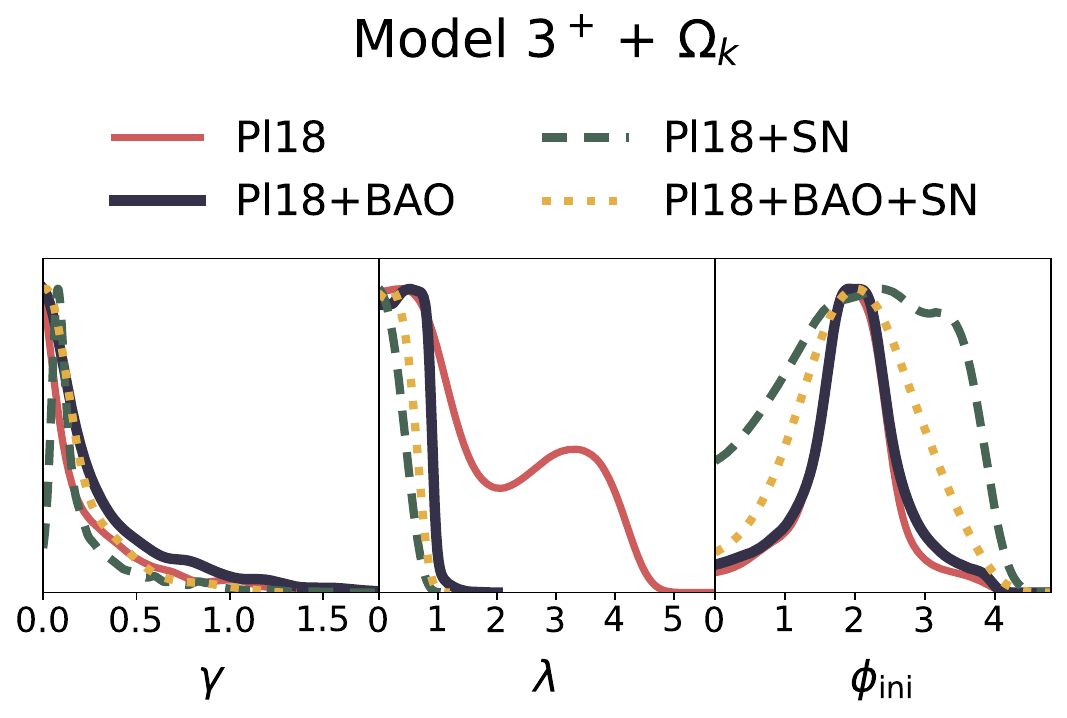}} \hfill
\subfloat{\includegraphics[width=0.48\textwidth]{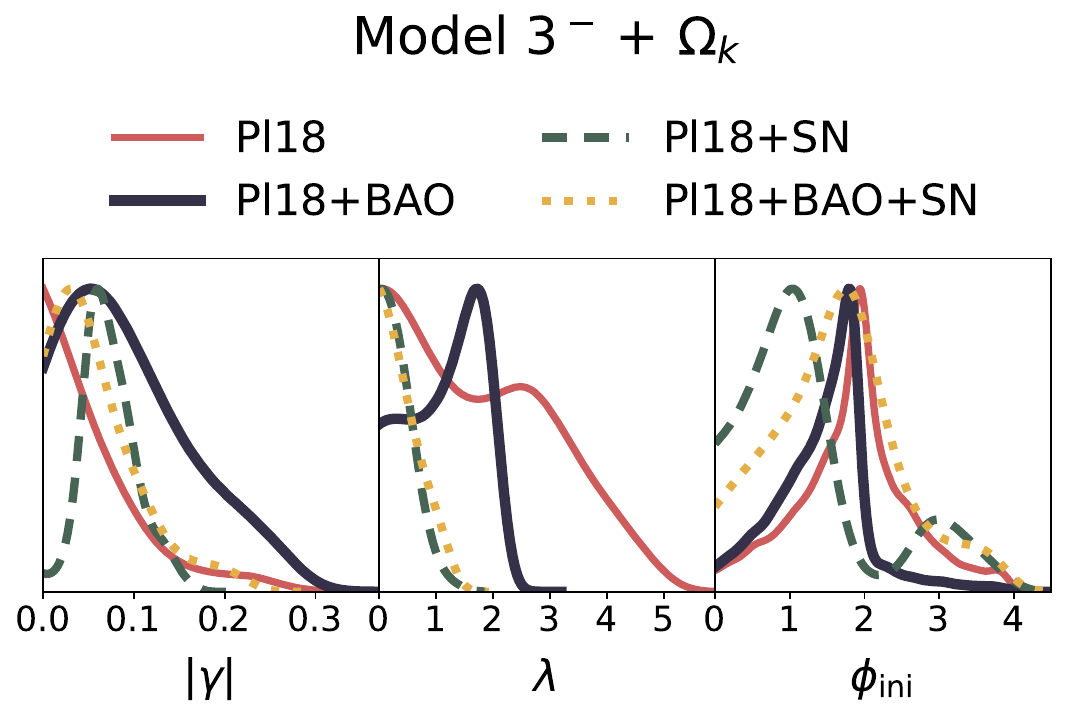}}
\caption[1D marginalised posterior distributions in the curved $\text{M}3$ models]{1D marginalised posterior distributions of the coupling parameter, $\gamma$, and slope of the potential, $\lambda$, in the $\text{M}3^{+}\, +\, \Omega_K$ model (upper panel) and $\text{M}3^{-}\, +\, \Omega_K$  model (lower panel) using the Pl18 (red), Pl18+BAO (blue), Pl18+SN, (green) and Pl18+BAO+SN (yellow) data set combinations.}     
\label{fig:1d_m3_curved} 
\end{figure*}

\begin{figure*}[ht!]     
\centering     
\subfloat{\includegraphics[width=\textwidth]{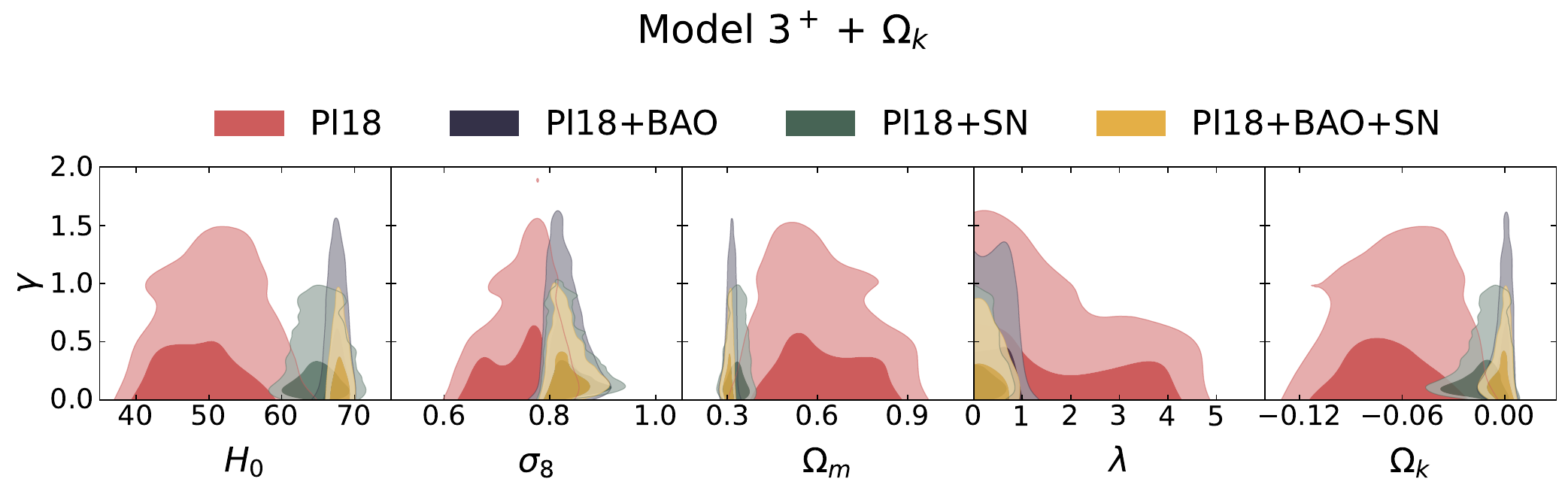}} \hfill
\subfloat{\includegraphics[width=\textwidth]{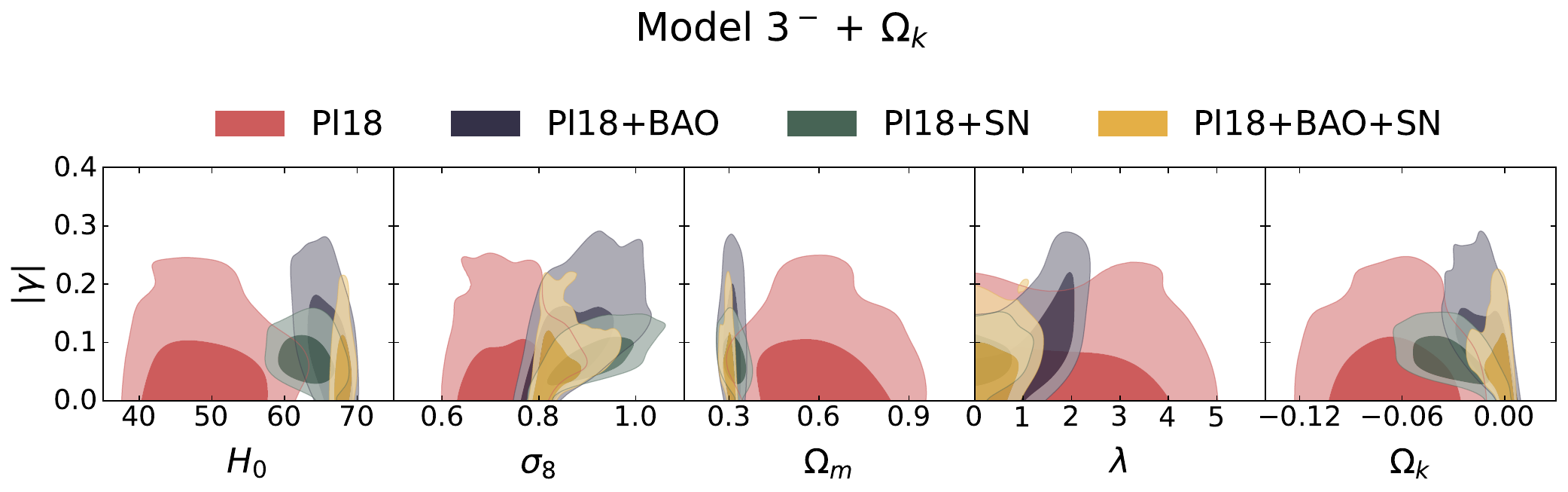}} 
\caption[2D marginalised posterior distributions in the curved $\text{M}3$ models]{2D marginalised posterior distributions of parameters in the $\text{M}3^{+}\, +\, \Omega_K$ model (upper panel) and $\text{M}3^{-}\, +\, \Omega_K$  model (lower panel) using the Pl18 (red), Pl18+BAO (blue), Pl18+SN, (green) and Pl18+BAO+SN (yellow) data set combinations. We plot the 2D marginalised posterior distributions of the conformal coupling parameter, $\gamma$, against the slope of the potential, $\lambda$, the Hubble constant in units ${\rm \: km \: s^{-1} \: {Mpc}^{-1}}$, $H_0$, the present-day mass fluctuation amplitude in spheres of radius $8h^{-1} {\rm Mpc}$,  $\sigma_8$, and the total matter density parameter, $\Omega_m$.  The shaded contours indicate the $1\sigma$ and $2\sigma$ confidence limits.}     
\label{fig:rectangle_m3_curved} 
\end{figure*}

In \cref{tab:cq_ok_m3p_curved,tab:cq_ok_m3n_curved}, we summarise the observational limits on both free and derived model parameters for the M$3 + \Omega_K$ models. \cref{fig:1d_m3_curved,fig:rectangle_m3_curved} display the 1D and 2D marginalised distributions for selected relevant parameters.

We find that when considering only the Pl18 data, there is a preference for a closed Universe at over 99\% CL. Namely, we find $\Omega_K = -0.067^{+0.052}_{-0.051}$ and $\Omega_K = -0.066^{+0.051}_{-0.057}$ at $99\%$ CL. The mean $H_0$ value in this scenario is lower than in the $\Lambda\text{CDM}$ case ($H_0 = 49.0^{+4.2}_{-5.7}\, \text{km/s/Mpc}$ and $H_0 = 49.0^{+4.8}_{-6.0}\, \text{km/s/Mpc}$ at 68\% CL), but the larger error bars ensure a similar value for the $H_0$ tension of approximately $4-5 \sigma$. The coupling and potential parameters are compatible with an uncoupled and cosmological constant-like scenario at $68\%$ CL).

When incorporating BAO data, the potential for a flat Universe is recovered at $68\%$ CL.
With the addition of SN data, we once more reaffirm the non-zero curvature at over $68\%$ CL for M$3^+ + \Omega_K$ ($\Omega_K =-0.033^{+0.028}_{-0.027}$) and $95\%$ CL for M$3^- + \Omega_K$ ($\Omega_K =-0.033^{+0.028}_{-0.027}$), with only a non-vanishing prediction for $\lambda$ being recovered with Pl18+BAO at $68\%$ CL ($\lambda = 0.47 \pm 0.28$ for M$3^+ + \Omega_K$ and $\lambda = 1.23 \pm 0.63$ for M$3^- + \Omega_K$). 
The coupling parameter $\gamma$ is more constrained in the M$3^- + \Omega_K$ case, with only non-zero values predicted at $1\sigma$ for Pl18+SN.
For the full joint analysis using Pl18+BAO+SN, the data break most of the degeneracies and slightly improve the agreement for $H_0$ ($H_0 = 68.00\pm 0.73$ and $H_0 = 68.00\pm 0.74$ km/s/Mpc at $68\%$ CL). Both $\gamma$ and $\lambda$ are consistent with non-dynamic behaviour at $95\%$ CL.

\begin{itemize}

    \item Initial Conditions: Just as in the flat case, the field dependence in the coupling introduces a minimum and a maximum of the effective coupling at $\phi_*=2\, \text{M}_{\text{Pl}}$ for $\text{M}3^{+} + \Omega_K$ and $\text{M}3^{-} + \Omega_K$, respectively. In the flat case, we came across an unphysical highly peaked posterior distribution for $\phi_{\text{ini}}$, which wants to sit at the minimum of the potential in $\text{M}3^{+}$. When the curvature is allowed to vary, this feature becomes less sharp, with the larger error bars and freedom introduced in the model implying a posterior distribution for $\phi_{\text{ini}}$ that is still peaked around $\phi_*$ but much less stringently. We see in \cref{fig:1d_m3_curved} that there is a seemingly symmetric distribution around the peak in the $\phi_\text{ini}$ posterior distribution for $\text{M}3^{+} + \Omega_K$. At the same time, the same peaked pattern but skewed to the right (mainly by the effect of SN) is recovered in $\text{M}3^{-} + \Omega_K$ just as in the flat case.

    \item Coupling: The non-trivial expression for the coupling in the M$3$ models makes this scenario more distinguishable from M$1$ and M$2$. In all the cases considered the uncoupled scenario $|\gamma| = 0$ is accommodated by the $68\%$ CL region at $1\sigma$, with the exception of the Pl18+SN combination in the model with a maximum, M$3^- + \Omega_K$, for which $\gamma = -0.077^{+0.035}_{0.023}$, as reported in \cref{tab:cq_ok_m3p_curved,tab:cq_ok_m3n_curved}. For just the Pl18 dataset, the constraints are upper limits, $\gamma < 0.357$ and $|\gamma| < 0.0741$, for the respective models. The coupling parameter's size is quite different in both cases, consistently more constrained towards zero for M$3^- + \Omega_K$. This feature is depicted in \cref{fig:1d_m3_curved}, in which there is a similar peak for the $\gamma$ marginalised posterior distribution across all the data sets for M$3^+ + \Omega_K$, with only the Pl18+SN peak being considerably deviated from zero. In contrast, the background data pushes the peak in the $|\gamma|$ posterior away from zero in M$3^- + \Omega_K$, with SN most effectively narrowing the lower-$\gamma$ tail for Pl18+SN, consistent with the bounded $68\%$ CL region derived in this case only. Adding BAO data preserves the peak in the distribution but reverses the narrowing of the left-hand side tail. \cref{fig:rectangle_m3_curved} reveals the 2D marginalised distributions of $\gamma$ against$\{H_0,\sigma_8$,$\Omega_m,\lambda,\Omega_K\}$. No significant correlations are identified for M$3^+ + \Omega_K$ due to the very peaked distribution around the minimum of the coupling and $|\Omega_K| \ll 1$ for the background data, apart from a small positive correlation in $\{|\gamma|, \sigma_8\}$ for Pl18. For M$3^- + \Omega_K$ we identify an overall positive correlation in the $\{|\gamma|, \sigma_8\}$ plane and in $\{|\gamma|, \lambda\}$ for the background data. A negative correlation is identified in $\{|\gamma|, \Omega_K\}$ and $\{|\gamma|, H_0\}$ when the background data is included, more pronounced for Pl18+SN in both cases.

    \item Potential: In both M$3 + \Omega_K$ models, the steepness of the potential, $\lambda$, is only upper-bounded at $68\%$ CL for all datasets, except for the Pl18+BAO combination, for which we find $\lambda = 0.47 \pm 0.28$ and $\lambda = 1.23 \pm 0.63$ for M$3^+ + \Omega_K$ and M$3^- + \Omega_K$, as outlined in \cref{tab:cq_ok_m3p_curved,tab:cq_ok_m3n_curved}. The tighter bounds are set by SN data, as corroborated in \cref{fig:1d_m3_curved}. We see that the integration of background data in M$3^+ + \Omega_K$ brings the marginalised posterior towards smaller values, with a minor non-zero peak for Pl18+BAO, while in M$3^- + \Omega_K$ larger deviations are allowed for the more pronounced peak in Pl18+BAO. Once again, the inclusion of background data shrinks the high-$\lambda$ tail present for the Pl18 dataset. When considering only Pl18, the upper limits for $\lambda$ are $2.52$ and $2.60$. In comparison, the full Pl18+BAO+SN sets reduce the limits down to $0.445$ and $0.604$, for M$3^+ + \Omega_K$ and M$3^- + \Omega_K$, respectively, at $1\sigma$.
    
    \item $H_0$ Tension: A minor mitigation of the $H_0$ tension occurs even though no clear correlation between $|\gamma|$ and $H_0$ has been identified. Even if the mean values reported for $H_0$ in the $\text{M}3 + \Omega_K$ models are lower than the $\Lambda\text{CDM} + \Omega_K$ case, the increased size of the CL regions has a minor influence on the tension, which decreases from around $ 5.3\sigma $ to approximately $ 4.8\sigma $ and $ 4.4\sigma $ in M$3^+ + \Omega_K$ and M$3^- + \Omega_K$, respectively. Consistently with what was found for the $\text{M}1 + \Omega_K$ and $\text{M}2 + \Omega_K$ cases, the addition of background data restores the value of $H_0$ to the order reported in the flat cases, coinciding with $\Omega_K$ vanishing in the $1\sigma$ region, and reflecting the incompatibility of the different data sets.

    \item $ S_8 $ Tension: Even though the prediction for a closed Universe for the Pl18-only data set brings the value of $\sigma_8$ significantly down, compared with both the flat case and the $\Lambda\text{CDM} + \Omega_K$ cases, this is accompanied by an increase in the mean value of $\Omega_m$, and the $S_8$ parameter actually becomes larger. For Pl18 this results in an $S_8$ value of $1.041^{+0.058}_{-0.043}$ at $1\sigma$ for the M$3^+ + \Omega_K$ model, and $1.027^{+0.056}_{-0.038}$ for M$3^- + \Omega_K$, the highest values of all the models considered, with a similar size for the CL region. 

    \item Curvature: We report firm evidence for a closed Universe at more than $99\%$ CL for the M$3 + \Omega_K$ models with CMB data and a less pronounced but still present support with the addition of SN. In particular, at $68\%$ CL we find $\Omega_K = -0.067\pm 0.021$ for M$3^+ + \Omega_K$ and $\Omega_K = -0.066^{+0.024}_{-0.021}$ for M$3^- + \Omega_K$, in contrast with $\Omega_K = -0.052^{+0.020}_{-0.017}$ for $\Lambda\text{CDM} + \Omega_K$, with similar sized CL regions. On the contrary, no significant prediction for curvature is present for the Pl18+BAO+SN counterparts, which yield $\Omega_K = -0.0020^{+0.0047}_{-0.0027}$ for M$3^+ + \Omega_K$ and $\Omega_K = -0.0035^{+0.0066}_{-0.0032}$ for M$3^- + \Omega_K$, in contrast with $\Omega_K = 0.0010 \pm 0.0019$ for $\Lambda\text{CDM} + \Omega_K$, all accommodating $\Omega_K = 0$ in the $68\%$ CL region. Data set inconsistencies are further confirmed, showing tensions at approximately the $ 3\sigma $ level, more precisely of $\sim 2.9\sigma$ for $\Lambda\text{CDM} + \Omega_K$, $\sim 3.0\sigma$ for M$3^+ + \Omega_K$ and $\sim 2.7\sigma$ for M$3^- + \Omega_K$, already after the data sets have been combined.

    \item Model Evidence: The Bayes factor $ \ln B_{\text{M3}, \Lambda {\rm CDM}} $ points to no support for the M$3 + \Omega_K$ models over the $ \Lambda\text{CDM} + \Omega_K $ model. Nevertheless, the considerably negative values of $ \Delta \chi^2_{\text{min}} $ indicate a better fit to the data in all cases.

    \end{itemize}

\begin{table*}[ht!]
\begin{center}
\renewcommand{\arraystretch}{1.5}
\resizebox{\textwidth}{!}{
\begin{tabular}{l c c c c c c c c c c c c c c c }
\hline\hline
\textbf{Parameter} & \textbf{ Pl18 } & \textbf{ Pl18 + BAO } & \textbf{ Pl18 + SN } & \textbf{ Pl18 + BAO + SN } \\ 
\hline\hline

$ \omega_{\rm b }  $ & $  0.02264\pm 0.00017 $ & $  0.02240\pm 0.00015 $ & $  0.02249\pm 0.00017 $ & $  0.02241\pm 0.00015 $ \\ 
$ \omega_{\rm cdm }  $ & $  0.1186^{+0.0022}_{-0.0036} $ & $  0.1187^{+0.0030}_{-0.0018} $ & $  0.1151^{+0.0047}_{-0.0030} $ & $  0.1177^{+0.0032}_{-0.0018} $ \\ 
$ 100\theta_{s }  $ & $  1.04203\pm 0.00032 $ & $  1.04186\pm 0.00031 $ & $  1.04191\pm 0.00031 $ & $  1.04189\pm 0.00030 $ \\ 
$ \tau_{\rm reio }  $ & $  0.0478^{+0.0087}_{-0.0074} $ & $  0.0556\pm 0.0077 $ & $  0.0547\pm 0.0079 $ & $  0.0556\pm 0.0078 $ \\ 
$ n_{s }  $ & $  0.9741^{+0.0050}_{-0.0056} $ & $  0.9691^{+0.0049}_{-0.0059} $ & $  0.9741^{+0.0061}_{-0.0074} $ & $  0.9698^{+0.0049}_{-0.0062} $ \\ 
$ \ln \left( 10^{10} A_s \right)  $ & $  3.027^{+0.018}_{-0.016} $ & $  3.048\pm 0.016 $ & $  3.045\pm 0.016 $ & $  3.048\pm 0.016 $ \\ 
$ \Omega_K  $ & $  -0.067\pm 0.021 $ & $  -0.00099^{+0.0042}_{-0.0023} $ & $  -0.0138^{+0.011}_{-0.0082} $ & $  -0.0020^{+0.0047}_{-0.0027} $ \\ 
\hline 
$ \lambda  $ & $  < 2.52 $ & $ 0.47\pm 0.28 $ & $ < 0.348 $ & $ < 0.445 $ \\ 
$ \gamma  $ & $ < 0.357 $ & $ < 0.384$ & $ < 0.215$ & $ < 0.229 $ \\ 
$ \phi_i  $ & $  1.90^{+0.61}_{-0.49} $ & $ 1.96^{+0.66}_{-0.59}$ & $2.15^{+1.4}_{-0.90} $ & $2.03\pm 0.84 $ \\ 
\hline
$ \sigma_8  $ & $  0.738^{+0.060}_{-0.039} $ & $  0.824^{+0.015}_{-0.027} $ & $  0.841^{+0.024}_{-0.036} $ & $  0.830^{+0.015}_{-0.027} $ \\ 
$ \Omega_{\rm m }  $ & $ 0.610^{+0.096}_{-0.16} $ & $  0.310^{+0.011}_{-0.0096} $ & $  0.329\pm 0.022 $ & $ 0.304^{+0.010}_{-0.0078}$ \\ 
$ S_8  $ & $ 1.041^{+0.058}_{-0.043} $ & $  0.837^{+0.014}_{-0.020}$ & $  0.880\pm 0.036 $ & $0.836^{+0.014}_{-0.021} $ \\ 
$ H_0  $ & $  49.0^{+4.2}_{-5.7} $ & $  67.64\pm 0.89 $ & $  64.9\pm 2.3 $ & $  68.00\pm 0.73 $ \\ 
\hline\hline 
$\Delta \chi^{2}_{\rm min} $ &  $-2.96$ & $-1.76$ & $-5.08$ & $-0.56$ \\
$\ln B_{M3^+, \Lambda {\rm CDM}}$ & $-2.79$ &  $-3.67$ & $-5.42$ & $-6.39$ \\
\hline \hline
\end{tabular} }
\end{center}
\caption[Observational constraints for the curved M$3^+$ model]{Observational constraints at $68 \%$ confidence level on the sampled and derived cosmological parameters for different data set combinations under the curved M$3^+\, +\, \Omega_K$ model, studied in \cref{sec:cq_ok_results_m3}.}
\label{tab:cq_ok_m3p_curved}
\end{table*}

\begin{table*}[ht!]
\begin{center}
\renewcommand{\arraystretch}{1.5}
\resizebox{\textwidth}{!}{
\begin{tabular}{l c c c c c c c c c c c c c c c }
\hline\hline
\textbf{Parameter} & \textbf{ Pl18 } & \textbf{ Pl18 + BAO } & \textbf{ Pl18 + SN } & \textbf{ Pl18 + BAO + SN } \\ 
\hline\hline
$ \omega_{\rm b }  $ & $  0.02262\pm 0.00017 $ & $  0.02243\pm 0.00017 $ & $  0.02254\pm 0.00016 $ & $  0.02240\pm 0.00015 $ \\ 
$ \omega_{\rm cdm }  $ & $  0.1152^{+0.0050}_{-0.0019} $ & $  0.1118^{+0.0090}_{-0.0054} $ & $  0.1022\pm 0.0083 $ & $  0.1157^{+0.0054}_{-0.0025} $ \\ 
$ 100\theta_{s }  $ & $  1.04206\pm 0.00030 $ & $  1.04194\pm 0.00030 $ & $  1.04200\pm 0.00031 $ & $  1.04192\pm 0.00029 $ \\ 
$ \tau_{\rm reio }  $ & $  0.0483\pm 0.0082 $ & $  0.0554\pm 0.0078 $ & $  0.0517\pm 0.0081 $ & $  0.0547^{+0.0069}_{-0.0078} $ \\ 
$ n_{s }  $ & $  0.9722\pm 0.0048 $ & $  0.9698^{+0.0050}_{-0.0059} $ & $  0.9747\pm 0.0054 $ & $  0.9679\pm 0.0048 $ \\ 
$ \ln \left( 10^{10} A_s \right)  $ & $  3.028\pm 0.017 $ & $  3.046\pm 0.016 $ & $  3.036\pm 0.017 $ & $  3.045\pm 0.016 $ \\ 
$ \Omega_K  $ & $  -0.066^{+0.024}_{-0.021} $ & $  -0.0104^{+0.013}_{-0.0080} $ & $  -0.032^{+0.013}_{-0.015} $ & $  -0.0035^{+0.0066}_{-0.0032} $ \\ 
\hline 
$ \lambda  $ & $ < 2.60 $ & $ 1.23\pm 0.63$ & $< 0.494$ & $  < 0.604 $ \\ 
$ \gamma  $ & $ >-0.0741$ & $ >-0.128 $ & $-0.077^{+0.035}_{-0.023}$ & $ > -0.0757 $ \\ 
$ \phi_i  $ & $  1.86\pm 0.76 $ & $ 1.46^{+0.55}_{-0.50} $ & $1.37^{+0.24}_{-1.1}$ & $ 1.66^{+0.70}_{-0.97}$ \\ 
\hline 
$ \sigma_8  $ & $  0.738\pm 0.059 $ & $  0.868^{+0.042}_{-0.079} $ & $  0.909\pm 0.053 $ & $  0.838^{+0.017}_{-0.041} $ \\ 
$ \Omega_{\rm m }  $ & $ 0.60^{+0.10}_{-0.16}$ & $  0.311\pm 0.020 $ & $ 0.316\pm 0.023 $ & $ 0.300^{+0.015}_{-0.0079}$ \\ 
$ S_8  $ & $ 1.027^{+0.056}_{-0.038} $ & $ 0.882^{+0.028}_{-0.070} $ & $  0.931^{+0.050}_{-0.041} $ & $ 0.838^{+0.012}_{-0.027} $ \\ 
$ H_0  $ & $  49.0^{+4.8}_{-6.0} $ & $  65.9^{+2.5}_{-1.7} $ & $  63.1^{+2.3}_{-2.6} $ & $  68.00\pm 0.74 $ \\ 
\hline\hline 
$\Delta \chi^{2}_{\rm min} $ & $-1.28$  & $-8.24$ & $-10.5$ & $-0.98$ \\
$\ln B_{M3^-, \Lambda {\rm CDM}}$ & $-4.32$ & $-5.28$  & $-6.98$ & $-6.76$ \\
\hline \hline
\end{tabular} }
\end{center}
\caption[Observational constraints for the curved M$3^-$ model]{Observational constraints at $68 \%$ confidence level on the sampled and derived cosmological parameters for different data set combinations under the curved M$3^-\, +\, \Omega_K$ model, studied in \cref{sec:cq_ok_results_m3}.}
\label{tab:cq_ok_m3n_curved}
\end{table*}

\section{Summary and Discussion} \label{sec:cqc_discussion}

\subsection{$H_0$ Tension}

\begin{figure*}[ht!]     
\centering     
\subfloat{\includegraphics[width=0.48\textwidth]{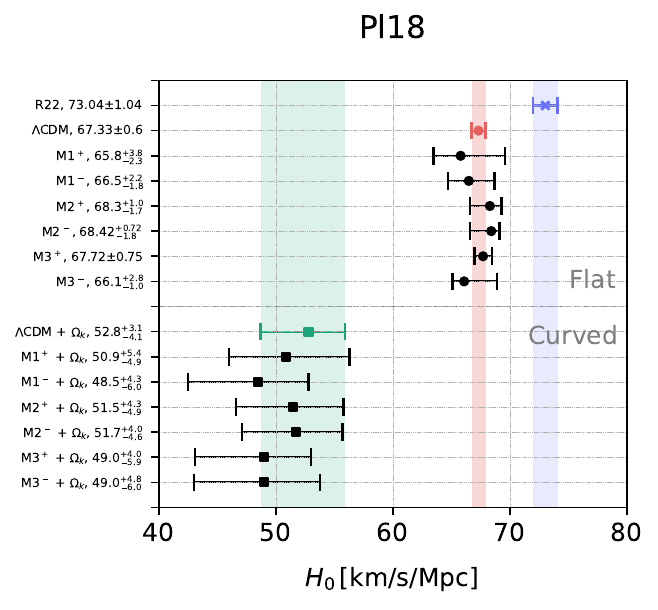}} \hfill
\subfloat{\includegraphics[width=0.48\textwidth]{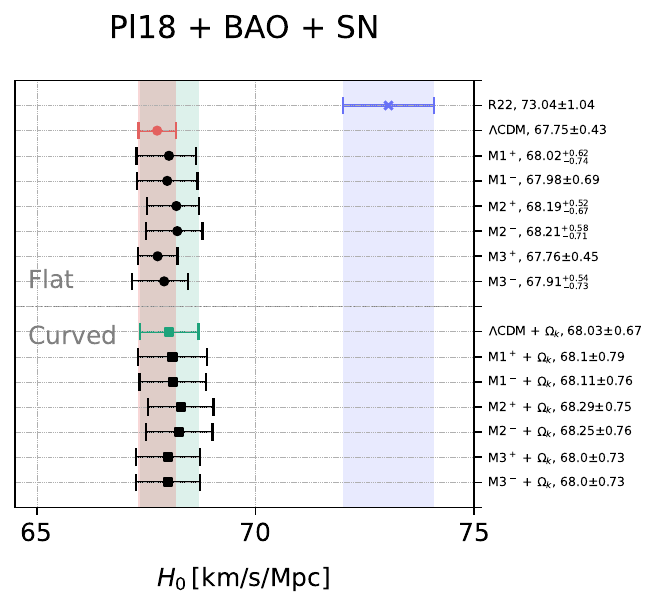}}
\caption[Whisker plot of the $68\%$ CL constraints on $H_0$ for the flat and curved models]{Whisker plot with the $68\%$ CL constraints on the Hubble parameter $H_0$ obtained for the cosmological models explored in this study for the Pl18 CMB data (left) and the combination Pl18 + BAO + SN (right), as detailed in \cref{sec:baseline}. Circle and squared markers denote the flat and curved models.  The red and green vertical bars illustrate to reference $\Lambda$CDM flat ($\Omega_K=0$) and curved $\Lambda$CDM + $\Omega_K$ values, respectively. The blue bar corresponds to the Kilo-Degree Survey (KiDS-1000) value \cite{KiDS:2020suj}, very close to one reported by the Dark Energy Survey (DES-Y3) \cite{DES:2021bvc}, as well, both in tension with the prediction for the six-parameter flat $\Lambda$CDM  from the \textit{Planck} collaboration 2018 data release \cite{Aghanim:2018eyx}, as explained in a \cref{sec:hubt}.}     
\label{fig:whisker_h0} 
\end{figure*}

The whisker plot in \cref{fig:whisker_h0} summarises the constraints derived on the Hubble constant $H_0$ at $68\%$ CL for the Pl18-only (left) and full Pl18+BAO+SN (right) for the various cosmological models considered in this chapter, showing the particular trends for how $H_0$ varies across the flat (top) and curved (bottom) cases. The first thing we can appreciate, discussed throughout the text, is the inherent degeneracy between $H_0$ and the curvature parameter, with a clear trend separating the flat and curved cases. In particular, a closed Universe drives the value of $H_0$ towards smaller values compared with the $\Lambda$CDM Pl18 case (red point and vertical bar) and, consequently, leads to an increase in the tension between the mean values and the R22 estimate (blue cross and vertical bar), which is superficially softened by the larger error bars.

For all the extensions to the $\Lambda\text{CDM} + \Omega_K$ model (green square and vertical bar), we reported an indication for a closed Universe at a confidence level exceeding 99\% when only Pl18 data is considered. The positive correlation between $H_0$ and $\Omega_K$ implies that, as $\Omega_K$ deviates from zero towards negative values, $H_0$ consequently decreases, exacerbating the tension with R22 as depicted in \cref{fig:whisker_h0}. This tension intensifies exclusively due to the Pl18-only evidence for a closed Universe. In the case of Pl18+BAO and Pl18+BAO+SN combinations, the $H_0$ estimates align closely with Planck's (flat) $\Lambda$CDM results as $\Omega_K$ approaches zero, and the mean value of $H_0$ is marginally higher, thus superficially easing the R22 tension. For Pl18+SN, a prediction for a closed Universe at more than $68\%$ for model M$3^+ + \Omega_K$ and $95\%$ for all the others bring the estimated $H_0$ values towards systematically lower values, and we do not observe any effective alleviation of the $H_0$ tension in any of these six extended scenarios.

The Pl18-only analysis also yields values of $H_0$ consistent at $1\sigma$ for all the flat and curved cases separately between themselves. There is an overall trend of models M$2$ yielding larger mean values of $H_0$, balanced by a reduction in the size of the $1\sigma$ regions, thereby not introducing significant benefits in addressing the tension. The larger tension reduction is present in the M$1$ flat models, predominantly due to the enlarged error bars.

Adding the background data to the Pl18 analysis brings the $H_0$ estimate in the curved models in agreement with the flat case at $1\sigma$, in line with the tension in BAO data for a closed Universe.

Finally, it is noteworthy that coupled quintessence models introduce new physics in comparison to $\Lambda$CDM, mostly at low-redshifts, once DE starts to dominate. This feature prevents these models from resolving the $H_0$ tension once BAO and SN data are considered, as evidenced by Refs.~\cite{Knox:2019rjx,Keeley:2022ojz,Arendse:2019hev}, unless other new exotic DE physics is considered~\cite{DiValentino:2020naf,Chudaykin:2022rnl}.

\subsection{$S_8$ Tension}

\begin{figure*}[ht!]     
\centering     
\subfloat{\includegraphics[width=0.48\textwidth]{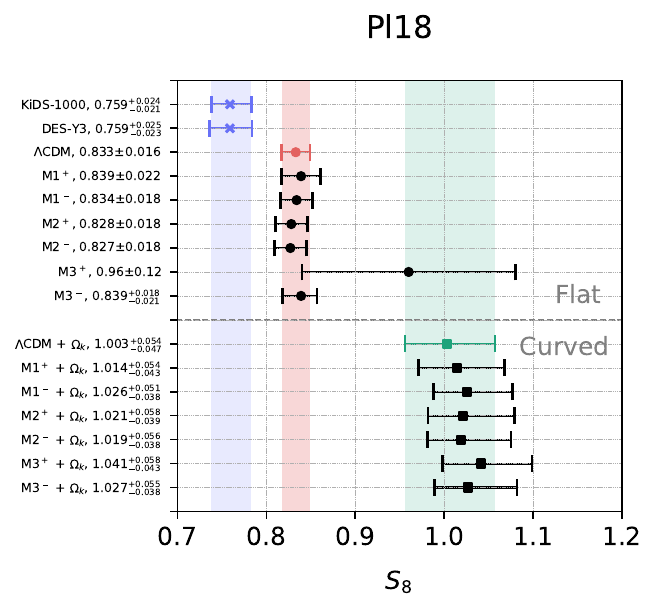}} \hfill
\subfloat{\includegraphics[width=0.48\textwidth]{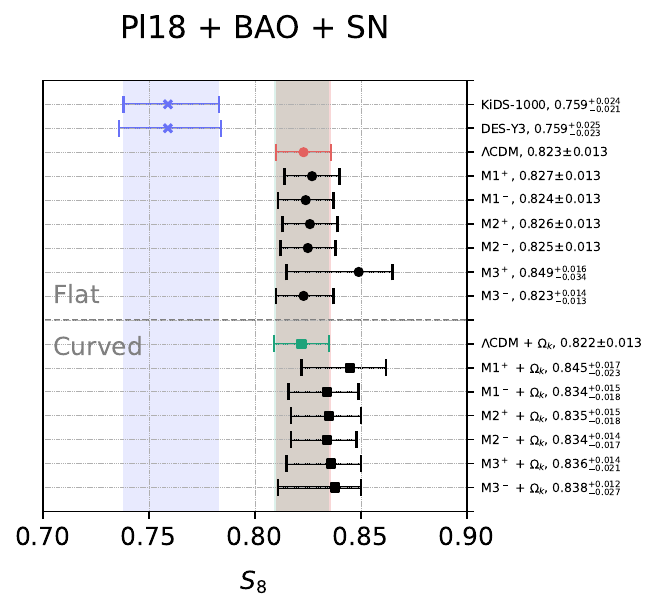}}
\caption[Whisker plot of the $68\%$ CL constraints on $S_8$ for the flat and curved models]{Whisker plot with the $68\%$ CL constraints on the parameter $S_8 \equiv \sigma_8 \sqrt{\Omega_m/0.3}$ derived for the cosmological models explored in this study for the Pl18 CMB data (left) and the combination Pl18 + BAO + SN (right), as detailed in \cref{sec:baseline}. Circle and squared markers denote the flat and curved models.  The red and green vertical bars illustrate the reference $\Lambda$CDM flat ($\Omega_K=0$) and curved $\Lambda$CDM + $\Omega_K$ values, respectively. The blue bar corresponds to the model-independent R22 value, reported in Ref.~\cite{Riess:2021jrx}, in tension with the prediction for the six-parameter flat $\Lambda$CDM  from the \textit{Planck} collaboration 2018 data release \cite{Aghanim:2018eyx}, as explained in \cref{sec:s8t}.}     
\label{fig:whisker_s8} 
\end{figure*}

In \cref{fig:whisker_s8}, the constraints on the $S_8$ parameter from the extended cosmological models are depicted at $68\%$ CL for Pl18-only (left) and Pl18+BAO+SN (right). For reference, we also show the similar estimated $S_8$ values from Kilo-Degree Survey (KIDS-1000)~\cite{Heymans:2020gsg} and the Dark Energy Survey (DES) Year 3 (DES-Y3)~\cite{DES:2021vln} (in blue), and the Planck~\cite{Aghanim:2018eyx} $\Lambda$CDM flat (red) and curved (green) cases with Pl18 data, all under the assumption of a baseline $\Lambda$CDM framework.

\cref{fig:whisker_s8} reveals two key observations. 
Firstly, for data involving just CMB, the $S_8$ values in all extended models deviate significantly from those of KIDS-1000, DES-Y3, and Planck when a standard $\Lambda$CDM model is assumed. For the curved scenarios, the estimated values of $S_8$ escalate sharply, largely due to the firm prediction of a closed Universe at more than $99\%$ CL. The only models that seem to move toward lower predictions of $S_8$ are the M$2$ flat models, even if marginally. Nevertheless, according to the discussion in \cref{sec:cosmotensions}, it should still be emphasised that any conclusions regarding the $S_8$ tension require a new analysis of the weak lensing data for the cosmological model in question. 

Secondly, estimations of $S_8$ using the background data sets closely resemble the prediction for the concordance $\Lambda$CDM with Planck. This might suggest that the $S_8$ tension is significantly enhanced in the extended coupled quintessence models that point toward a closed Universe. As was the case for the $H_0$ parameter, the apparent ease of the tension may occur in the curved models simply due to enlarged error bars and not through a more compatible mean value.

\subsection{Matter Density}

\begin{figure*}[ht!]     
\centering     
\subfloat{\includegraphics[width=0.48\textwidth]{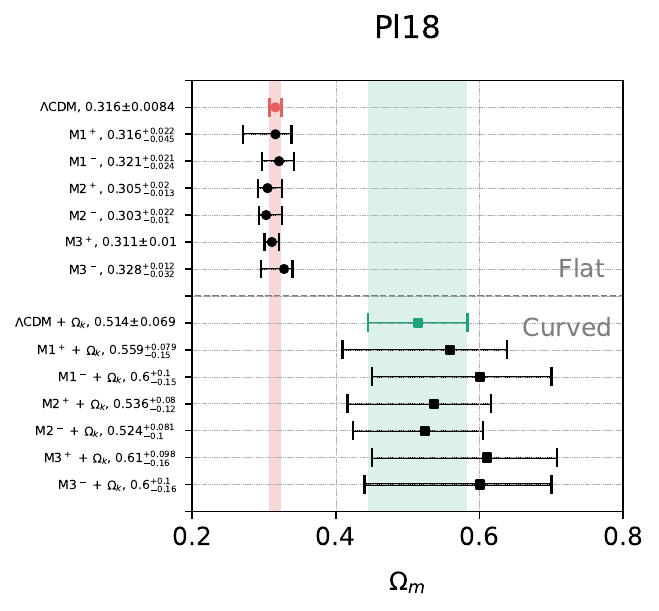}} \hfill
\subfloat{\includegraphics[width=0.48\textwidth]{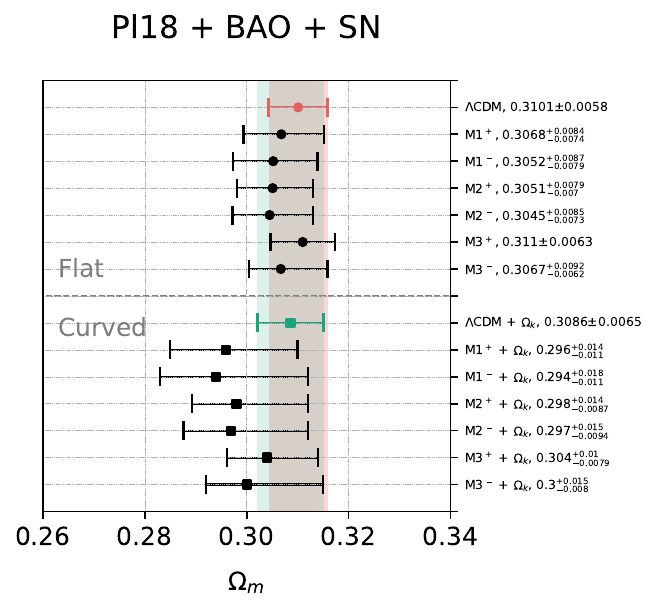}}
\caption[Whisker plot of the $68\%$ CL constraints on $\Omega_m$ for the flat and curved models]{Whisker plot with the $68\%$ CL constraints on the matter density parameter $\Omega_m$ obtained for the cosmological models explored in this study for the Pl18 CMB data (left) and the combination Pl18 + BAO + SN (right), as detailed in \cref{sec:baseline}. Circle and squared markers denote the flat and curved models.  The red and green vertical bars illustrate the reference $\Lambda$CDM flat ($\Omega_K=0$) and curved $\Lambda$CDM + $\Omega_K$ values, respectively.}     
\label{fig:whisker_om} 
\end{figure*}

For the matter density parameter $\Omega_m$ in \cref{fig:whisker_om} we compiled once more the finding of the cosmological models considered and for the Pl18 (left) and Pl18+BAO+SN (right) combinations of datasets. We depict both the flat cases (top) and the curved ones (bottom), for which a trend is readily identified. 

For the Pl18-only case, we see that the addition of spatial curvature brings $\Omega_m$ toward considerably large values. Even though all the models predict values consistent within their categories, we see that the curved coupled quintessence models have enlarged error bars compared with $\Lambda$CDM which, in some cases, allows to accommodate for more reasonable values. Once more, the addition of the background data brings the constraints closer to the flat prediction, in striking disagreement with the Pl18-case. The error bars for the curved coupled quintessence models ate considerably larger in comparison with both flat and curved $\Lambda$CDM. This gives more flexibility in predicting lower values of $\Omega_m$ which could yield a better fit to the data, at the cost of increasing the disagreement with the Pl18-only cases.


\subsection{Curvature}

\begin{figure*}[ht!]     
\centering     
\subfloat{\includegraphics[width=0.48\textwidth]{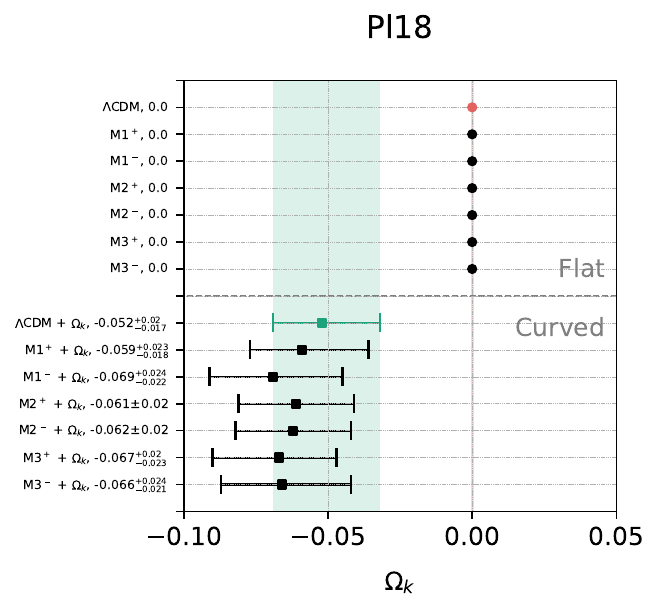}} \hfill
\subfloat{\includegraphics[width=0.48\textwidth]{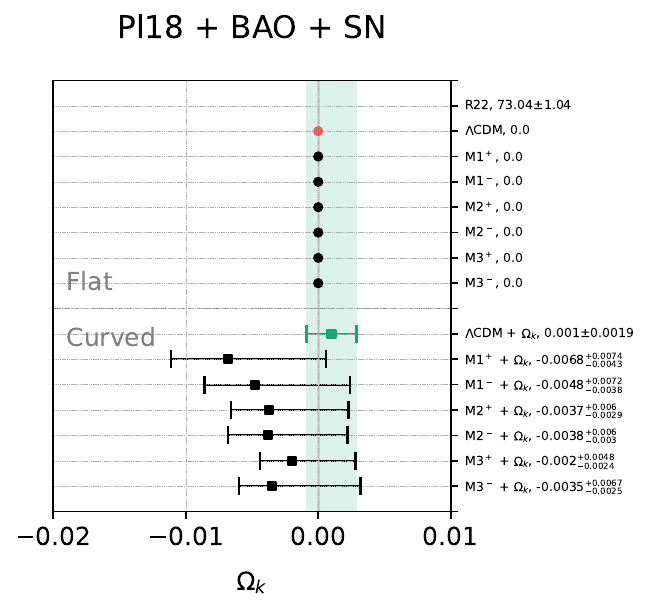}}
\caption[Whisker plot of the $68\%$ CL constraints on $\Omega_K$ for the flat and curved models]{Whisker plot with the $68\%$ CL constraints on the curvature parameter $\Omega_K$ obtained for the cosmological models explored in this study for the Pl18 CMB data (left) and the combination Pl18 + BAO + SN (right), as detailed in \cref{sec:baseline}. Circle and squared markers denote the flat and curved models. The red and green vertical bars illustrate to the reference $\Lambda$CDM flat ($\Omega_K=0$) and curved $\Lambda$CDM + $\Omega_K$ cases, respectively, illustrating the evidence for a closed Universe in the Planck data, as explained in \cref{sec:closed}.}     
\label{fig:whisker_ok} 
\end{figure*}

The findings related to the curvature density parameter, $\Omega_K$, are compiled in \cref{fig:whisker_ok} for the cosmological models considered and for the Pl18 (left) and Pl18+BAO+SN (right) combinations of datasets. We depict the flat cases (top) as dots in the plot for scale and comparison with the curved ones (bottom).

In the baseline $\Lambda$CDM model using Planck data, the indication for a closed Universe is evident, illustrated by the green bar, and persists across all the non-flat extended models examined in this work. However, this tendency varies in statistical significance due to the large error bars associated with expanding the parameter space or adopting alternative parametrisations for the potential and the coupling, even though this effect is not dramatic. This decrease in the constraining power is mainly attributed to the strong geometrical degeneracy between different parameters, most notably between the coupling parameter and the curvature density parameter $\Omega_K$, as seen by the lack of correlation in \cref{fig:rectangle_m1_curved,fig:rectangle_m2_curved,fig:rectangle_m3_curved}.

Breaking such degeneracies changes the predictions drastically, as it is evident for recovering the flatness condition in the full Pl18+BAO+SN combination. Introducing BAO large-scale structure data always wipes out the evidence for a closed Universe. Regardless of the coupled quintessence model under consideration, constraints on $\Omega_K$ tend to hover around zero, indicating spatial flatness within one standard deviation or slightly more. However, it is important to note that BAOs are generally in tension with Planck when curvature is a free parameter, making this data combination less robust.
On the other hand, for Pl18 combined with the distance moduli data from Pantheon (Pl18+SN), there is a systematic preference for a spatially closed geometry for coupled quintessence models.

\subsection{Discussion}

In this study, we explored whether the recently identified anomalies in CMB experiments, which are independent of Planck data, hold up when contrasted with other non-CMB observations, particularly when we extend the parameter space. In other words, when we shift to more comprehensive cosmological models allowing for additional variable parameters, does the evidence for a flat Universe with a cosmological constant and no interactions in the dark sector endure? To make the analysis more robust we have considered the full \textit{Planck} likelihood, with standard and nuisance parameters. Resolving these issues, as argued for in \cref{sec:cosmotensions}, is essential to give physical meaning to the origin of the cosmological discrepancies and formulate coherent alternative frameworks.



We find consistent discrepancies in the coupled quintessence models when compared to the concordance case which, while generally yielding a better fit to the data due to the increase parameter space, an analysis based on comparing the Bayesian evidence shows that the addition of the degrees of freedom is not justified. This leads us to conclude that the tensions in the data seem to genuinely favour the diverging patterns.

Given this, if we choose to maintain faith in the $\Lambda$CDM model, we must believe and prove that overlooked systematics could be skewing the data. Statistical fluctuations might explain the discrepancies related with the self-consistency between data collected based on the early-Universe (mostly CMB) and the distance ladder methods in the late Universe. Indeed, within the $\Lambda$CDM model, the CMB experiments are largely in agreement about the expansion rate, leaving the conflict with local measurements unresolved. The second scenario involves taking the data at face value, suggesting that solutions to current tensions could lie beyond the standard $\Lambda$CDM model. Given our limited understanding of the nature of the dark sector, this option deserves serious consideration. Yet, no combination of extra parameters entirely harmonises the conflicting data, hinting that a more radical overhaul of cosmological paradigms may be required.
While our analysis doesn't offer definitive answers about these tensions and anomalies, it highlights unresolved discrepancies in the standard model that remain present even for more accommodating models like coupled quintessence. The forthcoming generation of high-precision CMB and large-scale structure measurements might provide more conclusive constraints and insights on the origins of the cosmic tensions.

\cleardoublepage


 \chapter{Kinetically Coupled Dark Energy} \label{chap:kin}
 \setcounter{equation}{0}
\setcounter{figure}{0}


    \epigraph{Dava-se melhor com um irreal cotidiano, vivia em câmara leeeenta, lebre puuuuulando no aaaar sobre os ooooouteiros, o vago era o seu mundo terrestre, o vago era o de dentro da natureza.\blfootnote{\textit{She could deal better with her daily unreality, living in sloooow motion, hare leeeeaping through the aaaair over hiiiill and daaaale, vagueness was her earthly world, vagueness was the insides of nature.} --- \textsc{Clarice Lispector} in The Hour of the Star} \\ --- \textsc{Clarice Lispector}\ \small\textup{A Hora da Estrela}}

In this chapter, we study a particular extension in which dark energy is portrayed by a canonical scalar field, coupled to dark matter through an interaction term in the action, as proposed in Ref.~\cite{Barros:2019rdv}, and which we review in \cref{sec:kin_mot}. In \cref{sec:kin_numb,sec:kin_nump}, we perform a thorough numerical analysis of the dynamics of the model under consideration, both at the background and linear perturbative levels. The constraints for the cosmological and model-specific parameters according to different data sets are reported in \cref{sec:kin_const} from which we find that although the cosmic tensions persist in the \textit{best-fit} realisation of the model, there is a non-zero prediction for deviations from the standard model encoded by the coupling parameter. Computation of the Bayesian evidence indicates no significant preference for the Kinetic model. We conclude with a discussion of the results in \cref{sec:kin_dis}.
This work has been published in JCAP and can be found in Ref.~\cite{Teixeira:2022sjr}.

\section{Theoretical Motivation} \label{sec:kin_mot}

We begin by considering a phenomenological theory where dark energy is a dynamical quintessence field, denoted as $\phi$, minimally coupled to gravity. The DE source portrays a non-universal coupling to the dark matter component as expressed in the following action \cite{Barros:2019rdv}:
\begin{equation}
\mathcal{S} = \int \mathrm{d}^4 x \sqrt{-g} \left[ \frac{\text{M}_{\text{Pl}}^2}{2} R + X- V(\phi) + {f}(X)\tilde{\mathcal{L}}_c (\zeta,g_{\mu\nu} )+\mathcal{L}_\text{SM}(\psi_i,g_{\mu\nu}) \right] \mathperiod
\label{eq:kin_action}
\end{equation}
Here, $g$ represents the determinant of the metric tensor $g_{\mu\nu}$, $R$ is the curvature scalar, and $\text{M}_{\text{Pl}}^2=(8\pi G)^{-1}$ corresponds to the Planck mass in units where $c=1$, with $G$ being the Newtonian constant. The second and third terms in the action represent the Lagrangian of the scalar field, where $X=-g^{\mu\nu}\partial_{\mu}\phi\partial_{\nu}\phi/2$ is the kinetic term of $\phi$, and $V(\phi)$ is the self-interacting potential of the scalar field. The standard model is further extended by the introduction of a purely kinetic function $f(X)$ that multiplies the Lagrangian of cold dark matter \cite{Barros:2019rdv}, leading to a coupling between $\phi$ and the dark matter fields $\zeta$.

Variation of the action in \cref{eq:kin_action} with respect to the metric $g^{\mu\nu}$ yields the following field equations

\begin{equation}
    \label{eq:field}
    \text{M}_{\text{Pl}}^2 G_{\mu\nu} = T^{(\phi)}_{\mu\nu} + T^{(c)}_{\mu\nu} + T^{(b)}_{\mu\nu} + T^{(r)}_{\mu\nu} \mathcomma
\end{equation}

with $G_{\mu\nu}$ being the Einstein tensor and $T^{(i)}_{\mu\nu}$ the energy-momentum tensor for the $i$th species, defined as:

\begin{equation}
    T^{(i)}_{\mu\nu} = -\frac{2}{\sqrt{-g}}\frac{\delta \left( \sqrt{-g}\mathcal{L}_i \right)}{\delta g^{\mu\nu}} \mathcomma
\end{equation}

where $i=\phi,c,b,r$ and $c$ denotes the cold dark matter, $b$ the baryons, and $r$ the radiation. 
Let us note that, for the previous definition to be valid for all the fluids present in theory, we define an effective dark matter Lagrangian as follows \cite{Barros:2019rdv,Koivisto:2005nr,Kase:2019veo},

\begin{equation}
    \mathcal{L}_c \equiv f(X)\tilde{\mathcal{L}}_c \mathcomma
\end{equation}

incorporating the effect of the coupling.

The theory's matter components can be modelled as perfect fluids, with energy density $\rho_i$, pressure $p_i$, and \ac{eos} parameter $w_i=p_i/\rho_i$. Therefore, the energy-momentum tensor of each $i$th species becomes fully defined in terms of the fluid variables:

\begin{equation}
    \label{eq:emtensor}
    T^{(i)}_{\mu\nu}=\rho_i\left[\left(1+w_i\right)u^{(i)}_{\mu}u^{(i)}_{\nu} + w_i g_{\mu\nu}\right] \mathcomma
\end{equation}

with $u^{(i)}_{\mu}$ being the 4-velocity vector associated with the $i$th species, under the individual constraint $g^{\mu \nu} {u^{(i)}_{\mu}u^{(i)}_{\nu}=-1}$. Regarding the EoS parameter, we have $w_r=1/3$ for radiation and $w_b=w_c=0$ for baryons and cold dark matter, respectively. Given these considerations, the dark matter Lagrangian takes the particular form \cite{Koivisto:2005nr,Avelino:2018qgt},

\begin{equation}
    \label{L_c_rho_c}
    \mathcal{L}_c=-\rho_c \mathperiod
\end{equation}

The scalar field admits a perfect fluid description as well \cite{Faraoni:2012hn}, provided that

\begin{equation}
    u^{(\phi)}_{\mu}=-\frac{\partial_{\mu}\phi}{\sqrt{2X}} \mathcomma
\end{equation}

and $X>0$, where the energy density and pressure associated with the quintessence field are given by:

\begin{align}
    \rho_\phi&= X+V \mathcomma \label{rho_phi}\\
    p_\phi&=X-V \mathperiod \label{p_phi}
\end{align}

and the scalar field EoS parameter is $w_{\phi}=p_{\phi}/\rho_{\phi}$.

The equation of motion for the quintessence field, or simply the Klein-Gordon equation, is obtained through variation of the action in \cref{eq:kin_action} with respect to $\phi$ and reads:

\begin{equation}
    \label{eq:kleingordon}
    \square\phi - V_{,\phi} = - Q \mathcomma
\end{equation}

with $V_{,\phi} = \odv{V}{\phi}$. The term on the right-hand side of \cref{eq:kleingordon} includes the interaction in the dark sector in terms of $f(X)$ \cite{Barros:2019rdv}, and may be expressed generally in terms of $f$ and its derivatives

\begin{equation}
    \begin{aligned}
    \label{coupling}
    Q &= -\mathcal{L}_c\left\{ \frac{f_{,X}}{f}\left[ \square\phi + \partial^{\mu}\phi \left( \frac{\nabla_{\mu}\mathcal{L}_c}{\mathcal{L}_c}  + \frac{f_{,X}}{f} \, \partial_{\alpha}\phi\nabla_{\mu}\partial^{\alpha}\phi \right) \right] - \frac{f_{,XX}}{f} \, \partial^{\mu}\phi\partial_{\alpha}\phi\left( \nabla_{\mu}\partial^{\alpha}\phi \right) \right\} \mathperiod
    \end{aligned}
\end{equation}

where $f_{,X} \equiv \frac{df}{dX}$ and $f_{,XX} \equiv \frac{d^2f}{dX^2}$.
The uncoupled case ($Q=0$) is naturally recovered when $f$ is a constant function. Let us note that \cref{eq:kleingordon} could likewise be found through the contracted Bianchi identities, yielding the following conservation relations,

\begin{equation}
    \label{eq:consphim}
    \nabla_{\mu}T^{(c)\mu}{}_{\nu} = -\nabla_{\mu}T^{(\phi)\mu}{}_{\nu} = Q \nabla_{\nu}\phi \mathperiod
\end{equation}

These equations illustrate clearly the energy transfer between the scalar field and DM when $f$ is not a constant, meaning that the dark components are not individually conserved. However, since radiation and baryons remain non-interacting, \textit{i.e.},

\begin{equation}
    \label{eq:consradbar}
    \nabla_{\mu}T^{(r)\mu}{}_{\nu} = \nabla_{\mu}T^{(b)\mu}{}_{\nu} = 0 \mathcomma
\end{equation}

then, consistently, the overall energy-momentum tensor of the theory is conserved, rendering the total action covariant.

In particular, we focus on the most straightforward power-law kinetic interaction, as motivated in Ref.~\cite{Barros:2019rdv}, described by the function:
\begin{equation}
f(X) = \left(\text{M}_{\text{Pl}}^{-4}\, X\right)^{\alpha} \ \ \ \Longrightarrow\ \ \  Q = -\rho_c\frac{\alpha}{X}\left(\square\phi+\frac{\partial^{\mu}\phi\partial_{\nu}\phi\covd_{\mu}\partial^{\nu}\phi}{X}+\partial^{\mu}\phi\frac{\partial_{\mu}\rho_c}{\rho_c}\right) \mathcomma
\label{eq:kineticcoupling}
\end{equation}
where $\alpha$ is a dimensionless constant governing the strength of the coupling $Q$ in the dark sector, and $\rho_c$ represents the energy density of the cold dark matter. We assume a simple exponential potential for the field:
\begin{equation}
V(\phi) = V_0 \exp\left(-\frac{\lambda\phi}{\text{M}_{\text{Pl}}}\right) \mathcomma
\end{equation}
where $V_0$ represents the energy scale of the potential, and $\lambda$ characterises its steepness. These choices for the coupling function and potential are motivated by the desire to have a scaling regime \cite{Barros:2019rdv} at early times, followed by an accelerated expansion driven by the scalar field. The exponential potential drives the system out of the scaling solution and towards the late-time attractor, as described in \cref{sec:quin}.

It should be noted that the action described by \cref{eq:kin_action} is mathematically equivalent to a scalar-tensor theory in the Einstein frame with a conformal coupling in the DM action $\mathcal{S}_c\left[\tilde{g}_{\mu\nu}(X),\zeta \right]$. The coupling arises \textit{via} the conformal function $C(X) = f^2(X)$ in the metric transformation $\tilde{g}_{\mu\nu}=C(X)g_{\mu\nu}$, with a fundamentally different physical interpretation.


\section{Background Dynamics} \label{sec:kin_numb}

We assume cosmological dynamics in a flat Friedmann-Lemaître-Robertson-Walker (FLRW) background, expressed by the scale factor of the Universe $a(\tau)$ in conformal time $\tau$, as defined in \cref{eq:flrwchi}.
The equations governing the background evolution are obtained through variation of the action according to the metric and the scalar degree of freedom, namely the modified Friedmann equation and conservation relations:
\begin{align}
3 \text{M}_{\text{Pl}}^2 \mathcal{H}^2&=a^2(\rho_c+\rho_b+\rho_r+\rho_\phi)\label{eq:modfrid}\mathcomma\\
\phi'' + 2\mathcal{H}\phi'+a^2V_{,\phi} &= a^2Q \label{eq:phiII} \mathcomma\\ 
\rho_c'+3\mathcal{H}\rho_c &= -Q\phi' \mathcomma \label{eq:rhoII}\\
\rho'_b+3\mathcal{H}\rho_b &= 0 \mathcomma \label{eq:rhoIII}\\
\rho'_r+4\mathcal{H}\rho_r &= 0 \mathcomma \label{eq:rhoIV}
\end{align}
respectively, and with $' \equiv \dd/ \dd \tau$ and $\mathcal{H} = \frac{a'}{a}$ is the conformal Hubble rate. The coupling $Q$ given in \cref{eq:kineticcoupling} becomes:
\be
Q = 2\alpha\rho_c\frac{3\mathcal{H}\phi'+a^2V_{,\phi}}{2\alpha a^2\rho_c + \left( 1+2\alpha \right)\phi'^2} \mathperiod
\ee

We can further define the energy density and pressure of the $\phi$ field at the background level through \cref{rho_phi,p_phi}, as

\begin{align}
    \rho_{\phi} &= \frac{\phi'^2}{2a^2} + V, \label{eq:kin_rhoph} \\
    p_{\phi} &= \frac{\phi'^2}{2a^2} - V. \label{eq:pphi}
\end{align}

respectively. Therefore \cref{eq:phiII} can be written as:

\begin{equation}    
    \label{eq:contphi}
    \rho_\phi' + 3\mathcal{H}(1 + w_{\phi})\rho_\phi = Q\phi'.
\end{equation}

\cref{eq:rhoII,eq:contphi} imply that when $Q\phi'>0$, energy is transferred from the cold dark matter source to the scalar field, and \textit{vice versa} when $Q\phi'<0$.

At the classical level, the energy exchange in the dark sector may be interpreted as a mass variation for dark matter particles since $m_c = a^3\rho_c$, assuming conservation of the number of particles, \textit{i.e.} $N_c = N_{c}(\tau_0)$, with $\tau_0$ being the present conformal time. Integration of \cref{eq:rhoII} yields an expression for the total energy density of coupled dark matter,

\begin{equation} 
    \label{rho_c_an}
    \rho_c = \rho_c(\tau_0) a^{-3} \exp\left( 2\alpha \int_{\tau_0}^{\tau} Q \frac{\phi'}{\rho_c} \mathrm{d}\tau \right),
\end{equation}

that can be expressed equivalently in terms of the mass of the dark matter particles:

\begin{equation} 
    \label{eq:DMmass}
    m_{c}(\tau) = m_{c}(\tau_0) \exp\left( 2\alpha \int_{\tau_0}^{\tau} Q \frac{\phi'}{\rho_c} \mathrm{d}\tau \right).
\end{equation}

Finally, let us note that the modified Friedmann equation, \cref{eq:modfrid}, can be cast to the form of the well-known Friedmann constraint:

\begin{equation} 
    \label{eq:friedmannback}
    1 = \Omega_\phi + \Omega_m + \Omega_r,
\end{equation}

where we have defined a collective matter density $\rho_m = \rho_c + \rho_b$, and the fractional density parameter of the $i$-th species $\Omega_i = \frac{\rho_i a^2}{3 \text{M}_{\text{Pl}}^2 \mathcal{H}^2}$.
\cref{eq:friedmannback} can be rewritten in the form of a constraint on the present scalar field fractional density, $\Omega_{\phi}^0 = 1 - \Omega_m^0 - \Omega_r^0$, where ``0'' stands for quantities evaluated at present, $\Omega_i^0 = \frac{\rho_i^0}{3 \text{M}_{\text{Pl}}^2 H_0^2}$, and $H_0$ is the Hubble parameter. For numerical purposes, $V_0$, implicitly entering the definition of $\Omega_\phi^0$, is used to perform a shooting method that yields the fiducial value of $\Omega_\phi^0$ fulfilling the constraint relation in \cref{eq:friedmannback}, while simultaneously avoiding degeneracies. As such, $V_0$ will no longer be considered a free parameter of the model, leaving $\{\lambda, \alpha\}$ as the model free parameters.
The viable parameter space has been studied in Ref.~\cite{Teixeira:2022sjr} according to dynamical and stability conditions, along with motivation for the particular initial conditions for the scalar field used in the simulations.
We conduct numerical simulations using a modified version of the Einstein-Boltzmann solver \texttt{CLASS} \cite{Lesgourgues:2011re,Blas:2011rf,lesgourgues2011cosmic2} to study the predictions of the model over the expansion history for different $\{\lambda, \alpha\}$ combinations, using standard \textit{Planck} 2018 \cite{Aghanim:2018eyx} reference values for the cosmological parameters.

\begin{figure}[h]
      \subfloat{\includegraphics[width=0.5\linewidth]{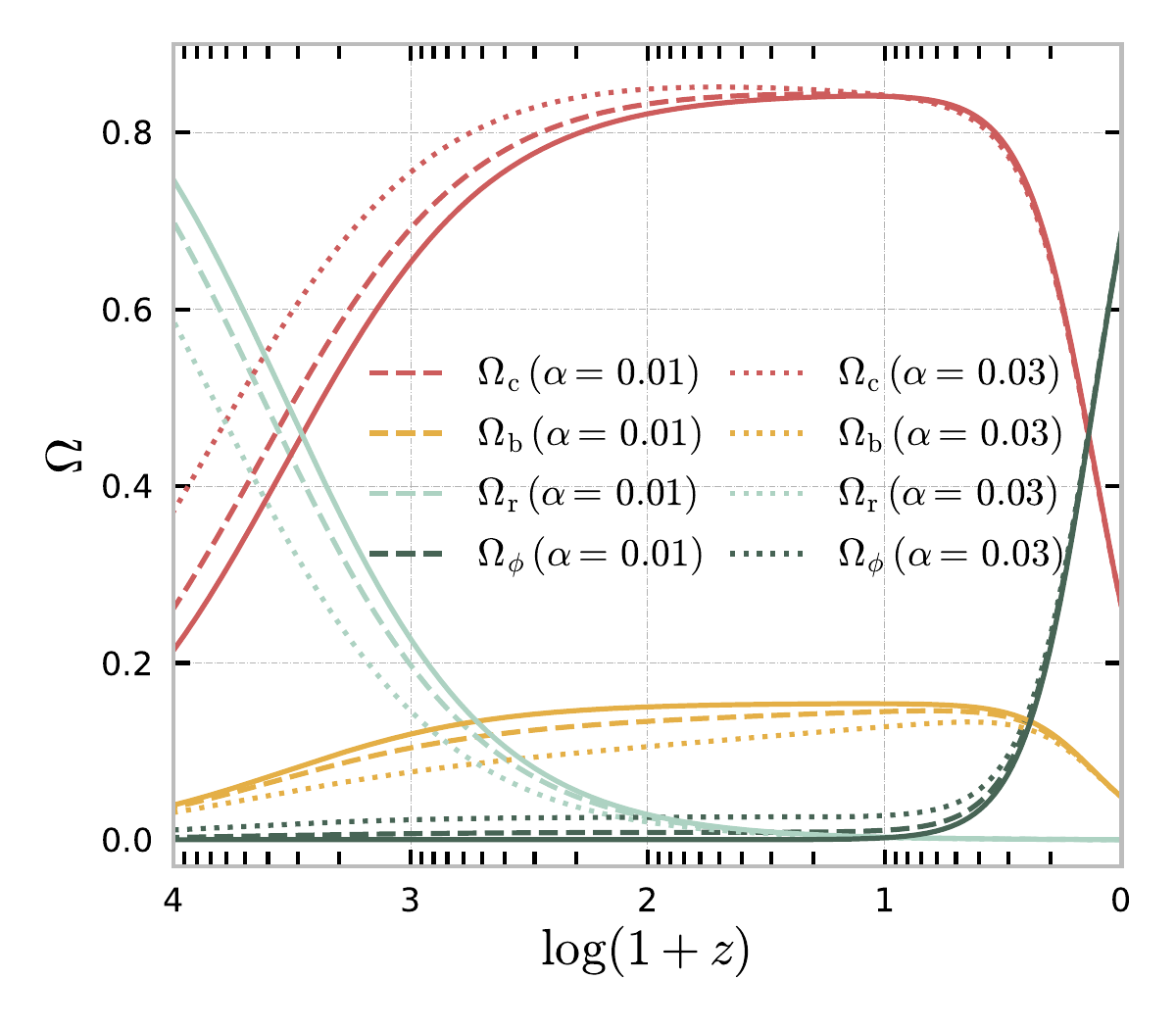}}
      \subfloat{\includegraphics[width=0.5\linewidth]{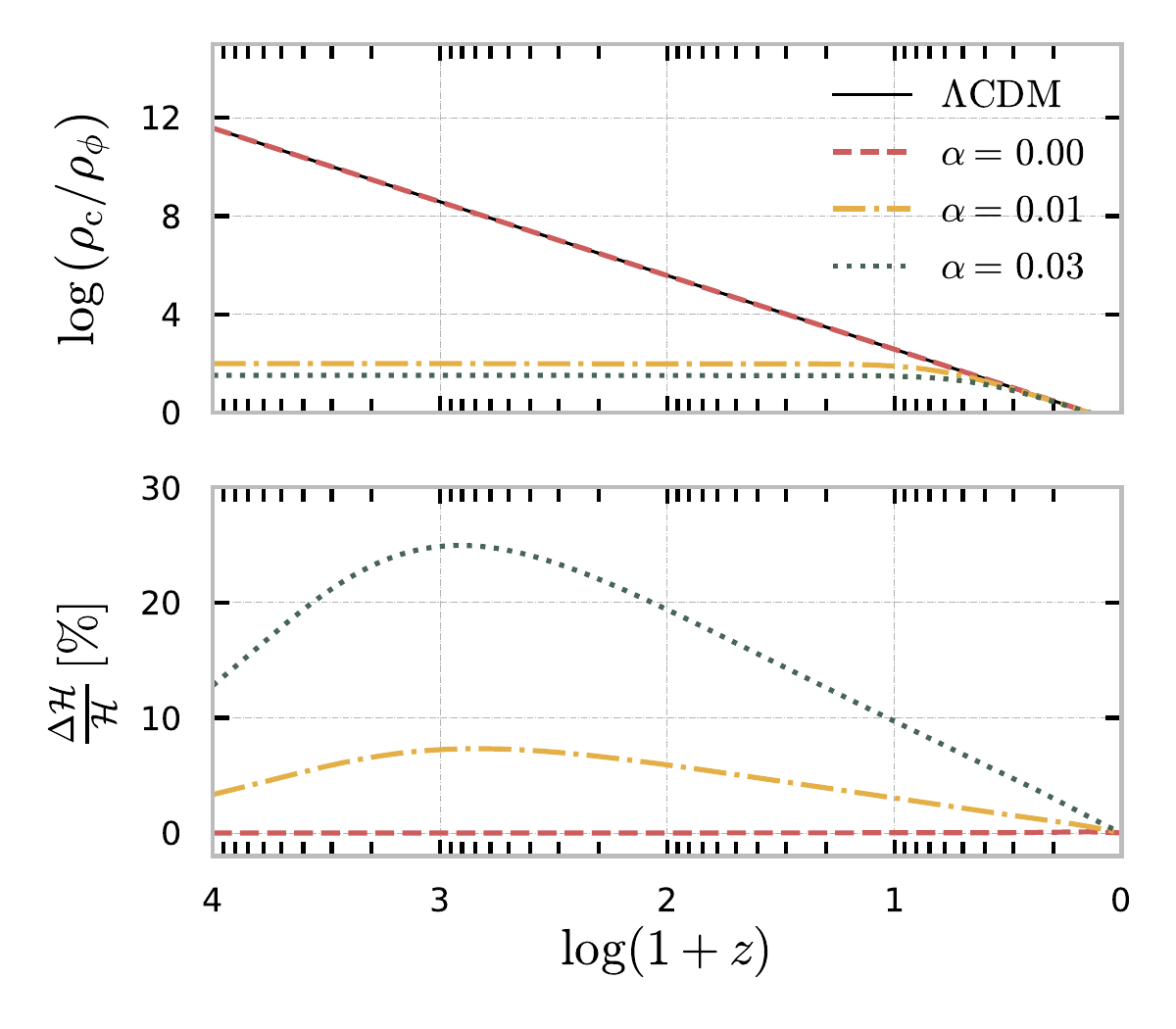}}
  \caption[Evolution of relative energy densities $\Omega_i$, the ratio $\rho_c/\rho_\phi$ and variations of the Hubble rate $\Delta \hub/\hub$]{\label{fig:fig1} \textit{Left panel}: Evolution of the relative energy densities  $\Omega_i$ with redshift, $1+z$, of the scalar field (green), dark matter (red), baryons (yellow) and radiation (grey) for $\Lambda$CDM (solid lines), ${\alpha =0.01}$ (dashed lines) and ${\alpha = 0.03}$ (dotted lines).  \textit{Upper right panel}: Ratio of the energy densities of cold dark matter and dark energy for $\Lambda$CDM (solid black), the uncoupled case $\alpha=0$ (red dashed), ${\alpha =0.01}$ (yellow dashed-dotted) and ${\alpha = 0.03}$ (green dotted). \textit{Lower right panel}: Fractional deviation of the conformal Hubble expansion rate for the same examples.}
\end{figure}

In the left panel of \cref{fig:fig1}, we present the evolution of the relative energy densities $\Omega_i = \rho_{i} a^2/ (3 \text{M}_{\text{Pl}}^2 \mathcal{H}^2)$ for each species with respect to redshift ($1+z$). As expected, the introduction of the coupling leads to the emergence of an early scaling regime during the radiation-dominated epoch, where the energy density of the scalar field is proportional to that of dark matter, approximately following the relation $\rho_c/\rho_{\phi}=1/\alpha$, as shown in the upper right panel of \cref{fig:fig1}. Eventually, the field exits this scaling regime and transitions towards the future attractor solution, causing its energy density to continuously dilute as $\rho_{\phi}\propto a^{-\lambda^2}$.
For the values considered, the coupling strength $Q\phi'/\rho_c$ remains positive at all redshifts, implying energy transfer from the dark matter fluid to the scalar field. It should be noted that by fixing the present-day values of fluid densities, the energy density of cold dark matter is relatively higher at earlier times due to its contribution in supplying energy to the scalar field at later times. This effect becomes more pronounced for larger values of $\alpha$. As the DM energy density decreases over time, the additional energy transferred to the scalar field compensates for this effect compared to the uncoupled case. This behaviour is illustrated in the left panel of \cref{fig:fig1}. Consequently, we also see in the figure that the matter-radiation equality occurs at earlier times as the value of $\alpha$ increases. Furthermore, the $\phi$ field starts acquiring energy at a rate proportional to its energy density, and the equality between matter and dark energy occurs earlier.
In the lower right panel of \cref{fig:fig1}, we depict deviations in the Hubble rate for the Kinetic model and the uncoupled case ($\alpha=0$) compared to $\Lambda$CDM. We remark that no significant variations in $\mathcal{H}$ are observed during the radiation-dominated epoch, as interactions between the dark and radiation sectors are excluded. However, as the matter contribution becomes significant, around ${z\approx 10^6}$, the Kinetic models exhibit an increased value of $\mathcal{H}$, which becomes more pronounced for higher values of $\alpha$.

\section{Evolution of Linear Perturbations} \label{sec:kin_nump}

We describe the perturbed FLRW metric in the Newtonian gauge according to the standard line element, as defined in \cref{eq:pmetric}. The linearised Einstein equations describe the evolution of perturbations for different scales in terms of independent Fourier modes and extra contributions from the dark sector coupling.


In particular, from \cref{rho_phi,p_phi}, we derive the perturbations for the energy density and pressure of the scalar field:
\begin{align}
\delta \rho_{\phi} &= \frac{\phi'}{a^2}\delta\phi' - \frac{\phi'^2}{a^2}\Psi + V_{,\phi}\delta\phi \mathcomma \label{eq:deltaphi1} \\
\delta p_{\phi} &= \frac{\phi'}{a^2}\delta\phi' - \frac{\phi'^2}{a^2}\Psi - V_{, \phi}\delta\phi \mathperiod \label{eq:deltaphi2}
\end{align}

The perturbations of the energy-momentum tensor, \cref{eq:emtensor}, for each species and at first order, reads
\begin{equation}
\delta T^{(i)}{}^{\mu}_{\nu} = (\delta\rho_i + \delta p_i)u^{(i)}{}^{\mu}u^{(i)}{}_{\nu} + \delta p_i \delta^{\mu}_{\nu} + (\rho_i + p_i)\left( \delta u^{(i)}{}^{\mu}u^{(i)}{}_{\nu} + u^{(i)}{}^{\mu} \delta u^{(i)}{}_{\nu} \right) \mathcomma
\end{equation}
where $\delta u^{(i)}{}_{\mu}$ is the perturbation on the four-velocity vector of the $i$th species, \textit{i.e.} $u^{(i)}{}_{\mu} = a(-1,v^{(i)}{}_j)$, with $v_j$ being the peculiar velocity.
As introduced in \cref{sec:peinst}, the perturbed Einstein equations assuming no shear are expressed as

\begin{align}
k^2 \Phi + 3\mathcal{H}\left( \Phi' + \mathcal{H}\Psi \right) &= -4\pi G a^2 \sum_i \delta \rho_i \label{eq:lEE1} \mathcomma \\
k^2\left( \Phi' + \mathcal{H}\Psi \right) &= 4\pi G a^2 \sum_i \rho_i (1 + w_i) \theta_i \label{eq:lEE2} \mathcomma \\
\Phi'' + \mathcal{H}\left( \Psi' + 2\Phi' \right) + \Psi\left( \mathcal{H}^2 + 2\mathcal{H}' \right) + \frac{k^2}{3}\left( \Phi - \Psi \right) &= 4\pi G a^2 \sum_i \delta p_i \label{eq:lEE3} \mathcomma \\
\Phi &= \Psi \label{eq:lEE4} \mathperiod
\end{align}

The first equation, corresponding to the time-time component, provides the constraint on the energy density. \cref{eq:lEE2}, which is derived from the time-space components of the perturbed Einstein equations, specifies the momentum constraint. Here, we adopt the definition of the velocity divergence as \(\theta_i = \nabla \cdot v^{(i)}\). The trace of the spatial components results in \cref{eq:lEE3}, and finally, \cref{eq:lEE4} relates to the propagation of shear in the absence of anisotropic stress. This is expected because of the absence of a non-minimal coupling in the action in \cref{eq:kin_action}.
The governing equations for the evolution of perturbations in each fluid can be obtained via the conservation relations, specifically \cref{eq:consphim,eq:consradbar}, at the first order of perturbation. For species that do not interact, namely baryons and radiation, these equations are as follows:

\begin{align}
\delta'_i + 3\mathcal{H}\left( \frac{\delta p_i}{\delta \rho_i} - w_i \right)\delta_i + (1+w_i)\left( \theta_i - 3\Phi' \right) &= 0 \label{eq:nonintdelta} \mathcomma \\
\theta_i' + \left[ \mathcal{H}(1-3w_i) + \frac{w_i'}{1+w_i} \right]\theta_i - k^2\left( \Psi + \frac{\delta p_i}{\delta \rho_i} \frac{\delta_i}{1+w_i} \right) &= 0 \mathperiod
\end{align}
%

For the coupled DM, these become
%
\begin{align}
\delta_c' + \theta_c -3\Phi' &= \frac{Q}{\rho_c} \left( \phi'\delta_c - \delta\phi' \right) - \frac{\phi'}{\rho_c}\delta Q \mathcomma \\
\theta'_c + \mathcal{H}\theta_c -k^2\Psi &= \frac{Q}{\rho_c}\left( \phi'\theta_c - k^2\delta\phi \right) \mathcomma
\end{align}
describing the evolution of the density contrast, denoted as $\delta_i = \delta \rho_i / \rho_i$ and the velocity divergence ${\theta_i = \covd \cdot v^{(i)}}$. The perturbed coupling term is defined as 
\begin{align}
\delta Q =& \frac{2\alpha\rho_c}{2\alpha a^2 \rho_c + (1+2\alpha)\phi'^2}\left\{ -3\Phi'\phi' -\phi'\theta_c + \left[ 3\mathcal{H}\phi' + a^2 (V_{,\phi}-Q) \right]\delta_c + \left( 2k^2 + a^2 V_{,\phi\phi} \right)\delta\phi \right. \nonumber \\
&\left. \hspace{3.8cm} -\left[ 3\mathcal{H}\phi' + 2a^2 (V_{,\phi}-Q) \right] \frac{\delta\phi'}{\phi'} +2a^2\Psi\left(Q-V_{,\phi}\right)\right\} \mathcomma\label{eq:deltaQ}
\end{align}
and we remark the explicit dependence of $\delta Q$ on the velocity potential $\theta_c$, which is unusual compared to other coupled dark energy models \cite{vandeBruck:2016hpz,vandeBruck:2020fjo}. Finally, the linearisation of the Klein-Gordon equation gives:
\be\label{eq:perturbedKG}
\delta\phi'' + 2\mathcal{H}\delta\phi' + \left( a^2V_{,\phi\phi} +k^2 \right)\delta\phi-\left(\Psi'+3\Phi'\right)\phi'+2a^2\Psi V_{,\phi} = a^2 \delta Q+2a^2Q\Psi \mathperiod
\ee

Including the coupling leaves an imprint on important cosmological observables that can be probed against different data, particularly the matter power spectrum and the temperature-temperature (TT) and lensing angular power spectra of the cosmic microwave background (CMB). Again, assuming standard \textit{Planck} 2018 values for the shape of the scalar primordial power spectrum \cite{Aghanim:2018eyx} as well, it is possible to single out the model-specific signatures by taking different sets of $\{\lambda,\alpha\}$ relevant for the scales under consideration. 

\begin{figure}[h]
      \subfloat{\includegraphics[width=0.5\linewidth]{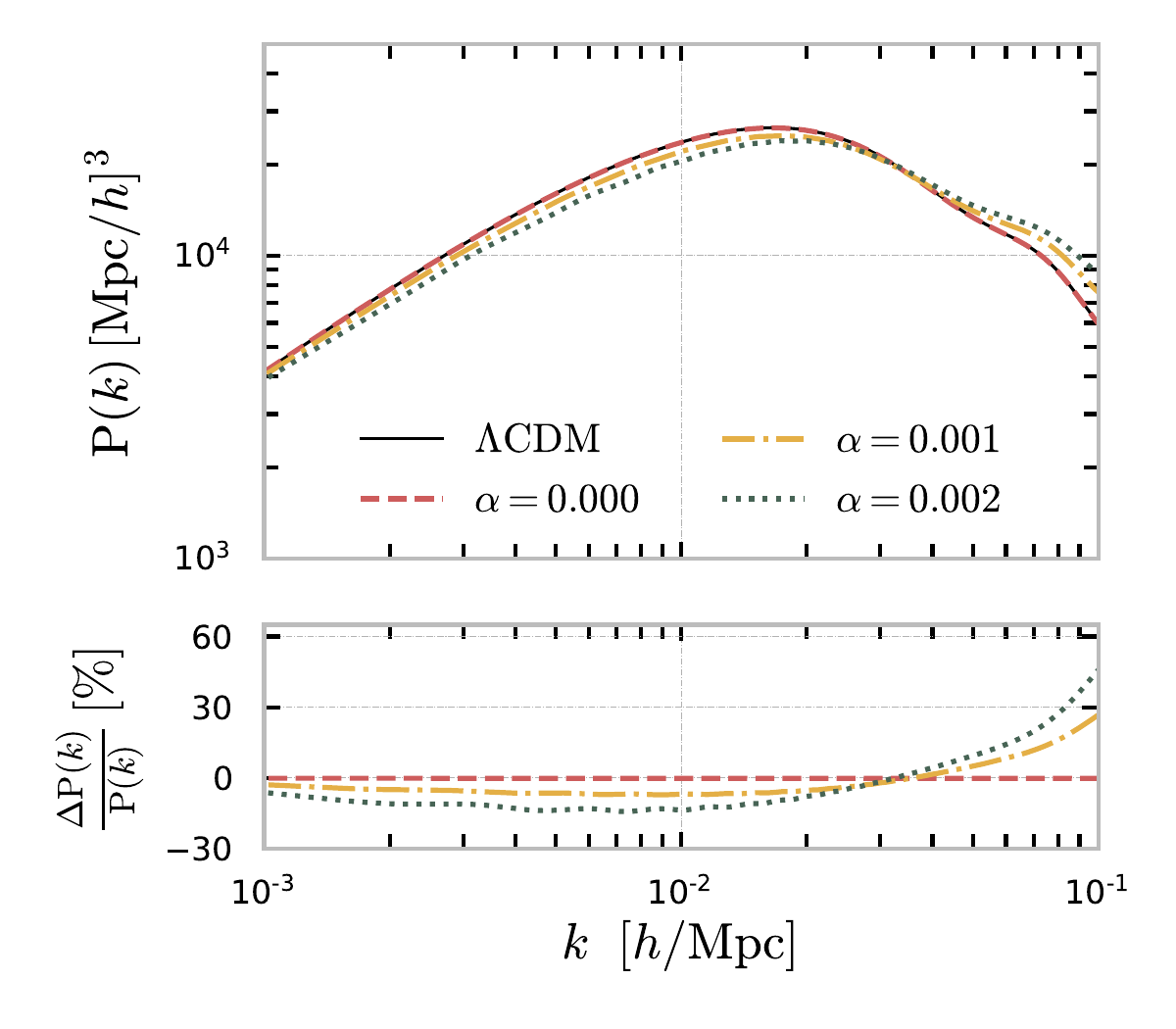}}
      \subfloat{\includegraphics[width=0.5\linewidth]{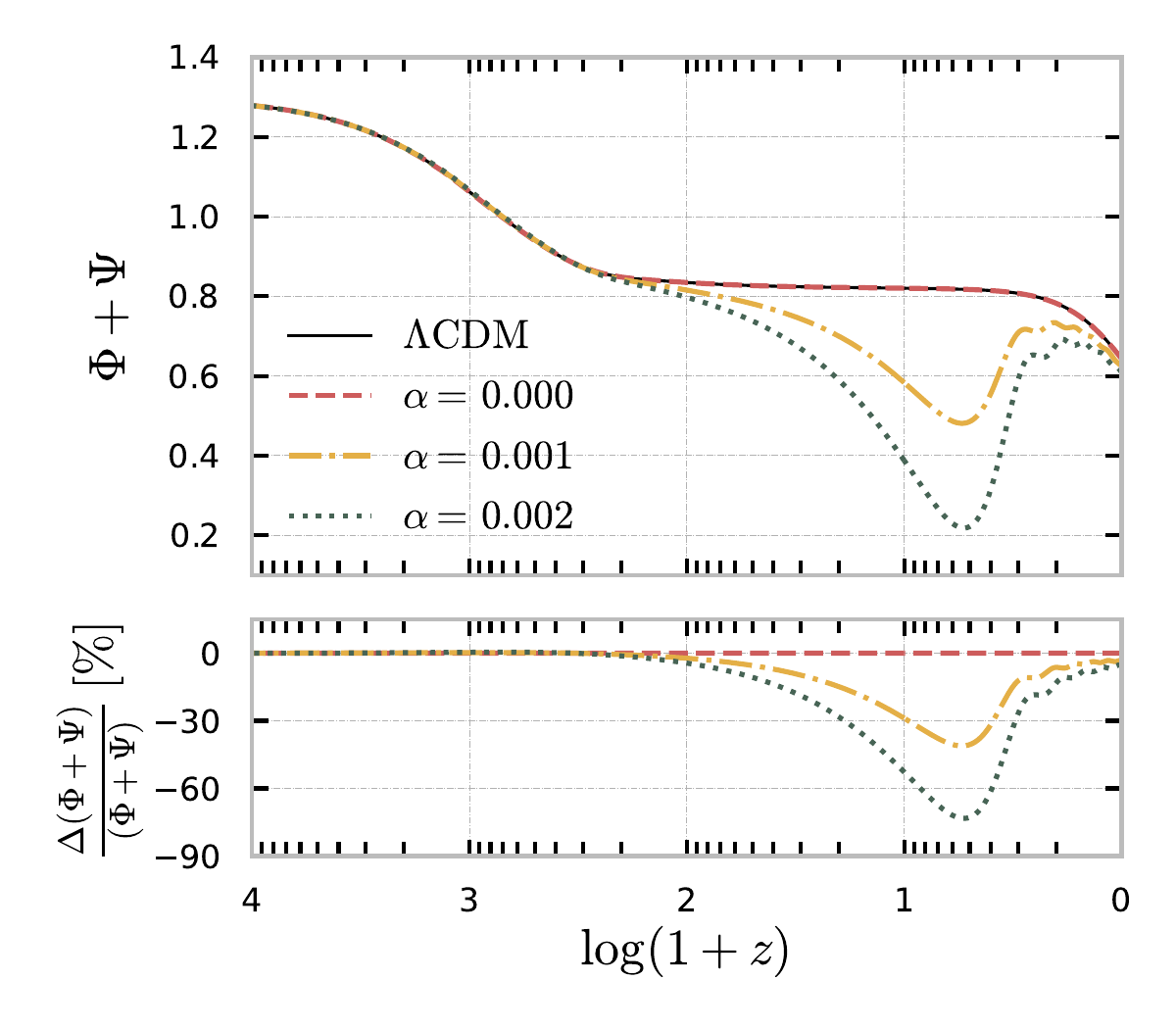}}
  \caption[The matter power spectrum and the lensing potential]{\label{fig:fig2} \textit{Left panel}: The matter power spectrum as function of $k$, for the uncoupled case (dashed red line), ${\alpha = 0.001}$ (yellow dot-dashed line), ${\alpha = 0.002}$ (green dotted line) and $\Lambda$CDM (black solid line), along with the percentage deviations from the $\Lambda$CDM case (lower panel). \textit{Right panel}: Evolution of the sum of the gravitational potentials as a function of the redshift at $k=0.01$ Mpc$^{-1}$ for the same examples.}
\end{figure}

In the left panel of \cref{fig:fig2}, we present the linear matter power spectrum at present up to the scale $k_{\rm max}= 0.1 h$ Mpc$^{-1}$ along with the fractional differences for the Kinetic model in contrast with $\Lambda$CDM. We identify an overall suppression at intermediate scales ($10^{-3} h$ Mpc$^{-1}\lesssim k \lesssim 3\times 10^{-2} h$ Mpc$^{-1}$), followed by an enhancement at smaller scales. This effect is primarily related to the global deviation of the turnover in the matter power spectrum to higher $k$ values due to the shift of the radiation-matter equality era to earlier times, as identified in the study of the background evolution. 
The positive exchange of energy from cold dark matter to dark energy at late times inevitably suppresses the growth of matter perturbations at intermediate scales and an enhancement at smaller scales. The highest deviations are observed for larger values of $\alpha$. Consequently, the amplitude of the matter power spectrum at present and a scale of 8 $h^{-1}$Mpc, denoted by $\sigma_8$, is expected to be larger for the Kinetic model.

\begin{figure}[h]
      \subfloat{\includegraphics[width=0.5\linewidth]{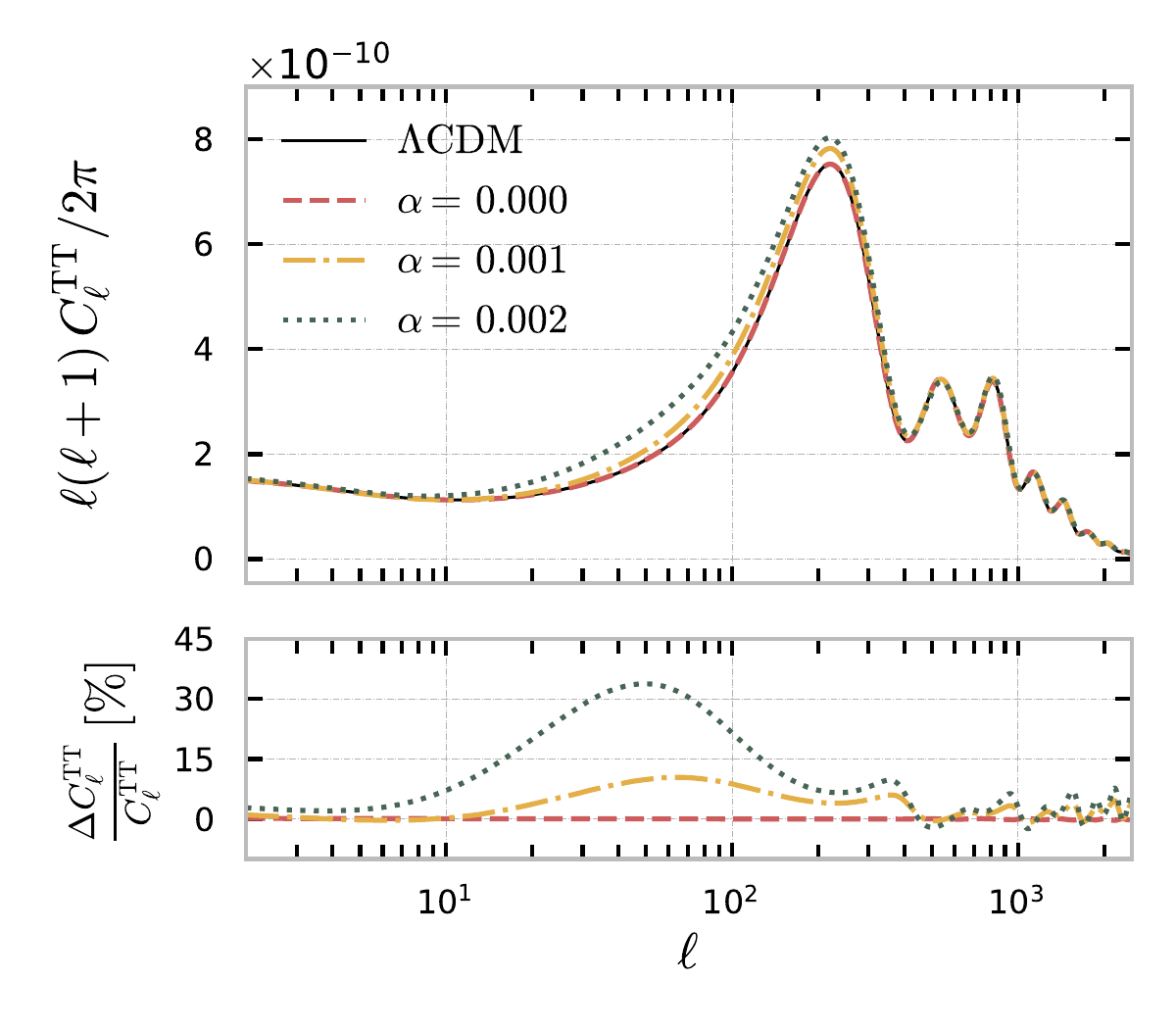}}
      \subfloat{\includegraphics[width=0.5\linewidth]{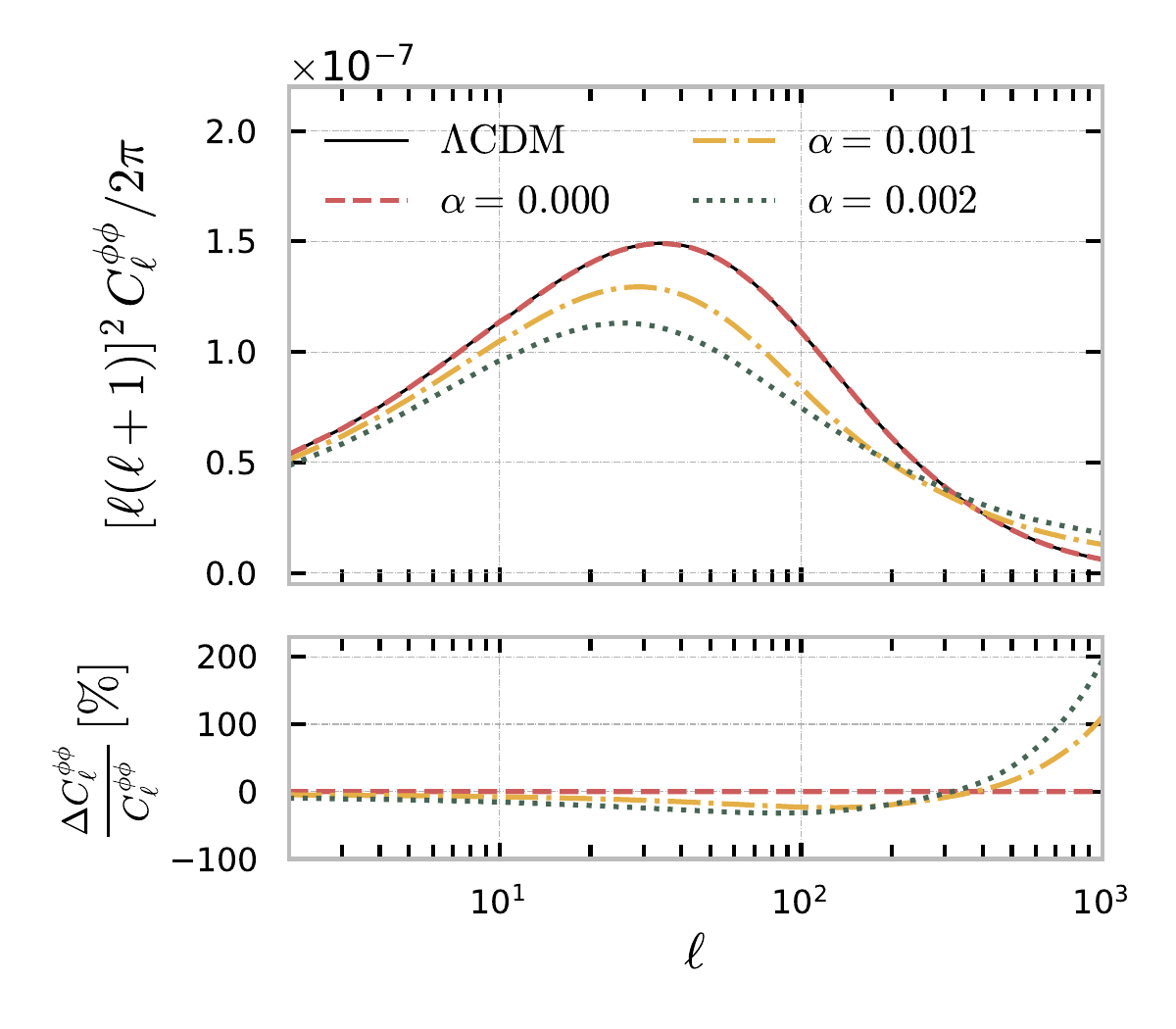}}
  \caption[The CMB temperature and lensing power spectra]{\label{fig:fig3} \textit{Left panel}: TT power spectrum of anisotropies as a function of the angular scale $\ell$, for the uncoupled case (dashed red line), ${\alpha = 0.001}$ (yellow dot-dashed line), ${\alpha = 0.002}$ (green dotted line) and $\Lambda$CDM (black solid line) for reference, along with percentage deviations with respect to $\Lambda$CDM (lower panel). \textit{Right panel}: Lensing angular power spectra for the same examples and relative difference between the predictions for each model and for $\Lambda$CDM.}
\end{figure}

The inclusion of the coupling is also reflected in the evolution of the gravitational potentials, parametrised in terms of the lensing potential $\phi_{\text{lens}}=(\Psi+\Phi)/2$ for particular scales, as depicted in the right panel of \cref{fig:fig2}. We identify an overall suppression in $\phi_{\text{lens}}$ which in turn leads to a suppression of the lensing power spectrum as well, as demonstrated in the right panel of \cref{fig:fig3}, with this effect becoming more pronounced for larger values of $\alpha$. On the other hand, the time variation of $\phi_{\text{lens}}$ is directly related to the integrated Sachs-Wolfe effect (ISW), imprinted in the shape of the temperature-temperature (TT) power spectrum as a contribution to the radiation transfer function, and is illustrated in the left panel of \cref{fig:fig3} as a function of the angular multipole $\ell$. This effect comprises two contributions: an early-time term occurring during the transition from the radiation- to matter-dominated epochs, shifted to earlier times in the Kinetic model, and a late-time term associated with the dynamics of the dark energy component. We observe an apparent overall enhancement compared to the reference case for $\ell \lesssim 300$, with milder differences around the plateau at $\ell <10$ and the most significant deviations around $\ell\sim 50$. Furthermore, there is an apparent increase in the amplitude of the first peak accompanied by a broadening of its shape and slight variations between the peaks and troughs at higher multipoles. 
The validity of these combined deviations can be assessed with cosmological data from background observations and the large-scale structure.

\section{Observational Constraints} \label{sec:kin_const}

\subsection{Data Sets}

The baseline data set considered is the CMB \textit{Planck} 2018 (Pl18) \cite{Aghanim:2018eyx} henceforth to which we add BAO data from the Sloan Digital Sky Survey (SDSS) DR7 Main Galaxy Sample \cite{Ross:2014qpa}, SDSS DR12 consensus release \cite{BOSS:2016hvq} and the 6dF Galaxy Survey \cite{Beutler:2011hx}, in combination with distance \textit{moduli} measurements of type Ia Supernova (SN) data from Pantheon \cite{Scolnic:2017caz}(Pl18+BAO+SN). Finally, we included CMB lensing potential data from \textit{Planck} 2018 ~\cite{Planck:2019nip,Planck:2018lbu} (Pl18len+BAO+SN). Further details are provided in \cref{sec:baseline}.
We use a set of free sampling parameters consisting of the baseline $\Lambda$CDM cosmological parameters $(\Omega_b h^2, \Omega_c h^2,z_{reio},\theta_s, A_{\rm s}, n_{\rm s})$ as detailled in \cref{sec:lcdm_param}, as well as the two free parameters associated with the Kinetic model $(\alpha,\lambda)$. 
We consider flat priors for all the parameters and provide the specific range of values in \cref{tab:kin_priors}.
Our analysis also yields derived constraints on $H_0$ and $S_8 =  \sigma_8 \sqrt{\Omega_m/0.3}$. The latter is also known to be in tension with cosmic shear measurements \cite{Heymans:2020gsg,DiValentino:2020vvd,Abdalla:2022yfr} for the standard model, with CMB data favouring higher values. Finally, to produce the Monte Carlo Markov Chain (MCMC) samples, we follow our modification of the Einstein Boltzmann solver \texttt{CLASS} \cite{Lesgourgues:2011re,Blas:2011rf,lesgourgues2011cosmic2} interfaced with the \texttt{MontePython} sampler \cite{Audren_2013,Brinckmann:2018cvx}, following the Metropolis-Hastings algorithm. Subsequently, we analyse the MCMC chains and produce the results reported in Ref.~\cite{Teixeira:2022sjr}, with the aid of the \texttt{GetDist} Python package \cite{Lewis:2019xzd}.

\begin{table}[]
\begin{center}
\begin{tabular}{c|c}
\hline
Parameter                    & Prior \\
\hline
$\Omega_b h^2$                & $[0.005,0.1]$ \\
$\Omega_c h^2$                & $[0.001,0.99]$ \\
$100\theta_{s}$                    & $[0.5,10]$ \\
$\tau_{reio}$                          & $[0.01,0.8]$ \\
$n_s$                      & $[0.7,1.3]$ \\
$\log \left(10^{10}A_{s} \right)$   & $[1.7, 5.0]$ \\
\hline
$\lambda$                           & $[0,2]$ \\
$\alpha$                            & $[0,1]$ \\
\hline 
\end{tabular}
\end{center}
\caption[Priors on the model parameters]{Flat priors on the cosmological and model parameters sampled in this work.}
\label{tab:kin_priors}
\end{table}

\subsection{Results}

\begin{figure}[t]
\begin{centering}
    \subfloat{\includegraphics[width=0.75\linewidth]{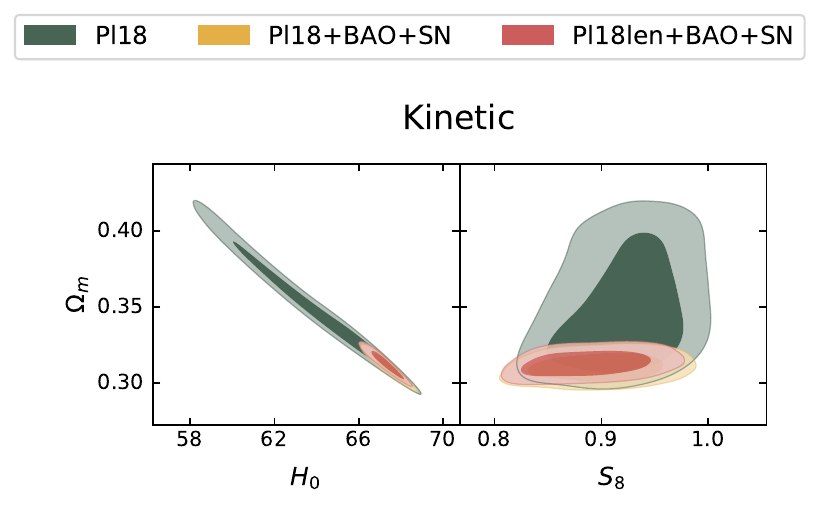}} \\
    \vspace{-5pt}\subfloat{\includegraphics[width=0.67\linewidth]{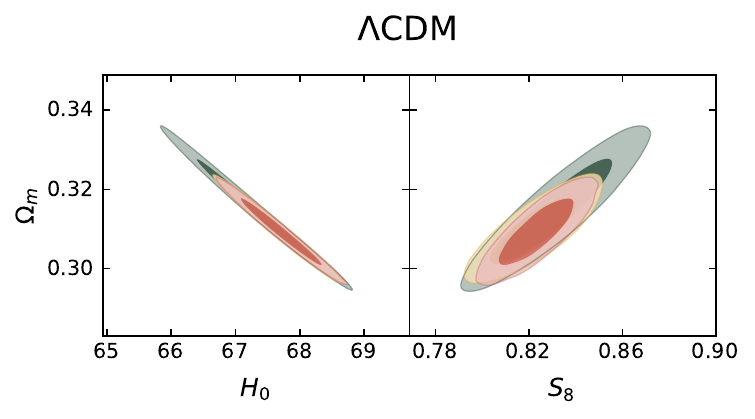}} \\
  \caption[2D marginalised posterior distributions in the $\Lambda$CDM and Kinetic models]{\label{fig:OmH0S8} 68\% and 95\% CL 2D contours derived for the parameter combinations $H_0$-$\Omega_m$ (left panels) and $S_8$-$\Omega_m$ (right panels) in the  Kinetic model (upper panels) and $\Lambda$CDM model (lower panels) for the {\it Planck} 2018 data (green), the {\it Planck} 2018, BAO and SN combination (yellow), and their combination with CMB lensing (red).}
\end{centering}
\end{figure}

In \cref{tab:Kineticbounds,tab:LCDMbounds}, we present the constraints on the sampled parameters for the Kinetic and $\Lambda$CDM models, respectively. These results are illustrated in \cref{fig:OmH0S8} where the corresponding 2D marginalised contour plots for all the considered combinations of data sets are displayed. 
Although the constraints on the cosmological parameters of the Kinetic model are compatible with those of the $\Lambda$CDM case within the uncertainties, the latter yields higher and lower mean values for $H_0$ and $\Omega_m$, respectively, across all the three data combinations considered. This tendency is consistent with the dominating effects of the coupling in the TT power spectrum as depicted in \cref{fig:fig3}, following the analysis provided in \cref{sec:lcdm_param}.
Furthermore, \cref{fig:OmH0S8} depicts contour plots for the constraints in the $H_0-\Omega_m$ and $S_8 - \Omega_m$ planes. 
The Kinetic model under the Pl18 data predicts $H_0 = 64.0^{+3.3}_{-1.8}$, with a lower mean value ($67.31$) than in $\Lambda$CDM, thereby apparently worsening the tension with late-time distance-ladder measurements, as detailed in \cref{sec:cosmotensions}. However, this discrepancy is attenuated by larger error bars ($0.61$ in $\Lambda$CDM), leading to an artificial reduction of the $H_0$ tension from $\sim 4.8 \sigma$ to $\sim 3.3 \sigma$. The compatibility with $\Lambda$CDM is restored when considering other data sets, indicating tensions between the BAO and SN data within this framework, just as was found in \cref{chap:cquint}. A similar situation in a Galileon model \cite{Frusciante:2019puu} suggests a potential bias towards $\Lambda$CDM-like models in the BAO data \cite{Carter:2019ulk}.
With only the Pl18 data, we find $S_8 = 0.921^{+0.044}_{-0.034}$ at a $68\%$ confidence level (CL) in the Kinetic model, with a higher mean value than in the corresponding $\Lambda$CDM case, but also accompanied by larger error bars, with the same tendency persisting across all the data set combinations. Even though a lower value is expected for mitigating the discordance with cosmic shear measurements in the standard model ($S_8 = 0.833 \pm 0.016$), a full reanalysis of the weak lensing data under the Kinetic model would be needed to evaluate the tension. 

\begin{figure}[t!]
      \subfloat{\includegraphics[width=1.\linewidth]{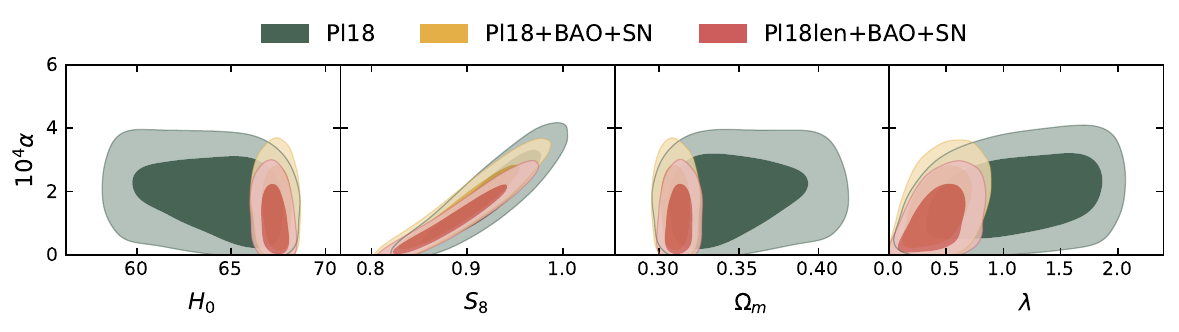}}
  \caption[2D marginalised posterior distributions in the Kinetic model]{\label{fig:const_kinetic} $68\%$ and $95\%$ CL contours obtained in the Kinetic model for the {\it Planck} 2018 data (green), the {\it Planck} 2018, BAO and SN combination (yellow), and their combination with CMB lensing (red).}
\end{figure}

The results for the specific parameters of the Kinetic model can be appreciated in the 2D contour plots of \cref{fig:const_kinetic}, where we see that the coupling parameter $\alpha$ is consistently constrained to be of the order $10^{-4}$, regardless of the data set combination. When considering only the \textit{Planck} data, a higher mean value of $\alpha$ is preferred, primarily to accommodate the TT likelihood better. However, incorporating BAO and SN data slightly reduces the mean value of $\alpha$. Furthermore, adding CMB lensing data shifts the peak of the posterior distribution for $\alpha$ to an even lower central value. This behaviour can be traced to the reported lensing excess by the \textit{Planck} collaboration \cite{Planck:2015mrs,Aghanim:2018eyx}. In the Kinetic model, the lensing power spectrum is always suppressed compared to the $\Lambda$CDM scenario, with higher values of $\alpha$ corresponding to lower amplitudes. Consequently, a lower mean value of $\alpha$ is favoured to match the CMB lensing data better.
Including BAO and SN data leads to narrower constraints on $\Omega_m$, resulting in tighter constraints on other parameters such as $H_0$, $S_8$, and $\lambda$. 

\begin{table*}[ht!]
\begin{center}
\renewcommand{\arraystretch}{1.5}
\resizebox{\textwidth}{!}{
\begin{tabular}{l c c c c c c c c c c c c c c c }
\hline\hline
\textbf{Parameter} & \textbf{ Pl18 } & \textbf{ Pl18 + BAO + SN } & \textbf{ Pl18len + BAO + SN }  \\ 
\hline\hline
$\Omega_b h^2$ & $0.02236\pm 0.00015$ &  $0.02242\pm 0.00013$ &  $0.02243\pm 0.00013$ \\
$\Omega_{cdm} h^2$ &$0.1202\pm 0.0014$ &  $0.11937\pm 0.00091$ &  $0.11935\pm 0.00092$ \\
$100 \theta_s$ &   $1.04188\pm 0.00030$ & $1.04195\pm 0.00028$ &  $1.04196\pm 0.00029$ \\
$\tau_{reio}$ &   $0.0542\pm 0.0078$ & $0.0554^{+0.0071}_{-0.0079}$  &  $0.0573\pm 0.0074$ \\
$n_s$  & $0.9652\pm 0.0044$ & $0.9672\pm 0.0036$  &  $0.9672\pm 0.0037$ \\
$\ln \left( 10^{10} A_s \right)$ & $3.046\pm 0.016$ & $3.046\pm 0.016     $ &  $3.050\pm 0.015$ \\
\hline 
$\sigma_8$ &$0.8115\pm 0.0075$ &   $0.8093\pm 0.0071$ &  $0.8110\pm 0.0061$ \\
$\Omega_m$ & $0.3149\pm 0.0085$ &   $0.3095\pm 0.0054$ &  $0.3093\pm 0.0055$  \\
$S_8$ &  $0.831\pm 0.016$ & $0.822\pm 0.012$ & $0.824\pm 0.011$  \\
$H_0$ &  $67.31\pm 0.61$ &   $67.68\pm 0.40$ &  $67.71\pm 0.42$ \\
\hline 
\hline
\end{tabular}}
\end{center}
\caption[Observational constraints for the $\Lambda$CDM model]{$68\%$ CL bounds on the cosmological parameters for the $\Lambda$CDM model for the three different combinations of data sets: {\it Planck} 2018, {\it Planck} 2018 combined with BAO and SN, and their full combination with CMB lensing.}
\label{tab:LCDMbounds}
\end{table*}

\begin{table*}[ht!]
\begin{center}
\renewcommand{\arraystretch}{1.5}
\resizebox{\textwidth}{!}{
\begin{tabular}{l c c c c c c c c c c c c c c c }
\hline\hline
\textbf{Parameter} & \textbf{ Pl18 } & \textbf{ Pl18 + BAO + SN } & \textbf{ Pl18len + BAO + SN }  \\ 
\hline\hline
$\Omega_b h^2$ & $0.02226\pm 0.00016$ &  $0.02234\pm 0.00015$ &  $0.02236\pm 0.00014$ \\
$\Omega_{cdm} h^2$ &$0.1195\pm 0.0015$ &  $0.1186\pm 0.0011$ &  $0.11911\pm 0.00098$ \\
$100 \theta_s$ &   $1.04196\pm 0.00030$ & $1.04203\pm 0.00028$ &  $1.04200\pm 0.00029$ \\
$\tau_{reio}$ &   $0.0534\pm 0.0079$ & $0.0555\pm 0.0079$  &  $0.0573\pm 0.0074$ \\
$n_s$  & $0.9667\pm 0.0047$ & $0.9688\pm 0.0040$  &  $0.9675\pm 0.0038$ \\
$\ln \left( 10^{10} A_s \right)$ & $3.039\pm 0.017$ & $3.041\pm 0.017$ &  $3.052\pm 0.015$ \\
\hline
$\lambda$                & $1.11\pm 0.48$            & $0.45^{+0.18}_{-0.21}$  & $0.43^{+0.18}_{-0.20}$\\
$10^{4} \alpha$          & $1.88\pm 0.95$     & $1.57^{+0.79}_{-1.00}$ & $1.14^{+0.55}_{-0.92}$ \\
\hline 
$\sigma_8$ &$0.858\pm 0.042$ &   $0.884^{+0.042}_{-0.038}$ &  $0.869^{+0.033}_{-0.043}$ \\
$\Omega_m$ & $0.347^{+0.019}_{-0.037}$ &   $0.3102\pm 0.0065$ &  $0.3128\pm 0.0061$  \\
$S_8$ &  $0.921^{+0.044}_{-0.034}$ & $0.899^{+0.043}_{-0.039}$ & $0.887^{+0.036}_{-0.045}$  \\
$H_0$ &  $64.0^{+3.3}_{-1.8}$ &   $67.41^{+0.59}_{-0.50}$ &  $67.26^{+0.55}_{-0.47}$ \\
\hline \hline 
$\Delta \chi^{2}_{\rm eff} $ & $ -0.9$  & $0.7$ & $1.0$ \\
$\ln B_{K, \Lambda {\rm CDM}}$  & $-4.2$  & $-5.4$ & $-6.8$  \\
\hline 
\hline
\end{tabular}}
\end{center}
\caption[Observational constraints for the Kinetic model]{$68\%$ CL bounds on the cosmological and model parameters for the Kinetic model for the three different combinations of data sets: \textit{Planck}, \textit{Planck} combined with BAO and SN, and their full combination with CMB lensing.}
\label{tab:Kineticbounds}
\end{table*}


\subsection{Model Selection Analysis}

Lastly, we aim to evaluate whether the Kinetic model is favoured over the $\Lambda$CDM case, using different statistical indicators for comparison purposes. First, we consider the effective $\chi^2$-statistics, corresponding to the maximum likelihood, denoted as $\chi_\text{eff}^2$, which allows us to determine how well the Kinetic model fits the data when compared to $\Lambda$CDM. This is accomplished by calculating $\Delta\chi_{\rm eff}^2=\chi^2_{\rm eff, Kinetic}-\chi^2_{\rm eff, \Lambda CDM}$, where a negative value indicates support for the Kinetic model, while a positive value suggests otherwise.
On the other hand, to statistically compare the level of support for one model against the other, we calculate the Bayes factor of the Kinetic model relative to $\Lambda$CDM, as introduced in \cref{eq:evdef}. The greater the evidence for the Kinetic model relative to $\Lambda$CDM, the larger the Bayes factor ratio $\ln B_{K, \Lambda {\rm CDM}} $ will be. The numerical Bayes factor may be translated into a qualitative statement about the strength of evidence for an extended model against $\Lambda$CDM through the Jeffreys scale in \cref{tab:jeff_scale}. 

In the last two rows of \cref{tab:Kineticbounds}, we present the values for both the $\Delta \chi_{\rm eff}^2$ and the $\ln B_{K, \Lambda {\rm CDM}}$, along with the corresponding individual values. From the analysis, we find that when considering only the Pl18 data, the Kinetic model shows a slightly better fit to the data compared to the $\Lambda$CDM model, with a decrease in the chi-squared value of $\Delta \chi^2=-0.9$. However, this preference disappears when other data sets are included. This is mainly due to the BAO and SN data affecting the fit to the temperature-temperature (TT) likelihood, which worsens after incorporating the CMB lensing data. The Kinetic model predicts a suppressed lensing amplitude, whereas the CMB lensing data indicates an excess of power. However, any support for the Kinetic model becomes insignificant when considering the Bayesian evidence, for which we report negative values across all the data sets. This is directly related to the increased complexity of the Kinetic model, including two additional parameters resulting in either reduced or no improvement in the data fitting. Therefore, we can conclude that this analysis does not affirm statistical evidence for the Kinetic model.

\section{Discussion} \label{sec:kin_dis}

This study investigated a Kinetic model consisting of a coupled quintessence theory with a power-law kinetic interaction. We explored its impact on the evolution of the background and linear perturbations in the Universe. We analysed the relevant cosmological observables and derived constraints on the model parameters using various data sets, namely CMB, CMB lensing, BAO, and SN data.

We first performed a numerical study of the model's predictions to assess the valid parameter space with physical interest. We used a modified version of the publicly available Einstein-Boltzmann code \texttt{CLASS}.
At the background level, we found that a non-zero value of $\alpha$ allows for a scaling regime during the radiation-dominated epoch, where the ratio of the densities of cold dark matter and the scalar field approximately scale with $\alpha$. We identified that only energy transfer from DM to DE can be realised in such scenarios. Additionally, the radiation-matter equality is shifted towards earlier times, with direct consequences on the matter power spectrum, namely a suppression of small-scale power and an enhancement of large-scale growth compared to the $\Lambda$CDM model. Consequently, we observed an overall suppression of the lensing potential and modifications in the ISW effect, which alter the shape of the temperature power spectrum of anisotropies at large angular scales.

Using an MCMC method for cosmological parameter extraction, we applied the theoretical insight gained in the numerical study to constrain the model. We found that even though there is an apparent reduction of the $H_0$ tension with the \textit{Planck} data only, this is due to the worse constraining power (larger error bars), and not an actual higher mean value of $H_0$. Regardless of the combination of data sets, the parameter $\alpha$ was consistently constrained to be of the order of $10^{-4}$. We also reported constraints on the other parameter of the Kinetic model, $\lambda$, with the most robust bounds obtained when combining BAO and SN data. This is primarily caused by the solid constraining power of BAO data on $\Omega_m$, which indirectly affects the bounds on $\lambda$. Finally, we presented a model selection study using the effective $\chi_{\rm eff}^2$ and the Bayesian evidence, with the latter indicating a statistical preference for $\Lambda$CDM over the Kinetic toy-model related to the increased dimension of the parameter space.
In conclusion, we highlight the importance of considering the Kinetic model and other variations for future investigations, especially with the availability of new probes from upcoming surveys. Further studies using high-precision data will help resolve tensions and establish a definitive preference for one model, possibly hinting at more complicated shapes for the kinetic function.

  \cleardoublepage

    \chapter{Forecasts on Coupled Quintessence with Standard Sirens} \label{chap:gwcons}
 \setcounter{equation}{0}
\setcounter{figure}{0}


 \epigraph{Esse não-saber pode parecer ruim mas não é tanto porque ela sabia muita coisa \\ assim como ninguém ensina cachorro a abanar o rabo e nem a pessoa a sentir fome; \\ nasce-se e fica-se logo sabendo. Assim como ninguém lhe ensinaria um dia a morrer: \\ na certa morreria um dia como se antes tivesse estudado de cor a representação \\ do papel de estrela. Pois na hora da morte a pessoa se torna brilhante estrela de cinema,\\ é o instante de glória de cada um e é quando como no canto coral se ouvem agudos sibilantes.\blfootnote{\textit{That not-knowing might seem awful but it’s not that bad because she knew lots of things in the way nobody teaches a dog to wag his tail or a person to feel hungry; you’re born and you just know. Just as nobody one day would teach her how to die: yet she’d surely die one day as if she’d learned the starring role by heart. For at the hour of death a person becomes a shining movie star, it’s everyone’s moment of glory and it’s when as in choral chanting you hear the whooshing shrieks.} --- \textsc{Clarice Lispector} in The Hour of the Star} \\ --- \textsc{Clarice Lispector}\ \small\textup{A Hora da Estrela}}

In the past few years, we have witnessed the rise of gravitational wave (GW) astronomy as a new independent probe of gravitational effects \cite{Holz:2005df}. An accurate redshift-luminosity relation can be constructed when GW events are combined with an electromagnetic (EM) counterpart multi-messenger signal. In 2015, the Laser Interferometer Gravitational-Wave Observatory (LIGO) and Virgo collaborations made the first direct detection of a gravitational wave signal from the inspiral, merger, and ringdown of a binary black hole system (GW150914) \cite{grav}. This inaugural measurement marked only the beginning, with the Virgo-LIGO catalogue of events growing ever since. Two years later, one of these events (GW170817) opened up a new door for multi-messenger astronomy by being matched with an electromagnetic counterpart for a binary neutron star merger event (GRB170817A), detected by the International Gamma-ray Astrophysics Laboratory (INTEGRAL)-Fermi collaboration \cite{LIGOScientific:2017vwq,LIGOScientific:2017zic}. Akin to BAO and SN data, conventionally used as standard rulers and candles, gravitational wave events were now established as \textit{standard sirens} \cite{Schutz:1986gp}. This single combined detection had a strong impact on the allowed modifications to the gravitational interaction by ruling out many proposals \cite{Creminelli:2017sry,Baker:2017hug,Ezquiaga:2017ekz,Creminelli:2018xsv,amendola:2017orw} with many other models further constrained \cite{LISACosmologyWorkingGroup:2019mwx,Belgacem:2018lbp,Allahyari:2021enz,Califano:2022syd,Ferreira:2022jcd}.  
The significance of these observations for cosmology is twofold. First, GW150914 confirmed the existence of gravitational waves, as predicted by GR. Second, GW170817 established that the tensor degrees of freedom seemingly propagate at the speed of light \cite{LIGOScientific:2017zic}, imposing severe constraints on modified gravity theories. Furthermore, the latter is why GW events are standard sirens for independent measurement of the Hubble parameter $H_0$ \cite{LIGOScientific:2017adf}. Although this method does not yet rival the traditional distance ladder measurements \cite{Chen:2017rfc}, it shows how future gravitational wave telescopes could be instrumental in addressing the present cosmological tensions. 

In this chapter, we wish to investigate the constraining power of combined distance-redshift measurements based on geometric, electromagnetic and gravitational sources on the classes of coupled quintessence models discussed throughout this thesis. These forecasts are conducted by simulating mock catalogues of gravitational wave standard sirens (SS) from next-generation detectors, namely the Einstein Telescope (ET) and the Laser Interferometer Space Antenna (LISA), along with current data from Type Ia supernovae (SN), and baryon acoustic oscillations (BAO). 
This study aims to provide insight into the ability of upcoming SS missions to constrain extended theories of gravity while simultaneously offering supplementary constraints on $H_0$, potentially shedding light on the cosmic tensions. 
We analyse four distinct models, each characterised by particular coupling functions between dark matter and dark energy, arising from a non-universal metric transformation, as discussed in \cref{sec:coupde}. More specifically, these are a conformally coupled quintessence model, featuring a constant coupling stemming from a conformal function, which is an exponential function of the scalar field (as explored in \cref{chap:cquint}); a kinetic model, with the conformal function as a power law of the kinetic term of the scalar field (as investigated in \cref{chap:kin}); a purely disformally coupled quintessence field, with a constant disformal function; and a mixed disformally coupled quintessence, extending the previous model with an exponential conformal coupling.

This chapter is organised as follows. We start by giving a brief introduction to the physics of standard sirens in \cref{sec:cq_ss_int}. The supplementary details on the simulation of the standard siren events developed for this study are provided in \cref{app:mock}. \cref{sec:cq_ss_method} provides an overview of the methodology used as well as a brief account of the data set combinations considered. We outline the criteria for particular catalogue choices, and discuss the sampling method employed for the forecasts. In \cref{sec:cq_ss_res} we  introduce each of the four models under study and present the results of our analysis, emphasising their significant implications. Lastly, in \cref{sec:cq_ss_sum}, we summarise our results and outline our concluding thoughts and future prospects.

This work has been published in Physical Review D and is available in \cite{Teixeira:2023zjt}.

\section{Standard Sirens} \label{sec:cq_ss_int}

Current GW detectors, (advanced) Virgo \cite{VIRGO:2014yos}, (advanced) LIGO \cite{LIGOScientific:2014pky}  and the Kamioka Gravitational Wave Detector (KAGRA) \cite{Somiya:2011np},
 are second generation (2G) ground-based detectors, with another one under planning (2030), the Indian Initiative in Gravitational-wave Observations (IndIGO) \cite{IndIGO}.
The increasing number of detectors will boost the capabilities of GW astronomy both in the number of confirmed events (a larger volume of the Universe is covered) and sky localisation (a better triangulation of the source), which will also aid in the search for a counterpart.  
However, 2G detectors are limited in their sensitivity and future ground-based detectors are designed to become more sensitive, precise and capable of probing a larger range of frequencies. Special emphasis should be given to the \ac{et}, a proposed ground-based triangular interferometer designated as a third-generation gravitational wave detector, which is expected to improve the current sensitivity by a factor of 10 \cite{Punturo:2010zz}. ET will also extend the redshift range, \textit{e.g.} $z \sim 5$ for binary black-holes (BBHs) compared to $z\sim 0.5$ for 2G detectors \cite{Sathyaprakash:2012jk}. The number of detectable multi-messenger events is expected to reach tens of thousands of standard sirens \cite{Maggiore:2019uih}. 
Contemporary terrestrial detectors, such as LIGO and Virgo, grapple with limitations at low frequencies due to seismic and thermal noise. In contrast, the ET will likely alleviate these through underground positioning and cryogenic cooling of its interferometer mirrors.
These enhancements will empower the ET and other third-generation detectors to make incredibly sensitive gravitational wave measurements, propelling standard siren detections into precision cosmology's forefront \cite{Hild:2010id}.

While these ground-based detectors will cover a frequency band in the range $1 \lesssim f\lesssim 10^3$ Hz \cite{Cai:2016sby}, the upcoming space-based 3G detectors, such as the \ac{lisa} \cite{LISA:2017pwj} will have a  peak sensitivity near $10^{-3}$ Hz and will be able to detect GW events beyond $z=20$, probing a wide range of targets. There are many proposals of 3G GW observatories, such as DECi-hertz Interferometer Gravitational wave Observatory (DECIGO) \cite{Kawamura:2011zz}. However, we have opted to focus our analysis on ET and LISA covering ground and space-based experiments. 


\section{Methodology and Data Sets}\label{sec:cq_ss_method}

Given the main objective of this study, we create simulated data that forecasts the potential future observations of standard siren events. Specifically, we focus on those that could be detected by ET and LISA. Below, we provide a concise overview of the samples we have generated along with the methodology and the data combinations used. The detailed account of the simulations of the mock data is found in \cref{app:mock}.


To assess the fit of the mock data to the coupled quintessence models explored in this study, we use the Markov Chain Monte Carlo (MCMC) technique, as described in \cref{sec:stat}. We generate the samples through our private modified branch of the Einstein-Boltzmann code \texttt{CLASS}, interfaced with the \texttt{MontePython} sampler \cite{Audren_2013, Brinckmann:2018cvx}. Moreover, we opt for the Nested Sampling algorithm implemented in MultiNest\footnote{\href{https://github.com/farhanferoz/MultiNest}{https://github.com/farhanferoz/MultiNest}} \cite{Feroz:2007kg,Feroz:2008xx,Feroz:2013hea} and PyMultiNest\footnote{\href{https://github.com/JohannesBuchner/PyMultiNest}{https://github.com/JohannesBuchner/PyMultiNest}} \cite{Buchner:2014nha}, instead of the conventional Metropolis-Hastings algorithm. This choice is due to the latter's limitations in dealing with complex degeneracies between parameters and handling multi-modal distributions, as discussed in \cref{sec:cq_ss_res}. For this reason, we recommend caution when employing more straightforward methods like Metropolis-Hastings and emphasise the importance of considering multiple sampling strategies. Additionally, this underscores the value of relying on various types of observations, which can break the degeneracy. The MCMC chains are then analysed and processed using the \texttt{GetDist}\footnote{\href{https://github.com/cmbant/getdist}{https://github.com/cmbant/getdist}} Python package \cite{Lewis:2019xzd}.

We integrate a new likelihood module into \texttt{MontePython} to assess the constraints imposed by the upcoming SS surveys. The luminosity distances for gravitational waves at each sampled point in the parameter space, $d_{SS} (z)$, are computed from \texttt{CLASS} according to \cref{eq:lumdist}.
More precisely, we constructed a likelihood function for our simulated dataset of SS events based on an effective Gaussian distribution:

\begin{equation}
    \ln \mathcal{L}_{SS} = - \frac{1}{2} \sum_{i=1}^n \left[ \frac{d_{SS}^{\text{(obs)}} (z_i) - d_{SS} (z_i)}{\sigma_{d_{L,i}}} \right]^2 \mathcomma
\end{equation}

where $d_{SS}^{\text{(obs)}} (z)$ is the observed luminosity distance, which in this case corresponds to the simulated data, with total error $\sigma_{d_{L}}$ as described in \cref{app:mock}, for $n$ observed events.

In summary, we investigate the potential of constraining coupled quintessence models with upcoming standard siren data probed by ET and LISA, either as a substitute for \textit{current} background data or in conjunction with them.
This allows for a direct assessment of the origin of the improvement in the constraints on $\{\Omega_m, H_0 \}$ and the model-specific parameters affecting the background evolution. In particular, we include BAO data from the Sloan Digital Sky Survey (SDSS) DR7 Main Galaxy Sample \cite{Ross:2014qpa}, SDSS DR12 consensus release \cite{BOSS:2016hvq} and the 6dF Galaxy Survey \cite{Beutler:2011hx}, in combination with distance \textit{moduli} measurements of 1048 SN data from Pantheon \cite{Scolnic:2017caz}. This combined data set is referred to throughout this chapter as SN+BAO.
Despite our focus on extended theories of gravity, we adopt flat $\Lambda \text{CDM}$ fiducial values for the parameters specified in \cref{sec:lcdm_param}, intentionally overlooking any gravity modifications, accounted for by the model parameters, for which we consider as fiducial value their $\Lambda$CDM limit. The models discussed in \cref{sec:cq_ss_res} reduce to $\Lambda$CDM in the following limits: $\lambda=0$ and $\beta=0$ for \cref{sec:cq_ss_casea}; $\lambda=0$ and $\alpha=0$ for \cref{sec:cq_ss_caseb}; $\lambda=0$ and $D_0=0$ for \cref{sec:cq_ss_casec}; $\lambda=0$, $\beta=0$ and $D_0=0$ for \cref{sec:cq_ss_cased}. We adopt flat priors for all the parameters, as detailed in \cref{tab:cq_gw_priorsgw}.

\begin{table}[ht!]
\begin{center}
\begin{tabular}{|c|c|c|}
\hline
Model & Parameter                    & Prior \\
\hline \hline
\multirow{4}{*}{All} & $\Omega_b h^2$                & $[0.018,0.03]$ \\
& $\Omega_c h^2$                & $[0.1,0.2]$ \\
& $h$                    & $[0.6,0.8]$ \\
& $\lambda$                           & $[0,2]$ \\
\hline \hline
\cref{sec:cq_ss_casea,sec:cq_ss_cased} & $\beta$                            & $[0,2]$ \\
\hline
\cref{sec:cq_ss_caseb} & $\alpha$                            & $[0,0.001]$ \\
\hline
\cref{sec:cq_ss_casec,sec:cq_ss_cased} & $D_0/\text{meV}^{-1}$                            & $[0,2]$ \\
\hline %
\end{tabular}
\end{center}
\caption[Priors on cosmological models]{Flat priors on the cosmological and model parameters sampled in \cref{sec:cq_ss_res}.}
\label{tab:cq_gw_priorsgw}
\end{table}

\section{Forecasts Results}\label{sec:cq_ss_res}

Following the methodology and data sets elaborated in \cref{sec:cq_ss_method}, we explore the predicted power of standard sirens probed by LISA and ET on constraining the cosmological parameters $\{\Omega_m,H_0\}$ and the model-specific conformal and disformal coupling parameters and the steepness of the self-interacting potential. Specifically, we discuss four distinct coupled quintessence models: a canonical coupled quintessence model in \cref{sec:cq_ss_casea}, as introduced in \cref{sec:cq_m1_flat}; a model with a kinetic coupling in \cref{sec:cq_ss_caseb}, previously explored in \cref{chap:kin}; a constant disformal model in \cref{sec:cq_ss_casec} and a hybrid conformal-disformal model in \cref{sec:cq_ss_cased}, both particular cases of the disformally coupled quintessence model \cite{vandeBruck:2015ida,Mifsud:2017fsy}. We provide a brief account of the theoretical framework of each scenario before presenting the forecast results, derived according to the theoretical and numerical specifications discussed in \cref{sec:cq_ss_method} and in \cref{app:mock}.

Each subsection also contains the derived 1D and 2D marginalised posterior distributions and contours for $\{H_0,\Omega_m\}$ and any of the particular model-dependent parameters of the model in question, as specified in \cref{tab:cq_gw_priorsgw}. These plots comprise both ET and LISA data, individually and combined, and feature an overlay of SN+BAO data and standalone SN+BAO results for reference. We provide a summary of the results in a table with the corresponding $1\sigma$ values, denoted in the text as $\{\sigma_{\text{p}}\}$, where $p$ is the corresponding model parameters. Additionally, we define
$\mathcal{F}_{\text{p}}^{(\text{i,j})} = \{\sigma_{\text{p}}^{(\text{j})}/\sigma_{\text{p}}^{(\text{i})}\} $, where $i$ and $j$ represent distinct data sets, to denote the effective change in the parameter $p$.

\subsection{Conformal Coupling} \label{sec:cq_ss_casea}

\begin{table*}[ht!]
\begin{adjustbox}{width=1\textwidth}
\begin{tabular}{|l|l|l|l|l|l|l|l|l|}
\hline
\multicolumn{9}{|c|}{Conformal Coupled Quintessence}  \\ \hline\hline
 Data sets &  $\Omega_m$ & $\sigma_{\Omega_m}$ & $H_0$ & $\sigma_{H_0}$ & $\beta$ & $\sigma_{\beta}$ & $\lambda$ & $\sigma_{\lambda}$\\ \hline \hline
 SN+BAO & $0.3019^{+0.0088}_{-0.0059}$ & $0.0074$ & $73.2^{+4.7}_{-3.5}$ & $4.1$ & $0.085^{+0.055}_{-0.043}$ & $0.049$ &  $0.42^{+0.20}_{-0.36}$ & $0.28$\\ 
 \hline\hline
ET & $0.307^{+0.011}_{-0.0050}$ & $0.0080$ & $67.49^{+0.39}_{-0.34}$ & $0.37$ & $0.115^{+0.060}_{-0.079}$ & $0.070$ & $0.50^{+0.26}_{-0.38}$ & $0.32$ \\
  ET+SN+BAO & $0.3046^{+0.0099}_{-0.0051}$ & $0.0075$ & $67.37\pm 0.36$ & $0.36$ & $0.063^{+0.033}_{-0.045}$ & $0.039$ & $0.49^{+0.26}_{-0.35}$ & $0.31$ \\
\hline
\hline 
   LISA  & $0.3039^{+0.0093}_{-0.0049}$ & $0.0071$ & $67.50^{+0.50}_{-0.44}$ & $0.47$ & $0.167^{+0.085}_{-0.11}$ & $0.098$ & $0.33^{+0.15}_{-0.32}$ & $0.24$ \\
   LISA+SN+BAO & $0.3028^{+0.0065}_{-0.0036}$ & $0.0051$ & $67.52\pm 0.37$ & $0.37$ & $0.048^{+0.025}_{-0.037}$ & $0.031$ & $0.33^{+0.15}_{-0.29}$ &  $0.22$ \\
  \hline
  \hline 
   ET+LISA  & $0.3079^{+0.0061}_{-0.0034}$ & $0.0048$ & $67.56\pm 0.26$ & $0.26$ & $0.178^{+0.099}_{-0.081}$ & $0.090$ & $0.30^{+0.13}_{-0.27}$ & $0.24$ \\
   ET+LISA+SN+BAO & $0.3044^{+0.0063}_{-0.0032}$ & $0.0048$ & $67.45\pm 0.28$ & $0.28$ & $0.052^{+0.028}_{-0.038}$ & $0.033$ & $0.35^{+0.17}_{-0.30}$ &  $0.24$ \\
  \hline
\end{tabular}
\end{adjustbox}
\caption[Observational constraints for conformal coupled quintessence]{Marginalised constraints on cosmological and model parameters for the Conformal Coupled Quintessence model at 68\% CL.}
\label{tab:cq_gw_boundscq}
\end{table*}

The first scenario we explore is the conformal coupling model introduced in \cref{sec:cq_m1_flat}. We recall that this model is characterised by the following conformal and potential functions $ C(\phi) $ and $ V(\phi) $:
\begin{equation}
    C(\phi) = e^{\frac{2 \beta \phi}{M_{\text{Pl}}}}, \qquad V(\phi) = V_0 e^{-\frac{\lambda \phi}{M_{\text{Pl}}}} \mathcomma
\end{equation}
with $ C(\phi) $ laying out the coupling in the dark sector according to \cref{eq:einconf}. The parameters $ \beta $ and $ \lambda $ are dimensionless constants, while $ V_0 $ is a constant with dimensions of (mass)$^4$ representing the energy scale of the potential\footnote{To avert degeneracies and for numerical robustness, $ V_0 $ is not a free parameter but rather serves as a shooting parameter to satisfy the flatness condition for $ \Omega_{\phi}^0 $.}. 

In these models, the mass of the dark matter particles becomes dependent on $ \phi $, as the DE field mediates a long-range force between DM particles. This translates into an effective gravitational coupling expressed as $ G_{\text{eff}} = G (1 + 2 \beta^2) $ \cite{Wetterich:1994bg,Farrar:2003uw,Amendola:2003wa}. For this analysis, we are mainly concerned with the slope of the potential $\lambda $ and the coupling constant $ \beta $ as free parameters. Constraints on this model have been previously reported in Ref.~\cite{vandeBruck:2016hpz} using only background data ($ H(z) $, BAO, and supernova Union 2.1), leading to upper limits $ \beta < 0.193 $ and $ \lambda < 1.27 $. Further stringent limits, $ \beta < 0.0298 $ and $ \lambda < 0.6 $ at $1\sigma$, have been derived in Ref.~\cite{vandeBruck:2017idm} using \textit{Planck}, BAO, and SN data, also in agreement with Ref.~\cite{Gomez-Valent:2020mqn} .

\begin{figure}[t!]
      \subfloat{\includegraphics[width=\linewidth]{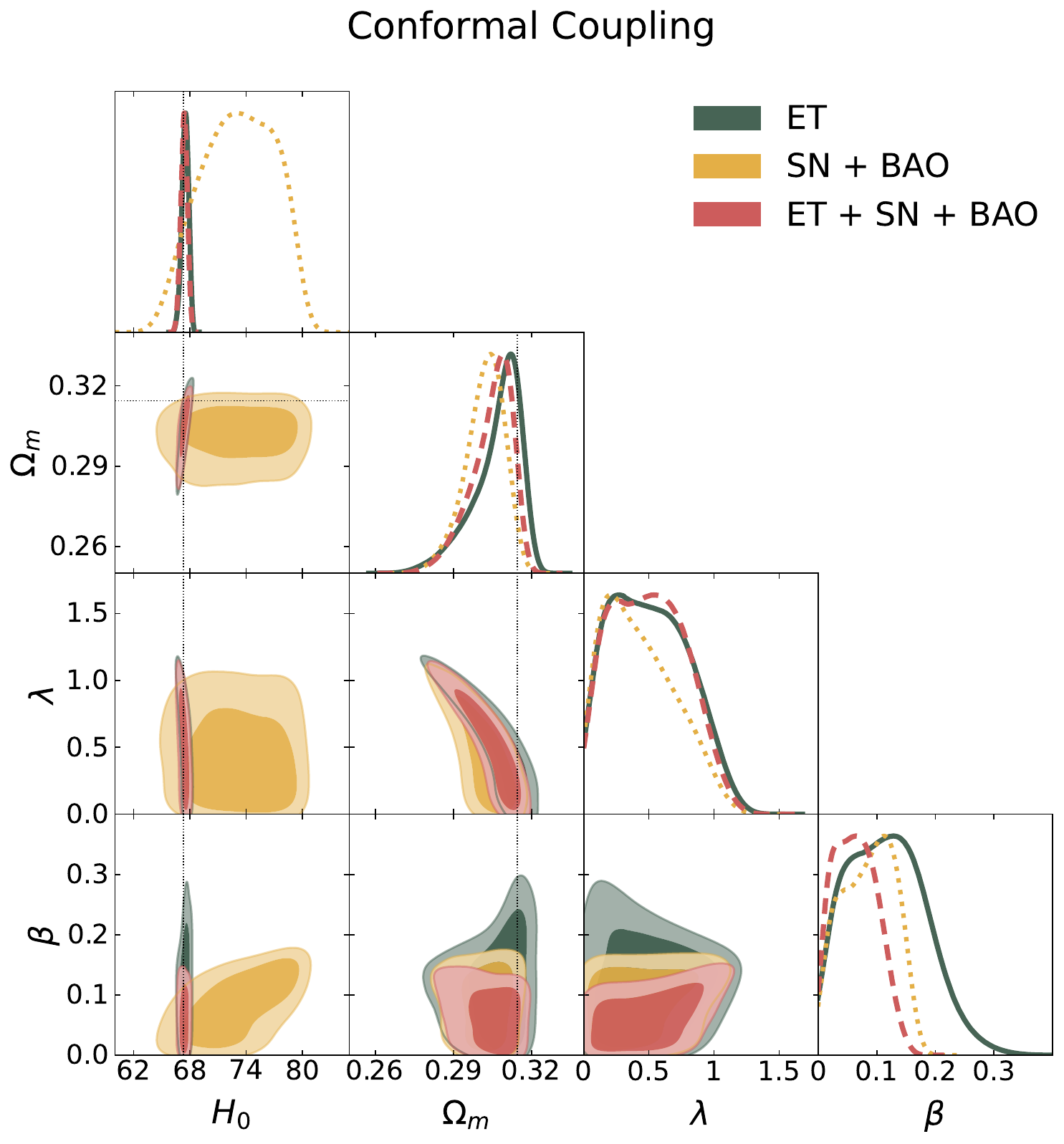}} 
  \caption[Triangle plot of 2D marginalised posterior distributions for conformal coupled quintessence with ET]{\label{fig:etCQ} 68\% and 95\% CL 2D contours and 1D marginalised posterior distributions for the parameters $\{H_0,\Omega_m, \lambda,\beta\}$ in the conformal coupled quintessence model with the ET mock data (green filled line), SN+BAO data (yellow dotted line) and their combination (red dashed line). The dotted lines depict the fiducial values for the mock data $\{\Omega_m,H_0 \} = \{0.3144, 67.32\}$.}
\end{figure}

\begin{figure}[t!]
      \subfloat{\includegraphics[width=\linewidth]{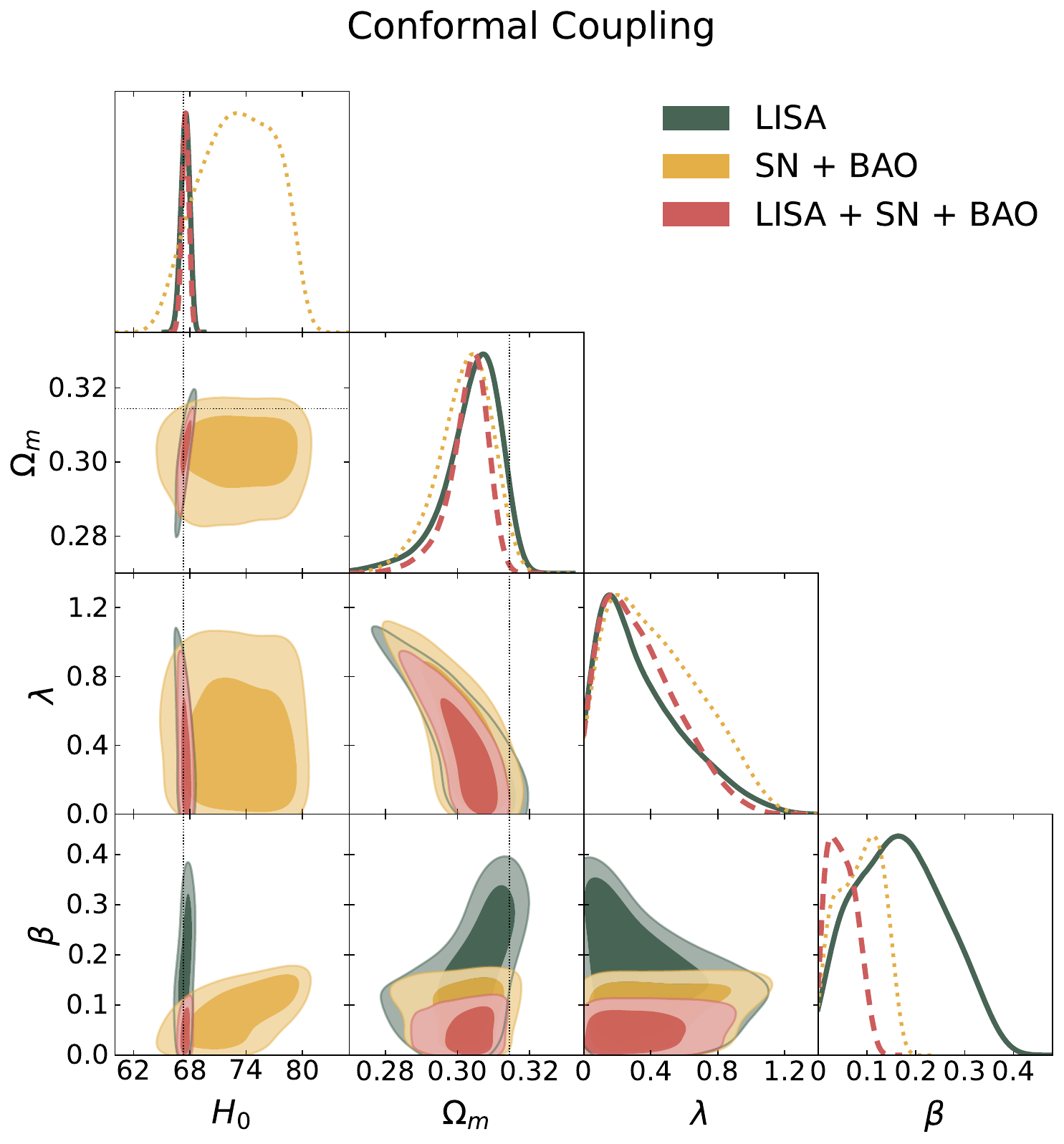}} 
  \caption[Triangle plot of 2D marginalised posterior distributions for conformal coupled quintessence with LISA]{\label{fig:lpCQ} 68\% and 95\% CL 2D contours and 1D marginalised posterior distributions for the parameters $\{H_0,\Omega_m, \lambda,\beta\}$ in the conformal coupled quintessence model with LISA mock data (green filled line), SN+BAO data (yellow dotted line) and their combination (red dashed line). The scale is the same as in \cref{fig:etCQ} for comparison purposes, with the SN+BAO contours standing as the reference. The dotted lines depict the fiducial values for the mock data $\{\Omega_m,H_0 \} = \{0.3144, 67.32\}$.}
\end{figure}

\begin{figure}[t!]
      \subfloat{\includegraphics[width=\linewidth]{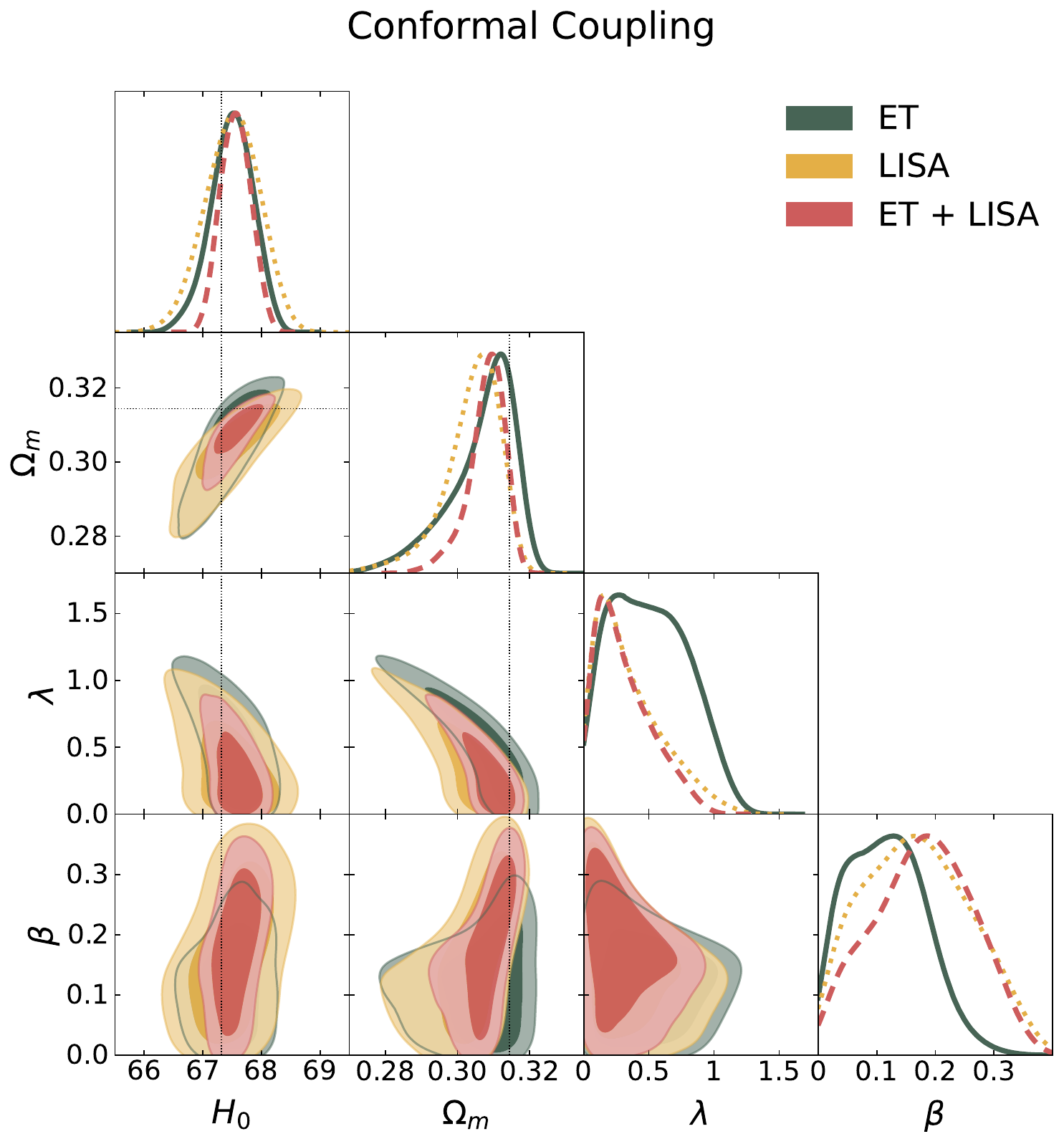}} 
  \caption[Triangle plot of 2D marginalised posterior distributions for conformal coupled quintessence with ET+LISA]{\label{fig:etlp2CQ} 68\% and 95\% CL 2D contours and 1D marginalised posterior distributions for the parameters $\{H_0,\Omega_m, \lambda,\beta\}$ in the conformal coupled quintessence model with ET mock data (green filled line), LISA mock data (yellow dotted line) and their combination (red dashed line). The dotted lines depict the fiducial values for the mock data $\{\Omega_m,H_0 \} = \{0.3144, 67.32\}$.}
\end{figure}

Based on the results illustrated in \cref{fig:etCQ,fig:lpCQ,fig:etlp2CQ} and summarised in \cref{tab:cq_gw_boundscq}, we discuss the parameter constraints from gravitational wave (GW) observations in comparison to SN+BAO data sets for the parameters $\{\Omega_m,H_0,\beta, \lambda\}$. Specifically, with ET standard sirens, the set of parameters is constrained at $ 1\sigma $ with accuracy $ \{0.0080, 0.37, 0.0070, 0.32\} $ for ET and $ \{0.0075, 0.36, 0.039, 0.31\} $ for ET+SN+BAO. This results in a change in error quantified by $ \mathcal{F}_{\Omega_m, H_0,\beta,\lambda}^{(\text{ET,ET+SN+BAO})} = \{0.94, 0.97, 0.56, 0.97\} $. Thus, adding the background data, compared to ET alone, increases accuracy in all parameters shown by the reduction in $\sigma$.
A similar trend is present for the LISA data set, with $ 1\sigma $ regions of $ \{0.0071, 0.47, 0.098, 0.24\} $, while for LISA+SN+BAO, they become $ \{0.0051, 0.37, 0.031, 0.22\} $. This leads to an overall error reduction given by $ \mathcal{F}_{\Omega_m, H_0,\beta,\lambda}^{(\text{LISA,LISA+SN+BAO})} = \{0.72, 0.79, 0.32, 0.92\} $.

For the standalone SN+BAO data set, the accuracy is $ \{0.0074, 4.1, 0.049, 0.28\} $. When compared to the data sets, including SS data, ET+SN+BAO and LISA+SN+BAO, a reduction in $ \sigma $ is observed for all the parameters except for $ \sigma_{\Omega_m} $, which nominally increases in the ET+SN+BAO case.

When evaluating the error performance of ET and LISA in comparison to SN+BAO, we report only minor variations in the constraining capability for $ \Omega_m $. Specifically, ET shows a slight degradation, while LISA demonstrates marginal improvement. A parallel trend occurs for the parameter $ \lambda $, where ET's performance is nominally worse, whereas LISA's is better than with the background data. Of particular note is the substantial reduction in $ \sigma_{H_0} $ when comparing ET and LISA against SN+BAO. This results in an error reduction by factors of $ \mathcal{F}_{H_0}^{(\text{SN+BAO,ET})} = 0.090 $ and $ \mathcal{F}_{H_0}^{(\text{SN+BAO,LISA})} = 0.11 $. Such forecasts suggest that GWs will be instrumental in addressing the Hubble tension. Conversely, the parameter $ \beta $ experiences an increase in error, quantified by $ F_{\beta}^{(\text{SN+BAO,ET})} = 1.4 $ and $ \mathcal{F}_{\beta}^{(\text{SN+BAO,LISA})} = 2.0 $. However, this degradation is alleviated when ET and LISA are combined with the background data, improving to $ F_{\beta}^{(\text{ET,ET+SN+BAO})} = 0.56 $ and $ \mathcal{F}_{\beta}^{(\text{LISA,LISA+SN+BAO})} = 0.32 $.

The results shown in \cref{fig:etlp2CQ} show that ET and LISA offer comparable constraining power over the cosmological parameters. A noteworthy feature observed is ET's superior constraining ability for $ H_0 $, ascribed to its abundance of low-redshift data points as depicted in \cref{fig:ETLISAmock}. 

By bringing the GW data from both LISA and ET together, which implies more data points also covering a broader redshift spectrum, we predict an improved constraining ability on $ \{H_0, \Omega_m\} $ relative to SN+BAO. Specifically, $ \mathcal{F}_{\Omega_m, H_0}^{(\text{SN+BAO,ET+LISA})} = \{0.65, 0.063\} $. For the model-specific parameters $ \beta $ and $ \lambda $, the constraints are modified to $ \mathcal{F}_{\beta, \lambda}^{(\text{SN+BAO,ET+LISA})} = \{1.8, 0.86\} $. By combining  ET+LISA with SN+BAO, only negligible changes in the constraining power for $ \Omega_m, H_0, \lambda $ arise. Only $ \beta $ shows a significant tightening in its constraint, reducing the $ 1\sigma $ error by nearly one-third.

In the context of our analysis, we find improvements in the $ 1\sigma $ upper limits of the model parameters when compared to the existing background constraints stated earlier for the following cases: $ \beta < 0.14 $ and $ \lambda < 0.62 $ (SN+BAO); $ \beta < 0.175 $ and $ \lambda < 0.76 $ (ET); $ \beta < 0.096 $ and $ \lambda < 0.75 $ (ET+SN+BAO); $ \lambda < 0.48 $ (both LISA and LISA+SN+BAO) and $ \beta < 0.073 $ (LISA+SN+BAO); $ \lambda < 0.43 $ (ET+LISA); $ \lambda < 0.52 $ and $ \beta < 0.08 $ (ET+LISA+SN+BAO).

\subsection{Kinetic Conformal Coupling} \label{sec:cq_ss_caseb}

In a more intricate example of a coupled quintessence model, we turn our attention to a conformal function that depends on the kinetic term of the scalar field $ \phi $, expressed as $ X = - \frac{\partial_{\mu} \phi \partial^{\mu} \phi}{2} $. This is called the kinetically coupled model, proposed in \cite{Barros:2019rdv} (more details in the references cited therein). Our discussion will focus on the specific instance of a power law form for $C(X)$, as explored in \cref{chap:kin}, based on Ref.~\cite{Teixeira:2022sjr}. Although originating from a Lagrangian formulation $ \mathcal{L}_{\text{DM}} \rightarrow \left( \frac{X}{{\text{M}_{\text{Pl}}}^4} \right)^{\alpha} \mathcal{L}_{\text{DM}} $, the model at the background level is equivalent to a kinetic-dependent conformal transformation $ \bar{g}_{\mu\nu} = C(X) g_{\mu\nu} $, where:

\begin{equation}
  C(X) = \left( \frac{X}{{\text{M}_{\text{Pl}}}^{4}} \right)^{2\alpha}, \quad \text{and} \quad V(\phi) = V_0 e^{- \frac{\lambda \phi}{\text{M}_{\text{Pl}}}} \mathperiod
\end{equation}

Here, $ \alpha $ is a dimensionless constant, and the assumption of a simple exponential potential remains in line with the previous case, implying similar considerations for $ \lambda $ and $ V_0 $.

In summary, an investigation based on the \textit{Planck} likelihood and SN+BAO background data, as outlined in Ref.~\cite{Teixeira:2022sjr}, reveals the substantial influence of BAO data in constraining $ \Omega_m $, intrinsically correlated to the steepness of the potential $ \lambda $. The coupling constant $ \alpha $ is constrained to the order $ 10^{-4} $. The results derived for the cosmological parameters were compatible with those of $ \Lambda $CDM within the error margins. Additionally, we highlight a direct correlation between $ H_0 $ and $ \Omega_m $. Initially ascribed to the behaviour of linear perturbations when $ \alpha $ is non-zero, this correlation persists even for the standard siren background data sets.

\begin{table*}[ht!]
\centering
\begin{adjustbox}{width=1\textwidth}
\begin{tabular}{|l|l|l|l|l|l|l|l|l|}
\hline
\multicolumn{9}{|c|}{Kinetic Coupled Quintessence}  \\ \hline\hline
 Data sets &  $\Omega_m$ & $\sigma_{\Omega_m}$ & $H_0$ & $\sigma_{H_0}$ & $10^4 \alpha$ & $\sigma_{10^4 \alpha}$ & $\lambda$ & $\sigma_{\lambda}$ \\ \hline \hline
 SN+BAO & $0.3016^{+0.0075}_{-0.0057}$ & $0.0066$ & $70.4\pm 3.1$ & $3.1$ & $5.1\pm 2.9$ & $2.9$ &  $0.34^{+0.16}_{-0.29}$ & $0.23$ \\ 
\hline
\hline
ET & $0.3067^{+0.0093}_{-0.0046}$ & $0.0070$ & $67.45\pm 0.36$ & $0.36$ & $4.8\pm 2.9$ & $2.9$ & $0.41^{+0.20}_{-0.31}$ & $0.26$ \\
  ET+SN+BAO & $0.3062^{+0.0074}_{-0.0043}$ & $0.0059$ & $67.36\pm 0.33$ & $0.33$ & $5.0\pm 2.9$ & $2.9$ & $0.37^{+0.19}_{-0.28}$ & $0.24$ \\
  \hline
\hline
   LISA  & $0.2997^{+0.0079}_{-0.0041}$ & $0.0060$ & $67.30\pm 0.39$ & $0.39$ & $4.9\pm 2.9$ & $2.9$ & $0.34^{+0.16}_{-0.30}$ & $0.23$ \\
   LISA+SN+BAO & $0.3024^{+0.0058}_{-0.0035}$ & $0.0047$ & $67.47\pm 0.36$ & $0.36$ & $5.0\pm 2.9$ & $2.9$ &  $0.29^{+0.13}_{-0.26}$ & $0.20$ \\
  \hline
  \hline
   ET+LISA & $0.3040^{+0.0058}_{-0.0031}$ & $0.0045$ & $67.42\pm 0.26$ & $0.26$ & $5.1\pm 2.9$ & $2.9$ & $0.31^{+0.15}_{-0.26}$ & $0.21$ \\
   ET+LISA+SN+BAO & $0.3040^{+0.0058}_{-0.0031}$ & $0.0045$ & $67.42\pm 0.27$ & $0.27$ & $4.9\pm 2.9$ & $2.9$ &  $0.29^{+0.14}_{-0.25}$ & $0.20$ \\
  \hline
\end{tabular}
\end{adjustbox}
\caption[Observational constraints for kinetic coupled quintessence]{Marginalised constraints on cosmological and model parameters for the Kinetic Model at 68\% CL. }
\label{tab:cq_gw_boundskcq}
\end{table*}

\begin{figure}[t!]
      \subfloat{\includegraphics[width=\linewidth]{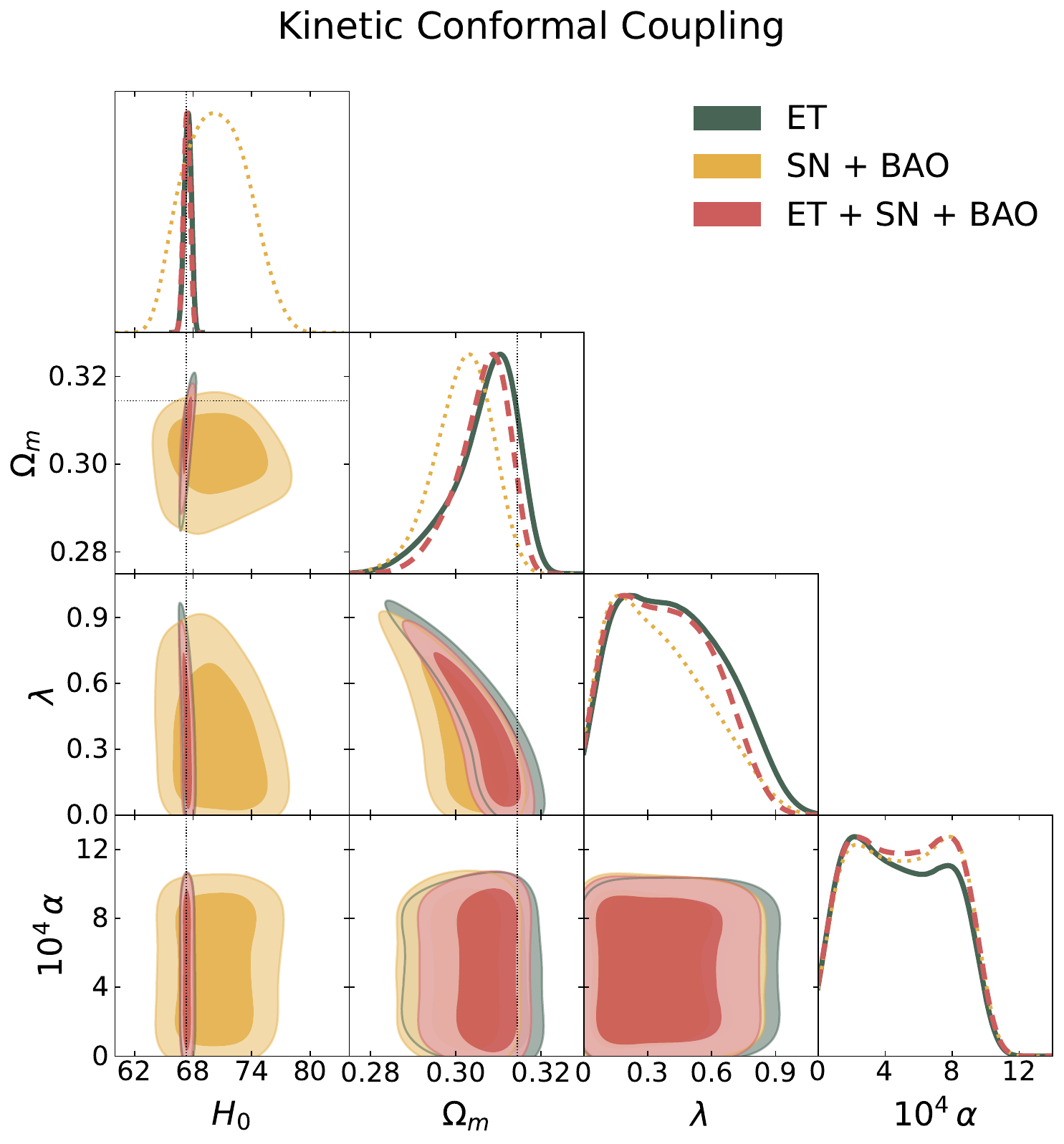}} 
  \caption[Triangle plot of 2D marginalised posterior distributions for kinetic coupled quintessence with ET]{\label{fig:etk}68\% and 95\% CL 2D contours and 1D posterior distributions for the parameters $\{H_0,\Omega_m,\lambda,10^4\alpha\}$ in the kinetic conformal coupled quintessence model with ET (green filled line), SN+BAO (yellow dotted line) data  and their combination (red dashed line). The dotted lines depict the fiducial values for the mock data $\{\Omega_m,H_0 \} = \{0.3144, 67.32\}$.}
\end{figure}

\begin{figure}[t!]
      \subfloat{\includegraphics[width=\linewidth]{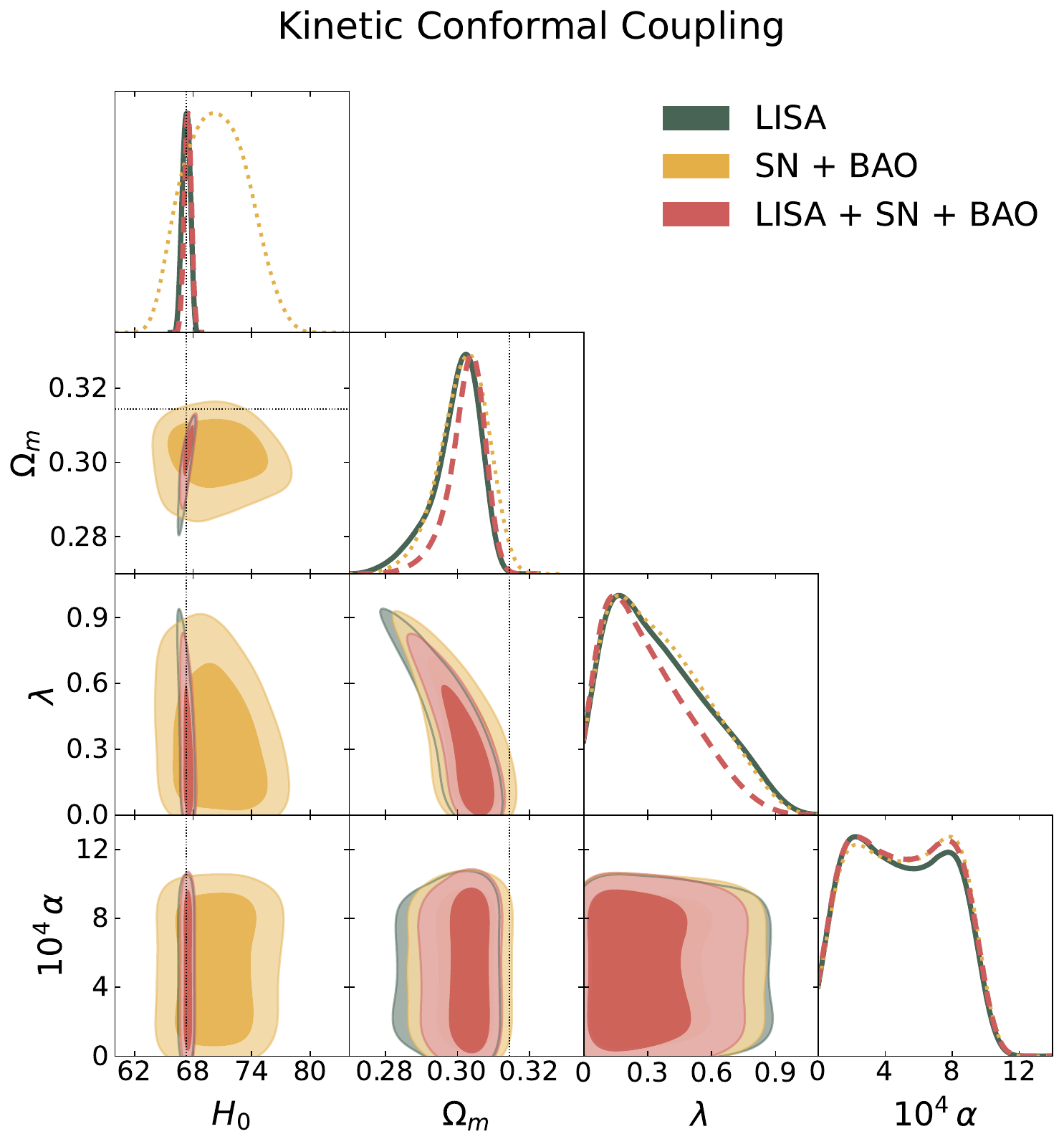}} 
  \caption[Triangle plot of 2D marginalised posterior distributions for kinetic coupled quintessence with LISA]{\label{fig:lpk} 68\% and 95\% CL 2D contours and 1D posterior distributions for the parameters $\{H_0,\Omega_m,\lambda,10^4\alpha\}$ in the kinetic conformal coupled quintessence model with LISA mock data (green filled line), SN+BAO (yellow dotted line) data  and their combination (red dashed line). The scale is the same as in \cref{fig:etk} for comparison purposes, with the SN+BAO contours standing as the reference. The dotted lines depict the fiducial values for the mock data $\{\Omega_m,H_0 \} = \{0.3144, 67.32\}$.}
\end{figure}

\begin{figure}[t!]
      \subfloat{\includegraphics[width=\linewidth]{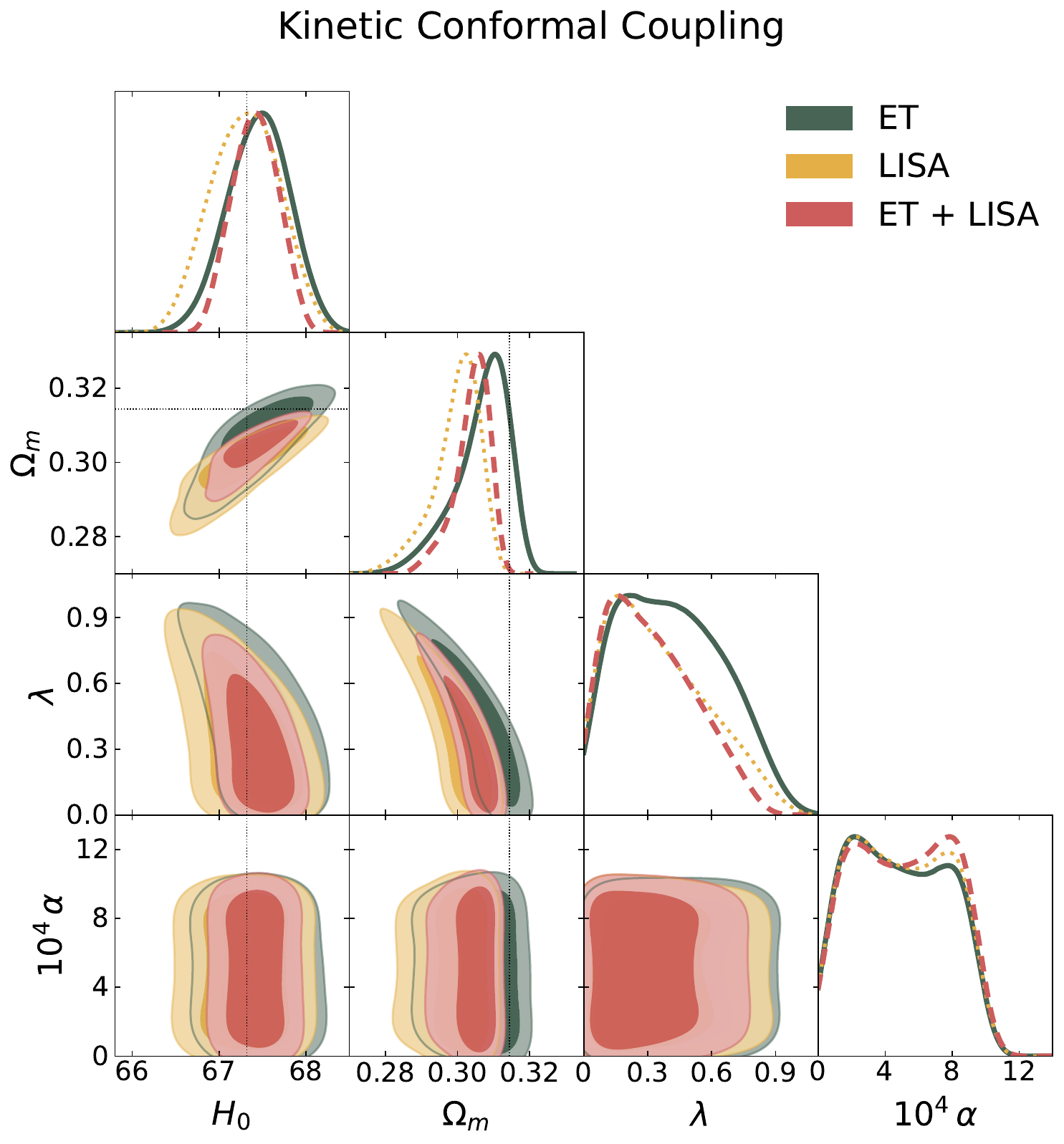}} 
  \caption[Triangle plot of 2D marginalised posterior distributions for kinetic coupled quintessence with ET+LISA]{\label{fig:etlp2k} 68\% and 95\% CL 2D contours and 1D marginalised posterior distributions for the parameters $\{H_0,\Omega_m, \lambda,10^4\alpha\}$ in the kinetic conformal coupled quintessence model with ET mock data (green filled line), LISA mock data (yellow dotted line) and their combination (red dashed line). The dotted lines depict the fiducial values for the mock data $\{\Omega_m,H_0 \} = \{0.3144, 67.32\}$.}
\end{figure}

From the findings presented in \cref{fig:etk,fig:lpk,fig:etlp2k}, and summarised in \cref{tab:cq_gw_boundskcq}, we investigate constraints on the parameters $ \{\Omega_m, H_0, \lambda, 10^4\alpha\} $ using to the same data sets as in the \cref{sec:cq_ss_casea}. Comparison of the errors obtained from the ET standard sirens catalogue to the SN+BAO data reveals that, with the exception of the $ H_0 $ parameter, ET's $ 1\sigma $ bounds are comparable in magnitude. The constraint in $ H_0 $ improves by a factor of ten, quantified by $ \mathcal{F}_{\Omega_m, H_0, \beta, \lambda}^{(\text{SN+BAO,ET})} = \{1.1, 0.12, 1.0, 1.1\} $. Combining the SS and background data sets (ET+SN+BAO), the $ 1\sigma $ constraints tighten for all parameters when set against ET alone, expressed by $ \mathcal{F}_{\Omega_m, H_0, \alpha, \lambda}^{(\text{ET, ET+SN+BAO})} = \{0.84, 0.92, 1.0, 0.92\} $.

For the LISA SS scenario, it becomes apparent that every cosmological and model parameter is either better or equally constrained when using LISA alone, compared to SN+BAO data, with $ \mathcal{F}_{\Omega_m, H_0, \alpha, \lambda}^{(\text{SN+BAO,LISA})} = \{0.91, 0.13, 1.0, 1.0\} $. Pairing LISA with SN+BAO improves upon constraints compared to SN+BAO alone, and LISA+SN+BAO even surpasses LISA alone in constraint power, particularly for $ \Omega_m $, demonstrated by $ \mathcal{F}_{\Omega_m, H_0, \alpha, \lambda}^{(\text{LISA,LISA+SN+BAO})} = \{0.78, 0.92, 1.0, 0.87\} $.

Regarding both ET and LISA, $ H_0 $ accuracy improves by an order of magnitude (0.36 for ET and 0.39 for LISA) relative to SN+BAO ($3.1$), corroborating the findings of \cref{sec:cq_ss_casea}. It is worth noting that the errors in the model-specific parameters remain predominantly unaltered across all data sets and combinations, with only minor changes reported for $ \lambda $ within the $ 1\sigma $ region, while $ \alpha $ remains effectively unaltered.

In comparing the constraining power of ET and LISA against their combination, ET+LISA, we highlight that the latter yields better results for the cosmological parameters relative to any of the data sets analysed. For model parameters, negligible variations in accuracy are observed between individual and combined ET and LISA data sets. However, compared to SN+BAO, ET+LISA exhibits better accuracy with $ \mathcal{F}_{\Omega_m, H_0, \alpha, \lambda}^{(\text{SN+BAO,ET+LISA})} = \{0.68, 0.084, 1.0, 0.91\} $. Incorporating all data sets, ET+LISA+SN+BAO, results in a negligible change in parameter constraints when compared against ET+LISA.

Finally, when assessing the constraints derived for the Kinetic model in Ref.~\cite{Teixeira:2022sjr} against CMB \textit{Planck} 2018, \textit{Planck} CMB lensing, BAO, and SN, we observe that the $ \alpha $ parameter is constrained more tightly by CMB and its combinations by an order of magnitude, relative to our data sets that rely solely on background evolution. Specifically, $ \sigma_{10^4\alpha} $ values are reduced in Ref.~\cite{Teixeira:2022sjr} to 0.95 (Pl18), 0.84 (Pl18+SN+BAO), and 0.7 (Pl18len+SN+BAO) as opposed to 2.9 in our cases. Future ET and LISA data will provide constraints on $ \lambda $ comparable to \textit{Planck} CMB ($ \sigma_\lambda = 0.48 $ with Pl18 and $ \sigma_\lambda = 0.2 $ with Pl18+SN+BAO and Pl18len+SN+BAO). Moreover, this analysis suggests that standard siren data will offer superior $ H_0 $ constraints by an order of magnitude compared to Pl18 ($ \sigma_{H_0} = 2.5 $). CMB lensing enhances this constraint by an order of magnitude, achieving $ \sigma_{H_0} = 0.6 $, comparable to ET and LISA, even though standard sirens yield better relative errors with $ \sigma_{H_0} < 0.4 $ for all considered combinations.

\subsection{Disformal Coupling} \label{sec:cq_ss_casec}

In what follows, we focus on a model featuring only a disformal coupling, given by

\begin{equation}
    C = 1 \mathcomma \quad D = D_0^4 \quad \text{and} \quad V(\phi) = V_0 e^{-\lambda \phi/M_{\text{Pl}}} \mathperiod
\end{equation}

In this scenario, the conformal contribution vanishes, leaving $D \equiv D_0^4$ as a constant with units of (mass)$^{-4}$ as per \cref{eq:disfphii}. The functional form of $V(\phi)$ is maintained relative to the previous scenarios. Prior studies on this model's constraints have been performed in \cite{vandeBruck:2016hpz} and \cite{vandeBruck:2017idm}. In particular, constraints derived solely from background data (including $H(z)$, BAO, and Supernova Union2.1 data) report $D_0 > 0.07$ meV$^{-1}$ and $\lambda < 1.56$ at $95\%$ confidence level \cite{vandeBruck:2016hpz}. When incorporating CMB data with lensing effects, along with BAO, SN, cosmic chronometers, cluster abundances, and $H_0$ priors, stringent upper limits for $D_0$ and $\lambda$ are set at $D_0 < 0.2500$ meV$^{-1}$ and $\lambda< 0.6720$ at $68\%$ CL \cite{vandeBruck:2017idm}.

\begin{table*}[ht!]
\centering
\begin{adjustbox}{width=1\textwidth}
\begin{tabular}{|l|l|l|l|l|l|l|l|l|}
\hline
\multicolumn{9}{|c|}{Constant Disformal Coupled Quintessence}  \\ \hline\hline
 Data sets &  $\Omega_m$ & $\sigma_{\Omega_m}$ & $H_0$ &  $\sigma_{H_0}$ & $D_0/\text{meV}^{-1}$ & $\sigma_{D_0}$ & $\lambda$ & $\sigma_{\lambda}$ \\ \hline \hline
 SN+BAO & $0.315\pm 0.017$ & $0.017$ & $70.5\pm 3.1$ & $3.1$ & $1.20^{+0.65}_{-0.38}$ & $0.52$ & $0.87^{+0.59}_{-0.76}$ & $0.68$ \\ 
 \hline
 \hline
ET & $0.290^{+0.011}_{-0.013}$ & $0.012$ & $67.58^{+0.36}_{-0.27}$ & $0.32$ & $1.06\pm 0.51$ & $0.51$ & $1.06\pm 0.58$ & $0.58$\\
  ET+SN+BAO & $0.298^{+0.011}_{-0.014}$ & $0.013$ & $67.45\pm 0.31$ & $0.31$ & $1.15^{+0.66}_{-0.44}$ & $0.55$ & $0.92\pm 0.58$ & $0.58$\\
  \hline
  \hline
   LISA  & $0.320\pm 0.012$ & $0.012$ & $67.43\pm 0.33$ & $0.33$ & $1.22^{+0.64}_{-0.38}$ & $0.51$ & $0.87\pm 0.58$ & $0.58$ \\
   LISA+SN+BAO & $0.317\pm 0.012$ & $0.012$ & $67.52\pm 0.34$ & $0.34$ & $1.24^{+0.64}_{-0.36}$ & $0.50$ &  $0.86^{+0.65}_{-0.77}$ & $0.71$\\
  \hline
   \hline
   ET+LISA  & $0.3094^{+0.0087}_{-0.0099}$ & $0.0093$ & $67.49\pm 0.22$ & $0.22$ & $1.23^{+0.63}_{-0.36}$ & $0.50$ & $0.88\pm 0.58$ & $0.58$ \\
   ET+LISA+SN+BAO & $0.3100^{+0.0092}_{-0.0100}$ & $0.0096$ & $67.47\pm 0.25$ & $0.25$ & $1.24^{+0.63}_{-0.36}$ & $0.50$ &  $0.88^{+0.68}_{-0.77}$ & $0.73$\\
  \hline
\end{tabular}
\end{adjustbox}
\caption[Observational constraints for disformal coupled quintessence]{Marginalised constraints on cosmological and model parameters for the Constant Disformal Coupled Quintessence Model at 68\% CL }
\label{tab:cq_gw_boundscdcq}
\end{table*}

\begin{figure}[t!]
      \subfloat{\includegraphics[width=\linewidth]{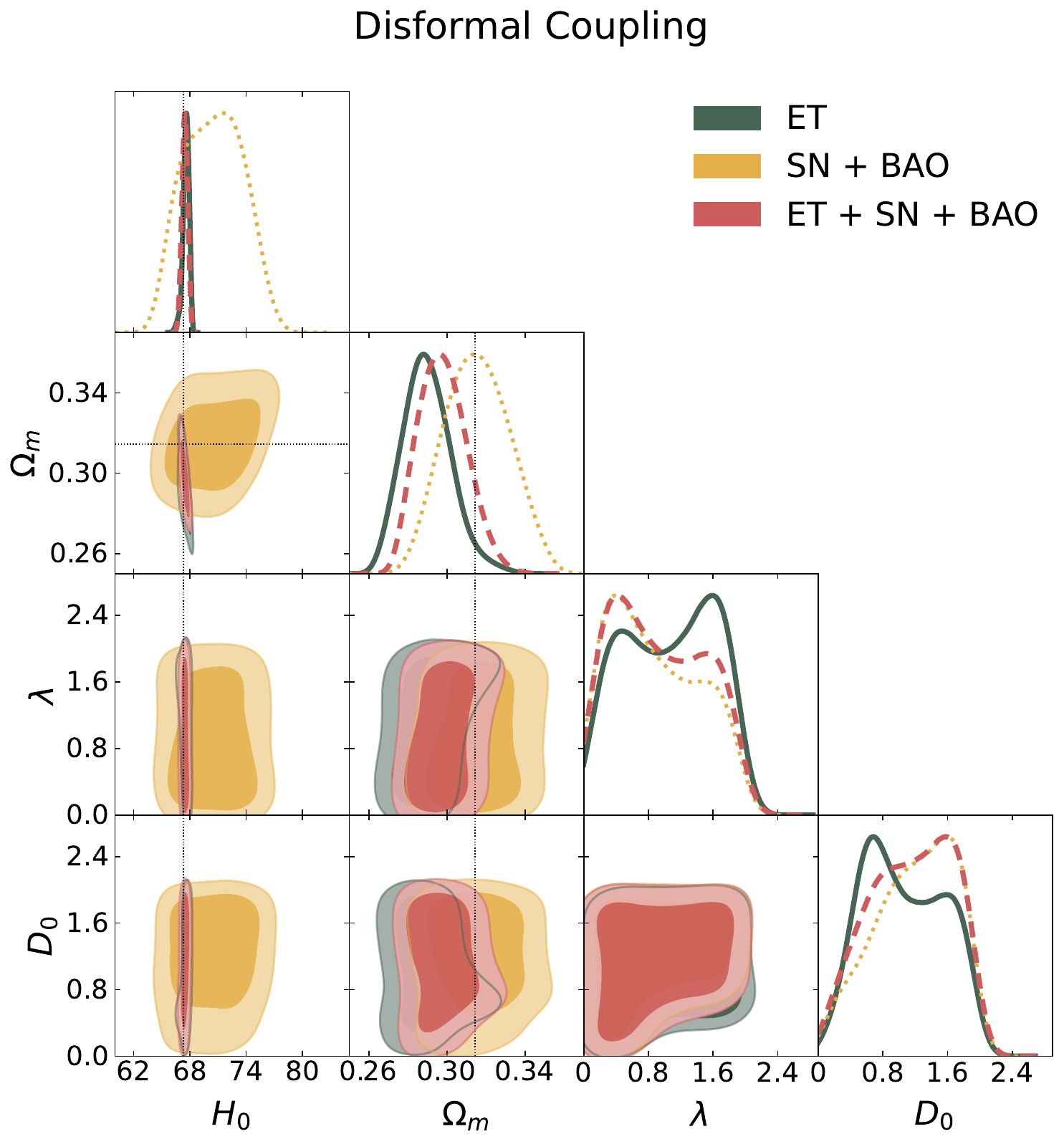}} 
  \caption[Triangle plot of 2D marginalised posterior distributions for disformal coupled quintessence with ET]{\label{fig:etdc} 
  68\% and 95\% CL 2D contours and 1D posterior distributions for the parameters $\{H_0,\Omega_m,\lambda, D_0\}$ in the constant disformal coupled quintessence model with ET (green filled line), SN+BAO (yellow dotted line) data  and their combination (red dashed line). The dotted lines depict the fiducial values for the mock data $\{\Omega_m,H_0 \} = \{0.3144, 67.32\}$.}
\end{figure}

\begin{figure}[t!]
      \subfloat{\includegraphics[width=\linewidth]{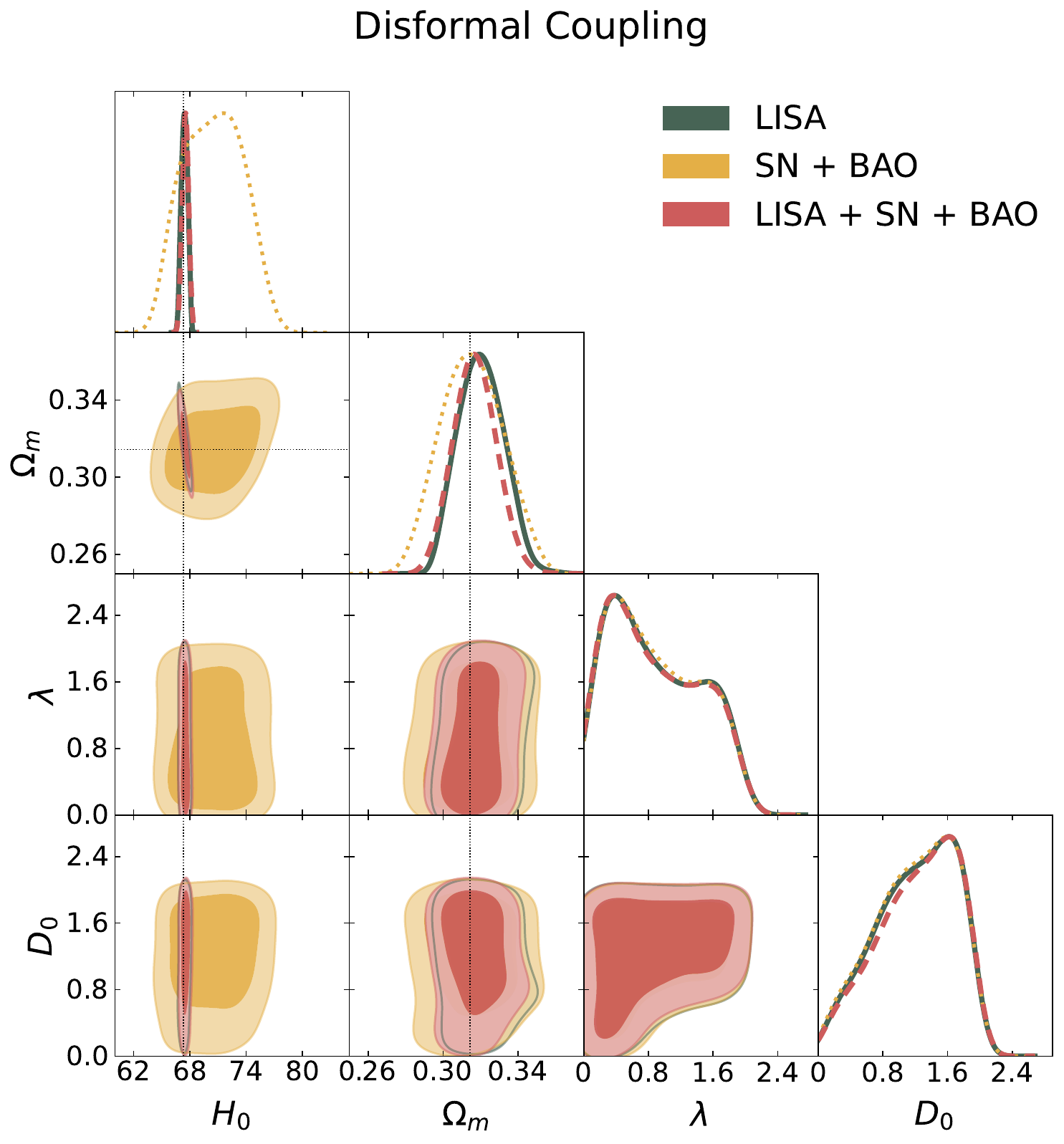}} 
  \caption[Triangle plot of 2D marginalised posterior distributions for disformal coupled quintessence with LISA]{\label{fig:lpdc} 68\% and 95\% CL 2D contours and 1D posterior distributions for the parameters $\{H_0,\Omega_m,\lambda, D_0\}$ in the constant disformal coupled quintessence model with LISA (green filled line), SN+BAO (yellow dotted line) data  and their combination (red dashed line). The scale is the same as in \cref{fig:etdc} for comparison purposes, with the SN+BAO contours standing as the reference. The dotted lines depict the fiducial values for the mock data $\{\Omega_m,H_0 \} = \{0.3144, 67.32\}$.}
\end{figure}

\begin{figure}[t!]
      \subfloat{\includegraphics[width=\linewidth]{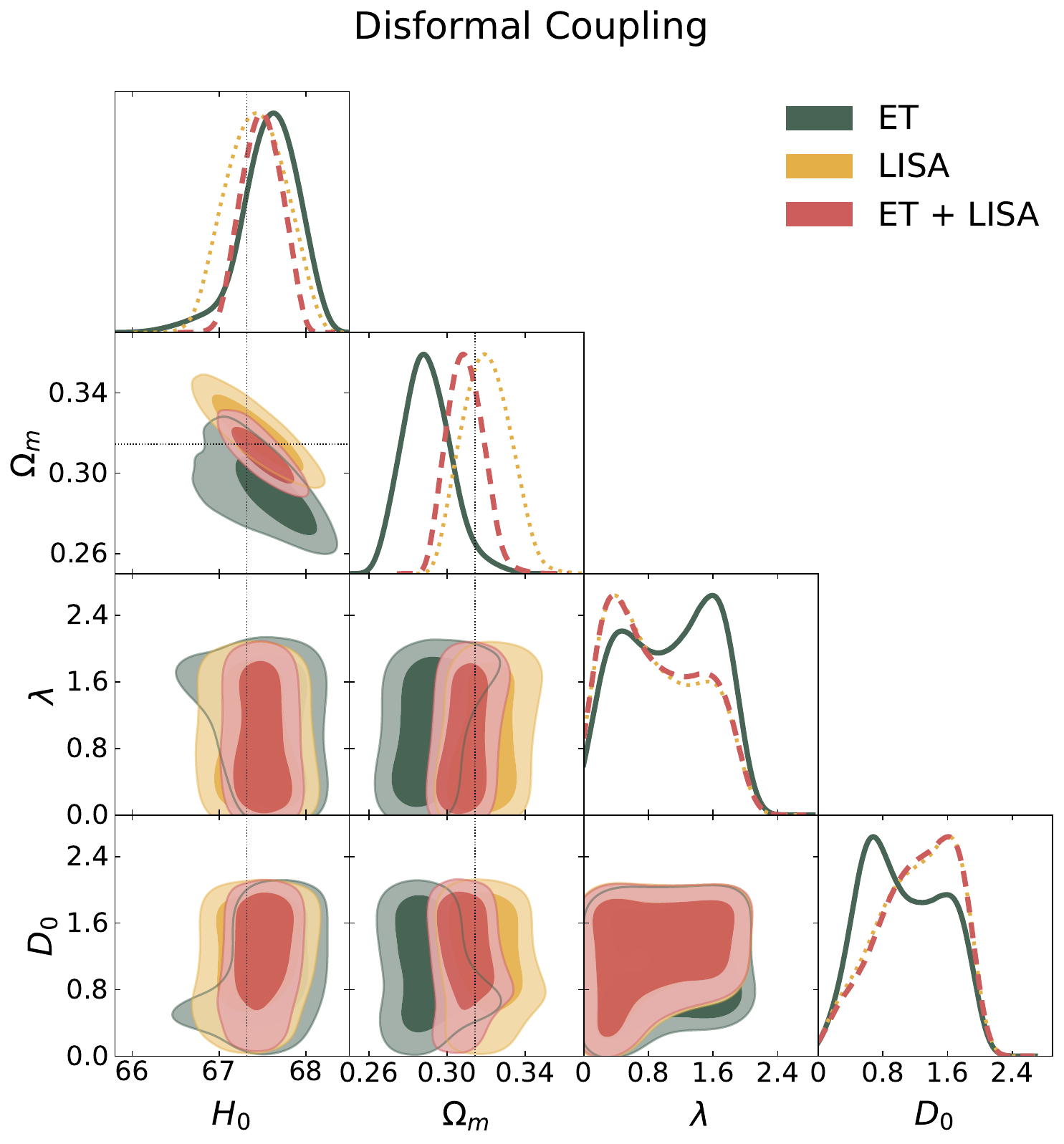}} 
  \caption[Triangle plot of 2D marginalised posterior distributions for disformal coupled quintessence with ET+LISA]{\label{fig:etlp2dc} 68\% and 95\% CL 2D contours and 1D marginalised posterior distributions for the parameters $\{H_0,\Omega_m,\lambda, D_0\}$ in the constant disformal coupled quintessence model with ET mock data (green filled line), LISA mock data (yellow dotted line) and their combination (red dashed line). The dotted lines depict the fiducial values for the mock data $\{\Omega_m,H_0 \} = \{0.3144, 67.32\}$.}
\end{figure}

In \cref{fig:etdc,fig:lpdc,fig:etlp2dc}, and in \cref{tab:cq_gw_boundscdcq}, we summarise the results for the same data sets as previously discussed in \cref{sec:cq_ss_casea,sec:cq_ss_caseb}. For the ET catalogue alone, improved accuracy is observed across all cosmological and model parameters $\{\Omega_m, H_0, D_0, \lambda\}$, relative to SN+BAO, namely $\mathcal{F}_{\Omega_m, H_0, D_0, \lambda}^{(\text{SN+BAO,ET})} = \{0.71,0.10, 0.98,0.85\}$. The joint ET+SN+BAO data set performs better than SN+BAO, although only modest changes in parameter accuracy are seen relative to ET alone, according to $\mathcal{F}_{\Omega_m, H_0, D_0, \lambda}^{(\text{ET,ET+SN+BAO})} = \{1.1,0.97, 1.1, 1.0\}$.

For the LISA standard sirens, we find an analogous trend in parameter accuracy compared to the background data, $\mathcal{F}_{\Omega_m, H_0, D_0, \lambda}^{(\text{SN+BAO,LISA})} = \{0.71,0.11, 0.98,0.85\}$. This extends to the LISA+SN+BAO combination, encapsulated by $\mathcal{F}_{\Omega_m, H_0, D_0, \lambda}^{(\text{LISA,LISA+SN+BAO})} = \{1.0,1.0, 0.98, 1.22\}$. The variation in parameter accuracy is minimal compared to LISA alone, except for $\lambda$, which displays a larger $1\sigma$ region with $\sigma_\lambda = 0.71$.

Regardless of the data combination used, the parameters $\{\Omega_m, D_0, \lambda\}$ are constrained at the same level, with $\lambda$ showing only a minor improvement in both ET and LISA scenarios ($\sigma_\lambda=0.58$ as opposed to $\sigma_\lambda=0.7$ for SN+BAO). The accuracy in the $H_0$ parameter improves by one order of magnitude for ET and LISA compared to SN+BAO.

\sloppy
The model parameters for ET and LISA exhibit no significant variations, leading to consistent constraints for ET+LISA. Nonetheless, there is an increased accuracy for the cosmological parameters. Following the observed trend for both ET and LISA against SN+BAO, the combined ET+LISA data set improves constraints further, according to $\mathcal{F}_{\Omega_m, H_0, D_0, \lambda}^{(\text{SN+BAO,ET+LISA})} = \{0.55,0.071, 0.96, 0.85\}$. There is negligible change in accuracy when combining all data sets (ET+LISA+SN+BAO), except for $\lambda$, which sees a decrease in accuracy as $\mathcal{F}_{\Omega_m, H_0, D_0, \lambda}^{(\text{ET+LISA,ET+LISA+SN+BAO})} = \{1.0,1.1,1.0,1.3\}$.

Contrasted with the results reported in Ref.~\cite{vandeBruck:2016hpz}, we highlight the ability of the SS data to impose constraints at the $68\%$ CL, and not only at $95\%$ CL, for all the model parameters. Particularly, the error in $H_0$ significantly reduces from $\sigma_{H_0} \approx 2.2$ to $\sigma_{H_0} \approx 0.3$ with standard sirens. When compared with results in Ref.~\cite{vandeBruck:2017idm} for CMB, CMB lensing and additional data, we find lower and upper bounds for both $\lambda$ and $D_0$ at $68\%$ CL, contrasting with prior works that reported only upper bounds in particular with more accommodating upper bounds, considering that this analysis includes only background data. Consequently, the error in $H_0$ is brought to the same order of magnitude with $\sigma_{H_0} \approx 0.9$, though still about three times larger than the one reported in this analysis.

\subsection{Mixed Conformal-Disformal Coupling}\label{sec:cq_ss_cased}

Lastly, we explore a mixed coupling, incorporating both conformal and disformal components. Specifically, the model is defined by

\begin{equation}
C(\phi) = e^{2 \beta \phi/ M_{\rm Pl}}, \quad D(\phi) = D_0^4  \quad \text{and}\quad V(\phi) = V_0 e^{-\lambda \phi/ M_{\rm Pl}} \mathperiod
\end{equation}

Similar to the disformal-only scenario in \cref{sec:cq_ss_casec}, constraints for this model have been presented in Refs.~\cite{vandeBruck:2016hpz} and \cite{vandeBruck:2017idm}. For the same background data, Ref.~\cite{vandeBruck:2016hpz} reported constraints of $D_0 > 0.102$, $\beta < 0.453$, and $\lambda < 1.59$ at the $95\%$ confidence level. In contrast, the work in \cite{vandeBruck:2017idm}, which incorporates CMB data, yields $\beta \lesssim 0.17$ and $\lambda \lesssim 0.35$ at $1\sigma$, with the exact constraints varying depending on the specific data sets employed. The disformal coupling $D_0$ is not always well constrained for this case, with lower limits of $D_0 \gtrsim 0.35$ found in particular data set combinations.

\begin{table*}[ht!]
\centering
\begin{adjustbox}{width=1\textwidth}
\begin{tabular}{|l|l|l|l|l|l|l|l|l|l|l|}
\hline
\multicolumn{11}{|c|}{Mixed Conformal-Disformal Coupled Quintessence}  \\ \hline\hline
 Data sets &  $\Omega_m$ & $\sigma_{\Omega_m}$ & $H_0$ & $\sigma_{H_0}$ & $\beta$ & $\sigma_{\beta}$ & $D_0$ & $\sigma_{D_0}$ & $\lambda$ & $\sigma_{\lambda}$ \\ \hline \hline
 SN+BAO & $0.308^{+0.021}_{-0.015}$ & $0.018$ & $71.2\pm 3.3$ & $3.30$ & $1.01\pm 0.57$ & $0.57$ & $1.23^{+0.59}_{-0.43}$ & $0.51$ &  $0.98\pm 0.57$ & $0.57$\\ 
 \hline\hline
ET & $0.286^{+0.010}_{-0.012}$ & $0.011$ & $67.65\pm 0.29$ & $0.29$ & $0.85\pm 0.58$ & $0.58$ & $1.27^{+0.58}_{-0.35}$ & $0.47$ & $1.03\pm 0.58$ & $0.58$ \\
  ET+SN+BAO & $0.294^{+0.011}_{-0.013}$ & $0.012$ & $67.50\pm 0.30$ & $0.30$ & $0.92^{+0.66}_{-0.76}$ & $0.71$ & $1.32^{+0.53}_{-0.35}$ & $0.44$ & $0.97\pm 0.58$ & $0.58$\\
   \hline\hline
   LISA  & $0.310^{+0.017}_{-0.0087}$ & $0.013$ & $67.55^{+0.27}_{-0.31}$ & $0.29$ & $0.97\pm 0.56$ & $0.56$ & $1.15^{+0.63}_{-0.44}$ & $0.54$ & $1.01\pm 0.56$ & $0.56$\\
   LISA+SN+BAO & $0.310^{+0.016}_{-0.010}$ & $0.013$ & $67.59\pm 0.33$ & $0.33$ & $1.01\pm 0.58$ & $0.58$ & $1.25^{+0.53}_{-0.43}$ & $0.48$ &  $0.98\pm 0.57$ & $0.57$\\
  \hline
  \hline
   ET+LISA & $0.302^{+0.0120}_{-0.0058}$ & $0.0089$ & $67.54\pm 0.20$ & $0.20$ & $0.92\pm 0.55$ & $0.55$ & $1.09^{+0.76}_{-0.42}$ & $0.59$ & $1.05^{+0.71}_{-0.56}$ & $0.64$\\
   ET+LISA+SN+BAO & $0.304^{+0.0120}_{-0.0089}$ & $0.011$ & $67.53\pm 0.24$ & $0.24$ & $0.98\pm 0.57$ & $0.57$ & $1.27^{+0.53}_{-0.41}$ & $0.47$ &  $0.97\pm 0.57$ & $0.57$\\
  \hline
\end{tabular}
\end{adjustbox}
\caption[Observational constraints for mixed conformal-disformal coupled quintessence]{Marginalised constraints on cosmological and model parameters for the Mixed Conformal-Disformal Coupled Quintessence Model at 68\% CL.}
\label{tab:cq_gw_boundsfdcq}
\end{table*}

\begin{figure}[t!]
      \subfloat{\includegraphics[width=\linewidth]{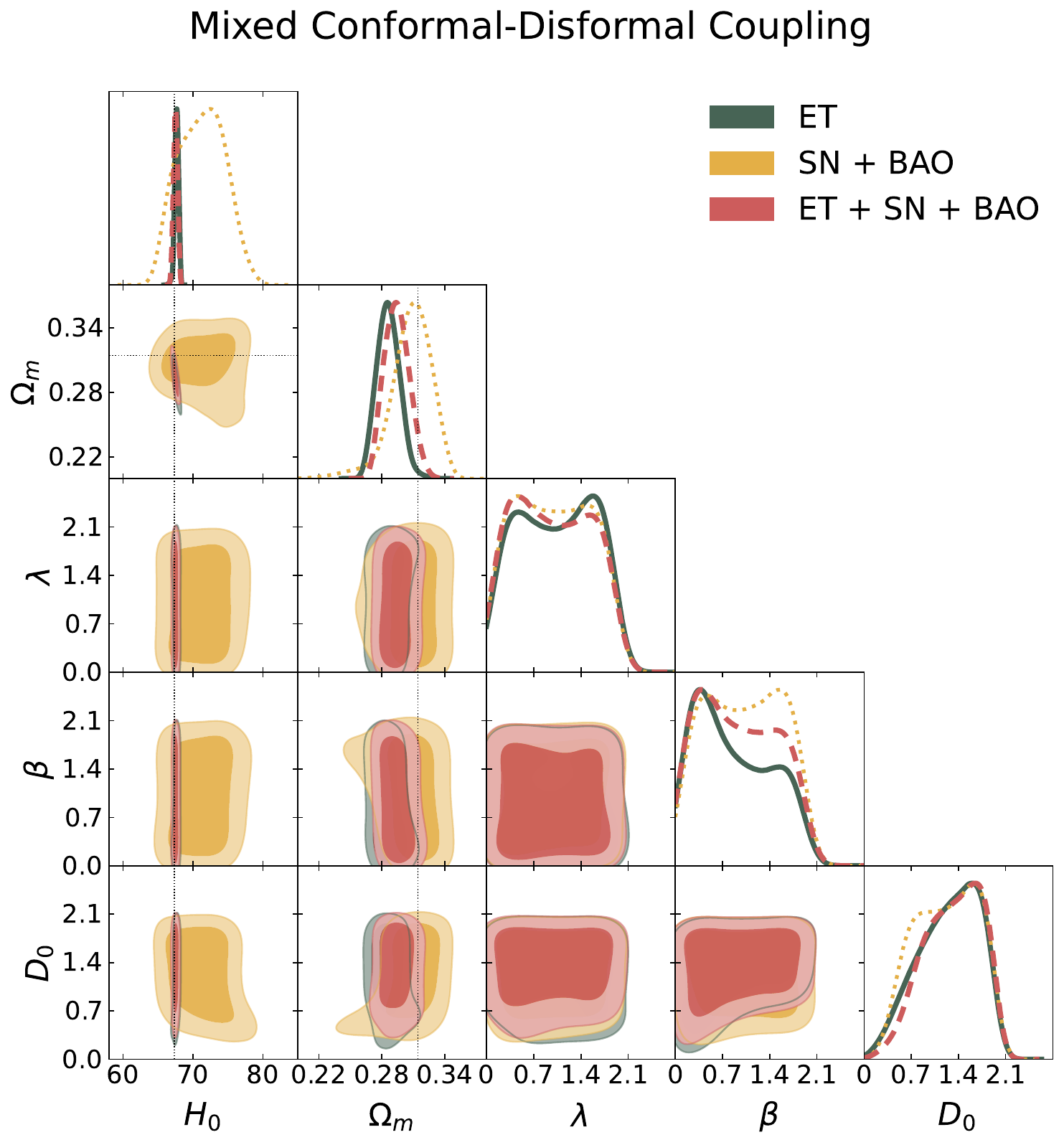}} 
  \caption[Triangle plot of 2D marginalised posterior distributions for mixed conformal-disformal coupled quintessence with ET]{\label{fig:etdf} 68\% and 95\% CL 2D contours and 1D posterior distributions for the parameters $\{ H_0,\Omega_m,\lambda, \beta, D_0\}$ in the mixed conformal-disformal coupled quintessence model with ET (green filled line), SN+BAO (yellow dotted line) data  and their combination (red dashed line). The dotted lines depict the fiducial values for the mock data $\{\Omega_m,H_0 \} = \{0.3144, 67.32\}$.}
\end{figure}

\begin{figure}[t!]
      \subfloat{\includegraphics[width=\linewidth]{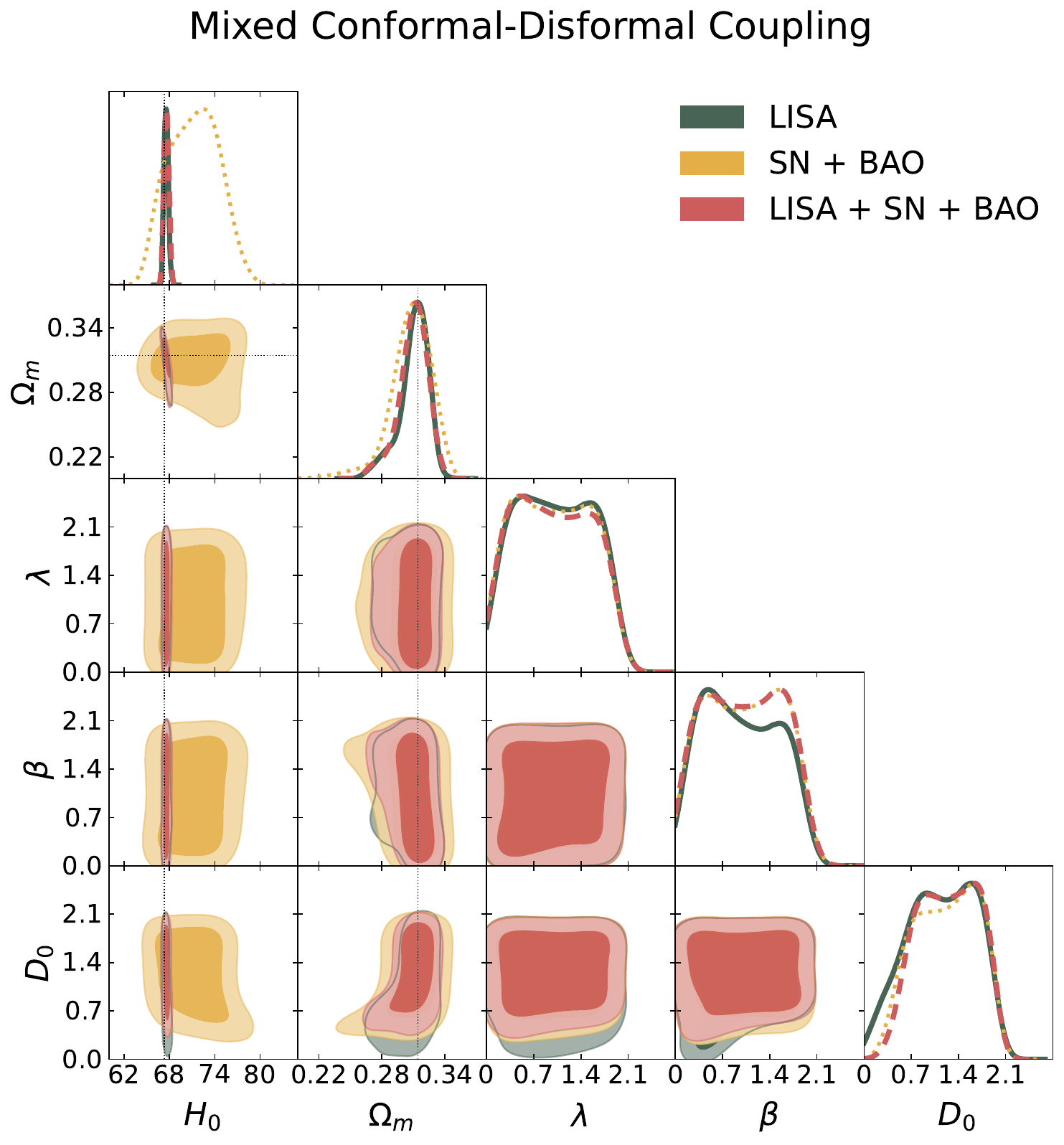}} 
  \caption[Triangle plot of 2D marginalised posterior distributions for mixed conformal-disformal coupled quintessence with LISA]{\label{fig:lpdf} 68\% and 95\% CL 2D contours and 1D posterior distributions for the parameters $\{ H_0,\Omega_m,\lambda, \beta, D_0\}$ in the mixed conformal-disformal coupled quintessence model with LISA (green filled line), SN+BAO (yellow dotted line) data  and their combination (red dashed line). The scale is the same as in \cref{fig:etdf} for comparison purposes, with the SN+BAO contours standing as the reference. The dotted lines depict the fiducial values for the mock data $\{\Omega_m,H_0 \} = \{0.3144, 67.32\}$.}
\end{figure}

\begin{figure}[t!]
      \subfloat{\includegraphics[width=\linewidth]{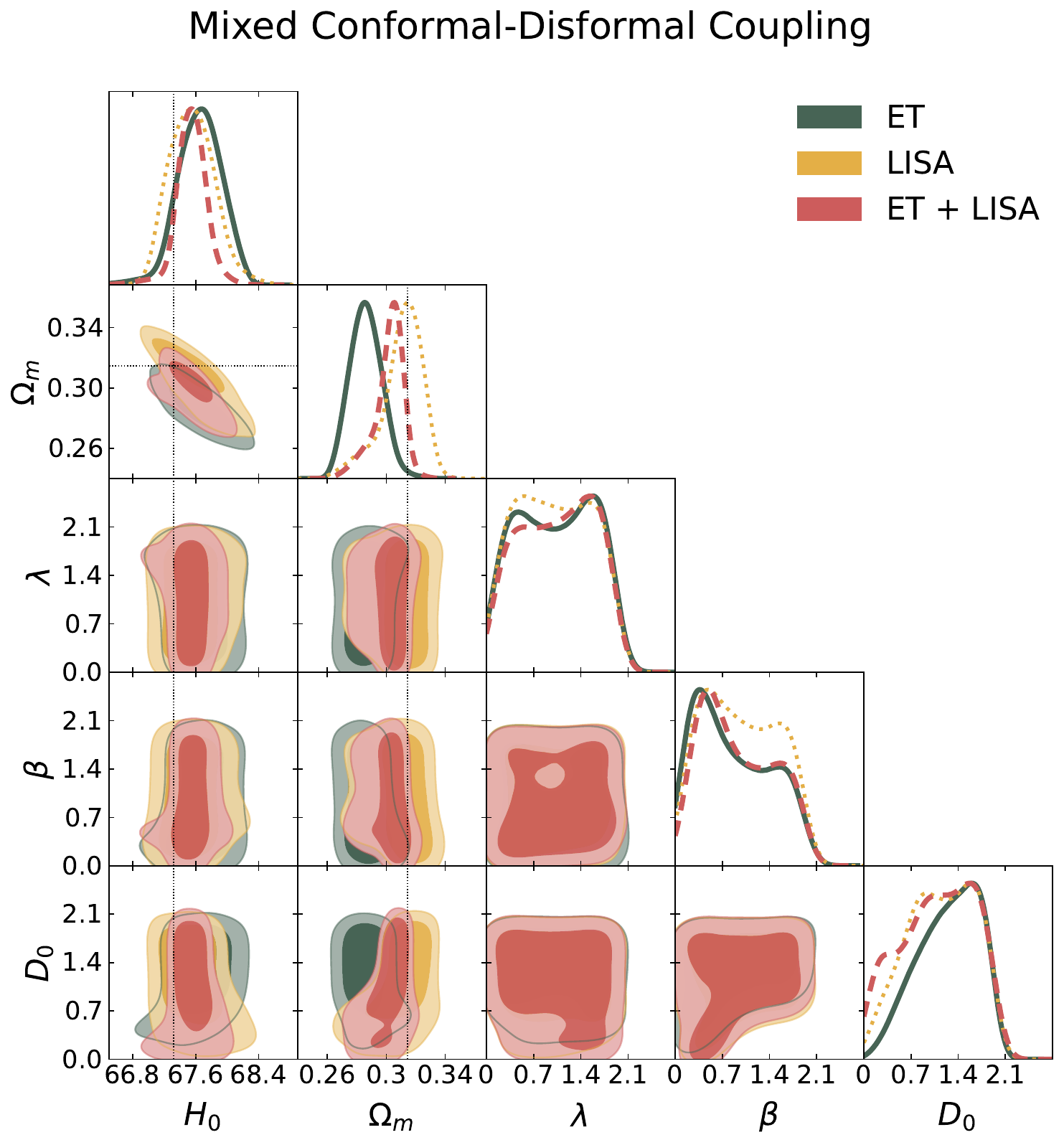}} 
  \caption[Triangle plot of 2D marginalised posterior distributions for mixed conformal-disformal coupled quintessence with ET+LISA]{\label{fig:etlp2df} 68\% and 95\% CL 2D contours and 1D marginalised posterior distributions for the parameters $\{ H_0,\Omega_m,\lambda, \beta, D_0\}$ in the mixed conformal-disformal coupled quintessence model with ET mock data (green filled line), LISA mock data (yellow dotted line) and their combination (red dashed line). The dotted lines depict the fiducial values for the mock data $\{\Omega_m,H_0 \} = \{0.3144, 67.32\}$.}
\end{figure}

In \cref{fig:etdf,fig:lpdf,fig:etlp2df}, and Table \cref{tab:cq_gw_boundsfdcq}, the estimated values of the parameters $\{\Omega_m, H_0, \beta, D_0, \lambda\}$ are presented using the same data sets as in \cref{sec:cq_ss_casea,sec:cq_ss_caseb,sec:cq_ss_casec}. Considering the ET catalogue alone, we note an improvement in accuracy in the cosmological parameters $\Omega_m$ and $H_0$ when compared to the SN+BAO data set; specifically, $\mathcal{F}_{\Omega_m, H_0}^{(\text{SN+BAO,ET})} = \{0.61,0.088\}$. Similar accuracy levels are maintained upon combining the ET, SN, and BAO data sets, with a marginal enhancement relative to the ET-only case. As for the model-specific parameters $\{\beta, D_0, \lambda\}$, the constraining ability of ET is nearly identical to that of SN+BAO, with $\mathcal{F}_{\beta, D_0, \lambda}^{(\text{SN+BAO,ET})} = \{1.0,0.92,1.0\}$. Nevertheless, it is worth mentioning that the combined data set yields an increase in the $\beta$-error, with $\sigma_\beta = 0.71$.

When considering LISA's standard sirens, there is an improved accuracy for the cosmological parameters, akin to the ET results, with $\mathcal{F}_{\Omega_m, H_0}^{(\text{SN+BAO,LISA})} = \{0.72,0.088\}$. However, an observable accuracy drop for $H_0$ is evident when combining LISA with other data sets, as given by $\mathcal{F}_{H_0}^{(\text{LISA, LISA+SN+BAO})} = \{1.1\}$.
Regarding model parameters $\{\beta, D_0, \lambda\}$, LISA, in contrast to ET, shows heightened accuracy over SN+BAO, except for $D_0$, represented by $\mathcal{F}_{\beta, D_0, \lambda}^{(\text{SN+BAO,LISA})} = \{0.98,1.1,0.98\}$. The composite data sets maintain a comparable level of accuracy to the LISA-only case, indicated by $\mathcal{F}_{\beta, D_0, \lambda}^{(\text{LISA,LISA+SN+BAO})} = \{1.0,0.89,1.0\}$.

Similarly to what was discussed in \cref{sec:cq_ss_casea,sec:cq_ss_casec}, the combined GW data sets, ET+LISA yield a marked improvement in the accuracy of $\Omega_m$ and $H_0$, when compared to SN+BAO alone, with $\mathcal{F}_{\Omega_m,H_0}^{(\text{SN+BAO, ET+LISA})} = \{0.49,0.061\}$. The model-specific parameters exhibit only minor changes, with both $D_0$ and $\lambda$ being somewhat less constrained, $\mathcal{F}_{\beta,D_0,\lambda}^{(\text{SN+BAO, ET+LISA})}=\{0.96,1.2,1.1\}$. When all data sets are considered, the cosmological parameters $\Omega_m$ and $H_0$ display enhanced constraints compared to SN+BAO, while the model parameters remain largely unchanged, $\mathcal{F}_{\Omega_m,H_0,\beta,D_0,\lambda}^{(\text{SN+BAO, ET+LISA+SN+BAO})}=\{0.58, 0.073,1.0,0.92,1.0\}$.

Regardless of the data set combinations, the constraints on $\{\Omega_m, \beta, D_0, \lambda\}$ closely resemble those derived from SN+BAO. Furthermore, the accuracy in constraining $H_0$ is improved by an order of magnitude for ET and LISA relative to SN+BAO.

Much like the comparison in \cref{sec:cq_ss_casec}, the main improvement predicted in our analysis in comparison to Ref.~\cite{vandeBruck:2016hpz} is the possibility of constraining all the model parameters at $68\%$ CL. The upper bounds on $\lambda$ for standard sirens at $1\sigma$ are consistent with the $2\sigma$ limits stated for earlier studies. Moreover, the precision of $H_0$ significantly improves from $\sigma_{H_0} \approx 2.1$ for the background data to $\sigma_{H_0} \approx 0.3$ in all the cases including SS data. Compared to Ref.~\cite{vandeBruck:2017idm}, which takes into account CMB, CMB lensing and additional data and which either does not provide constraints for $D_0$ or only finds a lower limit for $\lambda$ and $\beta$, the inclusion of standard sirens in our analysis yields solid constraints for all three model-parameters at $68\%$ CL. This represents a great achievement that could be improved by combining the background data with the CMB observations. Furthermore, including CMB data brings the error in $H_0$ to the same order of magnitude, $\sigma_{H_0} \approx 0.6$, although still approximately two times larger than the values reported in this analysis.

\section{Discussion} \label{sec:cq_ss_sum}

In this work, we have investigated the potential of upcoming gravitational wave detectors, specifically ET and LISA, to improve constraints on conformal and disformal couplings between dark energy and dark matter. Four theoretical frameworks were considered: a conformal coupled quintessence, a kinetic model, a constant disformal coupled quintessence, and a mixed conformal-disformal model. We also included a self-interacting exponential potential for each case and looked for constraints in its slope parameter $\lambda$.

We have simulated mock catalogues of standard siren events in accordance with the ET and LISA specifications, which were subsequently used to conduct an MCMC analysis. In particular, we compared the predictions for the individual SS data and also in combination with the current background SN and BAO data. Keeping in mind the assumptions that go into simulating data and performing forecasts, we summarise the main results for each model below:

\begin{itemize}
    \item Conformal Coupled Quintessence: Both LISA+SN+BAO and ET+SN+BAO combinations improve the constraints on $\lambda$ and the conformal coupling parameter $\beta$. Moreover, integrating ET+LISA with SN+BAO reduces the error in $\beta$ by one-third.
    
    \item Kinetic Model: The ET and LISA catalogues independently, when taken independently, fall short of improving the constraints on $\lambda$ and on the conformal exponential parameter $\alpha$. Combining LISA with SN+BAO yields a marginal improvement for $\lambda$ and the matter density $\Omega_m$.
    
    \item Constant Disformal Coupled Quintessence: All combined data sets effectively constrain the disformal parameter $D_0$ at the $1\sigma$ level and the same order of magnitude, with a small improvement for LISA+SN+BAO. ET, LISA and their combination perform better than SN+BAO in constraining $\lambda$. The combination of ET+LISA with SN+BAO allows for the error in $\Omega_m$ to be reduced.
    
    \item Mixed Conformal-Disformal Coupled Quintessence: ET and LISA, when analysed separately, do not significantly enhance the precision of the model parameters. Only a marginal reduction in the $1\sigma$ uncertainty of $D_0$ is noted for the full combination scenarios. The parameter $\Omega_m$ experiences a slight decrease in the associated error when combining ET and LISA.
\end{itemize}

Independent of the model considered, we find that the accuracy of the $H_0$ parameter is consistently enhanced by an order of magnitude at $1\sigma$, relative to the SN+BAO data set. This holds promise for providing further insight into addressing the $H_0$ tension and contributes to the overall improvement in model parameter estimates when the full combinations are considered, as we have just reviewed. Ultimately, our results show that upcoming third-generation gravitational wave detectors stand to enrich our understanding of dark energy-dark matter interactions substantially and may provide insights into the $H_0$ tension.

It should be noted that the forecast results presented in \cref{sec:cq_ss_res} reveal a considerable deviation in the constraints derived for the cosmological parameters in the extended models, from the expected $\Lambda$CDM fiducial values used to simulate the SS data. Even if the mean values found for $H_0$ and $\Omega_m$ differ from the ones used in the simulated cosmology, these are still expected to be recovered within the corresponding errors. Several factors might contribute to this bias observed in the forecasts. One plausible explanation is the presence of parameter degeneracies, where the variations in one parameter could compensate for changes in another, leading to deviations from the fiducial $\Lambda$CDM values. Additionally, the combination of \textit{real} BAO and SN data with \textit{simulated} GW data introduces complexities, especially when the fiducial cosmology for the GW simulations does not align exactly with the best-fit SN+BAO cosmology. Furthermore, the implicit dependence of the GW simulations on the cosmological parameters can introduce uncertainties that influence the forecasted values. 
In addition to the observed bias in the forecasted values away from the fiducial $\Lambda$CDM values, another noteworthy aspect of the results is the occurrence of certain parameters reaching their prior bounds. For instance, \cref{fig:etlp2df} demonstrates that the confidence regions for parameters such as $\lambda$ and $\beta$ touch on the specified low boundaries of the prior at zero. While the particular symmetries of the mixed conformal-disformal model should guarantee that the same dynamics, and therefore the same predictions, are expected for symmetric values of these parameters, this phenomenon further supports potential degeneracies or limitations in the parameter estimation process. This observation also underscores the importance of carefully selecting and defining prior distributions for parameters in cosmological analyses to ensure that the parameter space is adequately explored, and that results are not unduly influenced by the prior assumptions. 
Understanding and addressing these potential sources of bias is crucial for refining the accuracy and reliability of future cosmological predictions and parameter estimations.

 \cleardoublepage


 \chapter{Dark D-Brane Cosmology} \label{chap:dbi}
 \setcounter{equation}{0}
\setcounter{figure}{0}



  \epigraph{Céu é para baixo ou para cima? Pensava a nordestina. Deitada, não sabia.\blfootnote{\textit{Is the sky above or below? Wondered the northeastern girl. Lying there, she didn’t know.} \\ --- \textsc{Clarice Lispector} in The Hour of the Star} \\ --- \textsc{Clarice Lispector}\ \small\textup{A Hora da Estrela}}

Though the $\Lambda$CDM model continues to be considered the most plausible account of cosmological phenomena, the physical nature of the dark sector remains unfamiliar. Cold dark matter is generally posited as a particle, predicted by many extensions to the standard model, but has not been directly observed. Moreover, the magnitude of the cosmological constant remains incompatible with traditional quantum field theories, as detailed in \cref{sec:theoprob} \cite{Rugh:2000ji,Weinberg:2000yb,martin}. These theoretical considerations motivate the investigation of alternate paradigms, such as the quintessence field introduced in \cref{sec:quin} and which was the main focus of \cref{chap:cquint}. A scalar field can emulate cosmological constant-like behaviour and explain the apparent high coincidence in the present value of $\Lambda$ in a more effortless dynamical way \cite{Wetterich:1987fm,PhysRevD.37.3406,Peebles:1987ek,Caldwell:1997ii,Tsujikawa:2013fta}. Though relatively few studies consider a unified origin for both dark matter and dark energy \cite{Gao:2009me,Ansoldi:2012pi,Arbey:2020ldf,Brandenberger:2018xnf}, various coupled dark sector models have been proposed in the literature, which is also the main focus of this dissertation (literally in the title!) \cite{ELLIS1989264,Wetterich:1994bg,Amendola:1999er,Holden:1999hm,Copeland:2006wr,debook}. 

In this chapter, we will build on the theoretical framework presented in \cref{app:dbi}, in which dark matter and dark energy, despite being distinct entities, share a higher-dimensional geometrical origin, as proposed in \cite{Koivisto:2013fta}. In particular, the dark matter sector arises from the matter on a D-brane moving in a higher-dimensional spacetime. In this framework, dark energy parametrises the brane's position, whose higher-dimensional dynamics are encoded in the low-energy effective action's kinetic and potential energy terms. 
This scenario hinges on the concept of hidden sector branes in string theory, which experience no interaction with the D-branes accountable for the visible sector (the standard model of particle physics). This means that the setup imposes that dark matter can only interact gravitationally with standard model fields \cite{Cicoli:2023opf,McAllister:2007bg}, with gravity being the fundamental interaction propagating in the higher-dimensional bulk. Ultimately, in this framework, while still separate components, dark energy and dark matter emerge as geometrical properties of the D-brane, becoming inevitably coupled. Moreover, because the dark energy parametrises the dynamics of the \textit{Dark D-brane}, the kinetic term in action adopts a non-canonical Dirac-Born-Infeld (DBI) form, commonly considered in cosmology (see \cref{app:dbi}), particularly concerning inflationary \cite{Garriga:1999vw,Silverstein:2003hf,Alishahiha:2004eh,Panda:2005sg,Chimento:2007es} and dark energy models \cite{Abramo:2004ji,Martin:2008xw,Guo:2008sz,Gumjudpai:2009uy,Ahn:2009hu,Ahn:2009xd,Chimento_2009,Copeland:2010jt,Brax:2012jr,Kaeonikhom:2012xr,Burrage:2014uwa,Mahata:2015lja,Panpanich:2017nft}.
In this chapter, we focus on such a framework in which dark matter and dark energy are inherently coupled and, more precisely, due to differing metrics in the D-branes laying out the geodesics for particles of the standard model and dark matter~\cite{Will:2014kxa,Sakstein:2014isa,Ip:2015qsa,Wang:2016lxa}. This non-universal coupling of the scalar field to matter evades conflicts with Solar System tests and the stringent constraints on the speed of gravitational waves \cite{LIGOScientific:2017zic}.
The relation between these two metrics is actually a realisation of the disformal transformation \cite{Bekenstein:1992pj} introduced in \cref{sec:coup}, which we recall is generally described as

\begin{equation}
\bar{g}_{\mu \nu} = C(\phi) g_{\mu \nu} + D(\phi) \partial_{\mu} \phi \partial_{\nu} \phi \mathperiod
\label{trans}
\end{equation}

The conformal and disformal factors, $C(\phi)$ and $D(\phi)$, are related and encapsulate information about the curvature of the extra dimensions.

Disformal relations have been previously investigated in the context of brane-world cosmology~\cite{KOIVISTO:2013jwa,Koivisto:2014gia,Koivisto:2015vda,Cembranos:2016jun,Cicoli:2023opf} and have found various applications in a range of cosmological studies \cite{deRham:2010ik,Clayton:1998hv,Bettoni:2011fs,Deruelle:2014zza,Brax:2012hm,Olmo:2009xy,Sakstein:2014aca,Ezquiaga:2017ner}. They have also been instrumental in the formulation of inflationary settings \cite{Kaloper:2003yf,vandeBruck:2015tna}, and disformal quintessence \cite{Koivisto:2008ak,Zumalacarregui:2010wj}. Various models of disformally coupled dark energy have beem explored~\cite{Noller2012,Bettoni:2012xv,Koivisto:2012za,Zumalacarregui:2012us,Zumalacarregui:2013pma,Bettoni:2013diz,Sakstein:2014aca,Sakstein:2015jca,vandeBruck:2015ida,Bettoni:2015wla,vandeBruck:2016jgg,Teixeira:2019hil}, whose results can be contrasted with the work derived here.

In this chapter, we report on an extension to the work presented in Ref.~\cite{Koivisto:2013fta}, where the model was first formulated, along with an account of the background evolution through a dynamical systems analysis. We not only present a precise numerical exploration by evolving the equations in an Einstein-Boltzmann code but also derive and study the linearly perturbed equations, allowing for a direct comparison between this dark D-brane scenario and other theories of dark energy with disformal couplings~\cite{Zumalacarregui:2012us,vandeBruck:2015ida,vandeBruck:2017idm,Mifsud:2017fsy,Teixeira:2019hil}. Armed with this study we can derive and better understand constraints for the relevant cosmological parameters against current data sets.

This chapter is structured as follows. \cref{sec:modeld} provides a thorough exposition of the theoretical underpinnings and general equations of motion. \cref{sec:back} focuses on cosmological evolution within a flat FLRW spacetime. \cref{sec:pert} is devoted to discussing cosmological perturbations, where we present the governing equations in the Newtonian gauge and append the equations in synchronous gauge. We also evaluate the growth of the gravitational constant in this dark D-brane cosmological setting and provide predictions for CMB temperature fluctuations and matter power spectra. In \cref{sec:dbi_num}, we present the main cosmological observable features of the model through a numerical study. In \cref{sec:dbi_const}, we list the cosmological data constraints on the parameters of the model according to a Bayesian data analysis. Finally, we conclude in \cref{sec:concdbi}. 

This chapter is based on work published in Physical Review D \cite{vandeBruck:2020fjo} and ongoing work.

\section{The Model} \label{sec:modeld}

In the framework explored in this chapter, the dark sector emerges from a hidden D3-brane, moving in a higher-dimensional spacetime. This sector includes two distinct degrees of freedom: the matter fields that populate the brane and the brane's radial position \footnote{We restrict our investigation to a single species living on the hidden brane, presumed to be pressureless and thus serving as a suitable cold dark matter candidate.}. The cold dark matter is identified with the particles living on the so-called \textit{dark D-brane}, which does not intersect with the D-brane(s) from which the standard model fields emerge, explaining why the dark D-brane is said to be effectively \textit{hidden}. Consequently, any interactions between dark matter and the standard model fields are purely gravitational in the low-energy regime. Dark energy is ascribed to the scalar degree of freedom that parametrises the brane's position in the extra dimensions, naturally inducing the interaction in the dark sector. The geometry of this higher-dimensional spacetime is encoded in the warp factor, a function of the radial coordinate. For practical purposes, we will focus on AdS$_5 \times$ S$^5$ warped regions.

The theoretical structure behind the \ac{ddb} was proposed in \cite{Koivisto:2013fta}, constructed from a warped flux compactification of Type IIB String Theory. The corresponding low-energy 4D effective action takes the form (keeping with the convention in which $c = \hbar = 1$ and the metric signature is ($-+++$)):

\begin{align}
\label{action}
S &= \frac{1}{2 \kappa^2} \int d^4 x \sqrt{-g} R + \int d^4 x \sqrt{-g} \left[ h^{-1}(\phi) \left( 1 - \sqrt{1 + h(\phi) \partial^\mu \phi \partial_\mu \phi} \right) - V(\phi) \right] \nonumber \\
& \quad + \sum_i \int d^4 x \sqrt{-g} \mathcal{L}_\text{SM} \left( g_{\mu \nu}, \psi_i, \partial_\mu \psi_i \right) + \sum_j \int d^4 x \sqrt{-\bar{g}} \mathcal{L}_\text{c} \left( \bar{g}_{\mu \nu}, \chi_j, \partial_\mu \chi_j \right) \mathperiod
\end{align}

The first term is the conventional Einstein-Hilbert action, where $\kappa^2 = M_{\text{Pl}}^{-2} = 8 \pi G$ represents the reduced Planck mass, $G$ stands for Newton's gravitational constant, $g$ is the determinant of the metric tensor $g_{\mu \nu}$, and $R$ is the Ricci scalar. The metric $g_{\mu\nu}$ prescribes geodesics for the standard model fields. The second term is the Dirac-Born-Infeld action \cite{Silverstein:2003hf,Alishahiha:2004eh} for a D3-brane, in which the scalar $\phi$ parametrises a canonical normalisation of the radial position $r$ of the D3-brane in terms of the tension of the brane $T_3$: $\phi \equiv \sqrt{T_3} r$. 
The warp factor is also a function of $\phi$, namely $h(\phi) \equiv T_3^{-1} h(r)$, holding the geometrical information of the warped throat region in the compactified space. The term $V(\phi)$ is a potential function. The last two terms denote the actions for the standard model fields, $\psi_i$, and the dark D3-brane, where the matter fields $\chi_j$ reside. In the latter, these fields follow geodesics determined by the induced metric on the brane $\bar{g}_{\mu \nu}$, connected to $g_{\mu\nu}$ through a disformal transformation as given by \cref{trans}. For this study, we assume that the matter fields within the hidden D-brane are \ac{ddm}, inherently coupled to the scalar field $\phi$.  

Before moving on to the cosmological implications of the described setting, we remark on the assumptions underlying the low-energy effective action \cite{Koivisto:2013fta}. The dark D-brane was implicitly assumed to be a \textit{probe brane}, meaning that its presence does not induce back-reaction effects on the background geometry. This implies that any extra degrees of freedom that might emerge due to the propagation of the brane in the higher-dimensional space are neglected in this treatment. We will study the evolution of cosmological perturbations to the background cosmology within this model, which are minute and only treated in linear perturbation theory. In addition, the brane never enters a highly relativistic state, rendering any resulting back-reaction on the bulk geometry inconsequential. Hence, for this investigation, the action given in \cref{action} offers a satisfactory description of disformally coupled dark matter in this model. It is worth noting, however, that when studying the spectrum of non-linear cosmological perturbation, corrections to the action in \cref{action} may become indispensable due to emergent degrees of freedom.

In the following discussions, we will address the particular scenario emerging from this framework in which the conformal and disformal functions are not independent and depend on the warp factor $h(\phi)$. Nevertheless, for completeness, we present the general equations for $C(\phi)$ and $D(\phi)$. This will elaborate on how the metrics embody the phenomenological interactions between the DBI scalar field and the disformal dark matter, recalling that standard model particles remain decoupled from the dark sector.

From the action in \cref{action}, the Einstein equations must be derived as

\begin{equation}
G_{\mu \nu} \equiv R_{\mu \nu} - \frac{1}{2} g_{\mu \nu} R = \kappa^2 \left( T_{\mu \nu}^{\phi} + T_{\mu \nu}^{c} + T_{\mu \nu}^{S} \right) \mathcomma
\label{eins}
\end{equation}

where each component of the energy-momentum tensor is defined accordingly as

\begin{equation}
T_{\mu \nu}^{\phi} = - \frac{2}{\sqrt{-g}} \frac{\delta \left( \sqrt{-g} \mathcal{L}_{\phi} \right)}{\delta g^{\mu \nu}}\mathcomma \quad
T_{\mu \nu}^{c} = - \frac{2}{\sqrt{-g}} \frac{\delta \left( \sqrt{-\bar{g}} \mathcal{L}_{c} \right)}{\delta g^{\mu \nu}}\mathcomma \quad
T_{\mu \nu}^{S} = - \frac{2}{\sqrt{-g}} \frac{\delta \left( \sqrt{-g} \mathcal{L}_{S} \right)}{\delta g^{\mu \nu}}\mathperiod
\label{tmunu}
\end{equation}

Due to the coupling in the dark sector, the energy-momentum tensors for both the scalar field and dark matter are not individually conserved. Nevertheless, to preserve general covariance, the total energy-momentum tensor must still be conserved, as dictated by the Bianchi identities, expressed as

\begin{equation}
\nabla_{\mu} \left( T^{\mu \nu}_{\phi} + T^{\mu \nu}_{c} \right) = 0, \quad \nabla_{\mu} T^{\mu \nu}_{S} = 0 \mathperiod
\label{emtcons}
\end{equation}

The equation of motion for the scalar field, derived from variation of the action in \cref{action}, is generally given by

\begin{equation}
\nabla_{\mu} \left( \gamma \partial^{\mu} \phi \right) - V_{,\phi} + \frac{h_{,\phi}}{h^2} \frac{\gamma}{2} \left( \gamma^{-1} - 1 \right)^2 = \nabla_{\mu} \left[ \frac{D}{C} T_{c}^{\mu \alpha} \partial_{\alpha} \phi \right] - \frac{1}{2} \left[ \frac{C_{,\phi}}{C} T_{c} + \frac{D_{,\phi}}{C} T_{c}^{\mu \nu} \partial_{\mu} \phi \partial_{\nu} \phi \right] \mathcomma
\label{eqmogen}
\end{equation}

where the subscript $\phi$ indicates differentiation with respect to the scalar field, $T_{c} \equiv g_{\mu \nu} T_{c}^{\mu \nu}$ is the trace of the dark matter energy-momentum tensor, and

\begin{equation}
\gamma = \frac{1}{\sqrt{1 + h(\phi) \partial^{\mu} \phi \partial_{\mu} \phi}} \mathcomma
\end{equation}

is a Lorentz factor for the brane's motion, quantifying its relativistic character, which is always real. When $\gamma \to 1$, the canonical kinetic term is restored, and the scalar field $\phi$ resembles the quintessence field dynamics. On the other hand, the framework enters a purely relativistic regime in the limit $\gamma \to \infty$.

Until now, the equations were derived for general conformal and disformal functions. We will now converge to the scenario where the disformal metric in \cref{trans} is the induced metric on a probe D3-brane moving in a warped higher-dimensional space-time. In this context, the functions $C$ and $D$ in the transformation have been shown to be expressed in terms of the warp factor of the brane $ h(\phi)$ \cite{Koivisto:2013fta}:

\begin{equation}
C(\phi) = \left[ T_3 h(\phi) \right]^{-1/2}, \quad D(\phi) = \left[ h(\phi) / T_3 \right]^{1/2} \mathperiod
\label{transhcd}
\end{equation}

In this case, \cref{eqmogen} reduces to 

\begin{equation}
\nabla_{\mu} \left( \gamma \partial^{\mu} \phi \right) - V_{, \phi} + \frac{h_{, \phi}}{2h^2} \gamma \left( \gamma^{-1} -1 \right)^2 = \nabla_{\mu} \left[ h(\phi) T_{c}^{\mu \nu} \partial_{\nu} \phi \right] - \frac{T_{c}^{\mu \nu}}{4} \left[ -\frac{h_{, \phi}}{h} g_{\mu \nu} + h_{, \phi} \partial_{\mu} \phi \partial_{\nu} \phi \right] \mathperiod
\end{equation}

Under the assumption of a homogeneous and isotropic Universe, all the matter species in the theory can be accurately modelled as perfect fluids. For cold dark matter, this implies an energy-momentum tensor expressed as

\begin{equation}
T^{c}_{\mu \nu} = \rho_c u^c_{\mu} u^c_{\nu} \mathcomma
\label{tmunuddm}
\end{equation}

where $u^c_{\mu}$ is the four-velocity of the fluid for a comoving observer and $\rho_c$ is the energy density.

From \cref{tmunu}, defining the energy-momentum tensor for the scalar field, and from the DBI action in \cref{action}, we find

\begin{equation}
T_{\mu \nu}^{\phi} = \left( \frac{1-\gamma^{-1}}{h} - V \right) g_{\mu \nu} + \gamma \partial_{\mu} \phi \partial_{\nu} \phi \mathperiod
\label{tmunudef}
\end{equation}

For the dark energy fluid, the perfect fluid form implies

\begin{equation}
u_{\mu}^{\phi} = \frac{\partial_{\mu} \phi}{\sqrt{-\partial_{\nu} \phi \partial^{\nu} \phi}} \mathcomma
\end{equation}

for the scalar field's four-velocity, and

\begin{equation}
\rho_{\phi} = \frac{\gamma -1}{h} + V\mathcomma \quad \text{and} \quad p_{\phi} = \frac{1 - \gamma^{-1}}{h} - V \mathcomma
\label{rhopphi}
\end{equation}

for the energy density and pressure, respectively. Additionally, the \ac{eos} parameter for the scalar field is defined as

\begin{equation}
w_{\phi} = \frac{p_{\phi}}{\rho_{\phi}} = \frac{\left( \gamma - 1 \right) / h\gamma - V}{\left( \gamma - 1 \right) / h + V} \mathperiod
\label{wphi}
\end{equation}

The conservation relations given in \cref{emtcons} can be manipulated to yield an explicit expression for the coupling $Q$ between the dark sector fluids:

\begin{equation}
\nabla_{\mu} T^{\mu \nu}_{\phi} = Q \partial^{\nu} \phi \mathcomma
\label{cons}
\end{equation}

with

\begin{equation}
Q = \nabla_{\alpha} \left[ h(\phi) T_{c}^{\alpha \beta} \partial_{\beta} \phi \right] - \frac{h_{, \phi}}{4 h} T_{c}^{\alpha \beta} \left[ - g_{\alpha \beta} + h \partial_{\alpha} \phi \partial_{\beta} \phi \right] \mathperiod
\label{Qcov}
\end{equation}

The interaction term $Q$ accounts for the energy exchange between the dark energy and dark matter sectors and is central to the discussions in this work.

\section{Background FLRW Cosmology} \label{sec:back}

To study the cosmological implications of the \ac{ddb} model we consider a spatially flat Friedmann-Lema\^{i}tre-Robertson-Walker (FLRW) background, as defined in \cref{eq:frwsph}.


The scalar field $\phi$ is considered to be homogeneous, simply written as $\phi = \phi(\tau)$. In the dark D-brane disformal frame, in which the dark matter geodesics are defined, the line element is related to \cref{eq:frwsph} as

\begin{equation}
\odif{\bar{s}}^2 = C a^2(\tau) \left( - Z^2 \odif{\tau}^2 +  \odif{x}^2 + \odif{y}^2 + \odif{z}^2  \right) \mathcomma
\label{dst}
\end{equation}

where $Z$ represents the disformal scalar, related to the Jacobian of the metric transformation. In particular, with $C$ and $D$ given by \cref{transhcd}, the disformal scalar is found to be 

\begin{equation}
    Z \equiv \sqrt{1 - 2 X \frac{D}{C}} = \sqrt{1 - h(\phi) \frac{\phi'^2}{a^2}} = \frac{1}{\gamma} \mathcomma
    \label{zdef}
\end{equation}

where $X$ is the standard kinetic term for the scalar field, defined as $X = -\frac{1}{2}g^{\mu\nu} \partial_\mu\phi \partial_\nu \phi = \frac{\phi'^2}{2a^2}$. The implications of \cref{dst} are that while the conformal factor influences the entire line element and thus dilutes dark matter, the disformal factor impacts the time component only, altering the light cones of dark matter particles.

With $T^{\mu \nu}_{c}$ as provided in \cref{tmunuddm}, the coupling term in \cref{Qcov} can be recast as

\begin{equation}
Q = \frac{\rho_c}{2 C a^2} \left[ \phi'^2 D_{, \phi} + a^2 C_{, \phi} - 2 \phi'^2 \frac{D C_{, \phi}}{C} + 2D (\phi'' + \mathcal{H} \phi' + \frac{\rho_c'}{\rho_c} \phi') \right] \mathperiod
\label{Q0dc}
\end{equation}

For the particular definitions of the conformal and disformal functions for the dark D-brane in \cref{transhcd}, \cref{Q0dc} reduces to

\begin{equation}
Q = h \rho_c a^{-2} \left[ \frac{3}{4} \frac{\phi'^2 h_{, \phi}}{h} - \frac{a^2 h_{, \phi}}{4h^2} + \phi' \left(2 \mathcal{H} + \frac{\rho_c'}{\rho_c} \right) + \phi'' \right] \mathperiod
\label{Q0}
\end{equation}

The equation of motion for the scalar field, given by \cref{eqmogen}, in conjunction with \cref{Q0dc}, results in 

\begin{equation}
\phi'' - \mathcal{H} \left( 1 - 3 \gamma^{-2} \right) \phi' + \frac{a^2 h_{, \phi}}{2 h^2} \left( 1 - 3 \gamma^{-2} + 2 \gamma^{-3} \right) + a^2 \gamma^{-3} \left( V_{, \phi} + Q \right) = 0 \mathcomma
\label{phidd}
\end{equation}

where $Q$ is as specified in \cref{Q0}. Given that the brane's warp factor $h(\phi)$ is always non-negative, it is always warranted that $\gamma \geq 1$. The nonrelativistic regime is achieved when $a^{-2} h \phi'^2 \ll 1$ and $\gamma \to 1$. In this non-trivial regime, which depends on $\phi$ (through $h$) and $\phi'$, the DBI Lagrangian in \cref{action} simplifies to the standard quintessence case. 

It should also be emphasised that while the DBI action in \cref{action} can be thought of as a specific instance of the k-essence action presented in \cref{sec:kessence} \cite{PhysRevD.62.023511,ArmendarizPicon:2000ah}, the DBI scalar field's pressure can only become negative for a non-vanishing potential $V(\phi)$. In the limit $V(\phi) \to 0$, from \cref{wphi}, we note that $w_{\phi} \to 1/\gamma$, which is invariably non-negative. On the other hand, in the slow-roll limit where the kinetic term's influence is negligible, the scalar field mimics a cosmological constant, with $w_{\phi} \approx -1$. Thus, the coupled DBI model is a comprehensive framework that embodies various models previously explored, such as coupled quintessence (see \cref{chap:cquint}) \cite{vandeBruck:2015ida, vandeBruck:2017idm, Teixeira:2019hil} and coupled tachyonic dark energy \cite{Gumjudpai:2005ry,Teixeira:2019tfi}. This is reflected in the form of the equation governing the motion of the DBI scalar field in \cref{phidd}, which is considerably more complex than its canonical counterpart, to which it reduces to in the appropriate limit.

From the Einstein field equations, the Friedmann equations can be derived, describing the evolution of each $i$-th fluid component as

\begin{equation}
\mathcal{H}^2 = \frac{\kappa^2 a^2}{3} \sum_i \rho_i = \frac{\kappa^2 a^2}{3} \left( \rho_r + \rho_b + \rho_c + \rho_\phi \right) \mathcomma
\label{fried1}
\end{equation}

and

\begin{equation}
\mathcal{H}' + \mathcal{H}^2 = -\frac{\kappa^2 a^2}{6} \sum_i \left( \rho_i + 3 p_i \right) = -\frac{\kappa^2 a^2}{6} \left( \rho_r + 4 p_r + \rho_b + \rho_c + \rho_\phi + 3 p_\phi \right) \mathcomma
\end{equation}

where we include the standard model's matter sectors of the Universe: baryons and radiation, denoted by subscripts $b$ and $r$. As they are non-interacting, their evolution follows  the conventional conservation laws

\begin{equation}
\rho_r' + 4 \mathcal{H} \rho_r = 0 \mathcomma \quad \text{and}\quad \rho_b' + 3 \mathcal{H} \rho_b = 0 \mathcomma
\label{rhob}
\end{equation}

where $p_b=0$ and $p_r=\frac{1}{3} \rho_r$ for each respective fluid. Instead, from \cref{emtcons,cons}, the continuity equations for the interacting disformally coupled fluids are

\begin{equation}
\rho_\phi' + 3 \mathcal{H} \rho_\phi \left( 1 + w_\phi \right) = - Q \phi' \mathcomma
\label{rhophidbi}
\end{equation}

with $w_\phi$ as in \cref{wphi}, and 

\begin{equation}
\rho_c' + 3 \mathcal{H} \rho_c = Q \phi' \mathperiod
\label{rhocdm}
\end{equation}

In the non-relativistic regime, $\gamma \to 1 + \frac{h \phi'^2}{2a^2}$, and the EoS parameter for quintessence is recovered once more. Nevertheless, what distinguishes this setting is precisely the impact and signatures of the non-canonical relativistic behaviour. Using \cref{Q0,phidd}, the expression for $Q$ can be rewritten to include only first-order derivatives of the scalar field

\begin{equation}
Q = -\left[ \frac{h \left( V_{,\phi} + 3 a^{-2} \mathcal{H} \gamma \phi' \right) + \frac{h_{,\phi}}{h} \left( 1 - \frac{3}{4} \gamma \right) }{\gamma + h \rho_c} \right] \rho_c \mathperiod
\label{Qnod}
\end{equation}

The sign of $Q$ in \cref{rhophidbi,rhocdm} indicates the direction of energy transfer in the dark sector. The coupling yields distinct interpretations for each fluid. For the DBI scalar field, the coupling can be combined with the self-interacting potential to create an effective scalar field potential, $V_{\text{eff}} (\phi,\phi')$. The coupling can also be viewed as a local change in the geometry, encoded in $\bar{g}$, which governs the geodesics along which dark matter propagates. To facilitate the comparison with the canonical case, we introduce an effective coupling $\beta$, defined as 

\begin{equation}
\beta = - \frac{Q}{\kappa \rho_c} \mathcomma
\label{beta}
\end{equation}

such that, when $\phi'<0$, positive values of $\beta$ correspond to dark energy granting energy to dark matter and, conversely, when $\beta$ is negative, the dark energy field is being sourced by dark matter.

To compare with cosmological constraints assuming a non-interacting dark sector, we introduce an effective dark energy EoS parameter $w_{\phi, \text{eff}}$ \cite{Das:2005yj}. This can be thought of as a map from the coupled model to an uncoupled one at the background level, where the dark matter does not interact with dark energy, but all the coupling effects are transported into an effective dark energy fluid,

\begin{equation}
\rho_{\phi, \text{eff}} = \rho_\phi + \rho_c - \rho_{c,0} a^{-3} \mathcomma
\label{eq:rhophieff}
\end{equation}

where a subscript $0$ indicates present values.

The Friedmann equation can be reformulated according to \cref{eq:rhophieff} as

\begin{equation}
\mathcal{H}^2 = \frac{\kappa^2 a^2}{3} \left( \rho_{r,0} a^{-4} + \rho_{b,0} a^{-3} + \rho_{c,0} a^{-3} + \rho_{\phi, \text{eff}} \right) \mathcomma
\label{hubble}
\end{equation}

where $\rho_{r,0}$, $\rho_{b,0}$, and $\rho_{c,0}$ represent present-day energy densities for radiation, baryons, and dark matter, respectively. Taking the derivative of \cref{eq:rhophieff}, we obtain

\begin{equation}
\rho_{\phi, \text{eff}}' + 3 \mathcal{H} \rho_{\phi, \text{eff}} (1+ w_{\phi, \text{eff}}) = 0 \mathcomma
\end{equation}

consistent with a standard, uncoupled model. Comparison with \cref{rhophidbi,rhocdm} yields 

\begin{equation}
w_{\phi, \text{eff}} = \frac{p_\phi}{\rho_{\phi, \text{eff}}} \mathcomma
\label{weffphi}
\end{equation}

with $p_\phi$ as defined in \cref{rhopphi}. This formulation allows for a straightforward comparison between the background cosmological evolution of the disformally coupled dark sector in this model and empirical observations.


\subsection{Qualitative Dynamics and Initial Conditions}

In the present study, we focus on an AdS$_5$ throat model for the warp factor, endowed with a quadratic scalar field potential, expressed as

\begin{equation}
h(\phi) = \frac{h_0}{\phi^4}, \quad \text{and}\quad V(\phi) = \frac{V_0 \phi^2}{\kappa^2} \mathcomma
\label{hvmodel}
\end{equation}

where $h_0, V_0 > 0$ are dimensionless parameters in this definition. Four parameters govern the cosmological evolution in this model: the scales of the warp factor and the potential, namely $h_0$ and $V_0$, and the initial conditions for the scalar field and its rate of change, denoted by $\phi_{\text{i}}$ and $\phi'_{\text{i}}$. As discussed in \cite{Koivisto:2013fta}, the geometrical scales can be encapsulated in a single, dimensionless parameter

\begin{equation}
  \Gamma_0 \equiv \frac{V_0}{\td{h}_0}\mathcomma \quad \text{with} \quad \td{h}_0 = \frac{1}{h_0} \mathperiod
  \label{Gamma0}
\end{equation}

This combined parameter plays a crucial role in determining the dynamics near fixed-point solutions of the dynamical system of equations and has a significant impact on the ultimate fate of the Universe. The disformally coupled framework under consideration here has the same number of parameters as its standard DBI uncoupled counterpart \cite{Guo:2008sz,Gumjudpai:2009uy,Copeland:2010jt}. This implies that the conformal and disformal effects are intertwined, and the uncoupled scenario cannot be recovered from the limit values of these parameters.

A dynamical system analysis of the background cosmological evolution for different values of the parameters has been conducted in Ref.~\cite{Koivisto:2013fta} (see \cite{Bahamonde:2017ize} for a review on dynamical systems applied to cosmology). This methodology allows us to qualitatively understand the Universe's evolution through the fixed point solutions\footnote{Each realisation of the model maps to a specific trajectory in the phase space, encapsulating the Universe's evolutionary dynamics. The fixed points serve as the asymptotic scenarios of this evolution, illustrating distinct epochs in the cosmological history of the Universe.}, without the need for the full numerical evolution. 
As elaborated in \cref{chapter:sttheories}, a particularly compelling aspect of introducing couplings into the dark sector is the potential emergence of scaling fixed point solutions. In such solutions, both components dilute at an identical rate, resulting in a consistent ratio of their fractional energy densities. This offers a more intuitive explanation for the observed distribution of energy in the Universe.
The authors of Ref.~\cite{Koivisto:2013fta} reported that when $\Gamma_0 > 1$, interesting particular cosmological solutions appear, including a saddle point with a scaling behaviour, \cref{eq:scale}, between dark energy and dark matter. In this regime, dark matter could be driving the cosmic expansion, and the observed acceleration of the expansion could be initiated during the matter-dominated era, fuelled by the non-minimal coupling. Regardless of the initial conditions, owing to the repelling nature of the saddle point, the DDM starts diluting away, and the system eventually transitions to a standard accelerating solution dominated by the evolution of the DBI scalar field. 

The initial conditions for $\phi$ and $\phi'$ play an instrumental role in determining the onset and duration of the scaling and attractor regimes, respectively, more relativistic scenarios corresponding to a longer transition. However, if this scaling is happening at present, then the effects of $\phi'_{\text{i}}$ are not currently observable. In contrast, $\phi_{\text{i}}$ significantly influences the background and perturbation evolution. The system remains in a frozen latent state with $\gamma \approx 1$ and $w_\phi \approx -1$ during matter and radiation-dominated periods. Taking larger values of $\Gamma_0$, both the scaling and dark energy dominated solutions converge to a de Sitter-like state, with dynamics dominated by the potential, which leads to $w_{\phi} \rightarrow -1$, making the Universe increasingly similar to a $\Lambda$CDM cosmology near the fixed points, even though this may no longer hold at the level of the perturbations.
For this study, we limit our attention to scenarios where $\phi_{\text{i}} > 0$ and $\phi'_{\text{i}} < 0$ to maintain consistency with the extra-dimensional interpretation of the field $\phi$ as the radius as the brane moves down the throat, and to reproduce cosmologies that allow for the scaling regime.

In the following sections, we will present a numerical examination of the cosmological implications of this model, focusing on scenarios that yield a slowly evolving cosmological constant-like behaviour in the early Universe \cite{Pettorino:2013ia}. Since disformal dark matter can contribute to the accelerated expansion, we expect unique signatures compared to standard $\Lambda$CDM and coupled quintessence models.

\section{Linear Perturbations} \label{sec:pert}

In this section, we explore the dynamics of cosmological perturbations. Based on the background analysis, we anticipate that the dark D-brane model will exhibit a non-trivial and rich linear-level phenomenology, which we will connect with predictions according to existing observational data in \cref{sec:dbi_obs}. We show particular predictions for both the temperature anisotropy spectrum of the CMB and the matter power spectrum. The equations presented in this chapter are tailored for the Newtonian gauge and are specific to the model under discussion. Nevertheless, for a more comprehensive perspective, we provide the equations for the synchronous gauge, along with those for generic conformal and disformal functions, $C(\phi)$ and $D(\phi)$.

\subsection{Conformal Newtonian Gauge}

We first focus on scalar perturbations in the conformal Newtonian gauge, as described in \cref{sec:newtgauge}, where the perturbed line element takes the form

\begin{equation}
\odif{s}^2= a^2(\tau) \left[ - \left( 1 + 2 \Psi \right) \odif{\tau}^2 + \left( 1-2\Phi\right) \delta_{ij} \odif{x}^{i} \odif{x}^{j} \right] \mathcomma
\label{pertmet}
\end{equation}

where $\Psi (\tau, x^{i})$ and $\Phi (\tau, x^{i})$ are the scalar metric perturbations. From \cref{eins,pertmet}, the components of the perturbed Einstein equations are readily obtained and equated as

\begin{equation}
\delta G\indices{^{\mu}_{\nu}} = \kappa^2 \delta T\indices{^{\mu}_{\nu}} \mathcomma
\label{perteinst}
\end{equation}

where $\delta G\indices{^{\mu}_{\nu}}$ and $\delta T\indices{^{\mu}_{\nu}}$ are the perturbed Einstein and energy-momentum tensors, respectively. For each fluid, the distinct components of $\delta T\indices{^{\mu}_{\nu,f}}$ are given by

\begin{align}
    \delta T\indices{^{0}_{0,f}} &= - \delta \rho_f \mathcomma \label{dt1} \\
    \delta T\indices{^{0}_{i,f}} &= \left( \rho_f + p_f \right) \partial_i v_f \mathcomma \label{dt2} \\
    \delta T\indices{^{i}_{0,f}} &= - \left( \rho_f + p_f \right) \partial^{i} v_f \mathcomma \label{dt3} \\
    \delta T\indices{^{i}_{j,f}} &= \delta p_f \delta^{i}_{j} + \Pi^{i}_{~j,f} \mathcomma  \label{dt4}
\end{align}

where $f$ is an index running over each fluid. Consistently, $\delta \rho_f$, $\delta p_f$, $v_f$, and $\Pi^{i}_{j,f}$ denote the perturbed energy density, pressure perturbation, peculiar velocity potential, and anisotropic stress tensor for the fluid $f$, respectively. Given a specific matter source, the perturbed energy-momentum tensor is defined correspondingly. For the model in focus, the Fourier space representations of the perturbed Einstein equations are as follows:

\begin{align}
k^2 \Phi + 3 \mathcal{H} \left( \Phi'+ \mathcal{H} \Psi \right) &= -4 \pi G a^2 \sum_f \delta \rho_f \mathcomma
\label{eq:peq11} \\
k^2 \left( \Phi' + \mathcal{H} \Psi \right) &= 4 \pi G a^2 \sum_f \rho_f \left( 1 + w_f \right) \theta_f \mathcomma
\label{eq:peq22} \\
\Phi'' + \mathcal{H} \left( \Psi' + 2 \Phi' \right) + \Psi \left( \mathcal{H}^2 + 2 \mathcal{H}' \right) + \frac{k^2}{3} \left( \Phi - \Psi \right) &= 4 \pi G a^2 \sum_f \delta p_f \mathcomma
\label{eq:peq33} \\
k^2 \left( \Phi - \Psi \right) &= 12 \pi G a^2 \sum_f \rho_f \left( 1 + w_f \right) \sigma_f \mathperiod
\label{eq:peq44}
\end{align}

These equations establish a relationship between the scalar potentials $\Phi$ and $\Psi$ and the perturbations in the matter fluids. Fourier expansion has been applied to the metric potentials, effectively replacing spatial derivatives with Fourier modes characterised by the wave number $k$. Additionally, the velocity potential has been implicitly defined as $\theta_f = \nabla^i \nabla_i v_f$. For direct comparison purposed with the canonical case we introduce the anisotropic stress perturbation given by $\sigma_f$, according to \cite{Mifsud:2017fsy}.
As a first approximation, we focus on the scenario where $\sigma_f$ vanishes for all fluids. According to the fourth Einstein perturbation equation, \cref{eq:peq44}, this leads to $\Psi = \Phi$. Moving forward, we replace the perturbed energy density $\delta \rho_f$ with the density contrast $\delta_f = \frac{\delta \rho_f}{\rho_f}$.
We further assume that each fluid features an adiabatic speed of sound $c_{s,f}^2 = \frac{\delta p_f}{\delta \rho_f}$.

The perturbed conservation equations derive from the zero-divergence condition on the energy-momentum tensor, $\nabla_{\nu} T^{\mu\nu}_f = 0$, which must now be distinguished for uncoupled and coupled scenarios:

\begin{equation}
    \begin{aligned}
        \nabla_{\mu} \delta T^{\mu}_{\nu,u} + \delta \Gamma^{\mu}_{\mu \beta} T^{\beta}_{\nu, u} - \delta \Gamma^{\beta}_{\mu \nu} T^{\mu}_{\beta, u} = 0 \mathcomma
    \end{aligned}
    \label{contpert1}
\end{equation}

\begin{equation}
    \begin{aligned}
        \nabla_{\mu} \delta T^{\mu}_{\nu,c} + \delta \Gamma^{\mu}_{\mu \beta} T^{\beta}_{\nu, c} - \delta \Gamma^{\beta}_{\mu \nu} T^{\mu}_{\beta, c} = - Q \partial_{\nu} \phi \mathcomma
    \end{aligned}
    \label{contpert2}
\end{equation}

where $f = \{u, c\}$ was introduced to specify uncoupled and coupled fluids relative to the scalar field. 
The perturbations of the Christoffel symbols are denoted by $\delta \Gamma^{\mu}_{\nu \beta}$. From \cref{dt1,dt2,dt3,dt4}, we can rewrite \cref{contpert1,contpert2} for the evolution of the density contrast $\delta_f$ and the velocity potential $\theta_f$ for each fluid. Specifically, the baryonic and radiation fluids remain uncoupled from the scalar field and are described by the following conservation equations:

\begin{align}
\delta_u' + 3 \mathcal{H} (c_{s,u}^2-w_u) \delta_u &= (1+w_u) (3 \Phi' - \theta_u) \mathcomma \label{conseu} \\
\theta_u' + \mathcal{H} (1-3 w_u)\theta_u + \frac{w_u'}{1+w_u} \theta_u &= k^2 \left[ \frac{c_{s,u}^2}{1+w_u} \delta_u + \Psi \right] - k^2 \sigma_u \mathcomma \label{conseu2}
\end{align}

where $u=\{b,r\}$ and $w_u = p_u/\rho_u$ is the EoS parameter. The first equation represents the perturbed continuity equation, and the second one is the Euler equation, which emerges from the temporal and spatial components of the energy conservation equation, respectively.

For the disformal dark matter component, which is the only species coupled to the scalar field in this setting, the continuity and Euler equations transform as follows, considering $w_c=c_{s,c}=0$:

\begin{align}
\delta_c' &= - (\theta_c - 3 \Phi') - \frac{Q}{\rho_c} \phi' \delta_c + \frac{Q}{\rho_c} \delta \phi' + \frac{\delta Q}{\rho_c} \phi' \mathcomma \label{CDM1} \\
\theta_c' + \mathcal{H} \theta_c &= k^2 \Psi - \frac{Q \phi'}{\rho_c} \theta_c + k^2 \frac{Q}{\rho_c} \delta \phi \mathperiod \label{CDM2}
\end{align}

The perturbation of the DBI scalar field evolves according to the perturbed Klein-Gordon equation:

\begin{align}\label{PerturbKG}
&\delta \phi'' + \left[ \frac{3h_{,\phi}}{h} \left( 1- \gamma^{-1} \right) \phi'    - \mathcal{H} \left( 7 - 9 \gamma^{-2}  \right)   - 3 h  \left( V_{,\phi} + Q \right) \gamma^{-1} \phi' \right] \delta \phi' + \left[-\frac{3h_{,\phi}}{h} \mathcal{H} \left( 1 - \gamma^{-2} \right) \phi'  \right. \nonumber \\
& \left. + a^2 V_{,\phi \phi} \gamma^{-3} + \frac{h_{,\phi \phi}}{2h^2} a^2 \left( 1 - 3\gamma^{-2} + 2\gamma^{-3} \right) - \frac{3}{2} \frac{h_{,\phi}}{h} a^2 \left( V_{,\phi} + Q \right) \left( \gamma^{-1} - \gamma^{-3} \right) \right.  \\
&\left. + \frac{h_{,\phi}^2}{2 h^3} a^2 \left( 1 - 3\gamma^{-1} + 3\gamma^{-2} - \gamma^{-3} \right) \right] \delta \phi + \left[ 6 \mathcal{H} \left( 1- \gamma^{-2} \right) \phi'  - \frac{h_{,\phi}}{h^2} a^2  \left( 2 - 3 \gamma^{-1} + \gamma^{-3} \right) \right.  \nonumber \\
& \left. + a^2  \left( V_{,\phi} + Q \right) \left(3\gamma^{-1} -\gamma^{-3} \right) \right] \Psi - \phi'  \Psi' - 3 \gamma^{-2} \phi'  \Phi' - \gamma^{-2} \partial^i \partial_i \delta \phi+ a^2 \gamma^{-3} \delta Q=0 \mathperiod \nonumber
\end{align}

We will present the perturbed terms stemming from the introduction of the coupling in this setting, directly dependent on the warp factor $h(\phi)$. Nevertheless, we derive first the general expression of the perturbed disformal coupling $\delta Q$ for general conformal and disformal functions in the Newtonian gauge for completeness and as it may be relevant for extensions to this setting scenarios. 

\begin{equation}
\delta Q = \frac{a^{-2} \rho_c}{ C - \frac{D}{h} \left(1- \gamma^{-2} \right) + D \rho_c \gamma^{-3} } \left( \mathcal{Q}_1 \delta_c + \mathcal{Q}_2 \Phi' + \mathcal{Q}_3 \Psi + \mathcal{Q}_4 \delta \phi' + \mathcal{Q}_5 \delta \phi \right) \mathcomma
\label{dq}
\end{equation}

\noindent with

\begin{align}
\mathcal{Q}_1 = &\frac{1}{2} a^2 C_{,\phi} \left( 1 - 3 \frac{\delta p_c}{\delta \rho_c} \right) - 3 D \mathcal{H}  \frac{\delta p_c}{\delta \rho_c}  \phi' - \frac{D C_{, \phi}}{C} \phi'^2 +  \frac{D_{, \phi}}{2 }  \phi'^2  \\
&- a^2 \frac{D}{2} \frac{h_{, \phi}}{h^2} \left( 1 - \frac{3}{2} \gamma^{-2} \right) + a^2 D \frac{Q}{h \rho_c} \gamma^{-2} \mathcomma \nonumber \\
\mathcal{Q}_2 =& 3 D \left( \gamma^{-2} + w \right) \phi' \mathcomma \\
\mathcal{Q}_3 = &3 D \mathcal{H} \left( 1 +  \gamma^{-2} + 2w \right)  \phi' - a^2 \frac{D_{,\phi}}{h} \left( 1 - \gamma^{-2} \right) + \frac{3}{4} a^2 D \frac{h_{,\phi}}{h^2} \left( 1 - \gamma^{-2} \right) \nonumber  \\
& + 2 a^2 \frac{D}{h} \frac{C_{,\phi}}{C} \left( 1 - \gamma^{-2} \right) \nonumber + a^2 \frac{D}{h} \frac{Q}{\rho_c} \left( 1 - \gamma^{-2} \right) \mathcomma \\
\mathcal{Q}_4 =  &3D \mathcal{H} \left( 2 - 3 \gamma^{-2} - w_c \right) - 2 \frac{D C_{, \phi}}{C} \phi' + D_{, \phi} \phi' - 3 D \frac{h_{, \phi}}{h} \left( 1 - \gamma^{-1} \right) \phi'\\
&+ 3 D \mathcal{H} \gamma^{-1} \left( V_{, \phi} + Q \right) \phi' + 2 D \frac{Q}{\rho_c} \phi' \mathcomma \nonumber \\
\mathcal{Q}_5 = &- k^2 D \left( \gamma^{-2} + w_c \right)  + 3  \mathcal{H}  \left( \frac{D C_{, \phi}}{C} - D_{, \phi} \right) w_c  \phi' - \frac{a^2}{2} \frac{C_{, \phi}^2}{C} \left( 1 - 3 w_c \right) + 2  \frac{D C_{, \phi}}{C }  \phi'^2   \nonumber  \\
&  -  \frac{3}{2} \frac{C_{, \phi} D_{, \phi}}{C } \phi'^2 - \frac{3}{2} D \frac{h_{, \phi}}{h} \mathcal{H} \left( 1 - \gamma^{-2} \right) \phi'  + \frac{a^2}{2} \frac{D C_{, \phi}}{C} \frac{h_{, \phi}}{h^2} \left( 1 - \frac{3}{2} \gamma^{-2} \right)  \\
& - \frac{a^2}{2} D_{, \phi} \frac{h_{, \phi}}{h^2} \left( 1 - \frac{3}{2} \gamma^{-2}  \right) + \frac{a^2}{2} D \frac{h_{, \phi}^2}{h^3} \left( \frac{5}{4}  - \frac{21}{4} \gamma^{-2} + 4\gamma^{-3} \right) + \frac{1}{2} a^2 C_{, \phi \phi} \left(1 - 3 w_c \right)  \nonumber \\
&  -  \frac{D C_{, \phi \phi}}{C } \phi'^2 + \frac{D_{, \phi \phi}}{2}  \phi'^2  - \frac{a^2}{2} D \frac{h_{, \phi \phi}}{h^2} \left( 1 -3 \gamma^{-2} + 2 \gamma^{-3} \right) - a^2 D \gamma^{-3} V_{, \phi \phi}  \nonumber\\
&+  \frac{Q}{h \rho_c} \left[a^2 D_{, \phi} -  a^2\frac{D C_{, \phi}}{C } - \frac{3}{2} D h_{, \phi} \phi'^2 \right]  \nonumber \mathcomma
\end{align}

for general conformal and disformal functions, $C(\phi)$ and $D(\phi)$, respectively.

Finally, the perturbation of the coupling $Q$ as introduced in \cref{Qnod} for the dark D-brane model we are investigating, reduces to:

\begin{equation}
\delta Q = \frac{a^{-2} \rho_c}{\gamma^{-2} + h \rho_c \gamma^{-3} } \left( \mathcal{Q}_1 \delta_c + \mathcal{Q}_2 \Phi' + \mathcal{Q}_3 \Psi + \mathcal{Q}_4 \delta \phi' + \mathcal{Q}_5 \delta \phi \right),
\label{deltaQ}
\end{equation}

\noindent with the following perturbed coefficients:

\begin{align}
\mathcal{Q}_1 = & a^2 \frac{Q}{\rho_c}\gamma^{-2} + 3 h \frac{\delta p_c}{\delta \rho_c} \left(  a^2 \frac{h_{,\phi}}{4h^2} - \mathcal{H} \phi' \right) \mathcomma \\
\mathcal{Q}_2 = & 3 h \left( \gamma^{-2} + w \right) \phi' \mathcomma \\
\mathcal{Q}_3 = & 3 h \mathcal{H} \left( 1 +  \gamma^{-2} + 2w \right) \phi'  - a^2 \frac{3}{4} \frac{h_{,\phi}}{h} \left( 1 - \gamma^{-2} \right) + a^2 \frac{Q}{\rho_c} \left( 1 - \gamma^{-2} \right) \mathcomma \\
\mathcal{Q}_4 = & 3 h \mathcal{H} \left( 2 - 3 \gamma^{-2} - w \right) + 3 h^2 \left( V_{,\phi} + Q \right) \gamma^{-1} \phi' + 2 h \frac{Q}{\rho_c} \phi' - \frac{3}{2} h_{,\phi} \left( 1 - 2 \gamma^{-1} \right) \phi' \mathcomma \\
\mathcal{Q}_5 = & -k^2 h \left( \gamma^{-2} + w \right) + a^2 \frac{h_{,\phi}^2}{2 h^2} \left( \frac{3}{4} -\frac{15}{4} \gamma^{-2}  + 4 \gamma^{-3} - \frac{3}{2} w \right) + a^2 \frac{3}{4} \frac{h_{,\phi \phi}}{h} \left( \gamma^{-2} - \frac{4}{3} \gamma^{-3} + w \right) \nonumber  \label{q5} \\
& - \frac{3}{2} h_{,\phi} \mathcal{H} \left( 1 - \gamma^{-2} +2w \right) \phi' - a^2 h V_{,\phi \phi} \gamma^{-3} - a^2 \frac{h_{,\phi}}{2h} \frac{Q}{\rho_c} \left( 1 - 3 \gamma^{-2} \right) \mathperiod 
\end{align}

Our analysis confirms that in the limit $ h \phi'^2 \ll a^2 $ and $ \gamma \rightarrow 1 $, we retrieve the disformal quintessence scenario as previously investigated in \cite{vandeBruck:2015ida,Mifsud:2017fsy}. Additionally, the form of $ \delta Q $ illustrates that disformal couplings introduce a direct dependence on the Fourier scale $k $, namely through the first term of $ \mathcal{Q}_5 $ in \cref{q5}. This scale-dependence is a characteristic trait of disformal couplings \cite{Zumalacarregui:2012us, vandeBruck:2015ida, Mifsud:2017fsy}, and we will show how it reflects in the growth and distribution of perturbations through the matter power spectrum.

For the DBI scalar field, and focusing on adiabatic perturbations, the sound speed $ c_{s,\phi}^2 $ is defined according to

\begin{equation}
    c_{s,\phi}^2 \equiv \left( \frac{\partial p}{\partial X} \right) \left( \frac{\partial \rho}{\partial X} \right)^{-1} = \frac{1}{\gamma^2} \leq 1 \mathperiod
\end{equation}

This quantity is always positive, ensuring the stability of the perturbations. The fact that the DBI scalar field can accommodate for $ c_{s,\phi}^2 \neq 1 $ is correlated with its unique features, imprinted in the cosmic microwave background temperature fluctuations and the matter power spectrum - two key observables we will explore in the numerical study.


\subsection{Synchronous Gauge}

In this study, we have considered the equations in the synchronous gauge, as introduced in \cref{sec:sync} for the purpose of cross-checking the equations and the cosmological predictions. Therefore, to offer a comprehensive view, this section presents the perturbation equations in the synchronous gauge, covering general conformal $C(\phi)$ and disformal $D(\phi)$ coupling functions, before particularising to the scenario examined in this work, where both $C(\phi)$ and $D(\phi)$ are tied to the warp factor $h(\phi)$.

In a parallel manner to the derivation for the Newtonian gauge, we compute the perturbed Einstein equations:

\begin{equation}
k^2 \eta - \frac{1}{2} \mathcal{H} \mathpzc{h}' = -4 \pi G a^2 \sum \delta \rho_f \mathcomma
\end{equation}

\begin{equation}
k^2  \eta' = 4 \pi G a^2 \sum \rho_f \left( 1 + w_f \right) \theta_f \mathcomma
\end{equation}

\begin{equation}
\mathpzc{h}'' + 2\mathcal{H} \mathpzc{h}' - 2 k^2 \eta = - 24 \pi G a^2 \sum \delta p_f \mathcomma
\end{equation}

\begin{equation}
\mathpzc{h}'' + 6 \eta'' + 2 \mathcal{H} \left( \mathpzc{h}' + 6 \eta' \right) - 2 k^2 \eta = - 24 \pi G a^2 \sum \rho_f \left( 1 + w_f \right) \sigma_f \mathcomma
\end{equation}

with $\theta_f$, $\sigma_f$ and $v_f^i$ as previously defined.

The perturbed continuity and Euler equations for the uncoupled baryonic and radiation fluids become

\begin{equation}
\delta_u' + 3 \mathcal{H} \left( c_{s,u}^2-w_u \right) \delta_u = - \left(1 + w_u \right) \left( \frac{\mathpzc{h}'}{2} + \theta_u \right) \mathcomma
\end{equation}

\begin{equation}
\theta_u' + \left[ \mathcal{H} \left( 1 - 3 w_u \right) + \frac{w_u'}{1+w_u} \right] \theta_u = k^2 \frac{c_{s,u}^2}{1 + w_u} \delta_u - k^2 \sigma_u \mathcomma
\end{equation}

with $u=\{b,r\}$. For DDM, the coupled shear-free versions of these equations are: 

\begin{equation}
\delta_c' + 3 \mathcal{H} \left( c_{s,c}^2-w_c \right) \delta_c = - \left(1 + w_c \right) \left( \frac{\mathpzc{h}'}{2} + \theta_c \right) - \frac{Q}{\rho_c} \delta_c \phi' + \frac{Q}{\rho_c} \delta \phi' + \frac{\delta Q}{\rho_c} \phi' \mathcomma
\end{equation}

\begin{equation}
\theta_c' + \left[ \mathcal{H} \left( 1 - 3 w_c \right) + \frac{w_c'}{1+w_c} \right] \theta_c = k^2 \frac{c_{s,c}^2}{1 + w_c} \delta_c - \frac{Q}{\rho_c} \phi' \theta_c + k^2 \frac{Q}{\rho_c \left(1+w_c \right)} \delta \phi \mathperiod
\end{equation}




The perturbed Klein-Gordon equation in the synchronous gauge reads 



\begin{align}
&\delta \phi'' + \frac{\mathpzc{h}'}{2} \gamma^{-2} \phi' + \left[  2 \mathcal{H}  + \frac{3}{4} \frac{h_{, \phi}}{h} \phi' +3 \frac{Q}{\rho_c} \phi' \right] \delta \phi' + \left[ k^2 \gamma^{-2}  + \frac{3}{2} \frac{h_{, \phi}}{h} \mathcal{H} \left(1-\gamma^{-2} \right) \phi' \right. \nonumber \\
& \left.  - \frac{a^2}{2} \frac{h_{, \phi}^2}{h^3} \left(\frac{5}{4} + 4 \gamma^{-3} - \frac{21}{4} \gamma^{-2} \right) + \frac{a^2}{2}  \frac{h_{, \phi \phi}}{h^2} \left( 1 + 2 \gamma^{-3} - 3 \gamma^{-2} \right)+ a^{2} V_{, \phi \phi } \gamma^{-3} \right] \delta \phi + a^{2} \gamma^{-3} \delta Q = 0 \mathcomma
\end{align}

with the general perturbation of the coupling function $Q$ expressed as 

\begin{equation}
\delta Q = \frac{a^{-2} \rho_c}{C - \frac{D}{h} \left( 1- \gamma^{-2} \right) + D \gamma^{-3} \rho_c} \left( \mathcal{Q}_1 \delta_c + \mathcal{Q}_2 \mathpzc{h}' + \mathcal{Q}_3 \delta \phi' + \mathcal{Q}_4 \delta \phi \right) \mathcomma
\end{equation}

 with



\begin{align}
\mathcal{Q}_1 = & \frac{1}{2} a^2 C_{,\phi} \left( 1 - 3 \frac{\delta p_c}{\delta \rho_c} \right) - 3 D \mathcal{H}  \frac{\delta p_c}{\delta \rho_c}  \phi' - \frac{D C_{, \phi}}{C} \phi'^2 +  \frac{D_{, \phi}}{2 }  \phi'^2  - a^2 \frac{D}{2} \frac{h_{, \phi}}{h^2} \left( 1 - \frac{3}{2} \gamma^{-2} \right) \mathcomma \nonumber \\
&+ a^2 D \frac{Q}{h \rho_c} \gamma^{-2} + a^2 D \frac{Q}{h \rho_c} \gamma^{-2} \\
\mathcal{Q}_2 = &  - \frac{D}{2} \gamma^{-2} \left( 1 + w_c \right) \phi' \mathcomma \\
\mathcal{Q}_3 = &  3D \mathcal{H} \left( 2 - 3 \gamma^{-2} - w_c \right) - 2 \frac{D C_{, \phi}}{C} \phi' + D_{, \phi} \phi' - 3 D \frac{h_{, \phi}}{h} \left( 1 - \gamma^{-1} \right) \phi' \nonumber\\
&+ 3 D \mathcal{H} \gamma^{-1} \left( V_{, \phi} + Q \right) \phi' + 2 D \frac{Q}{\rho_c} \phi' \mathcomma \\
\mathcal{Q}_4 = &  - k^2 D \left( \gamma^{-2} + w_c \right)  + 3  \mathcal{H}  \left( \frac{D C_{, \phi}}{C} - D_{, \phi} \right) w_c  \phi' - \frac{a^2}{2} \frac{C_{, \phi}^2}{C} \left( 1 - 3 w_c \right) + 2  \frac{D C_{, \phi}}{C }  \phi'^2  \nonumber  \\
& -  \frac{3}{2} \frac{C_{, \phi} D_{, \phi}}{C } \phi'^2   - \frac{3}{2} D \frac{h_{, \phi}}{h} \mathcal{H} \left( 1 - \gamma^{-2} \right) \phi'  + \frac{a^2}{2} \frac{D C_{, \phi}}{C} \frac{h_{, \phi}}{h^2} \left( 1 - \frac{3}{2} \gamma^{-2} \right)  \\
&  - \frac{a^2}{2} D_{, \phi} \frac{h_{, \phi}}{h^2} \left( 1 - \frac{3}{2} \gamma^{-2}  \right)  + \frac{a^2}{2} D \frac{h_{, \phi}^2}{h^3} \left( \frac{5}{4}  - \frac{21}{4} \gamma^{-2} + 4\gamma^{-3} \right) + \frac{1}{2} a^2 C_{, \phi \phi} \left(1 - 3 w_c \right) \nonumber  \\
&  -  \frac{D C_{, \phi \phi}}{C } \phi'^2 + \frac{D_{, \phi \phi}}{2}  \phi'^2- \frac{a^2}{2} D \frac{h_{, \phi \phi}}{h^2} \left( 1 -3 \gamma^{-2} + 2 \gamma^{-3} \right) - a^2 D \gamma^{-3} V_{, \phi \phi} \nonumber\\
&+  \frac{Q}{h \rho_c} \left[a^2 D_{, \phi} -  a^2\frac{D C_{, \phi}}{C } - \frac{3}{2} D h_{, \phi} \phi'^2 \right] \mathcomma \nonumber
\end{align}

\noindent for general $C(\phi)$ and $D(\phi)$. If we now particularise to $C(\phi) = \left( T_3 h(\phi) \right)^{-1/2} $ and $D(\phi) = \left( h(\phi) / T_3 \right)^{1/2}$, these become

\begin{equation}
\delta Q = \frac{a^{-2} \rho_c}{\gamma^{-2} + h \rho_c \gamma^{-3} }\left( \mathcal{Q}_1 \delta_c + \mathcal{Q}_2 \mathpzc{h}' + \mathcal{Q}_3 \delta \phi' + \mathcal{Q}_4 \delta \phi \right) \mathcomma
\end{equation}

with



\begin{align}
\mathcal{Q}_1 = & a^2 \frac{Q}{\rho_c}\gamma^{-2} + 3 h \frac{\delta p_c}{\delta \rho_c} \left(  a^2 \frac{h_{,\phi}}{4h^2} - \mathcal{H} \phi' \right) \mathcomma \\
\mathcal{Q}_2 = & -\frac{h}{2} \left( \gamma^{-2} + w \right) \phi', \\
\mathcal{Q}_3 = & 3 h \mathcal{H} \left( 2 - 3 \gamma^{-2} - w \right) + 3 h^2 \left( V_{,\phi} + Q \right) \gamma^{-1} \phi' + 2 h \frac{Q}{\rho_c} \phi' - \frac{3}{2} h_{,\phi} \left( 1 - 2 \gamma^{-1} \right) \phi'\mathcomma \\
\mathcal{Q}_4 = & -k^2 h \left( \gamma^{-2} + w \right) + a^2 \frac{h_{,\phi}^2}{2 h^2} \left( \frac{3}{4} -\frac{15}{4} \gamma^{-2}  + 4 \gamma^{-3} - \frac{3}{2} w \right) + a^2 \frac{3}{4} \frac{h_{,\phi \phi}}{h} \left( \gamma^{-2} - \frac{4}{3} \gamma^{-3} + w \right) \nonumber \\
&  - \frac{3}{2} h_{,\phi} \mathcal{H} \left( 1 - \gamma^{-2} +2w \right) \phi' - a^2 h V_{,\phi \phi} \gamma^{-3} - a^2 \frac{h_{,\phi}}{2h} \frac{Q}{\rho_c} \left( 1 - 3 \gamma^{-2} \right) \mathperiod 
\end{align}

We will now proceed to investigate the phenomenology of these intricate equations.

\section{Numerical Study} \label{sec:dbi_num}

\subsection{Implementation and Background Results}


\begin{table*}[hbt!]
    \centering
    \begin{tabular}{|c|c|c|c|c|c|c|}
\hline
Model M$_{\phi_i/\Gamma_0}$ & $\Gamma_0$ & $\phi_i\, ({\rm M_{pl}})$ & $\gamma$ & $w_{\phi}$ & $G_{\rm eff}/G$ & $\sigma_8$ \\
\hline\hline
M$_{3/1.5}$ & 1.5 & 3 & 1.01 & -0.63 & 0.58 & 1.40 \\
M$_{3/5}$ & 5 & 3 & 1.01 & -0.70 & 0.46 & 1.24 \\ 
M$_{3/10}$ & 10 & 3 & 1.02 & -0.73 & 0.40 & 1.16 \\
\hline\hline
M$_{1.5/1.5}$ & 1.5 & 1.5 & 1.65 & -0.66 & 0.30 & 0.79 \\
M$_{1.5/5}$ & 5 & 1.5 & 1.49 & -0.76 & 0.41 & 0.77 \\ 
M$_{1.5/10}$ & 10 & 1.5 & 1.42 & -0.82 & 0.73 & 0.77 \\
\hline
\end{tabular}
\caption[Example solutions of the Dark D-Brane model]{\label{tabel1} Summary of parameter choices for the illustrative models examined in this study. All models use an initial scalar field velocity of $ \phi'_i = -10^{-25} \, \text{Mpl/Mpc} $ and all unspecified parameters are fixed to fiducial $\Lambda$CDM \textit{Planck} cosmological parameters \cite{Aghanim:2018eyx}, as detailed in \cref{tab:planckval}. The models are illustrative and may not be cosmologically viable. Key present-day quantities are indicated with a superscript $0$. The Lorentz factor $ \gamma $ quantifies current deviations from standard quintessence. The DBI EoS parameter $ w_{\phi} $ measures deviations from cosmological constant-like behaviour. The effective gravitational constant $ G_{\rm eff}^0/G $ and the predicted clustering amplitude $ \sigma_8 $ values are also included, with $ \sigma_8 \approx 0.85 $ for the reference $ \Lambda $CDM simulation.}
\end{table*}

We have conducted the numerical analysis using a modified version of the Boltzmann code \texttt{CLASS} \cite{Lesgourgues:2011re, Blas:2011rf} tailored to have dark energy portrayed by a DBI scalar field and a coupling to the dark matter component, as described in \cref{sec:cmbcodes}. From this modified patch, we can thoroughly investigate the cosmological implications of the theoretical model and infer the range of physically viable parameters. Moreover, the code is adapted to handle both the background and linear perturbations associated with a DBI scalar field. For the simulations, we employed \textit{Planck} 2018 cosmological parameters \cite{Aghanim:2018eyx} assuming a spatially flat $\Lambda$CDM cosmological background, listed in \cref{tab:planckval}. 

The scale of the potential $ V_0 $, defined in \cref{hvmodel}, is treated as a shooting parameter, which is adjusted for each instance of the code to match the Friedmann constraint, \cref{eq:omega0}, under the assumption that the system is currently near the scaling regime. This choice is motivated by the argument in Ref.~\cite{Koivisto:2013fta} that the true dynamics of the system is fully specified by the degenerate combination $\Gamma_0=h_0 V_0$.

\begin{figure}
\centering
      \subfloat{\includegraphics[width=0.5\linewidth]{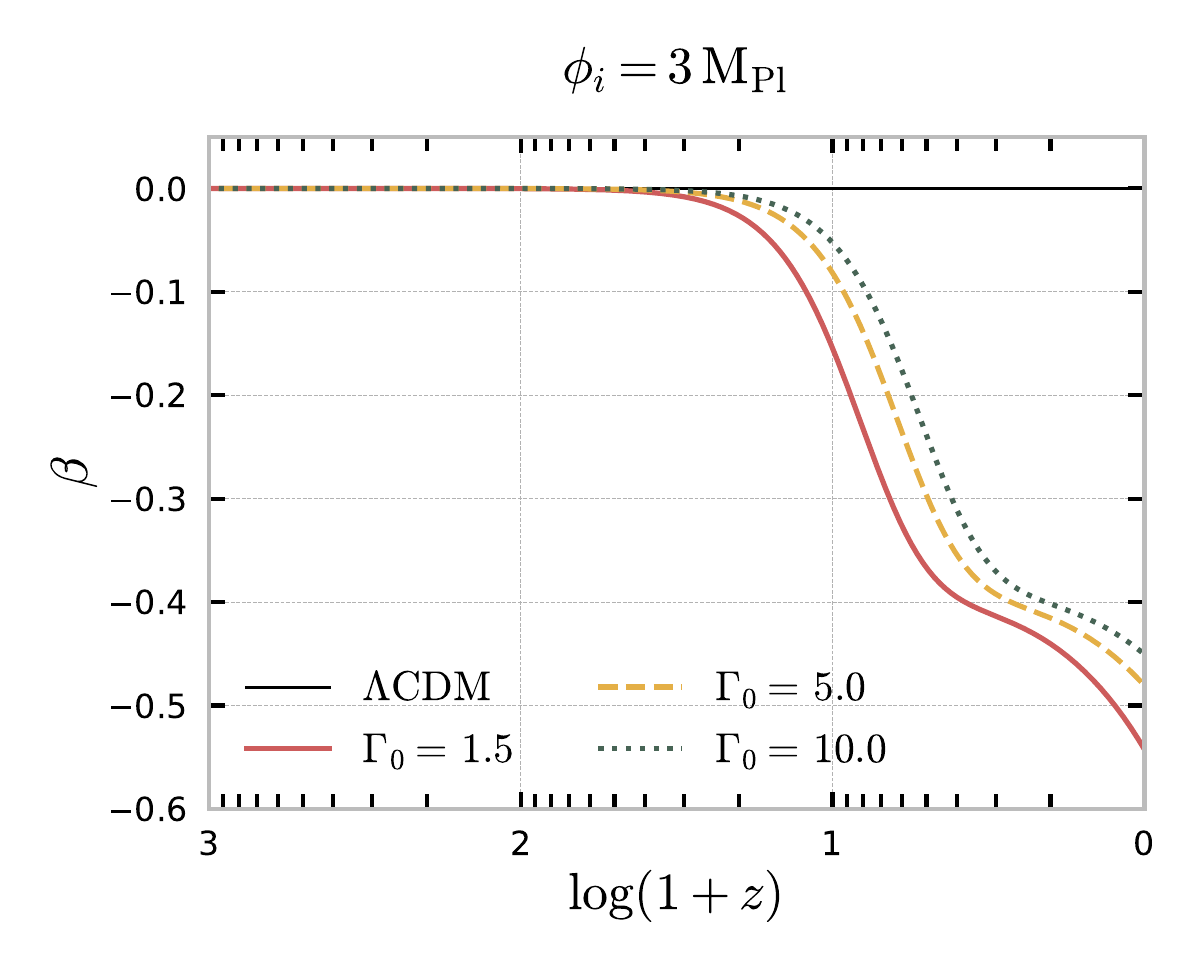}}
      \hfill
      \subfloat{\includegraphics[width=0.5\linewidth]{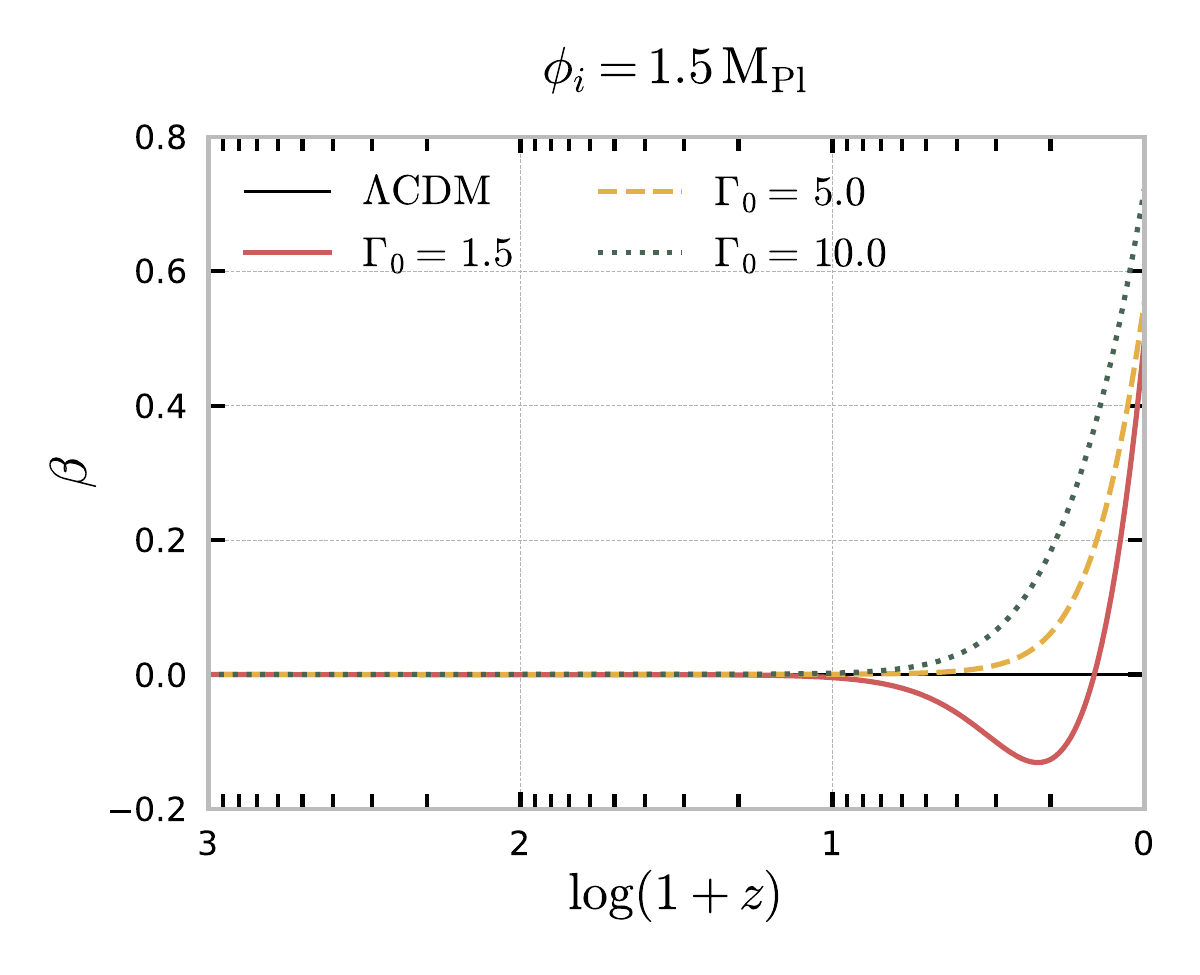}} 
  \caption[Evolution of the coupling parameter]{\label{fig:difgamma0beta} Background evolution of the effective coupling function $\beta$, defined in \cref{beta}, plotted as functions of the redshift $z$. The left and right panels correspond to Models M$_{3/1.5}$-M$_{3/10}$ and M$_{1.5/1.5}$-M$_{1.5/10}$ in \cref{tabel1}, respectively. We clearly identify two regimes of the theory: one in which the effective coupling is always negative throughout the cosmic evolution, depicted on the left panel, and another one for which the coupling may start out as being negative but eventually starts to increase towards positive values at present, pictured on the right panel.}
\end{figure}

Our study focuses on understanding different regimes of the model, differentiated by their initial conditions and $ \Gamma_0 $ values. The sets of parameters used for illustrative purposes are listed in \cref{tabel1}, along with some present-day quantities that show the impact of varying the model parameters. 
Two categories of coupled models are considered, each marked by a distinct initial scalar field value and thus different characteristics of the effective coupling, as illustrated in \cref{fig:difgamma0beta}. The first class, embodied by models M$_{3/1.5}$-M$_{3/10}$ in the upper section of \cref{tabel1}, assumes $ \phi_i = 3 \, \text{M}_{\text{pl}} $ and features a consistently negative effective coupling, with an earlier onset and larger absolute magnitude of the effective coupling for smaller $ \Gamma_0 $ values. Conversely, the second class, exemplified by models M$_{1.5/1.5}$-M$_{1.5/10}$ in the lower section of \cref{tabel1}, with $ \phi_i = 1.5 \, \text{M}_{\text{Pl}} $, maintains a positive coupling at present, even though it can start by being negative, as is the case in model M$_{1.5/1.5}$. The coupling value increases for higher $ \Gamma_0 $ values in this latter category.

\begin{figure}
\centering
       \subfloat{\includegraphics[width=0.5\linewidth]{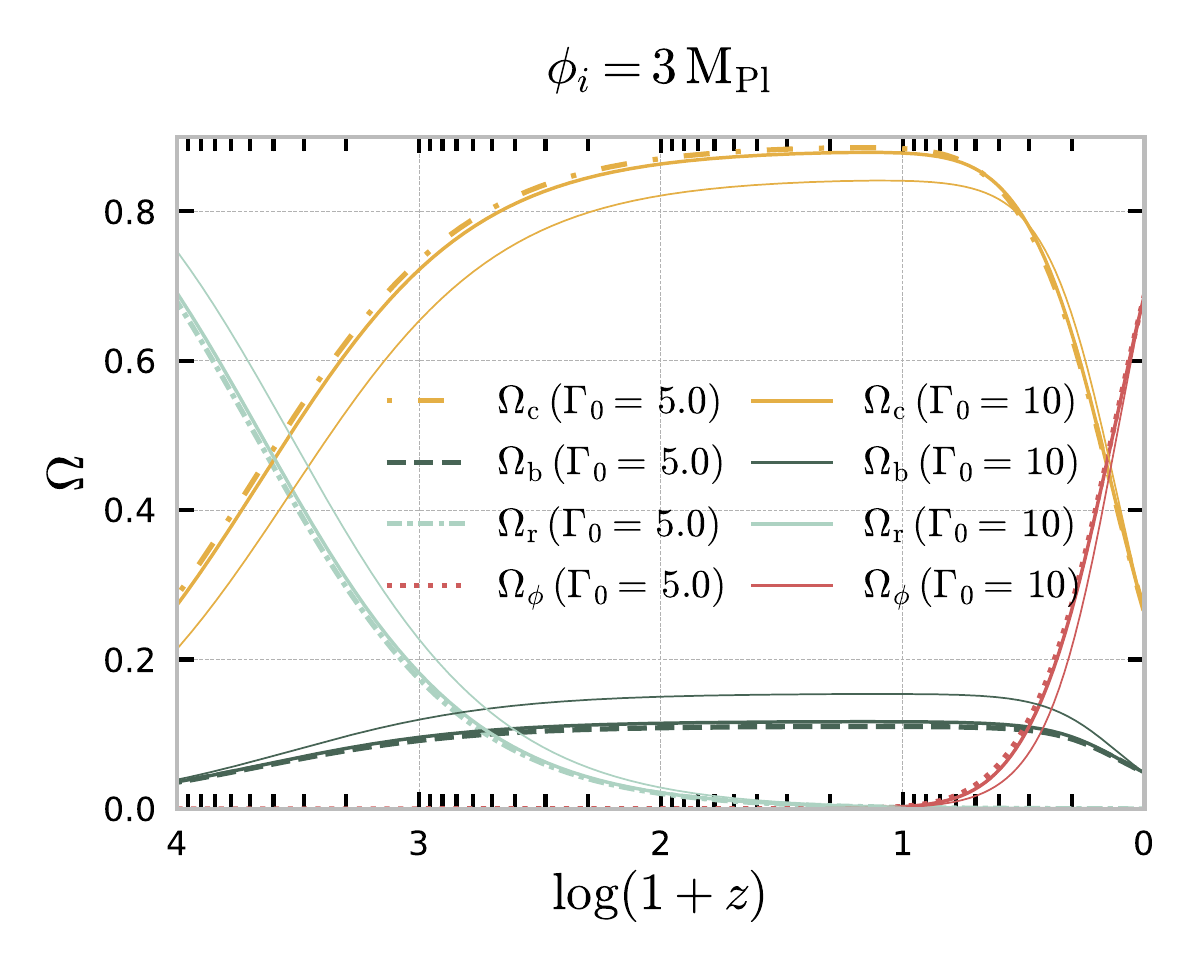}} 
       \hfill
      \subfloat{\includegraphics[width=0.5\linewidth]{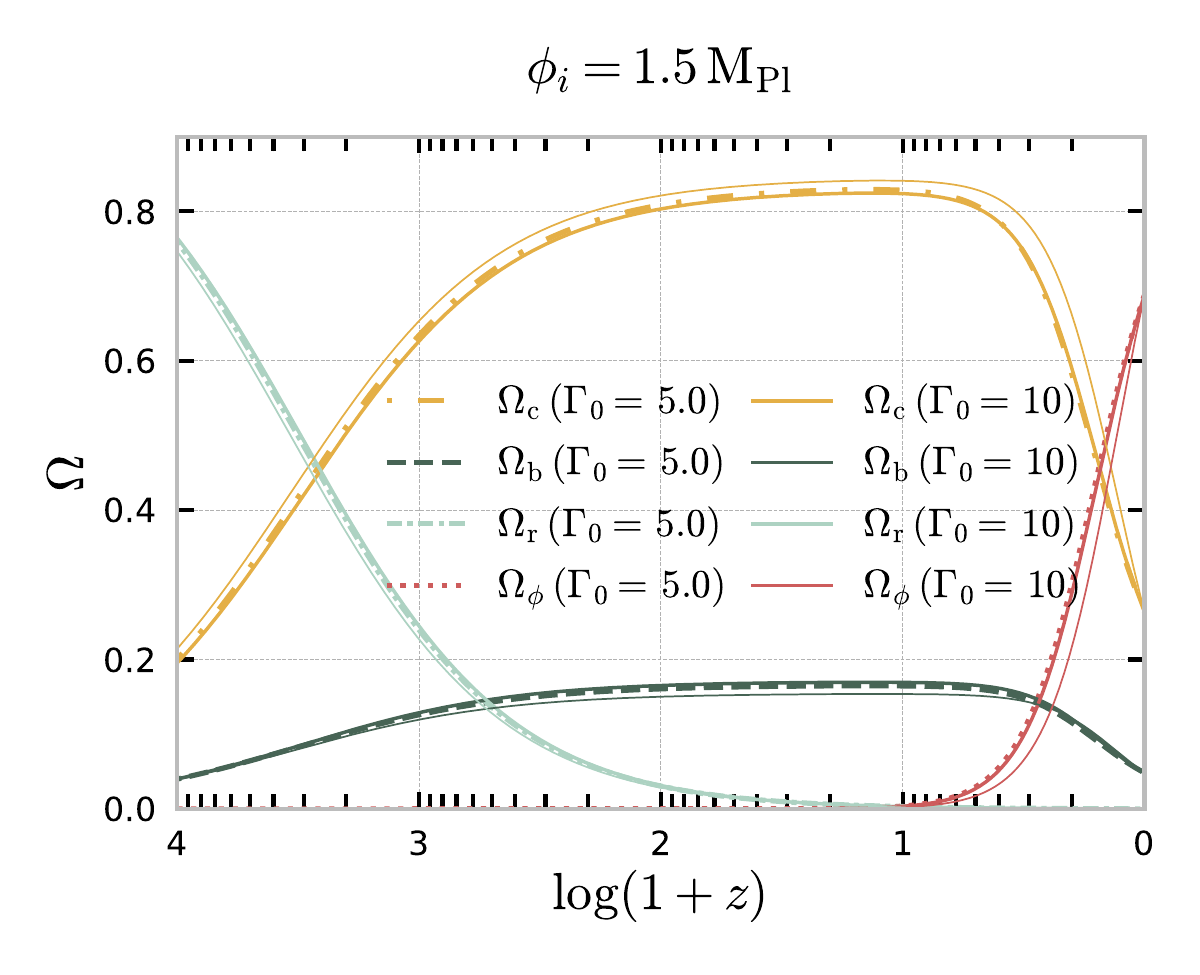}}
             \hfill
       \subfloat{\includegraphics[width=0.5\linewidth]{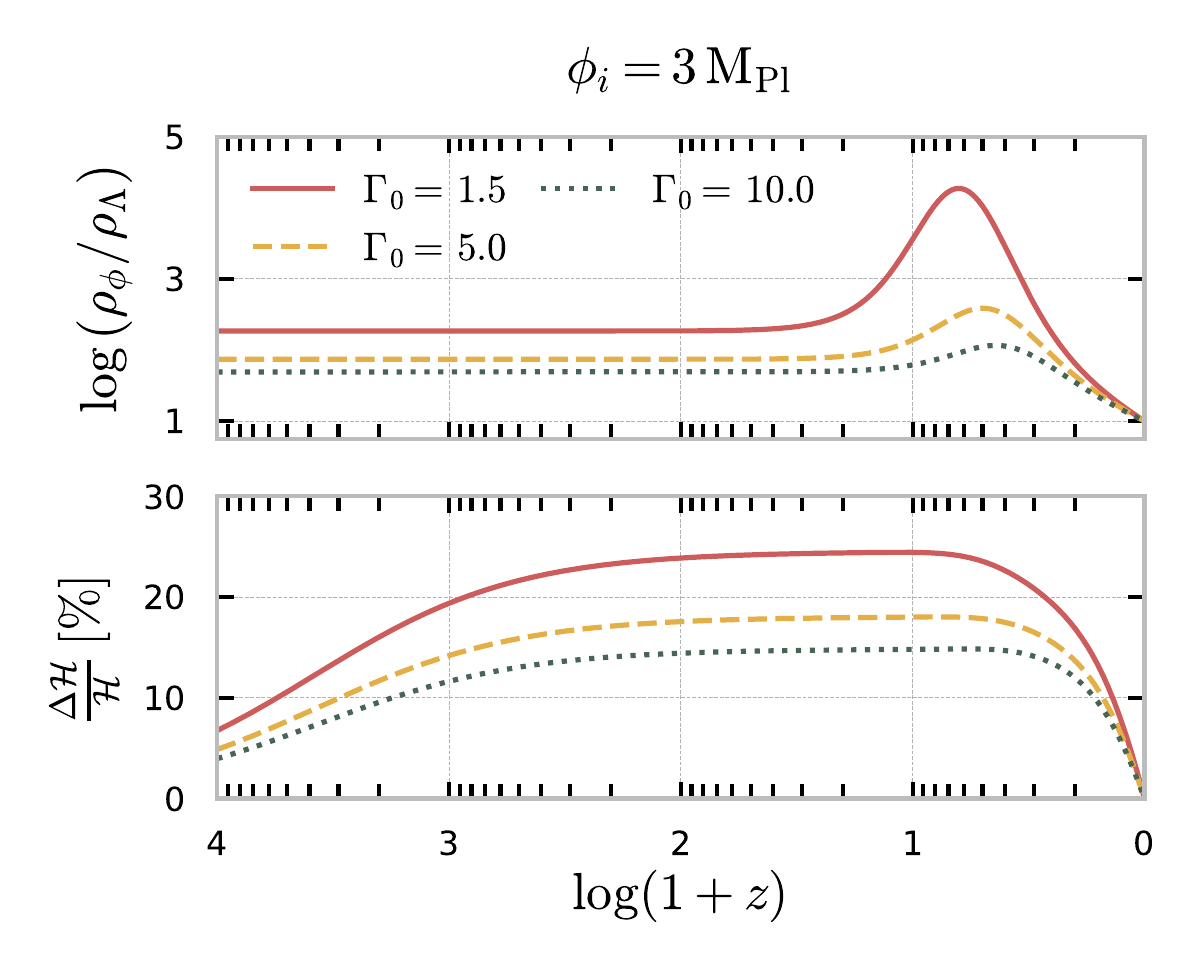}}
                          \hfill
             \subfloat{\includegraphics[width=0.5\linewidth]{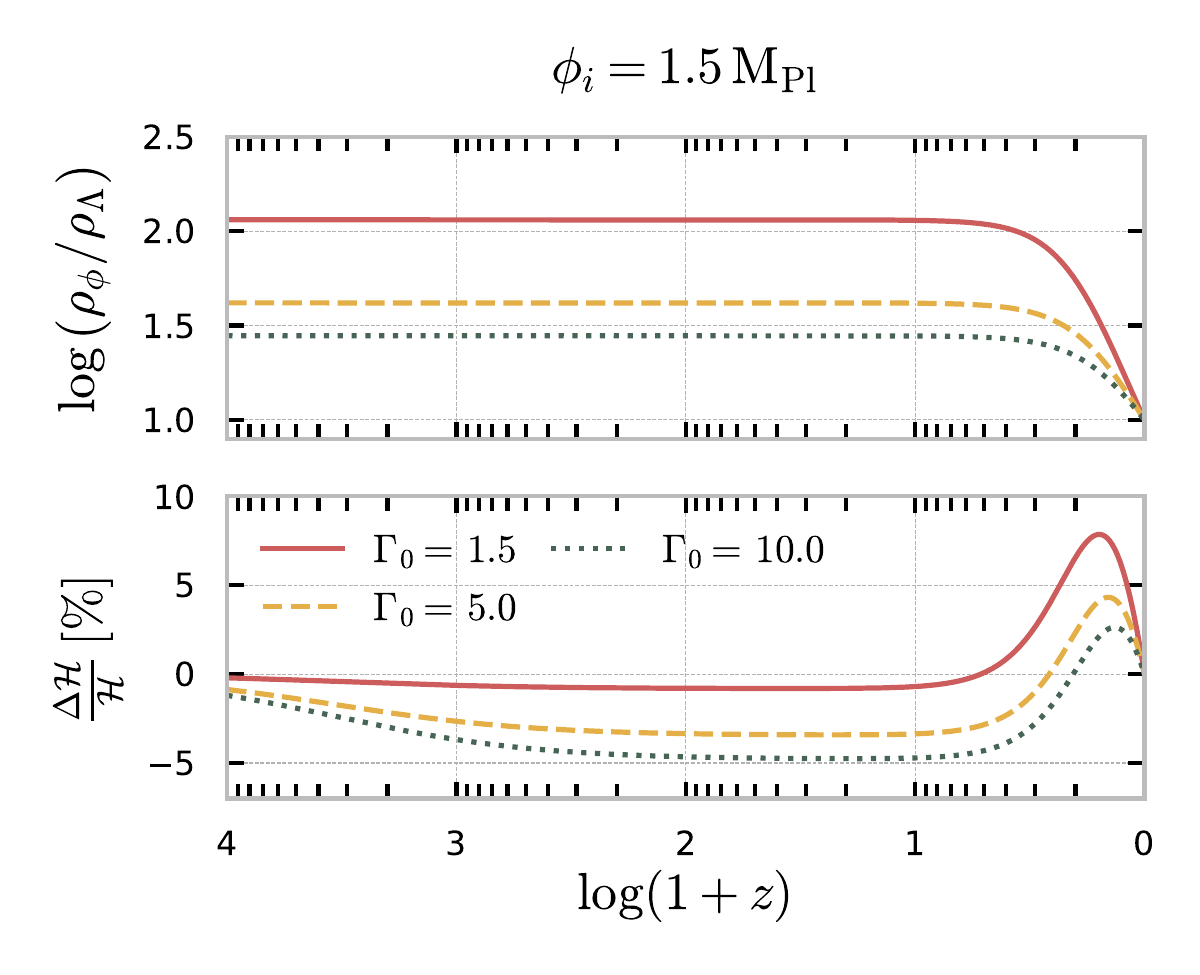}}
                       \hfill
      \subfloat{\includegraphics[width=0.5\linewidth]{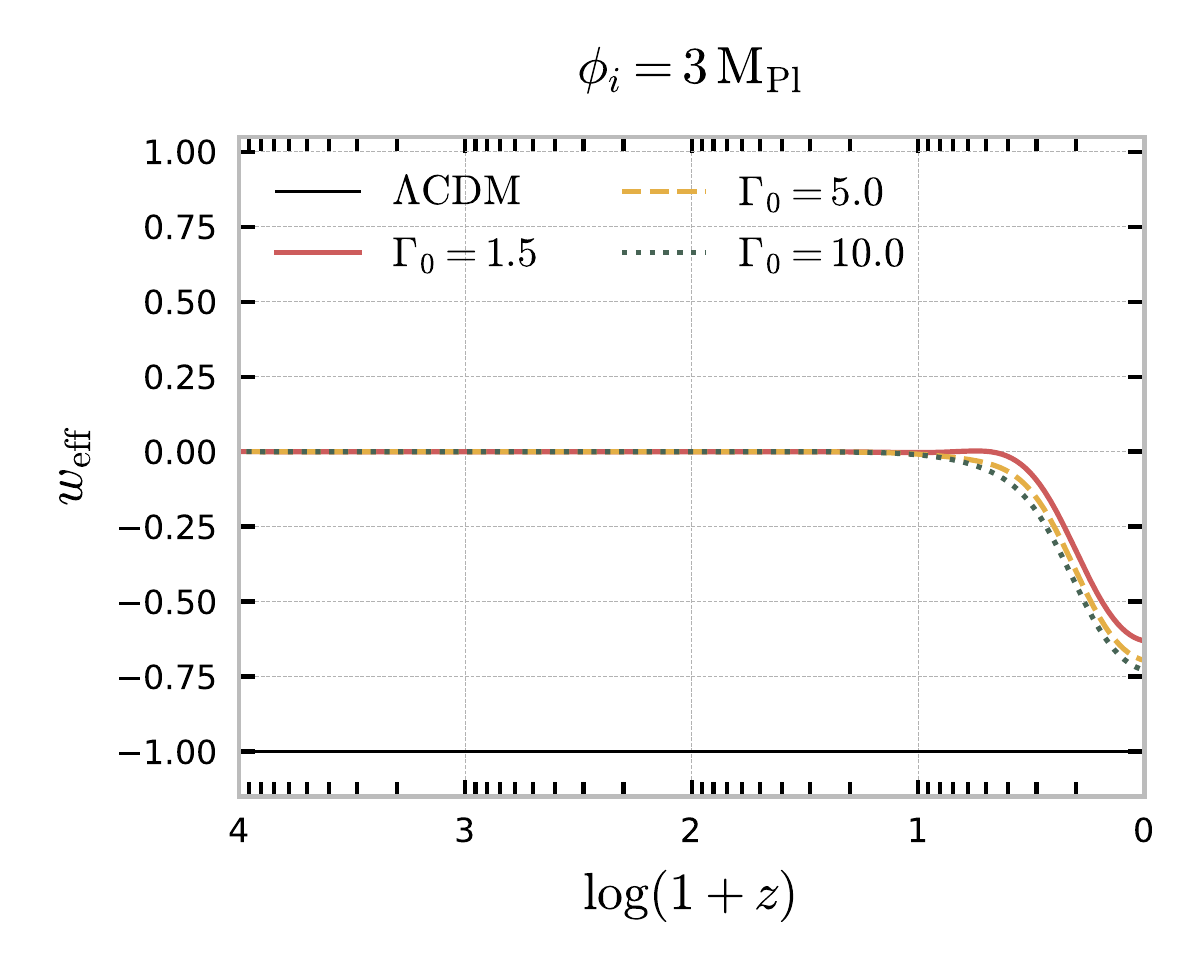}}
                         \hfill
            \subfloat{\includegraphics[width=0.5\linewidth]{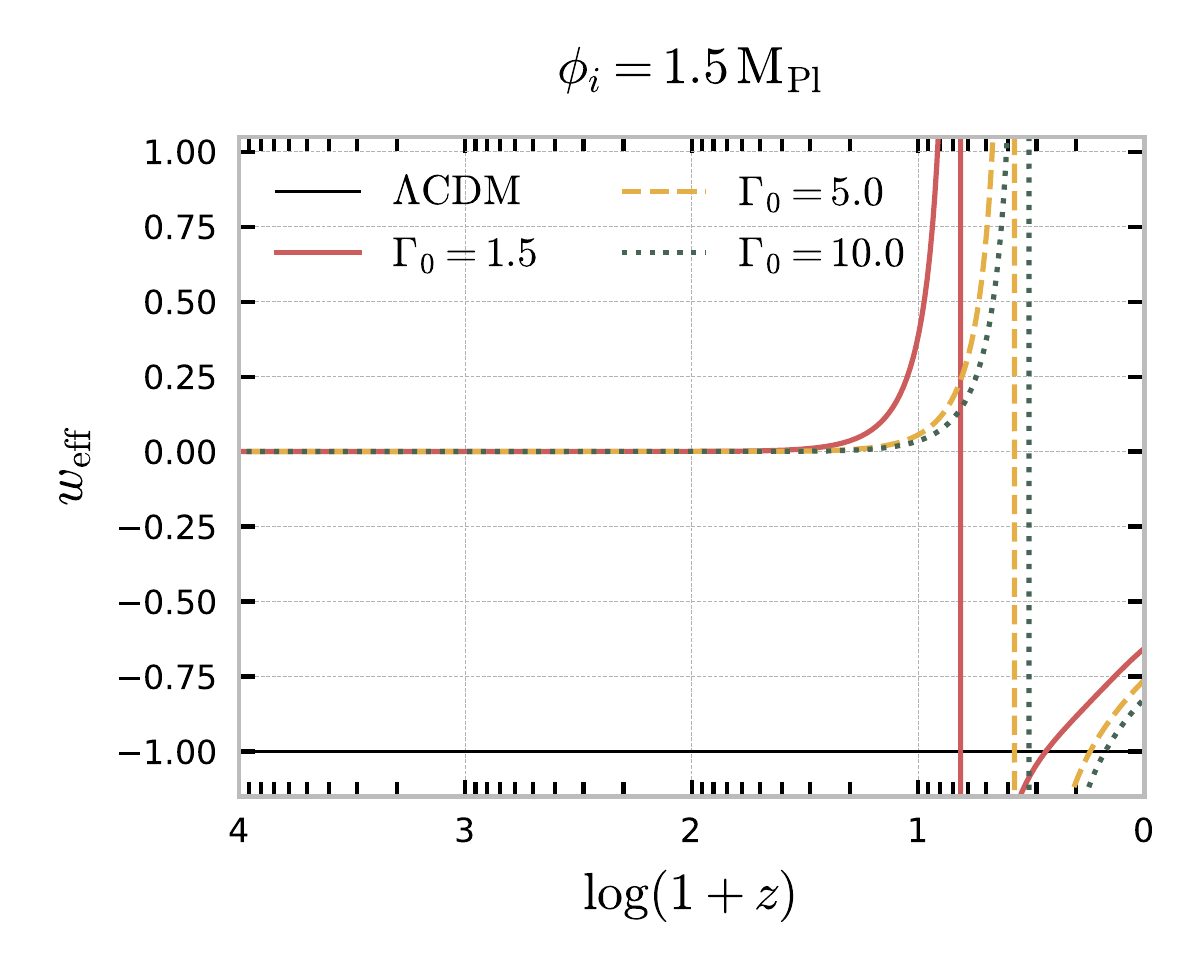}}
  \caption[Evolution of relative energy densities, the Hubble rate and the effective dark energy EoS]{\label{fig:difgamma0} Background evolution of the fractional energy densities $\Omega_i = k^2 \rho_i/3 H^2$, for each labelled $i$-th fluid (top), of the energy density of $\phi$ relative to $\Lambda$ and the Hubble expansion rate $H$ (middle), defined in \cref{hubble}, and the DBI effective equation of state parameters, as defined in \cref{weffphi} (bottom), as functions of the redshift, $z$. In the middle panel, we also present the deviations of the Hubble expansion rate in each model, $\Delta H$, when compared to the concordance model. The left and right panels correspond to Models M$_{3/1.5}$-M$_{3/10}$ and M$_{1.5/1.5}$-M$_{1.5/10}$ in \cref{tabel1}, respectively.}
\end{figure}

Our findings suggest a non-trivial relationship between $ \Gamma_0 $ and $ \phi_{\text{i}} $. Owing to the warp factor's inverse dependence on the scalar field value ($h(\phi) \propto \phi^{-4}$), higher $ \phi_{\text{i}} $ values correlate with larger $ h_0 $ values (and correspondingly lower $ V_0 $) to reproduce the fiducial cosmology at present maintaining the same $ \Gamma_0$.

We turn our attention to the redshift evolution of relevant background quantities in the six illustrative cosmological models: the effective coupling $ \beta $, as defined in \cref{beta}; the fractional energy densities $ \Omega_i $ for each fluid $ i $, defined as $ \Omega_i \equiv \kappa^2 \rho_i/3 H^2 $; the Hubble rate $ H $, according to \cref{hubble}, and the DBI and effective DE equations of state, articulated in \cref{wphi,weffphi}. These quantities are illustrated in \cref{fig:difgamma0beta,fig:difgamma0}. We emphasise that they encapsulate the Universe's background evolution and also hold implications for the linear growth rate of structures, deeply entangled with the background evolution.

The left panels of \cref{fig:difgamma0beta,fig:difgamma0} depict the models M$_{3/1.5}$-M$_{3/10}$ from \cref{tabel1}, with consistently negative effective coupling, corresponding to a transfer of energy from DM to DE. Nevertheless, as the system nears the scaling solution, characterised by $ 1/\gamma \to 0 $, we anticipate $ \beta $ turning positive. Conversely, the right panels depict models M$_{1.5/1.5}$-M$_{1.5/10}$, where $ \beta $ remains positive at late times, and energy is being transferred from dark energy to dark matter. Notably, model M$_{1.5/1.5}$ exhibits a unique behaviour for $ \beta $, which starts by being negative, where it peaks at $ z \approx 0.1 $ ($ \beta \approx -0.17 $) before rising to become positive at $ z=0 $ ($ \beta \approx 0.5 $). Such behaviour hints at scenarios where the dark sector coupling, albeit minimal today, could have been substantial in the past. Our simulations indicate that for larger $ \phi_{\text{i}} $ values (left panel), the coupling is activated sooner for the same value of $ \Gamma_0 $. On the other hand, keeping $ \phi_{\text{i}} $ fixed, for negative (positive) $ \beta $, higher $ \Gamma_0 $ values prompt later (earlier) onsets of $\beta$, culminating in lower (higher) absolute magnitudes for the coupling today. This also aligns with cosmologies nearing (or deviating from) a $ \Lambda $CDM-like background evolution during the matter-dominated epochs, underlining the distinction between the two regimes. Nevertheless, based on the dynamical systems study, we foresee that for models M$_{3/1.5}$-M$_{3/10}$, the coupling will eventually become positive, irrespective of the initial conditions, featuring a characteristic turning point akin to the one observed in  M$_{1.5/1.5}$.

The top panels of \cref{fig:difgamma0} illustrate the time evolution of fractional energy densities. The cosmological epochs occur in the sequence previously discussed: a radiation-dominant phase gives way to a DDM-dominant phase, culminating in dark energy becoming relevant in the current epoch as the system gravitates towards the scaling solution. We observe that the fact that the present cosmology has been fixed for all the simulations results in an artificial shift of the matter-radiation equality epoch. At later stages, we identify deviations from $ \Lambda $CDM in the fractional energy densities, with higher (lower) dark matter abundance in the presence of negative (positive) couplings, reflecting the energy flow between the dark fluids. Concerning $ \Omega_{\phi} $, we note how the curves corresponding to the highest value of $ \Gamma_0 $ (filled thicker lines) approximate $ \Omega_{\Lambda} $ in either regime most closely, as is also illustrated in the upper-middle panels of \cref{fig:difgamma0}.

The current value of the Hubble rate is fixed at the \textit{Planck} fiducial value. Nonetheless, different profiles for $ H(z) $ emerge across the examined models, an effect that is entangled with the differences in the evolution of the fractional energy densities of the uncoupled species, namely baryons and radiation. The variations to the evolution of $H(z)$ are depicted in the lower-middle panels of \cref{fig:difgamma0}, where generally enhanced (suppressed) expansion rates for most of the evolution are associated with negative (positive) effective couplings, increasingly closer to the $ \Lambda $CDM reference case for larger (smaller) $ \Gamma_0 $ values. Notably, at redshifts $ z \lesssim 0.5 $, we observe $ \Delta H(z) > 0 $ for all models, hinting at higher Hubble rates than $ \Lambda $CDM benchmarks. At higher redshifts, for the models with positive $\beta$, $ H(z) $ drops below the concordance model, whereas for negative $\beta$, it is the higher $ \Gamma_0 $ values that bring the evolution closer to the $ \Lambda $CDM curve. 
Variations in the evolution of $H(z)$, coupled with modifications in the effective gravitational interaction, give rise to distinct imprints in the linear growth of cosmic structures, as illustrated in \cref{fig:varying_all,fig:varying_pk_all}.


Overall, while the outcomes align with expectations from dynamical studies and qualitative considerations, we anticipate unique signatures at the linear cosmological perturbation level, on which we will focus next.

\subsection{The Growth of perturbations and the Effective Gravitational Constant}

In this section, we step aside for a moment to examine the evolution of the density contrast for DDM in the sub-horizon regime, where $ k \gg \hub $. We adopt the quasistatic framework where the temporal variations of the gravitational potential arise mainly from the matter and field fluctuations. Consequently, terms involving time derivatives, such as $ \Phi' $ in the Einstein equations or $ \delta\phi'' $ and $ \delta\phi' $ in the Klein--Gordon equation, are considered negligible relative to other terms. As a result, \cref{CDM1,CDM2} become

\begin{align} 
\label{one}
\delta_c' &\approx -\theta_c - \frac{Q}{\rho_c} \phi' \delta_c + \frac{\delta Q}{\rho_c} \phi'\mathcomma \\
\label{two}
\theta_c' &\approx - \mathcal{H} \theta_c + k^2 \Psi - \frac{Q \phi'}{\rho_c} + k^2 \frac{Q}{\rho_c} \delta\phi \mathperiod
\end{align}

We arrive at an approximate form for the Poisson-like equation governing the gravitational field from the perturbed Einstein equations. Here, we neglect baryons and radiation, which represent minor contributions at the current epoch. It follows that:

\begin{equation}
\label{three}
k^2 \Psi \approx - 4 \pi G \rho_c \delta_c \mathperiod
\end{equation}

Finally, the perturbed Klein--Gordon equation can be expressed as

\begin{equation}
\label{four}
\mathcal{A} \Psi + \left( k^2 \gamma^{-2} + a^2 m_{\text{eff}}^2 \right) \delta \phi + a^2 \gamma^{-3} \delta Q \approx 0 \mathcomma
\end{equation}

where $ \mathcal{A} $ was defined as 

\begin{align}
\mathcal{A} = \left[ 6 \mathcal{H} \left( 1- \gamma^{-2} \right) \phi'  - \frac{h_{,\phi}}{h^2} a^2  \left( 2 - 3 \gamma^{-1} + \gamma^{-3} \right) + a^2  \left( V_{,\phi} + Q \right) \left(3\gamma^{-1} -\gamma^{-3} \right) \right]\mathcomma
\end{align}

and the effective mass $ m_{\text{eff}} $ is given by

\begin{align}
a^2 m_{\rm eff}^2 =& -\frac{3h_{,\phi}}{h} \mathcal{H} \left( 1 - \gamma^{-2} \right) \phi' + \frac{h_{,\phi \phi}}{2h^2} a^2 \left( 1 - 3\gamma^{-2} + 2\gamma^{-3} \right) - \frac{3}{2} \frac{h_{,\phi}}{h} a^2 \left( V_{,\phi} + Q \right) \left( \gamma^{-1} - \gamma^{-3} \right)  \nonumber \\
+& a^2 V_{,\phi \phi} \gamma^{-3} + \frac{h_{,\phi}^2}{2 h^3} a^2 \left( 1 - 3\gamma^{-1} + 3\gamma^{-2} - \gamma^{-3} \right) \mathperiod
\end{align}

In applying these approximations to the expression for the coupling perturbation, \cref{deltaQ}, we obtain

\begin{equation}
\delta Q \approx \frac{a^{-2} \rho_c}{\gamma^{-2} + h \rho_c \gamma^{-3}} \left( \mathcal{Q}_1 \delta_c + \mathcal{Q}_5 \delta \phi \right) \mathperiod
\label{deltaqaprox}
\end{equation}

Considering the coefficient $ \mathcal{Q}_5 $, as described by \cref{q5}, we find that the $ k^2 $-term is the dominant component. By adopting the same analytical approach as for deriving \cref{deltaqaprox}, in the sub-horizon limit, we obtain the simplified relation

\begin{equation}
    \delta Q \simeq Q \delta_c \mathcomma
\end{equation}

 a result that has been confirmed through numerical analysis and also holds in other alternative theories featuring conformal and disformal couplings \cite{Zumalacarregui:2012us, vandeBruck:2015ida, Mifsud:2017fsy}.

Taking the time-derivative of \cref{one} and employing \cref{two,three,four}, we arrive at the following approximation to the evolution of $ \delta_c $:

\begin{equation}
\delta_c'' + \mathcal{H}_{\text{eff}} \delta_c' \approx 4 \pi G_{\text{eff}} \rho_c \delta_c \mathcomma
\end{equation}

where we have introduced the effective Hubble parameter as

\begin{equation}
    \mathcal{H}_{\text{eff}} = \mathcal{H} + \frac{Q \phi'}{\rho_c} \mathcomma
    \label{heff}
\end{equation}

and the effective gravitational constant, given by

\begin{equation}
G_{\text{eff}} = G \left(1 + 2 \beta^2 \frac{1}{\gamma (1 + a^2 m_{\text{eff}}^2 / k^2 \gamma^2 )} - A \frac{Q \gamma^2}{\rho_c k^2} \right) \mathperiod
\end{equation}

In the limits $ k^2 \gg a^2 m_{\text{eff}}^2 / \gamma^2 $ and $ k^2 \gg A Q \gamma^2 / \rho_c $, $ G_{\text{eff}} $ simplifies to

\begin{equation}
G_{\text{eff}} \approx G (1 + 2 \beta^2 / \gamma) \mathcomma
\label{geff}
\end{equation}

where $ \beta $ is as defined in \cref{beta}. $ G_{\text{eff}} $ represents the effective gravitational interaction between two dark matter particles, accounting for both standard gravitational and scalar field-mediated long-range forces. Our findings agree with Ref.~\cite{Tsujikawa:2007gd} for scalar-tensor gravity models with a conformal coupling, generalised here to include a disformal coupling and a distinct functional form for $ \beta $.

\begin{figure}[t!]
     \centering
 \subfloat{\includegraphics[width=0.5\linewidth]{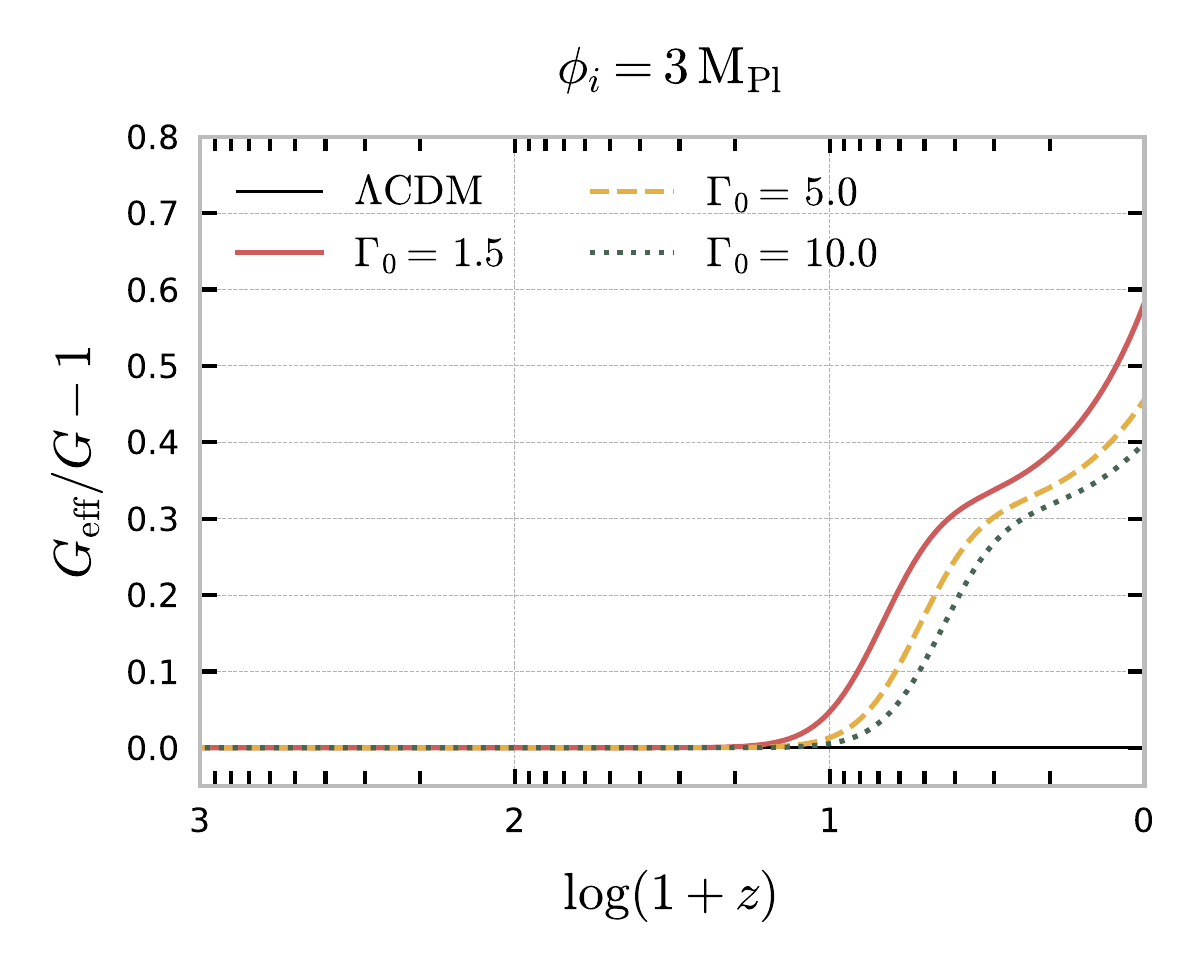}}
      \hfill
      \subfloat{\includegraphics[width=0.5\linewidth]{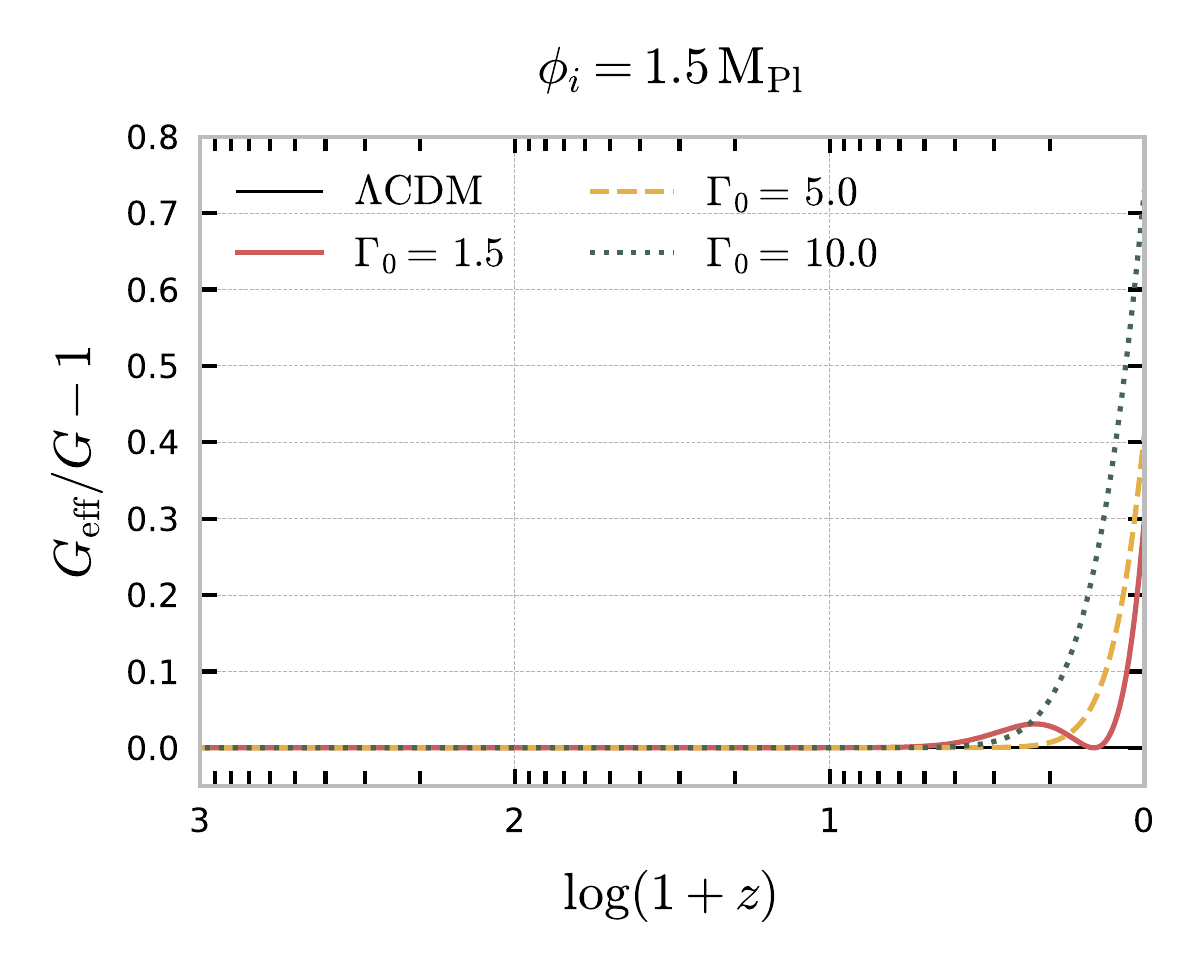}}
  \caption[Evolution of the effective gravitational constant]{\label{fig:Geff} Evolution of the effective gravitational constant, defined in \cref{geff}, as function of the redshift $z$. The left and right panels correspond to Models M$_{3/1.5}$-M$_{3/10}$ and M$_{1.5/1.5}$-M$_{1.5/10}$ in \cref{tabel1}, respectively, and the line styles used are the same as in the background evolution, depicted in \cref{fig:difgamma0beta,fig:difgamma0}.}
\end{figure}


The evolution of $ G_{\rm eff}/G$ for the models summarised in \cref{tabel1} is illustrated in \cref{fig:Geff}. We observe that deviations of $ G_{\rm eff} $ from $ G $ arise at lower redshifts, while it converges to $ G $ in the early Universe. This is consistent with the background study for $\beta$, showing that smaller initial values of $ \phi_i $ result in higher present values of $ G_{\rm eff} $. Interestingly, because the onset of the coupling is also delayed in these cases, we expect that this should be reflected as smaller variations in both CMB anisotropies and the matter power spectrum as compared to $ \Lambda $CDM. However, there should be a balance between both effects. This observation aligns with the findings in Ref.~\cite{vandeBruck:2017idm} for disformally coupled quintessence models. Nevertheless, the earlier activation of the deviations of $ G_{\rm eff} $ from $G$ presents a potential challenge to the model's consistency with observational data. However, we have also intentionally depicted a scenario where $ G_{\rm eff} $ initially rises (corresponding to a period of negative $\beta$) but ultimately declines (the $\Gamma_0 = 1.5$ case in the right panel of \cref{fig:Geff}), only to rise again at later times (as $\beta$ becomes positive). This suggests the existence of parameter combinations for which $ G_{\rm eff} \approx G $ today, yet the structure formation is distinctly impacted by the scalar-field mediated forces at intermediate redshifts. The varying dynamics for $ G_{\rm eff} $ across different models will inevitably result in diverse outcomes for the matter density fluctuations, which we will attempt to capture in the following section.

\subsection{Cosmological Observables} \label{sec:observables}

\begin{figure}
\centering
       \subfloat{\includegraphics[width=0.5\linewidth]{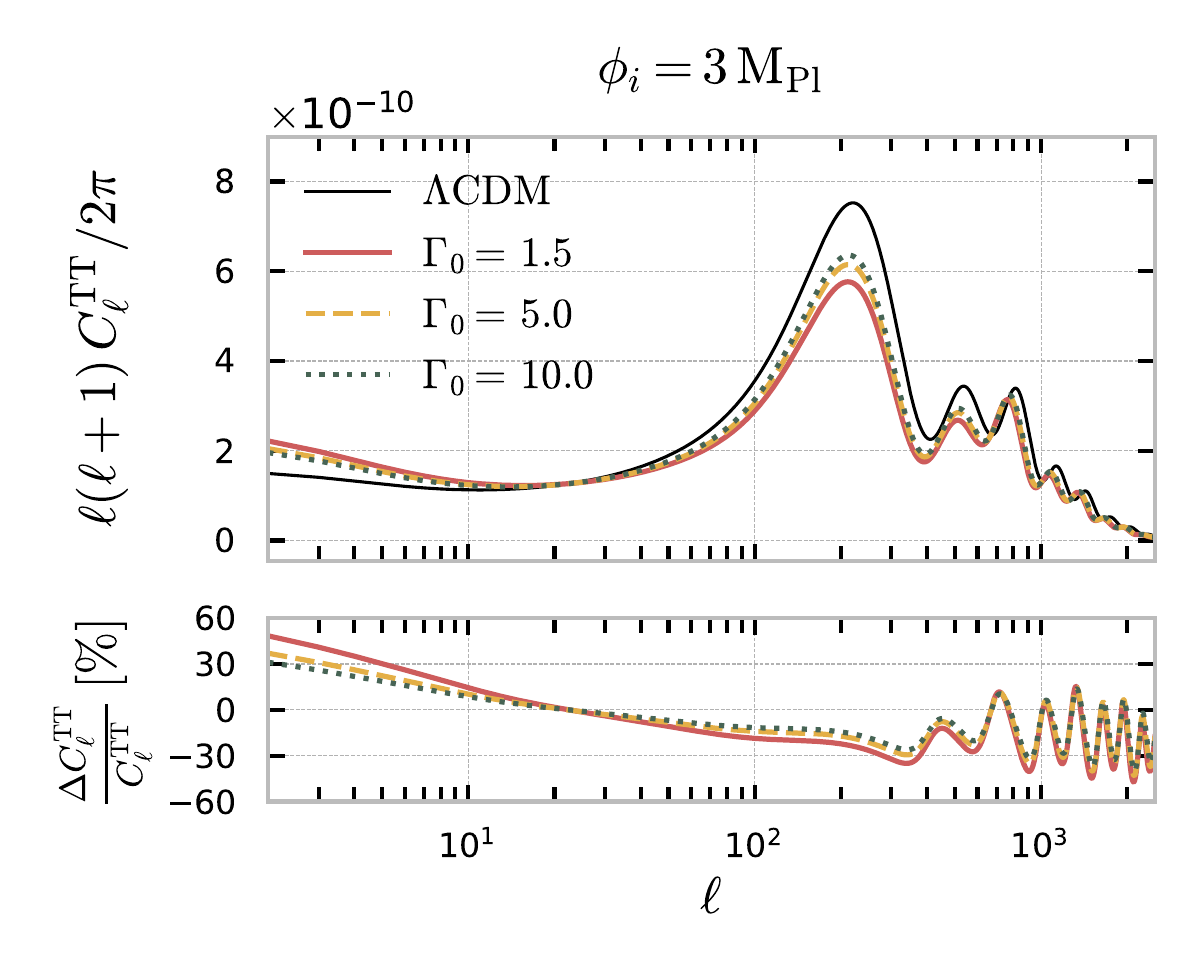}} 
       \hfill
       \subfloat{\includegraphics[width=0.5\linewidth]{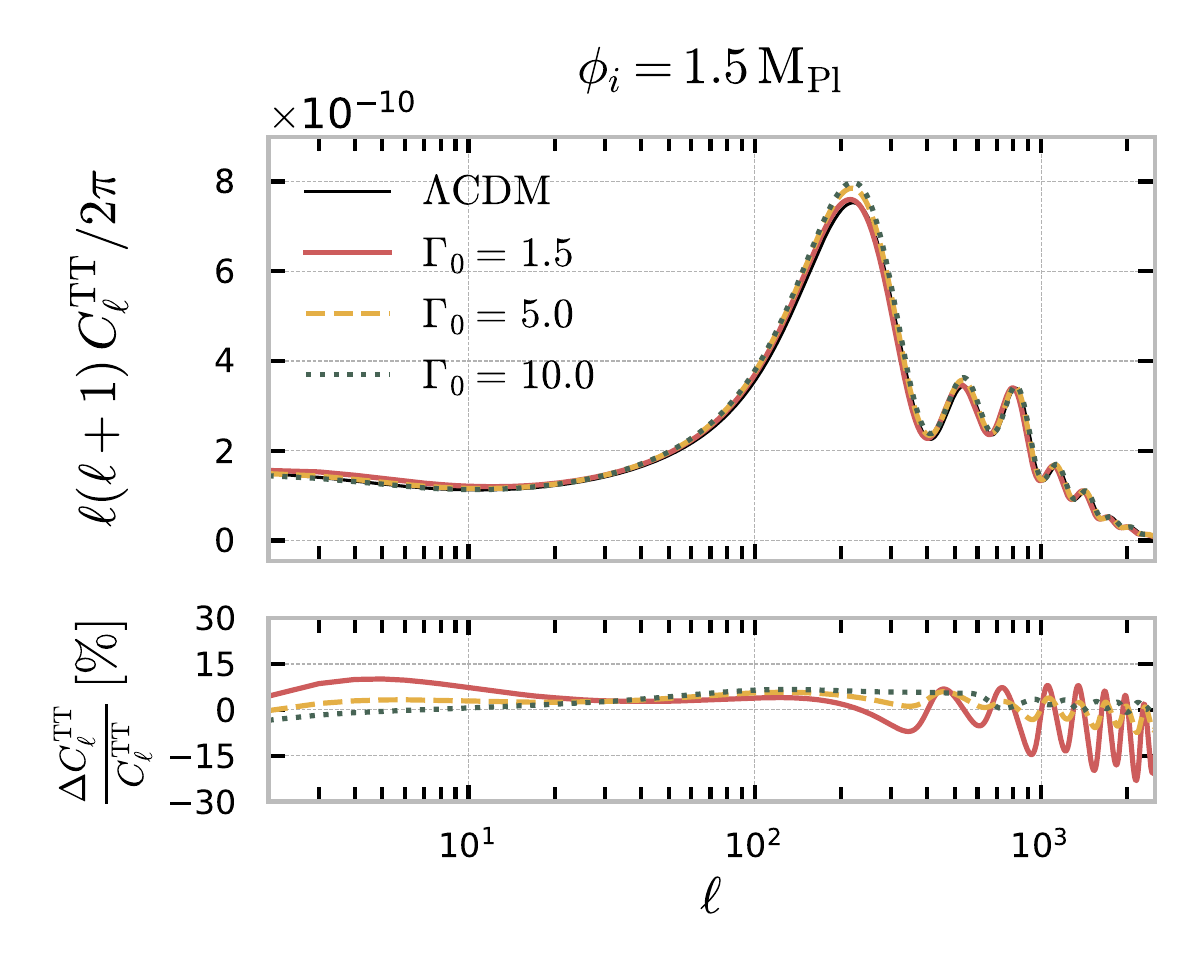}} 
       
  \subfloat{\includegraphics[width=0.5\linewidth]{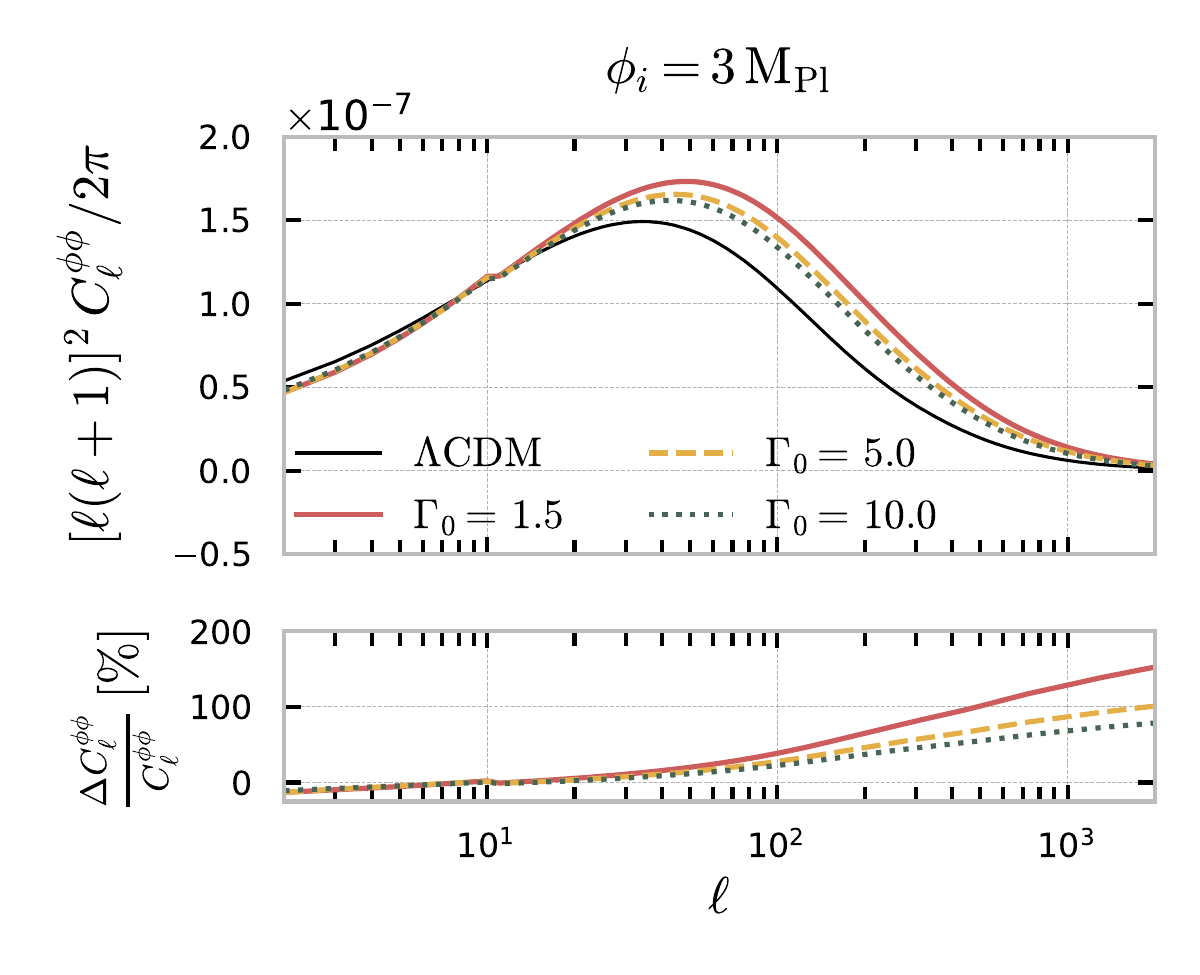}} 
       \hfill
       \subfloat{\includegraphics[width=0.5\linewidth]{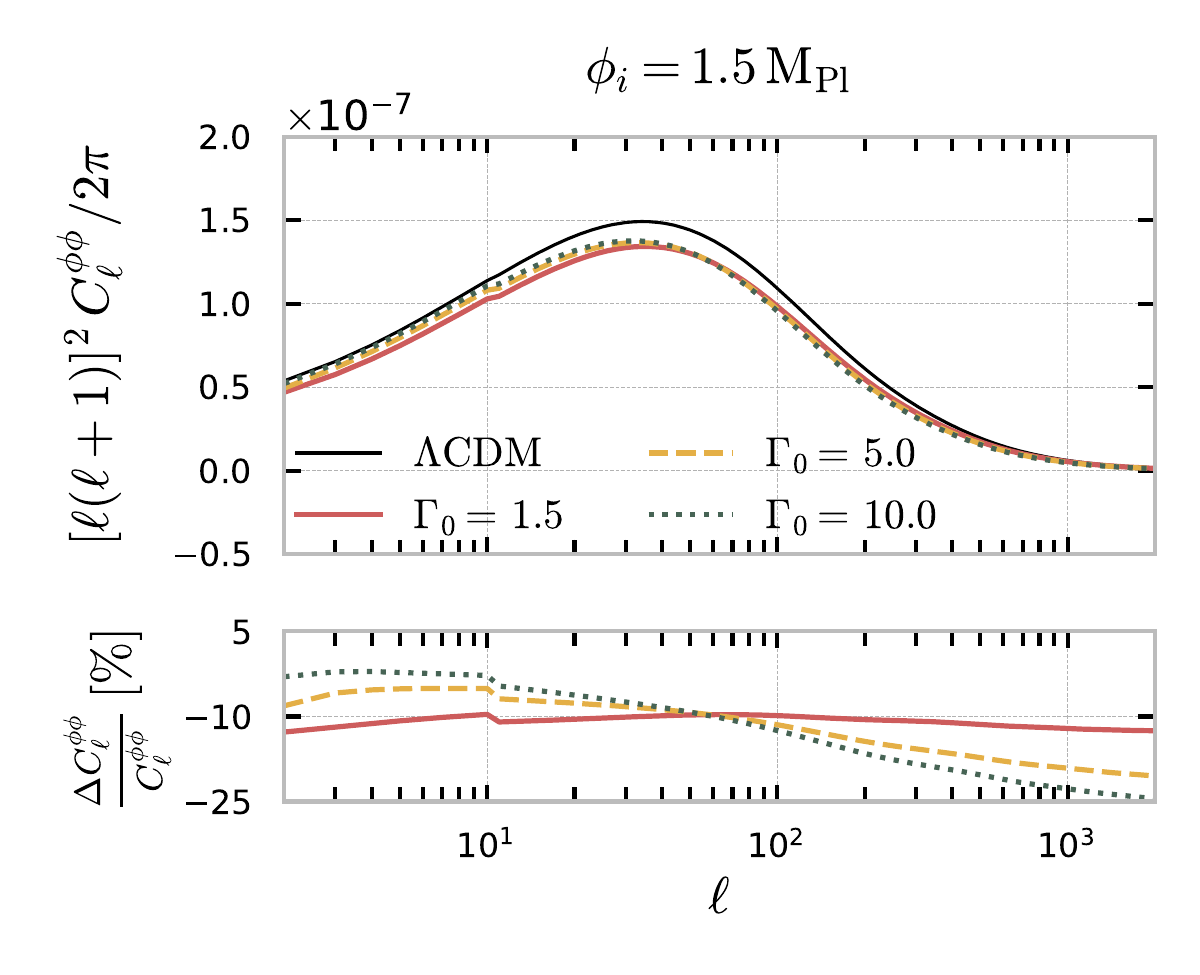}} 
       
      \subfloat{\includegraphics[width=0.5\linewidth]{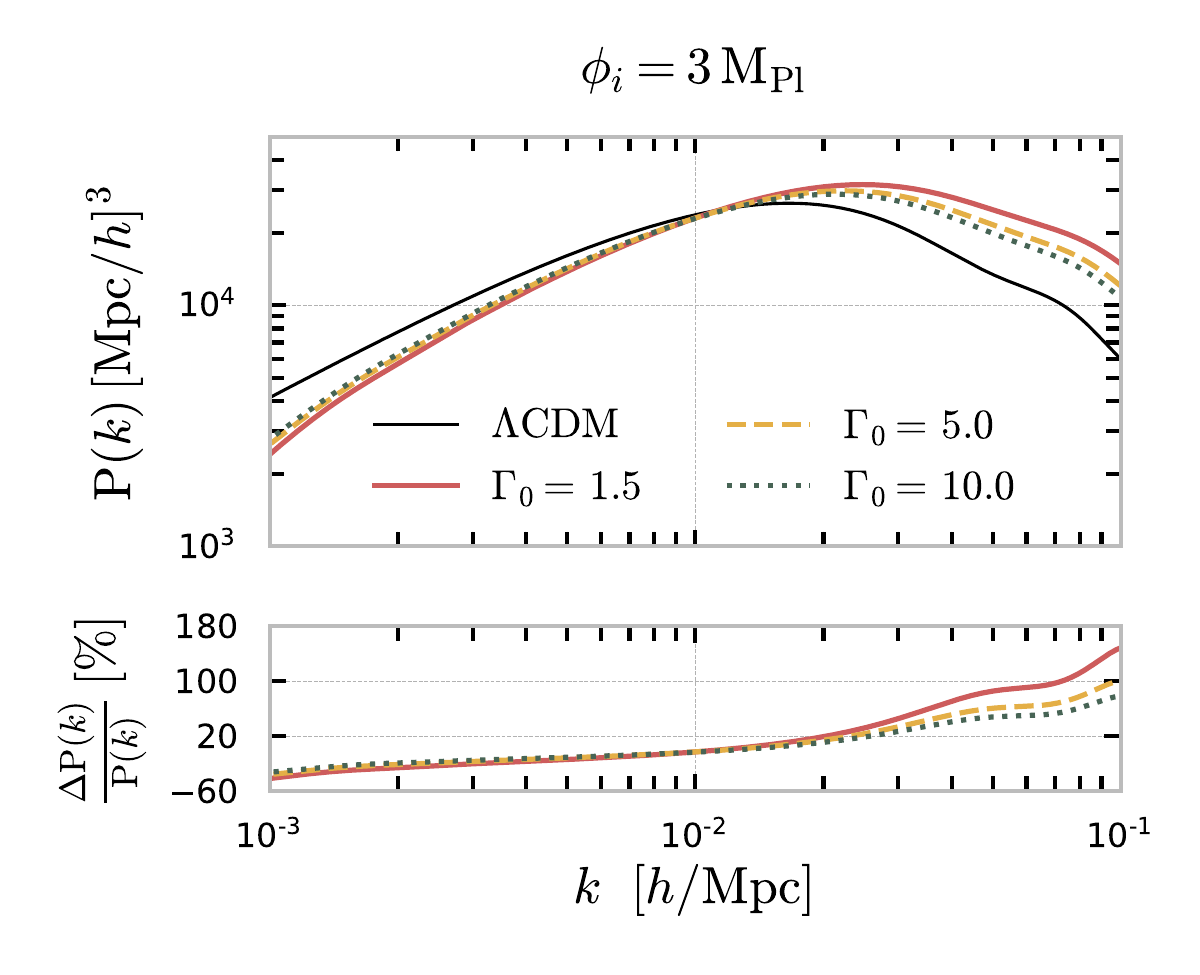}}
             \hfill
        \subfloat{\includegraphics[width=0.5\linewidth]{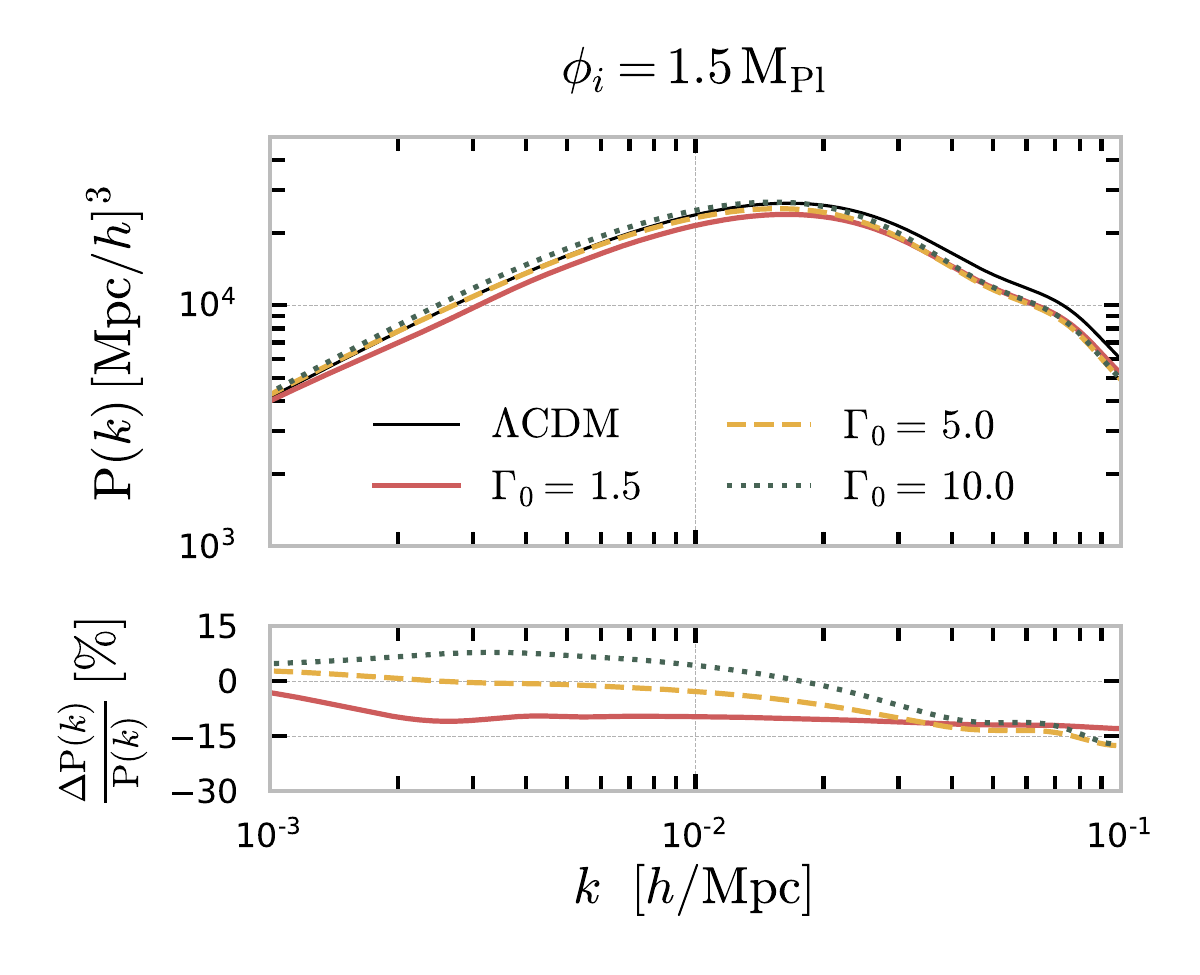}}
             \hfill     
  \caption[The CMB temperature and lensing angular power spectra and the matter power spectrum]{\label{fig:Powerspectra} The top and middle panels show the CMB temperature and lensing angular power spectra, $C_{\ell}^{TT}$ and $C_{\ell}^{\phi \phi}$, plotted as a function of the angle scale $\ell$. The bottom panels depict the matter power spectra $P_k$ for different Fourier scales (wavenumbers) $k$. For each observable, we also provide the $\Lambda$CDM predictions (black) and the relative deviation of each model from the standard model. The left and right panels correspond to Models M$_{3/1.5}$-M$_{3/10}$ and M$_{1.5/1.5}$-M$_{1.5/10}$ in \cref{tabel1}, respectively, and the line types used are the same as in \cref{fig:difgamma0beta,fig:difgamma0} for the background evolution.}
\end{figure}

The CMB anisotropy power spectrum and the matter power spectrum within the Dark D-brane framework can also be computed from the modified version of the CLASS code \cite{Lesgourgues:2011re, Blas:2011rf}. The results for the cases outlined in \cref{tabel1}, alongside the fiducial $\Lambda$CDM case for reference, are illustrated in \cref{fig:Powerspectra}. Here, the CMB temperature anisotropies are shown in the top panels, the lensing CMB power spectrum in the middle panels, and the matter power spectra for DDM and baryons at $ z = 0 $ are depicted in the bottom panels. We assumed standard adiabatic initial conditions, specifying an amplitude $ A_s = 2.215 \times 10^{-9} $ and a pivot scale $ k_{\text{pivot}} = 0.05 \, \text{Mpc}^{-1} $. In contrast, the initial perturbations for the scalar field are set to zero, $ \delta \phi_{\text{i}} = \delta \phi'_{\text{i}} = 0 $, without loss of generality. Mirroring the background cosmology, the evolution of the linearly perturbed quantities also exhibits a significant sensitivity to changes to the model parameters. This is corroborated in \cref{fig:Powerspectra}, where different values of $ \Gamma_0 $, with $ \phi_{\text{i}} $ held constant, lead to distinct shapes in the power spectra. On the other hand, in the left and right panels of \cref{fig:Powerspectra} we also show solutions with the same $ \Gamma_0 $ values: one with $ \phi_{\text{i}} = 3 \, \text{M}_{\text{Pl}} $ (models M$_{3/1.5}$-M$_{3/10}$ in the left panel) and another with $ \phi_{\text{i}} = 1.7 \, \text{M}_{\text{Pl}} $ (models M$_{1.5/1.5}$-M$_{1.5/10}$ in the right panel), both characterised by the same line styles for direct comparison. As anticipated, varying the initial condition for the scalar field yields distinct imprints on the linear perturbations, contrary to what was found in Ref.~\cite{Koivisto:2013fta} from a dynamical background study.

By contrasting the left and right panels of \cref{fig:Powerspectra}, we observe that the sign of the effective coupling, $ \beta $, which is directly related to the value of both $ \Gamma_0 $ and $ \phi_{\text{i}} $, markedly influences the evolution of the perturbations. In the context of the CMB temperature power spectrum, for $ \beta < 0 $ (upper-left panel), there is a marked enhancement for low multipoles $ \ell $ and suppression for medium and high $ \ell $ values, compared to the $ \Lambda $CDM prediction. On the other hand, for $\beta>0$ (upper-right panel), there is a mild enhancement at intermediate and large multipoles. Moreover, higher (lower) values of $ \Gamma_0 $ lead to fewer deviations from the $ \Lambda $CDM framework when $ \beta < 0 $ ($ \beta> 0 $). It is worth noting that although the present-day effective coupling (as demonstrated in \cref{fig:difgamma0beta}) in models M$_{3/1.5}$-M$_{3/10}$ in the left panel is of the same order of that in models M$_{1.5/1.5}$-M$_{1.5/10}$ in the right panel, the deviations from $ \Lambda $CDM are considerably more pronounced. This discrepancy is ascribed to the onset of the coupling at higher redshifts, resulting in an earlier departure from the standard model.

The predicted signatures for the CMB temperature anisotropies across the models can be explained on two fronts. Firstly, the background cosmology differs, as elaborated in \cref{sec:back}. Specifically, the ratio $ \Omega_b/\Omega_{\text{c}} $ at the time of recombination is generally not kept constant, and its high-redshift evolution depends on the chosen parameters. This ratio is the primary driver of the shifts in multipole and changes in amplitude of the peaks and valleys in the spectrum (see the discussion is \cref{sec:lcdm_param}); we can see that there is a tendency for the peaks to become narrower (wider) for models M$_{3/1.5}$-M$_{3/10}$ with $\beta<0$ (M$_{1.5/1.5}$-M$_{1.5/10}$ with $\beta>0$) relative to the $ \Lambda $CDM scenario. This introduces a degeneracy between the effective gravitational coupling and the Hubble parameter, as the latter has the greatest impact on the magnitude and position of the first peak. This non-trivial degeneracy is why we observe an overall enhancement of the low-$\ell$ tail of the spectrum in both cases considered, which depends on the balance between the effects.
Secondly, the time evolution of $ G_{\text{eff}} $, intimately linked to the coupling $ \beta $ and the $ \gamma $-factor, impacts the DDM fluid's perturbation growth. In models M$_{3/1.5}$-M$_{3/10}$, $ G_{\text{eff}} $ may start departing from zero as early as redshift $z \approx 1.1$ for the given $ \Gamma_0 $ values, while for models M$_{1.5/1.5}$-M$_{1.5/10}$, $ G_{\text{eff}} $ only starts growing at around a redshift of $z \approx 0.3$ or later. The late-time surge in the effective coupling affects the late-time Integrated Sachs--Wolfe (ISW) effect, as described in \cref{sec:cmb}. This effect alters the distance to the last scattering surface, suppressing (enhancing) the sound horizon at the baryon-drag epoch (see \cref{sec:lcdm_param}), shifting the peaks and troughs to the left (right) for $\beta<0$ in the left panel ($\beta>0$ in the right panel). This shift is more pronounced for the cases with the largest deviations from $\Lambda$CDM in the background.

\begin{figure}
\centering
      \subfloat{\includegraphics[width=0.5\linewidth]{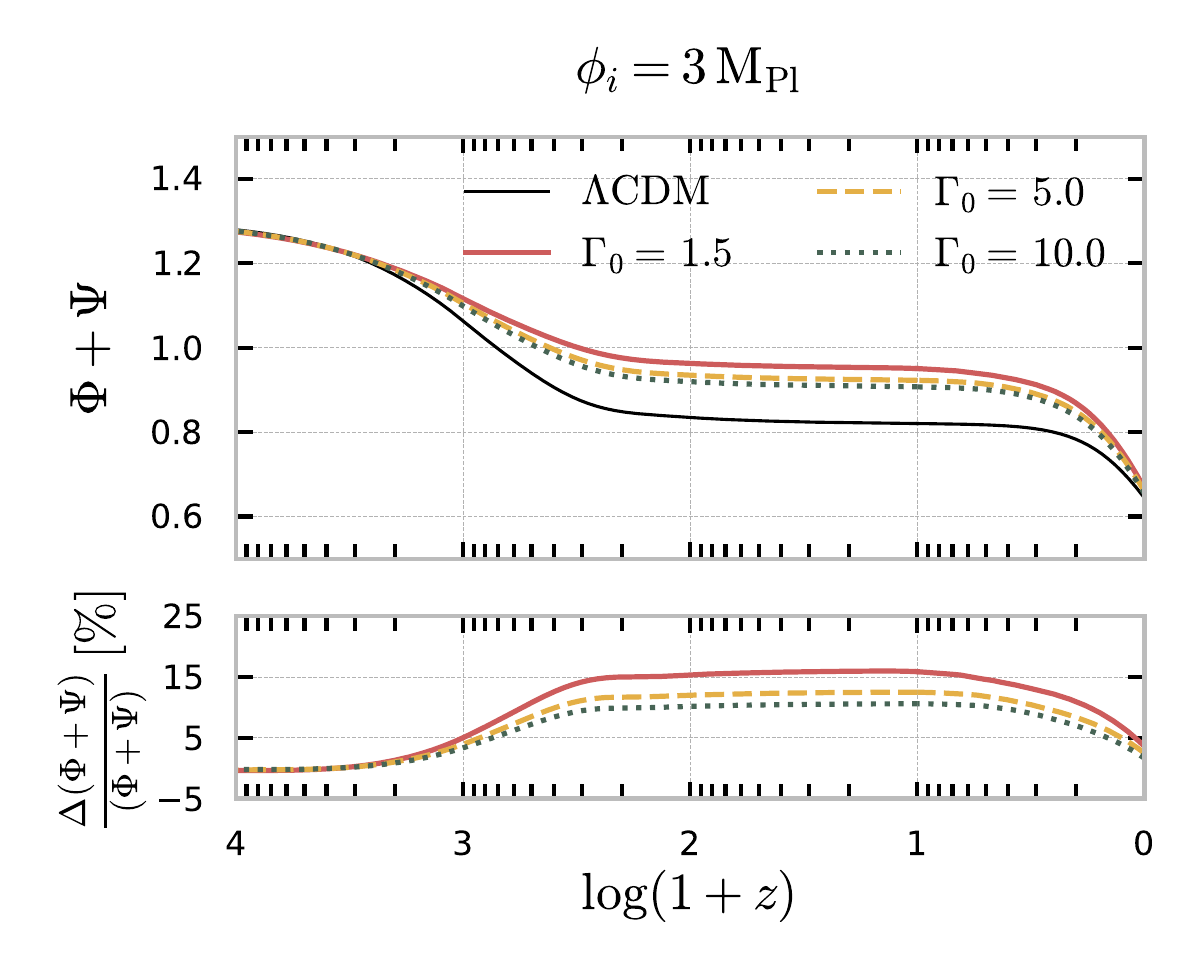}}
             \hfill
        \subfloat{\includegraphics[width=0.5\linewidth]{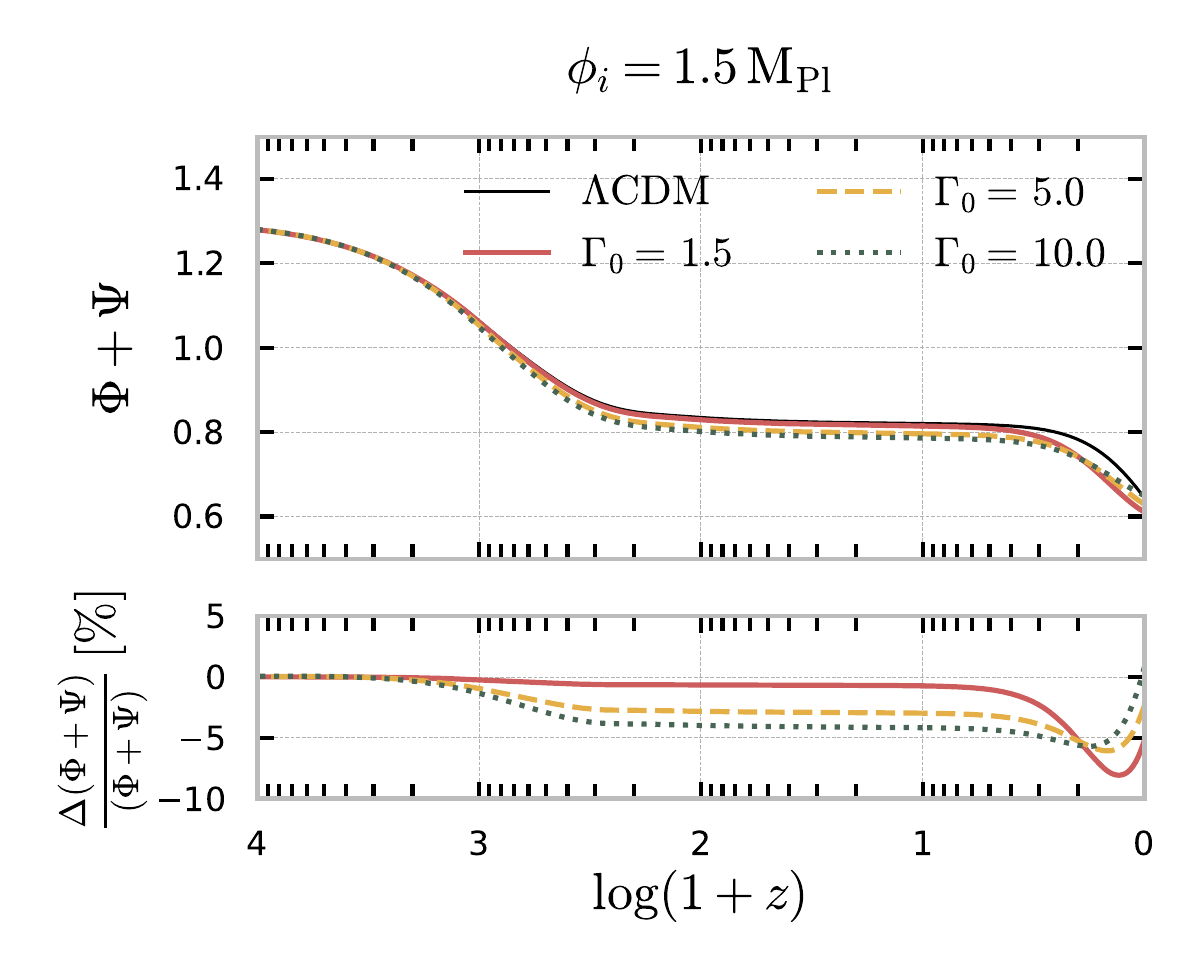}}
        \hfill
      \subfloat{\includegraphics[width=0.5\linewidth]{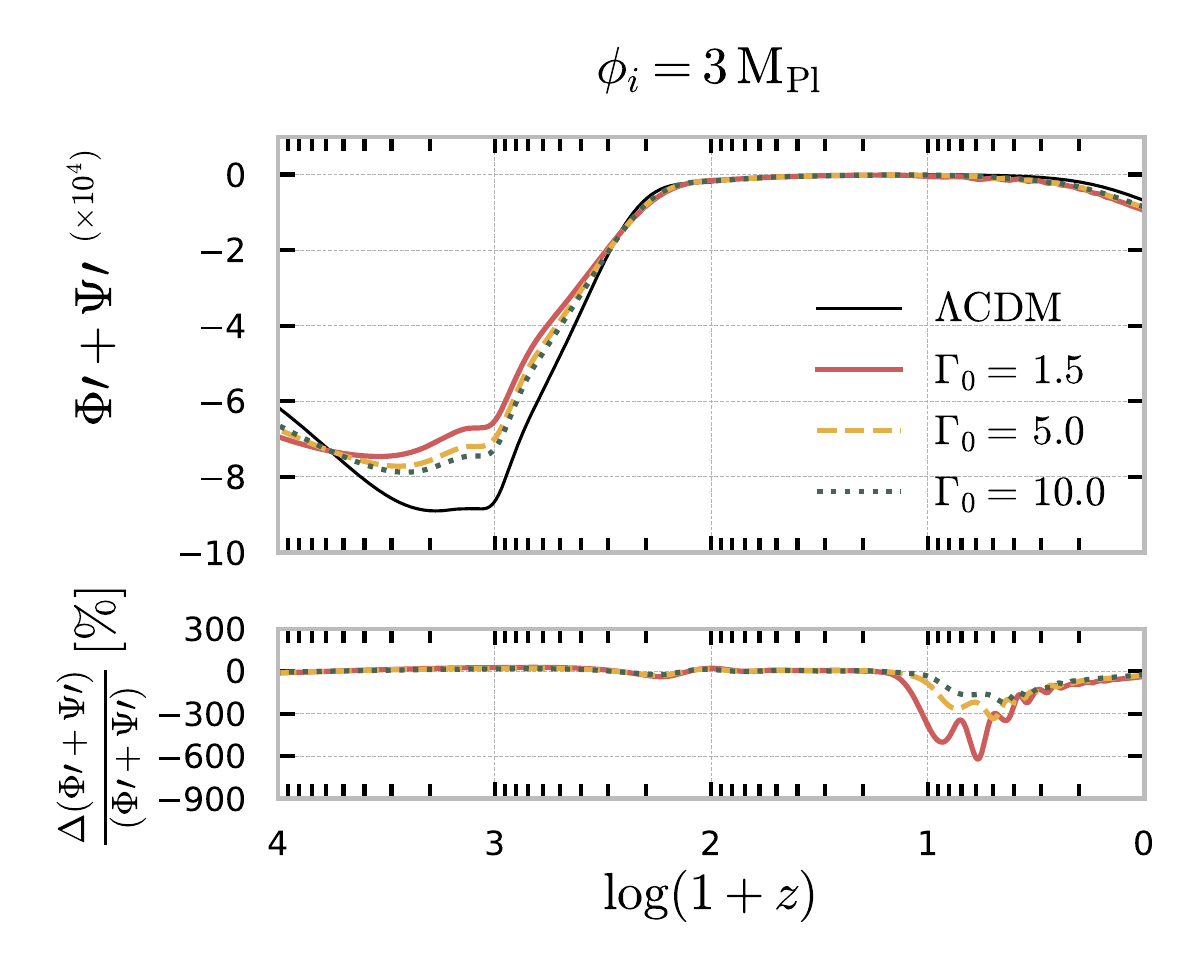}}
             \hfill
        \subfloat{\includegraphics[width=0.5\linewidth]{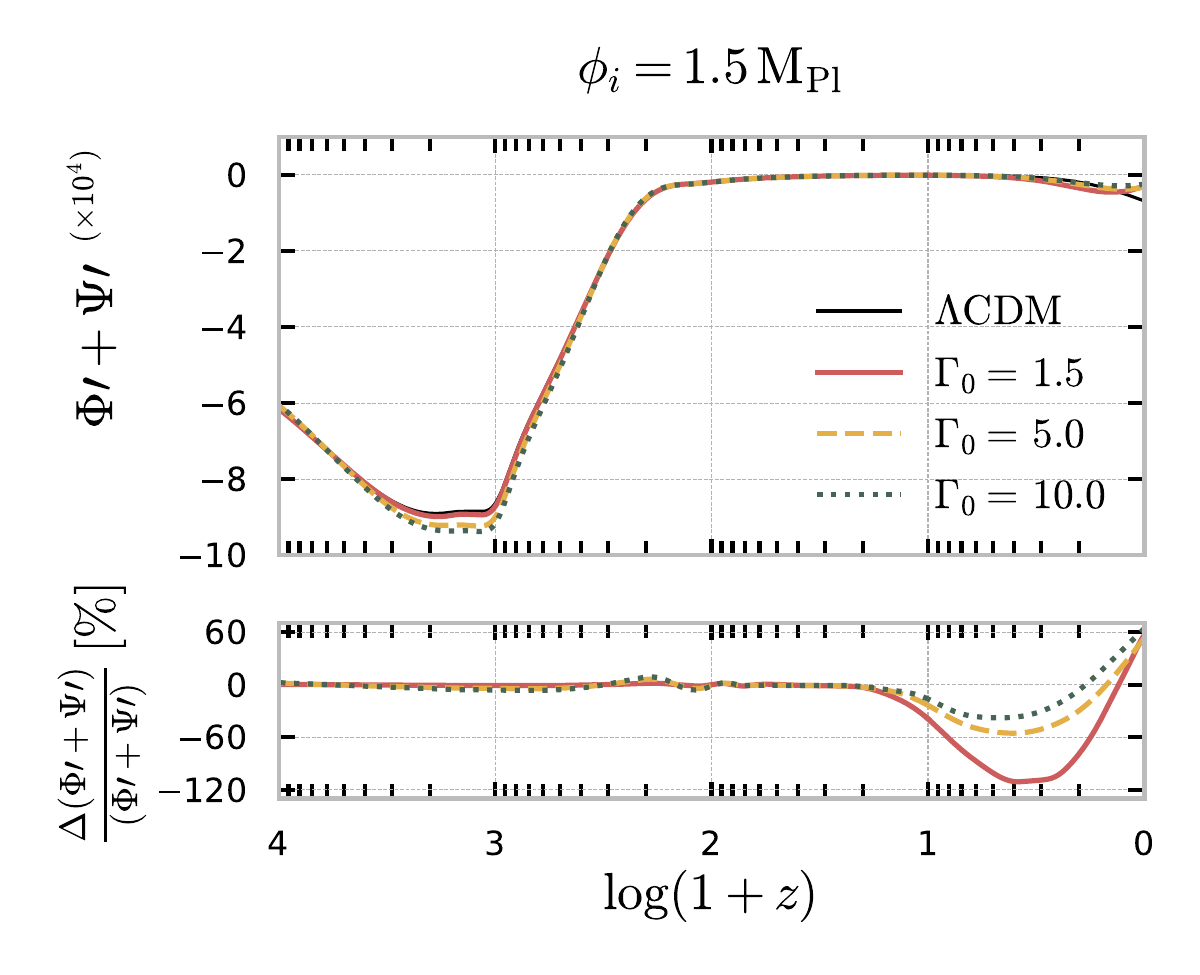}}
  \caption[Evolution of the lensing potential and its derivative]{\label{fig:lenspot} Evolution of the sum of the gravitational potentials, $\Phi+\Psi$, (\textit{top panels}) and the corresponding derivative with respect to conformal time, $\Phi'+\Psi'$ (\textit{lower panels}) for $ k = 0.01 $ Mpc$^{-1}$, plotted as a function of the redshift $z$. The left and right panels correspond to Models M$_{3/1.5}$-M$_{3/10}$ and M$_{1.5/1.5}$-M$_{1.5/10}$ in \cref{tabel1}, respectively, and the line styles used are the same as in \cref{fig:difgamma0beta,fig:difgamma0} for the background evolution.}
\end{figure}

In the middle panels of \cref{fig:Powerspectra}, we depict the lensing power spectrum, directly related to the evolution of the lensing potential $ \phi_{\text{len}} = \Phi + \Psi $, depicted in the top panels of \cref{fig:lenspot}, along with the corresponding time derivatives in the lower panel, at an intermediate scale of $ k = 0.01 $ Mpc$^{-1}$. The most pronounced deviations from $\Lambda$CDM manifest from around $ z \lesssim 10^3 $ during the matter-dominated epoch when changes in the disformal dark matter evolution become important. From this point on, there is an overall enhancement of $ \phi_{\text{len}}$ for the cases with $\beta<0$ (left panel) and a suppression for $\beta>0$ (right panel), with the largest departures identified for the lower and higher values of $\Gamma_0$ in each case, consistent with the considerations for the temperature power spectrum. The changes to the lensing potential arise from the energy flow in the dark sector and reflect directly in the lensing power spectrum $ C_{\ell}^{\phi \phi} $, depicted in the middle panels of \cref{fig:Powerspectra}, through their role in the line-of-sight integration presented in \cref{sec:cmb}. We observe a similar but opposite trend compared with $ C_{\ell}^{TT} $ for $ C_{\ell}^{\phi \phi} $, with an overall enhancement (suppression) of the spectrum at intermediate and high multipoles for $\beta<0$ in the left panel ($\beta>0$ in the right panel). The most significant deviations appear once again in the models with the earlier onset of the coupling, corresponding to the lowest (highest) values of $\Gamma_0$ for $\beta<0$ ($\beta>0$). This overall trend could help address the lensing excess reported in the \textit{Planck} temperature data of CMB anisotropies \cite{Aghanim:2018eyx,DiValentino:2019dzu}, as discussed in \cref{sec:Alens}.

The evolution of the lensing potential is also imprinted in the CMB temperature spectrum of anisotropies in the upper panels of \cref{fig:Powerspectra}. The largest contribution to the modifications comes from the integrated Sachs Wolfe (ISW) effect, which depends directly on $ \dot{\phi}_{\text{len}} $ (see \cref{sec:cmb}), depicted in the lower panels of \cref{fig:lenspot}. 
The contributions to the ISW effect can be separated into early and late-time components. In the early case, the ISW effect amplifies (reduces) the time derivatives of the gravitational potentials for $\beta<0$ ($\beta>0$), directly associated with an earlier (later) matter-radiation equality. Conversely, late-time ISW effects are primarily dictated by changes in the CMB lensing large-scale structure due to the modified dynamics of the dark sector. This culminates in a late-time attenuation of $\Phi' + \Psi'$, as demonstrated in the lower panels of \cref{fig:lenspot}.

The predictions for the matter power spectrum $P(k)$, depicted in the bottom panels of \cref{fig:Powerspectra} for for Fourier scales $ 10^{-3} h \, \text{Mpc}^{-1} < k < 10^{-1} h \, \text{Mpc}^{-1} $, are influenced by several factors. 
Besides the differing background evolution, the interplay between the effective gravitational coupling $ G_{\text{eff}} $, as defined in \cref{geff}, and the effective Hubble expansion rate, defined in \cref{heff}, play a crucial role in shaping the dark matter density contrast, as elaborated in the previous section. 
We observe that for models M$_{3/1.5}$-M$_{1.5/1.5}$ (M$_{1.5/5}$ and M$_{1.5/10}$), on large scales $ k \lesssim k_{\text{peak}} $, the dynamical contributions of the dark sector lead to a suppression (enhancement) in structure growth compared to the $ \Lambda $CDM model. On smaller scales, $ k \gtrsim k_{\text{peak}} $, the opposite effect is observed, except for model M$_{1.5/1.5}$ which still experiences a slight suppression due to the unique balance and sign change in the evolution of $\beta$. 
In agreement with the CMB power spectrum discussion, the deviations from the standard model are more accentuated in models M$_{3/1.5}$-M$_{3/10}$, owing to the earlier redshift onset of the coupling.


Furthermore, at least for particular periods in the evolution, the perturbation in the coupling, $ \delta Q $, is scale dependence \textit{via} the coefficient $ \mathcal{Q}_5 $, as presented in \cref{q5}, consistent with the change in the overall trend of suppression/enhancement before and after the peak of the spectra.
With the chosen normalisation of the power spectrum, for the models with $\beta<0$ ($\beta>0$), the amplitude of the primordial perturbations must be greater (lower) to match the current structure count. This is in line with the feature of these models exhibiting slightly enlarged dark energy density fractions at late times, compared to the $ \Lambda $CDM predictions (red curves in the upper-left panel of \cref{fig:difgamma0}), and explains why the peak of the spectrum is shifted to the right (left).
Deviations from $\Lambda$CDM surpass $100\%$ in the models depicted in the left panels of \cref{fig:Powerspectra}, peaking at $ k \approx 10^{-1} h \text{Mpc}^{-1} $, a point where the linear approximation is expected to break down, and non-linear effects take over. 

The ISW effect is correlated with the growth, with the behaviour at larger scales (smaller $k$) contributing to the enhancement/suppression trend in the low-$ \ell $ tail of the CMB anisotropy spectrum depicted in the top panels of \cref{fig:Powerspectra}. Another potential contributor to the ISW effect could be a phase of early dark energy. However, unlike other dark energy models with disformal couplings, such as in Ref.~\cite{Zumalacarregui:2012us}, we do not witness any relevant early dark energy signatures in the Dark D-brane framework, given the form of the potential adopted here.

We remark that, from \cref{tabel1}, the toy models M$_{3/1.5}$-M$_{3/10}$ with $\beta<0$ (M$_{1.5/1.5}$-M$_{1.5/10}$ with $\beta>0$) predict values for $ \sigma_8 $ that are higher (lower) than the $ \Lambda $CDM counterpart, $ \sigma_8^{\Lambda {\rm CDM}} \approx 0.85 $.

The DDM scenario shares certain traits with other disformal models discussed in the literature \cite{Zumalacarregui:2012us,vandeBruck:2015ida,Mifsud:2017fsy,vandeBruck:2017idm}. Specifically, the effective gravitational coupling between dark matter particles is not constant. In the Dark D-brane setting, the coupling is negligible in the early Universe, suppressed by the denominator in \cref{beta}. This aspect further motivates research exploring violations of the equivalence principle in the dark sector at later cosmological times. In the following section, we will focus on the impact of this time-evolving coupling on structure formation by testing the DDB model against existing cosmological data.

\section{Observational Constraints} \label{sec:dbi_const}

\subsection{Methodology and Data}

After reviewing the main predictions of the Dark D-Brane Model (DDB) at the background and linear perturbative level, we wish to study the constraints on the model's parameters imposed by incremental combinations of the data sets defined below. For this purpose, we rely once more on the \texttt{MontePython} code \cite{Audren_2013, Brinckmann:2018cvx}, an MCMC sampler of the cosmological parameter space based on our modification of the \texttt{CLASS} code, employing the Metropolis-Hastings algorithm with a Gelman-Rubin \cite{1992StaSc...7..457G} convergence criterion of $R - 1 < 10^{-2}$ for all the chains, as described in \cref{chapter:statistics}. From this method we derive the corresponding posterior distribution for the sampled parameters. In the following section, we discuss the main results of this analysis.

The baseline data set for this study is as described in \cref{sec:baseline}. More precisely, we consider the full TTTEEE+lowE CMB likelihood from the \textit{Planck} 2018 data release,composed of data on the CMB temperature (TT) and polarisation (EE) anisotropies, including their joint cross-correlations (TE), at both low and high multipoles. More precisely, this corresponds to the high-$\ell$ \texttt{Plik} likelihood for TT in the range $30 \leq \ell \leq 2508$ and TE and EE for $30 \leq \ell \leq 1996$, combined with the low-$\ell$ ($2 \leq \ell \leq 29$) TT and EE likelihoods based on the \texttt{Commander} algorithm and the \texttt{SimAll} likelihood. This data set is abbreviated as Pl18 throughout the text. We also investigate the effect of including the CMB lensing likelihood \cite{Planck:2018lbu}. We also study the combination of Planck with other background data sets. In this case, we include baryon acoustic oscillations (BAO) distance and expansion rate measurements from the 6dF Galaxy Survey \cite{Beutler:2011hx}, the Sloan Digital Sky Survey (SDSS) DR7 Main Galaxy Sample \cite{Ross:2014qpa} and SDSS DR12 \cite{BOSS:2016hvq}, as employed by the \textit{Planck} collaboration as well. Finally, we include Pantheon's 1048 supernova (SN) data sample \cite{Scolnic:2017caz}. The combination of this background data is referred to throughout this chapter as BAO+SN. 

For each data set combination, we sample the standard cosmological parameters, namely the reduced dark matter and baryon energy densities, $\Omega_{\rm c} h^2$ and $\Omega_{\rm b} h^2$; the ratio between the sound horizon and the angular diameter distance at decoupling, $\theta_s$; the amplitude and power of the scalar primordial power spectrum $A_s$ (expressed as $\ln \left( 10^{10} A_s \right)$) and $n_s$; and the reionisation redshift, $z_{\rm reio}$. In addition to these, we also consider the scalar field parameters $\td{h}_0 = 1/h_0$, as defined in \cref{Gamma0}, and the scalar field's initial condition $\phi_i$. We employ flat priors on the sampled parameters, including the nuisance parameters associated with each likelihood. The velocity of the scalar field is considered to be fixed and set to $\phi'_{i} = - 10^{-25} ({\rm Mpl\, Mpc}^{-1})$ as its influence on the dynamics was found to be negligible for this study. As briefly referred, the potential scale $V_0$ is used as a derived shooting parameter to account for degeneracies in the parameter space and ensure the Friedmann flatness condition. It can be constrained as a derived parameter and combined with $\td{h}_0$ to express the results regarding the physical parameter $\Gamma_0$. The priors for the sampled parameters are listed in \cref{tab:dbi_priors} and the results of the study are summarised in \cref{ tab:dbi_lcdm , tab:dbi_dbi } for the $\Lambda$CDM reference case and for the Dark D-Brane model.

The sampled MCMC chains obtained from \texttt{MontePython} are analysed using the \texttt{GetDist} Python package \cite{Lewis:2019xzd}. The statistical significance compared to the concordance model is inferred through the Bayes factor $\ln B_{ij}$ in \cref{eq:evdef}, an estimator of the Bayesian Evidence, computed from the publicly-available \texttt{MCEvidence} code \cite{Heavens:2017afc}. 
A negative value indicates preference over the concordance model, with the opposite holding for positive values and the level of support is assessed by the Jeffreys scale criteria presented in \cref{tab:jeff_scale}.

\begin{table}[hbt!]
\begin{center}
\begin{tabular}{c|c}
\hline
Parameter                    & Prior \\
\hline
$\Omega_b h^2$                & $[0.005,0.1]$ \\
$\Omega_c h^2$                & $[0.001,0.99]$ \\
$100\theta_{s}$                    & $[0.5,10]$ \\
$z_{reio}$                          & $[0.,20.]$ \\
$n_s$                      & $[0.7,1.3]$ \\
$\log \left(10^{10}A_{s} \right)$   & $[1.7, 5.0]$ \\
\hline
$\td{h}_0$                           & $[10^{-20},5]$ \\
$\phi_i$                            & $[0,20]$ \\
\hline 
\end{tabular}
\end{center}
\caption[Priors on the model parameters]{Flat priors on the cosmological and model parameters sampled in this work.}
\label{tab:dbi_priors}
\end{table}

\subsection{Results} \label{sec:dbi_obs}

\begin{figure}
\centering
      \subfloat{\includegraphics[width=1.\linewidth]{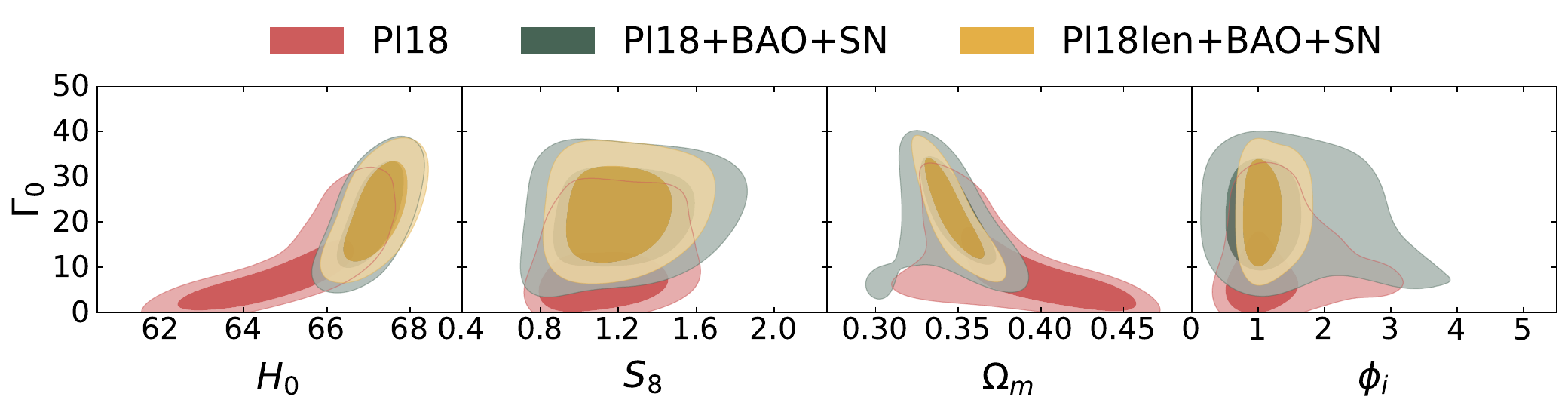}} \hfill
      \subfloat{\includegraphics[width=1.\linewidth]{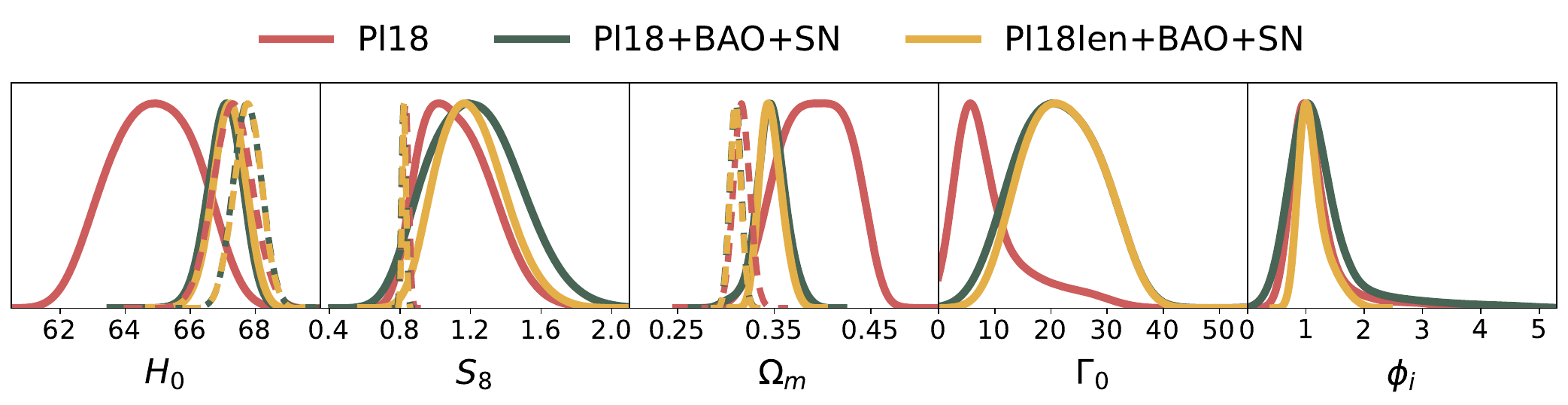}}
  \caption[Marginalised 1D and 2D constraints for the Dark D-Brane model]{\label{fig:const_g0} 2D marginalised $68\%$ and $95\%$ CL contours (top panel) and 1D marginalised curves (lower panel) obtained in the Dark D-Brane model under consideration for the {\it Planck} 2018 data (red), the {\it Planck} 2018, BAO and SN combination (green), and their combination with CMB lensing (yellow). The dashed lines correspond to the $\Lambda$CDM reference case.}
\end{figure}

\sloppy
 In \cref{ tab:dbi_lcdm , tab:dbi_dbi }, we report the statistical study results for the sampled cosmological and model parameters for the various data combinations considered for the Dark D-Brane and $\Lambda$CDM models at $68\%$ confidence level. This allows for a direct comparison of the results, even if the fact that the Dark D-Brane model does not possess a well-defined uncoupled and $\Lambda$CDM limits on the parameters makes this comparison non-trivial. As we have seen in the numerical analysis of the toy models, the case $\Gamma_0 \to \infty$ seems to represent a $\Lambda$CDM limit at the background level, explaining the apparent preference for high values of $\Gamma_0$ seen in \cref{ tab:dbi_dbi }.
 The results are also represented in \cref{fig:const_g0} as a 2D rectangular plot for marginalised distributions in the dark D-brane model and as the 1D marginalised distribution curves, including a comparison with $\Lambda$CDM for the cosmological parameters.

In the first column of \cref{ tab:dbi_lcdm , tab:dbi_dbi }, we present the constraints obtained considering Planck data only. We find a non-zero and bounded prediction at $1\sigma$ for $\Gamma_0$, roughly quantifying the deviation from the canonical case, even if this is a broader constraint compared to the combination with other data sets. On the other hand, the fact that we obtained non-vanishing constraints for $\phi_i$ at $68\%$ CL even with \textit{Planck} data alone highlights its importance as a cosmological parameter with impact over the dynamics, especially at the perturbative level. We verify that these considerations are maintained at $95\%$ and $99\%$ CLs, driven exactly by the model's unique features which cannot be trivially related to $\Lambda$CDM. 

We also look at the effect of adding the BAO+SN background data, followed by the lensing CMB data on top of the baseline Planck. We see that the inclusion of the CMB lensing data does not significantly impact the results, apart from a considerable reduction in the constraining power on $\phi_i$, which is related with a slight better convergence for the remaining parameters, illustrated by the yellow contours in \cref{fig:const_g0}. On the other hand, the addition of the SN and BAO data significantly improves the constraints on $\Omega_m$, leading to tighter constraints for the other background parameters as well, namely $\{H_0,\Gamma_0,\phi_i\}$, and bringing $\Gamma_0$ to considerably higher values, closer to the $\Lambda$CDM-like limit. The latter effect is also connected to the enhancement in the lensing power spectrum for higher values of $\Gamma_0$, which better accommodates the CMB lensing data under the lensing excess reported by \textit{Planck}. The addition of the background data sets also brings the value of $H_0$ considerably closer to the $\Lambda$CDM prediction. 
This discussion should be taken with a grain of salt as is a well-known effect, reported in other models as well \cite{Frusciante:2019puu,DiValentino:2022oon,DiValentino:2017gzb,Abdalla:2022yfr,DiValentino:2020leo,sym10110585,DiValentino:2020hov}, possibly related to a bias in the BAO data towards $\Lambda$CDM, which could require some corrections to be properly compared with other models. More importantly, with the inclusion of all the data sets, there is still evidence for a non-vanishing coupling (bounded $\Gamma_0$ and $\phi_i$) at more than $1\sigma$ without considerably worsening the tensions with the low-redshift data, present in the \textit{Planck} only case.

\begin{figure}[t]
   \centering
 \subfloat{\includegraphics[width=1.\linewidth]{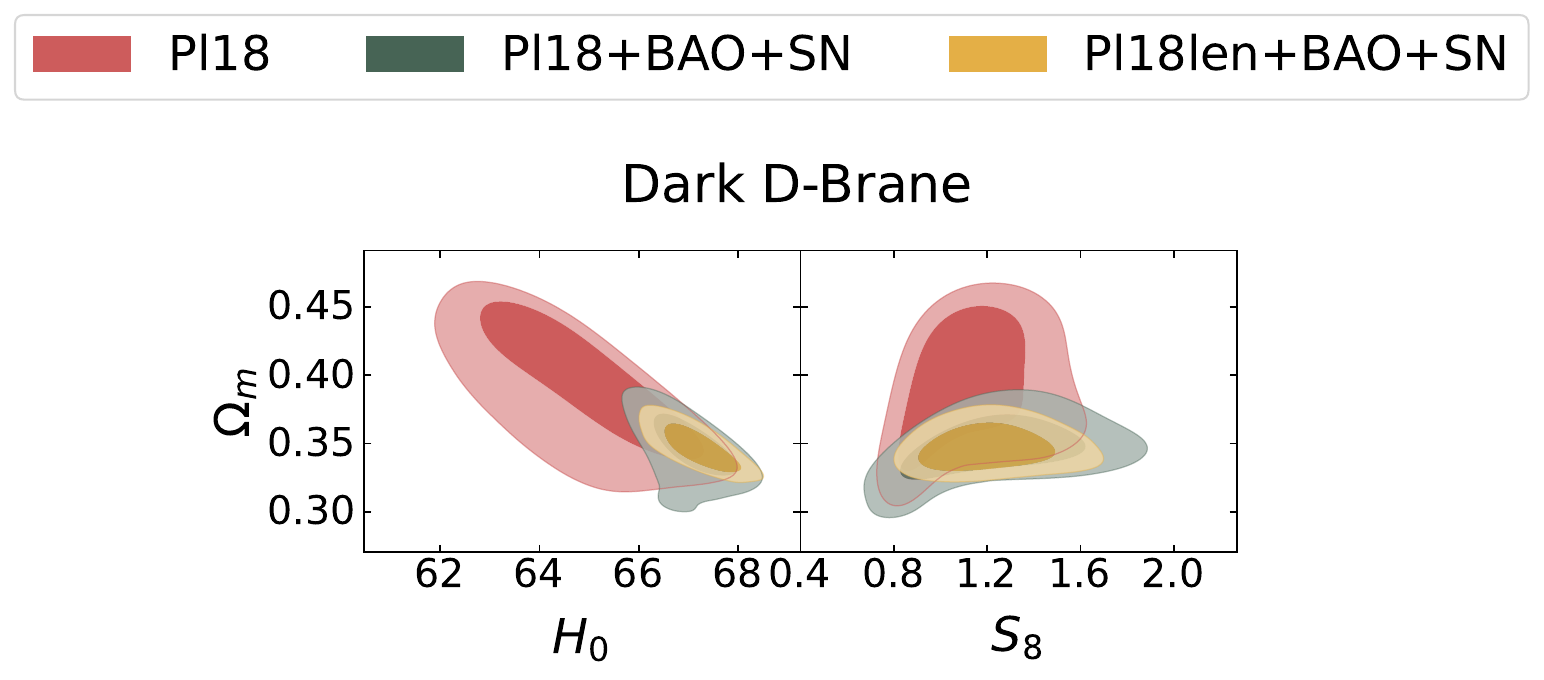}} \\
     \hspace{-5.25pt}\vspace{-15pt}\subfloat{\hspace{-10pt}\includegraphics[width=0.72\linewidth]{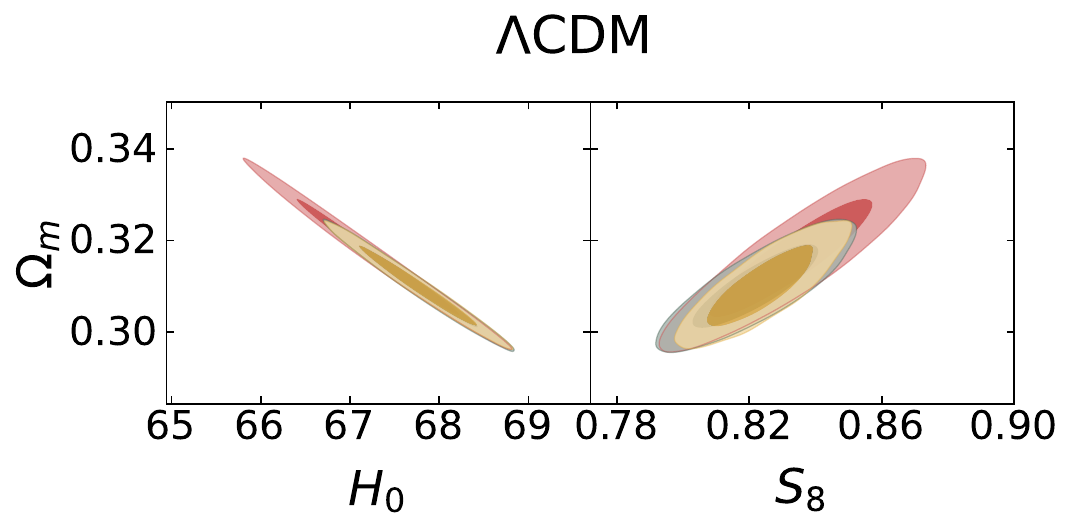}\hspace{10pt}} \\
  \caption[2D marginalised posterior distributions in the $\Lambda$CDM and Dark D-Brane models]{\label{fig:OmH0S8dbi} 68\% and 95\% CL 2D contours derived for the parameter combinations $H_0$-$\Omega_m$ (left panels) and $S_8$-$\Omega_m$ (right panels) in the Dark D-Brane model (upper panels) and $\Lambda$CDM model (lower panels) for the {\it Planck} 2018 data (red), the {\it Planck} 2018, BAO and SN combination (green), and their combination with CMB lensing (yellow).}
\end{figure}

In the upper panel of \cref{fig:const_g0} we see the 2D marginalised correlations between $\Gamma_0$ and the parameters $\{H_0,S_8,\Omega_m, \phi_i\}$. We find a distinct positive correlation between the coupling parameter $\Gamma_0$ and $H_0$ and a negative correlation between $\Gamma_0$ and $\Omega_m$. In particular, we see that the contours are more accommodating in the Pl18-only case. The introduction of the background data, namely the BAO data, greatly constrains $\Omega_m$ towards lower values, which results in larger values for $\Gamma_0$ and $H_0$ as well. We see that the addition of the lensing data contributes to an improvement on the constraints, especially in the value of $\phi_i$.
In the lower panel of \cref{fig:const_g0} we depict the 1D marginalised posterior distributions for $\{H_0,S_8,\Omega_m, \Gamma_0,\phi_i\}$ for the DDB model, including the reference $\Lambda$CDM curves for $\{H_0,S_8,\Omega_m\}$ as dashed lines. We see that the parameters are consistently constrained in $\Lambda$CDM, with $H_0$ taking higher values, while $\Omega_m$ and $S_8$ are constrained towards lower values. Although this seems to push the results in the opposite direction than what is necessary to address the cosmic tension in $H_0$ and $S_8$ in the standard model, we point to the enlarged error bars in the curves for those parameters, and emphasise once more that the weak lensing data should be consistently reanalysed in the context of the DDB model in order to quantify any tension in $S_8$. The $H_0$ tension of $\sim 4.8\sigma$ in $\Lambda$CDM when considering Pl18 data alone is maintained at $\sim 4.9 \sigma$ in the DDB model. 
While this could still hint at the need for new physics, \textit{e.g.} a Dark D-Brane formulation in a less simplified setting, we conclude that this particular framework alone does not lead to a significant ease of the tensions between early- and late-time probes. The comparison in the constraints for the Dark D-Brane and $\Lambda$CDM cases is given in \cref{fig:OmH0S8dbi} for the 2D marginalised contours for the parameters directly involved in the tensions reported above.

We find that $\phi_i$ is constrained to be of the order of unity, with similar and consistent results for all the data combinations at $68\%$ CL, considering that we have shown that its value greatly impacts the dynamics, even possibly altering the direction of energy flow in the dark sector interaction. For this range of values, the numerical study points at higher values of $\Gamma_0$, leading to a greater transfer of energy from DE to DM ($\beta>0$), associated with an overall enhancement of the CMB temperature-temperature angular power spectrum, as discussed over the previous sections and depicted in the right-upper panel of \cref{fig:Powerspectra}. This is consistent with the positive correlation between $\Gamma_0$ and $H_0$, illustrated in the upper panel of \cref{fig:const_g0}. It is worth noting that the contour plots show a clear saturation point for $\Gamma_0 \gtrsim 20$, corresponding to the case in which the cosmological realisations become too close to $\Lambda$CDM and the effect of $\Gamma_0$ becomes degenerate, with increasing values just approaching more closely the $\Lambda$CDM predictions. Nevertheless, we see that the constraints on $\phi_i$ are relatively tight and not significantly correlated with $\Gamma_0$, showing how well the data can constrain $\phi_i$ in that range of values independently of $\Gamma_0$. Indeed, in the previous sections we have confirmed how the two DBI parameters are not degenerate, given that they play considerably different roles in the cosmological evolution and in shaping the observables.

In the last two rows of \cref{ tab:dbi_dbi } we present the results for the effective $\Delta \chi^2$ and for the Bayesian evidence model comparison criteria, according to \cref{sec:modelsel,sec:modelcomp}. We see that the Bayesian evidence is considerably negative for all the data combinations, implying no support for the Dark D-brane model over $\Lambda$CDM. On the other hand, the non-negligible values of $ \Delta \chi^2_{\text{min}} $ indicate an improvement in the fit. However, this arises as a result of having introduced more parameters, making the model naturally more accommodating, with the Bayesian evidence confirming that the increased parameter space is not justified. Even if we were tempted to say that this might point to an arguable evidence for the Dark D-Brane model over $\Lambda$CDM, from an objective point of view, we have shown that, as for most interacting dark energy models with couplings proportional to the density of DM, this model does not address the cosmological tensions.


\begin{table*}
\begin{center}
\renewcommand{\arraystretch}{1.5}
\resizebox{\textwidth}{!}{
\begin{tabular}{l c c c c c c c c c c c c c c c }
\hline\hline
\textbf{Parameter} & \textbf{ P18 } & \textbf{ P18 + BAO + SN } & \textbf{ P18len + BAO + SN } \\ 
\hline
$ \omega_{\rm b}  $ & $  0.02236\pm 0.00015 $ & $  0.02243\pm 0.00014 $ & $  0.02245\pm 0.00014 $ \\ 
$ \omega_{\rm c}  $ & $  0.1202\pm 0.0014 $ & $  0.11921\pm 0.00098 $ & $  0.11923\pm 0.00094 $ \\ 
$ 100\theta_{s}  $ & $  1.04188\pm 0.00030 $ & $  1.04197\pm 0.00028 $ & $  1.04196\pm 0.00029 $ \\ 
$ \tau_{\rm reio}  $ & $  0.0542\pm 0.0080 $ & $  0.0558\pm 0.0078 $ & $  0.0565\pm 0.0074 $ \\ 
$ n_{s}  $ & $  0.9652\pm 0.0043 $ & $  0.9676\pm 0.0038 $ & $  0.9677\pm 0.0037 $ \\ 
$ \ln 10^{10}A_{s}  $ & $  3.045\pm 0.016 $ & $  3.046\pm 0.016 $ & $  3.048\pm 0.015 $ \\ 
\hline
$ \sigma_8  $ & $  0.8115\pm 0.0076 $ & $  0.8091\pm 0.0072 $ & $  0.8105\pm 0.0061 $ \\ 
$ S_8  $ & $  0.833\pm 0.016 $ & $  0.822\pm 0.012 $ & $  0.824\pm 0.011 $ \\ 
$ \Omega_{\rm m}  $ & $  0.3163\pm 0.0085 $ & $  0.3100\pm 0.0059 $ & $  0.3100\pm 0.0057 $ \\ 
$ H_0  $ & $  67.31\pm 0.61 $ & $  67.76\pm 0.44 $ & $  67.77\pm 0.43 $ \\ 
\hline \hline
\end{tabular} }
\end{center}
\caption[Observational constraints for the reference $\Lambda$CDM]{ Observational constraints at a $68 \%$ confidence level on the independent and derived cosmological parameters using different data set combinations for the $\Lambda$CDM model. }
\label{ tab:dbi_lcdm }
\end{table*}



\begin{table*}
\begin{center}
\renewcommand{\arraystretch}{1.5}
\resizebox{\textwidth}{!}{
\begin{tabular}{l c c c c c c c c c c c c c c c }
\hline \hline
\textbf{Parameter} & \textbf{ P18 } & \textbf{ P18 + BAO + SN } & \textbf{ P18len + BAO + SN } \\ 
\hline
$ \omega_{\rm b} $ & $  0.02241\pm 0.00015 $ & $  0.02251\pm 0.00014 $ & $  0.02253\pm 0.00014 $ \\ 
$ \omega_{\rm c} $ & $  0.142^{+0.012}_{-0.0086} $ & $  0.1327^{+0.0053}_{-0.0041} $ & $  0.1333^{+0.0030}_{-0.0043} $ \\ 
$ 100\theta_{s}  $ & $  1.04191\pm 0.00029 $ & $  1.04204\pm 0.00028 $ & $  1.04205\pm 0.00029 $ \\ 
$ \tau_{\rm reio}  $ & $  0.0533\pm 0.0080 $ & $  0.0553\pm 0.0080 $ & $  0.0531\pm 0.0081 $ \\
$ n_{s}  $ & $  0.9662\pm 0.0045 $ & $  0.9699\pm 0.0038 $ & $  0.9704\pm 0.0038 $ \\ 
$ \ln 10^{10}A_{s}  $ & $  3.043\pm 0.017 $ & $  3.043\pm 0.016 $ & $  3.038\pm 0.015 $ \\ 
$ \td{h}_0  $ & $  0.146^{+0.075}_{-0.14} $ & $  0.038^{+0.015}_{-0.029} $ & $  0.036^{+0.012}_{-0.022} $ \\ 
$ \phi_i  $ & $  1.14^{+0.16}_{-0.45} $ & $  1.24^{+0.18}_{-0.56} $ & $  1.11^{+0.13}_{-0.28} $ \\ 
\hline
$ \sigma_8  $ & $  0.98^{+0.11}_{-0.20} $ & $  1.14^{+0.19}_{-0.27} $ & $  1.12^{+0.14}_{-0.20} $ \\ 
$ S_8  $ & $ 1.12^{+0.16}_{-0.24} $ & $ 1.22^{+0.23}_{-0.28} $ & $ 1.20^{+0.16}_{-0.21}$ \\ 
$ \Omega_{\rm m}  $ &$  0.392^{+0.039}_{-0.035} $ & $  0.347\pm 0.016 $ & $  0.3467^{+0.0096}_{-0.013} $ \\ 
$ H_0  $ & $  64.9\pm 1.3 $ & $  67.08^{+0.55}_{-0.50} $ & $  67.20\pm 0.57 $ \\ 
$ \Gamma_0  $ & $ 8.9^{+2.5}_{-7.3} $ & $ 22\pm 8 $ & $ 23^{+7}_{-8} $ \\ 
\hline
$\Delta \chi^{2}_{\rm min} $ &  $-5.42$ & $-4.24$ & $-2.48$ &  \\
$\ln B_{i, \Lambda {\rm CDM}}$ & $-5.72$ & $-8.25$  & $-7.40$ &  \\
\hline \hline
\end{tabular} }
\end{center}
\caption[Observational constraints for the Dark D-Brane model]{ Observational constraints at a $68 \%$ confidence level on the independent and derived cosmological parameters using different data set combinations for the Dark D-Brane model. }
\label{ tab:dbi_dbi }
\end{table*}


\section{Discussion} \label{sec:concdbi}

In this Chapter, we have explored the dynamics and phenomenology of the dark D-brane model. As presented in previous works, it is well known that the Dark D-Brane model alleviates some of the theoretical problems associated with $\Lambda$CDM by introducing a theoretical motivation and geometrical interpretation for the dark sector. This study analysed a simple toy model to realise the DDB model in AdS$_5 \times$S$^5$ warped regions corresponding to a quadratic potential and an inverse quartic warp factor.

We performed a numerical study of the background and linearly perturbed equations. We found that the model has a non-trivial complex behaviour, with initial scalar field conditions significantly influencing the late-time cosmological evolution. Specifically, we categorised two distinct regimes based on the behaviour of the effective coupling at varying redshifts. These regimes yield different cosmological outcomes, highlighting the model's richness.

At the background level, we showed that the Universe evolves toward a scaling fixed point, where the fractional energy densities of DE and DM maintain a constant ratio. For higher values of the parameter $\Gamma_0$, the evolution closely resembles that of the $\Lambda$CDM model. We emphasised that the initial condition of the field $\phi_i$ plays a significant role, notably affecting the direction of energy exchange between dark energy and dark matter. The coupling is consistently negligible at earlier times and its onset redshift is determined by the DBI parameters $\Gamma_0$ and $\phi_i$. By fixing $\Gamma_0$ we gather that, by taking initial conditions for the field that lead to higher present values of $\beta$, the coupling starts to grow much earlier, during the matter dominated epoch, leaving distinct signatures at the level of the perturbations. Therefore, we conclude that the difference in the direction of the energy exchange between the dark fluids introduces distinct features at the level of the background, potentially shifting the matter-radiation equality and the time of transition from matter to dark energy dominated epochs.

On the perturbation front, we derived equations for both the Newtonian and synchronous gauges, uncovering diverse implications for the CMB and matter power spectrum. Particularly, we noted deviations from the $\Lambda$CDM model that are consistent with the coupling behaviour. 
Moreover, the numerical results in this work imply that the Dark D-brane model has the potential to produce unique observational features, especially in the CMB spectra. 

In our numerical study, we consistently considered the scalar field's initial value close to the \textit{Planck} mass. Within the framework of higher-dimensional theories, this corresponds to the D-brane being positioned far from the tip of the AdS throat. On the other hand, the study in Ref.~\cite{Koivisto:2013fta} examined an alternative limit where $\phi$ is much smaller than the Planck mass and the D-brane is close to the tip of the AdS throat. In this setting, to account for the present-day vacuum energy, the scalar field mass, $m_{\phi}$, is subject only to a lower bound—making it fundamentally different from conventional quintessence theories where a small mass is often assumed. We found that numerically simulating this case is non-trivial, primarily due to the reduced initial value needed for the scalar field and the need to consider high values of $\Gamma_0$. Future detailed studies on this specific scenario would be interesting, although we anticipate that the overarching physics presented in our current work should remain relevant in such contexts.

To test the model against observational data we have performed an MCMC analysis. We list below some of the main findings of the study:

\begin{itemize}
    \item Cosmic Tensions: The Dark D-Brane model gives a lower mean value of $H_0$ across all data sets, compared to $\Lambda$CDM. While having slightly enlarged error bars it does not address the tension between different $H_0$ measurements. The model accommodates for larger values of $\Omega_m$, consequently leading to higher $S_8$ values, associated with an increased $1\sigma$ region.

    \item Constraints on Model Parameters: The mean value of $\Gamma_0$ is always constrained at $68\%$ CL and predicted to be greater than one, while $\phi_i$ is on the order of $1\, \text{M}_\text{Pl}$ for all data combinations. There seems to be no correlation or degeneracy between the two DBI parameters, owing to their fundamentally different impact over the cosmological evolution as shown in the numerical study.

    \item Data Dependence: Inclusion of BAO and SN data results in tighter constraints on $\Omega_m$, and consequently on $H_0$, $S_8$, and $\phi_i$. These datasets also impact the degeneracies associated with $H_0$ and $S_8$ but do not clearly resolve the existing tensions. A slightly higher central value of $\Gamma_0$ is favoured with the lensing data, possibly accommodating for the lensing excess in the \textit{Planck} data. 

    \item Statistical Evidence: $\Delta\chi_{\rm eff}^2$ and $\ln B_{i, \Lambda {\rm CDM}}$ statistical tests were used to assess model preference. While the Bayesian evidence is significantly negative to show preference for $\Lambda$CDM, we find that the $\Delta\chi_{\rm eff}^2$ also takes considerable negative values across all the data set combinations, indicating a better fit to the data in the Dark D-Brane model. Nevertheless, when the two metrics are considered together, we conclude that the better fit to the data is not sufficient to justify the added complexity of the Dark D-Brane model.
    
\end{itemize}

In summary, while the Dark D-Brane model introduces intriguing dynamics and constraints, it does not conclusively resolve existing tensions in cosmological parameters. Future data and analyses are crucial for a more definitive understanding of these issues. Nevertheless, it is still a very rich phenomenological framework, which could be extended or considered at different regimes.

\cleardoublepage


 \chapter{A Hybrid Model for the Dark Sector} \label{chap:sfdm}
 \setcounter{equation}{0}
\setcounter{figure}{0}


    \epigraph{Quando eu era pequeno pensava que de um momento para outro eu cairia para fora do mundo. Por que as nuvens não caem, já que tudo cai? É que a gravidade é menor que a força do ar que as levanta. Inteligente, não é? Sim, mas caem um dia em chuva. É a minha vingança.\blfootnote{\textit{When I was a little boy I thought that from one minute to the next I could fall off the face of the earth. Why don’t clouds fall, since everything else does? Because gravity is less than the strength of the air that keeps them up there. Clever, right? Yes, but one day they fall as rain. That is my revenge.} --- \textsc{Clarice Lispector} in The Hour of the Star} \\ --- \textsc{Clarice Lispector}\ \small\textup{A Hora da Estrela}}

The mysterious components that dominate the present Universe constitute the dark sector of cosmology. Unveiling their intrinsic properties and origins is a key problem in the field. Numerous phenomenological dark matter candidates have been put forth, inspired by extensions to the standard model of particle physics. These range from weakly interacting massive particles to lighter scalar fields (\textit{e.g.}, \cite{Jungman:1995df,Bertone:2004pz,Holman:1982tb,Blumenthal:1984bp,KolbWimplzilla1,Hlozek:2014lca,Marsh:2015xka,Ballesteros:2016euj,Hlozek:2017zzf,Roszkowski:2017nbc}). Dark energy, however, is often treated as a separate matter from DM. In this study, we propose that both DE and DM originate from two coupled scalar fields governed by a common potential energy $V$ (see also \cite{DAmico:2016jbm,CarrilloGonzalez:2017cll,Benisty:2018qed,Brandenberger:2019jfh,Brandenberger:2020gaz,Johnson:2020gzn,Sa:2021eft,Johnson:2021wou}). We focus on the potential energy model commonly employed in hybrid inflation \cite{Lindehybridinflation} which introduces a mass hierarchy between DE and DM, and which will be detailed further in this chapter. Here, the heavier field is identified as DM, and its mass is determined by the DE field expectation value, which aligns with the flat direction of the potential. This framework strongly restricts the rate at which the DE field can vary, accommodating a constant-like behaviour but implying that the observed accelerated cosmic expansion is only temporary. Ultimately, both fields will settle at the minimum of the potential, and other players, such as spatial curvature, will determine the Universe's future evolution.

This chapter is structured as follows: the model is introduced in \cref{sec:model}, with the theoretical implications of its particular parameters detailed in \cref{sec:param}. A fluid approximation for the DM field is developed in \cref{sec:fluid}, followed by an account of the influence of the scalar-field parameters in the cosmic evolution and its implications for CMB anisotropies and large-scale structures in \cref{sec:eqs}. Finally, we discuss the results in \cref{sec:results} and present our conclusions in \cref{sec:conc}.

This work has been published in JCAP and can be found in Ref.~\cite{vandeBruck:2022xbk}.

\section{The Hybrid Model} \label{sec:model}

In this section, we motivate and introduce the scalar field scenario proposed and studied in this work. Due to its well-studied dynamics, the framework we consider here is an extension of the hybrid inflation model \cite{Lindehybridinflation}, populated by two scalar fields and with the conventional matter fields of the standard model. The action that describes this setting is

\begin{equation}
    S = \int d^4x \sqrt{-g} \left[ \frac{1}{2} \text{M}_{\text{Pl}}^2 R - \frac{1}{2} (\nabla \phi)^2 - \frac{1}{2} (\nabla \chi)^2 - V(\phi, \chi) \right] + S_{\text{SM}} \mathperiod
\end{equation}

\begin{figure*}
      \subfloat{\includegraphics[width=0.75\linewidth]{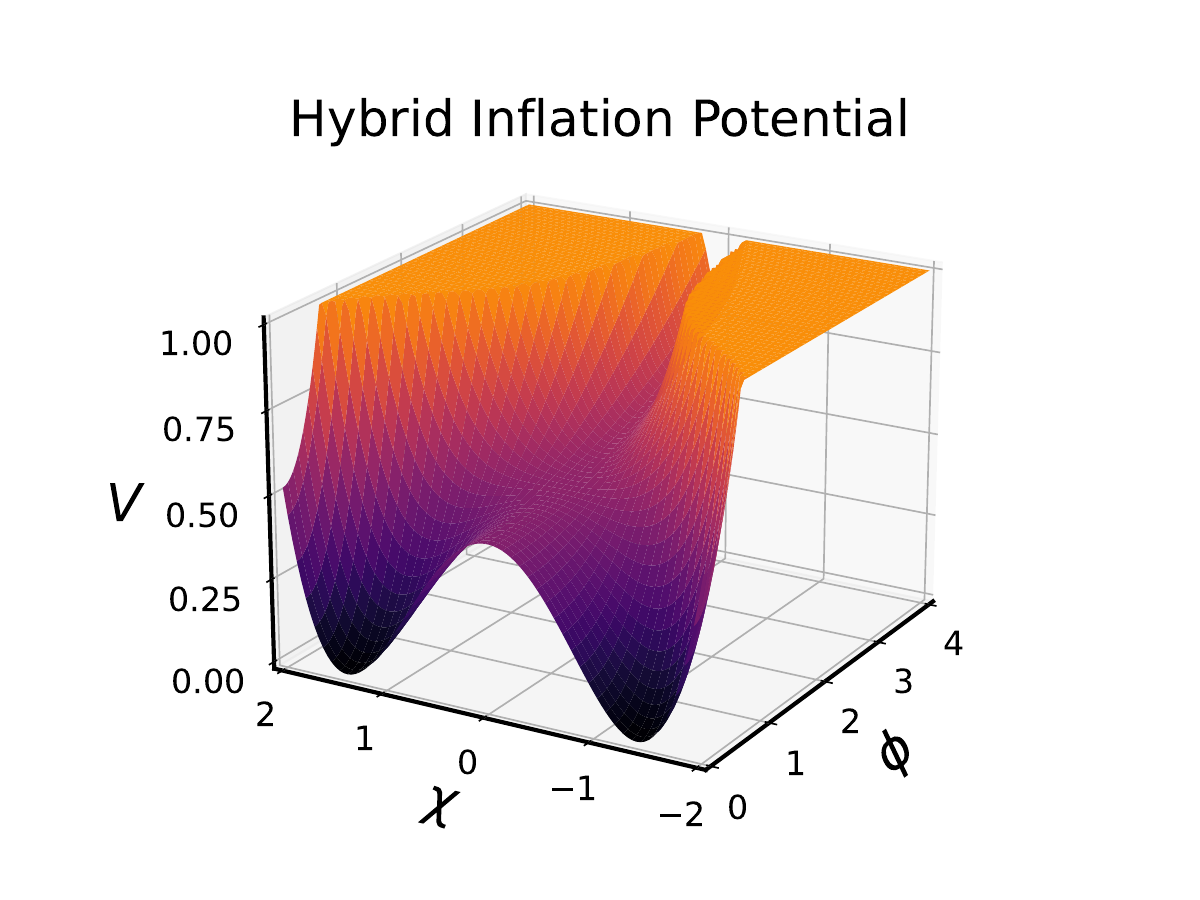}}
  \caption[Hybrid Inflation Potential]{\label{fig:hybpot} Illustrative example of the potential in the hybrid inflation model proposed in \cite{Lindehybridinflation}, following \cref{eq:potential2} with $m = 0.2$, $M = 1.4$, $\lambda = 0.5$ and $g = 0.8$.}
\end{figure*}

The field $\phi$ is identified as the DE field, while the field $\chi$ is the DM field, and the dynamics of the standard model fields are encapsulated in $ S_{\text{SM}} $. The potential term $ V(\phi, \chi) $ comprises the effective interaction in the dark sector and is akin to the potential in the hybrid inflation model, as is illustrated in \cref{fig:hybpot}, expressed as

\begin{equation}
     V(\phi,\chi) = \frac{\lambda}{4} (M^2 - \chi^2)^2 + \frac{g^2}{2} \phi^2 \chi^2 + \frac{\mu^2}{2} \phi^2 = V_0 - \frac{\lambda M^2 \chi^2}{2} + \frac{\lambda \chi^4}{4} + \frac{g^2 \phi^2 \chi^2}{2} + \frac{\mu^2 \phi^2}{2} \label{eq:potential2} \mathcomma
\end{equation}
   
where $ M $ and $ \mu $ are mass scales, $ g $ and $ \lambda $ are dimensionless coupling constants, and $ V_0 \equiv \lambda M^4/4 $ sets the potential scale. For $ \phi $ and $ \chi $ to be suitable candidates for DE and DM, the parameters must be carefully selected, and this question will be elaborated on in the next section. The potential has a global minimum at $ \chi = \pm M $ and $ \phi = 0 $, where the potential energy vanishes (see \cref{fig:hybpot}). It should be noted that another model that diverges from ours has been proposed in \cite{Axenides:2004kb}, with the reverse role for DM and DE, leading to different phenomenology, dynamics, and valid parameter choices.

The effective masses $ m_{\chi}^2 $ and $ m_{\phi}^2 $ for the DM and DE fields, $ \chi $ and $ \phi $, respectively, are given by the second derivatives of the potential as

\begin{equation}
    m_{\chi}^2 \equiv \frac{\partial^2 V}{\partial \chi^2} = g^2 \phi^2 - \lambda M^2 + 3 \lambda \chi^2 \mathcomma
\label{eq:DM_mass}
\end{equation}

and

\begin{equation}
\label{eq:DE_mass}
    m_{\phi}^2 \equiv \frac{\partial^2 V}{\partial \phi^2} = g^2 \chi^2 + \mu^2 \mathcomma
\end{equation}

accordingly.

We focus on a spatially flat cosmological model with the background given by the Friedmann--Lema\^{i}tre--Robertson--Walker (FLRW) metric. The line element in Cartesian coordinates is 

\begin{equation}
\odif{s}^2 = - \odif{t}^2 + a^2(t) \delta_{ij} \odif{x}^i \odif{x}^j \mathcomma
\end{equation}

where $ a(t) $ denotes the scale factor and, as in previous chapters, $ H = \dot{a}/a $ represents the cosmic expansion rate, and over-dots indicate derivatives of cosmic time $ t $. The field equations of motion for each scalar field are

\begin{align}
    \ddot{\phi} + 3H\dot{\phi} = -(g^2\chi^2 + \mu^2)\phi \mathcomma \quad \text{and} \quad \ddot{\chi} + 3H\dot{\chi} = -\lambda\chi^3 + (\lambda M^2 - g^2\phi^2)\chi \label{eq:KGchi} \mathperiod
\end{align}

The Friedmann equations take the form

\begin{align}
    H^2 &= \frac{\rho}{3 M_{\text{Pl}}^2} \mathcomma \quad \text{and} \quad \dot{H} = -\frac{1}{2 M_{\text{Pl}}^2} (\rho + p) \mathcomma 
\end{align}

where $ \rho $ and $ p $ denote the total energy density and pressure, encompassing contributions from both scalar fields, baryons, and radiation:

\begin{equation}
    \rho = \frac{1}{2}\dot{\phi}^2 + \frac{1}{2}\dot{\chi}^2 + V(\phi, \chi) + \rho_b + \rho_{r} \mathcomma \quad \text{and}\quad p = \frac{1}{2}\dot{\phi}^2 + \frac{1}{2}\dot{\chi}^2 - V(\phi, \chi) + p_{r} \mathperiod
\end{equation}

For practical purposes, we decompose the energy densities of each scalar field with different contributions from the hybrid potential:

\begin{align}
    \rho_{\chi} &= \frac{1}{2}\dot{\chi}^2 - \frac{\lambda M^2 \chi^2}{2} + \frac{\lambda \chi^4}{4} + \frac{g^2\phi^2\chi^2}{2} \mathcomma \label{eq:rhochi} \\
    \rho_{\phi} &= \frac{1}{2}\dot{\phi}^2 + V_0 + \frac{\mu^2\phi^2}{2} \mathperiod \label{eq:rhophi}
\end{align}

It should be noted that this decomposition is purely a matter of choice and does not impact the inherent physics of the model, provided the sum of both components matches the complete energy density of the scalar fields. The configuration in \cref{eq:rhochi} groups all oscillating terms (\textit{i.e.} terms involving $ \chi $) to achieve an effective pressureless behaviour required for structure formation in the matter domination era, as will become clear further on. While $ \phi $ is expected to act like a cosmological constant in the late Universe, its early evolution may differ significantly due to interactions with the $ \chi $-field.

In this framework, if $ \phi $ deviates significantly from the origin, $ \chi $ will experience oscillations around zero. A sudden shift in the potential shape occurs as the effective dark matter mass, specified in \cref{eq:DM_mass}, transitions from a positive to a negative value. According to \cref{eq:DM_mass}, and assuming $ \chi $ oscillates around 0, this transition occurs when $ \phi $ crosses a critical threshold, given by

\begin{equation} \label{phicrit}
    \left| \phi_c \right| \approx \frac{\sqrt{\lambda} M}{g} \mathperiod
\end{equation}

When $ \phi > \phi_c $, $ \chi $ acts as dark matter, and $ \phi $ is a dynamic dark energy component that slowly rolls down its potential. The dynamics of $ \phi $ are primarily dictated by the dominant constant scale in the potential, $ V_0 $, and its interaction with $ \chi $. However, as $ \phi $ nears the critical value $ \phi_c $, $ \chi $ drops abruptly and starts oscillating around $ \chi = \pm M $. Simultaneously, $ V(\phi, \chi) $ collapses to zero, signalling a rapid decay of dark energy and suggesting that the era of dark energy domination comes to a halt and, therefore, is just a transient phenomenon in this model.

\subsection{Conditions on Model Parameters} \label{sec:param}

Before moving on to the model dynamics, we must examine the conditions imposed by the model considerations discussed above. For that purpose, we must derive constraints for the free parameters $g$, $M$, and $\lambda$ such that the assumptions remain valid. For the $\phi$-field to act as DE, two conditions must be met: the field must evolve slowly, and its potential energy has to be of the order of the current DE density $\rho_{\mathrm{DE,0}}$. This translates into the requirement

\begin{equation}
V_0 = \frac{1}{4}\lambda M^4 \approx 10^{-47} \text{GeV}^4 \mathperiod
\end{equation}

Any contribution from the $\mu^2$-term in \cref{eq:potential2} should not exceed this value as it modifies the DE density. Hence, the mass parameter $M$ is of the order $10^{-3}\, \text{eV}$, as expected in standard DE models.

Conversely, for $\chi$ to mimic the dark matter behaviour, the field must oscillate in a quadratic potential from the early Universe onward \cite{turnerDM}. First, to avoid any damping effects in \cref{eq:KGchi} that smooth out the oscillations, it is required that $m_{\chi} \approx g \phi \gg H$. Second, for the quadratic term to dominate over the quartic term in \cref{eq:potential2}, the following condition must be fulfilled

\begin{equation}
g^2\phi^2 - \lambda M^2 \gg \frac{1}{2}\lambda \chi^2 \mathperiod
\end{equation}

The DE $\phi$-field is only changing very slowly throughout the cosmological evolution, and its present-day value is large ($\phi_0 \gtrsim 10~M^2_{\mathrm{Pl}}$), implying that the mass of $\chi$ must also be large unless $g$ is remarkably small, according to $m_{\chi} = g\phi$.

At some time in the early Universe, $t_i$, the Hubble rate will become of the same order as $ m_{\chi}$ ($H \approx m_{\chi}$), triggering rapid oscillations in the $\chi$-field as the expansion rate drops below its mass. To estimate the temperature of this transition, we assume a radiation-dominated epoch after inflation, such that \cite{Kolb:1990vq}

\begin{equation}
H^2 = \frac{1}{3M^2_{\mathrm{Pl}}}\frac{\pi^2}{30} g_*(T) T^4 \mathcomma
\end{equation}

where $g_*(T)$ is the effective number of relativistic degrees of freedom at temperature $T$ (of the order of several hundred in theories beyond the standard model). Therefore, an estimate for the oscillation temperature is

\begin{equation}
T \approx 10^{15} \left(\frac{g}{10^{-7}}\right)^{1/2} \left(\frac{\phi_i}{10 M^2_{\mathrm{Pl}}}\right)^{1/2} \left(\frac{g_*}{100}\right)^{-1/4} \text{GeV} \mathperiod
\end{equation}

This essentially confirms the assumption that the field will start to oscillate very early in the radiation-dominated epoch, almost immediately after a period of inflation in this framework. 
From here, we solve for the evolution of the $\chi$-field \cref{eq:chisolution} to estimate its initial amplitude $\chi_i$ in the early Universe. Considering that $\rho_{\mathrm{DM,0}} \approx g^2\phi_0^2\chi_0^2 \approx 4 \times 10^{-47} \text{GeV}^4$ and $\chi(t) = \chi_i (a_i/a)^{3/2} = \chi_i (T/T_i)^{3/2}$, we find

\begin{equation}
\frac{\chi_i}{\mathrm{GeV}} \approx 1.4 \times 10^6 \left(\frac{g}{10^{-7}}\right)^{-1/4} \left(\frac{\phi_0}{10 M^2_{\mathrm{Pl}}}\right)^{-1/4} \left(\frac{g_*}{100}\right)^{-3/8} \mathperiod
\end{equation}

This is the required initial amplitude for $\chi$ after inflation if it is to be responsible for the observed amount of dark matter (emphasising again that we assume that the field $\chi$ is responsible for all DM). 

As currently framed, the $\phi$-field remains light during inflation. The only requirement is a large field excursion, $\phi \gtrsim 10 M_{\mathrm{Pl}}$, ensuring that the mass of the $\chi$-field remains substantial in the radiation-dominated epoch, and the coupling between $\chi$ and $\phi$ remains sufficiently small. Being light during inflation, $\phi$ is susceptible to quantum fluctuations of magnitude $H_{\mathrm{inf}}/2\pi$, where $H_{\mathrm{inf}}$ is the inflationary expansion rate. Given that $\phi$ represents an (almost) flat direction and remains subdominant through the radiation and matter-dominated phases, these quantum fluctuations will not generate large isocurvature perturbations in the DE sector.

However, the situation concerning the $\chi$-field is less straightforward. If $g\phi < H_{\mathrm{inf}}$ during inflation, in which case $\chi$ is light, then the quantum fluctuations of $\chi$ are also of order $H_{\mathrm{inf}}/2\pi$, potentially resulting in substantial isocurvature modes. The amplitude of these modes is given by~\cite{Marsh:2014qoa}

\begin{equation}
A_I = \frac{(H_{\mathrm{inf}}^2/M^2_{\mathrm{Pl}})}{\pi^2 (\chi_{\mathrm{inf}}^2/M^2_{\mathrm{Pl}})} \mathcomma
\end{equation}

where $\chi_{\mathrm{inf}}$ is the field value during inflation, expected to be around $10^6\, \text{GeV}$ for $g \approx 10^{-7}$.

Two scenarios are considered as a mechanism to circumvent the isocurvature constraint (just like in the extensively studied axion-like fields \cite{Marsh:2015xka}). First, the $\chi$-field could be heavy during inflation ($g\phi > H_{\mathrm{inf}}$), which would suppress the isocurvature modes. The challenge here lies in setting an adequate field amplitude at the end of inflation for $\chi$ to behave like DM. Second, $\chi$ could exhibit non-standard dynamics during inflation, either through gravitational coupling as proposed in Ref.~\cite{Folkerts:2013tua} or by direct coupling to the inflaton field. For the remainder of the chapter, we focus on the post-inflation era, assuming that the isocurvature perturbations can be restrained.

For the ensuing numerical investigation, initial conditions are chosen at $z_i = 10^{14}$ such that $\phi_i \gg \phi_c$, or equivalently, $g\phi_i \gg \sqrt{\lambda}M$ as dictated by \cref{phicrit}. Given that $m_\chi \gg H$, the following constraint emerges

\begin{equation} \label{eq:phiconstraint}
    g\phi_i \gg H \mathcomma
\end{equation}

where the subscript $i$ denotes quantities evaluated at $z_i$ in the numerical simulations.
Moreover, the $\phi$-field must evolve slowly, requiring $m_\phi^2 \ll H^2$. Assuming that $\mu$ is small compared to $g\chi$, we obtain from \cref{eq:DE_mass}

\begin{equation} \label{eq:chiconstraint}
    g^2 \chi^2 \ll H^2 \mathperiod
\end{equation}

In the matter-dominated epoch, the $\chi$-field takes the leading role. As discussed in further detail, the relative energy density of dark matter will eventually start decreasing as the Universe keeps expanding. This dilution sets the stage for the $\phi$-field, which was slowly evolving until it may finally become the director of the cosmic evolution, driven by its potential energy. Initially, the condition

\begin{equation}
  \frac{1}{2}\mu^2 \phi^2 + V_0 \ll \frac{1}{2} g^2 \phi^2\chi^2 \mathcomma
\end{equation}

must be met to ensure a matter-dominated phase. This condition implies an energy density of $\chi$ evolving as

\begin{equation}
    \rho_{\chi} = \frac{1}{2}\dot{\chi}^2+\frac{1}{2}m^2_{\chi}\chi^2 \simeq m^2_{\chi} \chi^2 \mathcomma
    \label{eq:rho_chi}
\end{equation}

where we have implicitly assumed $m_{\chi} \simeq g \phi$ and that the $\chi$-field, exhibiting rapid oscillations, is essentially pressureless when averaged over multiple oscillation periods. Given how slowly $\phi$ evolves, the effective DM mass is also virtually constant. Solving \cref{eq:rho_chi} for $\chi^2$ and replacing into \cref{eq:chiconstraint}, we obtain

\begin{equation} \label{eq:condition}
    g^2 \frac{\rho_{\chi}}{m^2_{\chi}} \ll H^2 \mathperiod
\end{equation}

During the matter-dominated epoch, with $\chi$ as the primary contributor, the Friedmann equation simplifies to

\begin{equation}
    H^2 \simeq \frac{\rho_{\chi}}{3M^2_{\mathrm{Pl}}} \mathperiod
\end{equation}

Consequently, from \cref{eq:condition}, we derive

\begin{equation} \label{eq:phiconstraint2}
     1 \ll \frac{1}{3} \left(\frac{\phi}{M_{\mathrm{Pl}}} \right)^2 \mathcomma
\end{equation}

with $m_{\chi} \simeq g \phi$. To satisfy this inequality, $\phi$ must be \textit{trans-Planckian}, that is to say, that $\phi \gg \text{M}_{\mathrm{Pl}}$. Ultimately, to meet the constraints in \cref{eq:phiconstraint2}, $\phi$ must at least take on values comparable to the Planck mass. Once more, unless $g$ is remarkably small, this implies a substantially large DM mass, contrasting sharply with models with ultralight and light scalar field DM candidates~\cite{Hu:2000ke, Hui:2016ltb, Hlozek:2014lca}. On a more practical note, we remark that this is potentially consistent with the WIMPzilla framework as examined in~\cite{KolbWimplzilla1, Kolbwimpzilla2}.

One additional criterion for the viability of $\chi$ as a dark matter candidate is its stability over cosmological time scales. Phenomenologically, even though a decay of $\chi$ into $\phi$ is kinematically allowed, it is effectively negligible due to the disparity in their mass scales. The decay rate for $\chi \chi \rightarrow \phi \phi$ is quantified by~\cite{Kofman:1997yn}

\begin{equation}
    \Gamma (\chi \chi \rightarrow \phi \phi) = \frac{g^4 \langle \chi^2 \rangle}{8 \pi m_{\chi}} \mathcomma
\end{equation}

where $\langle \cdot \rangle$ denotes an average over one oscillation period. The decay rate, $\Gamma$, needs to be sub-dominant to the Hubble parameter, $H$, \textit{i.e.}, $\Gamma < H$.
As elaborated in the next section, the decay rate $\Gamma$ scales as $a^{-3}$ throughout the expansion history, owing to its direct dependence on the $\chi$-field coupled to the slow variation of $\phi$. In contrast, $H$ scales as $a^{-3/2}$ and $a^{-2}$ during matter- and radiation-dominated epochs, respectively. The overall result is that $\Gamma$ decreases more abruptly than $H$ as the Universe expands, thereby ensuring the DM field's stability for reasonable values of $g < 1$.

Finally, we address possible quantum corrections to the potential that might invalidate the considerations above. Generally, quantum corrections to the tree-level potential are expected at the order of $M_{\mathrm{Pl}}^2$, with $M_{\mathrm{Pl}}$ serving as the natural cut-off scale. Nonetheless, in the context of supersymmetric models, the corrections are logarithmic in nature, $\ln(\phi/M_{\mathrm{Pl}})$~\cite{Lyth:1998xn}. These logarithmic corrections can be suppressed if the coupling constants are relatively small, which is a natural assumption for the model under investigation. Consequently, the hybrid model shares the same challenges as other models of a similar kind.

\section{Fluid Approximation} \label{sec:fluid}

To mitigate the computational difficulties associated with the oscillations of the $ \chi $-field, we seek reasonable approximations that facilitate the study of its cosmological behaviour. Specifically, we reformulate the problem as an interacting quintessence model by adopting a fluid representation for the DM field $ \chi $. A key feature of this setup is that the mass of $ \chi $ evolves in parallel with the slow rolling of the DE field. Previous investigations into scalar fields oscillating in a quadratic potential have established that the dynamics can be approximated by an oscillating envelope with amplitude $ \mathcal{A}(t) \propto a^{-3/2} $ \cite{turnerDM}. Employing the WKB approximation and using the constraints derived previously ($ g\phi \gg H $ and $ \dot{\phi}/\phi \ll 1 $), we obtain a solution given by

\begin{equation}
\label{eq:chisolution}
\chi(t) = \chi_{i} \left( \frac{\phi_{i}}{\phi} \right)^{1/2} \left( \frac{a_i}{a} \right)^{3/2} \sin\left( g\phi \left( t-t_{i} \right) \right) \mathperiod
\end{equation}

Here, $ \chi_{i} $ is the initial amplitude of $ \chi $. Due to the slow variation of $ \phi $, the term $ \phi_{i}/\phi $ remains nearly constant, implying that $ \chi $ behaves like a pressureless fluid with $ \rho_{\chi} \propto \chi^2 \propto a^{-3} $.
Using \cref{eq:rho_chi} to replace $\rho_{\chi}$, the oscillation-averaged energy density of DM can be approximated to

\begin{equation}
\label{eq:fluidDMequation}
\langle \rho_{\chi} \rangle \approx \rho_{\chi,i} \left( \frac{\phi}{\phi_i} \right) \left( \frac{a_i}{a} \right)^3 \mathcomma
\end{equation}

where $ \rho_{\chi,i} = \frac{1}{2} g^2 \phi_i^2 \chi_i^2 $ represents the initial energy density of $ \chi $.

We will henceforth neglect the bracket notation for simplification, as the quantities under consideration are always oscillation-averaged. The averaged density in \cref{eq:fluidDMequation} shows a linear dependence on $ \phi $. This leads us to the following continuity equation for the oscillation-averaged interacting fluid:

\begin{equation}
\label{eq:coupling}
\dot{\rho_{\chi}} + 3H\rho_{\chi} = \frac{\dot{\phi}}{\phi} \rho_{\chi} \mathperiod
\end{equation}

Accordingly, the equation of motion for the DE field is reformulated as
\begin{equation}
    \ddot{\phi} + 3 H \dot{\phi} = -\frac{1}{\phi} \rho_{\chi} \mathperiod
    \label{eq:coupledphiKG}
\end{equation}
This equation is fully in line with the usual DE continuity equation, assuming a perfect fluid approximation for the field, given by $ \rho_{\phi} \approx \frac{\dot{\phi}^2}{2} + V_0 $:
\begin{equation}
    \dot{\rho_{\phi}} + 3H (\rho_{\phi} + P_{\phi}) = -\frac{\dot{\phi}}{\phi} \rho_{\chi} \mathcomma
\end{equation}
ensuring the conservation of the total energy density for both $ \phi $ and $ \chi $, as required by the general covariance of Einstein's equations. In line with the approximation in \cref{eq:fluidDMequation,eq:coupledphiKG} simplifies to
\begin{equation}
    \frac{1}{a^3} \odv{}{t} \left( a^3 \dot{\phi} \right) = -\frac{\rho_{\chi,i}}{\phi_i} \left( \frac{a_i}{a} \right)^3 \mathcomma
    \label{eq:slowrollKGphi}
\end{equation}
which can be integrated and solved for $ \dot{\phi} $:
\begin{equation}
    \dot{\phi} = \left( \frac{a_i}{a} \right)^3 \left( K_i - \frac{\rho_{\chi,i}}{\phi_i} t \right) \mathcomma
    \label{eq:phidotanalytical}
\end{equation}
where $ K_i \equiv \dot{\phi_i} + \frac{\rho_{\chi,i}}{\phi_i} t_i $ is an integration constant and $ \dot{\phi_i} $ denotes the initial velocity of the field, when $ a = a_i $. Provided that the relation between $ a $ and $ t $ is known, $ \phi $'s behaviour is completely determined. As demonstrated earlier, the fluid approximation is valid immediately after inflation, given that the $ \chi $-field starts oscillating around zero as soon as this era comes to a halt, and radiation becomes the dominant cosmic component at $ t = t_i $. Therefore, by solving \cref{eq:phidotanalytical} in the radiation-dominated epoch, during which $ a(t) \propto t^{1/2} $, we obtain
\begin{equation}
    \phi(t) = \phi_i + C_i - A_i \left( \frac{t}{t_i} \right)^{1/2} - B_i \left( \frac{t}{t_i} \right)^{-1/2} \mathcomma
    \label{eq:raddomphi}
\end{equation}
with $ C_i \equiv 2 \left( \frac{\rho_{\chi,i}}{\phi_i} t_i^2 + K_i t_i \right) $, $ A_i \equiv 2 \frac{\rho_{\chi,i}}{\phi_i} t_i^2 $, and $ B_i \equiv 2 K_i t_i $ as integration constants. During this epoch, $ \rho_{\phi} $'s scaling behaviour is dictated by
\begin{equation}
    \rho_{\phi} \propto \dot{\phi}^2 \propto a^{-1} \mathperiod
\end{equation}

\cref{eq:raddomphi} fully describes the time variation of the field up to the matter-radiation equality at $ t_{\rm eq} $. Upon entering the matter-dominated era, $ a(t) \propto t^{2/3} $, resulting in the following solution to \cref{eq:phidotanalytical}:
\begin{equation}
    \phi(t) = \phi_{\rm eq} + C_{\rm eq} - A_{\rm eq} \ln \left( \frac{t}{t_{\rm eq}} \right) - B_{\rm eq} \left( \frac{t}{t_{\rm eq}} \right)^{-1} \mathcomma
    \label{eq:matterdomphi}
\end{equation}
where $ C_{\rm eq} \equiv t_{\rm eq} K_{\rm eq} $, $ A_{\rm eq} \equiv t_{\rm eq}^2 \frac{\rho_{\chi, \rm eq}}{\phi_{\rm eq}} $, and $ B_{\rm eq} \equiv t_{\rm eq} K_{\rm eq} $ are constants that depend on the initial conditions specified at the point of matter-radiation equivalence, labelled by the subscript "eq". The constant $ K_{\rm eq} $ is conceptually analogous to $ K_i $ from \cref{eq:phidotanalytical}, with each term being evaluated at $ t_{\rm eq} $ instead of $ t_i $. Given the slow-rolling nature of the field, it is fair to consider $ \dot{\phi_{\rm eq}} \ll t_{\rm eq} \rho_{\chi, \rm eq} / \phi_{\rm eq} $, thus leading to $ B_{\rm eq} \approx t_{\rm eq} A_{\rm eq} $.

We note that \cref{eq:matterdomphi} suggests that $ \dot{\phi} \propto a^{-3/2} $ since $ t \propto a^{3/2} $ in the matter-dominated period. Furthermore, given that the coupling to dark matter is the main dynamical influence in the field dynamics and $ \dot{\phi}^2 \gg V_0 $, we derive
\begin{equation}
    \rho_{\phi} \propto \dot{\phi}^2 \propto a^{-3} \mathperiod
    \label{eq:scaling}
\end{equation}
In this regime, the dark energy component scales analogously to ordinary and cold dark matter. This property is not typical among interacting dark energy models with a constant potential. We have confirmed that this is, in fact, the only specific formulation that yields this unique behaviour, which is pivotal for tackling the cosmic coincidence problem in $ \Lambda $CDM described in \cref{sec:theoprob}, that concerns the remarkable similarity between the value of the energy densities of dark energy and cold dark matter at present times \cite{Ferreira:1997hj,Bahamonde:2017ize,vandeBruck:2016jgg,Teixeira:2019tfi}. In the next section, we will present numerical simulations that illustrate the dynamics within this regime.

The form of the interaction term on the right-hand side of \cref{eq:coupling} suggests that $ \phi $ should remain significantly large through the expansion history up until the present. This aligns with the considerations made in \cref{sec:param} and is an indispensable condition to avoid drastic deviations from the $ \Lambda $CDM paradigm. Though it may appear contradictory, the framework is sustained by the assumption that $\phi$ is slowly rolling, according to $ \frac{\dot{\phi}}{\phi H} \ll 1 $, a condition that has been verified numerically, and which corroborates that $ \phi > M_{\rm Pl} $ in agreement with \cref{eq:phiconstraint2}.

\subsection{Background Dynamics}

Mathematically speaking, the fluid approximation used in our analysis mimics a fifth force mediated by the dark energy scalar field. As reviewed through the course of \cref{chapter:sttheories}, this arises in theories with a non-universal conformal rescaling of the metric $ \tilde{g}_{\mu \nu} $, which in this case would dictate the geodesics of the dark matter particles, and is expressed as:
\begin{equation}
    \tilde{g}_{\mu \nu} = C(\phi) g_{\mu \nu} \mathcomma
\end{equation}
where $ g_{\mu \nu} $ is the gravitational metric, \textit{i.e.}, the one according to which the Einstein-Hilbert retains its form. For this approximation to hold in our setting, conformal factor $ C(\phi) $ must be given by:
\begin{equation}
    C(\phi) = \frac{\phi^2}{M_{\rm Pl}^2} \quad \mathrm{for} \quad |\phi| > |\phi_c| \mathperiod
\end{equation}
The transformation remains invertible as long $ |\phi| > |\phi_c| $. It should be noted that the fluid approximation will cease to be valid long before $ \phi $ could reach zero. Further, $ C(\phi) > 0 $ ensures that the Lorentzian nature of the metric is preserved, evading any instabilities related to metric singularities. Hence, from the considerations above, the functional form of the coupling term in the fluid approximation is reinstated as:
\begin{equation}
    Q = - \frac{C_{\phi}}{2C} \rho_{\chi} = - \frac{\rho_{\chi}}{\phi} \mathperiod
    \label{eq:qconf}
\end{equation}

Under this approximation, both components of the dark sector are considered perfect fluids, allowing us to recast the relevant equations for laying out the dynamics. For numerical purposes, we use conformal time, which we recall is defined as $ \mathrm{d}\tau = \frac{\mathrm{d}t}{a} $, with derivatives denoted by a prime and the Hubble rate scaled as $ \hub = a H $. The conservation equations for the DM and DE fluids are given by
\begin{equation}
    \rho'_{\chi} + 3 \mathcal{H} \rho_{\chi} = - Q \phi' = \frac{\phi'}{\phi} \rho_{\chi} \mathcomma
    \label{eq:conschi}
\end{equation}
and,
\begin{equation}
    \rho'_{\phi} + 3 \mathcal{H} \left( \rho_{\phi} + p_{\phi} \right) = Q \phi' = - \frac{\phi'}{\phi} \rho_{\chi} \mathperiod
    \label{eq:consphi}
\end{equation}
The interaction in the dark sector means that the right-hand side of these equations is no longer zero but composed of symmetric terms. The direction of the energy flow depends solely on the sign of $ \frac{\phi'}{\phi} $: when the ratio is positive, DE feeds the DM component, whereas a negative ratio indicates energy transfer from $ \phi $ to the DM fluid. Independent of the initial conditions, $ \phi $ and $ \phi' $ always have opposite signs, implying a unidirectional energy transfer from the $ \chi $-fluid to the $ \phi $-field. The same physics is encapsulated in the modified Klein-Gordon equation:
\begin{equation}
    \phi'' + 2 \mathcal{H} \phi' = - \frac{a^2}{\phi} \rho_{\chi} = a^2 Q \mathcomma
    \label{eq:kgphip}
\end{equation}
which can be numerically integrated to obtain specific realisations of the evolution under a given set of cosmological parameters.
For that purpose, the only free parameters specific to the model are the initial conditions for the DE field, $ \phi_i = \phi(\tau_i) $ and $ \phi'_i = \phi'(\tau_i) $, along with the scale of the hybrid potential, $ V_0 $. It should be emphasised that the remaining parameters in the potential, given in \cref{eq:potential2}, are irrelevant to the numerical solutions, so long as they meet the constraints outlined in \cref{sec:param}. We set $ \mu = 0 $ for simplicity as it does not affect current dynamics. As for all the cases studied in this thesis, $ V_0 $ is determined using a shooting method for a fiducial value of the present DE relative energy density $ \Omega_{\phi}^0 = \frac{\rho_{\phi}^0}{3 M_{\rm Pl}^2 H_0^2} $. Furthermore, as $ \phi'_i $ does not influence the dynamics due to the rapid stabilisation of the scalar field at the minimum of its potential in the radiation-dominated epoch, we set $ \phi'_i = 0 $ without loss of generality. This narrows our study to the effects of varying the single free model parameter, the initial condition $ \phi_i $. We exclusively consider scenarios where $ \phi_i > 0 $, based on the symmetry of the potential over the dynamics.

\subsection{Cosmological Perturbations} \label{sec:eqs}

We map the cosmological perturbations into an interacting dark energy framework by expanding upon the background evolution. The aim is to identify the changes in the gravitational interactions compared to the standard $\Lambda$CDM model and to evaluate the observational signatures left by the background fluid approximations. To this end, we consider once more the conformal Newtonian gauge for perturbations, as outlined in \cref{sec:newtgauge}, for which we present the metric once again
\begin{equation}
     \odif{s}^2 = a^2(\tau) \left[ -(1 + 2\Psi)\odif{\tau}^2 + (1 - 2\Phi)\delta_{ij}\odif{x}^i \odif{x}^j \right] \mathcomma
\end{equation}
where $\Psi(\tau, \vec{x})$ and $\Phi(\tau, \vec{x})$ are the Bardeen potentials and, accordingly, $\delta$ designates perturbed quantities and the correspondence $\nabla^2 \rightarrow -k^2$ holds in Fourier space. For the scalar field system, there is no anisotropic stress.
The perturbed variables $ \delta \phi $ and $ \delta \rho_{\chi} $ evolve according to
\begin{equation}
    \delta \phi'' + 2 \mathcal{H} \delta \phi' + k^2 \delta \phi = (\Psi' + 3\Phi')\phi' + 2a^2 Q \Psi + a^2 \delta Q  \mathcomma
\end{equation}
and
\begin{equation} \label{eq:pertrho}
    \delta_{\chi}' = - (\theta_{\chi} - 3\Phi') + \frac{Q}{\rho_{\chi}}\phi' \delta_{\chi} - \frac{Q}{\rho_{\chi}} \delta \phi' - \frac{\theta'}{\rho_{\chi}} \delta Q  \mathcomma
\end{equation}
where $ \delta_{\chi} = \delta \rho_{\chi} / \rho_{\chi} $ is the density contrast and $ \delta Q $ is defined as
\begin{equation}
    \delta Q = \frac{\rho_{\chi} \delta \phi - \phi \delta \rho_{\chi}}{\phi^2} \mathperiod
\end{equation}

In \cref{eq:pertrho}, both the equation of state $ w_{\chi} = p_{\chi}/\rho_{\chi} $ and the sound speed $ c_s^2 = \delta p_{\chi}/\delta \rho_{\chi} $ are set to zero. The background dynamics support the former assumption, while the latter follows naturally from the equation for the sound speed of an oscillating scalar field \cite{Hlozek:2014lca},
\begin{equation}
    c_s^2 = \frac{k^2 / (4 m_{\chi}^2 a^2)}{1 + k^2 / (4 m_{\chi}^2 a^2)} \mathperiod
    \label{eq:cs2}
\end{equation}
While this assumption holds for an uncoupled scalar field, it still captures the relevant physics, as the coupling is restricted to be very small. Since we have demanded $m_{\chi}$ to be considerably large for the background approximations to hold, and given that we are considering scales $k \ll 2 m_{\chi} a$, \cref{eq:cs2} implies $c_s^2 \approx 0$ which we nevertheless verified numerically. 

The impact of the coupling on the density matter perturbations can be primarily assessed by looking at the sub-horizon regime $ k \gg \mathcal{H} $ under the quasi-static approximation when the matter and field perturbations are the main drivers of the linear evolution. In practical terms, this consists of neglecting the time derivatives of the perturbations and metric potentials, greatly simplifying the equation of motion for $\delta_{\chi}$ in \cref{eq:pertrho} \cite{vandeBruck:2015ida,vandeBruck:2020fjo},
\begin{equation}
    \delta_{\chi}'' + \mathcal{H}_{\mathrm{eff}} \delta_{\chi}' \simeq 4\pi G_{\mathrm{eff}} \rho_{\chi} \delta_{\chi} \mathcomma
    \label{eq:deltapp}
\end{equation}
where the subdominant contribution of baryons has been neglected, and the effective Hubble term is defined as
\begin{equation}
    \mathcal{H}_{\mathrm{eff}} = \mathcal{H} \left( 1 + \frac{Q}{\rho_{\chi}} \frac{\phi'}{\mathcal{H}} \right) \mathcomma
    \label{eq:heff}
\end{equation}
consisting of an additional friction contribution that accounts for the changes in the coupled background evolution. On the other hand, the effective gravitational constant in the small-scale limit (large $k$) becomes important as it is expressed as
\begin{align}
G_{\rm eff} \simeq G \left(1 + 2 \beta^2 \right), \quad \text{with}\quad \beta \equiv \text{M}_{Pl} \frac{Q}{\rho_{\chi}} \mathcomma
\label{eq:geff}
\end{align}
as expected, considering the general results for scalar-tensor gravity models under a conformal transformation \cite{Mifsud:2017fsy,Tsujikawa:2007gd,Amendola:2003wa}.

\section{Observables and Phenomenology} \label{sec:results}

In this section, we investigate the evolution of the hybrid model for different realisations of the coupling in the dark sector and its impact on cosmological observables, compared with the standard $\Lambda$CDM model. As anticipated, the general features of the model are consistent with canonical coupled quintessence frameworks featuring stable, effective interactions \cite{Wetterich:1994bg,Amendola:1999er}, just as the ones considered in \cref{chap:cquint} (see also \cite{Barros:2018efl,Gomez-Valent:2020mqn,Teixeira:2019hil,Amendola:2000ub,Pettorino:2012ts,Pettorino:2013oxa,Xia:2013nua,vandeBruck:2016hpz,vandeBruck:2017idm,Planck:2015bue,Aghanim:2018eyx,Agrawal:2019dlm} for recent works). Nevertheless, this is not to say that there will not be any distinct signatures arising from the scalar field's slow-roll evolution and the $\phi$-dependent coupling.
For illustration purposes, we explore four scenarios, parameterised by different initial field values, namely $\phi_i/\text{M}_{\text{Pl}} = \{8,10,15,20\}$, recalling that $\phi_i$ sets the initial scalar field dynamics as well as the strength of the coupling in the dark sector, meaning that its effect over the dynamics can be non-trivial. We set the initial condition for the field velocity at $\phi'_i = 0$, as we demonstrate later that it has a negligible effect on the evolution. The cosmological parameters are kept at the \textit{Planck} 2018 fiducial values for a $\Lambda$CDM cosmology \cite{Aghanim:2018eyx}: $H_0 = 67.56\, \text{km/s/Mpc}$, $\Omega_b h^2 = 0.022$, and $\Omega_c h^2 = 0.12$, where $h = H_0/100$. The scale of the potential is fully fixed \textit{via} the Friedmann constraint for a flat cosmology.
The same applies to the analysis of the perturbations. We assume Gaussian adiabatic initial conditions and a scalar power spectrum characterised by an amplitude of curvature fluctuation of $A_s = 2.215 \times 10^{-9}$ at a pivot scale of $k_{\rm piv} = 0.05\, \text{Mpc}^{-1}$, and a spectral index $n_s = 0.962$. Without loss of generality, we also consider vanishing initial conditions for the perturbations in the scalar field and its corresponding velocity, i.e., $\delta\phi_i = \delta\phi'_i = 0$.
The background and cosmological perturbations are calculated using a version of the open-source \texttt{CLASS} code\footnote{\href{https://github.com/lesgourg/class\_public}{https://github.com/lesgourg/class\_public}} \cite{Lesgourgues:2011re,Blas:2011rf,lesgourgues2011cosmic2} modified for this purpose.

\subsection{Background Evolution}

\begin{figure*}
      \subfloat{\includegraphics[width=0.5\linewidth]{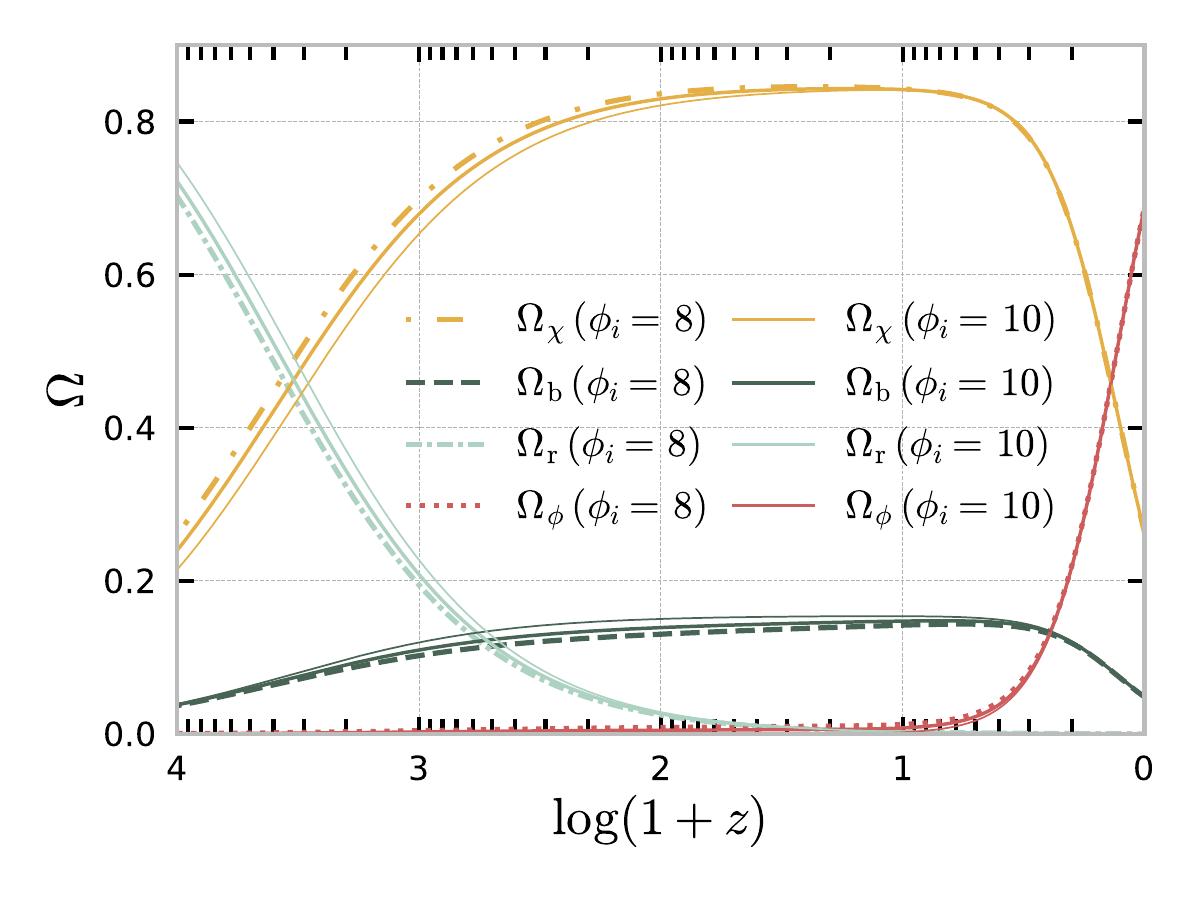}}
      \subfloat{\includegraphics[width=0.5\linewidth]{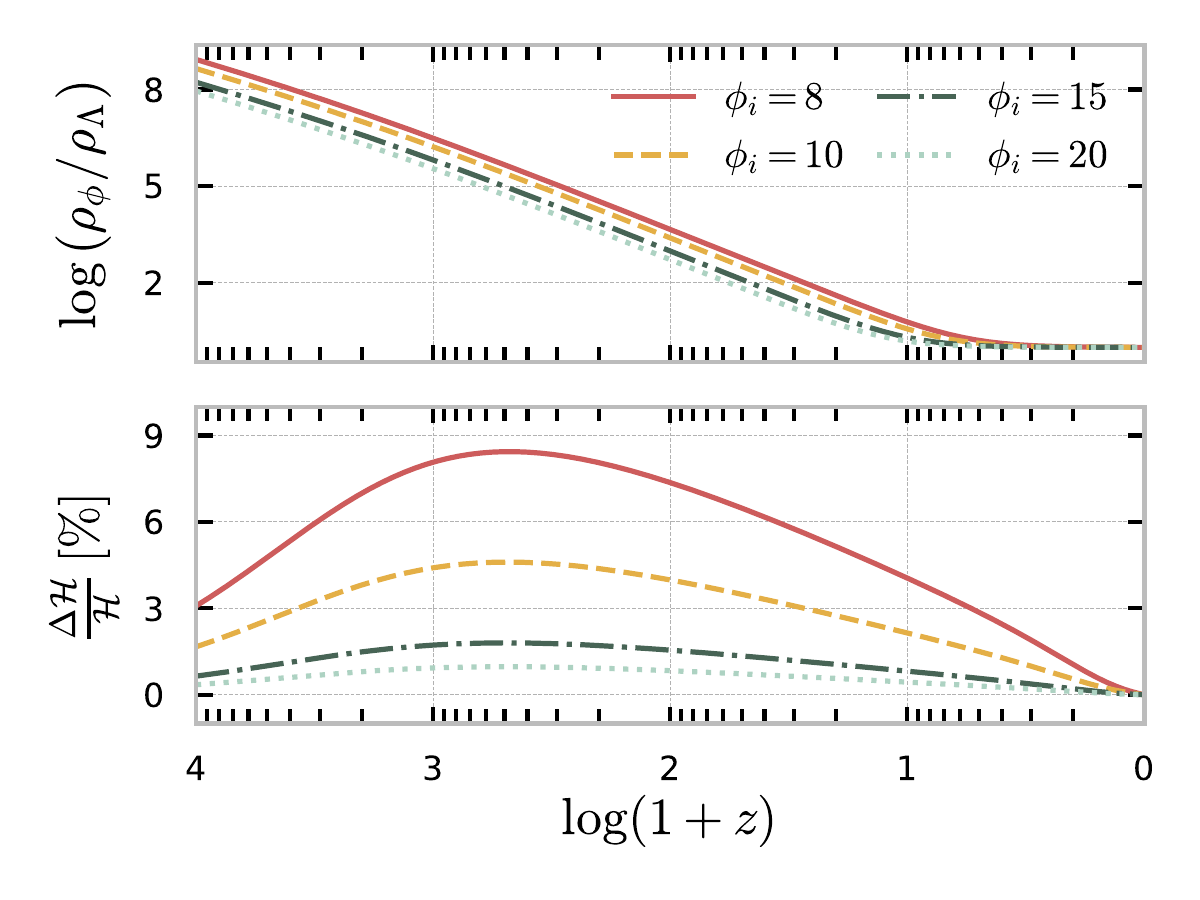}}
  \caption[Evolution of the relative energy densities, the ratio $\rho_{\phi}/\rho_{\Lambda}$ and the Hubble rate]{\label{fig:oms} \textit{Left panel:} Redshift evolution of the relative energy densities $\Omega_i$ of the dark matter fluid $\chi$ (yellow), baryons (green), radiation (sea blue) and the scalar field $\phi$ (red) for the $\Lambda$CDM model (thin solid lines), $\phi_i = 8$ M$_{\text{Pl}}$ (dashed/dotted-dashed lines) and $\phi_i = 10$ M$_{\text{Pl}}$ (thick solid lines). \textit{Right panel:} Ratio of the dark energy density (\textit{top panel}) and fractional deviations in the Hubble rate (\textit{bottom panel}) in the hybrid coupled model with respect to the standard model as a function of redshift $1+z$, for $\phi_i = \{8,10,15,20\}$ M$_{\text{Pl}}$ (solid red, dashed yellow, dotted-dashed green and dotted sea blue lines).}
\end{figure*}

\begin{figure*}
      \subfloat{\includegraphics[width=0.5\linewidth]{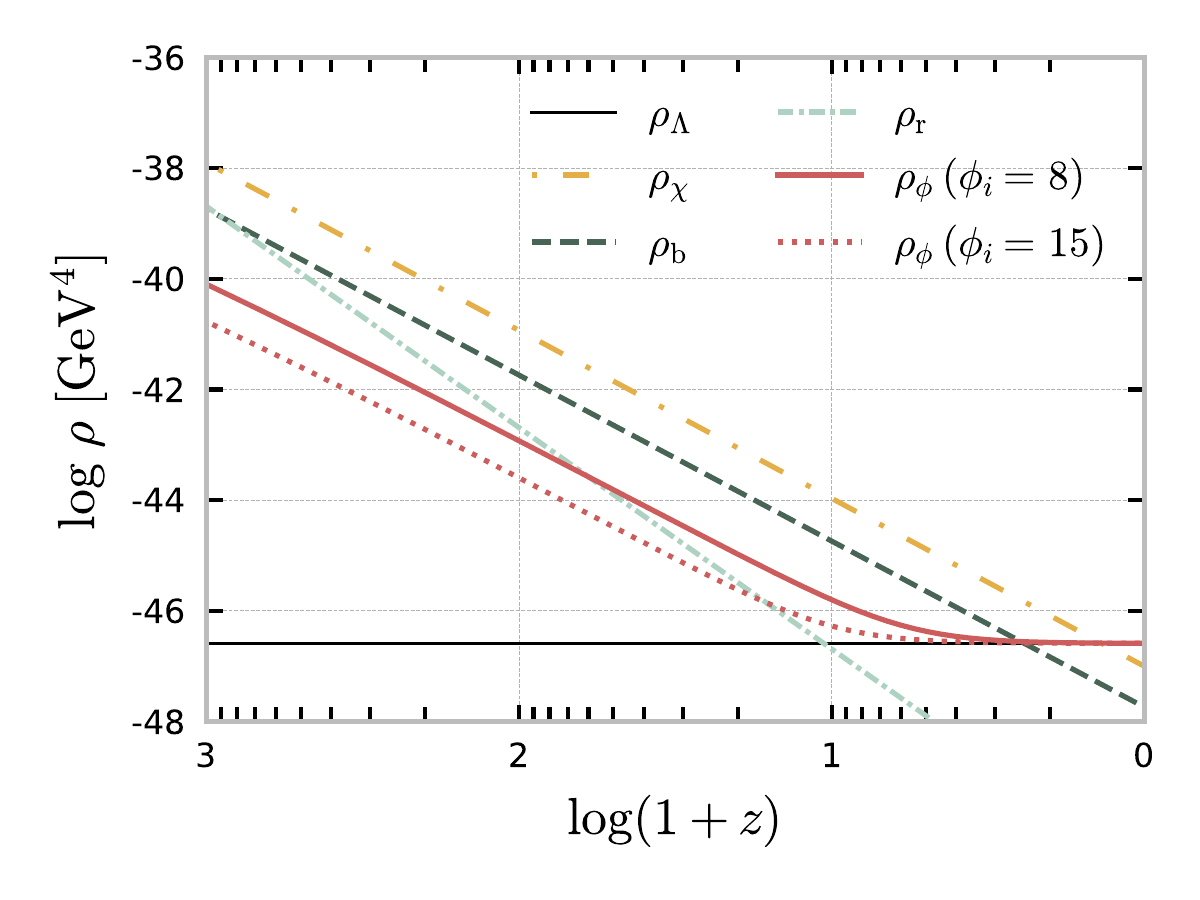}}
      \subfloat{\includegraphics[width=0.5\linewidth]{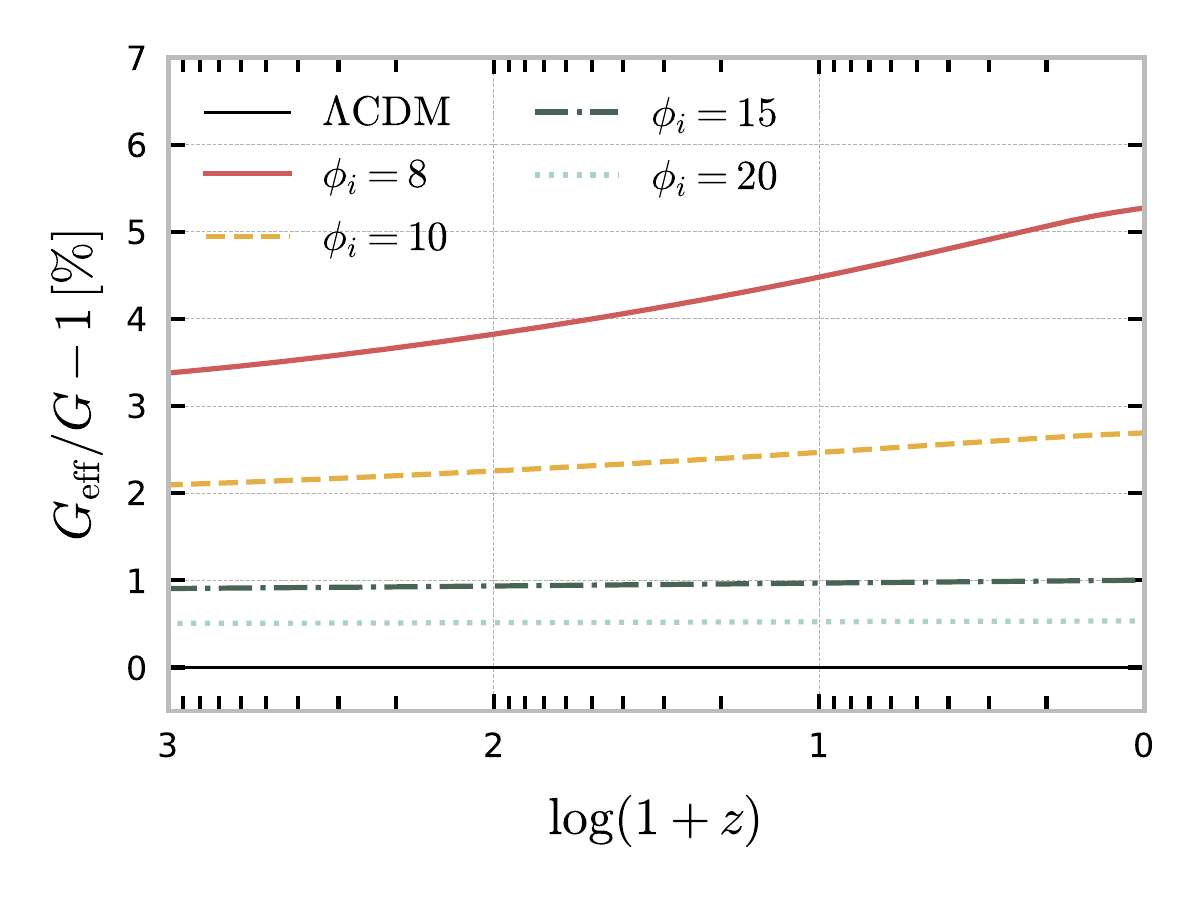}}
  \caption[Evolution of relative energy densities and effective gravitational constant]{\label{fig:back} \textit{Left panel:} Evolution of the energy densities $\rho$ of the dark matter fluid $\chi$ (yellow), baryons (green), radiation (sea blue) and the scalar field $\phi$ (filled red) for $\phi_i = 8$ M$_{\text{Pl}}$. To appreciate the differences, we also include $\rho_{\phi}$ for the $\phi_i = 15$ M$_{\text{Pl}}$ case (dotted-dashed red line) and $\rho_{\Lambda}$ for the standard model (thin black solid line) for completeness. \textit{Right panel:} Percentage deviations of the effective gravitational constant, as defined in \cref{eq:geff}, with respect to the standard $G$ (thin black solid line) for $\phi_i = \{8,10,15,20\}$ M$_{\text{Pl}}$ (solid red, dashed yellow, dotted-dashed green and dotted sea blue lines).}
\end{figure*}

In the fluid approximation, where the potential $V(\phi, \chi)$ remains constant, the field's dynamics would be indistinguishable from a cosmological constant in the absence of the coupling (which would also imply $\phi' \equiv 0$ or $\phi \rightarrow \infty$). The coupling with the dark matter fluid is the primary driver of the energy density of dark energy at early times, as depicted in the left panel of \cref{fig:back}. Once this coupling becomes significant after the radiation-dominated era and when DM starts playing a dominant role, 
$\phi$ begins to slowly evolve until its energy density mimics that of DM, the component to which it couples. During this period of the evolution, the scalar field acts as an \textit{effective} pressureless fluid, and this scaling phase ceases once the kinetic and potential energies become comparable and $\phi'^2 \lesssim a^2 V_0$.

The end of the scaling phase is directly influenced by the initial conditions set for $\phi_i$ and $\rho_{\chi}$ (see \cref{eq:phidotanalytical}). Higher $\phi_i$ values lead to late-time behaviour that resembles $\Lambda$CDM more closely, implying an earlier exit from the scaling regime, as seen in the right panel of \cref{fig:oms} and the left panel of \cref{fig:back}. In contrast, $\phi'$ quickly adjusts as the scalar field is driven towards the minimum of the effective potential, irrespective of its initial value, implying that $\phi'_i$ regulates the onset of the scaling phase., which starts earlier for the largest initial velocities. This effect is negligible as it occurs early in the radiation-dominated phase when the scalar field's contribution is insignificant. Similar behaviour has been reported in \cite{Pettorino:2008ez}, driven by a significant acceleration of the scalar field instead. When the $\phi$-field exits the matter-scaling regime, it settles at a cosmological constant-like attractor solution, where it will remain diluting with the expansion until the fluid approximation breaks down.

The coupling also manifests as an amplification of $\rho_\phi$, particularly in scenarios with stronger coupling (smaller $\phi_i$), which naturally induces an earlier matter-DE equality. This shift is an artificial result of fixing the present cosmology. As DM loses energy to DE, it must be more abundant early on, leading to higher initial amplitudes for the DM energy density to maintain equilibrium and match the present value. On the other hand, this effect goes in hand with a more significant contribution from the coupling to the scalar field dynamics, as evident from \cref{eq:kgphip}. This also reflects as an earlier matter-radiation equality with decreasing $\phi_i$ values, as depicted in the left panel of \cref{fig:oms}.

In the lower right panel of \cref{fig:oms}, we show the relative deviations in the Hubble rate $H(z)$ when comparing the hybrid coupled model with the standard $\Lambda$CDM model. Even though $H_0$ is fixed at the current value for both models, $H(z)$ is enhanced by up to $9\%$ during the matter-dominated phase for the model with the smallest $\phi_i$ while remaining negligible during radiation domination. This is correlated with an enhancement of $\rho_\phi$ and $\rho_\chi$ and will be instrumental for understanding the scale-dependent growth of matter perturbations, which we will discuss in the following section.

\subsection{Cosmological Perturbations}

\begin{figure*}
      \subfloat{\includegraphics[width=0.5\linewidth]{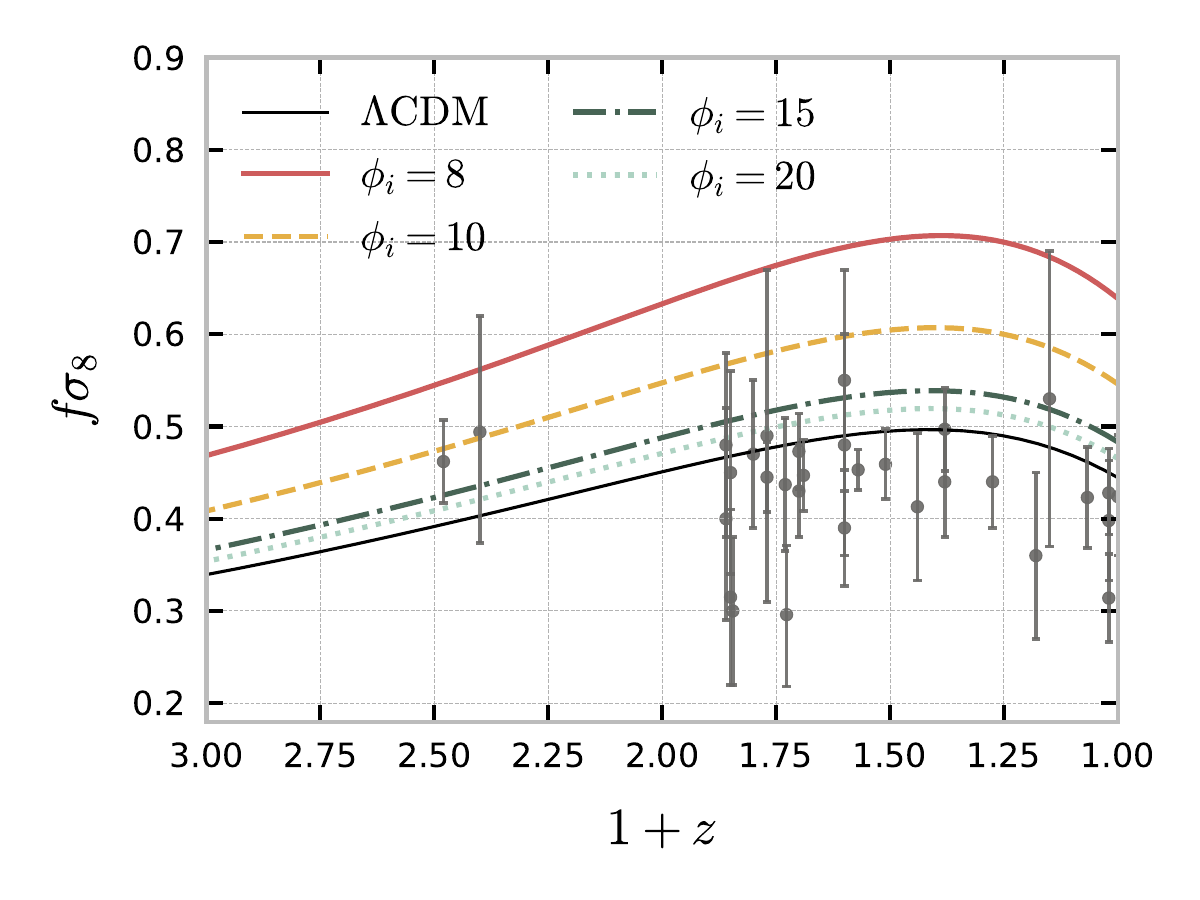}}
      \subfloat{\includegraphics[width=0.5\linewidth]{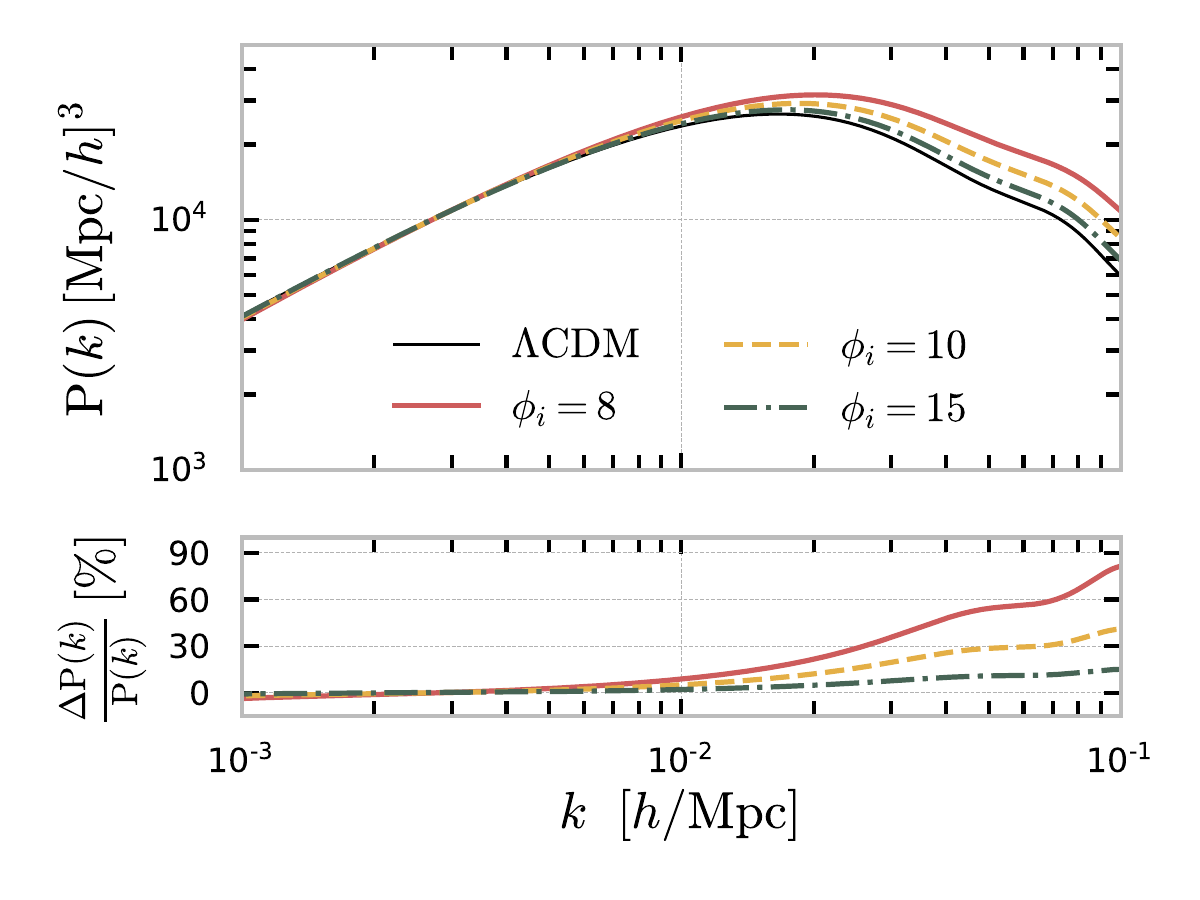}}
  \caption[Evolution of growth factor and matter power spectrum]{\label{fig:growth} \textit{Left panel:} Evolution of the cosmological observable $f \sigma_8$ (defined in \cref{eq:fs8}) with redshift $1+z$ for the hybrid coupled model with $\phi_i = \{8,10,15,20\}$ M$_{\text{Pl}}$ (solid red, dashed yellow, dotted-dashed green and dotted sea blue lines) and for the $\Lambda$CDM case (thin black solid line). The RSD data points and corresponding error bars (solid grey) correspond to the compilation presented in \cite{Marulli:2020uyy}. \textit{Right panel:} The matter power spectrum as a function of Fourier scales $k$ (\textit{top panel}) and corresponding percentage deviations (\textit{bottom panel}), for the hybrid coupled model with $\phi_i = \{8,10,15,20\}$ M$_{\text{Pl}}$ (solid red, dashed yellow and dotted-dashed green lines) with respect to the $\Lambda$CDM case (thin black solid line).}
\end{figure*}

The linear growth rate $ f(a) $ of the total matter perturbation - comprising both baryonic and dark matter and denoted as $ \delta_m $ - quantifies this phenomenon as
\begin{equation}
f(z, k) = \frac{1}{\mathcal{H}} \frac{ \delta_m'(z, k)}{ \delta_m(z, k)} \mathcomma
\end{equation}
with
\begin{equation}
\delta_m(z, k) = \frac{\Omega_b \delta_b + \Omega_{\mathrm{DM}} \delta_{\mathrm{DM}}}{\Omega_b + \Omega_{\mathrm{DM}}} \mathperiod
\end{equation}
The changes to the evolution of $ f(z) $, compared to the $ \Lambda $CDM model, correlate with the modifications to the onset of the matter-dominant era, at which point the dark sector coupling becomes substantial. The combined quantity $ f \sigma_8 $ is directly connected to observations, as was introduced in \cref{sec:rsd}, defined as


\begin{equation}
f \sigma_8 (z,k_{\sigma_8}) = \frac{\sigma_8 (0,k_{\sigma_8})}{\mathcal{H}} \frac{\delta_m' (z,k_{\sigma_8}) }{\delta_m (0,k_{\sigma_8}) } \mathcomma
\label{eq:fs8}
\end{equation}
where $\sigma_8$ is the root mean square mass fluctuation amplitude for spheres of size $8h^{-1}$ Mpc (or equivalently for scales $k_{\sigma_8} = 0.125 h$ Mpc$^{-1}$), generally used to set the amplitude of the matter power spectrum at present $\sigma_8^0 \equiv \sigma_8 (0,k_{\sigma_8})$. 

 In \cref{fig:growth} (left panel), we display the redshift evolution of $ f \sigma_8 $ for the examined models, as defined in \cref{eq:fs8}. We identify an overall enhancement in the linear growth of matter perturbations. The most notable deviations from $ \Lambda $CDM occur in the model with the minimal $ \phi_i $ as a direct consequence of changes to the expansion history. 
 The numerical analysis reveals that $ \sigma_8^0 $ increases across all models to match the same amplitude $ \mathcal{A}_s $, with higher values $ \phi_i $ progressively approaching the $ \Lambda $CDM prediction. We also incorporate RSD observational data\footnote{\href{https://gitlab.com/federicomarulli/CosmoBolognaLib/tree/master/External/Data/}{CosmoBolognaLib}} from the compilation in Ref.~\cite{Marulli:2020uyy} which encompasses measurements reported by multiple surveys. 
 The plots suggest an enhanced linear growth in the hybrid model, parametrised by a larger $ f \sigma_8 $, and associated with the higher $ \sigma_8 $ values as well when assuming the same spectrum of initial primordial perturbation amplitudes $ A_s $. We conclude that, in these specific conditions, the models appear not to resolve the $ S_8 $ tension \cite{DiValentino:2020vvd,Abdalla:2022yfr}, although a more rigorous investigation is needed to affirm this conclusion, as the data interpretation is based on $ \Lambda $CDM assumptions. 

In \cref{fig:growth} (right panel), we depict the power spectrum of matter density fluctuation $ P(k) $ and the corresponding relative variations for Fourier scales $ 10^{-3} h \, \text{Mpc}^{-1} < k < 10^{-1} h \, \text{Mpc}^{-1} $. We report a mild suppression for larger scales (low $ k $) and a significant amplification for intermediate to small scales (high $ k $). Deviations reach up to $ 81\% $ for the smallest value of $ \phi_i $ at $ 10^{-1} h \text{Mpc}^{-1} $, a point where the linear approximation is expected to no longer hold, as non-linear effects become dominant. This trend for large $ k $ was expected on the basis of the fifth force, which is more significant for lower values of $ \phi_i $, having a considerable influence on the growth of perturbations according to \cref{eq:geff}. We expect that this should play an important role when comparing the model with existing observational data.

The slight suppression for larger scales (smaller $ k $ values) is ascribed to the changes in the background cosmic expansion, specifically through the friction term in \cref{eq:heff}, which inhibits the growth of matter density perturbations. This effect overcomes the fifth-force only at the largest scales, peaking at $ 4\% $ for the models considered, and is essentially negligible for the models with larger $ \phi_i $, for which $ G_{\rm eff} \approx G $, as displayed in \cref{fig:back} (right panel). Furthermore, we note the shift in the turnover of the power spectrum towards smaller scales, in comparison with the $ \Lambda $CDM case, due to the change in the radiation-matter equality epoch towards larger redshifts.

\begin{figure*}
      \subfloat{\includegraphics[width=0.5\linewidth]{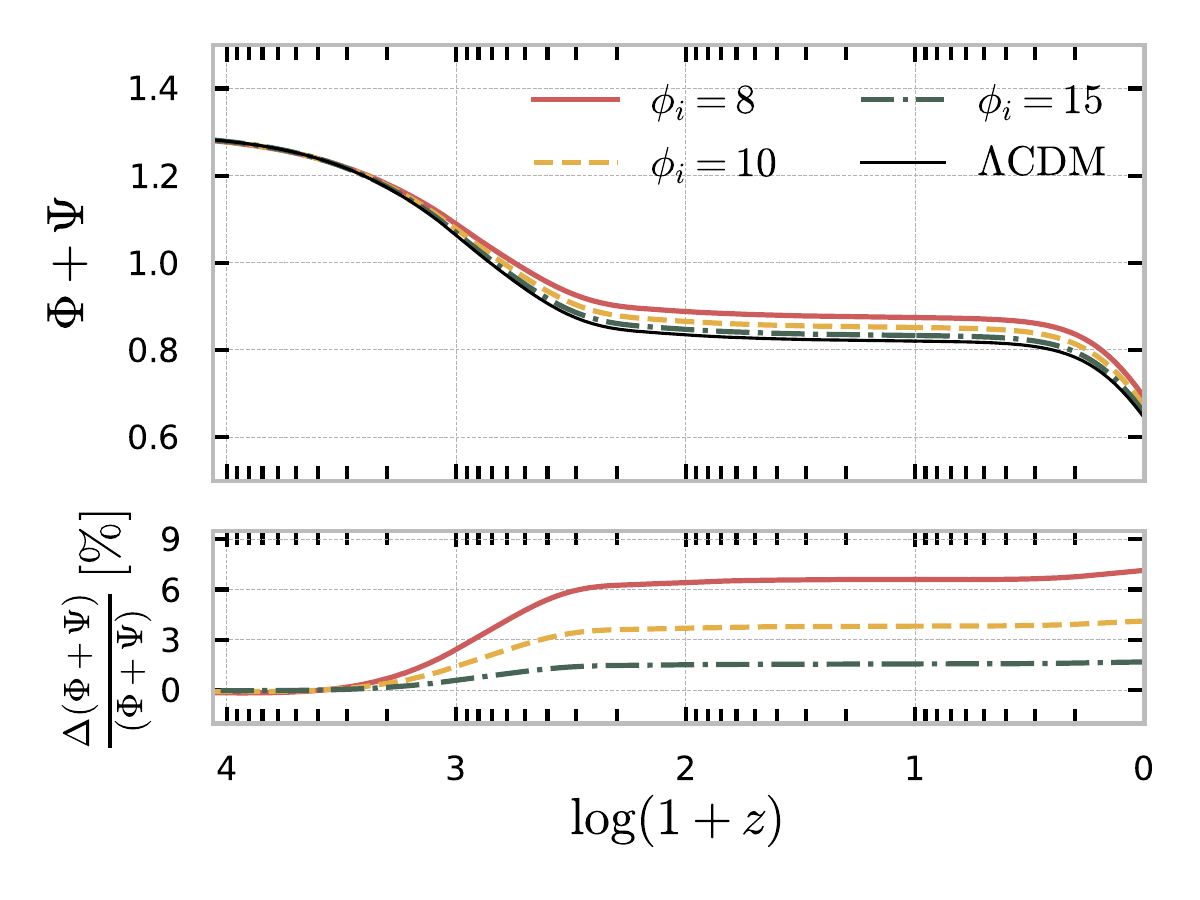}}
      \subfloat{\includegraphics[width=0.5\linewidth]{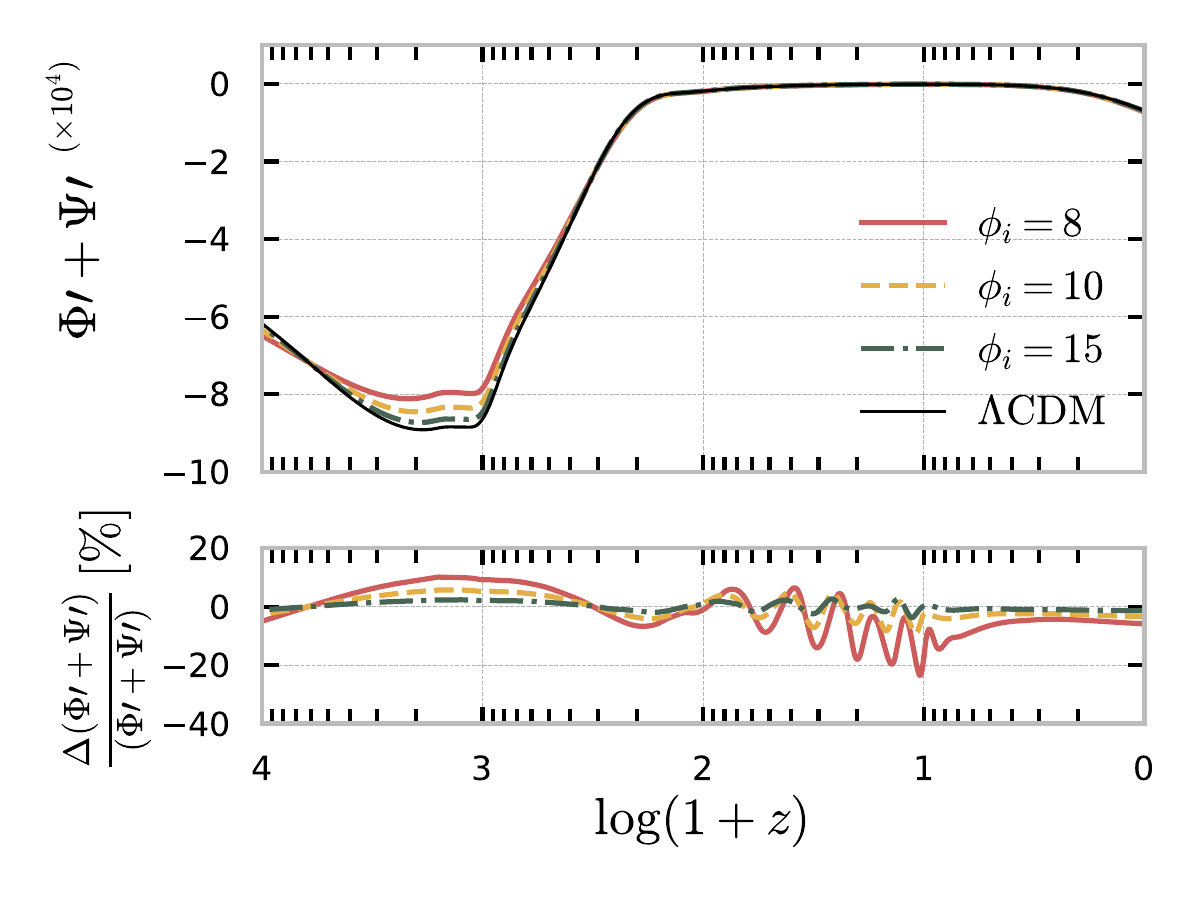}}
  \caption[Evolution of the lensing potential and its derivative]{\label{fig:lens} Redshift evolution of the sum of the gravitational potentials, $\Phi+\Psi$, (\textit{top left panel}) and the corresponding derivative with respect to conformal time, $\Phi'+\Psi'$ (\textit{top right panel}) for the hybrid coupled model with $\phi_i = \{8,10,15\}$ M$_{\text{Pl}}$ (solid red, dashed yellow and dotted-dashed green lines) and for the $\Lambda$CDM case (thin black solid line), including the percentage deviations from the standard model (\textit{bottom panels}). }
\end{figure*}

\begin{figure*}
      \subfloat{\includegraphics[width=0.5\linewidth]{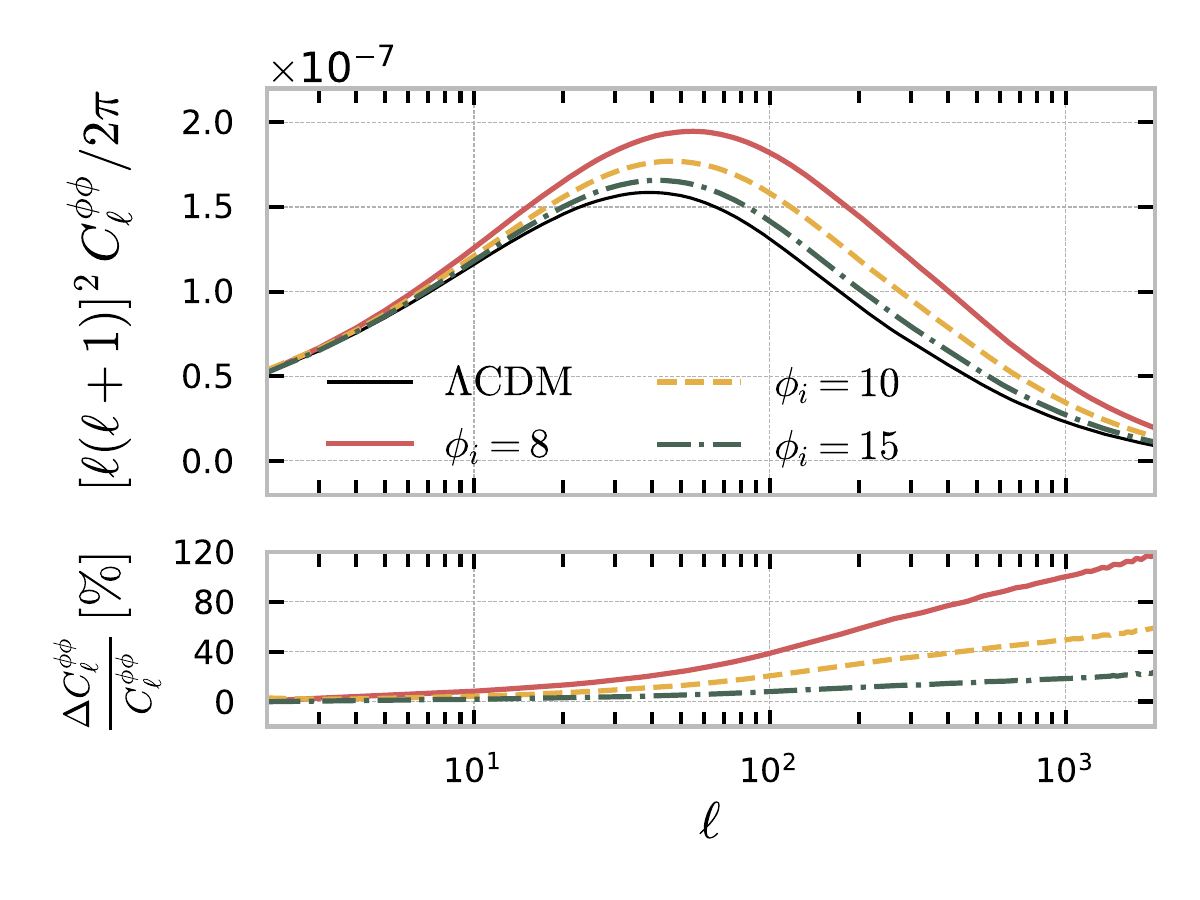}}
      \subfloat{\includegraphics[width=0.5\linewidth]{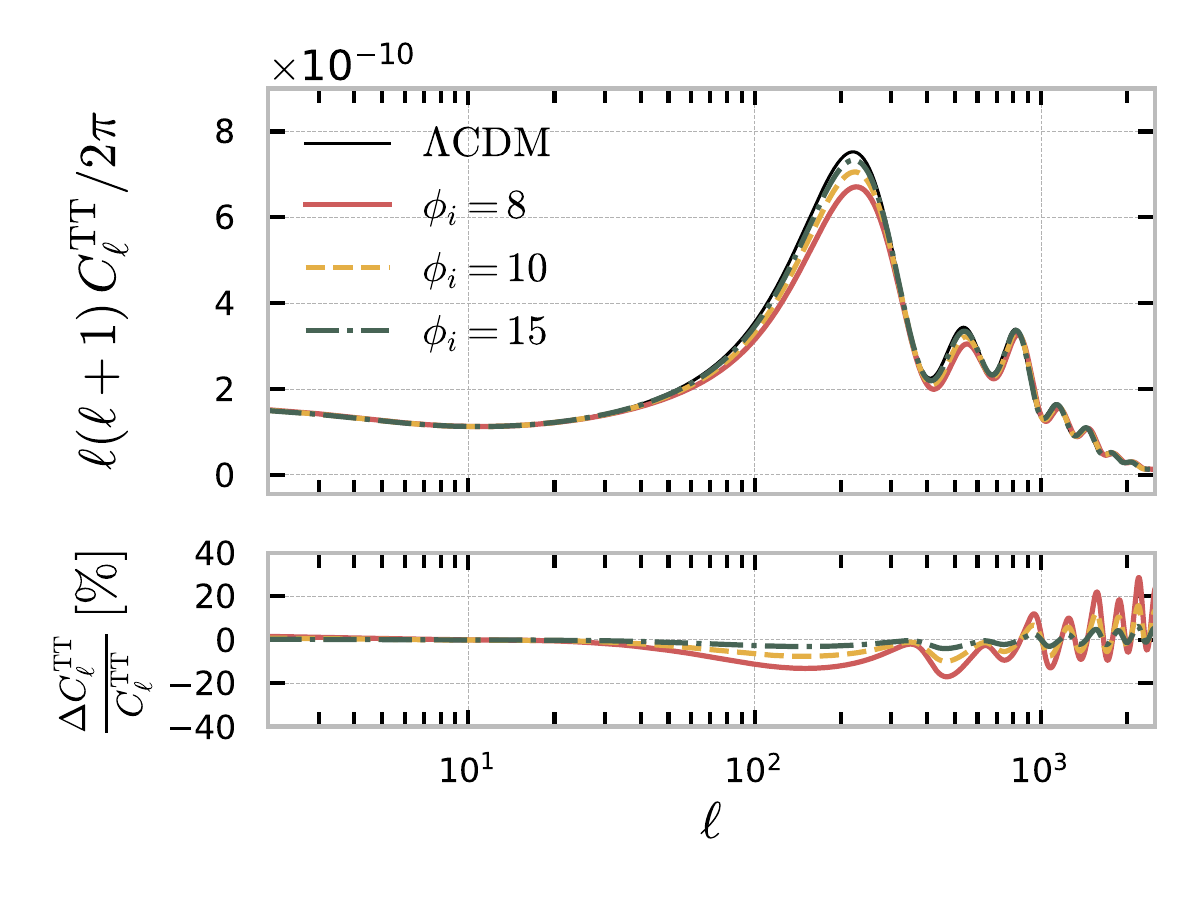}}
  \caption[The temperature and polarisation CMB angular power spectra]{\label{fig:cmbh} Lensing (\textit{top left panel}) and TT (\textit{top right panel}) power spectra as a function of the angular scale $\ell$ for the hybrid coupled model with $\phi_i = \{8,10,15,20\}$ M$_{\text{Pl}}$ (solid red, dashed yellow, dotted-dashed green and dotted sea blue lines) and for the $\Lambda$CDM case (thin black solid line), along with the fractional deviations from the standard model (\textit{bottom panels}).}
\end{figure*}

In the left panel of \cref{fig:lens}, we illustrate the evolution of the gravitational potentials $ \Phi $ and $ \Psi $ (top) at an intermediate scale of $ k = 0.01 $ Mpc$^{-1}$, along with the corresponding percentage differences relative to $ \Lambda $CDM. The most evident deviations occur at $ z \lesssim 10^3 $, in the matter-dominated epoch, where alterations in the DM evolution become significant, and the energy density of the scalar field is scaling with the matter. The most important variations are consistently associated with the smallest $ \phi_i $ values. The modifications to the lensing gravitational potential $ \phi_{\text{len}} = \Phi + \Psi $ are mainly ascribed to the energy transfer from DM to the DE sector. This potential is the relevant term in the line-of-sight integration for the lensing power spectrum $ C_{\ell}^{\phi \phi} $, as presented in \cref{sec:cmb}, and is depicted in the left panel of \cref{fig:cmbh}. We identify an overall enhancement of $ C_{\ell}^{\phi \phi} $ across all angular scales, with the most substantial deviations from $ \Lambda $CDM occurring for the lowest $ \phi_i $ case. This could potentially explain the lensing excess observed in the \textit{Planck} temperature data of CMB anisotropies \cite{Aghanim:2018eyx,DiValentino:2019dzu}, which was the focus of \cref{sec:Alens}. Likewise, this enhancement aligns with the amplified effective gravitational interaction for DM particles. We note that the matter density contrast $ \delta_m $ follows a broadly consistent trend across most of the Fourier scales depicted, mirroring the increased effective gravitational constant, as presented in \cref{fig:back}.

The same effects can also be dug out of CMB anisotropies' temperature-temperature (TT) power spectrum, presented in the right panel of \cref{fig:cmbh}. The most evident modifications arise from the integrated-Sachs-Wolfe (ISW) effect, proportional to $ \dot{\phi}_{\text{len}} $ (see \cref{sec:cmb}), depicted in the right panel of \cref{fig:lens}. This effect can be dissected into the early and late-time contributions to the ISW. The early ISW effect amplifies the time derivatives of the potentials due to an earlier transition from radiation- to matter-dominated eras. Most importantly, the late-time ISW effects stem from changes to the CMB lensing by large structures, owing to modified dynamics in the dark sector. This leads to a late-time suppression of $ \Phi' + \Psi' $, as shown in the right panel of \cref{fig:lens}. The most apparent changes in the TT power spectrum are a decrease in amplitude and a narrowing of the peaks and troughs, tied to a reduction in the $ \rho_{\text{b}}/\rho_{\text{DM}} $ ratio during recombination (see \cref{sec:lcdm_param}) \cite{Pettorino:2013oxa,Amendola:2011ie}. This induces a degeneracy between the effective coupling and the Hubble parameter, as the latter mainly affects the first peak's location and magnitude. 
Likewise, the shift in the acoustic peaks to higher multipoles is also linked to changes in the cosmic expansion history, which modify the distance to the last scattering surface and thus reduce the sound horizon at the baryon-drag epoch (see \cref{sec:lcdm_param}). The more this shift is pronounced, the more significant the deviations from $ \Lambda $CDM at the background level. From both panels of \cref{fig:cmbh}, we observe that an enhanced Hubble rate drives the CMB spectra towards smaller angular scales (higher multipoles). Lastly, the lensing power spectrum amplification correlates with an increased ISW tail at large angular scales, although this is typically a secondary effect.



\section{Discussion} \label{sec:conc}

In this study, we have introduced a hybrid model for the dark sector, in which DM and DE are ascribed to two interacting scalar fields. We started from a formulation based on an interacting potential frequently encountered in hybrid inflation, which we transported to the DM-DE system. We performed a thorough examination of the cosmology implications of such a setting. With the onset of a rapid oscillatory behaviour deep within the radiation-dominated epoch, the heavy scalar field essentially acts as pressureless DM. We have shown that, under appropriate approximations, the scalar fields can be treated approximately as a DM fluid coupled to a gradually evolving DE field through this oscillatory pattern.

In conclusion, we list the key predictions of the model proposed in this chapter:

\begin{itemize}
    \item The DE field is required to have a sizeable present value ($ \phi > \text{M}_{\text{Pl}} $) for the fluid description to hold. Consequently, the coupling in the dark sector is currently highly suppressed, and DM is considerably massive, akin to the WIMPZilla framework. The energy scales in the interacting potential are markedly below the Planck scale; for instance, the scale $ M $ is approximately on the eV scale, as it depends on the coupling constant $ \lambda $. The requirement for super-Planckian excursions of the field, reminiscent of inflationary models, poses challenges for model building in theories beyond the standard model. Nonetheless, it is noteworthy that the current model combines two seemingly disconnected mass scales: the minute mass scale $ M $ and the large field excursions for the DE field $ \phi $.
    
    \item Another exciting feature of the model is the transient nature of the dark energy-dominated epoch. The DM field is predicted to become light and cease to behave as a pressureless fluid. Both fields will ultimately settle at the actual minimum of the potential ($ \phi = 0 $ and $ \chi = \pm M $). The fate of the Universe will then hinge on its spatial curvature. Should it be closed, the expansion will cease, and the Universe will collapse, paving the way for a potential bounce far in the future.
\end{itemize}

As demonstrated, if the coupling is large enough, unique imprints are expected on the temperature-temperature power spectrum of CMB anisotropies and the growth of structures encoded in the matter power spectrum. Both existing and forthcoming observational datasets can test such characteristic signatures. Drawing upon diverse independent data sets and probing the remaining relevant cosmological variables, an in-depth analysis is ongoing work.

\cleardoublepage



\chapter{Conclusions}
\label{chapter:conclusions}

 \epigraph{Meu Deus, só agora me lembrei que a gente morre. Mas, mas eu também?! \\ Não esquecer que por enquanto é tempo de morangos. Sim...\blfootnote{\textit{My God, I just remembered that we die. But — but me too?! Don’t forget that for now it’s strawberry season. Yes...} --- Clarice Lispector in The Hour of the Star} \\ --- \textsc{Clarice Lispector}\ \small\textup{A Hora da Estrela}}

This project aimed to enhance our understanding of the properties of dark matter and dark energy in the context of cosmology, focusing specifically on the possibility of an interaction in the dark sector and its observational signatures. We have touched on the remarkable richness of independent probes, with particular emphasis on measurements of the physics of the cosmic microwave background anisotropies, invaluable archaeological sites of buried knowledge.
We have delved deeply into the enigmatic realms of coupled quintessence and interacting dark energy, much like cosmic archaeologists excavating layers of ancient expansion history, piecing together shards of pottery from various cosmological data in an attempt to understand the intricate nature of the dark sector. Nevertheless, it is fascinating to think that the more we uncover, the more we realise the scope of what remains hidden. With each model we explore, independent of the results, we enrich our understanding of the relics and artefacts that reveal something while hinting at larger structures yet undiscovered, constantly suggesting new avenues of research. We report on the main conclusions from each investigation included in this dissertation, which, while just a tiny overdensity contribution to the literature, may contribute to the gravitational clustering of the ceaseless quest to unravel the mysteries of our cosmic habitat.

\begin{itemize}
    \item In \cref{chap:cquint}, we have analysed the standard coupled quintessence model by assuming a flat cosmology and letting the spatial curvature vary as a free parameter. We found no significant improvement regarding the evidence for a closed Universe in the \textit{Planck} data assuming $\Lambda$CDM. Nevertheless, we were able to show that the disagreement between CMB data and other geometrical probes is maintained in other scenarios.

    \item In \cref{chap:kin}, we have studied a coupled quintessence model in which the interaction depends on the kinetic energy of the scalar field. We showed the particular observational signatures of such a framework. We found no statistically significant evidence for either of the models, which could be a hint that forthcoming data might be able to address these differences.

    \item In \cref{chap:gwcons}, we assessed the potential of constraining coupled quintessence models with next-generation gravitational wave detectors. We showed that standard sirens are a promising alternative to supernova catalogues and baryon acoustic oscillation observations. We found that the accuracy in the $H_0$ parameter is consistently enhanced by an order of magnitude at $1\sigma$, which is promising for providing further insight into addressing the $H_0$ tension.

    \item In \cref{chap:dbi}, we performed a thorough study of models motivated by string theory in which the dark sector has a fundamental geometrical higher-dimensional origin. Even though this is a more intricate scenario and a preliminary analysis does not seem to suggest evidence for the particular framework considered, this analysis motivates further studies with functions that might be more appropriate for a cosmological description.

    \item In \cref{chap:sfdm}, we introduced a hybrid model for the dark sector, in which DM and DE are ascribed to two interacting scalar fields. We have shown how the DM scalar field dynamics can be approximated to a fluid description that holds similarities with coupled quintessence models. We found exciting novel predictions for the nature of this DM source and a unique transient nature for the dark energy-dominated epoch. The numerical analysis of the observables shows promising cosmological signatures without introducing extra parameters, and comparing this setup against data is ongoing work.

\end{itemize}


    \cleardoublepage
    

    \appendix





\cleardoublepage

\chapter{Scalar Field Inflation} \label{app:infl}
\setcounter{equation}{0}
\setcounter{figure}{0}




Inflation refers to an epoch of accelerated expansion in the early Universe, needed in order to address the problems listed below:

\begin{itemize}
    \item The horizon problem: the observed homogeneity implies extraordinary smoothness of the early Universe even if most of the Universe does not appear to have been in causal contact in the hot Big Bang paradigm. In comoving coordinates, the size of a causally connected region of space is defined according to the maximum distance that light can travel and still be detected today. This threshold defines the \textit{particle horizon}:
\begin{equation} \label{eq:horizon}
    d_h (\tau) = \tau - \tau_i = \int_{t_i}^{t} \frac{\odif{t}}{a(t)} = \int_{a_i}^{a} \frac{\odif{a}}{a \dot{a}} = \int_{\ln a_i}^{\ln a} (a H)^{-1} \odif{\ln a} \mathcomma
\end{equation}
    %
with $\tau$ standing for the comoving time, which corresponds to the comoving particle horizon under the convention that $\tau_i = 0$ since $a_i \equiv 0$ corresponds to the Big Bang singularity. The last equality of \cref{eq:horizon} shows how causality relates to the \textit{comoving Hubble radius} 
\begin{equation} \label{eq:horizonhi}
d_{\hub} = \hub^{-1} =  (a H)^{-1} \mathperiod
\end{equation}

  The $\Delta \tau$ between the BB singularity and the last-scattering surface is much smaller than the present conformal age of the Universe, implying that most of the CMB is composed of non-causally connected patches. This can be shown with simple examples, such as in Ref.~\cite{baumann_2022}. Then the question of \textit{how can the CMB be homogeneous at scales much larger than the particle horizon at the time when the CMB photons were released} remains.

    \item The flatness problem: the fact that the spatial curvature of the Universe is so small implies a fine-tuning of the initial conditions. Relying on current constraints for the time-evolving curvature parameter of the Universe, $| \Omega_{k,0}| <0.005$ \cite{Aghanim:2018eyx}, it is possible to extrapolate bounds on the same parameter evaluated at the time-scales of the early Universe, resulting in even tighter constraints such as $| \Omega_{k,\text{BBN}}| < 10^{-16}$ \cite{baumann_2022}. This implies that the curvature scale at that epoch must have been larger than the Hubble radius by many orders of magnitude. As we have seen above, during the standard hot Big Bang, the particle horizon is well approximated by the Hubble radius, suggesting again a large fine-tuning degree over many causally disconnected patches of space. 

    \item The problem of superhorizon correlations: the Universe contains density fluctuations correlated over apparently acausal distances. For comoving coordinates, the wavelength of a fluctuation $\lambda$ remains fixed while the Hubble radius $(aH)^{-1}$ keeps increasing due to the expansion. This results in every fluctuation inside the Hubble radius today (\textit{subhorizon} regime) being outside the Hubble radius at sufficiently early times (\textit{superhorizon} regime). Scales larger than the particle horizon (approximately the same as the Hubble radius at early times) at recombination would not have been inside the horizon before the emission of the CMB photons. However, the CMB fluctuations are not random, containing characteristic correlations on scales much larger than this apparent horizon. This is akin to a modern version of the horizon problem, corroborated by the CMB map of temperature fluctuations - the CMB is not just homogeneous on apparently acausal scales, as stated by the horizon problem. However, its fluctuations are also strongly correlated with such scales.
\end{itemize}

To grasp the dynamics of inflation, we first define the \textit{slow-roll parameter} $\varepsilon$:

\begin{equation}
    \odv{}{t} \left[ (aH)^{-1} \right] = - \frac{1}{a} (1- \varepsilon)\quad \text{with}\quad \varepsilon \equiv - \frac{\dot{H}}{H^2} \mathperiod
    \label{eq:epsdef}
\end{equation}
Comparing this condition with \cref{eq:infcon}, it follows that $\varepsilon < 1$, or in other words, that the fractional change in the Hubble parameter ($\dot{H}/H$) per Hubble time ($H^{-1}$) must be small.
The limit $\varepsilon = 0$ such that $H = \text{const.} \Rightarrow a(t) = e^{Ht}$ corresponds to quasi-de Sitter expansion. While this approximation is true for the cosmological constant, inducing accelerated expanding behaviour at late times, it cannot hold for the inflationary epoch because any early period of accelerated expansion must eventually end in a finite amount of time. Nevertheless, to address the early Universe issues of the Hot Big Bang model, inflation must last for a sufficiently long period, typically measured in the number of e-folds, denoted as $N = \ln a$. This relies on $\varepsilon$ remaining small for a substantial number of Hubble times and is embodied by the \textit{second slow-roll parameter}:
\begin{equation}
 \eta \equiv \frac{\dot{\varepsilon}}{H \varepsilon} \mathperiod   
 \label{eq:etadef}
\end{equation}
Analogously, inflation proceeds while $|\eta| < 1$ and the fractional change in $\varepsilon$ per Hubble time is small.

 The simplest models of inflation incorporate the time-dependent dynamics of inflation into a scalar field $\varphi(t,\vec{x})$ called the inflaton, with a respective potential energy density $V(\varphi)$ and standard kinetic energy. In this way, it is possible to assess under which conditions the inflaton can drive inflation. The Lagrangian for such a scalar field minimally coupled to gravity is given by:
\begin{equation}
    \mathcal{L}_{\varphi} = \frac{1}{2} g^{\mu \nu} \partial_{\mu} \varphi \partial_{\nu} \varphi - V(\varphi) \mathperiod
\end{equation}
The standard assumption in the inflationary paradigm is that, at the start of inflation, the inflaton is displaced sufficiently far from the minimum of its potential, which is nearly flat, to maintain the slow-roll conditions.

The equation of motion for a homogeneous configuration of the field $\varphi(t)$ is
\begin{equation}
    \ddot{\varphi} + 3H \dot{\varphi} + \odv{V}{\phi} = 0 \mathperiod
\end{equation}
This is the familiar Klein-Gordon equation with an extra term, $3H \dot{\varphi}$, called the \textit{Hubble friction}, which plays a crucial role during inflation and reflects the influence of the expansion. Assuming that this scalar field dominates the Universe, we can infer how it impacts the expansion through the Friedmann equations sourced by a perfect fluid with energy density and pressure:
\begin{equation}
    \rho_{\varphi} = \frac{1}{2} \dot{\varphi}^2 + V(\varphi) \mathcomma \quad \text{and}\quad p_{\varphi} = \frac{1}{2} \dot{\varphi}^2 - V(\varphi) \mathcomma
\end{equation}
leading to
\begin{equation}
    H^2 = \frac{1}{3 \text{M}_{\text{Pl}}^2} \left[\frac{1}{2} \dot{\varphi}^2 + V(\varphi) \right] \mathcomma \quad \text{and} \quad \left( \frac{\ddot{a}}{a} \right) =  \frac{1}{3 \text{M}_{\text{Pl}}^2} \left[ V(\varphi) - \dot{\varphi}^2 \right] \mathcomma
\end{equation}
which in the limit $\dot{\varphi} \rightarrow 0$ becomes
\begin{equation}
    H^2 \approx  \frac{1}{3 \text{M}_{\text{Pl}}^2} V(\varphi) \approx \text{const.} \mathcomma
\end{equation}
corresponding to the exponential expansion, $a(t) \propto e^{Ht}$. If instead the field is slowly varying, such that $\dot{\varphi}^2 /2 \ll V(\varphi)$, the equation of motion can be approximated by
\begin{equation}
    3 H \dot{\varphi}  \sim -V'(\varphi) \mathcomma
    \label{eq:sr1}
\end{equation}
and the first Friedmann equation reads
\begin{equation}
      H^2 (t) \sim  \frac{V\left[\varphi(t)\right]}{3 \text{M}_{\text{Pl}}^2}  \mathperiod
      \label{eq:sr2}
\end{equation}
taking the field evolution to be encapsulated by the Hubble friction such that the scale factor evolves as $a(t) \propto e^{-N(t)}$. This will be verified as long as the potential is sufficiently flat, $V'(\varphi) \ll V(\varphi)$. It is possible to write the equation of state of the field in the slow roll-approximation as
\begin{equation}
    p = \left[\frac{2}{3} \varepsilon (\varphi) - 1 \right] \rho \mathcomma
\end{equation}
where now, substituting \cref{eq:sr1,eq:sr2} into the definition of the Hubble slow-rolling parameters in \cref{eq:epsdef,eq:etadef} yields the potential slow-rolling parameters
\begin{equation}
   \varepsilon_V \approx \varepsilon = \frac{\text{M}_{\text{Pl}}^2}{2} \left( \frac{V'(\varphi)}{V(\varphi)} \right)^2, \quad \text{and}\quad \eta_V \approx 2 \varepsilon - \frac{1}{2} \eta = \text{M}_{\text{Pl}}^2 \frac{V''(\varphi)}{V(\varphi)} \mathperiod
\end{equation}
and the slow roll approximation remains valid as long as both parameters are small, \textit{i.e.} $\varepsilon_V \ll 1$ and $|\eta_V| \ll 1$. The number of e-folds can be expressed as 
\begin{equation}
    N = \ln \frac{a(t_e)}{a(t_i)} = - \int_{a(t_i)}^{a(t_e)} H \odif{t} \sim \text{M}_{\text{Pl}}^2 \int_{\varphi_e}^{\varphi_i} \frac{V(\varphi)}{V'(\varphi)} \odif{\varphi} \mathcomma
\end{equation}
with the subscripts $i$ and $e$ labelling the beginning and end of inflation, respectively.
As previously mentioned, the minimum duration for inflation to resolve the HBB problems is around $60$ e-folds. However, inflation is typically so rapid that most models can accommodate this condition very easily. In summary, during inflation, the rapid exponential expansion of the Universe ensures flatness and homogeneity. As the potential becomes steeper, inflation eventually ends, and the scalar field starts oscillating around its vacuum state at the minimum of the potential.

To transition to a radiation-dominated Universe after inflation, the inflaton field must decay into standard model particles, a process commonly referred to as \textit{reheating} \cite{Dimopoulos2020}. The specifics of this process depend on the particular inflationary model. To ensure that the Universe becomes radiation-dominated and in thermal equilibrium during primordial nucleosynthesis, which occurs at temperatures around the MeV scale, it is often assumed that the reheating temperature falls within the range of $1\, \text{TeV}$ to $10^{16}\, \text{GeV}$ which sets the minimum bound of $N_{\text{min}} \sim \left[45, 60\right]$ e-folds of inflation \cite{Kolb:1990vq,Dimopoulos2020}.

\cleardoublepage



\chapter{Mock Data Set Simulation} \label{app:mock}
\setcounter{equation}{0}
\setcounter{figure}{0}


This Appendix details the methodology followed to create the mock catalogue of standard sirens used in \cref{chap:gwcons}.


The inspiral and merger events of compact objects - such as black holes and neutron stars - generate gravitational waves (GW) that propagate through spacetime. Interferometers are sensitive to the strain, $h(t)$ produced by a GW event, which in the transverse-traceless gauge is described as \cite{zhao:2010sz}  
\begin{align}
    h(t) = F_{\times}  (\theta_0, \phi_0, \psi) h_{\times}(t) + F_+(\theta_0, \phi_0, \psi) h_+(t)\mathperiod
\end{align}
The gravitational wave tensor perturbation $h_{\mu \nu}$ is decomposed into two independent propagating modes, $h_{\times}(t)$ and $h_{+}(t)$, which in general are functions of cosmic time $t$. The corresponding antenna beam pattern functions $F_{\times}$ and $F_+$ depend on the angular location of the source relative to the detector in polar coordinates, $\left( \theta_0, \phi_0 \right)$, and the polarisation angle, $\psi$. We adopt a random sampling method in the range $[0,2\pi]$ for $\theta_0$ and $[0,\pi]$ for both $\phi_0$ and $\psi$. The factors $F_{\times,+}$ are defined as: 
\begin{align}
    F_{\times}^{(1)} = \frac{\sqrt{3}}{2}&\left[\frac{1}{2}(1+\cos^2(\theta))\cos(2\phi)\cos(2\psi) + \cos(\theta)\sin(2\phi)\cos(2\psi)  \right]\mathcomma
    \\
    F_{+}^{(1)} = \frac{\sqrt{3}}{2}&\left[\frac{1}{2}(1+\cos^2(\theta))\cos(2\phi)\cos(2\psi)- \cos(\theta)\sin(2\phi)\cos(2\psi)  \right]\mathperiod
\end{align}
The superscript indicates the specific interferometer under consideration. For instance, LISA comprises only two interferometers, hence $F^{(3)}=0$. 
The detectors are spatially distributed in an equilateral triangle formation, implying that the other two antenna pattern functions are located with respect to $F_{\times,+}^{(1)}$ as
\begin{equation}
    F_{\times,+}^{(2)}(\theta, \phi, \psi)= F_{\times,+}^{(1)}(\theta, \phi+ \frac{2\pi}{3}, \psi)\quad \text{and}\quad F_{\times,+}^{(3)} = F_{\times,+}^{(1)}(\theta, \phi+ \frac{4\pi}{3}, \psi) \mathperiod
\end{equation}
LISA, sensitive to lower frequencies, can observe GWs from long-lasting merger events. The relative positions of the interferometer and the event change over time, a fact that is accommodated using the approach described in \cite{cai:2017aea}.
Since LISA can probe lower frequencies and equivalently larger masses, it can detect GWs from long-lasting merger events. The relative positions of the interferometer and the event change over time, a fact that is accommodated following the method described in \cite{cai:2017aea}, assuming a timescale of the event of
\begin{equation}
t=t_c-5(8\pi f)^{-8/3}M_c^{-5/3} \mathperiod
\end{equation}
Here, $t_c$ denotes the merger time, $t$ indicates the time at which LISA detects the merger, $f$ is the frequency of the GW, and $M_c$ is the chirp mass. The location angles are updated accordingly:
\begin{align}
\theta&=\cos^{-1}\left[\frac{1}{2}\cos(\theta_0) -\frac{\sqrt{3}}{2}\sin(\theta_0)\cos\left(\frac{2\pi t}{T}-\phi_0\right) \right]\mathcomma \\
\phi&=\frac{2\pi t}{T} -\tan\left[\frac{\sqrt{3}\cos(\theta_0)+\sin(\theta_0)\cos\left(\frac{2\pi t}{T}-\phi_0\right)}{2\sin(\theta_0)\cos\left(\frac{2\pi t}{T}-\phi_0\right)}	\right]\mathcomma
\nonumber
\end{align}
which, in turn, are used to update the beam pattern functions. Here, we have specified the period, $T$, as the orbit around the Sun.

While the modelling degeneracies prevent the distinction between the individual masses of the objects, GW detectors are sensitive to the chirp mass, a collective mass quantity related to the frequency evolution of the signal emitted before the merger, during the inspiral phase of the binary \cite{Hilborn:2017liy}, defined as
\begin{equation}
M_c=(1+z)\left(\frac{(m_1\, m_2)^3}{m_1+m_2}\right)^{1/5}\mathcomma
\end{equation}
where $(1+z)$ is a conversion redshift factor from the physical to the observational chirp mass.

The Fourier transform of the strain $h(t)$ based on the stationary phase approximation \cite{Li:2013lza}, which neglects changes in the orbital frequency averaged over one period during the inspiral, reads:
\begin{equation}
\mathcal{H}=\mathcal{A}f^{-7/6}e^{i\Psi(f)} \mathcomma
\label{eq:H}
\end{equation}
where $\Psi(f)$ is the phase of the waveform. Note that when replacing $\mathcal{H}$ into \cref{eq:SNR}, the exponential term will vanish, and the $\Psi(f)$ factor may be discarded in this analysis. $\mathcal{A}$ is the Fourier amplitude of the waveform signal, given by 
\begin{align}
\mathcal{A}= \frac{M_c^{5/6}}{d_L^{\text{GW}} (z)}\pi^{-2/3}\sqrt{\frac{5}{96}}\times\sqrt{F_+^2 \left(1+\cos^2(l) \right)^2 + 4 F_\times^2 \cos^2(l)}\mathcomma
\label{eq:gw_amp}
\end{align}
where $d_L^{\text{GW}} (z)$ is the luminosity distance from the merger, and $l$ is the inclination angle of the orbital angular momentum with respect to the line of sight, which we have sampled randomly between $[0^\circ,20^\circ]$, as that is the maximum detection inclination range. We can see from \cref{eq:gw_amp} that measuring the amplitude of gravitational wave signals allows the derivation of estimates of the luminosity distances of the associated mergers.

Designed for low-frequency detection, as low as $f_{min}=1\times10^{-4}\, \text{Hz}$, LISA is particularly promising for probing extreme mass ratio inspirals (EMRI) and binary massive black hole (BMBH) mergers. For the simulations of LISA, two quantities are responsible for determining the upper bound frequency: the structure of LISA itself and the last stable orbit of the merging system. LISA can detect frequencies up to $f_{max}=c\,(2\pi L)^{-1}$, where $L$ is the length of LISA's interferometer arm, taken to be $2.5\, \text{Gm}$ and $c$ is the speed of light. Moreover, the total mass of an orbiting system is inversely proportional to its measured frequency, implying that even though massive mergers give rise to large detection amplitudes, the frequency will fall below $f_{min}$. Therefore if the last stable orbit frequency, $f_{\text{LSO}}=(6^{3/2}2\pi M_{obs})^{-1} $, with $M_{obs}$ being the observed total mass, is lower than $f_{min}$, such an event is ignored. Otherwise, if it lies between $f_{min}$ and $f_{max}$, then $f_{\text{LSO}}$ defines the new maximum frequency for that event. 

\subsection{Simulated cosmology}

The simulation of the GW catalogue requires the following cosmological parameters: the redshift at which the merger occurred, $z$, the value of the Hubble rate at that redshift, $H(z)$, its comoving and luminosity distance, $d_c(z)$ and $d_L^{\text{GW}} (z)$ respectively, as well as the cosmic time between the merger and its observation, $t$.
For this purpose, we use the public Einstein-Boltzmann code \texttt{CLASS} code\footnote{\href{https://github.com/lesgourg/class_public}{https://github.com/lesgourg/class\_public}} \cite{Lesgourgues:2011re,Blas:2011rf,lesgourgues2011cosmic2}, which we have modified to accommodate a broad range of interacting dark energy models. 
With this new patch, we generate a \textit{mock Universe} adopting the flat $\Lambda$CDM as the fiducial model, based on the best-fit cosmological parameters of the \textit{Planck} 2018 data release \cite{Aghanim:2018eyx}. These include the current value of the Hubble parameter, $H_0 = 67.32\, \text{km/s/Mpc}$, the baryon density parameter, $\Omega_b h^2 = 0.022383$ (with $h=H_0/100$) and the density of cold dark matter, $\Omega_ch^2 = 0.12011$. Furthermore, we are also concerned with the derived quantity $\Omega_m = \Omega_b + \Omega_c$, fixed to $\Omega_m^0 = 0.3144$ for the fiducial \textit{Planck} case.

Armed with the background cosmology, we simulate the merger events to establish the redshift-luminosity relation. First, we construct a redshift distribution of events weighted by a probability distribution. The characteristics of these events, such as the chirp mass, are simulated based on a uniform distribution. Although each simulation run produces a unique data set, the overarching conclusions remain unaltered, as the fiducial parameters constrain them.
Once the mergers have been modelled, we emulate the observational process, accounting for errors associated with each event. Events are excluded if they yield a signal-to-noise ratio below a given threshold.

\subsection{Distribution of simulated merger events}

The Einstein Telescoped is engineered to explore a range of frequencies, $f$, akin to those of LIGO, thereby probing merger events of nearby compact objects. These include binary neutron stars (BNS) with mass ranges of $[1,2],[1,2]\, \text{M}\textsubscript{\(\odot\)}$, and black hole neutron star binaries (BHNS) with $[3,10], [1,2]\, \text{M}\textsubscript{\(\odot\)}$, respectively, with the $[\cdot,\cdot]$ notation indicating the uniformly distributed mass ranges considered. According to the advanced LIGO proposal, the ratio of BHNS to BNS mergers is approximately $0.03$ \cite{LIGOScientific:2010weo}. The probability distribution for the redshift of these events follows:
\begin{equation}
    P\propto\frac{4\pi d_c^2(z) R(z)}{(1+z)H(z)} \mathcomma
\end{equation}
with the comoving distance and the Hubble rate at various redshifts sourced from \texttt{CLASS}. $R(z)$ represents the rate of binary mergers, which, at a linear approximation level, is \cite{Nishizawa:2010xx}
\begin{align}R=
    \begin{cases}
    1+2z \hspace{0.5cm} &\text{if    } z<1\mathcomma\\
    \frac{3}{4}(5-z) &\text{if    } 1\leq z<5\mathcomma\\
    0 &\text{otherwise}\mathperiod
    \end{cases}
\end{align}

In contrast, LISA focuses on lower frequencies than other proposed 3G detectors,  making it more sensitive to higher mass binary systems since $f \propto M^{-1}$. Therefore, we consider simulations of detected events from extreme mass ratio inspirals (EMRI) and binary massive black holes (BMBH) in the ranges $[1-30],[10^4-10^8]\, \text{M}\textsubscript{\(\odot\)}$ \cite{Gair:2017ynp} and $[10^4-10^8],[10^4-10^8]\, \text{M}\textsubscript{\(\odot\)}$ \cite{Caprini:2016qxs}, respectively. The number of detected BMBH to EMRI events is assumed to follow a $2:1$ ratio according to the mission's proposal \cite{eLISA:2013xep,Amaro-Seoane:2012aqc}.

Although, in principle, LISA will also be able to probe mergers of binary intermediate-mass black holes (IMBH) and binary compact objects, we have excluded these from the simulations due to the lack of definitive observational proof of IMBH. Furthermore, events from binary compact objects are only expected to be observed at redshifts $z\approx 3$ \cite{Mapelli:2010ht} irrelevant for our cosmological aim since we are interested in the higher range of redshifts which LISA can reach. 

Focusing solely on BMBH events, we base their redshift probability distributions on the histogram outlined in the L6A2M5N2 mission specification \cite{Caprini:2016qxs}, which explores three different mechanisms for BMBH formation. Our study zeroes in on the light seed model, or the \textit{pop III model}, which suggests that BMBHs originate from the remnants of population III stars around a redshift range of $z=15-20$. In \cite{Caprini:2016qxs}, two other scenarios for massive black hole creation were discussed, namely \textit{delay} and \textit{no-delay} scenarios. These alternative routes involve the gravitational collapse of gas at the centre of a galaxy in the same redshift range, $z=15-20$, resulting in a black hole formed through a heavy seed mechanism. The critical difference between them is the timing - with and without delay - of the black hole's formation relative to the merger of its host galaxy. More details about these scenarios are given in Refs.~\cite{Klein:2015hvg}.

In our investigation, we provided mock data and obtained forecasts for both the delay and no delay cases but found no significant improvement in constraining power from variations in these models compared to the pop III case. Consequently, we restrict our focus to the pop III model, as it proves sufficient to forecast the constraining power of LISA.

\subsection{Simulation of measurements and errors}

We follow the methodologies of \cite{zhao:2010sz,cai:2017aea, Li:2013lza,Nishizawa:2010xx} to model the errors associated with the GW standard siren catalogue. 
An event is considered genuinely detected only if its combined signal-to-noise ratio (SNR), $\rho$, exceeds the particular threshold $\rho>8$. The SNR is calculated taking into account the specifications of the interferometer in use: 
\begin{equation}
\rho_{1,2,3}^2 =4 \int_{f_{\text{min}}}^{f_{\text{max}}} \odif{f} \frac{|\mathcal{H} (f)|^2}{S_h (f)}\mathcomma
\label{eq:SNR}
\end{equation}
where the numbers label the interferometer, the definition of $\mathcal{H}$ was provided in \cref{eq:H} and $S_h$ is the noise power spectral density. This SNR weighting function reflects the unique characteristics of the instruments in use. For ET, in particular, $S_h$ is according to
\begin{align}
    \begin{aligned}
        S^{(\text{ET})}_h= S_0\left( x^{p_1} + a_1 x^{p_2} + 
         a_2\frac{1 + \sum^{6}_{n=1}b_n x^n}{1 + \sum^{4}_{m=1}c_m x^m}  \right) \mathcomma
    \end{aligned}
    \label{eq:S_ET}
\end{align}
where, $x = f/200$ Hz$^{-1}$, $S_0 = 1.449\times 10^{-52}$ \text{Hz}, $p_1 = -4.05$,  $p_2 = -0.69$, $a_1 = 185.62$, $a_2 = 232.56$, $b_n = \left\{ 31.18, -64.72, 52.24, -42.16, 10.17, 11.53 \right\}$, $c_m = \left\{13.58, -36.46, 18.56, 27.43 \right\}$, assuming a lower cutoff at $f = 1\, \text{Hz}$. 

On the other hand, for LISA, $S_h$ relies on three noise components: the instrumental (or short) noise, $S_{\text{inst}}$, the noise from low-level acceleration, $S_{\text{acc}}$, and the confusion background noise, $S_{\text{conf}}$ \cite{Klein:2015hvg}:
\begin{align}
\begin{aligned}
  &S^{(\text{LISA})}_h =\frac{20}{3}\frac{4S_{\text{acc}} + S_{\text{inst}} + S_{\text{conf}}}{L^2} \left[1+ \left(\frac{fL}{0.81c} \right)  \right]  \mathcomma
\end{aligned}
\label{eq:S}
\end{align}
where $S_{\text{acc}}=9\times 10^{-30}/(2\pi f)^4( 1 +10^{-4}/f)$, $S_{\text{inst}}= 2.22\times10^{-23}$ and $S_{\text{conf}}=2.65\times10^{-23}$.

Therefore, combining \cref{eq:H} with either \cref{eq:S_ET} or \cref{eq:S} allows us to determine the total SNR contribution for ET and LISA respectively, as expressed by
\begin{equation}
\rho_{\text{tot}}=\sqrt{\rho_1^2+\rho_2^2 + \rho_3^2} \mathperiod
\end{equation}

The instrumentally induced error in the luminosity distance is computed using the Fisher Matrix approach,
\begin{equation}
\sigma_{d_L^{\text{GW},\text{inst}}} \approx \left\langle \frac{\partial \mathcal{H}}{\partial d_L^{\text{GW}}} , \frac{\partial \mathcal{H}}{\partial d_L^{\text{GW}}} \right\rangle^{-\frac{1}{2}} \mathcomma
\end{equation}
following \cite{cai:2017aea}. This expression simplifies due to the proportionality $\mathcal{H}\propto 1/d_L^{\text{GW}}$:
\begin{equation}
\sigma_{d_L^{\text{GW},\text{inst}}}\approx \frac{2d_L^{\text{GW}}}{\rho} \mathcomma
\label{eq:siginst}
\end{equation}
where the factor of $2$ accounts for the symmetry in the inclination angle, which ranges from $-20^\circ$ to $20^\circ$.
There is a correlation between the gravitational wave luminosity distance and the inclination of the source to the observer. For a single detector, $d_L^{\text{GW}}(z)$ and $l$ are completely degenerate with each other and the antenna patterns $F_{\times,+}$. However, this degeneracy can be broken with multiple detectors and sensitivity to both polarisations. The maximum effect of this degeneracy on the signal-to-noise ratio is a factor of two, between the source being face-on (inclination $l = 0$) and edge-on ($l = \pi/2$).

Gravitational lensing introduces another layer of error, represented by 
\begin{equation}
\sigma_{d_L^{\text{GW},\text{len}}}= \frac{d_L^{\text{GW}}}{2}\times 0.066\left[4(1- (1+z)^{1/4})  \right]^{1.8} \mathcomma
\label{eq:siglen}
\end{equation}
and it is reduced by half to account for the event's merger and ringdown.  

Being space-based, LISA is also subject to a peculiar velocity-associated error of the GW sources \cite{Tamanini:2016zlh}:
\begin{equation}
\sigma_{d_L^{\text{GW},\text{pec}}}=d_L^{\text{GW}}\frac{\sqrt{\langle v^2\rangle}}{c}\left[1+\frac{c(1+z)}{Hd_L^{\text{GW}}}\right] \mathcomma
\label{eq:sigpec}
\end{equation}
with an estimate of the peculiar velocity of the host galaxy with respect to the Hubble flow of $\sqrt{\langle v^2\rangle}=500\, \text{km}\, \text{s}^{-1}$.

The total error in luminosity distance combines all these individual errors in \cref{eq:siginst}-\cref{eq:sigpec}:
\begin{equation} \label{eq:tots}
\sigma_{d_L^{\text{GW}}}=\sqrt{(\sigma_{d_L^{\text{GW},\text{inst}}})^2+(\sigma_{d_L^{\text{GW},\text{len}}})^2+(\sigma_{d_L^{\text{GW},\text{pec}}})^2} \mathperiod
\end{equation}

The simulation allows us to interpolate any number of events over a continuous redshift distribution in the range $0<z\lesssim 5$ for ET and $0<z\lesssim 10$ for LISA. However, 
the number of mergers detected by the ET will depend on factors such as operational costs and the complementary with other experiments \cite{zhao:2010sz}. ET is anticipated to document over $10^4$ mergers per year. However, due to the scarcity of EM counterpart signals, the predicted number of detectable mergers with an actual EM counterpart over $10$ years is brought down to approximately $200$ \cite{Hou:2022rvk}.
According to \cite{Caprini:2016qxs}, LISA is expected to detect around 56 events over a 10-year mission.

\begin{figure}[t!]
\centering
\includegraphics[width=\textwidth]{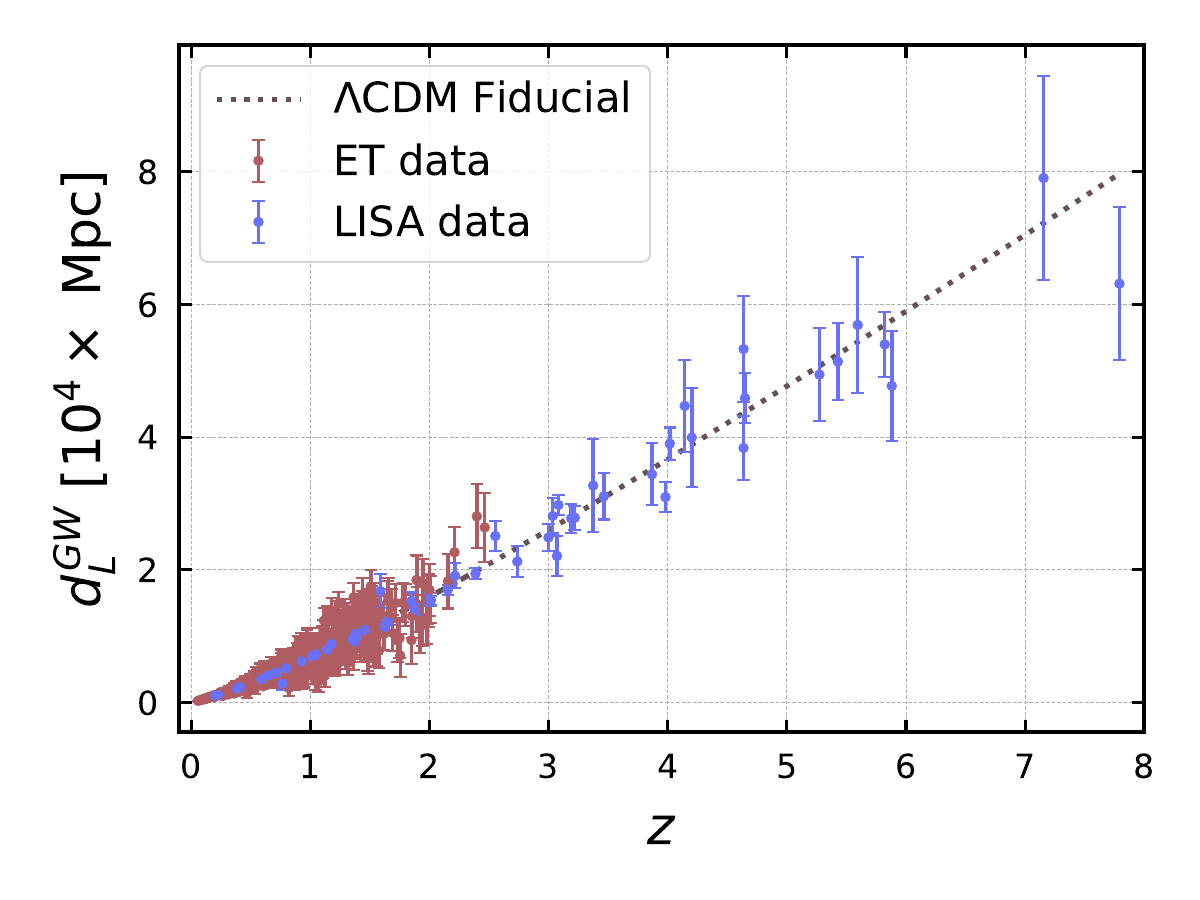}
 \caption[ET and LISA mock data]{\label{fig:ETLISAmock} Luminosity distance as a function of redshift for the mock data points from ET (red) and LISA (blue). The data is generated from the fiducial model, $\Lambda$CDM, shown by the black dotted line. The error bars represent the $1\sigma$ total errors in the luminosity distance.}
\end{figure}

To account for uncertainties in the luminosity distance of each merger, we employ a Gaussian distribution centred around the background cosmology, with a standard deviation determined by the calculated errors, $\sigma_{d_L^{\text{GW}}}$. This introduces an artificial randomness factor around each merger, leading to a higher deviation from $\Lambda$CDM for LISA than ET, essentially because LISA covers a more extensive redshift range with corresponding larger errors, resulting in a broader data spread. This is visually illustrated in \cref{fig:ETLISAmock}.

\chapter{String Theory and the DBI Action } \label{app:dbi}
\setcounter{equation}{0}
\setcounter{figure}{0}

This Appendix provides a brief overview of string theory applied to cosmology and the main concepts needed to understand the formulation of the Dark D-Brane Model studied in \cref{chap:dbi}. For more details on this and other cosmological applications of string theory, we refer the reader to the review in \cite{Cicoli:2023opf}.


Many cosmological models proposed in the literature have motivations based on string theory. Instead of treating particles as point-like entities, string theory posits that the basic building blocks of the Universe are one-dimensional \textit{strings} that vibrate at different frequencies. These vibrations give rise to various particle types and their properties, such as mass and charge. String theory also suggests the existence of more than the conventional four dimensions of GR, potentially providing a unified description of all four fundamental forces of physics in a single, coherent picture. The process of starting from a higher-dimensional framework and then obtaining a 4-dimensional effective theory is known as compactification and has been studied in great detail \cite{Cicoli:2023opf}. The extra dimensions are typically \textit{compactified}, \textit{i.e.} mathematically folded up so small that they are effectively invisible at the scales we can currently probe, which is why we do not readily observe them. 
This has profound implications for cosmology, particle physics, and our understanding of the fundamental nature of reality, although it is yet to be empirically verified.

These strings can be attached to branes, mathematical objects that generalise the notion of physical membranes to $n$ dimensions, which can propagate in space and time and can be characterised by physical properties, such as tension and charge. D-branes are an important class of branes that arise when one considers open (two-endpoints) strings whose endpoints must lie on a D-brane (with the D standing for Dirichlet boundary conditions). Therefore, D-branes can be regarded as surfaces (which could be 0-dimensional points, 1-dimensional strings, 2-dimensional membranes, and so on) to which open strings can attach their endpoints. While strings are 1-dimensional objects, D-branes can exist in various dimensions and are essential in describing non-gravitational forces like electromagnetism and nuclear forces in this context.

D-branes play a crucial role in various string theory dualities and phenomenology. A \textit{spacetime-filling D-brane} would be an object that fills all the dimensions of the spacetime in which it is embedded. The \textit{throat} refers to a region of this extra-dimensional space shaped like a funnel or a tube. The term \textit{warped} refers to the specific type of geometry that this throat can have, often characterised by a metric tensor that varies along the length of the throat. In this setup, the D-branes can move along the throat, often essential for various phenomena such as brane inflation. Therefore, we will be concerned with D-branes that are dynamically embedded within a region of that spacetime characterised by a warped geometry. 
Lastly, a D3-brane is a specific type of D-brane that is a 3-dimensional hypersurface embedded in a higher-dimensional spacetime. They provide a setting for the famous Anti-de-Sitter/Conformal Field Theory (AdS/CFT) correspondence \cite{Hubeny:2014bla}, a form of the holographic principle. This duality allows physicists to relate a theory of gravity in a higher-dimensional \textit{bulk} spacetime (described by the geometry around the D3-brane) to a conformal field theory on the brane itself.


This leads to the idea of brane cosmological models, where the visible, four-dimensional Universe is restricted to a D-brane (or a stack of branes) contained inside a compact higher-dimensional spacetime, called the \textit{bulk} or the hyperspace. The standard model particles and their interactions live on the \textit{visible branes}. There can also be\textit{ hidden dark sector D-branes}, from which the dark sector stems, and which can only interact with the visible sector branes gravitationally, with gravity being the force that propagates throughout the bulk. Interactions with the bulk, and possibly with other branes, may leave observational imprints on our visible brane. Hidden sector D-branes could be relevant for various theoretical constructions or even new forces and fields that have not yet been observed. They can interact with the visible sector via gravitational interactions or other weaker-than-Planckian couplings, making them candidates for explaining phenomena not accounted for by the standard model of particle physics. For example, if dark matter arises from the hidden sector, it would interact very weakly with the particles of the visible sector, making it difficult to detect directly.




The Dirac-Born-Infeld (DBI) action is a type of action functional in theoretical physics that describes D-branes' dynamics, including their interactions with other fields. Initially, the Born-Infeld action was formulated to address some of the infinities that arose in classical electrodynamics and the context of string theory, and it has been adapted to describe the dynamics of D-branes.
The DBI action is non-linear and includes higher-order terms in the field strengths. Unlike the standard actions in quantum field theory, which typically include quadratic terms in the field or its derivatives, the DBI action has a square root of the determinant of the induced metric on the brane, making it a non-canonical action. This results in a more complex set of equations of motion that capture a more comprehensive range of phenomena, including some that simpler models do not cover, encapsulated in the following action:

\begin{equation}
S = \int \odif[order=4]{x} \left[ T(\phi) \left(1 - \sqrt{1+ \frac{(\nabla \phi)^2}{T(\phi)}} \right) - V(\phi) \right] \mathcomma
\label{eq:dbiaco}
\end{equation}

where $\phi$ represents the D-brane's position modulus and $T(\phi)$ and $V(\phi)$ are its tension and potential, respectively. 
The functional form of $T(\phi)$ and $V(\phi)$ depend on the dimensionality of the brane and determine the phenomenology of the scenario \cite{Shandera:2006ax,Bean:2007hc,Durrer:2005nz}.
A phase of accelerated expansion can be achieved by the non-canonical form of the kinetic terms, given constraints on the speed of the moving D3-brane. However, constructing DBI inflation within a consistent string compactification is challenging \cite{Chen:2008hz}. Nonetheless, this scenario serves as an essential example of an inflationary mechanism relying on the symmetries of an ultraviolet theory, motivating further investigations of its phenomenology and leading to the observational search for specific cosmological signatures.

For D-branes moving in the internal compact space, the scalar field(s) associated with the open string representing, e.g. the radial position of a D-brane along a warped throat, couples disformally to matter living on such a brane \cite{KOIVISTO:2013jwa}. Indeed, the induced metric on the brane takes the form presented in \cref{eq:disfphii}.
Since a critical constraint on quintessence candidates is always their interactions with standard model matter resulting in fifth forces, a natural place to look for string theory quintessence candidates is from hidden sector D-branes, as they may be coupled to visible sector D-branes with weaker-than-Planckian couplings.
The simplest case is of a D$3$-brane moving in a 5-dimensional AdS space, describing the mid-region of a warped throat, $T(\phi) \propto \phi^4$.

This opens up a new avenue for exploring the dark Universe, providing a fresh perspective that goes beyond the limitations of the standard cosmological model. These dark D-branes can give rise to dark matter and dark energy candidates through their vibrations and interactions with the visible sectors.

For the late-time acceleration, quintessence candidates need to have highly suppressed interactions with standard matter (photons and baryons) to avoid any undetected fifth forces, highly constrained by local measurements in the Solar System \cite{Will:2014kxa,Ip:2015qsa,Sakstein:2014isa}. Further observational support for a non-universal or highly suppressed coupling to standard matter comes from the recent constraints on the speed of gravitational waves inferred from observations of the optical counterpart of a binary neutron star merger \cite{LIGOScientific:2017zic}. The cosmological viability of the DBI action in the sense of allowing for quintessence attractor tracker solutions for the scalar field representing the brane's position has been the focus of Refs. \cite{Martin:2008xw,Gumjudpai:2009uy}. In this scenario, 
the dynamics of the field are constrained by the causality of the gravity side of the AdS/CFT correspondence \cite{Maldacena:1997re}.
The DBI action effectively describes a probe D3-brane moving in the radial direction of the AdS$_5$ spacetime, with a string-inspired phenomenological action given by 
\begin{equation}
S =  \int \odif[order=4]{x}\, \left[ h(\phi)^{-1} \left(1 - \sqrt{1+ h(\phi) \partial_{\mu} \phi \partial^{\mu} \phi}  \right) - V(\phi) \right] \mathperiod
\label{eq:dbiac}
\end{equation}
The DBI field $\phi$ has mass dimensions and stands as the dark energy component. The brane's \textit{warp facto}r $h(\phi) \equiv T(\phi)^{-1}$ encodes the geometry of the higher-dimensional space, in terms of the DBI field and $V(\phi)$ represents the potential term arising from possible interactions with the bulk and other brane stacks. 


This cosmological context allows for accelerated expansion even when the Lorentz-like factor in the action in \cref{eq:dbiac}, $\gamma = 1/\sqrt{1-h(\phi) \dot{\phi}^2}$, is much larger than unity, meaning that $h(\phi) \dot{\phi}^2 \approx 1$. The $\gamma$ factor measures the relativistic motion of the D3-brane, and when it approaches unity, the canonical kinetic term is recovered, and $\phi$ behaves like a standard quintessence field. On the other hand, when $\gamma \rightarrow \infty$, the purely relativistic limit is approached. This factor acts as a speed limit, allowing the field to evolve relatively slowly, even in steep potentials. The application of DBI theories to dark energy has been seminally discussed in \cite{Martin:2008xw, Guo:2008sz}.

In \cref{chap:dbi}, we study a DBI-\textit{essence} scenario featuring a coupling to dark matter, which naturally arises in a stacked D-brane system, where dark energy emerges from open strings representing the radial position of a D-brane along a warped throat, and dark matter arises from open strings representing matter on the same hidden sector brane. The resulting theory, referred to as the \textit{Dark D-Brane Model}, was constructed in \cite{Koivisto:2013fta}, in which the action for the dark energy scalar takes the form of \cref{eq:dbiac}, while the action for the dark matter describes $N$ particles on the moving D3-brane at low energies. Further details are provided in \cref{chap:dbi}, in the particular case of an AdS warp factor, $h(\phi) \propto \phi^{-4}$, together with a mass term for the scalar potential, $V(\phi) \propto \phi^2$.


\cleardoublepage

\cleardoublepage

    \bibliography{MyThesis.bib}

\end{document}